%% file: review_doc.tex
\documentclass[10pt,a4paper, twoside]{book}

\usepackage{amsmath}
\usepackage{amsfonts}
\usepackage{amssymb}
\usepackage{graphicx}
\usepackage{epsfig}
\usepackage{color}
\usepackage{natbib}
\usepackage[hang,small,bf]{caption}
\usepackage{rotating}

\let\Oldcolor\color
\renewcommand{\color}[1]{\Oldcolor{black}}

\DeclareRobustCommand{\SkipTocEntry}[4]{}

\input{head}

\begin{document}
\title{{\bf Cosmology and fundamental physics\\with the Euclid satellite\footnote{Review of the Euclid Theory Working Group.}} \\[0.2cm]}


\author{
Luca Amendola\footnote{Corresponding authors. Please contact euclidtheoryreview@gmail.com for questions and comments.}, 
Stephen Appleby,
David Bacon,
Tessa Baker, \\
Marco Baldi, 
Nicola Bartolo,
Alain Blanchard,
Camille Bonvin, \\
Stefano Borgani,
Enzo Branchini,
Clare Burrage,
Stefano Camera$^2$, \\
Carmelita Carbone, 
Luciano Casarini, 
Mark Cropper,
Claudia de Rham, \\ 
Cinzia di Porto$^2$, 
Anne Ealet,
Pedro G. Ferreira$^2$,
Fabio Finelli, 
Juan Garcia-Bellido$^2$, \\
Tommaso Giannantonio, 
Luigi Guzzo,
Alan Heavens, 
Lavinia Heisenberg, \\
Catherine Heymans, 
Henk Hoekstra,
Lukas Hollenstein, 
Rory Holmes, \\
Ole Horst, 
Knud Jahnke,
Thomas D. Kitching$^2$, 
Tomi Koivisto, \\
Martin Kunz$^2$, 
Giuseppe La Vacca, 
Marisa March, 
Elisabetta Majerotto, \\
Katarina Markovic,
David Marsh, 
Federico Marulli, 
Richard Massey, \\
Yannick Mellier,
David F. Mota, 
Nelson J. Nunes,
Will Percival, \\
Valeria Pettorino$^2$,  
Cristiano Porciani$^2$,
Claudia Quercellini,  
Justin Read,  \\
Massimiliano Rinaldi, 
Domenico Sapone,  
Roberto Scaramella, \\
Constantinos Skordis,  
Fergus Simpson, 
Andy Taylor, 
Shaun Thomas, \\
Roberto Trotta$^2$, 
Licia Verde$^2$, 
Filippo Vernizzi,
Adrian Vollmer, \\
Yun Wang$^2$, 
Jochen Weller,
Tom Zlosnik}

\date{June 6, 2012}

\maketitle

\renewcommand\thefootnote{\roman{footnote}}
\addtocounter{footnote}{-1}

\null\vspace{7in}

\begin{center}
DISCLAIMER\\
{\bf This is not an official Euclid document and
its content reflects solely the views of the contributing authors.}

\end{center}

\newpage
\input{people}

\newpage
\tableofcontents


\include{acronyms}

\include{symbols}
\include{generalintro}

\include{de_mg/de_mg}

\include{dark_matter/dark_matter}

\include{ini_cond/ICmerged}
\include{structure/structure}

\include{methodology/methodology}

\include{ack}


\include{journals}

\bibliographystyle{apj}

{\footnotesize

\bibliography{bib_master}

}


\end{document}

%% file: head.tex
\usepackage{array}
\def\Mp{M_{\rm pl}}

\newcommand{\weff}{w_\text{eff}}

\def\ba{\begin{eqnarray}}
\def\ea{\end{eqnarray}}

\def\mn{_{\mu \nu}}

\def\({\left(}
\def\){\right)}
\def\p{\partial}

\def\vk{\vec{k}}
\def\fnl{f_{\rm NL}}

\newcommand{\data}{d}

\newcommand{\mdl}{\mathcal{M}}
\newcommand{\lsim}{\,\raise 0.4ex\hbox{$<$}\kern -0.8em\lower
  0.62ex\hbox{$\sim$}\,} 
\newcommand{\gsim}{\,\raise 0.4ex\hbox{$>$}\kern -0.7em\lower
  0.62ex\hbox{$\sim$}\,} 
\newcommand{\params}{{\Theta}}

\newcommand{\lnlike}{\ln \mathcal{L}}
\newcommand{\fid}{\params_*}
\newcommand{\dr}{\textrm{d}}
\newcommand{\ie}{{\it i.e.}}
\newcommand{\Uf}{{\mathcal U}}
\newcommand{\EU}{{\mathcal{EU}}}
\newcommand{\Df}{{D}_f}
\newcommand{\be}{\begin{equation}}
\newcommand{\ee}{\end{equation}}

\def\Mp{M_{\rm pl}}

\newcommand{\Vsur}{V_{\rm survey}}
\newcommand{\DA}{D\!_A(z)}
\newcommand{\hz}{H(z)}

\newcommand{\Veff}{V_{\rm eff}}

\newcommand{\metE}{\ensuremath{\tilde{g}}}
\newcommand{\metM}{\ensuremath{g}}
\newcommand{\RiemE}{\ensuremath{\tilde{R}}}
\newcommand{\volE}{\ensuremath{\sqrt{-\metE}}}
\newcommand{\volM}{\ensuremath{\sqrt{-\metM}}}
\newcommand{\metS}{\ensuremath{\hat{g}}}
\newcommand{\connE}{\ensuremath{\tilde{\nabla}}}

\newcommand{\Vp}{\ensuremath{\frac{dV}{d\mu}}}

\newcommand{\kmax}{k_{\rm max}}
\newcommand\bcite[1]{\citeauthor{#1} \citeyear{#1}}

\newcommand{\er}{\mbox{\boldmath $\hat{e}$}_r}
\newcommand{\et}{\mbox{\boldmath $\hat{e}$}_t}

\newcommand{\CD}{{\cal D}}
\newcommand{\CQ}{{\cal Q}}
\newcommand{\CR}{{\cal R}}
\newcommand{\average}[1]{\left\langle #1 \right\rangle_\CD}

\usepackage{hyperref}
\newcommand{\symbolref}[1]{p. \pageref{#1}}

%% file: people.tex
\section*{Credits version}
\subsubsection*{Euclid Theory Working Group Editorial Board (2012):}

Valeria Pettorino (editor in chief)\\
Tessa Baker\\
Stefano Camera\\
Elisabetta Majerotto\\
Marisa March\\
Cinzia Di Porto\\
Martin Kunz (Euclid Theory Working Group Coordinator)\\
Luca Amendola (Euclid Theory Working Group Coordinator)

\subsubsection*{Corresponding authors (2012):}

Luca Amendola\\
Stefano Camera\\
Cinzia Di Porto\\
Pedro G.\ Ferreira\\
Juan Garc{\'{\i}}a-Bellido\\
Thomas D.\ Kitching\\
Martin Kunz\\
Valeria Pettorino\\
Cristiano Porciani\\
Roberto Trotta\\
Licia Verde\\
Yun Wang

\vspace{1cm}

\subsubsection*{Contributing Authors (2012):}

Luca Amendola, Stephen Appleby, David Bacon, Tessa Baker, Marco Baldi, Nicola Bartolo, Alain Blanchard, Camille Bonvin, Stefano Borgani, Enzo Branchini, Clare Burrage, Stefano Camera, Carmelita Carbone, Luciano Casarini, Mark Cropper, Claudia de Rham, Cinzia Di Porto, Anne Ealet, Pedro G. Ferreira, Fabio Finelli, Juan Gar{\'{\i}}a-Bellido, Tommaso Giannantonio, Luigi Guzzo, Alan Heavens, Lavinia Heisenberg, Catherine Heymans, Henk Hoekstra, Lukas Hollenstein, Rory Holmes, Ole Horst, Knud Jahnke, Thomas D. Kitching, Tomi Koivisto, Martin Kunz, Giuseppe La Vacca, Marisa March, Elisabetta Majerotto, Katarina Markovic, David Marsh, Federico Marulli, Richard Massey, Yannick Mellier, David F. Mota, Nelson J. Nunes, Will Percival, Valeria Pettorino, Cristiano Porciani, Claudia Quercellini, Justin Read, Massimiliano Rinaldi, Domenico Sapone, Roberto Scaramella, Constantinos Skordis, Fergus Simpson, Andy Taylor, Shaun Thomas, Roberto Trotta, Licia Verde, Filippo Vernizzi, Adrian Vollmer, Yun Wang, Jochen Weller, Tom Zlosnik.

\vspace{1cm}

%% file: acronyms.tex

\chapter*{List of acronyms}

\begin{table}[htb]
\centering
\begin{tabular}{|l|l|}
\hline
 AGN & Active Galactic Nucleus\tabularnewline
\hline
 ALP & Axio-like Particle\tabularnewline
\hline
 BAO & Baryonic Acousitc Oscillations\tabularnewline
\hline
 BBKS & Bardeen-Bond-Kaiser-Szalay	\tabularnewline 
\hline
 BOSS & Baryon Oscillation Spectroskopic Survey	\tabularnewline 
\hline
 BPol & B-Polarization Satellite	\tabularnewline 
\hline
 BigBOSS & Baryon Oscillation Spectroskopic Survey	\tabularnewline 
\hline
 CAMB & Code for Anisotropies in the Microwave Background \tabularnewline 
\hline
 CDE & Coupled Dark Energy\tabularnewline
\hline
 CDM & Cold Dark Matter\tabularnewline
\hline
 CDMS & Cryogenic Dark Matter Search	\tabularnewline 
\hline
 CL & Confidence Level\tabularnewline
\hline
 CMB & Cosmic Microwave Background\tabularnewline
\hline
 COMBO-17 & Classifying Objects by Medium-Band Observations \tabularnewline 
\hline
 COSMOS & Cosmological Evolution Survey	\tabularnewline 
\hline
 CPL & Chevallier-Polarski-Linder	\tabularnewline 
\hline
 CQ & Coupled Quintessence\tabularnewline
\hline
 CRESST & Cryogenic Rare Event Search with Superconducting
Thermometers\tabularnewline
\hline
 DE & Dark Energy\tabularnewline
\hline
 DES & Dark Energy Survey \tabularnewline 
\hline
 DETF & Dark Energy Task Force\tabularnewline
\hline
 DGP & Dvali-Gabadadze-Porrati	\tabularnewline 
\hline
 DM & Dark Matter\tabularnewline
\hline
 EBI & Eddington-Born-Infeld	\tabularnewline 
\hline
 EDE & Early Dark Energy\tabularnewline
\hline
 EROS & Exp\'{e}rience pour la Recherche d'Objets Sombres	\tabularnewline 
\hline
 eROSITA & Extended ROentgen Survey with an Imaging Telescope Array \tabularnewline 
\hline
 FCDM & Fuzzy Cold Dark Matter\tabularnewline
\hline
 FFT & Fast Fourier Transform\tabularnewline
\hline
 FLRW & Friedmann-Lema\^{i}tre-Robertson-Walker	\tabularnewline 
\hline
 FOM & Figure of Merit\tabularnewline
\hline
 FoG & Fingers of God\tabularnewline
\hline
 GEA & Generalized Einstein-Aether\tabularnewline
\hline
 GR & General Relativity\tabularnewline
\hline
 HETDEX & Hobby-Eberly Telescope Dark Energy Experiment \tabularnewline 
\hline
 ICM & Intracluster Medium\tabularnewline
\hline
 IH & Inverted Hierarchy\tabularnewline
\hline
 IR & Infrared\tabularnewline
\hline
\end{tabular}
\end{table}

\begin{table}
\centering
\begin{tabular}{|l|l|}
\hline
 ISW & Integrated Sachs-Wolfe 	\tabularnewline 
\hline
 KL & Kullback-Leibler divergence	\tabularnewline 
\hline
 LCDM & Lambda Cold Dark Matter\tabularnewline
\hline
 LHC & Large Hadron Collider\tabularnewline
\hline
 LRG & Luminous Red Galaxy\tabularnewline
\hline
 LSB & Low Surface Brightness\tabularnewline
\hline
 LSS & Large Scale Structure\tabularnewline
\hline
 LSST & Large Synoptic Survey Telescope	\tabularnewline 
\hline
 LTB & Lema\^{i}tre-Tolman-Bondi	\tabularnewline 
\hline
 MACHO & MAssive Compact Halo Object\tabularnewline
\hline
 MCMC & Markov Chain Monte Carlo \tabularnewline 
\hline
 MCP & Mini-Charged Particles\tabularnewline
\hline
 MF & Mass Function\tabularnewline
\hline
 MG & Modified Gravity\tabularnewline
\hline
 MOND & MOdified Newtonian Dynamics\tabularnewline
\hline
 MaVaNs & Mass Varying Neutrinos \tabularnewline
\hline
 NFW & Navarro-Frenk-White	\tabularnewline 
\hline
 NH & Normal Hierarchy\tabularnewline
\hline
 PCA & Principal Component Analysis \tabularnewline 
\hline
 PDF & Probability Distribution Function \tabularnewline 
\hline
 PGB & Pseudo-Goldstein Boson\tabularnewline
\hline
 PKDGRAV & Parallel K-D tree GRAVity code	\tabularnewline 
\hline
 PPF & Parameterized Post-Friedmann\tabularnewline
\hline
 PPN & Parameterized Post-Newtonian\tabularnewline
\hline
 PPOD & Predictive Posterior Odds Distribution\tabularnewline
\hline
 PSF & Point spread function	\tabularnewline 
\hline
 QCD & Quantum ChromoDynamics \tabularnewline 
\hline
 RDS & Redshift Space Distortions \tabularnewline 
\hline
 RG & Renormalization Group\tabularnewline
\hline
 SD & Savage-Dickey\tabularnewline
\hline
 SDSS & Sloan Digital Sky Survey\tabularnewline
\hline
 SIDM & Self Interacting Dark Matter\tabularnewline
\hline
 SN & Supernova\tabularnewline
\hline
 TeVeS & Tensor Vector Scalar\tabularnewline
\hline
 UDM & Unified Dark Matter\tabularnewline
\hline
 UV & Ultra Violett\tabularnewline
\hline
 WDM & Warm Dark Matter\tabularnewline
\hline
 WFXT & Wide-Field X-Ray Telescope\tabularnewline
\hline
 WIMP & Weakly Interacting Massive Particle\tabularnewline
\hline
 WKB & Wentzel-Kramers-Brillouin	\tabularnewline 
\hline
 WL & Weak Lensing\tabularnewline
\hline
 WLS & Weak Lensing Survey\tabularnewline
\hline
 WMAP & Wilkinson Microwave Anisotropy Probe\tabularnewline
\hline
 XMM-Newton & X-ray Multi-Mirror Mission	\tabularnewline 
\hline
 vDVZ & van Dam-Veltman-Zakharov \tabularnewline 
\hline
\end{tabular}
\end{table}

%% file: symbols.tex
\chapter*{List of symbols}

\begin{table}[htb]
\centering
\begin{tabular}{|l|l|l|}
\hline
$c_a$ & Adiabatic sound speed & \symbolref{symbol:ad.sound.speed}  \tabularnewline
\hline
$D_A(z)$ & Angular diameter distance  & \symbolref{eq:ang} \tabularnewline
\hline
$\;\raise1.0pt\hbox{$/$}\hskip-6pt\partial$ & Angular spin raising operator
& \symbolref{eq:spinraising} \tabularnewline
\hline
$\Pi^i_j$ & Anisotropic stress perturbation tensor  & \symbolref{symbol:stressperturbation} \tabularnewline
\hline
$\sigma$ &Uncertainty  & {} \tabularnewline
\hline
$B$ & Bayes factor  & \symbolref{eq:bayesfactor} \tabularnewline
\hline
$b$ & Bias (ratio of galaxy to total matter perturbations)  &
\symbolref{symbol:bias} \tabularnewline
\hline
$B_\Phi(k_1,k_2,k_3)$ & Bispectrum of the Bardeen's  potential  & \symbolref{bispectrum} \tabularnewline
\hline
$g(X)$ & Born-Infeld kinetic term  & \symbolref{UDML} \tabularnewline
\hline
$\mathbf b$ & Bulleticity  & \symbolref{symbol:bulleticity} \tabularnewline
\hline
$\zeta$ & Comoving curvature perturbation  & \symbolref{eq:zeta} \tabularnewline
\hline
$r(z)$ & Comoving distance & {} \tabularnewline
\hline
$\mathcal H$ & Conformal Hubble parameter, $\mathcal H = aH$  & \symbolref{symbol:conformaltime}  \tabularnewline
\hline
$\eta,\tau$ & Conformal time  & \symbolref{symbol:conformaltime} \tabularnewline
\hline
$\kappa$ & Convergence  & \symbolref{eq:kappadef} \tabularnewline
\hline
$t$ & Cosmic time  & \symbolref{CQ_euler} \tabularnewline
\hline
$\Lambda$ & Cosmological constant  & {} \tabularnewline
\hline
$\mathbf\Theta$ & Cosmological parameters  & \symbolref{eq:likelihood_i} \tabularnewline
\hline
$r_c$ & Cross over scale  & \symbolref{symbol:rc} \tabularnewline
\hline
$\square$ & d'Alembertian, $\square=\mathbf \nabla^2$  & {} \tabularnewline
\hline
$F$ & Derivative of $f(R)$  & \symbolref{symbol:FR} \tabularnewline
\hline
$\theta$ & Divergence of velocity field  & \symbolref{ij_EM} \tabularnewline
\hline
$\mu$ & Direction cosine & \symbolref{eq:Pgini} \tabularnewline
\hline
$\pi$ & Effective anisotropic stress  & \symbolref{eq:real-sigma} \tabularnewline
\hline
$\eta(a,k)$ & Effective anisotropic stress parameterisation  & \symbolref{mod_constr} \tabularnewline
\hline
$\rho$ & Energy density  & {} \tabularnewline
\hline
$T_{\mu\nu}$ & Energy momentum tensor  & \symbolref{EMT} \tabularnewline
\hline
$w$ & Equation of state  & \symbolref{phi_bkg} \tabularnewline
\hline
$F_{\alpha\beta}$ & Fisher information matrix  & \symbolref{eq:Falphabeta} \tabularnewline
\hline

$\sigma_8$ & Fluctuation amplitude at 8 km/s/Mpc  & {} \tabularnewline
\hline
$u^\mu$ & Four-velocity  & \symbolref{EMT} \tabularnewline
\hline
$\Omega_m$ & Fractional matter density  & {} \tabularnewline
\hline
$f_\mathrm{sky}$ & Fraction of sky observed  & \symbolref{eq:3-fm} \tabularnewline
\hline
$\Delta_M$ & Gauge invariant comoving density contrast  & \symbolref{symbol:DeltaM} \tabularnewline
\hline
$\tau(z)$ & Generic opacity parameter  & \symbolref{symbol:tau} \tabularnewline
\hline
$\varpi$ & Gravitational slip parameter  & \symbolref{symbol:varpi} \tabularnewline
\hline
$G(a)$ & Growth function/Growth factor  & \symbolref{eq:growthdef} \tabularnewline
\hline
\end{tabular}
\end{table}
\clearpage
\begin{table}[htb]
\centering
\begin{tabular}{|l|l|l|}
\hline
$\gamma$ & Growth index/Shear  &
\symbolref{gamma_Delta}/\symbolref{eq:kappadef} \tabularnewline
\hline
$f_g$ & Growth rate  & \symbolref{def-growth-rate} \tabularnewline
\hline
$b_\mathrm{eff}$ & Halo effective linear bias factor  & \symbolref{symbol:halobias} \tabularnewline
\hline
$h$ & Hubble constant in units of 100 km/s/Mpc  & {} \tabularnewline
\hline
$H(z)$ & Hubble parameter  & {} \tabularnewline
\hline
$\mathrm \xi_i$ & Killing field   & \symbolref{eq:killing} \tabularnewline
\hline
$\delta_{ij}$ & Kronecker delta  & {} \tabularnewline
\hline
$f(R)$ & Lagrangian in Modified Gravity  & \symbolref{fRaction} \tabularnewline
\hline
$P_l(\mu)$ & Legendre polynomials  & \symbolref{symbol:legendre} \tabularnewline
\hline
$\mathcal L(\mathbf\Theta)$ & Likelihood function  & \symbolref{eq:likelihood_i} \tabularnewline
\hline
$\beta(z)$ & Linear redshift-space distortion parameter  & \symbolref{symbol:beta} \tabularnewline
\hline
$D_L(z)$ & Luminosity distance  & \symbolref{symbol:luminosity} \tabularnewline
\hline
$Q(a,k)$ & Mass screening effect  & \symbolref{mod_constr} \tabularnewline
\hline
$\delta_m$ & Matter density perturbation  & {} \tabularnewline
\hline
$g_{\mu\nu}$ & Metric tensor  & \symbolref{symbol:metric} \tabularnewline
\hline
$\mu$ & Modified gravity function: $\mu=Q/\eta$  & \symbolref{eq:muQeta} \tabularnewline
\hline
$C_\ell$ & Multipole power spectrum  & \symbolref{symbol:Cell} \tabularnewline
\hline
$G$ & Newton's gravitational constant  & {} \tabularnewline
\hline
$N$ & Number of e-folds, $N=\ln a$  & \symbolref{symbol:efolds} \tabularnewline
\hline
$P(k)$ & Matter power spectrum  & {} \tabularnewline
\hline
$p$ & Pressure  & {} \tabularnewline
\hline
$\delta p$ & Pressure perturbation  & {} \tabularnewline
\hline
$\chi(z)$ & Radial, dimensionless comoving distance  & \symbolref{chiz} \tabularnewline
\hline
$z$ & Redshift  & {} \tabularnewline
\hline
$R$ & Ricci scalar  & {} \tabularnewline
\hline
$\phi$ & Scalar field  & \symbolref{symbol:phi} \tabularnewline
\hline
$A$ & Scalar potential  & \symbolref{pert_0_ds} \tabularnewline
\hline
$\Psi,\Phi$ & Scalar potentials  & \symbolref{symbol:scalarpotentials} \tabularnewline
\hline
$n_s$ & Scalar spectral index  & \symbolref{eq:index_ns} \tabularnewline
\hline
$a$ & Scale factor  & {} \tabularnewline
\hline
$f_a$ & Scale of Peccei-Quinn symmetry breaking  & \symbolref{symbol:f_a} \tabularnewline
\hline
$\ell$ & Spherical harmonic multipoles  & \tabularnewline
\hline
$c_s$ & Sound speed  & \symbolref{symbol:c_s} \tabularnewline
\hline
$\Sigma$ & Total neutrino mass/Inverse covariance matrix/PPN parameter  &
\symbolref{symbol:Sigma}/\symbolref{symbol:invcov}/\symbolref{Sigma-wl-isw} \tabularnewline
\hline
$H_T^{ij}$ & Trace-free distortion  & \symbolref{symbol:H_T} \tabularnewline
\hline
$T(k)$ & Transfer function  & \symbolref{symbol:Tk} \tabularnewline
\hline
$B_i$ & Vector shift  & \symbolref{symbol:vectorshift} \tabularnewline
\hline
$\mathbf k$ & Wavenumber  & {} \tabularnewline
\hline
\end{tabular}
\end{table}

%% file: generalintro.tex
\chapter*{\centering Abstract}
\addcontentsline{toc}{chapter}{Abstract}

Euclid is a European Space Agency medium class mission selected for
launch in 2019 within the Cosmic
Vision 2015-2025 programme. The main goal of Euclid is to
understand the origin of the accelerated expansion of the Universe.
Euclid will explore the expansion
history of the Universe and the evolution of cosmic structures  by
measuring shapes and red-shifts of
galaxies as well as the distribution
of clusters of galaxies  over a large fraction of
the sky.

Although the main driver for Euclid is the nature of dark energy,
Euclid science covers a  vast range of topics, from cosmology to
galaxy evolution to planetary research.
In this review we focus on  cosmology and fundamental physics, with a strong
emphasis on science beyond the current standard models. We discuss
five broad topics:
dark energy and modified gravity, dark matter, initial conditions,
basic assumptions and questions of methodology
in the data analysis.

This review has been planned and carried out within Euclid's Theory
Working Group and is meant to provide
a  guide to the scientific themes that will underlie the activity of
the group during the preparation of the Euclid mission.

\chapter*{Introduction}
\addcontentsline{toc}{chapter}{Introduction}

Euclid\footnote{Continuously updated information on Euclid is available on http://www.euclid-ec.org.} (\cite{ECredbook:2011mu} \cite{2009ExA....23...17R}, \cite{2009ExA....23...39C}) is an ESA medium class mission selected
for the second launch slot (expected for 2019) of the Cosmic
Vision 2015-2025 programme. The main goal of Euclid is to
understand the physical origin of the accelerated expansion of the Universe. Euclid
is a satellite equipped with a 1.2 m telescope and
three imaging and spectroscopic instruments working in the visible and
near-infrared wavelength domains. These instruments will explore the expansion
history of the Universe and the evolution of cosmic structures
 by measuring shapes and redshifts of galaxies over a large fraction of
the sky. The satellite will be launched by a Soyuz ST-2.1B rocket and
transferred to the L2 Lagrange point for a six-year mission that will
cover at least 15,000 square degrees of sky. Euclid plans to image
a billion of galaxies and measure nearly 100 millions of galaxy redshifts.

These impressive numbers will allow Euclid to realize a detailed
reconstruction of
 the clustering of galaxies out to a redshift 2 and the pattern of
light distortion from weak lensing to redshift 3. The two main probes, redshift
clustering and weak lensing,
are complemented by a number of additional cosmological probes: cross
correlation between
the cosmic microwave background and the large scale structure;
luminosity distance through
supernovae Ia; abundance and properties of galaxy clusters and strong
lensing. To extract
the maximum of information also in the non-linear regime of
perturbations, these probes will require accurate high-resolution
numerical simulations.
Besides cosmology, Euclid will provide an exceptional dataset for
galaxy evolution, galaxy structure, and
planetary searches. All Euclid data will be publicly released after a
relatively short proprietary period
and will constitute for many years the ultimate survey database for
astrophysics.

A huge enterprise like Euclid requires a highly considered planning in terms
not only of technology
but also for the scientific exploitation of future data. Many ideas and
models that today seem to be
abstract exercises for theorists will in fact finally become testable with
the Euclid surveys.
The main science driver of Euclid is clearly the nature of dark
energy, the enigmatic
substance that is driving the accelerated expansion of the Universe.
As we will discuss
in detail in  Chapter \ref{dark-energy}, under the label ``dark energy''
we include a wide
variety of hypotheses, from extradimensional physics to higher-order
gravity, from new fields and new
forces to large violations of homogeneity and isotropy. The simplest
explanation, Einstein's
famous cosmological constant, is still currently acceptable from the
observational point of view,
but is not the only one, nor necessarily the most satisfying, as
we will argue in the following.
It is therefore important to identify the main observables that will
help distinguish the cosmological
constant from the alternatives and to forecast Euclid's performance in
testing the various models.

Since clustering and weak lensing also depend on the properties of
dark matter, Euclid is a dark matter
probe as well. In Chapter \ref{dark-matter} we focus on the models of dark
matter that can be tested with Euclid data,
from massive neutrinos to ultra-light scalar fields. We show that
Euclid can measure the neutrino mass to a very
high precision, making it one of the most sensitive neutrino experiments
of its time, and it can help identifying new light fields in the cosmic fluid.

The evolution of perturbations depends not only on the fields and forces
active during the cosmic eras, but also
on the initial conditions. By reconstructing the initial conditions we open
a window on the inflationary physics that created the
perturbations, and allow ourselves the
chance of determining
whether a single inflaton drove the expansion or a mixture of fields.
In Chapter \ref{ini-cond} we review the choices of initial conditions and their
impact on Euclid science. In particular we discuss deviations from
simple scale invariance,
mixed isocurvature-adiabatic initial conditions,
non-gaussianity, and the combined forecasts of Euclid and CMB experiments.

Practically all of cosmology is built on the Copernican Principle, a very fruitful idea
postulating a homogeneous and isotropic background. Although this assumption has been
confirmed time and again since the beginning of modern cosmology,
Euclid's capabilities can push
the test to new levels. In Chapter \ref{testing} we challenge some of
the basic cosmological assumptions and
predict how well Euclid can constrain them. We explore the basic
relation between luminosity and angular
diameter distance that holds in any metric theory of gravity if
the Universe is transparent to light,
and the existence of
large violations of homogeneity and isotropy, either due to local
voids or to the cumulative stochastic
effects of perturbations, or to intrinsically anisotropic vector
fields or space-time geometry.

Finally, in Chapter \ref{statistical} we review some of the
statistical methods that are used to forecast the performance
of probes like Euclid,
and we discuss some possible future developments.

This review has been planned and carried out within Euclid's Theory
Working Group and is meant to provide
a  guide to the scientific themes that will underlie the activity of
the group during the preparation of the mission.
At the same time, this review will help us and the community at large
to identify the areas
that deserve closer attention, to improve  the development of Euclid
science and to offer new
scientific challenges and opportunities.

%% file: de_mg/de_mg.tex
\chapter{Dark Energy}\label{dark-energy}


\section{Introduction}

\input{de_mg/introduction}

\section{Background evolution}


\input{de_mg/quintessence_parametrizations_v2}

\section{Perturbations  \label{sec:dof}}

\input{de_mg/mg_params}

\section{Models of dark energy and modified gravity}\label{models-of-modified-gravity}

\input{de_mg/intro-models}

\input{de_mg/defmodgrav}

\input{de_mg/coupled} \label{cde_eq}

\input{de_mg/crossing}

\input{de_mg/fr-general}

\input{de_mg/ModifiedGravity} 

\input{de_mg/GEA}

\input{de_mg/TeVeS}

\section{Generic properties of dark energy and modified gravity models} \label{genericproperties}

\input{de_mg/lessons}

\input{de_mg/smoking}

\input{de_mg/consistency}

\section{Nonlinear aspects}\label{non-linear-aspects}

\input{de_mg/nonlinear_aspects}

\input{de_mg/spherical_collapse}

\section{Observational properties of dark energy and modified gravity}\label{observational-properties-of-modified-gravity}

\input{de_mg/obs-intro}

\input{de_mg/magnification}

\input{de_mg/redshift-survey}

\input{de_mg/bulk}



\section{Forecasts for Euclid}


\input{de_mg/obs_prospects}

\input{de_mg/euclid-surveys}

\input{de_mg/gamma-param}

\input{de_mg/hg-weaklensing}

\input{de_mg/weaklensing-nbody}

\input{de_mg/soundspeed}

\input{de_mg/fRforecastconstraints}

\input{de_mg/cdeforecast}

\input{de_mg/extra_priors}

\input{de_mg/outlook}










%% file: de_mg/introduction.tex
With the discovery of  cosmic acceleration  at the end of the 1990s, and its
possible explanation in terms of a cosmological constant,  cosmology has {\color{red} returned to its roots in
Einstein's famous 1917 paper that simultaneously} inaugurated modern cosmology and the history of the constant $\Lambda$.
Perhaps cosmology is approaching a robust and all-encompassing standard model, like its cousin, 
the very successful standard model of particle physics. 
{\color{red}In this scenario}, the cosmological standard model could essentially close the search
for a broad picture of cosmic evolution, leaving to future generations only
the task of filling in a number of important, but not crucial, details.

The cosmological constant is still in remarkably good agreement with
almost all cosmological data more than ten years after the observational discovery of the
{\color{red}accelerated expansion rate of the universe}. However, our
knowledge of the universe's evolution is so incomplete that  it would be totally premature
to claim that we are close to understanding the ingredients of the cosmological standard model. 
If we ask ourselves what we {\color{red}know for certain} about
the expansion rate at redshifts larger than unity, or the growth rate of
matter fluctuations, or about the properties of gravity on large scales and at early times, or
about the influence of extra dimensions (or their absence) on our four dimensional world,
the answer would be surprisingly disappointing. 

{\color{red} Our present knowledge can be succinctly summarised as follows: we live in a universe that is consistent  with the presence of a cosmological constant in the field equations of General Relativity, and as of 2012, the value of this constant corresponds to a fractional energy density today of $\Omega_\Lambda\approx 0.73$.
Far from being disheartening}, this current lack of knowledge points however the way to an exciting future. 
A decade of research on dark energy has
taught to many cosmologists that this ignorance can be overcome by the same tools
that revealed it, together with many more which have been developed in recent years.

Why then is the cosmological constant not the end of the story as far as the cosmic
acceleration is concerned? There are at least three reasons. The first is that we have no simple
way to explain its small but non-zero value. In fact, its value is {\color{red}unexpectedly}
small with respect
to any physically meaningful scale, except the {\em current} horizon scale. The second {\color{red}reason} is that this value is not only small, but also
surprisingly close to another unrelated quantity, the {\em present} matter-energy density. That this
happens just by coincidence is hard to accept, as the matter density is diluted rapidly with the expansion
of space. {\color{red}Why is it that we happen to live at the precise, fleeting epoch when the energy densities of matter and the cosmological constant are of comparable magnitude?} Finally, observations of coherent acoustic
oscillations in the Cosmic Microwave Background (CMB)
have turned the notion of accelerated expansion in the very early universe (inflation) into an integral 
part of the cosmological standard model. Yet the simple truth that we exist {\color{red}as observers} demonstrates that this early accelerated expansion {\color{red}was of a finite duration, and hence cannot be ascribable to a true, constant
$\Lambda$; this sheds} doubt on the nature of the current accelerated expansion.
The very fact that
we know so little about the past dynamics of the universe forces us to enlarge
the theoretical parameter space and to consider also phenomenology that a simple
cosmological constant cannot accommodate.

These motivations {\color{red}have} led many scientists to challenge one of the most basic tenets
of physics: Einstein's law of gravity. Einstein's theory of General Relativity (GR)
 is a supremely successful theory on
scales {\color{red}ranging from the size of our Solar System down to micrometres},  the shortest {\color{red}distances at which GR has been probed in the laboratory
so far}. Although specific predictions about such
diverse phenomena as the gravitational redshift of light, energy loss from
binary pulsars, the rate of precession of the perihelia of bound orbits, and
light deflection by the Sun are not unique to General Relativity, it must be
regarded as highly significant that General Relativity is  consistent with each
of these tests and more. {\color{red}We can securely state that GR has been tested
to high accuracy \textit{at these distance scales}}.

The success of General Relativity on larger scales is less clear. On astrophysical
and cosmological scales, tests of General Relativity are complicated by the existence of 
invisible components like dark matter and by the effects of spacetime
geometry. 
{\color{red} We do not know whether the physics underlying the apparent cosmological constant originates from modifications to General Relativity (ie. an extended theory of gravity), or from a new fluid or field in our universe that we have not yet detected directly. The latter phenomena are generally referred to as `dark energy' models.} 

{\color{red} If we only consider observations of the expansion rate of the universe we cannot discriminate between a theory of modified gravity and a dark energy model. However, it is likely that these two alternatives will cause perturbations around the `background' universe to behave differently. Only by improving our knowledge of the growth of structure in the universe
can we hope to progress towards breaking the degeneracy between dark energy and modified gravity.} The first Chapter of this Review
is dedicated to this effort.
We begin with a review of the background and linear perturbation equations in a general
setting, defining quantities that will be employed throughout. We then explore the 
non-linear effects of dark energy, making use of analytical tools such as the spherical collapse model, perturbation theory and numerical $N$-body simulations.
We discuss a number of competing models proposed in literature and demonstrate what the Euclid survey
will be able to tell us about them.

%% file: de_mg/quintessence_parametrizations_v2.tex
Most of the calculations in this review are performed in the Friedmann-Lema\^\i tre-Robertson-Walker (FLRW) metric
\begin{equation}
ds^2=-dt^2+ {a(t)^2}(\frac {dr^2}{1-kr^2}+r^2d\theta^2+r^2\sin^2\theta d\phi^2 )
\end{equation}
where $a(t)$ is the scale factor and $k$ the spatial curvature. The usual symbols for
the Hubble function $H=\dot a/a$ and the density fractions $\Omega_x$, where $x$ stands
for the component, are employed. We characterise the components with the subscript $M$ or $m$ for matter,
$\gamma$ or $r$ for radiation, $b$ for baryons, $K$ for curvature and $\Lambda$ for the cosmological constant.
Whenever necessary for clarity, we append a subscript $0$ to denote the present epoch, e.g. $\Omega_{M,0}$.
Sometimes the conformal time $\eta=\int dt/a$ and the conformal Hubble
function ${\mathcal H}=aH= da/(ad\eta)  $ \label{symbol:conformaltime}
 are employed. Unless otherwise stated, we denote with a dot derivatives wrt cosmic time $t$ (and sometimes
we employ the dot for derivatives wrt conformal time $\eta$) while we use a prime for
derivatives with respect to $\ln a$.

The energy density due to a cosmological constant with $p=-\rho$ is obviously constant over time. This can be easily seen from the covariant
conservation equation $T_{\mu;\nu}^\nu=0$ for the homogeneous and isotropic
FLRW metric,
\begin{equation}
\dot{\rho} + 3 H (\rho+p) = 0 .
\end{equation}
However since we also observe radiation with $p=\rho/3$ and non-relativistic
matter for which $p\approx 0$, it is natural to assume that the dark energy is
not necessarily limited to a constant energy density, but that it could be
dynamical instead.

One of the simplest models that explicitly realises such a dynamical dark
energy scenario is described by a minimally coupled canonical scalar field evolving in a
given potential. For this reason, the very concept of dynamical dark energy
is often associated to this scenario, and in this context it is called `quintessence'
\citep{Wetterich_1988, Ratra:1987rm}. In the following, the scalar field will be indicated with $\phi$.
Although in this simplest framework the dark energy does not
interact with other species and influences space-time only through its energy density
and pressure, this is not the only possibility and we will encounter more general models later on. 
The homogeneous energy density and pressure of the scalar field $\phi$ are
defined as \begin{equation}
 \label{phi_bkg} \rho_{\phi} = \frac{{\dot\phi}^2}{2 } + V(\phi)  \,\,\, , \,\,\,
p_{\phi} = \frac{{\dot\phi}^2}{2 } - V(\phi)   \,\,\,, \,\,\, w_{\phi} =
\frac{p_{\phi}}{\rho_{\phi}} \,\,\,, \end{equation}
and $w_\phi$ is called the equation-of-state parameter. 
Minimally coupled dark energy models can allow for attractor solutions \citep{Copeland:1997et, Liddle:1998xm,
Steinhardt:1999nw}: if an attractor exists, depending on the potential $V(\phi)$ in which
dark energy rolls, the trajectory of the scalar field in
the present regime converges to the path given by the attractor, though starting from a wide set
of different initial conditions for $\phi$ and for its first derivative $\dot\phi$.
Inverse power law and exponential potentials are typical examples of potential
that can lead to attractor solutions.
As constraints on $w_\phi$ become tighter \citep[e.g.][]{Komatsu:2010fb}, the allowed
range of initial conditions to follow into the attractor solution shrinks, so
that minimally coupled quintessence is actually constrained to have very flat
potentials. The flatter the potential is, the more minimally-coupled quintessence mimics a
cosmological constant, the more it suffers from the same fine-tuning and
coincidence problems which affect a $\Lambda$CDM scenario
\citep{Matarrese:2004xa}. 

However, when General Relativity is modified or when an
interaction with other species is active, dark energy may very well have a non-negligible
contribution at early times. It is therefore important, already at the background level, 
to understand the best way to characterize the main features of the evolution of quintessence and dark energy in general, pointing out which parametrisations are more suitable
and which ranges of parameters are of interest to disentangle quintessence or modified gravity from a cosmological constant scenario.

In the following we discuss briefly how to describe the cosmic expansion rate
in terms of a small number of parameters. This will set the stage for the
more detailed cases discussed in the subsequent sections. Even within specific physical models it is often 
convenient to reduce the information to a few phenomenological parameters.

Two important points are left for later: from Eq.~(\ref{phi_bkg}) we can easily see that $w_\phi\geq-1$ as
long as $\rho_\phi>0$, i.e. uncoupled canonical scalar field dark energy never crosses $w_\phi=-1$. However, this is
not necessarily the case for non-canonical scalar fields or for cases where GR is modified. We postpone the discussion of how to parametrise
this so-called `phantom crossing' to avoid singularities to section \ref{sec:crossing}, as it requires also the study
of perturbations.

The second deferred part on the background expansion concerns a basic statistical question: 
what is a sensible precision target for a measurement of dark energy, e.g. of its equation 
of state? In other words, how close to $w_\phi=-1$ should we go
before we can be satisfied and declare that dark energy is the cosmological constant? We will address this question in section \ref{genericproperties}.

\subsection{Parametrisation of the background evolution}\label{parametrization-of-the-background-evolution}

If one wants to parametrise the equation of state of dark energy, two general 
approaches are possible. 
The first is to start from a set of dark energy models 
given by the theory and to find parameters describing their $w_\phi$ as accurately as 
possible. Only later, one can try and include as many theoretical models as possible 
in a single parametrisation.
In the context of scalar-field dark energy models (to be discussed in Section \ref{quintessence}),  \cite{Crittenden:2007yy} 
parametrise the case of slow-rolling fields, \cite{Scherrer:2007pu} study
thawing quintessence, \cite{Hrycyna:2007mq} and \cite{Chiba:2010cy} include 
non-minimally coupled fields,  \cite{Setare:2008sf} quintom quintessence,
\cite{Dutta:2008qn} parametrise hilltop quintessence, \cite{Chiba:2009nh} 
extend the quintessence parametrisation to a class of $k$-essence models,  
\cite{Huang:2010zr} study a common parametrisation for quintessence and phantom fields.
Another convenient way used to parametrise the presence of a non-negligible homogenous dark
energy component at early times (usually labelled as EDE) was presented in
\cite{Wetterich:2004pv}. We recall it here because we will refer to this example in {\color{red} Section \ref{quintessence_ede}}. 
In this case the equation of state is parametrised as:
\begin{equation} {w}_X (z) = \frac{{ w}_0}{1+b \ln{(1+z)}} \,\,\,\,
\label{w_ede_par} \end{equation}
where $b$ is a constant related to the amount of dark energy at early
times, i.e.
\begin{equation}
b = - \frac{3 {\bar
w}_0}{\ln{\frac{1-\Omega_{\text{X},e}}{\Omega_{\text{X},e}}} +
\ln{\frac{1-\Omega_{m,0}}{\Omega_{m,0}}}}.
\end{equation}
Here the subscripts `$0$' and `$e$' refer to quantities calculated today or
early times, respectively. With regard to the latter parametrisation, we note that concrete 
theoretical and realistic models involving a non-negligible energy component
at early times are often accompanied by further important
modifications (as in the case of interacting dark energy), not always included
in a  parametrisation of the sole equation of state such as (\ref{w_ede_par})
(for further details see also Sec. \ref{non-linear-aspects} on non linear aspects of dark energy
and modified gravity).

The second approach is to start from a simple expression of $w$ without 
assuming any specific dark energy model (but still 
checking afterwards whether known theoretical dark energy models can be represented).
This is what has been done by \cite{Huterer:2000mj}, \cite{Maor:2000jy}, 
\cite{Weller:2000pf} (linear and logarithmic parametrisation in $z$), 
\cite{chevallier01}, \cite{Linder03} (linear and power law 
parametrisation in $a$), \cite{Douspis:2006rs}, \cite{Bassett:2004wz} 
(rapidly varying equation of state).

The most common parametrisation, widely employed also in this review,
is the linear equation of state 
\citep{chevallier01,Linder03}
\begin{equation}
\label{CPL}
w_X(a)=w_0+w_a (1-a)\;,
\end{equation}
where the subscript $X$ refers to the generic dark energy constituent.
While this parametrization is useful as a toy model
in comparing the forecasts for different dark energy projects,
{\color{red} it should not be taken as all-encompassing. In general a dark energy model can introduce further significant terms in the effective $w_X(z)$ that cannot be mapped onto the simple form of eq.(\ref{CPL})}.

An alternative  to make model-independent constraints
is measuring the dark energy density $\rho_X(z)$
(or the expansion history $H(z)$) as a free function
of cosmic time \citep{WangGarnavich01,Tegmark02,Daly03}.
Measuring $\rho_X(z)$ has advantages over measuring the dark energy
equation of state $w_X(z)$ as a free function; $\rho_X(z)$ is more
closely related to observables, hence is more tightly 
constrained for the same number of redshift bins 
used \citep{WangGarnavich01,WangFreese04}.
Note that $\rho_X(z)$ is related to $w_X(z)$ as follows \citep{WangGarnavich01}:
\begin{equation}
\frac{\rho_X(z)}{\rho_X(0)} = \exp\left\{ \int_0^z {\rm d}z'\, \frac{3
    [1+w_X(z')]}{1+z'} \right\}\;. 
\end{equation} 
Hence parametrising dark energy with $w_X(z)$ implicitly assumes that 
$\rho_X(z)$ does not change sign in cosmic time.
This precludes whole classes of dark energy models in which $\rho_X(z)$ 
becomes negative in the future (``Big Crunch'' models, see 
\cite{WangLinde04} for an example) \citep{WangTegmark04}.

Note that the measurement of $\rho_X(z)$ is straightforward
once $H(z)$ is measured from baryon acoustic oscillations, and $\Omega_m$ is constrained
tightly by the combined data from galaxy clustering, weak lensing, and cosmic microwave background data
-- although strictly speaking this requires a choice of perturbation evolution for the dark energy as well, and in addition one that
is not degenerate with the evolution of dark matter perturbations, see \cite{Kunz:2007rk}.

Another useful possibility is to adopt the principal component approach 
\citep{Huterer:2002hy}, which avoids any assumption about the 
form of $w$ and assumes it to be constant or linear in redshift bins, 
then derives which combination of parameters is best 
constrained by each experiment.

For a cross-check of the results using more complicated parametrisations,
one can use simple polynomial parametrisations of
$w$ and $\rho_{DE}(z)/\rho_{DE}(0)$ \citep{Wang:2008zh}.

%% file: de_mg/mg_params.tex
This section is devoted to a discussion of linear perturbation theory
in dark energy models. Since we will discuss a number of non-standard models
in later sections, we present here the main equations in a general form that
can be adapted to various contexts. This section will identify
which perturbation functions the Euclid survey \cite{euclidredbook} will try to measure and how they can help
us characterising the nature of dark energy and the properies of gravity.

\subsection{Cosmological perturbation theory}\label{sec:cosmo_perts}

Here we provide the perturbation equations in a dark energy dominated Universe 
for a general fluid, focusing  on 
scalar perturbations.

For simplicity, we consider a flat  Universe containing only (cold dark) 
matter and dark energy, so that the Hubble parameter is given by
\begin{equation}
H^2 =\left( \frac{1}{a} \frac{{\rm d}a}{{\rm d}t} \right) ^2 = 
H_{0}^{2}\left[\Omega_{m_0} a^{-3}+\left( 1- \Omega_{m_0} \right)\exp\left( -3\int_1^a\frac{1+w(a')}{a'}{\rm d}a \right) \right]\,.
\end{equation}
We will consider linear perturbations on a spatially-flat background
model, defined by the line of element
\begin{equation}
{\rm d}s^{2} = a^{2} \left[ -\left( 1+2A\right) {\rm d}\eta^{2}+2B_{i}{\rm d}{\eta}{\rm d} x^{i}+\left( \left( 1+2H_{L}\right) \delta_{ij}+2H_{Tij} \right){\rm d} x_{i}{\rm d} x^{j} \right]
\label{pert_0_ds}
\end{equation}
where $A$ is the scalar potential; $B_{i}$\label{symbol:vectorshift} a vector shift; $H_{L}$ is the
scalar perturbation to the spatial curvature; $H_{T}^{ij}$ is the trace-free
distortion to the spatial metric; ${\rm d}\eta= {\rm d}t/a$ is the
conformal time. \label{symbol:H_T}

We will assume that the Universe is filled with perfect fluids only,
so that the energy momentum tensor takes the simple form
\begin{equation}
T^{\mu\nu}=\left( \rho+p\right) u^{\mu}u^{\nu} +p~g^{\mu\nu} +\Pi^{\mu\nu}
\label{EMT}
\end{equation}
where $\rho$ and $p$ are the density and the pressure of the 
fluid respectively, $u^{\mu}$ is the four-velocity and $\Pi^{\mu\nu}$ 
is the anisotropic-stress perturbation tensor which represents 
the traceless component of the $T_{j}^{i}$.
\label{symbol:stressperturbation}

The components of the perturbed energy momentum tensor can be written as:
\begin{eqnarray}
T_{0}^{0} &=& - \left( \bar\rho + \delta\rho \right) \\
T_{j}^{0} &=& \left( \bar\rho + \bar{p} \right) \left( v_{j} - B_{j} \right) \\
T_{0}^{i} &=& \left( \bar\rho + \bar{p} \right) v^{i} \\
T_{j}^{i} &=& \left( \bar{p} + \delta{p} \right) \delta_{j}^{i} + \bar{p}~\Pi_{j}^{i}.
\end{eqnarray}
Here $\bar\rho$ and $\bar p$ are the energy density and pressure of the
homogeneous and isotropic background Universe,
$\delta\rho$ is the density perturbation, $\delta p$ is the pressure perturbation, $v^{i}$ is the velocity {\color{red} vector}. 
Here, we want to investigate only the scalar modes of the perturbation equations.
So far the treatment of the matter and metric is fully general and applies 
to any form of matter and metric. We now choose the Newtonian gauge (also known as the longitudinal gauge), characterized by zero non-diagonal metric terms (the shift vector $B_{i}=0$ and $H_{T}^{ij}=0$) and by two \label{symbol:scalarpotentials}
scalar potentials $\Psi$ and $\Phi$; the metric Eq.~(\ref{pert_0_ds}) then becomes
\begin{equation}
{\rm d} s^{2} = a^{2} \left[ -\left( 1+2\Psi \right) {\rm d}\eta^{2} + \left( 1-2\Phi\right) {\rm d} x_{i}{\rm d} x^{i} \right]  \,\, .
\label{pert_newton_ds}
\end{equation}
The advantage of using the Newtonian gauge is that the metric tensor $g_{\mu\nu}$ is diagonal
and\label{symbol:metric} 
this simplifies the calculations. This choice does not only simplify the calculations 
but it is also the most intuitive one as the observers are attached to the points in 
the unperturbed frame; as a consequence, they will detect a velocity field 
of particles falling into the clumps of matter and will measure their gravitational potential, represented directly by $\Psi$; $\Phi$ corresponds to the perturbation to the spatial curvature. 
Moreover, as we will see later, the Newtonian gauge is the 
best choice for observational tests (i.e. for perturbations smaller than the horizon).

In the conformal Newtonian gauge, and in Fourier space, the first-order perturbed Einstein
equations give \citep[see][for more details]{Ma:1995ey}:
\begin{eqnarray}
k^2\Phi + 3\frac{\dot{a}}{a} \left( \dot{\Phi} + \frac{\dot{a}}{a}\Psi\right) &=& -4\pi G a^2 \sum_{\alpha}\bar{\rho}_{\alpha}\delta_{\alpha} \,,\label{ein-cona}\\
k^2 \left( \dot{\Phi} + \frac{\dot{a}}{a}\Psi \right)&=& 4\pi G a^2 \sum_{\alpha}(\bar{\rho}_{\alpha}+\bar{p}_{\alpha}) \theta_{\alpha}\,,\label{ein-conb}\\
\ddot{\Phi} + \frac{\dot{a}}{a} (\dot{\Psi}+2\dot{\Phi})+\left(2\frac{\ddot{a}}{a} - \frac{\dot{a}^2}{a^2}\right)\Psi+ \frac{k^2}{3} (\Phi-\Psi)
&=& 4\pi G a^2 \sum_{\alpha}\delta p_{\alpha}\,,\label{ein-conc}\\
k^2(\Phi-\Psi) &=& 12\pi G a^2 \sum_{\alpha}\left(\bar{\rho}_{\alpha}+\bar{p}_{\alpha}\right)\pi_{\alpha}\,,\label{ein-cond}
\end{eqnarray}
where a dot denotes $d/d\eta$, {\color{red}$\delta_\alpha=\delta\rho_\alpha/\bar{\rho}_\alpha$}, the {\color{red} index $\alpha$ indicates a sum over all matter} components in the universe and 
where $\pi$ is related to $\Pi_{j}^{i}$ through:
\begin{equation}
\left(\bar{\rho}+\bar{p}\right)\pi = -\left(\hat{k}_i\hat{k}_j-\frac{1}{3}\delta_{ij}\right)\Pi_{j}^{i}.
\end{equation}
The energy-momentum tensor components in the Newtonian gauge become:
\begin{eqnarray}
T_{0}^{0} &=& -\left( \bar\rho + \delta\rho \right) \label{00_EM}\\
ik_i T_{0}^{i} &=& -ik_i T_{i}^{0} = \left(\bar\rho + \bar{p} \right) \theta  \label{0i_EM}\\
T_{j}^{i} &=& \left( \bar p + \delta p \right) \delta_{j}^{i} +\bar{p}\Pi_{j}^{i} \label{ij_EM}
\end{eqnarray}
where we have defined the variable $\theta=ik_j v^j$ which represents the divergence
of the velocity field.

Perturbation equations for a single fluid are obtained taking
the covariant derivative of the perturbed energy momentum tensor, 
i.e. $T_{\nu;\mu}^{\mu}=0$. We have
\begin{eqnarray}
\dot\delta &=& -\left( 1+w \right) \left( \theta - 3\dot\Phi \right)
-3\frac{\dot a}{a} \left( \frac{\delta p}{\bar\rho} - w\delta \right)~~~~~~~~~~~~~~~~~~~{\rm for}~~~~\nu=0 \label{d_pert}\\
\dot\theta &=& -\frac{\dot a}{a} \left( 1-3w \right) \theta -
\frac{\dot{w}}{1+w}\theta +k^{2}\frac{\delta{p}/\bar\rho}{1+w} + k^{2}\Psi - k^2\pi~~~~~{\rm for}~~~~\nu=i \label{t_pert}.
\end{eqnarray}
The equations above are valid for any fluid. The evolution of the 
perturbations depends on the characteristics of the fluids considered, i.e. we need to 
specify the equation of state parameter $w$, the pressure perturbation $\delta p$ and the 
anisotropic stress $\pi$. 
For instance, if we want to study how matter perturbations evolve, we simply 
substitute $w=\delta p = \pi = 0$ (matter is pressureless) in the above equations. 
However, Eqs.~(\ref{d_pert})-(\ref{t_pert}) depend on the 
gravitational potentials $\Psi$ and $\Phi$ which in 
turn depend on the evolution of the perturbations of the other fluids. 
For instance, if we assume that the universe is filled by dark matter and dark energy 
then we need to specify also $\delta p$ and $\pi$ for the dark energy. 

The problem here is not only to parameterise the pressure perturbation and the 
anisotropic stress for the dark energy (there is not a unique way to do it, see in the 
next sections, especially \ref{sec:crossing} for what to do when $w$ crosses -1) 
but rather that we need to run the perturbation equations 
for each model we assume, making predictions and compare the results with observations. 
Clearly, this approach takes too much time. 
In the following section we show a general approach to understand the observed late-time 
accelerated expansion of the universe through the evolution of the 
matter density contrast. 

In the following, whenever there is no risk of confusion, we remove the overbars
from the background quantities.

\subsection{Modified Growth Parameters}
\label{mg_growth_params}

Even if the expansion history, $H(z)$, of the
Friedmann-Lema\^itre-Robertson-Walker (FLRW) background  has been measured {\color{red}(at least up to redshifts $\sim 1$ by supernova data)}, it
is not possible yet to identify the physics causing the recent acceleration of
the expansion of the Universe. Information on the growth of structure at
different scales and different redshifts is needed to discriminate between
models of dark energy (DE) and modified gravity (MG). 
A definition of what we mean by DE and MG will be postponed to the next section.

An alternative to testing predictions of specific theories is to parameterise
the possible departures from a fiducial model. Two conceptually different
approaches are widely discussed in the literature:
\begin{itemize}
\item \emph{Model parameters} capture the degrees of freedom of DE/MG and modify
the evolution equations of the energy-momentum content of the fiducial model.
They can be associated with physical meanings and have uniquely predicted
behaviour in specific theories of DE and MG.
\item \emph{Trigger relations} are derived directly from observations and only
hold in the fiducial model. They are constructed to break down if the fiducial
model does not describe the growth of structure correctly.
\end{itemize}
As the current observations favour the concordance cosmology, the fiducial model
is typically taken to be spatially flat FLRW in GR with cold dark matter and a
cosmological constant, hereafter referred to as $\Lambda$CDM.

For a large-scale structure and weak lensing survey the crucial quantities are
the matter density contrast and the gravitational potentials and we therefore focus on scalar perturbations in the Newtonian gauge with the metric (\ref{pert_newton_ds}).

We describe the matter perturbations using the gauge-invariant comoving density
contrast $\Delta_M\equiv\delta_M+3aH \theta_M/k^2$ where $\delta_M$ and
$\theta_M$ are the\label{symbol:DeltaM}
matter density contrast and the divergence of the fluid velocity for matter, respectively. 
The discussion can be generalised to include multiple fluids.

In $\Lambda$CDM, after radiation-matter equality there is no anisotropic stress
present and the Einstein constraint equations at ``sub-Hubble scales'' $k\gg aH$ become
\begin{equation}
  -k^2 \Phi = 4\pi G a^2 \rho_M \Delta_M \ ,\qquad\qquad  \Phi=\Psi \,.
  \label{lcdm_eeq}
\end{equation}
These can be used to reduce the energy-momentum conservation of matter simply to
the second order growth equation
\begin{equation}
  \Delta_M''+\left[2+(\ln H)'\right]\Delta_M' = \frac{3}{2}\Omega_M(a)\Delta_M
\,.
  \label{lcdm_geq}
\end{equation}
Primes denote derivatives with respect to $\ln a$ and we define the time
dependent fractional matter density as $\Omega_M(a)\equiv8\pi
G\rho_M(a)/(3H^2)$.  Notice that
the evolution of $\Delta_M$ is driven by $\Omega_M(a)$ and is scale-independent
throughout (valid on sub- and super-Hubble scales after radiation-matter
equality). We define the growth factor $G(a)$  as  $\Delta=\Delta_0G(a)$. This is
very well approximated by the expression
\begin{equation} G(a)\approx 
\exp\left\{ \int_1^a
\frac{da'}{a'}\left[\Omega_M(a')^\gamma\right] \right\} \label{def_gf}
\end{equation}
and
\begin{equation}
f_g\equiv \frac{d\log G}{d\log a}\approx \Omega_M(a)^\gamma\label{def-growth-rate}
\end{equation}
defines the growth rate and the growth index  $\gamma$ that is found to be $\gamma_{\Lambda}\simeq 0.545$
for the $\Lambda$CDM solution, \citep[see][]{wang98, Linder:2005in,
Huterer:2006mva,Ferreira:2010sz}.

Clearly, if the actual theory of structure growth is not the $\Lambda$CDM
scenario, the constraints (\ref{lcdm_eeq}) will be modified, the growth equation
(\ref{lcdm_geq}) will be different, and finally the growth factor (\ref{def_gf})
is changed, i.e.~the growth index is different from $\gamma_\Lambda$ and may
become time and scale dependent. Therefore, the inconsistency of these three
points of view can be used to test the $\Lambda$CDM paradigm.

\subsubsection{ Two new degrees of freedom}

Any generic modification of the dynamics of scalar perturbations with respect to
the simple scenario of a smooth dark energy component that only alters the
background evolution of $\Lambda$CDM can be represented by introducing two new
degrees of freedom in the Einstein constraint equations. We do this by replacing
(\ref{lcdm_eeq}) with
\begin{equation}
  -k^2 \Phi = 4\pi G Q(a,k) a^2 \rho_M \Delta_M \ ,\qquad\qquad 
\Phi=\eta(a,k)\Psi \,.
  \label{mod_constr}
\end{equation}
Non-trivial behaviour of the two functions $Q$ and $\eta$ can be due to a
clustering dark energy component or some modification to GR. In MG models the
function $Q(a,k)$ represents a mass screening effect due to local modifications
of gravity and effectively modifies Newton's constant. In dynamical DE models $Q$
represents the additional clustering due to the perturbations in the DE. On the
other hand, the function $\eta(a,k)$ parameterises the effective anisotropic
stress introduced by MG or DE, which is absent in $\Lambda$CDM.

Given a MG or DE theory, the scale- and time-dependence of the functions $Q$ and
$\eta$ can be derived and predictions projected into the $(Q,\eta)$ plane. This
is also true for interacting dark sector models, although in this case the
identification of the total matter density contrast (DM plus baryonic matter)
and the galaxy bias become somewhat contrived, \citep[see e.g.][for
an overview of predictions for different MG/DE models]{Song:2010rm}.

Using the above defined modified constraint equations (\ref{mod_constr}), the
conservation equations of matter perturbations can be expressed in the following
form (see \cite{Pogosian:2010tj})
\begin{eqnarray}
  \Delta_M' &=& -\frac{1/\eta-1+(\ln Q)'}{x_Q^2+\frac{9}{2}\Omega_M}\,
\frac{9}{2}\Omega_M \Delta_M
    - \frac{x_Q^2-3(\ln H)'/Q}{x_Q^2+\frac{9}{2}\Omega_M}\, \frac{\theta_M}{aH}
\nonumber \\
  \theta_M' &=& -\theta_M - \frac{3}{2}aH\Omega_M
\frac{Q}{\eta}\Delta_M
\end{eqnarray}
where we define $x_Q\equiv k/(aH\sqrt{Q})$. Remember $\Omega_M=\Omega_M(a)$ as defined
above. Notice that it is $Q/\eta$ which modifies the source term of the
$\theta_M$ equation and therefore also the growth of $\Delta_M$. Together with
the modified Einstein constraints (\ref{mod_constr}) these evolution equations
form a closed system for $(\Delta_M,\theta_M,\Phi,\Psi)$ which can be solved for
given $(Q,\eta)$.

The influence of the Hubble scale is modified by $Q$, such that now the size of
$x_Q$ determines the behaviour of $\Delta_M$; on ``sub-Hubble'' scales, $x_Q\gg
1$, we find
\begin{equation}
  \Delta_M''+\left[2+(\ln H)'\right]\Delta_M' = \frac{3}{2}\Omega_M(a)
\frac{Q}{\eta} \Delta_M \,.
\end{equation}
and $\theta_M=-aH\Delta_M'$. The growth equation is only modified by the factor
$Q/\eta$ on the RHS with respect to $\Lambda$CDM (\ref{lcdm_geq}). On
``super-Hubble'' scales, $x_Q\ll 1$, we have
\begin{eqnarray}
  \Delta_M' &=& -\left[1/\eta-1+(\ln Q)'\right] \Delta_M + \frac{2}{3\Omega_M}\frac{(\ln
H)'}{aH}\frac{1}{Q} \theta_M
\nonumber \\
  \theta_M' &=& -\theta_M - \frac{3}{2}\Omega_M\,aH
\frac{Q}{\eta}\Delta_M \,.
\end{eqnarray}
$Q$ and $\eta$ now create an additional drag term in the $\Delta_M$ equation,
except if $\eta>1$ when the drag term could flip sign. \cite{Pogosian:2010tj}
also showed that the metric potentials evolve independently and
scale-invariantly on super-Hubble scales as long as $x_Q\to 0$ for $k \to 0$.
This is needed for the comoving curvature perturbation, $\zeta$,  to be constant
on super-Hubble scales.

Many different names and combinations of the above defined functions $(Q,\eta)$
have been used in the literature, some of which are more closely related to
actual observables and are less correlated than others in certain situations,
\citep[see e.g.][]{Amendola:2007rr, Mota:2007sz,Song:2010rm, Pogosian:2010tj, Daniel:2010ky,
Daniel:2010yt, Ferreira:2010sz}.

For instance, as observed above, the combination $Q/\eta$ modifies the source
term in the growth equation. Moreover, peculiar velocities are following
gradients of the Newtonian potential, $\Psi$, and therefore the comparison of
peculiar velocities with the density field is also sensitive to $Q/\eta$. So we
define
\begin{equation}\label{eq:muQeta}
  \mu \equiv Q\,/\,\eta \qquad \Rightarrow \qquad
  -k^2 \Psi = 4\pi G a^2 \mu(a,k) \rho_M \Delta_M \,.
\end{equation}

Weak lensing and the ISW effect, on the other hand, are measuring
$(\Phi+\Psi)/2$ which is related to the density field via 
\begin{equation}
  \Sigma \equiv \frac{1}{2}Q(1+1/\eta) = \frac{1}{2}\mu(\eta+1)
   \qquad \Rightarrow \qquad
  -k^2 (\Phi+\Psi) = 8\pi G a^2 \Sigma(a,k) \rho_M \Delta_M \,.
\label{Sigma-wl-isw}
\end{equation}
A summary of different other variables used was given by \cite{Daniel:2010ky}.
For instance, the gravitational slip parameter introduced by \label{symbol:varpi}
\cite{Caldwell:2007cw} and widely used is related through $\varpi\equiv
1/\eta-1$. Recently \cite{Daniel:2010yt} used $\{{\cal G}\equiv\Sigma,\
\mu\equiv Q,\ {\cal V}\equiv\mu\}$, while \cite{Bean:2010zq} defined
$R\equiv1/\eta$. All these variables reflect the same two degrees of freedom
additional to linear growth of structure in $\Lambda$CDM. 

Any combination of two variables out of $\{Q,\eta,\mu,\Sigma,\ldots\}$ is a
valid alternative to $(Q,\eta)$. It turns out that the pair $(\mu,\Sigma)$ is
particularly well suited when CMB, WL and LSS data are combined as it is less
correlated than others, \citep[see][]{Zhao:2010dz, Daniel:2010yt,Axelsson:2011gt}.

\subsubsection{Parameterisations and non-parametric approaches}

So far we defined two free functions which can encode any departure of the
growth of linear perturbations from $\Lambda$CDM. However, these free
functions are not measurable but have to be inferred via their impact on the
observables. Therefore, one needs to specify a parameterisation of e.g.
$(Q,\eta)$ such that departures from $\Lambda$CDM can be quantified.
Alternatively, one can use non-parametric approaches to infer the time- and
scale-dependence of the modified growth functions from the observations.

Ideally, such a parameterisation should be able to capture all relevant physics
with the least number of parameters. Useful parameterisations can be motivated
by predictions for specific theories of MG/DE, \citep[see][]{Song:2010rm}, and/or by
pure simplicity and measurability, \citep[see][]{Amendola:2007rr}. For instance,
\cite{Zhao:2010dz} and \cite{Daniel:2010ky} use scale-independent
parameterisations which model one or two smooth transitions of the modified
growth parameters as a function of redshift. \cite{Bean:2010zq} also add a
scale-dependence to their parameterisation while keeping the time-dependence a simple power law:
\begin{eqnarray}
  Q(a,k) &\equiv& 1 + \left[ Q_0e^{-k/k_c} +
Q_\infty(1-e^{-k/k_c})-1\right]\,a^s
\nonumber \\
  \eta(a,k)^{-1} &\equiv& 1 + \left[ R_0e^{-k/k_c} +
R_\infty(1-e^{-k/k_c})-1\right]\,a^s
\end{eqnarray}
with constant $Q_0$, $Q_\infty$, $R_0$, $R_\infty$, $s$ and $k_c$. Generally,
the problem with any kind of parameterisation is that it is difficult---if not impossible---for it to be flexible enough to describe all possible modifications.

\cite{Daniel:2010ky} and \cite{Daniel:2010yt} investigate the modified growth
parameters binned in $z$ and $k$. The functions are taken constant in each bin.
This approach is simple and only mildly dependent on the size and number of the
bins. However, the bins can be correlated and therefore the data might not be
used in the most efficient way with fixed bins. Slightly more sophisticated than simple binning is a principal component
analysis (PCA) of the binned (or pixelised) modified growth functions. In PCA
uncorrelated linear combinations of the original pixels are constructed. In the
limit of a large number of pixels the model dependence disappears. At the moment
however, computational cost limits the number of pixels to only a few.
\cite{Zhao:2009fn, Zhao:2010dz} employ a PCA in the $(\mu,\eta)$ plane and find
that the observables are more strongly sensitive to the scale-variation of the
modified growth parameters rather than the time-dependence and their average
values. This suggests that simple, monotonically or mildly varying
parameterisations as well as only time-dependent parameterisations are poorly
suited to detect departures from $\Lambda$CDM.


\subsubsection{Trigger relations}

A useful and widely popular trigger relation is the value of the growth index
$\gamma$ in $\Lambda$CDM. It turns out that the value of $\gamma$ can be fitted
also for simple DE models and sub-Hubble evolution in some MG models, \citep[see
e.g.][]{Linder:2005in, Huterer:2006mva, Linder:2007hg, Linder:2009kq,Nunes:2004wn, Ferreira:2010sz}. For
example, for a non-clustering perfect fluid DE model with equation of state
$w(z)$ the growth factor $G(a)$ given in (\ref{def_gf}) with the fitting formula
\begin{equation}\label{eq:growthdef}
  \gamma = 0.55 + 0.05\left[1+w(z=1)\right]
\end{equation}
is accurate to the $10^{-3}$ level compared with the actual solution of the
growth equation (\ref{lcdm_geq}). Generally, for a given solution of the growth
equation the growth index can simply be computed using
\begin{equation}
  \gamma(a,k) = \frac{\ln(\Delta_M')-\ln\Delta_M}{\ln \Omega_M(a)} \,.
  \label{gamma_Delta}
\end{equation}
The other way round, the modified gravity function $\mu$ can be computed for a
given $\gamma$
\begin{equation}
  \mu = \frac{2}{3}\Omega_M^{\gamma-1}(a) \left[ 
  \Omega_M^\gamma(a) +2 +(\ln H)' -3\gamma +\gamma'\ln\gamma \right]
\end{equation}
\citep{Pogosian:2010tj}.

The fact that the value of $\gamma$ is quite stable in most DE models but
strongly differs in MG scenarios means that a large deviation from
$\gamma_\Lambda$ signifies the breakdown of GR, a substantial DE clustering
or a breakdown of another fundamental hypothesis like near-homogeneity. 
Furthermore, using the
growth factor to describe the evolution of linear structure is a very simple and
computationally cheap way to carry out forecasts and compare theory with data.
However, several drawbacks of this approach can be identified:
\begin{itemize}
\item As only one additional parameter is introduced, a second parameter, such as
$\eta$, is needed to close the system and be general enough to capture all
possible modifications.

\item The growth factor is a solution of the growth equation on sub-Hubble
scales and, therefore, is not general enough to be consistent on all scales.

\item The framework is designed to describe the evolution of the matter density
contrast and is not easily extended to describe all other energy-momentum
components and integrated into a CMB-Boltzmann code.
\end{itemize}

%% file: de_mg/intro-models.tex
In this section we review a number of popular models of dynamical DE and MG. This section is  more
technical than the rest and it is meant to provide a quick but self-contained
review of the current research in the 
theoretical foundations
of DE models. The selection of models is of course somewhat arbitrary
but we tried
to cover the most well-studied cases and those that introduce new and
interesting observable phenomena.

\subsection{Quintessence}\label{quintessence}

In this review we refer to scalar field models with canonical kinetic energy 
in Einstein's gravity as ``quintessence models". 
Scalar fields are  obvious candidates for dark energy, as they are for the inflaton, for many reasons: they are the simplest fields since {\color{red}they} lack internal degrees of freedom,  do not introduce
preferred directions, are typically weakly clustered (as discussed later on), and can easily drive an accelerated expansion.
If the kinetic energy has a canonical form, the only degree
of freedom is then provided by the field potential (and of course by the initial conditions).
The typical requirement is that the potentials are flat
enough to lead to the slow-roll inflation today with an energy scale
$\rho_{{\rm DE}}\simeq10^{-123}m_{{\rm pl}}^{4}$ and a mass scale
$m_{\phi}\lesssim10^{-33}$\,eV. 

Quintessence models are the protoypical DE models \citep{1998PhRvL..80.1582C}
and as such are the most studied ones. Since they have been
explored in many reviews of DE,
 we limit ourselves here to a few remarks \footnote{This subsection is based on   \cite{amendola_book}}.

 The quintessence model is  described by
the action \begin{eqnarray}
S=\int{\rm d}^{4}x\sqrt{-g}\,\left[\frac{1}{2\kappa^{2}}R+{\cal L}_{\phi}\right]+S_{M}\,,\qquad{\cal L}_{\phi}=-\frac{1}{2}g^{\mu\nu}\partial_{\mu}\phi\partial_{\nu}\phi-V(\phi)\,,\label{action}\end{eqnarray}
 where $\kappa^{2}=8\pi G$ and $R$ is the Ricci scalar and  $S_{M}$ is the matter action.
The fluid satisfies the continuity equation
 \begin{eqnarray}
\dot{\rho}_{M}+3H(\rho_{M}+p_{M})=0\,.\label{rhomeq}\end{eqnarray}
 The energy-momentum tensor of quintessence
is \begin{eqnarray}
T_{\mu\nu}^{(\phi)} & = & -\frac{2}{\sqrt{-g}}\frac{\delta(\sqrt{-g}{\cal L}_{\phi})}{\delta g^{\mu\nu}}\\
 & = & \partial_{\mu}\phi\partial_{\nu}\phi-g_{\mu\nu}\left[\frac{1}{2}g^{\alpha\beta}\partial_{\alpha}\phi\partial_{\beta}\phi+V(\phi)\right]\,.\end{eqnarray}
As we have already seen, in a FLRW background, the energy density $\rho_{\phi}$
and the pressure $p_{\phi}$ of the field are \begin{eqnarray}
\rho_{\phi}=-{T_{0}^{0}}^{(\phi)}=\frac{1}{2}\dot{\phi}^{2}+V(\phi)\,,\quad p_{\phi}=\frac{1}{3}{T_{i}^{i}}^{(\phi)}=\frac{1}{2}\dot{\phi}^{2}-V(\phi)\,,\end{eqnarray}
 which give the equation of state \begin{eqnarray}
w_{\phi}\equiv\frac{p_{\phi}}{\rho_{\phi}}=\frac{\dot{\phi}^{2}-2V(\phi)}{\dot{\phi}^{2}+2V(\phi)}\,.\label{wphiqui}\end{eqnarray}
 {}In the flat universe, Einstein's equations give
 the following equations of motion: \begin{eqnarray}
 &  & H^{2}=\frac{\kappa^{2}}{3}\left[\frac{1}{2}\dot{\phi}^{2}+V(\phi)+\rho_{M}\right]\,,\label{Heq1}\\
 &  & \dot{H}=-\frac{\kappa^{2}}{2}\left(\dot{\phi}^{2}+\rho_{M}+p_{M}\right)\,,\label{Heq2}\end{eqnarray}
 where $\kappa^{2}=8\pi G$. The variation of the action (\ref{action})
with respect to $\phi$ gives \begin{eqnarray}
\ddot{\phi}+3H\dot{\phi}+V_{,\phi}=0\,,\label{Klein}\end{eqnarray}
 where $V_{,\phi}\equiv{\rm d}V/{\rm d}\phi$. 

During radiation or matter dominated epochs, the energy density $\rho_{M}$
of the fluid dominates over that of quintessence, i.e. $\rho_{M}\gg\rho_{\phi}$.
If the potential is steep so that the condition $\dot{\phi}^{2}/2\gg V(\phi)$
is always satisfied, the field equation of state is given by $w_{\phi}\simeq1$
from Eq.~(\ref{wphiqui}). In this case the energy density of the
field evolves as $\rho_{\phi}\propto a^{-6}$, which decreases much
faster than the background fluid density.

The condition $w_{\phi}<-1/3$ is required to realize the late-time
cosmic acceleration, which translates into the condition $\dot{\phi}^{2}<V(\phi)$.
Hence the scalar potential needs to be shallow enough for the field
to evolve slowly along the potential. This situation is similar to
that in inflationary cosmology and it is convenient to introduce the
following slow-roll parameters \citep{2006RvMP...78..537B}
\begin{eqnarray}
\epsilon_{s}\equiv\frac{1}{2\kappa^{2}}\left(\frac{V_{,\phi}}{V}\right)^{2}\,,\qquad\eta_{s}\equiv\frac{V_{,\phi\phi}}{\kappa^{2}V}\,.\label{slowroll}\end{eqnarray}
 If the conditions $\epsilon_{s}\ll1$ and $|\eta_{s}|\ll1$ are satisfied,
the evolution of the field is sufficiently slow so that $\dot{\phi}^{2}\ll V(\phi)$
and $|\ddot{\phi}|\ll|3H\dot{\phi}|$ in Eqs.~(\ref{Heq1}) and (\ref{Klein}).

{}From Eq.~(\ref{Klein}) the deviation of $w_{\phi}$ from $-1$
is given by \begin{eqnarray} \label{eq:pot_w}
1+w_{\phi}=\frac{V_{,\phi}^{2}}{9H^{2}(\xi_{s}+1)^{2}\rho_{\phi}}\,,\end{eqnarray}
 where $\xi_{s}\equiv\ddot{\phi}/(3H\dot{\phi})$. This shows that
$w_{\phi}$ is always larger than $-1$ for a positive potential and energy density.
In the slow-roll limit, $|\xi_{s}|\ll1$ and $\dot{\phi}^{2}/2\ll V(\phi)$,
we obtain $1+w_{\phi}\simeq2\epsilon_{s}/3$ by neglecting the matter
fluid in Eq.~(\ref{Heq1}), i.e. $3H^{2}\simeq\kappa^{2}V(\phi)$.
The deviation of $w_{\phi}$ from $-1$ is characterized by the slow-roll
parameter $\epsilon_{s}$. It is also possible to consider Eq.~(\ref{eq:pot_w})
as a prescription for the evolution of the potential given $w_\phi(z)$ and to
reconstruct a potential that gives a desired evolution of the equation of
state (subject to $w\in[-1,1]$). This was used for example in \cite{Bassett:2002qu}.

However, in order to study the evolution of the perturbations of a quintessence field
it is not even necessary to compute the field evolution explicitly. Rewriting the perturbation
equations of the field in terms of the perturbations of the density contrast $\delta_\phi$ and
the velocity $\theta_\phi$ in the conformal Newtonian gauge, one finds \cite[see e.g.][Appendix A]{Kunz:2006wc} that they
correspond precisely to those of a fluid, (\ref{d_pert}) and (\ref{t_pert}), with $\pi=0$ and 
\mbox{$\delta p = c_s^2 \delta\rho + 3 a H (c_s^2-c_a^2) (1+w) \rho \theta/k^2$} with $c_s^2=1$. {\color{red}The adiabatic sound speed, $c_a$, is defined in eq.(\ref{c_a_def}).}
The large value of the sound speed $c_s^2$, equal to the speed of light, means that
quintessence models do not cluster significantly inside the horizon \citep[see][and Sec.~\ref{soundspeed} for a detailed analytical discussion of quintessence clustering
and its detectability with future probes, for arbitrary $c_s^2$]{Sapone:2009,Sapone:2010}.

Many quintessence potentials have been proposed in the literature. A simple
crude classification dives them into two classes, (i) {}``freezing models''
and (ii) {}``thawing'' models
\citep{2005PhRvL..95n1301C}. In the class (i) the field was rolling along the
potential in the past, but the movement gradually slows down after
the system enters the phase of cosmic acceleration. The representative
potentials that belong to this class are

\vspace{0.2cm}
 \underline{{\bf (i) Freezing models}} 
\begin{itemize}
\item $V(\phi)=M^{4+n}\phi^{-n}\quad(n>0)$\,, 
\item $V(\phi)=M^{4+n}\phi^{-n}\exp(\alpha\phi^{2}/m_{{\rm pl}}^{2})$\,. 
\end{itemize}
The former potential does not possess a minimum and hence the field
rolls down the potential toward infinity.
This appears, for example, in the fermion condensate model as a dynamical
supersymmetry breaking \citep{1999PhRvD..60f3502B}. The latter potential has
a minimum at which the field is eventually trapped (corresponding
to $w_{\phi}=-1$). This potential can be constructed in the framework
of supergravity \citep{1999PhLB..468...40B}.

\vspace{0.3cm}
In thawing models (ii) the field (with mass $m_{\phi}$) has been frozen
by Hubble friction (i.e. the term $H\dot{\phi}$ in eq.(\ref{Klein})) until recently and
then it begins to evolve once $H$ drops below $m_{\phi}$. The equation
of state of DE is $w_{\phi}\simeq-1$ at early times, which
is followed by the growth of $w_{\phi}$. The representative potentials
that belong to this class are

\vspace{0.2cm}
 \underline{{\bf (ii) Thawing models}} 
\begin{itemize}
\item $V(\phi)=V_{0}+M^{4-n}\phi^{n}\quad(n>0)$\,, 
\item $V(\phi)=M^{4}\cos^{2}(\phi/f)$\,. 
\end{itemize}
The former potential is similar to the one of chaotic inflation ($n=2,4$)
used in the early universe (with $V_{0}=0)$ \citep{1983PhLB..129..177L},
while the mass scale $M$ is very different. The model with
$n=1$ was  proposed 
by \cite{2003JCAP...10..015K} in connection with the possibility to allow
for negative values of $V(\phi)$. The universe will collapse in the
future if the system enters the region with $V(\phi)<0$. The latter
potential appears as a potential for the Pseudo-Nambu-Goldstone Boson
(PNGB)\index{Pseudo-Nambu-Goldstone Boson (PNGB)}. This was introduced
by  \cite{1995PhRvL..75.2077F} in response to the first
tentative suggestions that the universe may be dominated by the cosmological
constant. In this model the field is nearly frozen at the potential
maximum during the period in which the field mass $m_{\phi}$ is smaller
than $H$, but it begins to roll down around the present ($m_{\phi}\simeq H_{0}$).

Potentials can also be classified in several other ways, e.g. on the basis
of the existence of special solutions. For instance,
tracker solutions have approximately constant $w_{\phi}$ and $\Omega_{\phi}$ along
 special attractors. A wide range of initial conditions converge
to a common, cosmic evolutionary tracker. Early DE models
contain instead solutions in which DE was not negligible even during the last scattering.
{\color{red}While in the specific Euclid forecasts section we will not explicitly consider these models, it is worth to note that 
the combination of  observations of the CMB and of large scale structure (such as Euclid) can dramatically constrain these models  drastically improving the  inverse area figure of merit compared to current constraints, as discussed in \cite{Huterer:2006mv}.}

{\color{red}\subsection{K-essence}
In a quintessence model it is the potential energy of a scalar field which leads to the late time acceleration of the expansion of the universe; the alternative where it is the kinetic energy of the scalar field which dominates is known as k-essence.  Models of k-essence are characterized by an action for the scalar field of the following form
\begin{equation}
S=\int d^4 x\;\sqrt{-g}p(\phi,X)\;,
\end{equation}
where $X=(1/2)g^{\mu\nu}\nabla_{\mu}\phi\nabla_{\nu}\phi$.  The energy density of the scalar field is given by 
\begin{equation}
\rho_{\phi}=2X\frac{dp}{dX}-p\;,
\end{equation}
and the pressure is simply $p_{\phi}=p(\phi,X)$.  Treating the k-essence scalar as  perfect fluid, this means that k-essence has equation of state
\begin{equation}
w_{\phi}=\frac{p_{\phi}}{\rho_{\phi}}=-\frac{p}{p-2X p,_{X}}\;,
\end{equation}
where the subscript $,_{X}$ indicates a derivative with respect to $X$.  Clearly with a suitably chosen
$p$ the scalar can have an appropriate equation of state to allow it to act as dark energy

The dynamics of the k-essence field are given by a continuity equation
\begin{equation}
\dot{\rho}_{\phi}=-3H(\rho_{\phi}+p_{\phi})\;,
\end{equation}
or equivalently by the scalar equation of motion
\begin{equation}
G^{\mu\nu}\nabla_{\mu}\nabla_{\nu}\phi+2X\frac{\partial^2p}{\partial X \partial \phi}-\frac{\partial p}{\partial \phi}=0\;,
\end{equation}
where
\begin{equation}
G^{\mu\nu}=\frac{\partial p}{\partial X}g^{\mu\nu}+\frac{\partial^2 p}{\partial X^2}\nabla^{\mu}\phi\nabla^{\nu}\phi\;.
\end{equation}
For this second order equation of motion to be hyperbolic, and hence physically meaningful, we must impose
\begin{equation}
1+2X\frac{p,_{XX}}{p,_X}>0\;.
\end{equation}
k-essence was first proposed by \cite{Armendariz-PiconMukhanovSteinhardt2000, 2001PhRvD..63j3510A}, where it was also shown that tracking solutions to this equation of motion, which are attractors in the space of solutions, exist during the radiation and matter dominated eras for k-essence in a similar manner to quintessence.

The speed of sound for k-essence fluctuation is 
\begin{equation}
c_s^2 =\frac{p,_{X}}{p,_X+2Xp,_{XX}}\;.
\end{equation}
So that whenever the kinetic terms for the scalar field are not linear in $X$, the speed of sound of fluctuations differs from unity.  It might appear concerning that superluminal fluctuations are allowed in k-essence models, however it was shown in \cite{Babichev:2007dw} that this does not lead to any causal paradoxes.}

%% file: de_mg/defmodgrav.tex
\subsection{A definition of modified gravity}

In this review we often make reference to DE and MG models. Although in an increasing number of publications a similar dichotomy
is employed, there is currently no consensus on where to draw the line between the
two classes. Here we will introduce an operational definition for
the purpose of this document.

Roughly speaking, what most people have in mind when talking about
standard dark energy are models of minimally coupled scalar fields with standard kinetic
energy in 4-dimensional Einstein gravity, the only
functional degree of freedom being the scalar potential.
Often, this class of model is refered to simply as ``quintessence''.
When we depart from
this picture, however, a simple classification is not easy to draw.
One problem is that, as we have seen in the previous chapters, both at
background
and at the perturbation level, different models can have the same observational
signatures \citep{Kunz:2006ca}. This problem is not due to the use of perturbation theory:
any modification to Einstein's equations can be interpreted as
standard Einstein gravity
with a modified ``matter'' source, containing an arbitrary mixture of
scalars, vectors and tensors \citep{Hu:2007pj,Kunz:2008wt}.

The simplest example can be discussed by looking at Eqs.
(\ref{mod_constr}). One can
modify gravity and obtain a modified Poisson equation, and therefore
$Q\ne 1$, or one can
introduce a clustering dark energy (for example a $K$-essence model with
small sound speed) that also induces the same $Q\ne
1$ (see Eq.   \ref{mod_constr}). This extends to the anisotropic stress
$\eta$: there is in general a one-to-one relation at first order
between a fluid with arbitrary equation of state, sound speed,
and anisotropic stress and a modification of the Einstein-Hilbert Lagrangian.

We could therefore simply abandon any attempt to distinguish between DE and MG, and just analyse different models, comparing their properties and phenomenology.
There is however a possible classification that helps us setting targets
for the observations, which is often useful to communicate in few words
the results of complex arguments.  In this review, we will use the following
notation: 
\begin{itemize}
\item {\em Standard dark energy:} These are models in which dark energy lives in 
standard Einstein gravity {\it and} does not
cluster appreciably on sub-horizon scales. As already noted, the
prime example of a standard dark energy model is a
minimally coupled scalar field with standard kinetic energy, for which
the sound speed
equals the speed of light. 
\item {\em Clustering dark energy:} 
In clustering dark energy models, there
is an additional contribution to the Poisson equation due to the 
dark energy perturbation, which induces $Q \neq 1$. However, in this class 
we require $\eta=1$,
i.e. no extra effective anisotropic stress is induced by the extra dark component.
A typical example is a $K$-essence model with a low
sound speed, $c_s^2\ll 1$. 
\item {\em Explicit modified gravity models:} 
These are models where from the start the Einstein equations
are modified, for example scalar-tensor and $f(R)$ type theories, DGP as
well as interacting dark energy, in which effectively a fifth force is
introduced in addition to gravity. Generically they change the clustering and/or induce
a non-zero anisotropic stress. Since our definitions are based
on the phenomenological parameters, we also add dark energy models that 
live in Einstein's gravity
but that have
non-vanishing anisotropic stress into this class since they cannot be distinguished
by cosmological observations.
\end{itemize}

Notice that both clustering dark energy and explicit modified gravity models
lead to deviations from what is often called `General Relativity' (or, like here, standard dark energy) in the literature when
constraining extra perturbation parameters like the growth index $\gamma$.
For this reason we generically call both of these classes MG models. 
In other words, in this review we use the simple and by now extremely
popular (although admittedly somewhat
misleading) expression ``modified gravity'' to denote models in which
gravity is modified and/or dark energy clusters or interacts with other fields.
Whenever we feel useful, we will  remind the reader of the actual meaning of the
expression ``modified gravity'' in this review.

On sub-horizon scales and at first order in perturbation theory, therefore,
our definition of MG is
straightforward: models with $Q=\eta=1$ (see Eq.   \ref{mod_constr}) are standard DE,
otherwise they are MG models. In this sense the definition above
is rather convenient: we can use it to quantify, for instance, how well
Euclid will distinguish between standard dynamical dark energy and
modified gravity
by forecasting the errors on $Q,\eta$ or on related quantities like
the growth index $\gamma$.

On the other hand, it is clear that this definition is only a
practical way to group
different models and should not be taken as a fundamental one. We do not try to
set a precise threshold on, for instance, how much dark energy should
cluster before we call it
modified gravity: the boundary between the classes is therefore left
undetermined but we think
this will not harm the understanding of this document.

%% file: de_mg/coupled.tex
\subsection{Coupled dark energy models} \label{mg:cde}

A first class of models in which Dark Energy shows dynamics, in
connection to the presence of a fifth force different from gravity, is the
case of `interacting Dark Energy': we consider the
possibility that Dark Energy, seen as a dynamical scalar field, may
interact with other components in the Universe. This class of models
effectively enters in the ``explicit modified gravity models" in the
classification above, because the gravitational attraction between dark matter particles is modified by the presence of a fifth force. However, we note that the anisotropic stress for DE is still zero in the Einstein frame while is in general non zero in the Jordan frame. In some cases (when a universal coupling is present) such an interaction can be explicitly recast in a non-minimal coupling to gravity, after a redefinition of the metric and matter fields (Weyl scaling).
We would like to identify
whether interactions (couplings) of dark energy with matter fields, neutrinos or
gravity itself can affect the universe in an observable way.

In this subsection we give a general description of the following main interacting
scenarios:
\begin{enumerate}
 \item  couplings between dark energy and baryons;
 \item \label{de_dm}  couplings between dark energy and dark matter (coupled
quintessence);
 \item \label{de_nu} couplings between dark energy and neutrinos (growing
neutrinos, MaVaNs);
 \item  universal couplings with all species (scalar-tensor theories and $f(R)$). 
\end{enumerate}
In all these cosmologies the coupling introduces a fifth force, in addition to standard
gravitational attraction. 
The presence of a new force, mediated by the DE scalar field (sometimes called
the `cosmon' \cite{Wetterich_1988}, seen as the mediator of a cosmological interaction) has several
implications and can significantly modify the process of structure formation.
We will discuss cases (\ref{de_dm}) and (\ref{de_nu}) also in 
section \ref{dark-matter}.

In these scenarios the presence of the additional interaction couples the evolution of components 
that in the standard $\Lambda$-FLRW would evolve independently. 
The stress-energy tensor
${T{^{\mu}}}_{\nu}$ of each species is in general not conserved - only
the total stress-energy tensor is. 
Usually, at the level of the Lagrangian, the coupling is introduced by allowing
the mass $m$ of matter fields to depend on a scalar field $\phi$ via a function
$m(\phi)$ whose choice specifies the interaction.\label{symbol:phi}
This wide class of cosmological models can be described by the following action:
\begin{equation}
  \label{mg:cde:action} {\cal S} = \int{d^4x \sqrt{-g} \left[-\frac{1}{2}\partial^\mu 
\phi \partial_\mu \phi - U(\phi) - m(\phi)\bar{\psi}\psi + {\cal
L}_{kin}[\psi]\right] }
\end{equation}
where $U(\phi)$ is the potential in which the scalar field $\phi$ rolls, $\psi$
describes matter fields, and $g$ is defined in the usual way as the determinant
of the metric tensor, whose background expression is $g_{\mu \nu} = \mathrm{diag}[-a^2,
a^2, a^2, a^2]$.

For a general treatment of background and perturbation equations we refer to
\cite{Kodama:1985bj, Amendola:1999er, Amendola:2003wa, Pettorino:2008ez}. Here
the coupling of the dark energy scalar field to a generic matter component
({\color{red} denoted by index} $\alpha$) is treated as an external source $Q_{(\alpha)\mu}$ in the Bianchi
identities:
\begin{equation}
\nabla_{\nu}T_{(\alpha)\mu}^{\nu}=Q_{(\alpha)\mu}\,,\label{tensor_conserv_alpha}
\end{equation}
with the constraint \begin{equation}
\sum_{\alpha}Q_{(\alpha)\mu}=0\label{Q_conserv_total} ~~~.\end{equation}

The zero component of (\ref{tensor_conserv_alpha}) gives the background conservation equations:
\begin{eqnarray}
 \label{cons_phi} \frac{d\rho_{\phi}}{d\eta} = -3 {\cal H} (1 + w_\phi) \rho_{\phi} +
\beta(\phi) \frac {d \phi}{d\eta} (1-3 w_{\alpha}) \rho_{\alpha} ~~~, \\
\label{cons_gr} \frac{ d \rho_{\alpha}}{d\eta} = -3 {\cal H} (1 + w_{\alpha}) \rho_{\alpha} -
\beta(\phi) \frac{d\phi}{d\eta} (1-3 w_{\alpha}) \rho_{\alpha}
~~~,
\end{eqnarray}
{\color{red} for a scalar field $\phi$ coupled to one single fluid $\alpha$ with a function $\beta(\phi)$ which in general may not be constant.} 
The choice of the mass function $m(\phi)$ corresponds to a choice of
$\beta(\phi)$ and equivalently to a choice of the source $Q_{(\alpha)\mu}$ and
specifies the strength of the coupling according to the following relations:
\begin{equation} \label{mass_def}
Q_{(\phi)\mu}=\frac{\partial\ln{m(\phi)}}{\partial\phi}
T_{\alpha}\,\partial_{\mu}\phi ~~~~~,~~~~~~
m_\alpha=\bar{m}_\alpha ~ e^{-{\beta(\phi)}{\phi}}
\end{equation}
{\color{red} where $\bar{m}_\alpha$ is the constant Jordan-frame bare mass.}
The evolution of dark energy is related to the trace
$T_{\alpha}$ and, as a consequence, 
to density and pressure of the species $\alpha$. We note that a description of
the coupling via an action such as (\ref{mg:cde:action}) 
is originally motivated by the wish to modify general relativity with extension
such as scalar-tensor theories.
In general, one of more couplings can be active \citep{Brookfield:2007au}.

As for perturbation equations, it is possible to include the
coupling in a modified Euler equation: 
\begin{eqnarray}
& &\frac {d{\bf{{v}_\alpha}}}{d\eta} + \left({\cal H} - {\beta(\phi)} \frac{d \phi}{d\eta} \right) {\bf
{v}_\alpha} - {\bf{\nabla }} \left[\Phi_\alpha + \beta \phi \right] = 0
\,.
\end{eqnarray}
The Euler equation in cosmic time (${\rm d}t = a\, {\rm d}\tau$) can also be
rewritten in the form of an acceleration equation for
particles at position ${\bf{r}}$:
\begin{equation}
\label{CQ_euler}
\dot{{\bf{v}}}_{\alpha} = -\tilde{H}{\bf{v}}_{\alpha} - {\bf{\nabla
}}\frac{\tilde{G}_{\alpha}{m}_{\alpha}}{r} \,.
\end{equation}
The latter expression contains explicitely all main ingredients which affect
dark energy interactions:
\begin{enumerate}
 \item a fifth force ${\bf{\nabla }} \left[\Phi_\alpha + \beta \phi
\right]$ with an effective $\tilde{G}_{\alpha} = G_{N}[1+2\beta^2(\phi)]$ ;
\item a velocity dependent term $\tilde{H}{\bf{v}}_{\alpha} \equiv H \left(1 -
{\beta(\phi)} \frac{\dot{\phi}}{H}\right) {\bf{v}}_{\alpha}$
\item a time-dependent mass for each particle $\alpha$, evolving according to
(\ref{mass_def}).
\end{enumerate} 

The relative {\color{red} significance} of these key ingredients can lead to a variety of
potentially observable effects, expecially on structure formation. We will recall some of
them in the following subsections as well as, in more detail, for two specific
couplings in the Dark Matter Section (\ref{dms:de_dm}, \ref{dms:de_nu}) of this
report. 

\subsubsection{Dark energy and baryons}
A coupling between dark energy and baryons is active when the baryon mass is a
function of the dark energy scalar field: $m_b = m_b(\phi)$. 
Such a coupling is
constrained to be very small: main bounds come from tests
of the equivalence principle and solar system constraints
\citep{Bertotti:2003rm}. 
More in general, depending on the coupling, bounds on the variation of fundamental constants over cosmological time-scales
may have to be considered (\cite{Marra:2005ma, Dent:2008gx, Dent:2008vd, Martins:2010gu} and references therein). 
It is presumably very difficult to have significant
cosmological effects due to a coupling to baryons only. Uncoupled baryons can
however still play a role in presence of a coupling to dark matter (see section (\ref{non-linear-aspects}) 
on non-linear aspects). 

\subsubsection{Dark energy and dark matter}
An interaction between dark energy and dark matter (CDM) is active when CDM mass
is a function of the dark energy scalar field: $m_c = m_c(\phi)$. In this case
the coupling is not affected by tests on the equivalence principle and solar
system constraints and can therefore be stronger than the one with baryons.
{\color{red} One may argue that dark matter particles are themselves coupled to baryons, which leads, through quantum
corrections, to direct coupling between dark energy and baryons. The strength of such couplings can still be small and was discussed in \cite{Dent:2008vd} for the case of neutrino - dark energy couplings. Also, quantum corrections are often recalled to spoil the flatness of a quintessence potential. 
However, it may be misleading to calculate quantum corrections up to a cutoff scale, as contributions above the cutoff can possibly compensate terms below the cutoff, as discussed in \cite{2008PhRvD..77j3505W}. 
}
Typical values of $\beta$ presently allowed by observations (within current CMB 
data) are within the range $0< \beta < 0.06$ (at 95\% CL for a constant coupling and an exponential potential) 
\citep{Bean:2008ac, amendola_etal_2003, Amendola:2003wa, amendola_quercellini_2003}, or possibly more 
\citep{LaVacca:2009yp, Kristiansen:2009yx} if neutrinos are taken into account or for
 more realistic time-dependent choices of the coupling. This 
framework is generally referred to as `coupled quintessence' (CQ).
Various choices of couplings have been investigated in literature, including
constant and varying $\beta(\phi)$ \citep{Amendola:1999er, Mangano:2002gg,
Amendola:2003wa, Koivisto:2005nr, Guo:2007zk, Quartin:2008px, quercellini_etal_2008, Pettorino:2008ez,Gannouji:2010fc}.

The presence of a coupling (and therefore, of a fifth force acting among dark
matter particles) modifies the background expansion and linear perturbations \citep{amendola1999, Amendola:1999er, Amendola:2003wa}, therefore affecting CMB and cross-correlation of CMB and LSS \citep{amendola_quercellini_2003, Amendola:2003wa, amendola_etal_2003, amendola_quercellini_2004, Bean:2008ac, LaVacca:2009yp, Kristiansen:2009yx, Xia:2009zzb, Mainini:2010ng, Amendola:2011ie}.

Furthermore, structure formation itself is modified \citep{maccio_etal_2004, Manera:2005ct, Koivisto:2005nr, Mainini:2006zj, Sutter:2007ky, Abdalla:2007rd,Mota:2008ne, Bertolami:2007tq, Wintergerst:2010ui, Baldi_etal_2010, Baldi:2010vv, Baldi_Pettorino_2010, Baldi:2010ks, Li:2010zw, Li:2010eu, Baldi:2010pq, Zhao:2010dz, Marulli:2011jk}.

An alternative approach, also investigated in literature \citep{Mangano:2002gg,Valiviita:2008iv,Valiviita:2009nu,Majerotto:2009np,
Gavela:2009cy,Gavela:2010tm,CalderaCabral:2008bx,Schaefer:2008ku,CalderaCabral:2009ja}, where the authors consider as a starting point 
eq.(\ref{tensor_conserv_alpha}): the coupling is then introduced by choosing directly a covariant stress 
energy tensor on the right hand side of this equation, treating dark energy as a fluid and 
in absence of a starting action. The advantage of this
approach is that a good parametrisation allows us to investigate several models of dark energy 
at the same time. Problems connected to instabilities of some parametrisations or to the definition
of a physically motivated speed of sound for the density fluctuations can be found in \cite{Valiviita:2008iv}.
It is also possible to both take a covariant form for the coupling and a
quintessence dark energy scalar field, starting again directly from eq.(\ref{tensor_conserv_alpha}). This has been done e.g. in \cite{Boehmer:2008av},
\cite{Boehmer:2009tk}.
At the background level only, \cite{Chimento:2003iea}, \cite{Chimento:2005xa}, \cite{delCampo:2006vv} and \cite{Olivares:2006jr} have also considered
which background constraints can be obtained when starting from a fixed present ratio of Dark Energy and Dark Matter. The disadvantage of this approach is that it is not clear how to perturb a coupling which has been defined as a background quantity. 

A Yukawa-like interaction was investigated \citep{Farrar:2003uw, Das:2006cc}, pointing out that coupled dark energy behaves as a fluid with an effective equation of state $w \lesssim -1$, though staying well-defined and without the presence of ghosts \citep{Das:2006cc}. 

For an illustration of observable effects related to dark energy - dark matter interaction see also section
(\ref{dms:de_dm}) of this report.

\subsubsection{Dark energy and neutrinos}
A coupling between dark energy and neutrinos can be even stronger than the one
with dark matter and as compared to gravitational strength. Typical values of
$\beta$ are order 50-100 or even more such that even the small fraction of
cosmic energy density in neutrinos can have a substantial influence on the time
evolution of the quintessence field. In this scenario neutrino masses change in
time, depending on the value of the dark energy scalar field $\phi$. 
Such a coupling has been investigated within MaVaNs \citep{Fardon:2003eh, Peccei:2004sz, Bi:2004ns, Afshordi:2005ym,
Weiner:2005ac, Das:2006ht, Takahashi:2006jt, Spitzer:2006hm, Bjaelde:2007ki, Brookfield:2005td, Brookfield:2005bz} and more recently within growing neutrino 
cosmologies \citep{Amendola2008b, Wetterich:2007kr, Mota:2008nj, Wintergerst:2009fh, Wintergerst:2010ui,
Pettorino:2010bv, Brouzakis:2010md, Baldi:2011es}.
In this latter case, DE properties are related to the neutrino mass and to a cosmological event, i.e. neutrinos becoming non-relativistic.
This leads to the formation of stable neutrino lumps \citep{Mota:2008nj, Wintergerst:2009fh,Baldi:2011es} at very large scales only ($\sim 100$ Mpc and beyond) as well as to signatures in the CMB spectra \citep{Pettorino:2010bv}. For an illustration of observable effects related to this case see section (\ref{dms:de_nu}) of this report.

\subsubsection{Scalar-tensor theories}
Scalar-tensor theories \citep{Wetterich_1988, Hwang:1990re, Hwang:1990jh, Damour_Gibbons_Gundlach_1990, casas_etal_1991, casas_etal_1992, Wetterich:1994bg,
Uzan:1999ch, Perrotta:1999am, Faraoni:2000wk, Boisseau:2000pr, Riazuelo:2001mg,
Perrotta:2002sw, Schimd:2004nq, Matarrese:2004xa, Pettorino:2004zt,
Pettorino:2005pv, Capozziello:2007iu, Appleby:2010dx} extend general relativity by introducing a non-minimal
coupling between a scalar field (acting also as dark energy) and the metric
tensor (gravity); they are also sometimes referred to as `extended
quintessence'. 
We include scalar-tensor theories among `interacting cosmologies' because, via a
Weyl transformation, they are equivalent to a general relativity framework
(minimal coupling to gravity) in which the dark energy scalar field $\phi$ is
coupled (universally) to all species \citep{Wetterich_1988, Maeda:1988ab, Wands:1993uu, EspositoFarese:2000ij, Pettorino:2008ez, Catena:2006bd}. In other words, these theories correspond
to the case where, in action (\ref{mg:cde:action}), the mass of all species
(baryons, dark matter, ...) is a function $m=m(\phi)$ with the same coupling for
every species $\alpha$. Indeed, a description of the coupling via an action such
as (\ref{mg:cde:action}) is originally motivated by extentions of general relativity
such as scalar-tensor theories.
Typically the strength of the scalar mediated interaction is required to be
orders of magnitude weaker than gravity (\cite{Lee:2010zy}, \cite{Pettorino:2004zt} and references therein for recent constraints).  
It is possible to tune this coupling
to be as small as is required---for example by choosing a suitably flat potential $V(\phi)$ for the scalar field. This leads back however to naturalness and fine-tuning problems.

In the next subsection we will discuss in more detail a number of ways in which new scalar degrees of freedom can
naturally couple to standard model fields, whilst still being in agreement with
observations. We mention here only that the presence of
  chameleon mechanisms \citep{Brax:2004qh, Mota:2010uy, Mota:2006fz,Brax:2008hh, Hui:2009kc,
Brax:2010kv, Davis:2011qf} can, for example, modify the coupling depending on the environment. In
this way, a small (screened) coupling in high density regions, in agreement with
observations, is still compatible with a bigger coupling ($\beta \sim 1$) active
in low density regions. In other words, a dynamical mechanism ensures that the effects of the coupling are
screened in laboratory and solar system tests of gravity.

Typical effects of scalar-tensor  theories on CMB and structure
formation include:
\begin{itemize}
 \item enhanced ISW \citep{Pettorino:2004zt, Giannantonio:2009zz, Zhao:2010dz}:;
 \item violation of the equivalence principle: extended objects such as galaxies
do not all fall at the same rate \citep{amendola_quercellini_2004, Hui:2009kc}.
\end{itemize}

It is important to remark, however, that screening mechanisms are meant to protect the scalar field in high density regions (and therefore allow for bigger couplings in low density environments) but they do not address problems related to self-acceleration of the DE scalar field, which still usually require some fine-tuning to match present observations on $w$.
So-called $f(R)$ theories, which can be mapped into a subclass of scalar-tensor theories, 
will be discussed in more detail in the next subsection.

%% file: de_mg/crossing.tex
\subsection{Phantom crossing\label{sec:crossing}}

In this section we pay attention to the evolution of the perturbations of a general 
dark energy fluid with an evolving equation of state parameter $w$.
Current limits on the equation of state parameter $w=p/\rho$
of the dark energy indicate that $p\approx -\rho$, and so do not
exclude $p<-\rho$, a region of parameter space often called
{\em phantom energy}. Even though the region for which $w<-1$ may be unphysical at the
quantum level, it is still important to probe
it, not least to test for coupled dark energy and alternative theories of gravity or
higher dimensional models which can give rise to an effective or apparent
phantom energy. 

Although there is no problem to consider $w<-1$ for the background evolution, there
are apparent divergencies appearing in the perturbations 
when a model tries to cross the limit $w=-1$. This is a potential headache for
experiments like Euclid that probe directly the perturbations through measurements
of the galaxy clustering and weak lensing. To analyse the Euclid data, we need to
be able to consider models that cross the phantom divide $w=-1$ at the level
of first order perturbations (since the only dark energy model that has
no perturbations at all is the cosmological constant).

However, at the level of cosmological first-order perturbation 
theory, there is no fundamental limitation that prevents an effective
fluid from crossing the phantom divide.

As $w \rightarrow -1$ the terms in Eqs.~(\ref{d_pert}) and (\ref{t_pert}) 
containing $1/(1+w)$ will generally
diverge. This can be avoided by replacing $\theta$ with a new variable
$V$ defined via {\color{red}$V=\rho\left( 1+w \right) \theta$}. This corresponds to rewriting
the $0$-$i$ component of the energy momentum tensor as 
$ik_j T_{0}^{j}= V$ which avoids problems if $T_{0}^{j}\neq0$ when
$\bar{p}=-\bar\rho$. 
Replacing the time derivatives by a derivative with respect
to the logarithm of the scale factor $\ln a$ (denoted by a prime), we obtain 
\citep{Ma:1995ey,Hu:2004xd, Kunz:2006wc}:
\begin{eqnarray}
\delta' &=& 3(1+w) \Phi' - \frac{V}{H a} 
- 3 \left(\frac{\delta p}{\bar\rho}-w \delta \right) \label{eq:delta} \\
V' &=& -(1-3w) V+ \frac{k^2}{H a} \frac{\delta p}{\bar\rho}
+(1+w) \frac{k^2}{H a}\left( \Psi -\pi\right)\,.  \label{eq:v}
\end{eqnarray}
In order to solve Eqs.~(\ref{eq:delta}) and (\ref{eq:v}) 
we still need to specify the expressions
for $\delta p$ and $\pi$, quantities which characterise the physical, intrinsic nature 
of the dark energy fluid at first oder in perturbation theory. 
While in general the anisotropic stress 
plays an important role as it gives a measure of how 
the gravitational potentials $\Phi$ and $\Psi$ differ, we will set it
in this section to zero, $\pi=0$. We will therefore focus on the
form of the pressure perturbation. There are two important special
cases: barotropic fluids which have no internal degrees of freedom
and for which the pressure perturbation is fixed by the evolution of
the average pressure, and non-adiabatic fluids like e.g. scalar fields
for which internal degrees of freedom can change the pressure
perturbation.

\subsubsection{Parameterising the pressure perturbation}

\noindent{\it Barotropic fluids.}

We define a fluid to be barotropic if the pressure $p$ depends strictly
only on the energy density $\rho$: $p=p(\rho)$. These fluids 
have only adiabatic perturbations, so that they are often called adiabatic.
We can write their pressure as
\begin{equation}
p(\rho) = p(\bar{\rho}+\delta\rho) 
= p(\bar{\rho}) + \left.\frac{{\rm d}p}{{\rm d}\rho}\right|_{\bar{\rho}} \delta\rho
+ O\left[(\delta\rho)^2\right].
\label{eq:baro_exp}
\end{equation}
Here $p(\bar\rho) = \bar{p}$ is the pressure of the isotropic and homogeneous
part of the fluid. The second term
in the expansion (\ref{eq:baro_exp}) can be re-written as
\begin{equation}
\label{c_a_def}
\left.\frac{dp}{d\rho}\right|_{\bar{\rho}}
= \frac{\dot{\bar{p}}}{\dot{\bar{\rho}}} = w - \frac{\dot{w}}{3 aH(1+w)} 
\equiv c_a^2
\end{equation}
where we used the equation of state and the conservation
equation for the dark energy density in the background.
We notice that the adiabatic sound speed $c_a^2$ \label{symbol:ad.sound.speed}will necessarily diverge
for any fluid where $w$ crosses $-1$. 

However, for a perfect barotropic fluid the adiabatic sound speed $c_{a}^2$ turns out to
be the physical propagation speed of perturbations, therefore it should never 
be larger than the speed of light---otherwise our theory becomes acausal---and 
it should never be negative ($c_{a}^{2}<0$)---otherwise classical, and possible quantum, instabilities 
appear. 
Even worse, the pressure perturbation
\begin{equation}
\delta p = c_{a}^{2} \delta\rho = \left( w - \frac{\dot{w}}{3 aH(1+w)} \right) \delta\rho
\end{equation}
will necessarily diverge if $w$ crosses $-1$ and $\delta\rho\neq0$.
Even if we find a way to stabilize the pressure perturbation, 
for instance an equation of state parameter {\color{red} that crosses the $-1$ limit} with zero slope ($\dot{w}$), there will be always 
the problem of a negative speed of sound that prevents these
models from being viable dark energy candidates.

\vspace{0.3cm}
\noindent{\it Non-adiabatic fluids \label{sec:non_ad}}

To construct a model that can cross the phantom divide, we therefore
need to violate the constraint that $p$ is a unique function of $\rho$. 
At the level of first-order perturbation theory, this amounts
to changing the prescription for $\delta p$ which now becomes an arbitrary
function of $k$ and $t$.
One way out to this problem is to choose an appropriate gauge 
where the equations are simple; one choice is for instance the rest 
frame of the fluid where the pressure perturbation reads (in this frame) 
\begin{equation}
\hat{\delta p} = \hat{c}_{s}^{2}\hat{\delta\rho}\label{eq:dp-rest}
\end{equation}
where now the $\hat{c}_{s}^{2}$ is the speed with which fluctuations 
in the fluid propagate, ie. the sound speed. 
We can write Eq.~(\ref{eq:dp-rest}), with an appropriate gauge transformation, 
in a form suitable for the Newtonian frame, ie. for Eqs.(\ref{eq:delta}) and (\ref{eq:v}). 
We find that the pressure perturbation is {\color{red}given by \citep{2002PhRvL..88l1301E, Bean:2003fb, 2003PhRvD..68j3501C}}
\begin{equation}
\delta p = \hat{c}_s^2 \delta\rho + 3 a H\left(a\right)\left(\hat{c}_s^2 - c_{a}^{2}\right) \bar{\rho}\frac{V}{k^2} .
\label{eq:dp_rest}
\end{equation}
The problem here is the presence of $c_{a}^2$ which goes to 
infinity at the crossing and it is impossible that this term
stays finite except if $V\rightarrow 0$ fast enough 
or $\dot{w}=0$, but this is not in general the case.

This divergence appears because for $w=-1$ the energy momentum tensor Eq.~(\ref{EMT}) 
reads: $T^{\mu\nu}=pg^{\mu\nu}$. 
Normally the four-velocity $u^{\mu}$ is the time-like 
eigenvector of the energy-momentum tensor, but now all vectors are 
eigenvectors. So the problem of fixing a unique rest-frame is no longer 
well posed. Then, even though the pressure perturbation looks fine for  {\em the 
observer in the rest-frame}, because it does not
diverge, the badly-defined gauge transformation to the Newtonian frame does, as it contains also 
$c_{a}^{2}$.

\subsubsection{Regularising the divergencies}

We have seen that neither barotropic fluids nor canonical scalar fields, for which the
pressure perturbation is of the type (\ref{eq:dp_rest}), can cross the phantom divide. 
However, there is a simple model (called the quintom model, \citealt{Feng:2004ad,Hu:2004kh})
consisting of two fluids of the same type as in the previous subsection but with
a constant $w$ on either side of $w=-1$. The combination of the two fluids then
effectively crosses the phantom divide if we start with $w_{\rm tot}>-1$, as the energy density in the 
fluid with $w<-1$ will grow faster, so that this fluid will eventually dominate and 
we will end up with $w_{\rm tot}<-1$.

The perturbations in this scenario were analysed in detail in \citet{Kunz:2006wc}, 
where it was shown that in addition to the rest-frame contribution, one also has
relative and non-adiabatic perturbations. All these contributions apparently diverge
at the crossing, but their sum stays finite. When parametrising the perturbations
in the Newtonian gauge as
\begin{equation}
\delta p(k,t) = \gamma(k,t)\, \delta\rho(k,t) \label{d_p_gamma}
\end{equation}
the quantity $\gamma$ will in general have a complicated time and scale
dependence. The conclusion of the analysis is that
indeed single canonical scalar fields with pressure perturbations of the type (\ref{eq:dp_rest})
in the Newtonian frame cannot cross $w=-1$, but that this is not the most general
case. More general models have a priori no problem crossing the phantom divide,
at least not with the classical stability of the perturbations.

\citeauthor{Kunz:2006wc} found that a good approximation to the quintom model
behaviour can be found by regularising the adiabatic sound speed in the gauge
transformation with
\begin{equation}
c_a^2 = w - \frac{\dot{w}(1+w)}{3 H a [(1+w)^2 + \lambda]} \label{eq:caeff}
\end{equation}
where $\lambda$ is a tuneable parameter which determines how close to $w=-1$ the regularisation kicks in. A value of $\lambda \approx 1/1000$ should work reasonably well. However, the final
results are not too sensitive on the detailed regularisation prescription.

This result appears also related to the behaviour found for coupled dark energy models (originally 
introduced to solve the coincidence problem) where 
dark matter and dark energy interact not only through gravity \citep{Amendola:1999er}. The
effective dark energy in these models can also cross the phantom divide without divergencies 
\citep{Huey:2004qv,Das:2005yj,Kunz:2007rk}.

The idea is to insert (by hand) a term in the continuity equations 
of the two fluids
\begin{eqnarray} 
&&\dot{\rho}_{M}+3H\rho_{M}=\lambda\\
&&\dot{\rho}_{x}+3H\left(1+w_{x}\right)\rho_{x}=-\lambda
\end{eqnarray}
where the subscripts $m,x$ refer to dark matter and dark energy, respectively.
In this approximation, the adiabatic sound speed $c_{a}^{2}$ reads
\begin{equation} 
c_{a,x}^{2}= \frac{\dot{p}_{x}}{\dot{\rho}_{x}} = w_{x}-\frac{\dot{w_{x}}}{3aH\left(1+w_{x}\right)+
\lambda/\rho_{x}}
\end{equation}
which stays finite at crossing as long as $\lambda\neq 0$.

However in this class of models there are other instabilities 
arising at the perturbation level regardless of the coupling 
used, \citep[cfr.][]{Valiviita:2008iv}.

\subsubsection{A word on perturbations when $w=-1$}

Although a cosmological constant has $w=-1$ and no perturbations, the converse is not
automatically true: $w=-1$ does not necessarily imply that there are no perturbations.
It is only when we set from the beginning (in the calculation): 
\begin{eqnarray}
p &=& -\rho\\
\delta p &=& -\delta\rho\\
\pi &=& 0
\end{eqnarray}
ie. $T^{\mu\nu} \propto g^{\mu\nu}$, then we have as a solution $\delta = V =0$. 

For instance, if we set $w=-1$ and $\delta p = \gamma\delta\rho$ 
(where $\gamma$ can be a generic function)
in Eqs.~(\ref{eq:delta}) and (\ref{eq:v}) we have 
$\delta\neq 0$ and $V\neq 0$. 
However, the solutions are decaying modes due to the 
$-\frac{1}{a}\left(1-3w\right)V$ term so they are not important  
at late times, but it is interesting to notice that they are 
in general not zero.

As another example, if we have a non-zero anisotropic stress $\pi$ then the 
Eqs. (\ref{eq:delta}) - (\ref{eq:v}) will have a source term that will influence the growth 
of $\delta$ and $V$ in the same way as $\Psi$ does (just because they appear in the same way). 
The $\left(1+w\right)$ term in front of $\pi$ should not worry us as we can always define 
the anisotropic stress through
\begin{equation}
\rho\left(1+w\right) \pi = -\left(\hat{k}_{i}\hat{k}_{j}-\frac{1}{3}\delta_{ij}\right)
\Pi^{i}_{\,j},
\label{eq:real-sigma}
\end{equation}
where $\Pi^{i}_{\,j}\neq 0$ when $i\neq j$ is the {\em real} traceless 
part of the energy momentum tensor, probably the quantity we need to look at:
as in the case of $V=(1+w) \theta$, there is no need for $\Pi \propto (1+w)\pi$ to
vanish when $w=-1$.

It is also interesting to notice that when $w = -1$ the 
perturbation equations tell us 
that dark energy perturbations are not influenced
through $\Psi$ and $\Phi'$ (see Eq.(\ref{eq:delta}) and (\ref{eq:v})). Since
$\Phi$ and $\Psi$ are the quantities directly entering the metric, they must
remain finite, and even much smaller than $1$ for perturbation theory to hold.
Since, in the absence of direct couplings, the dark energy only feels the other
constituents through the terms $(1+w)\Psi$ and $(1+w)\Phi'$, it decouples
completely in the limit $w=-1$ and just evolves on its own. But its perturbations
still enter the Poisson equation and so the dark matter perturbation will
feel the effects of the dark energy perturbations.

Although this situation may seem contrived, it might be that the acceleration of the universe is just an 
observed effect as a consequence of a modified theory of gravity. As was shown in \cite{Kunz:2006ca}, any 
modified gravity theory can be described as an effective fluid both at 
background and at perturbation level; in such a situation it is imperative to 
describe its perturbations properly as this effective fluid may manifest unexpected behaviour.

%% file: de_mg/fr-general.tex
\subsection{$f(R)$ gravity \label{fr-general}}

\def\rd{\rm d}
In parallel to models with extra degrees of freedom in the matter sector, such 
as interacting quintessence (and K-essence, not treated here), another promising approach to the
late-time acceleration enigma is to modify the left-hand side of the Einstein
equations and invoke new degrees of freedom, belonging this time to the
gravitational sector itself.
One of the simplest and most popular extensions of general relativity and known example of 
modified gravity models is the so-called $f(R)$
gravity\index{f(R) gravity@$f(R)$ gravity} in which the 4-dimensional
action is given by some generic function $f(R)$ of the Ricci scalar
$R$\index{Ricci scalar} (for an introduction see e.g. \cite{Amendola2010}): 
\begin{equation}
S=\frac{1}{2\kappa^{2}}\int{\rm d}^{4}x\sqrt{-g}f(R)+S_{m}(g_{\mu\nu},\Psi_{m})
\,,\label{fRaction}\end{equation}
 where as usual $\kappa^{2}=8\pi G$, and $S_{m}$ is a matter action
with matter fields $\Psi_{m}$. Here $G$ is a \emph{bare} gravitational
constant\index{gravitational constant!bare}: we will see that the 
observed value will in general be different. As mentioned in the previous section, it is possible to show that 
 $f(R)$ theories can be mapped into a subset of scalar-tensor theories and therefore to 
a class of interacting scalar field dark energy models universally coupled to all species. When seen 
in the Einstein frame \citep{Wetterich_1988, Maeda:1988ab, Wands:1993uu, EspositoFarese:2000ij, Pettorino:2008ez, Catena:2006bd},
action (\ref{fRaction}) can therefore be related to the action (\ref{mg:cde:action}) shown in the previous section. 
Here we describe $f(R)$ in the Jordan frame: the matter fields in
$S_{m}$ obey standard conservation equations and therefore the metric
$g_{\mu\nu}$ corresponds to the physical frame (which here is the
Jordan frame).  
 
There are two approaches to derive field equations
from the action (\ref{fRaction}). 
\begin{itemize}
\item \textbf{(I) The metric formalism}

The first approach is the so-called metric formalism\index{metric formalism}
in which the connections $\Gamma_{\beta\gamma}^{\alpha}$ are the
usual connections defined in terms of the metric $g_{\mu\nu}$. The
field equations can be obtained by varying the action (\ref{fRaction})
with respect to $g_{\mu\nu}$: \begin{eqnarray}
F(R)R_{\mu\nu}(g)-\frac{1}{2}f(R)g_{\mu\nu}-\nabla_{\mu}\nabla_{\nu}F(R)+g_{
\mu\nu}\square F(R)=\kappa^{2}T_{\mu\nu}\,,\label{fREin}\end{eqnarray}
 where $F(R)\equiv\partial f/\partial R$ (we also use the notation
$f_{,R}\equiv\partial f/\partial R,\, f_{,RR}\equiv\partial^{2}f/\partial
R^{2}$), \label{symbol:FR}
and $T_{\mu\nu}$ is the matter energy-momentum tensor.
The trace\index{trace} of Eq.~(\ref{fREin}) is given by 
\begin{equation}
3\,\square F(R)+F(R)R-2f(R)=\kappa^{2}T\,,\label{trace}
\end{equation}
where $T=g^{\mu\nu}T_{\mu\nu}=-\rho+3P$. Here $\rho$ and $P$ are
the energy density and the pressure of the matter, respectively.

\item \textbf{(II) The Palatini formalism}

The second approach is the so-called Palatini
 formalism\index{Palatini formalism}, where $\Gamma_{\beta\gamma}^{\alpha}$
and $g_{\mu\nu}$ are treated as independent variables. Varying the
action (\ref{fRaction}) with respect to $g_{\mu\nu}$ gives \begin{eqnarray}
F(R)R_{\mu\nu}(\Gamma)-\frac{1}{2}f(R)g_{\mu\nu}=\kappa^{2}T_{\mu\nu}\,,
\label{fRpala}\end{eqnarray}
 where $R_{\mu\nu}(\Gamma)$ is the Ricci tensor corresponding to
the connections $\Gamma_{\beta\gamma}^{\alpha}$. In general this
is different from the Ricci tensor $R_{\mu\nu}(g)$ corresponding
to the metric connections. Taking the trace of Eq.~(\ref{fRpala}),
we obtain
\begin{equation}
F(R)R-2f(R)=\kappa^{2}T\,,\label{trace2}
\end{equation}
where $R(T)=g^{\mu\nu}R_{\mu\nu}(\Gamma)$ is directly related to
$T$. Taking the variation of the action (\ref{fRaction}) with respect
to the connection, and using Eq.~(\ref{fRpala}), we find 
\begin{eqnarray}
R_{\mu\nu}(g)-\frac{1}{2}g_{\mu\nu}R(g) & = & 
\frac{\kappa^{2}T_{\mu\nu}}{F}-\frac{FR(T)-f}{2F}g_{\mu\nu}
+\frac{1}{F}(\nabla_{\mu}\nabla_{\nu}F-g_{\mu\nu}\square F)\nonumber \\
 &  & -\frac{3}{2F^{2}}\left[\partial_{\mu}F\partial_{\nu}F
 -\frac{1}{2}g_{\mu\nu}(\nabla F)^{2}\right]\,.\label{equationpala}
\end{eqnarray}

\end{itemize}
In General Relativity\index{General Relativity} we have $f(R)=R-2\Lambda$ 
and $F(R)=1$, so that the term $\square F(R)$
in Eq.~(\ref{trace}) vanishes. 
In this case both the metric and the Palatini formalisms give the relation
$R=-\kappa^{2}T=\kappa^{2}(\rho-3P)$, which means that the Ricci
scalar $R$ is directly determined by the matter (the trace $T$).

In modified gravity models where $F(R)$ is a function of $R$, the
term $\square F(R)$ does not vanish in Eq.~(\ref{trace}).
This means that, in the metric formalism, there is a propagating scalar
degree of freedom, $\psi\equiv F(R)$. The trace equation (\ref{trace})
governs the dynamics of the scalar field $\psi$---dubbed
{}``scalaron''\index{scalaron}
\cite{Sta80}. In the Palatini formalism the kinetic term $\square F(R)$
is not present in Eq.~(\ref{trace2}), which means that the scalar-field
degree of freedom does not propagate freely (\cite{Amarzguioui:2005zq,Li:2007xw,Li:2008bma,Li:2008fa}).

The de Sitter point\index{de Sitter!point} corresponds to a vacuum
solution at which the Ricci scalar is constant. Since $\square F(R)=0$
at this point, we get \begin{eqnarray}
F(R)R-2f(R)=0\,,\label{fRdeSitter}\end{eqnarray}
 which holds for both the metric and the Palatini formalisms. Since
the model $f(R)=\alpha R^{2}$ satisfies this condition, it possesses
an exact de Sitter solution \citep{Sta80}.

It is important to realise that the dynamics of $f(R)$ dark energy models is
different depending
on the two formalisms. Here we confine ourselves to  the metric case only.

Already in the early 1980s it was known that the model $f(R)=R+\alpha R^{2}$
can be responsible for inflation\index{inflation} in the early universe
\citep{Sta80}. This comes from the fact that the presence of the quadratic
term $\alpha R^{2}$ gives rise to an asymptotically exact de Sitter
solution. Inflation ends when the term $\alpha R^{2}$ becomes smaller
than the linear term $R$. Since the term $\alpha R^{2}$ is negligibly
small relative to $R$ at the present epoch, this model is not suitable
to realise the present cosmic acceleration.

Since a late-time acceleration requires modification for small $R$,
models of the type $f(R)=R-\alpha/R^{n}$ ($\alpha>0,n>0$) were proposed
as a candidate for dark energy \citep{Capo1,CDTT,Nojiri03}.
While the late-time cosmic acceleration is possible in these models,
it has become clear that they do not satisfy local gravity constraints
because of the instability associated with negative values of $f_{,RR}$
\citep{Chiba:2003ir,Dolgov,Woodard,Olmo,Faraoni}. Moreover a standard matter
epoch is not present because of a large coupling between the Ricci
scalar and the non-relativistic matter \citep{APT07}.

Then, we can ask what are the conditions for the viability of $f(R)$ dark energy  
models in the metric formalism. In the following we first present 
such conditions and then explain step by step why they are required.
\begin{itemize}
\item (i) $f_{,R}>0$ for $R\ge R_{0}~(>0)$, where $R_{0}$ is the Ricci
scalar at the present epoch. Strictly speaking, if the final attractor
is a de Sitter point with the Ricci scalar $R_{1}~(>0)$, then the
condition $f_{,R}>0$ needs to hold for $R\ge R_{1}$.

This is required to avoid a negative effective gravitational constant.

\item (ii) $f_{,RR}>0$ for $R\ge R_{0}$.

This is required for consistency with local gravity tests\index{local gravity
constraints}
\citep{Dolgov,Olmo,Faraoni,Navarro}, for the presence of the matter-dominated
epoch\index{matter-dominated epoch} \citep{APT07,AGPT}, and for the
stability of cosmological perturbations \citep{Carroll06,Song07,Bean07,Faulkner}.

\item (iii) $f(R)\to R-2\Lambda$ for $R\gg R_{0}$.

This is required for consistency with local gravity tests
\citep{2008PhLB..660..125A,Hu07,Star07,Appleby,Tsuji08}
and for the presence of the matter-dominated epoch \citep{AGPT}.

\item (iv) $0<\frac{Rf_{,RR}}{f_{,R}}(r=-2)<1$ at $r=-\frac{Rf_{,R}}{f}=-2$.

This is required for the stability of the late-time de Sitter point
\citep{Muller,AGPT}.

\end{itemize}
For example, the model $f(R)=R-\alpha/R^{n}$ ($\alpha>0$, $n>0$)
does not satisfy the condition (ii).

Below we list some viable $f(R)$ models that satisfy the above conditions.
\begin{eqnarray}
 &  & {\rm (A)}~f(R)=R-\mu R_{c}(R/R_{c})^{p}\qquad{\rm
with}~~0<p<1,~~\mu,R_{c}>0\,,\label{Amodel}\\
 &  & {\rm (B)}~f(R)=R-\mu
R_{c}\frac{(R/R_{c})^{2n}}{(R/R_{c})^{2n}+1}\qquad{\rm
with}~~n,\mu,R_{c}>0\,,\label{Bmodel}\\
 &  & {\rm (C)}~f(R)=R-\mu
R_{c}\left[1-\left(1+R^{2}/R_{c}^{2}\right)^{-n}\right]\qquad{\rm
with}~~n,\mu,R_{c}>0\,,\label{Cmodel}\\
 &  & {\rm (D)}~f(R)=R-\mu R_{c}{\rm tanh}\,(R/R_{c})\qquad{\rm
with}~~\mu,R_{c}>0\,.\label{Dmodel}\end{eqnarray}
 The models (A), (B), (C), and (D) have been proposed in \citet{AGPT},
\citet{Hu07}, \citet{Star07}, and \citet{Tsuji08}, respectively. A
model similar to (D) has been also proposed in \citet{Appleby},
while a generalised model encompassing (B) and (C) has been studied
in  ~\cite{mir09}. In the model (A), the power $p$ needs to
be close to 0 to satisfy the condition (iii). In the models (B) and
(C) the function $f(R)$ asymptotically behaves as $f(R)\to R-\mu
R_{c}[1-(R^{2}/R_{c}^{2})^{-n}]$
for $R\gg R_{c}$ and hence the condition (iii) can be satisfied even
for $n={\cal O}(1)$. In the model (D) the function $f(R)$ rapidly
approaches $f(R)\to R-\mu R_{c}$ in the region $R\gg R_{c}$. These
models satisfy $f(R=0)=0$, so the cosmological contant vanishes in
the flat spacetime.

Let us consider cosmological dynamics of $f(R)$ gravity in the metric
formalism. It is possible to carry out a general analysis without
specifying the form of $f(R)$. In the flat FLRW spacetime the Ricci
scalar is given by \begin{eqnarray}
R=6(2H^{2}+\dot{H})\,,\end{eqnarray}
 where $H$ is as usual the Hubble parameter. As a matter action $S_{m}$
we take into account non-relativistic matter and radiation, which
satisfy the usual conservation equations $\dot{\rho}_{m}+3H\rho_{m}=0$
and $\dot{\rho}_{r}+4H\rho_{r}=0$ respectively. {}From Eqs.~(\ref{fREin})
and (\ref{trace}) we obtain the following equations \begin{eqnarray}
3FH^{2} & = &
\kappa^{2}\,(\rho_{m}+\rho_{r})+(FR-f)/2-3H\dot{F}\,,\label{FRWfR1}\\
-2F\dot{H} & = &
\kappa^{2}\left[\rho_{m}+(4/3)\rho_{r}\right]+\ddot{F}-H\dot{F}\,.\label{FRWfR2}
\end{eqnarray}
 We introduce the dimensionless variables: \begin{eqnarray}
x_{1}\equiv-\frac{\dot{F}}{HF}\,,\quad x_{2}\equiv-\frac{f}{6FH^{2}}\,,\quad
x_{3}\equiv\frac{R}{6H^{2}}\,,\quad
x_{4}\equiv\frac{\kappa^{2}\rho_{r}}{3FH^{2}}\,,\end{eqnarray}
 together with the following quantities \begin{eqnarray}
\Omega_{m}\equiv\frac{\kappa^{2}\rho_{m}}{3FH^{2}}=1-x_{1}-x_{2}-x_{3}-x_{4}\,,
\qquad\Omega_{r}\equiv x_{4}\,,\qquad\Omega_{{\rm DE}}\equiv
x_{1}+x_{2}+x_{3}\,.\label{fromedef}\end{eqnarray}
 It is straightforward to derive the following differential equations
\citep{AGPT}: \begin{eqnarray}
 x_{1}' & = &
-1-x_{3}-3x_{2}+x_{1}^{2}-x_{1}x_{3}+x_{4}~,\label{x1fR}\\
 x_{2}' & = &
\frac{x_{1}x_{3}}{m}-x_{2}(2x_{3}-4-x_{1})~,\label{x2fR}\\
 x_{3}' & = &
-\frac{x_{1}x_{3}}{m}-2x_{3}(x_{3}-2)~,\label{x3fR}\\
x_{4}' & = &
-2x_{3}x_{4}+x_{1}x_{4}\,,\label{x4fR}\end{eqnarray}
 where the prime denotes $d/d\ln a$ and \begin{eqnarray}
m & \equiv & \frac{\rd\ln F}{\rd\ln
R}=\frac{Rf_{,RR}}{f_{,R}}\,,\label{eq:charact}\\
r & \equiv & -\frac{\rd\ln f}{\rd\ln
R}=-\frac{Rf_{,R}}{f}=\frac{x_{3}}{x_{2}}\,.\label{mdef}\end{eqnarray}
 {}From Eq.~(\ref{mdef}) one can express $R$ as a function of
$x_{3}/x_{2}$. Since $m$ is a function of $R$, it follows that
$m$ is a function of $r$, i.e. $m=m(r)$. The $\Lambda$CDM model,
$f(R)=R-2\Lambda$, corresponds to $m=0$. Hence the quantity $m$
characterises the deviation\index{deviation parameter} from the $\Lambda$CDM
model. Note also that the model, $f(R)=\alpha R^{1+m}-2\Lambda$,
gives a constant value of $m$. The analysis using Eqs.~(\ref{x1fR})-(\ref{x4fR})
is sufficiently general in the sense that the form of $f(R)$ does
not need to be specified.

The effective equation of state\index{effective equation of state!for $f(R)$
models}
of the system (i.e. $p_{tot}/\rho_{tot}$) is \begin{eqnarray}
w_{{\rm eff}}=-\frac{1}{3}(2x_{3}-1)\,,\label{ldef}\end{eqnarray}

The dynamics of the full system can be investigated by analysing the stability
properties of the critical phase-space points as in e.g. \cite{AGPT}. 
The general conclusions is that only models with a characteristic function $m(r)$ positive and
close to $\Lambda$CDM, i.e. $m\ge 0$, are cosmologically viable. That is, only for these
models one finds a sequence of a long decelerated matter epoch followed by a stable
accelerated attractor.

The perturbation equations have been derived in e.g. \cite{Hwang:2001qk,Tsujikawa:2007tg}.
Neglecting the contribution of radiation one has
\begin{align}
\delta_{m}'' & +\left(x_{3}-\frac{1}{2}x_{1}\right)\delta_{m}'-\frac{3}{2}(1-x_{1}-x_{2}-x_{3})\delta_{m}\nonumber \\
 & =\frac{1}{2}\biggl[\left\{ \frac{k^{2}}{x_{5}^{2}}-6+3x_{1}^{2}-3x_{1}'-3x_{1}(x_{3}-1)\right\} \delta\tilde{F}\nonumber \\
 & ~~~~+3(-2x_{1}+x_{3}-1)\delta\tilde{F}'+3\delta\tilde{F}''\biggr]\,,\\
\delta\tilde{F}'' & +(1-2x_{1}+x_{3})\delta\tilde{F}'\nonumber \\
 & +\left[\frac{k^{2}}{x_{5}^{2}}-2x_{3}+\frac{2x_{3}}{m}-x_{1}(x_{3}+1)-x_{1}'+x_{1}^{2}\right]\delta\tilde{F}\nonumber \\
 & ~~~~=(1-x_{1}-x_{2}-x_{3})\delta_{m}-x_{1}\delta_{m}'\,,\end{align}
 where $\delta\tilde{F}\equiv\delta F/F$, and the new variable $x_{5}\equiv aH$
satisfies \begin{eqnarray}
x_{5}'=(x_{3}-1)\, x_{5}\,.\label{x5eq}\end{eqnarray}
The perturbation $\delta F$ can be written as $\delta F=f_{,RR}\delta R$ and therefore $\delta\tilde{F}=m\delta R /R$.
These equations can be integrated numerically to derive the behavior of $\delta_m$ at all scales.
At sub-Hubble scales they however can be simplified and the following expression for the two MG function $Q,\eta$ of Eq. (\ref{mod_constr})
can be obtained:
\begin{eqnarray}
Q &=& 1-{ \frac{k^{2}}{ 3(a^2M^2 + k^{2})}
}\nonumber\\
\eta &=&   1- {\frac{2k^{2} }{3a^2M^2 +
4k^{2}} }
\end{eqnarray}
where
\begin{equation}\label{frr-mass} M^{2} = {\frac{1}{3 f_{,RR}} } \end{equation}
Note that in the $\Lambda$CDM limit $f_{,RR}\to 0$ and $Q,\eta\to 1$.

These relation can be straightforwardly generalised to more general cases. In \cite{2010PhRvD..82b3524D}
the perturbation equations for  $f(R)$ Lagrangian  have been extended to include coupled scalar fields 
and their kinetic energy $X\equiv -\phi_{,\mu}\phi^{\mu}/2$, 
resulting in a $f(R,\phi,X)$-theory. In the slightly simplified case in which
$f(R,\phi,X)=f_1(R,\phi)+f_2(\phi,X)$, with arbitrary functions $f_1,2$, one obtains
\begin{eqnarray}
Q &=&
-\frac{1}{F} \frac{(1+2r_1)(f_{,X}+2r_2)+2F_{,\phi}^2/F}
{(1+3r_1)(f_{,X}+2r_2)+3F_{,\phi}^2/F}\,.\nonumber\\
\eta &=&  \frac{(1+2r_1)(f_{,X}+2r_2)+2F_{,\phi}^2/F}
{(1+4r_1)(f_{,X}+2r_2)+4F_{,\phi}^2/F}\,.
\end{eqnarray}
where the notation $f_{,X}$ or $F_{,\phi}$ denote differentiation wrt $X$ or $\phi$, respectively,
and where
$
r_1 \equiv \frac{k^2}{a^{2}} \frac{m}{R}
$ and
$
r_2 \equiv \frac{a^2}{k^2}M_\phi^2\,
$, $M_{\phi}=-f_{,\phi\phi}/2$ being the scalar field effective mass. In the same paper
\cite{2010PhRvD..82b3524D} an extra term proportional to $X\Box\phi$ in the Lagrangian is also taken into account.

Euclid forecasts for the $f(R)$ models will be presented in Sec. \ref{fRforecastconstraints}

%% file: de_mg/ModifiedGravity.tex
\subsection{Massive gravity and higher-dimensional models}
Instead of introducing new scalar degrees of freedom such as in $f(R)$ theories, another philosophy in modifying gravity is to modify the graviton itself.
In this case the new degrees of freedom belong to the
gravitational sector itself; examples include massive gravity and higher-dimensional
frameworks, such as the Dvali-Gabadadze-Porrati (DGP) model \citep{dvali00} and its extensions. The new degrees of freedom can be responsible for a late-time
speed-up of the Universe, as is summarised below for a choice of selected models. 
We note here that whilst such self-accelerating solutions are interesting in their own right, they
do not tackle the old Cosmological Constant problem:
why the observed cosmological constant is so
much smaller than expected in the first place.
Instead of answering this question directly, an alternative approach is the
idea of
degravitation \citep[see][]{Degravitation1,Degravitation2,Degravitation3,Degravitation4}, where the cosmological
constant could be as large as expected from standard field theory,
but would simply gravitate very little {\color{red}(see subsection below)}.

%
%

\subsubsection{Self-acceleration}
\label{sec:Self-Acceleration}
\paragraph{DGP.}
\label{sec:DGP}

 The DGP model is one of the important infrared modified theories of gravity.
From a four-dimensional point of view this corresponds effectively to a theory
where the graviton acquires a soft mass $m$. In this braneworld model our
visible Universe is confined to a brane of 4 dimensions embedded into a 5
dimensional bulk.  At small distances the 4 dimensional gravity is recovered due
to an intrinsic Einstein Hilbert term sourced by the brane curvature causing a
gravitational force law which scales as $r^{-2}$. At large scales the
gravitational force law asymptotes to an $r^{-3}$ behavior.
 The cross over scale $r_c=m^{-1}$ is given by the ratio of the Planck masses in
4 ($M_4$) and 5 ($M_5$) dimensions.\label{symbol:rc}
 One can study perturbations around flat space-time and compute the
gravitational exchange amplitude between two conserved sources, which does not
reduce to the GR result even in the limit m$\to0$. However, the successful
implementation of the Vainshtein mechanism for decoupling the additional modes
from gravitational dynamics at sub-cosmological scales makes these theories
still very attractive \citep{Vainshtein:1972sx}. Hereby, the Vainshtein effect
is realised through the non-linear interactions of the helicity-0 mode $\pi$,
as will be explained in further detail {\color{red}below}.
 Thus, this vDVZ discontinuity does not appear  close to an astrophysical source
where the $\pi$ field becomes non-linear and these non-linear effects of $\pi$
restore predictions to those of GR. This is most easily understood in the limit
where $M_4, M_5\to\infty$ and $m\to 0$ while keeping the strong coupling scale
$\Lambda=(M_4m^2)^{1/3}$ fixed. This allows us to treat the usual helicity-2
mode of gravity linearly while and treating the helicity-0 mode $\pi$
non-linearly. The resulting effective action is then
 \ba
 \label{eq.DGP_decoupling}
 \mathcal{L}_\pi=3 \pi \Box \pi -\frac{1}{\Lambda^3}(\partial \pi)^2 \Box \pi\,,
 \ea
where interactions  already become important at the scale $\Lambda \ll M_{\rm
Pl}$ \citep{Luty:2003vm}.

Furthermore, in this model, one can recover an interesting range of cosmologies,
in particular a modified Friedmann equation with a self-accelerating solution.
The  Einstein equations thus obtained reduce to the following modified Friedmann
equation in a homogeneous and isotropic metric \citep{Deffayet:2001pu}
\ba
\label{eq:Friedmann_DGP}
H^2\pm m H=\frac{8\pi G}{3}\rho\,,
\ea
such that at higher energies one recovers the usual 4-dimensional behaviour,
$H^2\sim \rho$, while at later time corrections from the extra dimensions kick
in.  As is clear in this Friedmann equation, this braneworld scenario holds  two
branches of cosmological solutions with distinct properties. The
self-accelerating branch (minus sign) allows for a de Sitter behaviour $H={\rm
const}=m$ even in the absence of any cosmological constant $\rho_{\Lambda}=0$
and as such it has attracted a lot of attention. Unfortunately, this branch suffers
from a ghost-like instability. The normal branch (the plus sign) instead slows
the expansion rate but is stable. In this case a cosmological constant is
still required for late-time acceleration, but it provides significant
intuition for the study of degravitation.
%
%
\paragraph{The Galileon.}
{\color{red}Even though the DGP model is interesting for several reasons like giving the
Vainshtein effect a chance to work, the self-acceleration solution unfortunately introduces
extra ghost states as outlined above. However, it has been generalized to a ``Galileon'' model,
which can be considered as an effective field theory for the helicity-0 field $\pi$.  Galileon models are invariant under shifts of the field $\pi$ and shifts of the  gradients of $\pi$ (known as the Galileon symmetry), meaning that a Galileon model is invariant under the transformation
\begin{equation}
\pi \rightarrow \pi + c +v_{\mu}x^{\mu}\;,
\label{eq:Galileonsymmetry}
\end{equation}
for arbitrary constant $c$ and $v_{\mu}$.
 In induced gravity braneworld models, this symmetry is naturally inherited from
the 5-dimensional Poincar\'{e} invariance \cite{deRham:2010eu}. The Galileon theory relies strongly on this symmetry to
constrain the possible structure of the effective $\pi$ Lagrangian, and insisting that the effective field theory for $\pi$  bears no ghost-like instabilities further restricts the possibilities \cite{Nicolis:2008in}. It can be shown that there
exists only five derivative interactions which preserve the Galilean symmetry without introducing
ghosts. These interactions are symbolically of the form $\mathcal{L}_{\pi}^{(1)}=\pi$ and $\mathcal{L}_{\pi}^{(n)}=(\partial \pi)^2(\partial\partial \pi)^{n-2}$, for $n = 2,\ldots  5$. A general Galileon Lagrangian can be constructed as a linear combination of these Lagrangian operators.  The effective action for the DGP scalar \ref{eq.DGP_decoupling} can be seen to be a combination of $\mathcal{L}_{\pi}^{(2)}$ and $\mathcal{L}_{\pi}^{(3)}$. Such interactions have been shown to naturally arise from Lovelock invariants in
the bulk of generalized braneworld models \cite{deRham:2010eu}, however the Galileon does not necessarily require a higher-dimensional origin and can be consistently treated as a four-dimensional effective field theory.
 
 As shown in \cite{Nicolis:2008in}, such theories can allow for self-accelerating de Sitter solutions without any ghosts unlike
in the DGP model. In the presence of compact sources, these solutions can support spherically
symmetric, Vainshtein-like non-linear perturbations that are also stable against small fluctuations.
This is however constrained to the subset of the third order Galileon, which contains only $\mathcal{L}_{\pi}^{(1)}$, $\mathcal{L}_{\pi}^{(2)}$ and $\mathcal{L}_{\pi}^{(3)}$ \cite{Mota:2010bs}.

The Galileon terms described above  form a subset of the so called ``generalized Galileons''.  A generalized Galileon model allows non-linear derivative interactions of the scalar field $\pi$ in the Lagrangian while insisting that the equations of motion remain at most second order in derivatives thus removing any ghost-like instabilities.  However unlike the pure Galileon models generalized Galileons do not impose the symmetry of Equation (\ref{eq:Galileonsymmetry}).  These theories were first written down by Horndeski \cite{Horndeski:1974wa} and later rediscoved by Deffayet, Gao,  Steer, and Zahariade \cite{Deffayet:2011gz}.  They are a linear combination of Lagrangians constructed by multiplying the Galileon Lagrangians $\mathcal{L}_{\pi}^{(n)}$ by an arbitrary scalar function of the scalar $\pi$ and its first derivatives.  Just like the Galileon, generalized Galileons can give rise to cosmological acceleration and to Vainshtein screening.  However as they lack the Galileon symmetry these theories are not protected from quantum corrections.  Many other theories can also be found within the spectrum of generalized Galileon models, including k-essence.}

%
%

\paragraph{Degravitation.}
\label{sec:Degravitation}

The idea behind degravitation is to modify gravity in the IR, such that the
vacuum energy could have a weaker effect on the geometry, and therefore
reconcile a natural value for the vacuum energy as expected from particle
physics with the observed late time acceleration.  Such modifications of gravity
typically arise in models of massive gravity \citep{Degravitation1,Degravitation2,Degravitation3,Degravitation4}, i.e. where gravity is mediated by a massive spin-2 field.
The extra-dimensional DGP scenario presented previously, represents a
specific model of soft mass gravity, where gravity weakens down at large
distance, with a force law going as $1/r$. Nevertheless, this weakening is too
weak to achieve degravitation and tackle the cosmological constant problem.
However an obvious way out is to extend the DGP model to higher
dimensions, thereby diluting gravity more efficiently at large distances. This
is achieved in models of Cascading gravity, as is presented below. An
alternative to cascading gravity is to work directly with theories of constant
mass gravity (hard mass graviton).

\paragraph{Cascading Gravity.}
Cascading gravity is an explicit realisation of the idea of degravitation,
where gravity behaves as a high-pass filter, allowing sources with
characteristic wavelength (in space and in time) shorter than a characteristic
scale $r_c$ to behave as expected from General Relativity, but weakening the
effect of sources with longer wavelengths.
This could explain why a large cosmological constant does not backreact as much
as
anticipated from standard General Relativity. Since the DGP model does not
modify gravity enough in the IR, ``Cascading gravity" relies on the
presence of at least two infinite extra dimensions while our world
is confined on a four-dimensional brane \citep{deRham:2007rw}. Similarly as in
DGP, four-dimensional gravity is recovered at short distances thanks to an
induced Einstein-Hilbert term on the brane with associated Planck scale $M_4$.
The brane we live in is then embedded in a five-dimensional brane which bears a
five-dimensional Planck scale $M_5$, itself embedded in six dimensions (with
Planck scale $M_6$). From a four-dimensional perspective, the relevant scales
are the 5d and 6d masses $m_4=M_5^3/M_4^2$ and $m_5=M_6^4/M_5^3$ which
characterise the transition from the 4d to 5d and 5d to 6d  behaviour
respectively.

Such theories embedded in more-than-one extra dimensions involve at least one
additional scalar field that typically enters as a ghost. This ghost is
independent of the ghost present in the self-accelerating branch of DGP but is
completely generic to any
codimension-two and higher framework with brane localised kinetic terms. There
are however two ways to cure the ghost, both of which are natural when
considering a
realistic higher codimensional scenario, namely smoothing out the brane, or
including a brane tension \citep{deRham:2007rw,deRham:2007xp,deRham:2010rw}.

When properly taking into account the issue associated with the
ghost, such models give rise to a theory of massive gravity (soft mass graviton)
composed of one
helicity-2 mode, helicity-1 modes that decouple and 2 helicity-0
modes. In order for this theory to be consistent with standard
general relativity in four dimensions, both helicity-0 modes should
decouple from the theory. As in DGP, this decoupling does not happen in a
trivial way, and relies on a phenomenon of strong coupling. Close
enough to any source, both scalar modes are strongly coupled and therefore
freeze.

The resulting theory appears as a theory of a massless spin-2 field in
four-dimensions,
in other words as General Relativity.
If $r\ll m_5$ and for $m_6\le m_5$, the respective Vainshtein scale or strong
coupling scale, \ie the distance from the source $M$
within which each mode is strongly coupled is $r_{i}^3=M/m_i^2
M_4^2$, where $i=5,6$. Around a source $M$, one recovers
four-dimensional gravity for $r\ll r_{5}$, five-dimensional
gravity for $r_{5}\ll r \ll r_{6}$ and finally six-dimensional
gravity at larger distances $r\gg r_{6}$.

\paragraph{Massive Gravity.}

While laboratory experiments, solar systems tests and cosmological
observations have all been in complete agreement with General Relativity
for now almost a century, these bounds do not eliminate the
possibility for the graviton to bear a small hard mass $m\lesssim
6.10^{-32}$eV \citep{Goldhaber:2008xy}. The question of whether or not gravity
could be mediated by a hard-mass graviton
is not only a purely fundamental but abstract one. Since the degravitation
mechanism is also expected to be present if
the graviton bears a hard mass, such models can play an important role for late-time cosmology, and more precisely when the age of the Universe becomes of the
order of the graviton Compton wavelength.

Recent progress has shown that theories of hard massive gravity can be free of
any ghost-like pathologies in the decoupling limit where $M_{\rm Pl}\to \infty$
and  $m\to 0$ keeping the scale $\Lambda_3^3=M_{\rm Pl} m^2$ fixed
\citep{deRham:2010ik,deRham:2010kj}. The absence of pathologies in the decoupling limit does
not guarantee the stability of massive gravity on cosmological backgrounds, but
provides at least a good framework to understand the implications of a small
graviton mass. Unlike a massless spin-2 field which only bears two
polarisations, a massive one bears five of them namely 2 helicity-2 modes, 2
helicity-1 modes which decouple, and one helicity-0 mode (denoted as $\pi$). As
in the braneworld models presented previously, this helicity-0 mode behaves as a
scalar field with specific derivative interactions of the form
\ba
\mathcal{L}_\pi=h^{\mu\nu}\(X^{(1)}\mn+\frac{1}{\Lambda_3^3}
X^{(2)}\mn+\frac{1}{\Lambda_3^6}X^{(3)}\mn\)\,.
\label{eq:massiveGravity}
\ea
Here, $h_{\mu\nu}$ denotes the canonically normalised (rescaled by  $M_{\rm
pl}$)
tensor field perturbation (helicity-2 mode), while  $X^{(1)}\mn,X^{(2)}\mn,$ and
 $X^{(3)}\mn$ are
respectively, linear, quadratic and cubic in the helicity-0 mode $\pi$.
Importantly, they are all transverse (for instance, $X^{(1)}\mn  \propto \eta_
{\mu\nu}\square\pi - \partial_\mu \partial_\nu \pi$). Not only do these
interactions automatically satisfy the Bianchi identity,  as they should to
preserve diffeomorphism invariance, but they are also at most second order in
time derivatives. Hence, the interactions (\ref{eq:massiveGravity}) are  linear
in the helicity-2 mode, and are free of any ghost-like pathologies.  Such
interactions are therefore very similar in spirit to the Galileon ones, and bear
the same internal symmetry (\ref{eq:shift_symmetry},
\ref{eq:galileon_symmetry}), and present very similar physical properties. When
$X^{(3)}\mn$ {\color{red}is absent}, one can indeed recover an Einstein frame picture for which the
interactions are of the form
\ba
\label{galgen}
\mathcal{L}=\frac{M_{\rm Pl}^2}{2}\sqrt{-g}R +\frac 32 \pi \Box \pi
+\frac{3\beta}{2\Lambda_3^3}(\p \pi)^2 \Box \pi
+\frac{\beta^2}{2\Lambda_3^6}(\p \pi)^2\((\partial_\alpha\partial_\beta
\pi)^2-(\Box \pi)^2\)
+\mathcal{L}_{\rm mat}[\psi, \tilde g\mn]\,,
\ea
where $\beta$ is an arbitrary constant and matter fields $\psi$ do not couple to
the metric $g\mn$ but to $\tilde g\mn=g\mn+\pi\eta\mn+\frac{\beta}{\Lambda_3^3}
\partial_\mu \pi \partial_\nu \pi$.
Here again, the recovery of GR in the UV is possible via a strong coupling
phenomena, where the interactions for $\pi$ are already important at the scale
$\Lambda_3\ll M_{\rm Pl}$, well before the interactions for the usual helicity-2
mode. This strong coupling, as well as the peculiar coupling to matter sources,
have distinguishable features in cosmology as is explained below
\citep{Afshordi:2008rd,Jain:2010ka}.

\subsubsection{Observations}
All models of modified gravity presented in this section have in 
common the presence of at least one additional helicity-0 degree of freedom
which is not an arbitrary scalar, but descends from a full-fledged spin-two
field. As such it has no potential and enters the Lagrangian
via very specific derivative terms fixed by symmetries. However tests of gravity
severely
constrain the presence of additional scalar degrees of freedom. As is well
known, in theories of massive gravity
the helicity-0 mode can evade fifth force constrains  in the vicinity of matter
if the helicity-0 mode interactions
are important enough to freeze out the field fluctuations
\citep{Vainshtein:1972sx}. This Vainshtein mechanism is similar in spirit but
different in practise to the chameleon and symmetron mechanisms  presented in
detail in the next section. One key difference relies on the presence of
derivative interactions rather than a specific potential. So rather than
becoming massive in dense regions, in the Vainshtein mechanism the helicity-0
mode becomes weakly coupled to matter (and light, \ie sources in general) at
high energy. This screening of scalar mode can yet have distinct signatures in
cosmology and in particular for structure formation.

\subsubsection{Screening Mechanisms}
\label{sec:Screening}
While quintessence introduces a new degree of freedom to explain the late time acceleration of the Universe, the idea behind modified gravity is instead to tackle the core of the cosmological constant problem and its tuning issues as well as screening any fifth forces that would come from the introduction of extra degrees of freedom.
As we have mentioned at the end of the previous subsection,
the strength with which these new degrees of freedom can
couple to the fields of the standard model is very tightly constrained by
searches for fifth forces and violations of the weak equivalence principle.
Typically the strength of the scalar mediated interaction is required to be
orders of magnitude weaker than gravity. It is possible to tune this coupling
to be as small as is required, leading however to additional naturalness
problems. Here we discuss in more detail a number of ways in which new scalar degrees of freedom can 
naturally couple to standard model fields, whilst still being in agreement with
observations, because a dynamical mechanism ensures that their effects are
screened in laboratory and solar system tests of gravity.  This is done by
making some property of the field dependent on the background environment under
consideration.  These models typically fall into two classes;  either the field
becomes massive in a dense environment so that the scalar force is suppressed
because the Compton wavelength of the interaction is small, or the coupling to
matter becomes weaker in dense environments to ensure that the effects of the
scalar are suppressed.  Both types of behaviour require the presence on
non-linearities.

\paragraph{Density Dependent Masses: The Chameleon.}

The chameleon \citep{Khoury:2003rn} is the archetypal model of a scalar field
with a mass that depends on its environment, becoming heavy in dense
environments and light in diffuse ones.  The ingredients for construction of a
chameleon model are a conformal coupling between the scalar field and the matter
fields of the standard model, and a potential for the scalar field which
includes relevant self-interaction terms.

In the presence of non-relativistic matter these two pieces conspire to give
rise to an effective potential for the scalar field
\begin{equation}
V_{eff}(\phi) = V(\phi)+\rho A(\phi),
\end{equation}
where $V(\phi)$ is the bare potential, $\rho$ the local energy density and
$A(\phi)$ the conformal coupling function.  For suitable choices of $A(\phi)$
and $V(\phi)$ the effective potential has a minimum and the position of the
minimum depends on $\rho$.  Self-interaction terms in $V(\phi)$ ensure that the
mass of the field in this minimum also depends on $\rho$  so that the field
becomes more massive in denser environments.

The environmental dependence of the mass of the field allows the chameleon to
avoid the constraints of fifth force experiments through what is known as the
thin-shell effect.  If a dense object is embedded in a diffuse background the
chameleon is massive inside the object.  There, its Compton wavelength is small.
If the Compton wavelength is smaller than the size of the object then scalar
mediated force felt by an observer at infinity is sourced, not by the entire
object, but instead only by a thin shell of matter (of depth the Compton
wavelength) at the surface.  This leads to a natural suppression of the force
without the need to fine tune the coupling constant.

\subsubsection{Density Dependent couplings.}

{\it The Vainshtein Mechanism.}
In models such as DGP and the Galileon the effects of the scalar field are
screened by the Vainshtein mechanism \citep{Vainshtein:1972sx,Deffayet:2001uk}.
This occurs when nonlinear, higher derivative operators are present in the
Lagrangian for a scalar field, arranged in such a way that the equations of
motion for the field are still second order, such as the interactions presented
in (\ref{eq.DGP_decoupling}).

In the presence of a massive source  the non-linear terms force the suppression
of the scalar force in the vicinity of a massive object.  The radius within
which the scalar force is suppressed is known as the Vainshtein radius.  As an
example in the DGP model the Vainshtein radius around a massive object of mass
$M$ is
\begin{equation}
r_{\star}\sim \left( \frac{M}{4\pi M_{\rm Pl}}\right)^{1/3} \frac{1}{\Lambda},
\end{equation}
where $\Lambda$ is the strong coupling scale introduced in section
\ref{sec:DGP}. For the Sun, if $m\sim 10^{-33}$eV, or in other words,
$\Lambda^{-1}=1000$km, then the Vainshtein radius is $r_{\star} \sim 10^2 \mbox{
pc}$.

Inside the Vainshtein radius, when the non-linear, higher derivative terms
become important they cause the kinetic terms for scalar fluctuations to become
large.  This can be interpreted as a relative weakening of the coupling between
the scalar field and matter.   In this way the strength of the interaction is
suppressed in the vicinity of massive objects.

{\it The Symmetron.}
The symmetron model \citep{Hinterbichler:2010es} is in many ways similar to the
chameleon model discussed above.  It requires a conformal coupling between the
scalar field and the standard model and a potential of a certain form.  In the
presence of non-relativistic matter this leads to an effective potential for the
scalar field
\begin{equation}
V_{\mathrm{eff}}(\phi)
=-\frac{1}{2}\left(\frac{\rho}{M^2}-\mu^2\right){\color{red}\phi^2}+\frac{1}{4}\lambda \phi^4,
\end{equation}
where $M$, $\mu$ and $\lambda$ are parameters of the model, and $\rho$ is the
local energy density.

In sufficiently dense environments, $\rho>\mu^2M^2$, the field sits in a minimum
at the origin.  As the local density drops the symmetry of the field is
spontaneously broken and the field falls into one of the two new minima with a
non-zero vacuum expectation value. In high density symmetry-restoring environments, the scalar field
vacuum expectation value should be near zero and fluctuations of the field should not couple to
matter.  Thus the symmetron force in the exterior of a massive object is
suppressed because the field does not couple to the core of the object.

\paragraph{The Olive-Pospelov Model.}
The Olive-Pospelov model \citep{Olive:2007aj} again uses a scalar conformally
coupled to matter.  In this construction both the coupling function and the
scalar field potential are chosen to have quadratic minima.  If the background
field takes the value that minimises the coupling function then fluctuations of
the scalar field decouple from matter.  In non relativistic environments the
scalar field feels an effective potential which is a combinations of these two
functions.  In high density environments the field is very close to the value
which minimises the form of the coupling function.    In low density
environments the field relaxes to the minimum of the bare potential.  Thus the
interactions of the scalar field are suppressed in dense environments. 

%% file: de_mg/GEA.tex
\subsection{Einstein Aether and its Generalizations}
In 1983 it was suggested by Milgrom \citep{Milgrom1983a} that the
emerging evidence for the presence of dark matter in galaxies
could follow from a modification either to how `baryonic' matter
responded to the Newtonian
gravitational field it created
or to how the gravitational field was related to the baryonic matter density.
Collectively these ideas are referred to as MOdified Newtonian Dynamics
(henceforth MOND).  By way of illustration, MOND may be considered as a
modification to the nonrelativistic Poisson equation:
\begin{equation}
\label{eq:mond}
\nabla\cdot\left(\mu\left(\frac{|\nabla\Psi|}{a_{0}}\right)\nabla\Psi\right)=
4\pi G\rho 
\end{equation} 
where $\Psi$ is the gravitational potential, $a_{0}$ is a number with dimensions
Length$^{-1}$ and $\rho$ is the baryonic matter density. The number $a_{0}$ is
determined by looking at the dynamics
of visible matter in galaxies (\cite{SandersMcGaugh2002}). The function $\mu(x)$
would simply be equal to unity in Newtonian gravity. In MOND, the functional
form is only fixed at its limits: $\mu \rightarrow 1$ as $x \rightarrow \infty$
and $\mu \rightarrow x$ as $x \rightarrow 0$.

We are naturally interested in a relativistic version of such a proposal. The building
block is the perturbed space time metric already introduced in Eq. \ref{pert_newton_ds}
\begin{eqnarray} 
\label{eq:metric}
ds^{2}= -(1+2\Psi)dt^{2}+(1-2\Phi)a^{2}(t)(dR^{2}+R^{2}d\Omega^{2})
\end{eqnarray} 
A simple approach is to introduce a dynamical clock field, which we
will call $A^{\mu}$. If it has solutions aligned with the {\color{red}time-like coordinate} $t^{\mu}$ then
it will be sensitive to $\Psi$.  The dynamical 
nature of the field implies that it should have an action which will contain gradients
of the field  and thus potentially scalars formed from gradients of $\Psi$, as
we seek.
A family of covariant actions for the clock field is as follows
(\cite{Zlosnik:2006zu}): 
\begin{eqnarray}
\label{geaact}
\nonumber I [g^{ab},A^{a},\lambda] &=&  \frac{1}{16\pi G} \int d^4x \sqrt{-g}
\left[\frac{1}{\ell^2}  F(K) + \lambda \left(A^a A_a + 1 \right) \right] 
\end{eqnarray}
where 
\begin{equation}
  K = \ell^2 K^{\mu\nu\gamma\delta} \nabla_\mu A_\nu \nabla_\gamma A_\delta
\end{equation}
with
\begin{equation}
 K^{\mu\nu\gamma\delta} = c_1 g^{\mu\gamma} g^{\nu\delta} + c_2 g^{\mu\nu}
g^{\gamma\delta} + c_3 g^{\mu\delta} g^{\nu\delta} 
\end{equation}
The quantity $\ell$ is a number with dimensions of length, the $c_{A}$ are
dimensionless constants, the Lagrange multiplier field $\lambda$ enforces the
unit-timelike constraint on $A^{a}$, 
and $F$ is a function. These models
have been termed Generalized Einstein-Aether (GEA) theories, emphasising the
coexistence of general covariance and a `preferred' state of rest
in the model-i.e. keeping time with $A^{\mu}$.

Indeed, when the geometry is of the form (\ref{eq:metric}), anisotropic
stresses are negligible and $A^{\mu}$ is aligned with the flow of time
$t^{\mu}$, then one can find appropriate values of the $c_{A}$ and $\ell$ such
that $K$ is dominated by 
a term equal to
$|\nabla\Psi|^{2}/a_{0}^{2}$. This influence then leads to a modification to the
time-time component of Einstein's equations: instead of reducing to Poisson's
equation,
one recovers an equation of the form (\ref{eq:mond}). Therefore the models are
successful covariant realizations of MOND.

Interestingly, in the FLRW limit $\Phi,\Psi\rightarrow 0$, the time-time
component of Einstein's equations in the GEA model becomes a modified Friedmann
equation:
\begin{eqnarray}
\label{modfr}
\beta\left(\frac{H^{2}}{a_{0}^{2}}\right)H^{2} = \frac{8\pi G\rho}{3}
\end{eqnarray} 
where the function $\beta$ is related to $F$ and its derivatives with respect to
$K$.
The dynamics in galaxies prefer a value $a_{0}$ of the order the Hubble
parameter today $H_{0}$ (\cite{SandersMcGaugh2002}) and so one typically 
gets  a modification to the background expansion with a characteristic scale
$H_{0}$ i.e. the scale associated with modified gravity
models that produce dark energy effects. Ultimately the GEA model is a
phenomenological one and as such there currently lack deeper reasons to favour
any 
particular form of $F$. However, one may gain insight into the possible
solutions of (\ref{modfr}) by looking at simple forms for $F$. In
(\cite{Zuntz:2010jp})
the monomial case $F\propto K^{n_{ae}}$ was considered where the kinetic index
$n_{ae}$ was allowed to vary. Solutions with accelerated expansion were found that
could mimic dark energy.

Returning to the original motivation behind the theory, the next step is to look
at the theory on cosmological scales and see whether the GEA
models are realistic alternatives to dark matter. As emphasized, the additional
structure in spacetime is dynamical and so possesses
independent degrees of freedom. As the model is assumed to be uncoupled to other
matter, the gravitational field equations would regard
the influence of these degrees freedom as a type of dark matter (possibly
coupled non-minimally to gravity, and not necessarily `cold').

The possibility that the model may then be a viable alternative to the dark
sector in background cosmology and linear cosmological 
perturbations has been explored in depth in
(\cite{ZlosnikFerreiraStarkman2008}), (\cite{Li:2007vz})
and (\cite{Zuntz:2010jp}). As an alternative to dark matter,
it was found that the GEA models could replicate some but not all of the
following features of cold dark matter: 
influence on background dynamics of the universe;
negligible sound speed of perturbations;
growth rate of dark matter `overdensity';
absence of anisotropic stress and contribution to the cosmological Poisson
equation;
effective minimal coupling to the gravitational field.
When compared to the data from large scale structure and the CMB, the model
fared significantly less well than the Concordance Model
and so is excluded. If one relaxes the requirement that the vector field is
responsible for the effects of cosmological dark matter, one can 
look at the model as one responsible only for the effects of dark energy. It was
found  (\cite{Zuntz:2010jp}) that the current most stringent constraints
on the model's success as dark energy were from constraints on the size of large
scale CMB anisotropy. Specifically, possible variation in $w(z)$ of the
`dark energy' along with new degrees of freedom sourcing anisotropic stress in
the perturbations was found to lead to new, 
non-standard time variation of the potentials $\Phi$ and $\Psi$. These time
variations source large scale anisotropies via the integrated
Sachs-Wolfe effect, and the parameter space of the model is constrained in
avoiding the effect becoming too pronounced. 

In spite of this, given the status of current experimental bounds it is 
conceivable that a more successful alternative to the dark sector may share some
of
these points of departure from the Concordance Model and yet fare significantly
better at the level of the background and linear perturbations.

%% file: de_mg/TeVeS.tex
\subsection{The Tensor-Vector-Scalar Theory of Gravity}

Another proposal for a theory of modified gravity arising from Milgrom's observation is the Tensor-Vector-Scalar Theory of Gravity, or TeVeS.
TeVeS theory is {\it bimetric} with two frames:  the 
"Geometric frame" for the gravitational fields, and the "Physical frame", for the matter fields.  The three gravitational fields are the metric $\metE_{ab}$ (with connection $\tilde{\nabla}_a$) 
that we refer to as the Geometric metric, 
the  vector field $A_a$ and the scalar field $\phi$.
The action for all matter fields, uses a single physical metric 
$g_{ab}$ (with connection $\nabla_a$). The two metrics are related via an algebraic, disformal relation~\citep{Bekenstein1993} as
\begin{equation}
   \metM_{ab} = e^{-2\phi}\metE_{ab} - 2\sinh(2\phi)A_a A_b.
   \label{eq:metric_relation}
\end{equation}
Just like in the Generalized Einstein-Aether Theories, the  vector field is further enforced to be unit-timelike with respect to the geometric metric, i.e.
\begin{equation}
 \metE^{ab} A_a  A_b = A^a A_a = -1
\label{eq:A_unit}
\end{equation}
The theory is based on an action $S$, which is split as $S = S_{\metE} + S_A + S_{\phi}+S_m$
where
\begin{equation}
   S_{\metE} = \frac{1}{16\pi G}\int d^4x \; \volE \; \RiemE,
\label{eq:S_EH}
\end{equation}
where $\metE$ and $\RiemE$ are \label{def_detE} \label{def_RE}  the determinant and scalar curvature of $\metE_{\mu\nu}$ respectively and
$G$ \label{def_Gbare} is the bare gravitational constant, 
\begin{equation}
	S_A = -\frac{1}{32\pi G}  \int d^4x \; \volE \; \left[ K F^{ab}F_{ab}   - 2\lambda (A_a A^a + 1)\right],
\end{equation}
where  $F_{ab} = \nabla_a A_b - \nabla_b A_a$ leads to a Maxwellian kinetic term
and $\lambda$ \label{def_lambda} is a Lagrange multiplier ensuring the unit-timelike constraint on $A_a$ and $K$
is a   dimensionless constant (note that  indices on $F_{ab}$ are raised using the geometric metric, i.e. $F^a_{\;\;b} = \metE^{ac} F_{cb}$) and
\begin{equation}
    S_{\phi} = -\frac{1}{16\pi G} \int d^4x  \volE \left[ 
    \mu \; \metS^{ab}\connE_a\phi \connE_b\phi +   V(\mu) \right]
\end{equation}
 where  $\mu$ \label{def_mu} is a non-dynamical dimensionless scalar field, $\metS^{ab} = \metE^{ab} - A^a A^b$
  and $V(\mu)$ is an arbitrary  function which typically depends on a scale $\ell_B$. 
  The matter is coupled only to the physical metric $\metM_{ab}$ and  defines the matter stress-energy tensor $T_{ab}$ through
 $\delta S_m = -\frac{1}{2} \int d^4x \volM \; T_{ab} \; \delta\metM^{ab}$. The TeVeS action can be written entirelly in the physical 
frame~\citep{ZlosnikFerreiraStarkman2006,Skordis2009a} or in a diagonal frame~\citep{Skordis2009a} where the scalar and vector fields decouple.

In a Friedman Universe, the cosmological evolution is governed by the Friedmann equation
\begin{equation}
3\tilde{H}^2 = 8\pi G e^{-2\phi} \left( \rho_\phi + \rho\right)
\label{eq_beke_friedmann}
\end{equation}
where $\tilde{H}$ is the Hubble rate in terms of the geometric scale factor, $\rho$ is the physical matter density
 which obeys the energy conservation equation with respect to the physical metric 
and where the scalar field energy density is
\begin{equation}
\rho_\phi = \frac{e^{2\phi}}{16\pi G}\left( \mu \Vp + V \right) 
\end{equation}
Exact analytical and numerical solutions with the Bekenstein free function  have been found in~\cite{SkordisEtAl2006}
and in ~\cite{DodelsonLiguori2006}. It turns out that  
energy density tracks the matter fluid energy density.  
The ratio of the energy density of the scalar field to that of ordinary matter is approximately constant, so that the scalar field  exactly tracks the matter dynamics.
In realistic situations, the radiation era tracker is almost never realized, as has been noted by Dodelson and Liguori, but rather $\rho_\phi$ is subdominant
and slowly-rolling and $\phi \propto a^{4/5}$.
~\cite{BourliotEtAl2006} studied more general free functions which have the Bekenstein function as a special case and found a whole range of behaviour, from tracking and accelerated expansion to finite time singularities. ~\cite{Diaz-RiveraSamushiaRatra2006} have 
studied cases where the cosmological TeVeS  equations lead to inflationary/accelerated expansion solutions. 

Although no further studies of accelerated expansion in TeVeS have been performed, it is very plausible that certain choices of function will
inevitably lead to acceleration. It is easy to see that the scalar field action has the same form 
as a k-essence/k-inflation~\citep{Armendariz-PiconMukhanovSteinhardt2000} action which has been considered as
a candidate theory for acceleration.
It is unknown in general whether this has similar features as the uncoupled k-essence, although Zhao's study
indicates that this a promising research direction~\citep{Zhao2006a}. 

In TeVeS, cold dark matter is absent.  Therefore in order to get acceptable values for the physical Hubble constant today (i.e. around $H_0 \sim 70 Km/s/Mpc$) , 
we have to supplement the absence of CDM with something else.  Possibilities include the scalar field itself, massive neutrinos~\citep{SkordisEtAl2006,FerreiraSkordisZunkel2008}
and a cosmological constant. At the same time, one has to get the right angular diameter distance to  recombination~\citep{FerreiraSkordisZunkel2008}.
These two requirements can place severe constraints on the allowed free functions.

\begin{figure}
\includegraphics[width=6.5cm]{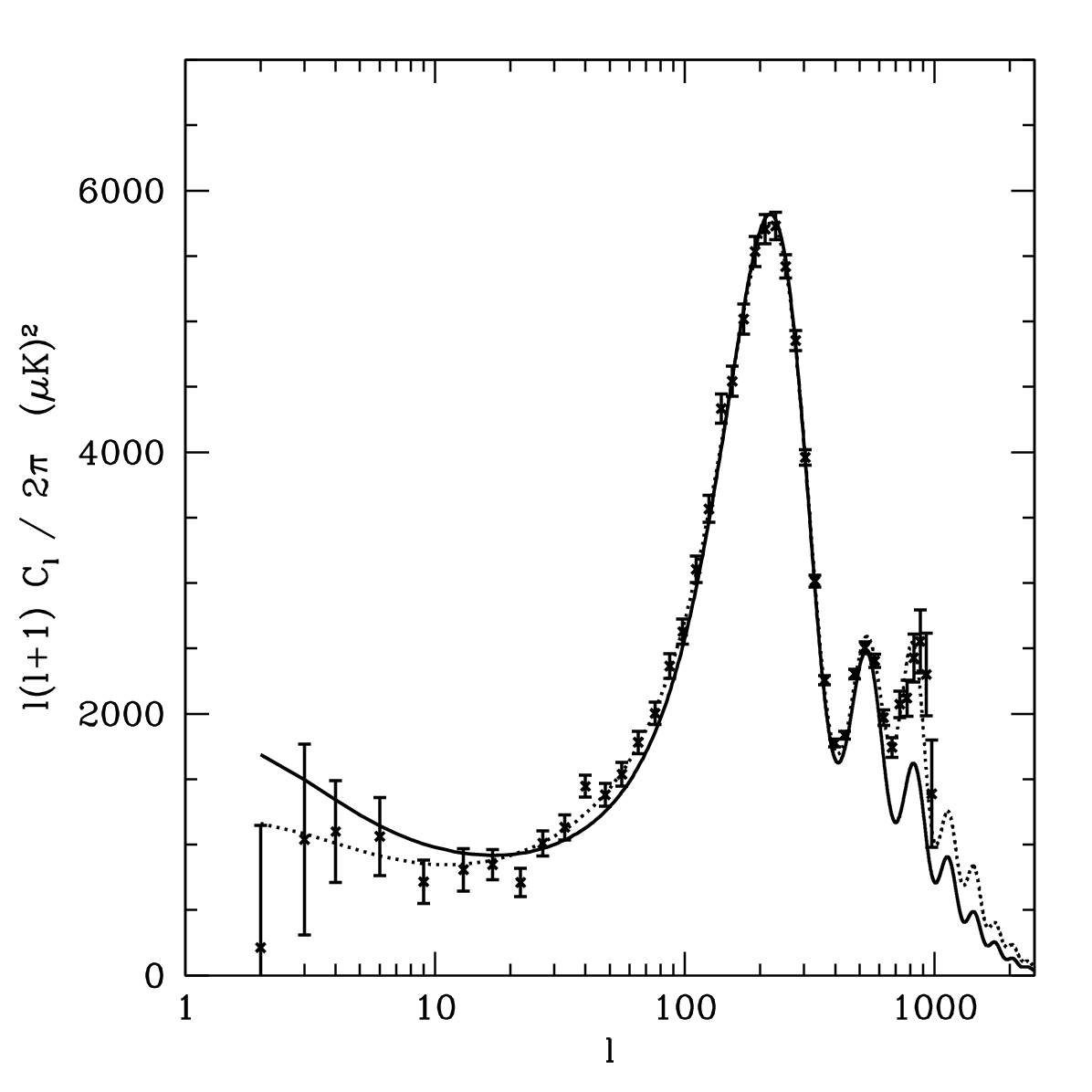}%
\includegraphics[width=6.5cm]{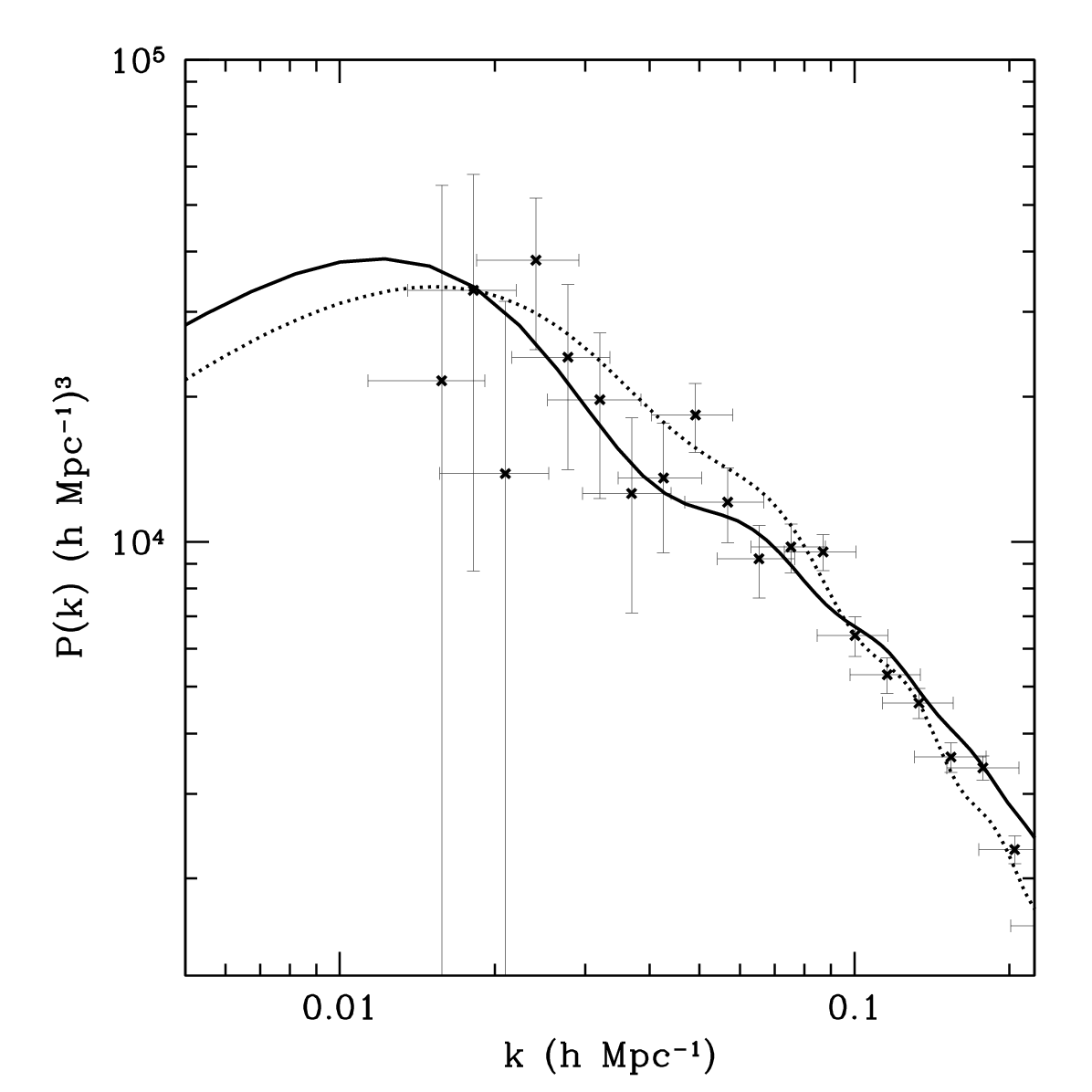}%
\caption{
LEFT: the Cosmic Microwave Background angular power spectrum $l(l+1)C_l/(2\pi)$ for TeVeS (solid) and $\Lambda$CDM (dotted) with 
WMAP 5-year data~\citep{NoltaEtAl2008}.
\newline
RIGHT: the matter power spectrum $P(k)$ for TeVeS (solid) and $\Lambda$CDM (dotted) plotted with SDSS data.
\label{fig_Pk}
 }
\end{figure}
Until TeVeS was proposed and studied in detail, MOND-type theories were assumed to be fatally flawed:  their lack of a dark matter component would necessarily prevent the formation of  large scale structure
compatible with current observational data.  In the case of an Einstein Universe,
it is well known that, since baryons are coupled to photons before recombination they do not have enough time to grow into structures on their own. 
In particular, on scales smaller than the diffusion damping scale perturbations in such a Universe are exponentially
damped due to the Silk-damping effect. CDM solves all of these problems because it does not couple to photons and therefore can start creating potential
wells early on, to which the baryons fall into. 

TeVeS contains two additional fields, which change the structure of the equations significantly. 
The first study of TeVeS predictions for Large Scale Structure observations was conducted in  ~\cite{SkordisEtAl2006}.
They found that TeVeS can indeed form large scale structure compatible with observations depending
on the choice of TeVeS parameters in the free function. In fact the form of the matter power spectrum $P(k)$ in TeVeS looks  quite similar to that in
$\Lambda$CDM. Thus TeVeS can produce matter power spectra that cannot be distinguished from $\Lambda$CDM by current observations. 
One would have to turn to other observables to distinguish the two models. The power spectra for TeVeS and $\Lambda$CDM are plotted on the right panel of Figure \ref{fig_Pk}. \cite{DodelsonLiguori2006} provided an analytical explanation of the growth of structure seen numerically by~\cite{SkordisEtAl2006} and found that the growth in TeVeS is due to the
vector field perturbation.

It is premature to claim (as
in~\cite{SlosarMelchiorriSilk2005,SpergelEtAl2006}) 
that only a theory with CDM can fit CMB observations; a prime example to the
contrary is the EBI theory~\cite{BanadosFerreiraSkordis2008}.
Nevertheless, in the case of TeVeS \cite{SkordisEtAl2006}
numerically solved the linear Boltzmann equation in the case of TeVeS and
calculated the CMB angular power
spectrum for TeVeS. By using initial conditions close to adiabatic the spectrum
thus found provides very poor fit as compared to
the $\Lambda$CDM model (see the left panel of Figure \ref{fig_Pk}). The  CMB
seems to put TeVeS into trouble, at least for the Bekenstein free function.
The result of \cite{DodelsonLiguori2006} has a further direct consequence. The
difference $\Phi -\Psi$, sometimes named the gravitational slip (see  Sec. 
\ref{mg_growth_params}),
 has additional contributions coming from the 
perturbed vector field $\alpha$. 
Since the vector field is required to grow in order to drive structure
formation, it will inevitably  lead to a growing $\Phi - \Psi$. 
If the difference $\Phi - \Psi$
 can be measured observationally, it can provide a substantial test of TeVeS
that can distinguish TeVeS from $\Lambda$CDM.

%% file: de_mg/lessons.tex
This section explores some generic issues that are not connected to particular models (although we use some 
specific models as examples).
First, we ask ourselves  to which precision we should measure $w$ in order
to make a significant progress in understanding dark energy. Second, we 
discuss the role of the anisotropic stress in  distinguishing between dark energy and modified gravity models. 
Finally, we present some general consistency relations among the perturbation variables that all
models of modified gravity should fulfill.

\subsection{To which precision should we measure $w$ ?} \label{twpswmw}

Two crucial questions that are often asked in the context of
dark energy surveys:
\begin{itemize}
\item Since $w$ is so close to $-1$, do we not already know that
the dark energy is a cosmological constant?
\item To which precision should we measure $w$? Or equivalently,
why is the Euclid target precision of about $0.01$ on $w_0$ and $0.1$
on $w_a$ interesting?
\end{itemize}

In this section we will attempt to answer these questions at least
partially, in two different ways. We will start by examining whether
we can draw useful lessons from inflation, and then we will look at
what we can learn from arguments based on Bayesian model comparison.

{\color{red}  In the first part we will see that for single field slow-roll inflation models we effectively measure $w \sim -1$ with percent-level accuracy (see Fig.\ref{fig:w_k}); however, the deviation from a scale invariant spectrum means that we nonetheless observe a dynamical evolution and thus a deviation from an exact and constant equation of state of w=-1. 
Therefore, we know that inflation was not due to a cosmological constant; we also know that we can see no deviation from  a de Sitter expansion for a precision smaller than the one Euclid will reach.}

{\color{red}  In the second part we will consider the Bayesian evidence in favour of a true cosmological constant if we keep finding w=-1; we will see that for priors on $w_0$ and $w_a$ of order unity, a precision like the one for Euclid is necessary to favour a true cosmological constant decisively. We will also discuss how this conclusion changes depending on the choice of priors.}

\subsubsection{Lessons from inflation}

In all probability the observed late-time acceleration of the Universe
is not the first period of accelerated expansion that occurred during its evolution: the current standard
model of cosmology incorporates a much earlier phase with $\ddot{a}>0$,
called inflation. Such a period provides a natural explanation for several
late-time observations:
\begin{itemize}
\item Why is the Universe very close to being spatially flat?
\item Why do we observe homogeneity and isotropy on scales that were
naively never in causal contact?
\item What created the initial fluctuations?
\end{itemize}
In addition, inflation provides a mechanism to get rid of unwanted relics
from phase transitions in the early universe, like monopoles, that arise
in certain scenarios (e.g. grand-unified theories).

While there is no conclusive proof that an inflationary phase took place
in the early universe, it is surprisingly difficult to create the observed
fluctuation spectrum in alternative scenarios that are strictly causal and
only act on sub-horizon scales \citep{Spergel:1997vq,Scodeller:2009iu}.

If however inflation took place, then it seems natural to ask the question
whether its observed properties appear similar to the current knowledge about
the dark energy, and if yes, whether we can use inflation to learn something
about the dark energy. The first lesson to draw from inflation is that
it was not due to a pure cosmological constant. This is immediately clear
since we exist: inflation ended. We can go even further: if Planck confirms 
the observations of a deviation from a scale invariant initial spectrum
($n_s\neq 1$) of WMAP \citep{Komatsu:2010fb} then this excludes an exactly exponential
expansion during the observable epoch and thus also a temporary, effective
cosmological constant.

If there had been any observers during the observationally accessible
period of inflation, what would they have been seeing? Following the
analysis in \citet{Ilic:2010zp}, we notice that
\begin{equation}
1+w = - \frac{2}{3} \frac{\dot{H}}{H^2} = \frac{2}{3} \varepsilon_H \;,
\end{equation}
where $\epsilon_H \equiv 2 M^2_{\rm Pl}(H'/H)^2$ and here the prime denotes a derivative with respect to the inflaton field.
Since therefore the tensor-to-scalar ratio is linked to the equation
of state parameter through $r \sim 24(1+w)$ we can immediately conclude
that no deviation of from $w=-1$ during inflation has been observed so
far, just as no such deviation has been observed for the contemporary
dark energy. At least in this respect inflation and the dark energy
look similar. However, we also know that
\begin{equation}
\frac{d \ln(1+w)}{d N} = 2 (\eta_H - \varepsilon_H) \label{eq:dw}
\end{equation}
where $\eta_H \equiv 2 M^2_{\rm Pl} H''/H $ is related to the scalar spectral index by $2\eta_H = (n_s-1)+4\varepsilon_H$. Thus if $n_s\neq1$ we have
that either $\eta_H\neq0$ or $\varepsilon_H\neq0$, and consequently
either $w\neq-1$ or $w$ is not constant. 

\begin{figure}
\begin{centering}
\scalebox{0.7}{\includegraphics{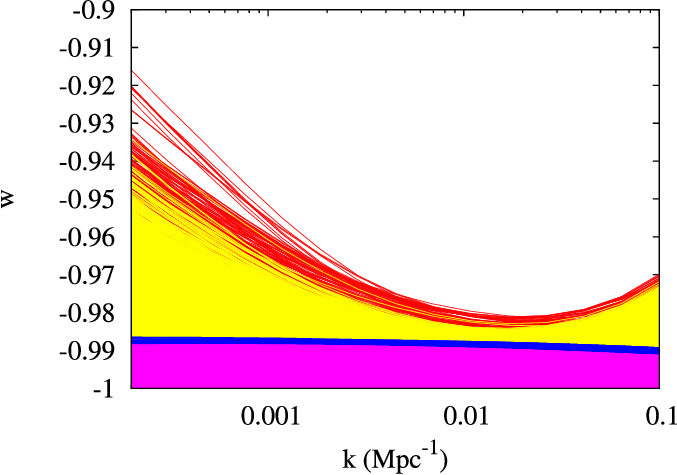}}
\caption{The evolution of $w$ as a function of the comoving scale $k$, using only the 5-year WMAP CMB data. Red and yellow are the 95\% and 68\% confidence regions for the LV formalism. Blue and purple are the same for the flow-equation formalism.
From the outside inward, the colored regions are red, yellow, blue, and purple. [Figure from Ilic et al 2010]
}
\label{fig:w_k}
\end{centering}
\end{figure}

As already said earlier, we conclude that inflation
is not due to a cosmological constant. However, an observer back then
would nonetheless have found $w\approx -1$. Thus, observation of
$w\approx -1$ (at least down to an error of about 0.02, see Figure~\ref{fig:w_k})
does not provide a very strong reason to believe that we are dealing with a
cosmological constant.

We can rewrite Eq.~(\ref{eq:dw}) as
\begin{equation}
(1+w) = - \frac{1}{6} (n_s-1) + \frac{\eta_H}{3}
\approx 0.007 +  \frac{\eta_H}{3} .
\end{equation}
Naively it would appear rather fine-tuned if $\eta_H$ precisely cancelled
the observed contribution from $n_s-1$. Following this line of reasoning,
if $\varepsilon_H$ and $\eta_H$ are of about the same size, then we would
expect $1+w$ to be about $0.005$ to $0.015$, well within current experimental
bounds and roughly at the limit of what Euclid will be able to observe.

\begin{figure}
\begin{centering}
\scalebox{0.8}{\includegraphics{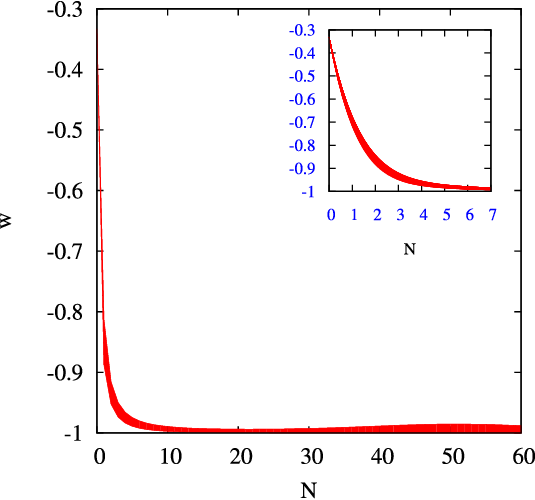}}
\caption{The complete evolution of $w(N)$, from the flow-equation results accepted by the CMB likelihood. Inflation is made to end at $N=0$ where $w(N=0)=-1/3$ corresponding to $\epsilon_H(N=0)=1$. For our choice of
priors on the slow-roll parameters at $N=0$, we find that $w$ decreases
rapidly towards $-1$ (see inset) and stays close to it during the period when the observable
scales leave the horizon ($N\approx 40 - 60$). [Figure from Ilic et al 2010]}
\label{fig:w_N}
\end{centering}
\end{figure}

However, this last argument is highly speculative, and at least for inflation
we know that there are classes of models where the cancellation is indeed natural, 
which is why one cannot give a lower limit for the amplitude of primordial gravitational
waves. On the other hand, the observed period of inflation is probably in the middle
of a long slow-roll phase during which $w$ tends to be close to $-1$ (cf Fig.~\ref{fig:w_N}), while
near the end of inflation the deviations become large. Additionally, inflation happened
at an energy scale somewhere between 1 MeV and the Planck scale, while the energy scale
of the late time accelerated expansion is of the order of $10^{-3}$ eV. At least in this
respect the two are very different.

\subsubsection{Higgs-Dilaton Inflation: a connection between the Early and Late Universe Acceleration}

{\color{red} Despite previous arguments, it is natural to ask for a connection between the two known
acceleration periods. In fact, in the last few years there has been a renewal of model building 
in inflationary cosmology by considering the fundamental Higgs as the inflaton field~\cite{Bezrukov_etal_2007}.
Such an elegant and economical model can give rise to the
observed amplitude of CMB anisotropies when we include a large non-minimal coupling of the 
Higgs to the scalar curvature. In the context of quantum field theory, the running of the Higgs
mass from the electroweak scale to the Planck scale is affected by this non-minimal coupling in
such a way that the beta function of the Higgs' self-coupling vanishes at an intermediate scale
($\mu\sim 10^{15}$ GeV), if the mass of the Higgs is precisely 126 GeV, as measured at the LHC. 
This partial fixed point (other beta functions do not vanish) suggests an enhancement of symmetry 
at that scale, and the presence of a Nambu-Goldstone boson (the dilaton field) associated with the 
breaking of scale invariance~\cite{Shaposhnikov_etal_2009}. In a subsequent paper~\cite{bellido_etal_2011}, the Higgs-Dilaton scenario 
was explored in full detail. The model predicts a bound on the scalar spectral index, $n_s<0.97$, 
with negligible associated running, $-0.0006 < d\ln n_s/d\ln k < 0.00015$, and a scalar to tensor 
ratio, $0.0009 < r < 0.0033$, which could be within reach of the Planck satellite mission. Moreover,
the model predicts that, after inflation, the dilaton plays the role of a thawing quintessence field,
whose slow motion determines a concrete relation between the early universe fluctuations and the 
equation of state of dark energy, $3(1+w) = 1-n_s > 0.03$, which could be within reach of Euclid
satellite mission~\cite{bellido_etal_2011}. Furthermore, within the HDI model, there is also a relation between 
the running of the scalar tilt and the variation of $w(a)$, $d\ln n_s/d\ln k = 3w_a$, a prediction that 
can easily be ruled out with future surveys.}

{\color{red} These relationships between early and late universe acceleration parameters constitute a
fundamental physics connection within a very concrete and economical model, where the Higgs
plays the role of the inflaton and the dilaton is a thawing quintessence field, whose dynamics 
has almost no freedom and satisfies all of the present constraints~\cite{bellido_etal_2011}.}

\subsubsection{When should we stop: Bayesian model comparison}

In the previous section we saw that inflation provides an argument why an observation
of $w \approx -1$ need not support a cosmological constant strongly. Let us now investigate
this argument more precisely with Bayesian model comparison. One model, $M_0$, posits that
the accelerated expansion is due to a cosmological constant. The other models
assume that the dark energy is dynamical, in a way that is well parametrised  either by 
an arbitrary constant $w$ (model $M_1$) or by a linear fit $w(a)=w_0+(1-a) w_a$ (model $M_2$).
Under the assumption that no deviation from $w=-1$ will be detected in the future, at
which point should we stop trying to measure $w$ ever more accurately? The relevant
target here is to quantify at what point we will be able to rule out an entire class of theoretical 
dark energy models (when compared to $\Lambda$CDM) at a certain threshold for the strength of evidence. 

Here we are using the constant and linear parametrisation of $w$ because on the one
hand we can consider the constant $w$ to be an effective quantity, averaged over redshift with the appropriate weighting
factor for the observable, see \citealt{Simpson:2006bd},
and on the other hand because the
precision targets for observations are conventionally phrased in terms of the figure
of merit (FoM) given by $1/\sqrt{|{\rm Cov}(w_0,w_a)|}$. We will therefore find a
direct link between the model probability and the FoM. It would be an interesting
exercise to repeat the calculations with a more general model, using e.g. PCA, although
we would expect to reach a similar conclusion.

Bayesian model comparison aims to compute the relative model probability
\begin{equation}
\frac{P(M_0|d)}{P(M_1|d)} = \frac{P(d|M_0)}{P(d|M_1)} \frac{P(M_0)}{P(M_1)} 
\end{equation}
where we used Bayes formula and where $B_{01}\equiv P(d|M_0)/P(d|M_1)$ is called
the Bayes factor. The Bayes factor is the amount by which our relative
belief in the two models is modified by the data, with $\ln B_{01} >
(<0)$ indicating a preference for model 0 (model 1). Since the model $M_0$ is nested in $M_1$ at the point
$w=-1$ and in model $M_2$ at $(w_0=-1,w_a=0)$, we can use the Savage-Dickey (SD) density ratio \citep[e.g.][]{Trotta:2005ar}.
Based on SD, the Bayes factor between the two models is just the ratio of
posterior to prior at $w=-1$ or at $(w_0=-1,w_a=0)$, marginalised over all other parameters.

Let us start by following Trotta (2009) and consider the Bayes factor $B_{01}$ between a cosmological
constant model $w=-1$ and a free but constant effective $w$.
If we assume that the data are
compatible with $\weff=-1$ with an uncertainty $\sigma$, then the
Bayes factor in favour of a cosmological constant is given by
 \begin{equation} \label{eq:B}
 B = \sqrt{\frac{2}{\pi}}\frac{\Delta_{+} + \Delta_{-}}{\sigma}
 \left[\text{erfc}\left(-\frac{\Delta_+}{\sqrt{2}\sigma}\right)
- \text{erfc}\left(\frac{\Delta_-}{\sqrt{2}\sigma}\right)
  \right]^{-1},
 \end{equation}
where for the evolving dark energy model we have adopted a flat
prior in the region $-1 - \Delta_{-} \leq \weff \leq -1+\Delta_+$
and we have made use of the Savage--Dickey density ratio formula
\citep[see][]{Trotta:2005ar}. The prior, of total width $\Delta =
\Delta_+ + \Delta_-$, is best interpreted as a factor describing
the predictivity of the dark energy model under consideration. For
instance, in a model where dark energy is a fluid with a negative
pressure but satisfying the strong energy condition we have that
$\Delta_+ = 2/3, \Delta_- = 0$. On the other hand, phantom models
will be described by $\Delta_+ = 0, \Delta_- > 0$, with the latter
being possibly rather large. A model with a large $\Delta$ will be more generic and
less predictive, and therefore is disfavoured by the Occam's razor
of Bayesian model selection, see Eq.~\eqref{eq:B}. According to the Jeffreys' scale for the strength
of evidence, we have a moderate (strong) preference for the
cosmological constant model for $2.5 < \ln B_{01} < 5.0$ ($\ln B_{01}>5.0$),
corresponding to posterior odds of $12:1$ to $150:1$ (above
$150:1$). 
\begin{figure}[tb]
\centering
\includegraphics[width=.5 \linewidth]{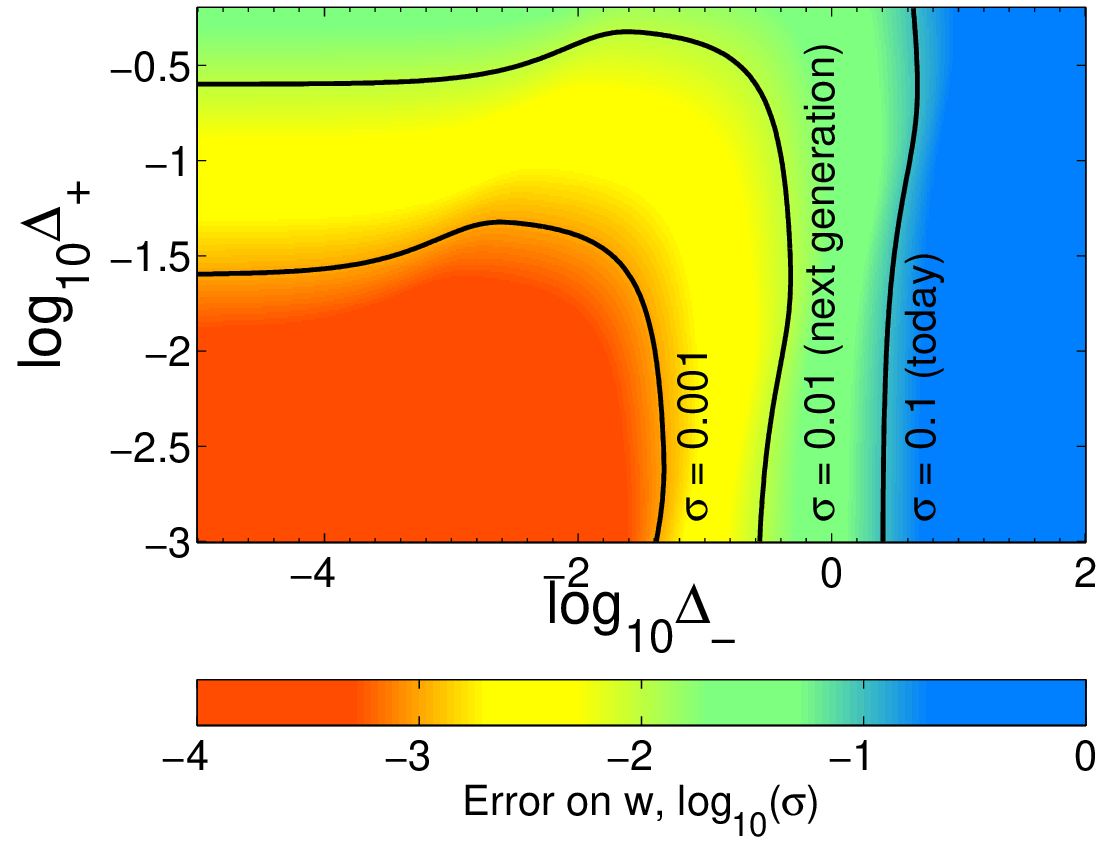}
\caption{Required accuracy on $\weff = -1$ to obtain strong
evidence against a model where $-1 - \Delta_{-} \leq \weff \leq
-1+\Delta_+$ as compared to a cosmological constant model, $w=-1$.
For a given $\sigma$, models to the right and above the contour
are disfavoured with odds of more than $20:1$.  } \label{fig:sigma}
\end{figure}
\begin{table}
\centering
\begin{tabular}{l l|l }
 Model & $(\Delta_+ , \Delta_- )$ & $\ln B$ today ($\sigma = 0.1$)
 \\\hline
 Phantom & $( 0, 10)$      & $4.4$ (strongly disfavoured)\\
 Fluid--like & $(2/3,  0)$ & $1.7$ (slightly disfavoured) \\
 Small departures & $( 0.01, 0.01)$ & $0.0$ (inconclusive) \\
\end{tabular}
\caption{Strength of evidence disfavouring the three benchmark
models against a cosmological constant model, using an indicative
accuracy on $w=-1$ from present data, $\sigma \sim 0.1$.
\label{table:evidence}}
\end{table}

We plot in Fig.~\ref{fig:sigma} contours of constant observational
accuracy $\sigma$ in the model predictivity space
$(\Delta_-,\Delta_+)$ for $\ln B = 3.0$ from Eq.~\eqref{eq:B},
corresponding to odds of 20 to 1 in favour of a cosmological
constant (slightly above the ``moderate'' threshold. The figure can be interpreted as giving the space of
extended models that can be significantly disfavoured with respect
to $w=-1$ at a given accuracy. The results for the 3 benchmark models mentioned above
(fluid--like, phantom or small departures from $w=-1$) are
summarized in Table~\ref{table:evidence}.  Instead, we can ask the question 
which precision needs to reached to support $\Lambda$CDM at a given
level. This is shown in table \ref{table:sigma} for odds 20:1 and 150:1. We see that
to rule out a fluid-like model, which also covers the parameter space expected for canonical
scalar field dark energy, we need to reach a precision comparable to the one that
the Euclid satellite is expected to attain.

\begin{table}
\centering
\begin{tabular}{l l|l c }
 Model & $(\Delta_+ , \Delta_- )$ & \multicolumn{2}{c}{Required $\sigma$ for odds} \\
       &                          & $>20:1$  & $>150:1$  \\\hline
 Phantom & $( 0, 10)$      & $0.4$           & $5\cdot10^{-2}$\\
 Fluid--like & $(2/3,  0)$ & $3\cdot10^{-2}$ & $3\cdot10^{-3}$\\
 Small departures & $( 0.01, 0.01)$ & $4\cdot10^{-4}$ & $5\cdot10^{-5}$\\
\end{tabular}
\caption{Required accuracy for future surveys in order to
disfavour the three benchmark models against $w=-1$ for two different strengths of evidence.
\label{table:sigma}}
\end{table}

By considering the model $M_2$ we can also provide a direct link with the target DETF FoM:
Let us choose (fairly arbitrarily) a flat probability distribution for the prior, of width
$\Delta w_0$ and $\Delta w_a$ in the dark energy parameters, so that the value of the prior is $1/(\Delta w_0 \Delta w_a)$
everywhere. Let us assume that the likelihood is Gaussian in $w_0$ and $w_a$ and centred
on $\Lambda$CDM (i.e. the data fully supports $\Lambda$ as the dark energy). 

As above, we need to distinguish different cases depending on the width of the prior.
If you accept the argument of the previous section that we expect only a small deviation from $w=-1$,
and set a prior width of order $0.01$ on both $w_0$ and $w_a$, then the posterior is dominated by the
prior, and the ratio will be of order 1 if the future data is compatible with $w=-1$. Since the precision of the experiment is
comparable to the expected deviation, both $\Lambda$CDM and evolving dark energy are equally probable (as
argued above and shown for model $M_1$ in table \ref{table:evidence}), 
and we have to wait for a detection of $w\neq-1$ or a significant further 
increase in precision (cf the last row in table \ref{table:sigma}).

However, one often considers a much wider range for $w$, for example the fluid-like model with $w_0\in [-1/3,-1]$
and $w_a \in [-1,1]$ with equal probability (and neglecting some subtleties near $w=-1$). If the likelihood
is much narrower than the prior range, then the value of the normalised posterior at $w=-1$ will be
$2/(2\pi\sqrt{|{\rm Cov}(w_0,w_a)|}={\rm FoM}/\pi$ (since we excluded $w<-1$, else it would half this value).
The Bayes factor is then given by
\begin{equation}
B_{01} = \frac{\Delta w_0 \Delta w_a {\rm FoM}}{\pi} .
\end{equation}
For the prior given above, we end up with $B_{01} \approx 4 {\rm FoM}/(3\pi) \approx 0.4 {\rm FoM}$. In order
to reach a ``decisive'' Bayes factor, usually characterised as $\ln B > 5$ or $B > 150$, we thus need a figure of
merit exceeding 375. Demanding that Euclid achieve a FoM $> 500$ places us therefore on the safe side and
allows to reach the same conclusions (the ability to favour $\Lambda$CDM decisively {\em if} the data is in full
agreement with $w=-1$) under small variations of the prior as well.

A similar analysis could be easily carried out to
compare the cosmological constant model against departures from
Einstein gravity, thus giving some useful insight into the
potential of future surveys in terms of Bayesian model selection.

 To summarize,  we used inflation as a dark energy prototype to show that the current experimental
bounds of $w \approx -1.0 \pm 0.1$ are not yet sufficient to significantly favour a cosmological constant
over other models.  In addition, even when expecting a deviation of $w=-1$
of order unity, our current knowledge of $w$ does not allow us to favour $\Lambda$ strongly in a Bayesian
context. Here we showed that we need to reach a percent level accuracy both
to have any chance of observing a deviation of $w$ from $-1$ if the dark energy is similar to inflation, and because it is at this
point that a cosmological constant starts to be favoured decisively for prior widths of order $1$.
{\color{red}In either scenario, we do not expect to be able to improve much our knowledge with a lower precision
measurement of $w$. The dark energy can of course be quite different from the inflaton and may lead to larger deviations from $w=-1$. This indeed would be the preferred situation for Euclid, as then we will be able to investigate much more easily the physical origin of the accelerate expansion. We can however have departures from $\Lambda$CDM even if w is very close to -1 today.
In fact most present models of modified gravity and dynamical dark energy have a value of $w_0$ which is asymptotically close to -1 (in the sense that large departures from this value is already excluded). In this sense, for example, early dark energy parametrizations ($\Omega_e$) test the amount of dark energy in the past, which can still be non negligible (ex. \cite{pettorino_etal_2013}). Similarly, a fifth force can lead to a background similar to LCDM but different effects on perturbations and structure formation \cite{Baldi_etal_2010}.
}

%% file: de_mg/smoking.tex
\subsection{The effective anisotropic stress as evidence for modified
gravity} 

\label{subsection:anisotropic_stress}

As discussed in Sec.\ref{models-of-modified-gravity}, all dark energy and modified gravity
models can be described with the same effective metric degrees of freedom. This
makes it impossible in principle to distinguish clearly between the two possibilities with
cosmological observations alone. But while the cleanest tests would come from
laboratory experiments, this may well be impossible to achieve. We would expect
that model comparison analyses would still favour the correct model as it should
provide the most elegant and economical description of the data. However, we may
not know the correct model a priori, and it would be more useful if we could
identify generic differences between the different classes of explanations,
based on the phenomenological description that can be used directly to analyse
the data.

Looking at the effective energy momentum tensor of the dark energy sector, we
can either try to find a hint in the form of the pressure perturbation $\delta
p$ or in the effective anisotropic stress $\pi$. {\color{red}Whilst all} scalar field dark
energy affects $\delta p$ (and for multiple fields with different sound speeds in
potentially quite complex ways), they generically have $\pi = 0$. The opposite
is also true, modified gravity models have generically $\pi\neq0$
\citep{Kunz:2006ca}. {\color{red}Radiation and neutrinos will contribute to anisotropic stress on cosmological scales, but their contribution is safely negligible in the late-time universe.} 
In the following sections we will first look at models with
single extra degrees of freedom, for which we will find that $\pi\neq0$ is a
firm prediction. We will then consider the $f(R,G)$ case as an example for
multiple degrees of freedom \citep{Saltas:2010tt}.

\subsubsection{Modified gravity models with a single degree of freedom}

In the prototypical scalar-tensor theory, where the scalar $\varphi$ is coupled
to R through $F(\varphi) R$, we find that $\pi \propto (F'/F) \delta\varphi$.
This is very similar to the $f(R)$ case for which $\pi\propto (F'/F) \delta R$ (where
now $F=df/dR$). In both cases the generic model with vanishing anisotropic stress
is given by $F'=0$, which corresponds to a constant coupling (for scalar-tensor)
or $f(R) \propto R + \Lambda$. In both cases we find the GR limit. The other
possibility, $\delta\varphi=0$ or $\delta R=0$, imposes a very specific
evolution on the perturbations that in general does not agree with observations.

Another possible way to build a theory that deviates from GR is to use a function of the second-order
Lovelock function, the Gauss-Bonnet term 
$G \equiv R^2 - 4 R_{\mu\nu} R^{\mu\nu} + R_{\alpha\beta\mu\nu}R^{\alpha\beta\mu\nu}$.
The Gauss-Bonnet term by itself is a topological invariant in 4 space-time dimensions
and does not contribute to the equation of motion.
It is useful here since it avoids an Ostrogradski-type instability \citep{Woodard07}.
In $R+f(G)$ models, the situation is slightly more complicated than for the
scalar-tensor case, as 
\be
\pi \sim \Phi - \Psi =  4H \dot{\xi}\Psi  - 4\ddot{\xi}\Phi + 4 \left( H^2 + \dot{H}
\right)\delta \xi
\label{eq:fG}
\ee
where the dot denotes derivative with respect to ordinary time and $\xi=df/dG$ 
(see e.g. \citeauthor{Saltas:2010tt}). An obvious choice to force
$\pi=0$ is to take $\xi$ constant, which leads to $R+G+\Lambda$ in the action,
and thus again to GR in four space-time dimensions . 
There is no obvious way to exploit the extra $\xi$ terms in Eq.
(\ref{eq:fG}), with the exception of curvature dominated evolution and on small
scales (which is not very relevant for realistic cosmologies). 

Finally, in DGP one has, with the notation of \cite{Amendola:2007rr},
\be
\Phi - \Psi = \frac{2 H r_c-1}{1+H r_c (3 H r_c -2 )} \Phi . \label{eq:pi_dgp}
\ee
This expression vanishes for $H r_c = 1/2$ (which is never reached in the usual
scenario in which $H r_c \rightarrow 1$ from above) and for $H r_c \rightarrow
\infty$ (for large $H r_c$ the expression in front of $\Phi$ in (\ref{eq:pi_dgp})  
vanishes like $1/(H r_c)$). In the DGP
scenario the absolute value of the anisotropic stress grows over time and approaches
the limiting value of $\Phi-\Psi = \Phi/2$. {\color{red}The only way to avoid this limit is to set the crossover scale to be unobservably large, \mbox{$r_c \propto
M_4^2/M_5^3 \rightarrow \infty$}}. In this situation the five-dimensional part of the
action is suppressed and we end up with the usual 4D GR action.

In all of these examples only the GR limit has consistently no effective
anisotropic stress in situations compatible with observational results (matter
dominated evolution with a transition towards a state with $w \ll -1/3$).

\subsubsection{Balancing multiple degrees of freedom}

In models with multiple degrees of freedom it is at least in principle possible
to balance the contributions in order to achieve a net vanishing $\pi$.
\cite{Saltas:2010tt} explicitly study the case of $f(R,G)$ gravity (please refer
to this paper for details). The general equation,
\be
\Phi  -  \Psi  = \frac{1}{F}  \left[  \delta F +  4H \dot{\xi}\Psi  -
4\ddot{\xi}\Phi + 4\left( H^2 + \dot{H} \right)\delta \xi\right] ,
\label{eq:noanisogeneral}
\ee
is rather complicated, and generically depends e.g. on scale of the
perturbations (except for $\xi$ constant, which in turn requires $F$ constant
for $\pi=0$ and corresponds again to the GR limit). Looking only at small
scales, $k\gg aH$, one finds
\be
f_{RR} + 16 (H^2+\dot{H}) (H^2 + 2 \dot{H}) f_{GG} +4 (2 H^2+3 \dot{H}) f_{RG} =
0 .  \label{eq:noanisoH}
\ee
It is in principle possible to find simultaneous solutions of this equation and
the modified Friedmann (0-0 Einstein) equation,  for a given $H(t)$. As an
example, the model $f(R,G) = R + G^n R^m$ with 
\be
n = \frac{1}{90} \left( 11 \pm \sqrt{41} \right) \; , \quad m = \frac{1}{180}
\left( 61 \pm 11 \sqrt{41} \right) \label{eq:mattertoyExponents}
\ee
allows for matter dominated evolution, $H=2/(3 t)$, and has no anisotropic
stress. It is however not clear at all how to connect this model to different
epochs and especially how to move towards a future accelerated epoch with
$\pi=0$ as the above exponents are fine-tuned to produce no anisotropic stress
specifically only during matter domination. Additionally, during the transition
to a de Sitter fixed point one encounters generically severe instabilities.

In summary, none of the standard examples with a single extra
degree of freedom discussed above allows for a viable model with $\pi=0$. While
finely balanced solutions can be constructed for models with several degrees of
freedom, one would need to link the motion in model space to the evolution of
the universe, in order to preserve $\pi=0$. This requires even more fine tuning,
and in some cases is not possible at all, most notably for evolution to a de
Sitter state. The effective anisotropic stress appears therefore to be a very
good quantity to look at when searching for generic conclusions on the nature of
the accelerated expansion from cosmological observations.

%% file: de_mg/consistency.tex
 \subsection{Parameterised Frameworks for Theories of Modified Gravity}

As explained in earlier sections of this report, modified gravity models cannot be distinguished
from dark energy models by using solely the FLRW background equations. But by comparing the background expansion rate of the universe with observables that depend on linear perturbations of an FRW spacetime we can hope to distinguish between these two categories of explanations. An efficient way to do this is via a parameterised, model-independent framework that describes cosmological perturbation theory in modified gravity. We present here one such framework, the Parameterised Post-Friedmann formalism (\cite{Baker2013:vx}) \footnote{Not to be confused with a different formalism of the same name by other authors (\cite{Hu:2007pj}).} that implements possible extensions to the linearised gravitational field equations.

The Parameterised Post-Friedmann approach (PPF) is inspired by the Parameterised Post-Newtonian (PPN) formalism (\cite{WillNordtvedt1972,Will1971}), which uses a set of parameters to summarise leading-order deviations from the metric of General Relativity. PPN was developed in the 1970s for the purpose of testing of alternative gravity theories in the Solar System or binary systems, and is valid in weak-field, low-velocity scenarios. 
PPN itself cannot be applied to cosmology, because we do not know the exact form of the linearised metric for our Hubble volume. Furthermore, PPN can only test for constant deviations from General Relativity, whereas the cosmological data we collect contain inherent redshift dependence.

For these reasons the PPF framework is a parameterisation of the gravitational field equations (instead of the metric) in terms of a set of functions of redshift. A theory of modified gravity can be analytically mapped onto these PPF functions, which in turn can be constrained by data.\newline

We begin by writing the perturbed Einstein field equations for spin-0 (scalar) perturbations in the form:
\begin{equation}
\label{abc}
\delta G_{\mu\nu} \;=\; 8\pi G\,\delta T_{\mu\nu}+\delta U_{\mu\nu}^{\mathrm{metric}}+\delta U_{\mu\nu}^{\mathrm{d.o.f}}+\mathrm{gauge\,invariance\,fixing\,terms}
\end{equation}
where $\delta T_{\mu\nu}$ is the usual perturbed stress-energy tensor of all cosmologically-relevant fluids. The tensor $\delta U_{\mu\nu}^{\mathrm{metric}}$ holds new terms that may appear in a modified theory, containing perturbations of the metric (in GR such perturbations are entirely accounted for by $\delta G_{\mu\nu}$). $\delta U_{\mu\nu}^{\mathrm{d.o.f.}}$ holds perturbations of any new degrees of freedom that are introduced by modifications to gravity. A simple example of the latter is a new scalar field, such as introduced by scalar-tensor or Galileon theories. However, new degrees of freedom could also come from spin-0 perturbations of new tensor or vector fields, St$\ddot{\mathrm{u}}$ckelberg fields, effective fluids and actions based on curvature invariants (such as $f\left(R\right)$ gravity).

In principle there could also be new terms containing matter perturbations on the RHS of eq.(\ref{abc}). However, for theories that maintain the weak equivalence principle - ie. those with a Jordan frame where matter is uncoupled to any new fields - these matter terms can be eliminated in favour of additional contributions to $\delta U_{\mu\nu}^{\mathrm{metric}}$ and $\delta U_{\mu\nu}^{\mathrm{d.o.f.}}$.

The tensor $\delta U_{\mu\nu}^{\mathrm{metric}}$ is then expanded in terms of two gauge-invariant perturbation variables $\hat\Phi$ and $\hat\Gamma$. $\hat\Phi$ is one of the standard gauge-invariant Bardeen potentials, whilst $\hat\Gamma$ is the following combination of the Bardeen potentials: \mbox{$\hat\Gamma=1/k (\dot{\hat\Phi}+{\cal H}\hat\Psi)$}. We use $\hat\Gamma$ instead of the usual Bardeen potential $\hat\Psi$ because $\hat\Gamma$ has the same derivative order as $\hat\Phi$ (whereas $\hat\Psi$ does not). We then deduce that the only possible structure of $\delta U_{\mu\nu}^{\mathrm{metric}}$ that maintains the gauge-invariance of the field equations is a linear combination of $\hat\Phi$, $\hat\Gamma$ and their derivatives, multiplied by functions of the cosmological background (see eqs.\ref{FE1}-\ref{FE4} below). 

$\delta U_{\mu\nu}^{\mathrm{d.o.f.}}$ is similarly expanded in a set of gauge-invariant potentials $\{\hat\chi_i\}$ that contain the new degrees of freedom. \cite{Baker2013:vx} presented an algorithm for constructing the relevant gauge-invariant quantities in any theory.

For concreteness we will consider here a theory that contains only one new degree of freedom and is second-order in its equations of motion (a generic but not watertight requirement for stability, see \cite{Woodard2006:cp}). Then the four components of eq.(\ref{abc}) are:
 \begin{align}
-a^2\delta G^0_0&=8\pi a^2 G\,\rho_M\delta_M+A_0 k^2\hat\Phi+F_0k^2\hat\Gamma+\alpha_0k^2\hat\chi+\alpha_1k\dot{\hat\chi}+k^3 M_{\Delta}(\dot\nu+2\epsilon)\label{FE1}\\ 
-a^2\delta G^0_i&=\nabla_i\left[8\pi a^2 G\,\rho_M (1+\omega_M)\theta_M+B_0 k\hat\Phi+I_0k\hat\Gamma+\beta_0 k\hat\chi+\beta_1\dot{\hat\chi}+k^2M_{\Theta}(\dot\nu+2\epsilon)\right]\label{FE2}\\
a^2\delta G^i_i&=3\,8\pi a^2 G\,\rho_M\Pi_M+C_0 k^2\hat\Phi+C_1 k\dot{\hat\Phi}+J_0k^2\hat\Gamma+J_1 k\dot{\hat\Gamma}+\gamma_0 k^2\hat\chi+\gamma_1 k \dot{\hat\chi}+\gamma_2 \ddot{\hat\chi}+k^3M_P (\dot\nu+2\epsilon)\label{FE3}\\
a^2\delta \hat{G}^i_j&=8\pi a^2 G\,\rho_M (1+\omega_M)\Sigma_M+ D_0\hat\Phi+\frac{D_1}{k} \dot{\hat\Phi}+K_0\hat\Gamma+\frac{K_1}{k}\dot{\hat\Gamma}+\epsilon_0\hat\chi+\frac{\epsilon_1}{k}\dot{\hat\chi}+\frac{\epsilon_2}{k^2} \ddot{\hat\chi}  \label{FE4}
\end{align}
where \mbox{$\delta \hat{G}^i_j=\delta G^i_j-\frac{\delta^i_j}{3}\delta G^k_k$}. Each of the lettered coefficients in eqs.(\ref{FE1})-(\ref{FE4}) is a function of cosmological background quantities, ie. functions of time or redshift; this dependence has been suppressed above for clarity. Potentially the coefficients could also depend on scale, but this dependence is not arbitrary (\cite{Silvestri2013:rt})). These PPF coefficients are the analogy of the PPN parameters; they are the objects that a particular theory of gravity `maps onto', and the quantities to be constrained by data. Numerous examples of the PPF coefficients corresponding to well-known theories are given in \cite{Baker2013:vx}.

The final terms in eqs.(\ref{FE1})-(\ref{FE3}) are present to ensure the gauge invariance of the modified field equations, as is required for any theory governed by a covariant action. The quantities $M_\Delta$, $M_\Theta$ and $M_P$ are all pre-determined functions of the background. $\epsilon$ and $\nu$ are off-diagonal metric perturbations, so these terms vanish in the conformal Newtonian gauge. The gauge-fixing terms should be regarded as a piece of mathematical book-keeping; there is no constrainable freedom associated with them. \newline

One can then calculate observable quantities -- such as the weak lensing kernel or the growth rate of structure $f(z)$ -- using the parameterised field equations (\ref{FE1})-(\ref{FE4}). Similarly they can be implemented in an Einstein-Boltzmann solver code such as CAMB (\cite{CAMB}) to utilise constraints from the CMB. If we take the divergence of the gravitational field equations (i.e. the unperturbed equivalent of eq.(\ref{abc})), the LHS vanishes due to the Bianchi identity, whilst the stress-energy tensor of matter obeys its standard conservation equations (since we are working in the Jordan frame). Hence the $U$-tensor must be separately conserved, and this provides the necessary evolution equation for the variable $\hat\chi$:
\begin{align}
\label{Uconsv}
\delta\left(\nabla^\mu\left[U_{\mu\nu}^{\mathrm{metric}}+U_{\mu\nu}^{\mathrm{d.o.f.}}\right]\right)&=0
\end{align}  
Eq.(\ref{Uconsv}) has two components. If one wishes to treat theories with more than two new degrees of freedom, further information is needed to supplement the PPF framework.

The full form of the parameterised equations (\ref{FE1})-(\ref{FE4}) can be simplified in the `quasistatic regime', that is, significantly sub-horizon scales on which the time derivatives of perturbations can be neglected in comparison to their spatial derivatives (\cite{Hu:2007pj}). Quasistatic lengthscales are the relevant stage for weak lensing surveys and galaxy redshift surveys such as those of Euclid. A common parameterisation used on these scales has the form:
\begin{align}
2\nabla^2\Phi&=8\pi a^2G\,\mu(a,k)\,{\bar \rho}_M\Delta_M \label{Poisson}\\
\frac{\Phi}{\Psi}&=\gamma(a,k) \label{slip}
\end{align}
where $\{\mu,\gamma\}$ are two functions of time and scale to be constrained. This parameterisation has been widely employed (\cite{Bertschinger:2008zb, Daniel:2010yt, linder07,Bean:2010zq, Pogosian:2010tj, Zhao:2010dz, Dossett:2011vu,Hojjati:2011df, Hojjati:2011vp}). It has the advantages of simplicity and somewhat greater physical transparency: $\mu (a,k)$ can be regarded as describing evolution of the effective gravitational constant, whilst $\gamma (a,k)$ can, to a certain extent, be thought of as acting like a source of anisotropic stress (see section \ref{subsection:anisotropic_stress}). The quasistatic limit of PPF is discussed in \cite{Baker2013}.\newline

Let us make a comment about the number of coefficient functions employed in the PPF formalism. One may justifiably question whether the number of unknown functions in eqs.(\ref{FE1})-(\ref{FE4}) could ever be constrained. In reality, the PPF coefficients are not all independent. The form shown above represents a fully agnostic description of the extended field equations. However, as one begins to impose restrictions in theory space (even the simple requirement that the modified field equations must originate from a covariant action), constraint relations between the PPF coefficients begin to emerge. These constraints remove freedom from the parameterisation.

Even so, degeneracies will exist between the PPF coefficients. It is likely that a subset of them can be well-constrained, whilst another subset have relatively little impact on current observables and so cannot be tested. In this case it is justifiable to drop the untestable terms. Note that this realisation, in itself, would be an interesting statement - that there are parts of the gravitational field equations that are essentially unknowable. \newline

Finally we note that there is also a completely different, complementary approach to parameterising modifications to gravity. Instead of parameterising the linearised field equations, one could choose to parameterise the perturbed gravitational action. This approach has been used recently to apply the standard techniques of effective field theory to modified gravity; see \cite{Battye2012:kd,Bloomfield2012:as,Gubitosi2013:pv} and references therein.

%% file: de_mg/nonlinear_aspects.tex
In this Section we discuss how the nonlinear evolution of cosmic structures in
the context of
different non-standard cosmological models can be studied by means of numerical
simulations
based on N-body algorithms and of analytical approaches based on the spherical collapse model.

\subsection{N-body simulations of Dark Energy and Modified Gravity}
\label{Nbody_sims}

Here we discuss the numerical methods presently available
for this type of 
analyses, and
we review the main results obtained so far for different classes of
alternative cosmologies. {\color{red} These can be grouped into models where structure formation is
affected only through
a modified expansion history (such as Quintessence and Early Dark Energy models, section \ref{quintessence})
and models where particles experience modified gravitational forces, either for individual particle
species (interacting dark energy models
and growing neutrino models, section \ref{cde_eq}) or for all types of particles in
the Universe (modified gravity models).}

\subsubsection{Quintessence and Early Dark Energy models}
\label{quintessence_ede}

In general, in the context of flat FLRW cosmologies, any dynamical evolution of the 
dark energy density ($\rho _{{\rm DE}}\ne {\rm const.} = \rho _{\Lambda }$) determines a modification of the cosmic expansion history with respect to
the standard $\Lambda $CDM cosmology. 
In other words, if the dark energy is a dynamical quantity, i.e. if its equation of state parameter $w\ne -1$ {\color{red}exactly}, 
for any given set of cosmological parameters ($H_{0}$,$\Omega _{{\rm CDM}}$,$\Omega _{{\rm b}}$,$\Omega _{{\rm DE}}$,$\Omega _{{\rm rad}}$),
the redshift evolution of the Hubble function $H(z)$ will differ from the standard $\Lambda $CDM case $H_{\Lambda }(z)$.

Quintessence models of dark energy \citep{Wetterich_1988,Ratra_Peebles_1988} 
based on a classical scalar field $\phi $ subject to a self-interaction potential $V(\phi )$ have an energy density $\rho _{\phi } \equiv \dot{\phi }^{2}/2 + V(\phi )$ that evolves in time according to the 
dynamical evolution of the scalar field, which is governed by the homogeneous Klein-Gordon equation:
\begin{equation}
\label{kg}
\ddot{\phi } + 3H\dot{\phi } + \frac{{\rm d}V}{{\rm d}\phi } = 0 \,.
\end{equation}
Here the dot denotes derivation wrt ordinary time $t$.

For a canonical scalar field, the equation of state parameter $w_{\phi }\equiv
\rho _{\phi}/p_{\phi }$
where $p_{\phi}\equiv \dot{\phi }^{2}/2 - V(\phi ),  $
will in general be larger than $-1$, and the density of dark energy $\rho _{\phi
}$ will consequently be larger than $\rho _{\Lambda }$
at any redshift $z > 0$.
Furthermore, for some simple choices of the
potential function {\color{red} such as those discussed in section ~\ref{quintessence} (e.g. an exponential
potential $V\propto {\rm exp}(-\alpha \phi /M_{{\rm Pl}})$ or an inverse-power
potential $V\propto (\phi /M_{{\rm Pl}})^{-\alpha }$)},
scaling solutions for the evolution of the system can be analytically derived.
In particular, for an exponential potential, a scaling solution exists
where the dark energy scales as the dominant cosmic component, with a fractional
energy density
\begin{equation}
\label{EDE_1}
\Omega _{\phi }\equiv \frac{8\pi G \rho _{\phi }}{3H^{2}} = \frac{n}{\alpha
^{2}}
\end{equation}
with $n=3$ for matter domination and $n=4$ for radiation domination.
This corresponds to a relative fraction of dark energy at high redshifts which
is in general not negligible, whereas during matter and radiation domination
$\Omega _{\Lambda }\sim 0$, and therefore represents a phenomenon of an early
emergence of dark energy as compared to $\Lambda $CDM where
dark energy is for all purposes negligible until $z\sim 1$.

Early Dark Energy (EDE hereafter) is therefore a common prediction of scalar
field models of dark energy, and observational constraints put firm bounds on
the
allowed range of $\Omega _{\phi }$ at early times, and consequently on the
potential slope $\alpha $.

As we have seen in Sec.  \ref{parametrization-of-the-background-evolution}, a completely phenomenological parametrization of EDE, independent from any
specific model of dynamical dark energy has been proposed by
\citet{Wetterich_2004}
as a function of the present dark energy density $\Omega _{{\rm DE}}$, its value
at early times $\Omega _{{\rm EDE}}$, and the present value of the equation of
state parameter
$w_{0}$. 
From eqn.~\ref{w_ede_par}, the full expansion history of the corresponding EDE model
can be derived.
\ \\

A modification of the expansion history indirectly influences also the growth of density perturbations and ultimately the formation of cosmic structures.
While this effect can be investigated analytically for the linear regime, N-body simulations are required to extend the analysis to the nonlinear stages
of structure formation. For standard Quintessence and EDE models, the only modification that is necessary to implement into standard N-body algorithms is the 
computation of the correct Hubble function $H(z)$ for the specific model under investigation, since this is the only way in which these non standard cosmological models can alter
structure formation processes. 

This has been done by the independent studies of \citet{Grossi_2008} and \citet{Francis_Lewis_Linder_2008},
 where a modified expansion history consistent with
EDE models described by the parametrization of eq.\ref{w_ede_par} has been implemented in the widely used N-body code {\small GADGET-2} \citep{gadget-2} and 
the properties of nonlinear structures forming in these EDE cosmologies have been analyzed.
Both studies have shown that the standard formalism for the computation of the halo mass function still holds for EDE models at $z=0$. In other words, both the
standard fitting formulae for the number density of collapsed objects as a function of mass, and their key parameter $\delta _{c} = 1.686$ representing the linear
overdensity at collapse for a spherical density perturbation, remain unchanged also for EDE cosmologies.

The work of \citet{Grossi_2008}, however, 
investigated also the internal properties of collapsed halos in EDE models, finding a slight increase of halo concentrations due to the earlier onset of structure formation
and most importantly a significant increment of the  line-of-sight velocity dispersion of massive halos. The latter effect could mimic a higher $\sigma _{8}$ normalization
for cluster mass estimates based on galaxy velocity dispersion measurements, and therefore represents a potentially detectable signature of EDE models.

\subsubsection{Interacting Dark Energy models}
\label{cDE}

Another interesting class of non standard dark energy models, as introduced in Sec.~\ref{mg:cde}, 
is given by coupled dark energy where a direct interaction
is present between a Quintessence scalar field $\phi $ and other cosmic components, in the 
form of a source term in the background continuity equations:
\begin{eqnarray}
\label{coupled_phi} \frac{d\rho_{\phi}}{d\eta} &=& -3 {\cal H} (1 + w_\phi) \rho_{\phi} +
\beta(\phi) \frac {d \phi}{d\eta} (1-3 w_{\alpha}) \rho_{\alpha} ~~~, \\
\label{cons_species} \frac{ d \rho_{\alpha}}{d\eta} &=& -3 {\cal H} (1 + w_{\alpha}) \rho_{\alpha} -
\beta(\phi) \frac{d\phi}{d\eta} (1-3 w_{\alpha}) \rho_{\alpha}
\end{eqnarray}
{\color{red} where $\alpha$ represents a single cosmic fluid coupled to $\phi$.}

While such direct interaction with baryonic particles ($\alpha=b$) is tightly constrained
by observational bounds, and while it is suppressed for relativistic particles ($\alpha=r$) by symmetry reasons ($1-3w_{r}=0$), 
a selective interaction with Cold Dark Matter (CDM hereafter) 
or with massive neutrinos is still observationally viable (see Sec.~\ref{mg:cde}).

Since the details of interacting dark energy models have been discussed in Sec.~\ref{mg:cde}, here we simply recall the main features of these models
that have a direct relevance for nonlinear structure formation studies.
For the case of interacting dark energy, in fact, the situation is much more complicated than for the simple EDE scenario discussed above.
The mass of a coupled particle changes in time due to the energy exchange with the dark energy scalar field $\phi $ according to the equation:
\begin{equation}
\label{mass_var}
m(\phi ) = m_{0}e^{-\int \beta (\phi '){\rm d}\phi '}
\end{equation}
where $m_{0}$ is the mass at $z=0$.
Furthermore, the Newtonian acceleration of a coupled particle (subscript $c$) gets modified as:
\begin{equation}
\label{mod_accel}
\dot{\vec{v}}_{c} = -\tilde{H}\vec{v}_{c} - \vec{\nabla }\tilde{\Phi }_{c} - \vec{\nabla }\Phi _{nc} \,.
\end{equation}
where $\tilde{H}$ contains a new velocity-dependent acceleration:
\begin{equation}
\tilde{H}\vec{v}_{c} = H\left( 1-\beta _{\phi }\frac{\dot{\phi }}{H}\right) \vec{v}_{c}\,,
\end{equation}
and where a fifth-force acts only between coupled particles as
\begin{equation}
\label{phitilde}
\tilde{\Phi }_{c} = (1 + 2\beta ^{2})\Phi _{c}\,,
\end{equation}
while $\Phi _{nc}$ represents the gravitational potential due to all massive particles with no coupling to the dark energy that
exert a standard gravitational pull.

As a consequence of these new terms in the Newtonian acceleration equation the growth of density perturbations will
be affected, in interacting dark energy models, not only by the different Hubble expansion due to the
dynamical nature of dark energy, but also by a direct modification of the effective gravitational interactions at subhorizon scales.
Therefore, linear perturbations of coupled species will grow with a higher rate in these cosmologies
In particular, for the case of a coupling to CDM, a different amplitude of the matter power spectrum
will be reached at $z=0$ with respect to $\Lambda $CDM if a normalization in accordance with CMB measurements at high redshifts is assumed.

Clearly, the new acceleration equation (\ref{mod_accel}) will have an influence also on the formation and evolution of nonlinear structures, and
a consistent implementation of all the above mentioned effects into an N-body algorithm is required in order to investigate this regime.

For the case of a coupling to CDM (a coupling with neutrinos will be discussed in the next section) 
this has been done e.g. by \citet{maccio_etal_2004,Sutter_Ricker_2007} with 1D or 3D grid-based field solvers, and more recently
by means of a suitable modification by \citet{Baldi_etal_2010} of the TreePM hydrodynamic N-body code {\small GADGET-2} \citep{gadget-2}.

Non-linear evolution within coupled
quintessence cosmologies has been addressed using various methods of
investigation, such as spherical collapse
\citep{Mainini:2006zj,Wintergerst:2010ui,Manera:2005ct,Koivisto:2005nr,
Sutter:2007ky,Abdalla:2007rd,Bertolami:2007tq} and alternative semi-analytic
methods \citep{Saracco_etal_2010,amendola_quercellini_2004}. $N$-body and
hydro-simulations have also been done
\citep{maccio_etal_2004,Baldi_etal_2010,Baldi:2010vv,Baldi_Pettorino_2010,
Baldi:2010ks,Li:2010zw,Li:2010eu,Baldi:2010pq,Zhao:2010dz}.
We list here briefly the main observable features typical of this class of
models:
\begin{itemize}
\item The suppression of power at small scales in the power spectrum of interacting dark energy models as compared to $\Lambda $CDM;
\item The development of a gravitational bias in the amplitude of density perturbations of uncoupled baryons and coupled CDM particles
defined as $P_{b}(k)/P_{c}(k)<1$, which determines
a significant decrease of the baryonic content of massive halos at low redshifts in accordance with a large number of observations \citep{Baldi_etal_2010,Baldi:2010pq};;
\item The increase of the number density of high-mass objects at any redshift as compared to $\Lambda $CDM \citep[see][]{Baldi_Pettorino_2010};
\item An enhanced ISW effect
\citep{Amendola:1999er,Amendola:2003wa,Mainini:2010ng}; such effects may be
partially reduced when taking into account non-linearities, as described in
\citet{Pettorino:2010bv};
\item A less steep inner core halo profiles (depending on the interplay between
fifth force and velocity-dependent terms)
\citep{Baldi_etal_2010,Baldi:2010vv,Li:2010zw,Li:2010eu,Baldi:2010pq};
\item A lower concentration of the halos
\citep{Baldi_etal_2010,Baldi:2010vv,Li:2010eu};
\item Voids are emptier when a coupling is active \citep{Baldi:2010ks}.
\end{itemize}
Subsequent studies based on Adaptive Mesh Refinement schemes for the solution of the local scalar field equation \citep{Li_Barrow_2010} have 
broadly confirmed these results.

The analysis has been extended to the case of non-constant coupling functions $\beta (\phi )$ by \citet{Baldi:2010vv}, and has shown how in the presence
of a time evolution of the coupling some of the above mentioned results no longer hold:
\begin{itemize}
\item Small scale power can be both suppressed and enhanced when a growing coupling function
is considered, depending on the magnitude of the coupling time derivative  ${\rm d}\beta (\phi )/{\rm d}\phi $
\item The inner overdensity of CDM halos, and consequently the halo concentrations, 
can both decrease (as always happens for the case of constant couplings) or increase, again depending on the rate of change of the 
coupling strength;
\end{itemize}

All these effects represent characteristic features of interacting dark energy models and could provide a direct way to observationally test these scenarios.
Higher resolution studies would be required in order to quantify the impact of a DE-CDM interaction on the statistical properties of halo substructures and on
the redshift evolution of the internal properties of CDM halos.

As discussed in subsection \ref{Nbody_sims}, when a variable coupling $\beta (\phi )$ 
is active the relative balance of the fifth-force and other dynamical effects depends on
 the specific time evolution of the coupling strength. Under such conditions, certain cases
 may also lead to the opposite effect of larger halo inner overdensities and higher concentrations, 
as in the case of a steeply growing coupling function \citep[see][]{Baldi:2010vv}. Alternatively, 
the coupling can be introduced by choosing directly a covariant stress-energy tensor, treating dark 
energy as a fluid in the absence of a starting action \citep{Mangano:2002gg,Valiviita:2008iv,CalderaCabral:2008bx,Schaefer:2008ku,Valiviita:2009nu,
Majerotto:2009np,Gavela:2009cy,CalderaCabral:2009ja,Gavela:2010tm}.

\subsubsection{Growing Neutrinos}
\label{cnu}

In case of a coupling between the dark energy scalar field $\phi $ and the relic fraction of massive neutrinos, all the above basic 
equations (\ref{mass_var}-\ref{phitilde}) still hold. However, such models are found to be cosmologically viable only for large negative values of the 
coupling $\beta $ \citep[as shown by][]{Amendola2008b}, that according to Eq.~\ref{mass_var} determines 
a neutrino mass that grows in time (from which these models have been dubbed ``Growing Neutrinos").
An exponential growth of the neutrino mass implies that cosmological bounds on the neutrino mass are 
no longer applicable
and that neutrinos remain relativistic much longer than in the standard scenario,
 which keeps them effectively uncoupled until
recent epochs, according to Eqs.~(\ref{coupled_phi},\ref{cons_species}). However, as soon as neutrinos become nonrelativistic at redshift $z_{{\rm nr}}$
due to the exponential
growth of their mass, the pressure terms $1-3w_{\nu }$ in Eqs.~(\ref{coupled_phi},\ref{cons_species}) no longer vanish
and the coupling with the DE scalar field $\phi $ becomes active.

Therefore, while before $z_{{\rm nr}}$ the model behaves as a standard $\Lambda $CDM scenario, after $z_{{\rm nr}}$ the nonrelativistic
massive neutrinos obey the modified Newtonian equation (\ref{mod_accel}) and a fast growth of neutrino density perturbation
takes place due to the strong fifth force described by Eq.~(\ref{phitilde}).

The growth of neutrino overdensities in the context of Growing Neutrinos models has been studied in the linear regime by
\citet{Mota:2008nj}, predicting the formation of very large neutrino lumps at the scale of superclusters and above (10-100 Mpc/h)
at redshift $z\approx 1$.

The analysis has been extended to the nonlinear regime in \citet{Wintergerst:2009fh} by following the spherical collapse of a neutrino
lump in the context of Growing Neutrino cosmologies. This study has witnessed the onset of virialization processes in the nonlinear evolution of
the neutrino halo at $z\approx 1.3$, and provided a first estimate of the associated gravitational potential at virialization being of the order of
$\Phi _{\nu }\approx 10^{-6}$ for a neutrino lump with radius $R \approx 15$ Mpc.

An estimate of the potential impact of such very large nonlinear structures onto the CMB angular power spectrum through 
the Integrated Sachs-Wolfe effect has been attempted by \citet{Pettorino:2010bv}. This study has shown that the linear
approximation fails in predicting the global impact of the model on CMB anisotropies at low multipoles,
and that the effects under consideration are very sensitive to the details of the transition between the linear and nonlinear regimes 
and of the virialization processes of nonlinear neutrino lumps, 
and that also significantly depend on possible backreaction effects of the evolved neutrino density field onto the local scalar filed evolution.

A full nonlinear treatment by means of specifically designed N-body simulations is therefore required 
in order to follow in further detail the evolution of a cosmological sample of neutrino lumps beyond virialization,
and to assess the impact of Growing Neutrinos models onto potentially observable quantities as the low-multipoles
CMB power spectrum or the statistical properties of CDM large scale structures.

\subsubsection{Modified Gravity}
\label{modgrav}

Modified gravity models, presented in Sec.~\ref{models-of-modified-gravity}, represent a different perspective to account for the nature of
the dark components of the Universe. Although most of the viable modifications of General Relativity are
constructed in order to provide an identical cosmic expansion history to the standard $\Lambda $CDM model,
their effects on the growth of density perturbations could lead to observationally testable predictions capable of
distinguishing modified gravity models from standard General Relativity plus a cosmological constant.

Since a modification of the theory of gravity would affect all test masses in the Universe, i.e. including the standard baryonic matter,
an asymptotic recovery of General Relativity for solar system environments, where deviations from GR are tightly 
constrained, is required for all viable modified gravity models. Such mechanism, often referred to as the ``Chameleon effect",
represents the main difference between modified gravity models and the interacting dark energy scenarios discussed above, by 
determining a local dependence of the modified gravitational laws in the Newtonian limit.

While the linear growth of density perturbations in the context of modified gravity theories can be studied 
\citep[see e.g.][]{Hu_Sawicki_2007,Motohashi_etal_2010,Amarzguioui:2005zq,Appleby:2010dx} by parametrizing
the scale dependence of the modified Poisson and Euler equations in Fourier space (see the discussion in Sec.~\ref{sec:dof}),
 the nonlinear 
evolution of the ``Chameleon effect" makes the implementation of these theories into nonlinear N-body algorithms
much more challenging. For this reason, very little work has been done so far in this direction. 
A few attempts to solve the modified gravity interactions
in the nonlinear regime by means of mesh-based iterative relaxation schemes have been carried out by 
\citet{Oyaizu_2008,Oyaizu_etal_2008,Schmidt_etal_2009,Khoury_Wyman_2009,Zhao_Li_Koyama_2010,Davis:2011pj,Winther:2011qb}
and showed an enhancement of the power spectrum amplitude at intermediate and small scales. 
These studies also showed that this nonlinear enhancement of small scale power cannot be accurately reproduced
by applying the linear perturbed equations of each specific modified gravity theory to the standard nonlinear
fitting formulae \citep[as e.g.][]{Smith2003}.

Higher resolution simulations and new numerical approaches will be necessary in order to extend these first results
to smaller scales and to accurately evaluate the deviations of specific models of modified gravity
from the standard GR predictions to a potentially detectable precision level.

%% file: de_mg/spherical_collapse.tex
\subsection{The spherical collapse model}\label{the-spherical-collapse-model}

A popular analytical approach to study non-linear clustering of dark matter
without recurring to N-body simulations is the spherical collapse model, first
studied by \cite{Gunn/Gott:1972}. In this approach, one studies the collapse of
a spherical overdensity and determines its critical overdensity for collapse as
a function of redshift. Combining this information with the extended
Press-Schechter theory (\cite{Press1974a,Bond/etal:1991}; see
\cite{Zentner:2007} for a review) one can provide a statistical model for the
formation of structures which allows to predict the abundance of virialized
objects as a function of their mass. Although it fails to match the details of
N-body simulations, this simple model works surprisingly well and can give
useful insigths into the physics of structure formation. Improved models
accounting for the complexity of the collapse exist in the literature and offer
a better fit to numerical simulations. For instance, \cite{Sheth/Tormen:1999}
showed that a significant improvement can be obtained by considering an
ellipsoidal collapse model. Furthermore, recent theoretical developments and new
improvements in the excursion set theory have been undertaken by
\cite{Maggiore/Riotto:2010} and other authors (see e.g. (\cite{Shaw:2007tr})).

The spherical collapse model has been generalized to include a cosmological
constant by \cite{Peebles:1984,Weinberg:1987}. \cite{Lahav/etal:1991} have used
it to study the observational consequences of a cosmological constant on the
growth of perturbations. The case of standard quintessence, with speed of sound
$c_s = 1$, have been studied by \cite{wang98}.  In this case,
scalar fluctuations propagate at the speed of light and sound waves maintain
quintessence homogeneous on scales smaller than the horizon scale. In the
spherical collapse pressure gradients maintain the same energy density of
quintessence between the inner and outer part of the spherical overdensity, so
that the evolution of the overdensity radius is described by
\begin{equation}
\label{uc_sp}
\frac{\ddot R}{R} = -\frac{4 \pi G}{3} (\rho_m + \bar \rho_Q + 3 \bar p_Q)\;,
\end{equation}
where $\rho_m$ denotes the energy density of dark matter while $\bar \rho_Q$ and
$\bar p_Q$ denote the homogeneous energy density and pressure of the
quintessence field.
Note that,
although this equation looks like one of the Friedmann equations, the dynamics
of $R$ is not the same
as for a FLRW universe. Indeed, $\rho_m$ evolves following the scale factor $R$,
while the quintessence follows the external scale factor $a$, according to the
continuity equation $\dot{\bar \rho}_Q + 3 (\dot a /a) (\bar \rho_Q + \bar p_Q)
=0$. 

In the following we will discuss the spherical collapse model in the contest of
other dark energy and modified gravity models.

\subsubsection{Clustering dark energy}

In its standard version, quintessence is described by a minimally-coupled
canonical field, with speed of sound $c_s=1$. As mentioned above, in this case
clustering can only take place on scales larger than the horizon, where sound
waves have no time to propagate. However, observations on such large scales are
strongly limited by cosmic variance and this effect is difficult to observe. A
minimally coupled scalar field with fluctuations characterized by a practically
zero speed of sound can cluster on all observable scales. There are several
theoretical motivations to consider this case. In the limit of zero sound speed
one recovers the Ghost Condensate theory proposed by
\cite{Arkani-Hamed/etal:2004} in the context of modification of gravity, which
is invariant under shift symmetry of the field $\phi \to \phi + {\rm constant}$.
Thus, there is no fine tuning in assuming that the speed of sound is very small:
quintessence models with vanishing speed of sound should be thought of as
deformations of this particular limit where shift symmetry is recovered.
Moreover, it has been shown that minimally coupled quintessence with an equation
of state $ w<  -1$ can be free from ghosts and gradient instabilities only if
the speed of sound is very tiny, $| c_s |  \lesssim 10^{-15}$. Stability can be
guaranteed by the presence of higher derivative operators, although their effect
is absent on cosmologically relevant scales
(\cite{Creminelli/etal:2006,Cheung/etal:2008,Creminelli/etal:2009}).

The fact that the speed of sound of quintessence may vanish opens up new
observational consequences. Indeed, the absence of quintessence pressure
gradients allows instabilities to develop on all scales, also on scales where
dark matter perturbations become non-linear. Thus, we expect quintessence to
modify the growth history of dark matter not only through its different
background evolution but also by actively participating to the structure
formation mechanism, in the linear and non-linear regime, and by contributing to
the total mass of virialized halos.

Following \cite{Creminelli/etal:2010}, in the limit of zero sound speed pressure
gradients are negligible and, as long as the fluid approximation is valid,
quintessence follows geodesics remaining comoving with the dark matter (see also
\cite{Lim/Sawicki/Vikman:2010} for a more recent model with identical
phenomenology). In particular, one can study the effect of quintessence with
vanishing sound speed on the structure formation in the nonlinear regime, in the
context of the spherical collapse model. The zero speed of sound limit
represents the natural counterpart of the opposite case $c_s = 1$. Indeed, in
both cases there are no characteristic length scales associated to the
quintessence clustering and the spherical collapse remains independent of the
size of the object  (see \cite{EggersBjaelde/Wong:2010,Mota:2004pa, Nunes:2004wn} for a study of the
spherical collapse when $c_s$ of quintessence is small but finite).

Due to the absence of pressure gradients quintessence follows dark matter in the collapse and the evolution of the overdensity radius is described by
\begin{equation}
\frac{\ddot R}{R} = -\frac{4 \pi G}{3} (\rho_m + \rho_Q + \bar p_Q)\;,
\end{equation}
where the energy density of quintessence $\rho_Q$ has now a different value
inside and outside the overdensity, while the pressure remains unperturbed. In
this case 
the quintessence inside the overdensity evolves following the internal scale
factor $R$, $\dot{ \rho}_Q + 3 (\dot R /R) (\rho_Q + \bar p_Q) =0$ and
the comoving regions behave as closed FLRW universes. $R$ satisfies the
Friedmann equation and the spherical collapse can be solved exactly
(\cite{Creminelli/etal:2010}). 

\begin{figure}[h!!!]
\begin{center}
\includegraphics[scale=1]{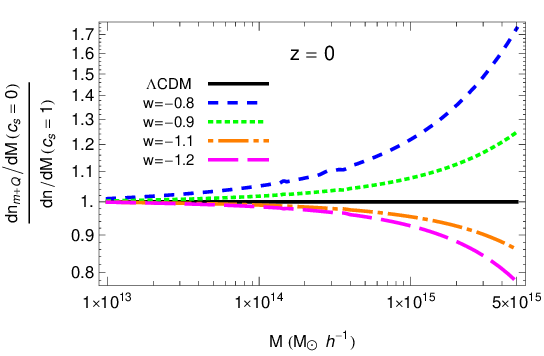}  

\vspace{0.5cm}
\includegraphics[scale=1]{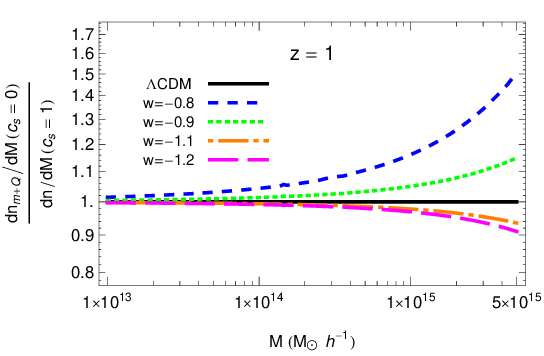} 
\end{center}
\caption{\label{fig:qmassfunctionratio} Ratio of the total mass functions, which
include the quintessence contribution, for $c_s=0$ and $c_s=1$ at $z=0$ (above)
and $z=1$ (below). [Figures from \cite{Creminelli/etal:2010}.]}
\end{figure}
Quintessence with zero speed of sound modifies dark matter clustering with
respect to the smooth quintessence case through the linear growth function and
the linear threshold for collapse. Indeed, for $w >-1$ ($w < -1$), it enhances
(diminishes) the clustering of dark matter, the effect being proportional to
$1+w$.
The modifications to the critical threshold of collapse are small and the
effects on the dark matter mass function are dominated by the modification on
the linear dark matter growth function. Besides these conventional effects there
is a more important and qualitatively new phenomenon: quintessence mass adds to
the one of dark matter, contributing to the halo mass by a fraction of order
$\sim (1 + w) \Omega_Q/ \Omega_m$. Importantly, it is possible to show that the
mass associated to quintessence stays constant inside the virialized object,
independently of the details of virialization. Moreover the ratio between the
virialization and the turn-around radii is approximately the same as the one for
$\Lambda$CDM computed by \cite{Lahav/etal:1991}.
In figure \ref{fig:qmassfunctionratio} we plot the ratio of the mass function
including the quintessence mass contribution, for the $c_s=0$ case to the smooth
$c_s=1$ case. 
The sum of the two effects is rather large: for values of $w$ still compatible
with the present data and for large masses the difference between the
predictions of the $c_s = 0$ and the $c_s = 1$ cases is of order one.

\subsubsection{Coupled Dark Energy}
We now consider spherical collapse within coupled dark energy cosmologies. 
The presence of an interaction that couples the cosmon dynamics to another species
introduces a new force acting between particles (CDM or neutrinos in the examples
mentioned in section \ref{mg:cde}) and mediated by dark energy fluctuations. Whenever such a coupling is active, spherical collapse,
whose concept is intrinsically based on gravitational attraction via the Friedmann equations, has to be suitably modified in order to account for other external forces. As shown in \cite{Wintergerst:2010ui} the inclusion of the fifth force within the spherical collapse picture deserves particular caution. Here we summarize the main results on this topic and
we refer to \cite{Wintergerst:2010ui} for a detailed illustration of spherical collapse in presence of a fifth force. 

\begin{figure}[ht]
\begin{center}
\includegraphics[width=85mm,angle=0.]{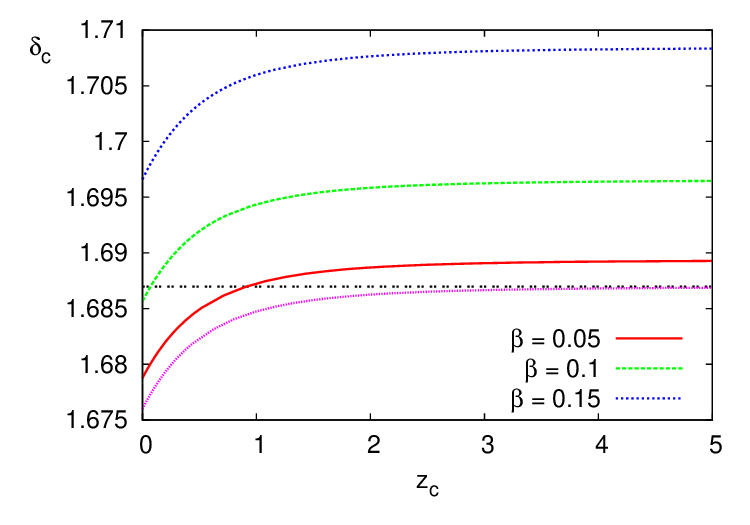}
\end{center}
\caption{Extrapolated linear density contrast at collapse for coupled quintessence models with different coupling strength $\beta$. 
For all plots we use a constant $\alpha = 0.1$. We also depict $\delta_c$ for reference $\Lambda$CDM (dotted, pink) and EdS (double-dashed, black) models. [Figure from \cite{Wintergerst:2010ui}]}
\label{fig:delta_c_ccdm}
\vspace{0.5cm}
\end{figure}

If CDM is coupled to a quintessence scalar field as described in secs.~\ref{mg:cde} and \ref{dms:de_dm} of the present document, the full nonlinear evolution equations within the Newtonian limit read:
\begin{eqnarray} \label{eq:sph_com_ns1} \dot\delta_m &=& -{\bf v}_m\,\nabla\delta_m - (1 + \delta_m)\,\nabla\cdot{\bf v}_m \\
\dot{\bf v}_m &=& -(2{\bar H} - \beta\,\dot{\bar\phi})\,{\bf v}_m - ({\bf v}_m\,\nabla){\bf v}_m \nonumber \\
\label{eq:sph_com_ns2} && - a^{-2}\,\nabla(\Phi - \beta\,\delta\phi) \\
\label{eq:sph_com_poisson} \Delta\delta\phi &=& -\beta\,a^2\,\delta\rho_m \\
\label{eq:sph_com_grav_pot} \Delta\Phi &=& -\frac{a^2}{2}\,\sum_{\alpha} \delta\rho_{\alpha} \end{eqnarray}
These equations can be derived from the nonrelativistic Navier-Stokes equations and from the Bianchi identities written in presence of an external source of the type:
\begin{equation}
\label{eq:ps_cons}\nabla_{\gamma}T_{\mu}^{\gamma} = Q_{\mu} = -\beta T_{\gamma}^{\gamma}  \partial_{\mu}\phi ~~~,
\end{equation}
where $T^{\gamma}_{\mu}$ is the stress energy tensor of the dark matter fluid and we are using comoving spatial coordinates $\bf x$ and cosmic time $t$. Note that ${\bf v}_m$ is the comoving velocity, related to the peculiar velocities by ${\bf v}_m = {\bf v}_{pec}/a$.  They are valid for arbitrary quintessence potentials as long as the scalar field is sufficiently light, i.e. $m_\phi^2 \delta\phi = V''(\phi)\delta\phi \ll \Delta\delta\phi$ for the scales under consideration. For a more detailed discussion see \cite{Wintergerst:2010ui}.
Combining the above equations yields to the following expression for the evolution of the matter perturbation $\delta_m$:
\begin{equation} \label{eq:sph_gen_del} \ddot\delta_m = -(2{\bar H}-\beta\,\dot{\bar\phi})\,\dot\delta_m 
  + \frac{4}{3}\frac{\dot\delta_m^2}{1 + \delta_m} + \frac{1 + \delta_m}{a^2} \,\Delta\Phi_{\text{eff}} \,\,\,, \end{equation} 
Linearization leads to: 
\begin{equation} \label{eq:sph_gen_del_lin} \ddot\delta_{m,L} = -(2{\bar H}-\beta\,\dot{\bar\phi})\,\dot\delta_{m,L} + a^{-2}\,\Delta\Phi_{\text{eff}} ~~~. \end{equation}
where the effective gravitational potential follows the modified Poisson equation:
\begin{equation} \Delta {\Phi_{\text{eff}}} = -\frac{a^2}{2} {\bar\rho}_m \delta_m \left(1+2 \beta^2\right) \, ~~~. \end{equation}
Equations (\ref{eq:sph_gen_del}) and (\ref{eq:sph_gen_del_lin}) are the two main equations which correctly describe the nonlinear and linear evolution for a coupled dark energy model. 
They can be used, among other things,
 for estimating the extrapolated linear density contrast at collapse $\delta_c$ in the presence of
a fifth force.
It is possible to reformulate Eqs. (\ref{eq:sph_gen_del}) and (\ref{eq:sph_gen_del_lin}) into 
an effective spherical collapse: 
\begin{equation} \label{eq:imp_sc_f2b} \frac{\ddot{R}}{R} = -\beta\,\dot{\phi}\left( H - \frac{\dot{R}}{R}\right) - \frac{1}{6} \sum_{\alpha} \left[{\rho}_{\alpha}(1 + 3 { w}_{\alpha})\right] 
 - \frac{1}{3}\,\beta^2\,\delta\rho_m ~~~. \end{equation}
Equation (\ref{eq:imp_sc_f2b}) (\cite{Mainini:2006zj, Wintergerst:2010ui}), describes the general evolution of
the radius of a spherical overdense region within coupled quintessence. 
Comparing with the standard case (\ref{uc_sp}) we notice the presence of two additional terms: a `friction' term 
and the coupling term $\beta^2\,\delta\rho_m$, the latter being responsible for the additional attractive fifth force.
Note that the ``friction'' term is actually velocity dependent and its effects on collapse depend, more realistically, on the
direction of the velocity, information which is not contained within a \emph{spherical} collapse picture and can be treated within simulations (\cite{Baldi_Pettorino_2010,Li:2010zw, Baldi:2010vv, Li:2010eu, Baldi:2010pq}).
We stress that it is crucial to include these additional terms in the equations, as derived from the non-linear equations, in order to correctly account for the presence of a fifth force.
The outlined procedure can easily be generalized to include uncoupled components, for example baryons. 
In this case, the corresponding evolution equation for $\delta_b$, will be fed by $\Phi_{\text{eff}} = \Phi$. 
This yields an evolution equation for the uncoupled scale factor $R_{uc}$ that is equivalent to the standard Friedmann equation.
In Fig.~\ref{fig:delta_c_ccdm} we show the linear density contrast at collapse $\delta_c(z_c)$ for three coupled quintessence
 models with $\alpha = 0.1$ and $\beta = 0.05$, $0.1$, $0.15$.
An increase of $\beta$ results in an increase of $\delta_c$.
As shown in \cite{Wintergerst:2010ui}, $\delta_c(\beta)$ is well described by a simple quadratic fitting formula,
\begin{equation} \delta_c(\beta) = 1.686(1 + a\beta^2)\,,a = 0.556 ~~~, \end{equation}
valid for small $\beta \lsim 0.4$ and $z_c \geq 5$. We recall that a non-linear analysis beyond the spherical collapse method can be addressed by means of the 
time-renormalization-group method, extended to the case of couple quintessence in (\cite{Saracco:2009df}).

\begin{figure}[ht]
\begin{center}
\includegraphics[width=85mm,angle=0.]{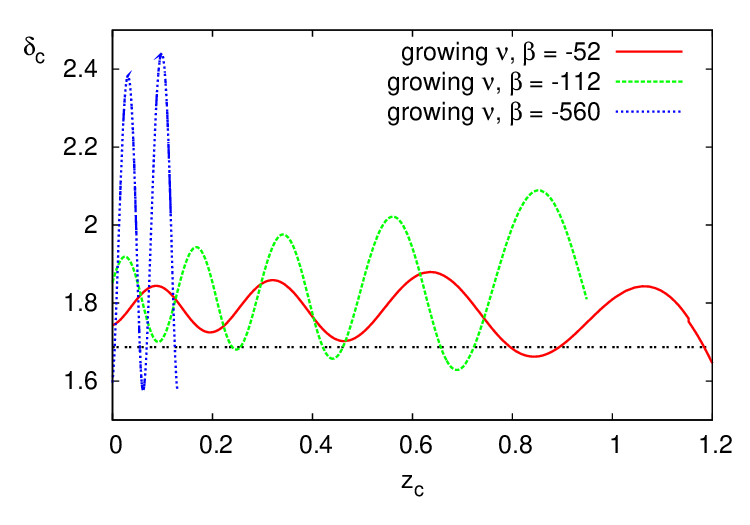}
\end{center}
\caption{Extrapolated linear density contrast at collapse $\delta_c$ vs. collapse redshift $z_c$ for growing neutrinos with $\beta = -52$ (solid, red), $\beta = -112$ (long-dashed, green) and $\beta = -560$ (short-dashed, blue). A reference EdS model (double-dashed. black) is also shown. [Figure from \cite{Wintergerst:2010ui}].}
\label{fig:delta_c_cnu}
\vspace{0.5cm}
\end{figure}

If a coupling between dark energy and neutrinos is present, as described in secs.~\ref{mg:cde} and \ref{dms:de_nu}, bound neutrino structures may form within these models (\cite{Brouzakis:2007aq}). It was shown in \cite{Mota:2008nj} that their formation will only start after neutrinos become nonrelativistic. A nonlinear treatment of the evolution of neutrino densities is thus only required for very late times, and one may safely neglect neutrino pressure as compared to their density. The evolution equations (\ref{eq:sph_gen_del}) and (\ref{eq:sph_gen_del_lin}) can then also be applied for the nonlinear and linear neutrino density contrast.
The extrapolated linear density at collapse $\delta_c$ for growing neutrino quintessence reflects in all respects 
the characteristic features of this model and results in a $\delta_c$ which looks quite different from standard dark energy cosmologies.
We have plotted the dependence of $\delta_c$ on the collapse redshift $z_c$ in Fig.~\ref{fig:delta_c_cnu} for three values of the coupling. 
The oscillations seen are the result of the oscillations of the neutrino mass caused by the coupling to the scalar field: the latter 
has characteristic oscillations as it approaches the minimum of the effective potential in which it rolls, given by a combination of
the self-interaction potential $U(\phi)$ and the coupling contribution $\beta(1-3{ w}_\nu){\rho}_\nu$. 
Furthermore, due to the strong coupling $\beta$, the average value of $\delta_c$ is found to be substantially higher than $1.686$, corresponding to the Einstein de Sitter value, shown in black (double-dashed) in Fig.~\ref{fig:delta_c_cnu}.
Such an effect can have a strong impact on structure formation and on CMB (\cite{Pettorino:2010bv}).
For the strongly coupled models, corresponding to a low present day neutrino mass $m_\nu(t_0)$, the critical density at collapse is only available for $z_c \lsim 0.2$, $1$ for $\beta = -560$, $-112$, respectively. This is again a reflection of the late transition to the nonrelativistic regime. 
Nonlinear investigations of single lumps beyond the spherical collapse picture was performed in (\cite{Wintergerst:2009fh, Brouzakis:2010md}), the latter showing the influence of the gravitational potentials induced by the neutrino inhomogeneities on the acoustic oscillations in the baryonic and dark-matter spectra.
\subsubsection{Early dark energy}\label{sec:early_results}
\begin{figure}[ht]
\begin{center}
\includegraphics[width=85mm,angle=0.]{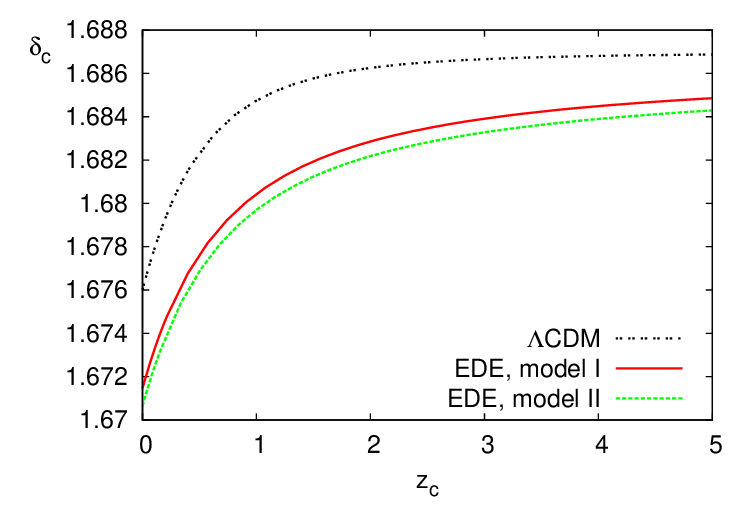}
\end{center}
\caption{Extrapolated linear density contrast at collapse $\delta_c$ vs. collapse redshift $z_c$ for EDE models I (solid, red) and II (long-dashed, green), as well as $\Lambda$CDM (double-dashed, black). [Figure from \cite{Wintergerst:2010ui}].}
\label{fig:delta_c_ede}
\vspace{0.5cm}
\end{figure}

A convenient way to parametrize the presence of a nonnegligible homogenous dark energy component at early times was presented in \cite{Wetterich:2004pv} and has been illustrated in sec.~\ref{parametrization-of-the-background-evolution} of the present review. 
If we specify the spherical collapse equations for this case, the nonlinear evolution of the density contrast follows the evolution equations (\ref{eq:sph_gen_del}) and (\ref{eq:sph_gen_del_lin}) without the terms related to the coupling. As before, we assume relativistic components to remain homogenous.
In Fig.~\ref{fig:delta_c_ede} we show $\delta_c$ for two models of early dark energy, namely model I and II, corresponding to the choices ($\Omega_{m,0} = 0.332\,,\,\,\,w_0 = -0.93\,,\,\,\,\Omega_{\text{DE},e}  = 2\cdot10^{-4}$) and ($\Omega_{m,0} = 0.314\,,\,\,\,w_0 = -0.99\,,\,\,\,\Omega_{\text{DE},e} = 8\cdot10^{-4}$) respectively.
Results show $\delta_c(z_c = 5) \sim 1.685$ ($\sim 5\cdot10^{-2} \%$) (\cite{Francis:2008ka}, \cite{Wintergerst:2010ui}).

%% file: de_mg/obs-intro.tex
Both scalar field dark energy models and modifications of gravity can in
principle lead to any desired expansion history $H(z)$, or equivalently any
evolution of the effective dark energy equation of state parameter $w(z)$. For
canonical scalar fields, this can be achieved by selecting the appropriate
potential $V(\varphi)$ along the evolution of the scalar field $\varphi(t)$, as
was done e.g. in \cite{Bassett:2002qu}. For modified gravity models, the same
procedure can be followed for example for $f(R)$ type models
\citep[e.g.][]{Pogosian:2007sw}. The evolution history on its own can thus not
tell us very much about the physical nature of the mechanism behind the
accelerated expansion (although of course a clear measurement showing that $w
\neq -1$ would be a sensational discovery). A smoking gun for modifications of
gravity can thus only appear at perturbation level.

In the next
 subsections we explore how dark energy or modified gravity effects  can be detected through  weak
lensing
and redshift surveys.

\subsection{General remarks}

Quite generally, cosmological observations fall into two categories: geometrical
probes and structure formation probes. While the former provide a measurement of
the Hubble function, the latter are a test of the gravitational theory in an
almost Newtonian limit on subhorizon scales. Furthermore, possible effects on
the geodesics of test particles need to be derived: naturally, photons follow
null-geodesics while massive particles, which constitute the cosmic large-scale
structure, move along geodesics for nonrelativistic particles.

In some special cases,  modified gravity models predict a strong  deviation from the standard
Friedmann equation as  in, e.g., DGP, (\ref{eq:Friedmann_DGP}).
While the Friedmann equation is not know explicitly in more general models of
massive gravity (Cascading gravity or hard mass gravity), similar modifications
are expected to arise and provide characteristic features, (see for instance
~\cite{Afshordi:2008rd,Jain:2010ka}) that could distinguish these models
from other scenarios of modified gravity or with additional dynamical degrees of
freedom. 

In general however the most interesting signatures of modified gravity
models are to be found in the perturbation sector.  For instance, in DGP,  growth functions 
differ from those in dark energy models by a few percent for identical Hubble
functions, and for that reason, an observation of both the Hubble and the growth
function gives a handle on constraining the gravitational theory,
\citep{Lue:2004rj}. The growth function can be estimated both through weak lensing
and through galaxy clustering and redshift distortions.

Concerning the interactions of light with the cosmic large-scale structure, one
sees a modified coupling in general models and a difference between the metric potentials.
These effects are present in the anisotropy pattern of the CMB, as shown in
\cite{Sawicki:2005cc}, where smaller fluctuations were found on large angular
scales, which can possibly alleviate the tension between the CMB and the
$\Lambda$CDM model on small multipoles where the CMB spectrum acquires smaller
amplitudes due to the ISW-effect on the last-scattering surface, but provides a
worse fit to supernova data. An interesting effect unexplicable in GR is the
anticorrelation between the CMB temperature and the density of galaxies at high
redshift due to a sign change in the integrated Sachs-Wolfe effect.
Interestingly, this behaviour is very common in modified gravity theories.

A very powerful probe of structure growth is of course weak lensing, but to evaluate the
lensing effect it is important to understand the nonlinear structure formation
dynamics as a good part of the total signal is generated by small structures.
Only recently  has it been possible to perform structure formation simulations
in modified gravity models, although still without a mechanism in which GR is recovered on
very small scales, necessary to be in accordance with local tests of gravity.

In contrast, the number density of collapsed objects relies only little on
nonlinear physics and can be used to investigate modified gravity cosmologies.
One needs to solve the dynamical equations for a spherically symmetric matter
distribution. Modified gravity theories show the feature of lowering the
collapse threshold for density fluctuations in the large-scale structure,
leading to a higher comoving number density of galaxies and clusters of
galaxies.
 This probe is degenerate with respect to dark energy cosmologies,
which generically give the same trends.

%% file: de_mg/magnification.tex
\subsection{Observing modified gravity with weak lensing}
The magnification matrix is a $2{\rm x}2$ matrix that relates the true shape of a galaxy to its image. It contains two distinct parts: the convergence, defined
as the trace of the matrix, modifies the size of the image, whereas the shear,
defined as the symmetric traceless part, distorts the shape of the image. At
small scales the shear and the convergence are not independent. They satisfy a
consistency relation, and they contain therefore the same information on matter
density perturbations. More precisely, the shear and the convergence are both
related to the sum of the two Bardeen potentials, $\Phi+\Psi$, integrated along
the photon trajectory. At large scales however, this consistency relation does
not hold anymore. Various relativistic effects contribute to the convergence,
see \cite{Bonvin:2008ni}. Some of these effects are generated along the photon
trajectory, whereas others are due to the perturbations of the galaxies
redshift. These relativistic effects provide independent information on the two
Bardeen potentials, breaking their degeneracy. The convergence is therefore a
useful quantity that can increase the discriminatory power of weak lensing.   

The convergence  can be measured through its effect on the galaxy number
density, see e.g.~\cite{Broadhurst:1994qu}. The standard method extracts the
magnification from correlations of distant quasars with foreground clusters,
see~\cite{Scranton:2005ci, Menard:2009yb}. Recently \cite{Zhang:2005pu,
Zhang:2005eb} designed a new method that permits to accurately measure
auto-correlations of the magnification, as a function of the galaxies redshift.
This method potentially allows measurements of the relativistic effects in the
convergence.

\subsubsection{Magnification matrix}

We are interested in computing the magnification matrix $\mathcal{D}_{ab}$ in a perturbed Friedmann
Universe. The magnification matrix relates the true shape of a galaxy to its image, and describes
therefore the deformations encountered by a light bundle along its trajectory.
$\mathcal{D}_{ab}$ can be computed by solving Sachs equation,  
see \cite{Sachs:1961zz}, that governs propagation of light in a generic geometry.
The convergence $\kappa$ and the shear $\gamma\equiv \gamma_1+i \gamma_2$
are then defined respectively as the trace and the symmetric traceless part
of $\mathcal{D}_{ab}$
\begin{equation}
\mathcal{D}_{ab}=\frac{\chi_S}{1+z_S}\left(\begin{array}{cc}
1-\kappa-\gamma_1&-\gamma_2\\
-\gamma_2&1-\kappa+\gamma_1 \end{array}\right)\; . \label{eq:kappadef}
\end{equation}
Here $z_S$ is the redshift of the source and $\chi_S$ is a time coordinate related to 
conformal time $\eta_S$ through $\chi_S=\eta_O-\eta_S$. 

We consider a spatially flat ($K=0$) Friedmann Universe with scalar perturbations.
We start from the usual  longitudinal (or Newtonian) gauge where the metric is given by
\begin{equation}
\label{eq:metricmag}
g_{\mu\nu}dx^\mu dx^\nu = a^2\left[ -(1+2\Psi)d\eta^2 +
(1-2\Phi)\delta_{ij}dx^idx^j\right]\; .
\end{equation}
We compute $\mathcal{D}_{ab}$ at linear order in $\Phi$ and $\Psi$ and then
we extract the shear and the convergence. We find, see
\cite{Bonvin:2008ni, Bernardeau:2009bm}
\begin{eqnarray}
\label{eq:gamma}\gamma&=&\frac{1}{2}\int_0^{\chi_S}d\chi
\frac{\chi_S-\chi}{\chi\chi_S}\;\raise1.0pt\hbox{$/$}
\hskip-6pt\partial^2(\Phi+\Psi)\label{gamma}\; ,  \\
\label{eq:kappa}\kappa&=& \frac{1}{2}\int_0^{\chi_S}d\chi
\frac{\chi_S-\chi}{\chi\chi_S}\;\overline{\raise1.0pt\hbox{$/$}
\hskip-6pt\partial}\;\raise1.0pt\hbox{$/$}\hskip-6pt\partial(\Phi+\Psi)
+ \Phi_S-\int_0^{\chi_S}\frac{d\chi}{\chi_S}(\Phi+\Psi)\label{kappa}\\
&+& \left(\frac{1}{\mathcal{H}_S \chi_S}-1\right)\left(\Psi_S+\mathbf{n}\cdot
\mathbf{v}_S-\int_0^{\chi_S}d\chi(\dot\Phi+\dot\Psi) \right)\; ,\nonumber 
\end{eqnarray}
where $\mathbf{n}$ is the direction of observation and $\mathbf{v}_S$ is the peculiar velocity of the source.
Here we are making use of
the angular spin raising $\;\raise1.0pt\hbox{$/$}\hskip-6pt\partial$ and
lowering $\;\overline{\raise1.0pt\hbox{$/$}\hskip-6pt\partial}$ operators (see
e.g.~\cite{Lewis:2001hp} for a review of the properties of these operators)
defined as
\begin{equation}
\;\raise1.0pt\hbox{$/$}\hskip-6pt\partial \; {}_s X   \equiv - \sin^s \theta
(\partial_\theta + i \csc \theta \partial_\varphi)  (\sin^{-s} \theta) \; {}_s X
\;, 
\qquad \;\overline{\raise1.0pt\hbox{$/$}\hskip-6pt\partial} \; {}_s X   \equiv -
\sin^{-s} \theta (\partial_\theta - i \csc \theta \partial_\varphi) (\sin^{s}
\theta) \; {}_s X \; ,
\label{eq:spinraising}
\end{equation} 
where ${}_s X$ is an arbitrary field of spin $s$ and $\theta$ and $\varphi$ are
spherical coordinates.

Eq.~(\ref{gamma}) and the first term in eq.~(\ref{kappa}) are the standard contributions of the shear and the
convergence, but expressed here with the full-sky transverse operators
\begin{eqnarray}
\frac{1}{\chi^2}\;\raise1.0pt\hbox{$/$}\hskip-6pt\partial^2&=&
\frac{1}{\chi^2}\left(\partial_\theta^2-\cot\theta\partial_\theta -\frac{1}{\sin^2\theta}\partial_\varphi\right)
+\frac{2 \rm{i}}{\chi^2\sin\theta}\Big( \partial_\theta\partial_\varphi-\cot\theta\partial_\theta \Big)~,\\
\frac{1}{\chi^2}\;\raise1.0pt\hbox{$/$}\hskip-6pt\partial\;\overline{\raise1.0pt\hbox{$/$}\hskip-6pt\partial}&=&
\frac{1}{\chi^2}\left(\partial_\theta^2+\cot\theta\partial_\theta +\frac{1}{\sin^2\theta}\partial_\varphi\right)~.
\end{eqnarray}
In the flat-sky approximation, where $\theta$ is very small, 
$\frac{1}{\chi^2}\;\raise1.0pt\hbox{$/$}\hskip-6pt\partial\;\overline{\raise1.0pt\hbox{$/$}\hskip-6pt\partial}$
reduces to the 2D Laplacian $\partial_x^2+\partial_y^2$ and one recovers the standard expression for the convergence.
Similarly, the real part of $\frac{1}{\chi^2}\;\raise1.0pt\hbox{$/$}\hskip-6pt\partial^2$ that corresponds to $\gamma_1$
reduces to $\partial_y^2-\partial_x^2$ and the imaginary part that corresponds to $\gamma_2$ becomes
$\partial_x\partial_y$.

The other terms in eq.~(\ref{kappa}) are relativistic corrections to the convergence,
that are negligible at small scales but may become relevant at large scales.
The terms in the first line are intrinsic corrections, generated respectively by the curvature
perturbation at the source position and the Shapiro time-delay. The terms in the
second line are due to the fact that we measure the convergence at a fixed redshift of the source $z_S$
rather that at a fixed conformal time $\eta_S$. Since in a perturbed universe, the observable redshift
is itself a perturbed quantity, this transformation generates additional contribution to the convergence. 
Those are respectively the Sachs-Wolfe contribution, the Doppler contribution and the integrated Sachs-Wolfe
contribution. 
Note that we have neglected the contributions at the observer position since they only give rise
to a monopole or dipole term. The dominant correction to the convergence is due
to the Doppler term.  Therefore in the following we are interested in comparing
its amplitude with the amplitude of the standard contribution. To that end we
define $\kappa_{\rm st}$ and $\kappa_{\rm vel}$ as 
\begin{eqnarray}
\label{eq:kp} \kappa_{\rm st}&=&\int_{0}^{\chi_S}d\chi
\frac{\chi_S-\chi}{2\chi\chi_S}\;\raise1.0pt\hbox{$/$}
\hskip-6pt\partial\;\overline{\raise1.0pt\hbox{$/$}\hskip-6pt\partial}
(\Phi+\Psi)\; ,\\ 
\label{eq:kv} \kappa_{\rm vel}&=&\left(\frac{1}{\mathcal{H}_S
\chi_S}-1\right)\mathbf{n}\cdot \mathbf{v}_S\; .
\end{eqnarray}

\subsubsection{Observable quantities}

The convergence is not directly observable. However it can be measured through
the modifications that it induces on the galaxy number density. Let us introduce
the magnification 
\begin{equation}
\mu=\frac{1}{\det \mathcal{D}}\simeq  1+2\kappa,
\hspace{0.4cm}\mbox{when}\hspace{0.5cm} |\kappa|, |\gamma|  \ll 1\; .
\end{equation}  
The magnification modifies the size of a source: $d\Omega_O=\mu d\Omega_S $,
where $d\Omega_S$ is the true angular size of the source and  $d\Omega_O$ is the
solid angle measured by the observer, i.e. the size of the image. The
magnification has therefore an impact on the observed galaxy number density. Let
us call $\bar{n}(f)df$ the number of unlensed galaxies per unit solid angle, at
a redshift $z_S$, and with a flux in the range $[f,f+df]$. The magnification
$\mu$ modifies the flux measured by the observer, since it modifies the observed
galaxy surface. It affects also the solid angle of observation and hence the
number of galaxies  per unit of solid angle. These two effects combine to give a
galaxy number overdensity, see~\cite{Broadhurst:1994qu, Scranton:2005ci}
\begin{equation}
\label{eq:overdens}
\delta^\mu_g=\frac{n(f)-\bar{n}(f)}{\bar{n}(f)}
\simeq1+2\big(\alpha-1\big)(\kappa_{\rm st}+\kappa_{\rm vel})\; .
\end{equation}
Here $\alpha\equiv -N'(>f_c)f_c/N(f_c)$, where $N(>f_c)$ is the number of
galaxies brighter than $f_c$ and $f_c$ is the flux limit adopted. Hence $\alpha$
is an observable quantity, see e.g.~\cite{Zhang:2005pu, Scranton:2005ci}. Recent
measurements of the galaxy number overdensity $\delta^\mu_g$ are reported
in~\cite{Scranton:2005ci, Menard:2009yb}. The challenge in those measurements is
to eliminate intrinsic clustering of galaxies, which induces an overdensity
$\delta_g^{cl}$ much larger than $\delta_g^\mu$. One possibility to separate
these two effects is to correlate galaxy number overdensities at widely
separated redshifts. One can then measure $\langle
\delta_g^\mu(z_S)\delta_g^{cl}(z_{S'})\rangle$, where $z_S$ is the redshift of
the sources and $z_{S'}<z_S$ is the redshift of the lenses. Another possibility,
proposed by \cite{Zhang:2005pu, Zhang:2005eb}, is to use the unique dependence
of $\delta^\mu_g$ on galaxy flux (i.e. on $\alpha$) to disentangle
$\delta^\mu_g$ from $\delta_g^{cl}$. This method, combined with precise
measurements of the galaxies redshift, allows to measure auto-correlations of
$\delta^\mu_g$, i.e. $\langle \delta_g^\mu(z_S)\delta_g^{\mu}(z_{S'})\rangle$,
either for $z_S\neq z_{S'}$ or for $z_S=z_{S'}$. The velocity contribution,
$\kappa_{\rm vel}$, has only an effect on  $\langle
\delta_g^\mu(z_S)\delta_g^{\mu}(z_{S'})\rangle$. The correlations between
$\delta_g^{cl}(z_{S'})$ and $\mathbf{v}_S$ are indeed completely negligible and
hence the source peculiar velocity does not affect $\langle
\delta_g^\mu(z_S)\delta_g^{cl}(z_{S'})\rangle$. In the following we study in
detail the contribution of peculiar motion to
$\langle\delta_g^\mu(z_S)\delta_g^{\mu}(z_S)\rangle$.

The two components of the convergence $\kappa_{\rm st}$ and $\kappa_{\rm vel}$
(and consequently the galaxy number overdensity)
are functions of redshift $z_S$ and direction of observation $\mathbf{n}$. We
can therefore determine the angular power spectrum
\begin{equation}
\langle \delta^\mu_g(z_S,\mathbf{n})\delta^\mu_g(z_S,\mathbf{n}')\rangle=
\sum_{\ell}\frac{2\ell+1}{4\pi}C_\ell(z_S)P_\ell(\mathbf{n}\cdot \mathbf{n}')\;
.
\end{equation}
The angular power spectrum $C_\ell(z_S)$ contains two contributions, generated
respectively by  $\langle \kappa_{\rm st} \kappa_{\rm st}\rangle$ and $\langle
\kappa_{\rm vel} \kappa_{\rm vel}\rangle$. The cross-term $\langle \kappa_{\rm
vel} \kappa_{\rm st}\rangle$  is negligible since $\kappa_{\rm st}$ contains
only Fourier modes with a wave vector $\mathbf{k}_\perp$ perpendicular to the
line of sight (see eq.~(\ref{eq:kp})), whereas $\kappa_{\rm vel}$ selects modes
with wave vector along the line of sight (eq.~(\ref{eq:kv})).  

So far the derivation has been completely generic. Eq.~(\ref{eq:gamma}) and (\ref{eq:kappa}) are valid in any theory of gravity whose metric can be written as in eq.~(\ref{eq:metricmag}). To evaluate the angular power spectrum we now have to be more specific. In the following we assume general relativity, with no anisotropic stress such that $\Phi=\Psi$. We use the Fourier transform convention 
\begin{equation}
\mathbf{v}(\mathbf{x},\chi)=\frac{1}{(2\pi)^3}\int d^3k
\mathbf{v}(\mathbf{k},\chi)e^{i\mathbf{k}\mathbf{x}}\; .
\end{equation}
The continuity equation, see e.g. \cite{Dodelson:2003ft}, allows us to express the peculiar velocity as
\begin{equation}
\label{eq:velocity}
\mathbf{v}(\mathbf{k},\chi)=-i\frac{\dot{G}(a)}{G(a)}\frac{\mathbf{k}}{k^2}
\delta(\mathbf{k},a)\; ,
\end{equation}
where $\delta(\mathbf{k},a)$ is the density contrast, $G(a)$ is the growth
function, and $\dot{G}(a)$ its derivative with respect to $\chi$. With this we can
express the angular power spectrum as
\begin{equation}
\label{cv}
C_\ell^{\rm
vel}(z_S)=\frac{16\pi\delta_H^2(\alpha_S-1)^2\dot{G}(a_S)^2}{H_0^4G^2(a=1)}
\left(\frac{1}{\mathcal{H}_S \chi_S}-1\right)^2 \int dk k T^2(k)j_\ell'(k
\chi_S)^2\; .
\end{equation}
Here $\delta_H$ is the density contrast at horizon and $T(k)$ is the transfer
function defined through, see e.g. \cite{Dodelson:2003ft}
\begin{equation}
\Psi(\mathbf{k},a)=\frac{9}{10}\Psi_p(\mathbf{k})T(k)\frac{G(a)}{a}\; .
\end{equation}
We assume a flat power spectrum, $n_s=1$, for the primordial potential $\Psi_p(\mathbf{k})$. We want to compare
this contribution with the standard contribution  
\begin{equation}
\label{eq:clstand}
C_\ell^{\rm
st}(z_S)=\frac{36\pi\delta_H^2(\alpha_S-1)^2\Omega_m^2\ell^2(\ell+1)^2}{G^2(a=1)
}\int\frac{dk}{k}T^2(k)
\left[\int_0^{\chi_S}d\chi
\frac{\chi_S-\chi}{\chi\chi_S}\frac{G(a)}{a}j_\ell(k\chi)\right]^2\; .
\end{equation}

\begin{figure}[t]
\begin{minipage}[!t]{7.cm}
\centerline{\epsfig{figure=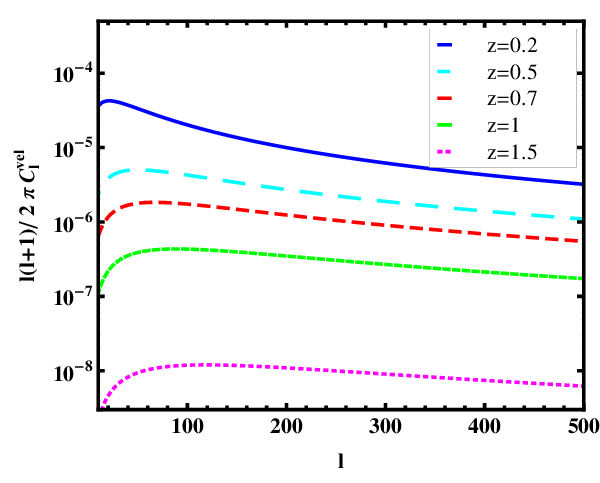,height=5.8cm}}
\caption{ \label{fig:clv} {\small The velocity contribution $C_\ell^{\rm vel}$
as a function of $\ell$ for various redshifts.}}
\end{minipage}
\hspace{1cm}\begin{minipage}[!t]{7.cm}
\centerline{\epsfig{figure=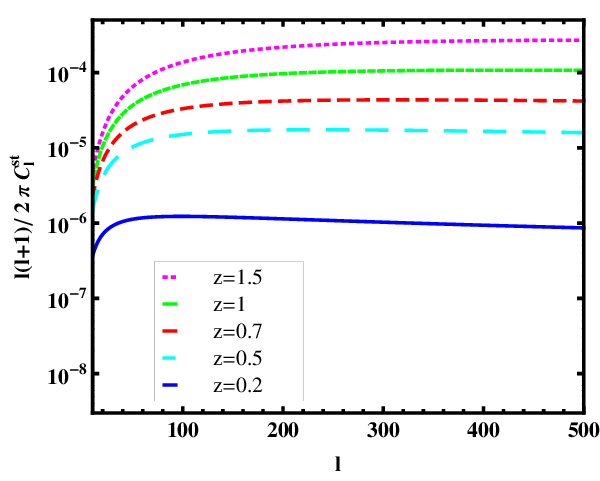,height=5.8cm}}
\caption{\label{fig:lens} {\small The standard contribution $C_\ell^{\rm st}$ as
a function of $\ell$ for various redshifts. }}
\end{minipage}
\end{figure}

We evaluate $C_\ell^{\rm vel}$ and $C_\ell^{\rm st}$ in a $\Lambda$CDM Universe
with $\Omega_m~=~0.25$, $\Omega_\Lambda~=~0.75$ and
$\delta_H=5.7\cdot 10^{-5}$.  We approximate the transfer function with the BBKS formula,
see~\cite{Bardeen:1985tr}.  In Fig.~\ref{fig:clv} and \ref{fig:lens}, we plot
$C_\ell^{\rm vel}$ and $C_\ell^{\rm st}$ for various source redshifts.  The
amplitude of $C_\ell^{\rm vel}$ and $C_\ell^{\rm st}$ depends on $(\alpha-1)^2$,
which varies with the redshift of the source, the flux threshold adopted, and
the sky coverage of the experiment. Since $(\alpha-1)^2$ influences $C_\ell^{\rm
vel}$ and $C_\ell^{\rm st}$ in the same way  we do not  include it in our plot.
Generally, at small redshifts, $(\alpha-1)$ is smaller than 1 and consequently
the amplitude of both $C_\ell^{\rm vel}$ and $C_\ell^{\rm st}$ is slightly
reduced, whereas at large redshifts $(\alpha-1)$ tends to be larger than 1 and
to amplify $C_\ell^{\rm vel}$ and $C_\ell^{\rm st}$, see
e.g.~\cite{Zhang:2005eb}. However, the general features of the curves and more
importantly the ratio between $C_\ell^{\rm vel}$ and $C_\ell^{\rm st}$ are not
affected by $(\alpha-1)$. 

Fig. \ref{fig:clv} shows that $C_\ell^{\rm vel}$ peaks at rather small $\ell$,
between $30$ and $120$ depending on the redshift. This correponds to rather
large angle $\theta\sim 90-360\,$arcmin. This behavior differs from the standard
term (Fig.~\ref{fig:lens}) that peaks at large $\ell$. It is therefore important
to have large sky surveys to detect the velocity contribution. The relative
importance of $C_\ell^{\rm vel}$ and $C_\ell^{\rm st}$ de\-pends stron\-gly on
the redshift of the source.  At small redshift, $z_S=0.2$, the velocity
contribution is about $4\cdot 10^{-5}$ and is hence larger than the standard
contribution which reaches $10^{-6}$.  At redshift $z_S=0.5$, $C_\ell^{\rm vel}$
is about $20\,\%$ of $C_\ell^{\rm st}$, whereas at redshift $z_S=1$, it is about
$1\,\%$ of $C_\ell^{\rm st}$.  Then at redshift $z_S=1.5$ and above,
$C_\ell^{\rm vel}$ becomes very small with respect to $C_\ell^{\rm st}$:
$C_\ell^{\rm vel} \leq 10^{-4}\,C_\ell^{\rm st}$. The enhancement of
$C_\ell^{\rm vel}$ at small redshift together with its fast decrease at large
redshift are due to the prefactor $\left(\frac{1}{\mathcal{H}_S
\chi_S}-1\right)^2$ in eq.~(\ref{cv}). Thanks to this enhancement we see that if the magnification can be measured with
an accuracy of $10\,\%$, then the velocity contribution is observable up to
redshifts $z\leq 0.6$. If the accuracy reaches  $1\,\%$ then the velocity
contribution becomes interesting up to redshifts of order 1.

The shear and the standard contribution in the convergence are not independent.
One can easily show that their angular power spectra satisfy the consistency
relation, see \cite{Hu:2000ee}
\begin{equation}
C_\ell^{\kappa\, {\rm st}}=\frac{\ell(\ell+1)}{(\ell+2)(\ell-1)}C_\ell^\gamma\;
.
\end{equation}
This relation is clearly modified by the velocity contribution. Using that the
cross-correlation between the standard term and the velocity term is negligible,
we can write a new consistency relation that relates the observed convergence
$C_\ell^{\kappa\, {\rm tot}}$ to the shear
\begin{equation}
\label{eq:cgamma}
\frac{\ell(\ell+1)}{(\ell+2)(\ell-1)}C_\ell^\gamma=C_\ell^{\kappa\, {\rm
tot}}-C_\ell^{\kappa\, {\rm vel}}\; .
\end{equation}
Consequently, if one measures both the shear $C_\ell^\gamma$ and the
magnification $C_\ell^{\kappa\, {\rm tot}}$ as functions of the redshift, eq.
(\ref{eq:cgamma}) allows to extract the peculiar velocity contribution
$C_\ell^{\kappa\, {\rm vel}}$. This provides a new way to measure peculiar
velocities of galaxies.

Note that in practice, in weak lensing tomography, the angular power spectrum is computed in redshift bins and therefore the square bracket in eq. (\ref{eq:clstand}) has to be integrated over the bin 

\begin{equation}
\int_0^\infty d\chi n_i(\chi)\int_0^{\chi}d\chi' \frac{\chi-\chi'}{\chi\chi'}\frac{G(\chi')}{a(\chi')}j_\ell(k\chi')~,
\end{equation}
where $n_i$ is the galaxy density for the i-th bin, convolved with a gaussian around the mean redshift of the bin. The integral over $\chi'$ is then simplified using Limber approximation, i.e.
\begin{equation}
\int_0^\chi d\chi' F(\chi')J_\ell(k\chi')\simeq \frac{1}{k}F\left(\frac{\ell}{k} \right)\theta(k\chi-\ell)~,
\end{equation}
where $J_\ell$ is the bessel function of order $\ell$. 
The accuracy of Limber approximation increases with $\ell$. 
Performing a change of coordinate such that $k=\ell/\chi$, eq.~(\ref{eq:clstand}) 
can be recast in the usual form used in weak lensing tomography, see e.g. 
eq.~(\ref{weak-lensing-non-parametric}).

%% file: de_mg/redshift-survey.tex
\subsection{Observing modified gravity with
redshift surveys}\label{dark-energy-and-redshift-surveys}

Wide-deep galaxy redshift surveys have the power to yield information on both
$H(z)$ and $f_{g}(z)$ through measurements of Baryon Acoustic Oscillations 
(BAO) and redshift-space distortions.
In particular, if gravity is not modified and matter is not interacting other
than gravitationally, then a detection of the expansion rate is directly linked
to a unique prediction of the growth rate. Otherwise galaxy redshift surveys
provide a unique and crucial way to make a combined analysis of $H(z)$ and
$f_{g}(z)$ to test gravity. 
As a wide-deep survey, Euclid allows us to measure $H(z)$ directly from BAO, but
also indirectly through the angular diameter distance $D_A(z)$ (and possibly
distance ratios from weak lensing). Most importantly,
Euclid survey enables us to measure the cosmic growth history using two
independent methods: $f_g(z)$ from galaxy clustering, and $G(z)$ from weak
lensing. In the following we discuss the estimation of
$[H(z), D_A(z)$ and $f_g(z)]$ from galaxy clustering.

\label{sec:rs-BAO}
From the measure of BAO in the matter power spectrum or in the 2-point
correlation function one can infer information on the expansion rate of the
universe. In fact, the sound waves imprinted in the CMB can be also detected in
the clustering of galaxies, thereby completing an important test of our theory
of gravitational structure formation. 
\begin{figure}[ht]
\centering
 \includegraphics[width=9cm]{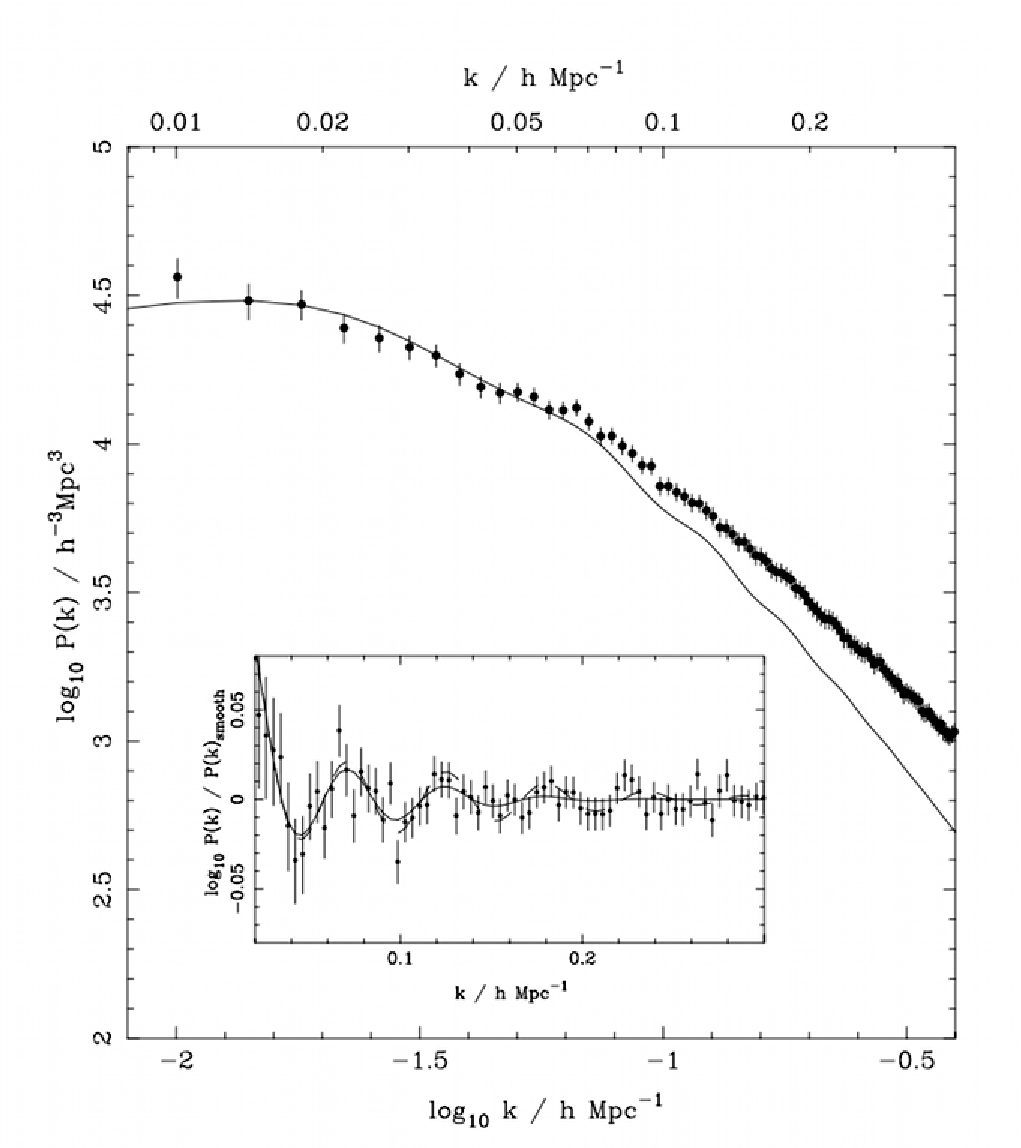}
\caption{Matter power spectrum form measured from SDSS
\citep{2007ApJ...657...51P}}
\label{fig:Pk} 
\end{figure}
The BAO in the radial and tangential directions offer a way to measure the
Hubble parameter and angular diameter distance, respectively. 
In the simplest Friedmann-Lemaitre-Robertson-Walker universe the basis to define
distances  is the dimensionless, radial, comoving distance: 
\begin{equation} 
\chi(z) \equiv\int_0^z \frac{dz'}{E(z')}\,. 
\label{chiz}
\end{equation}
{\color{red}The dimensionless version of the comoving distance
(defined in the previous section by the same symbol $\chi$) is:
\begin{equation} 
E^2(z) = \Omega_m^{(0)}(1+z)^3 + (1-\Omega_m^{(0)})
\exp\left[ \int_0^z
\frac{3(1+w(\tilde z))}{1+\tilde z}\mathrm d\tilde z \right].
\end{equation}}
The standard cosmological distances are related to $\chi(z)$ via
\begin{equation} 
D_A(z) = \frac{c}{H_0 (1+z)
\sqrt{-\Omega_k}}\sin\left(\sqrt{-\Omega_k}\chi(z)\right) 
\label{eq:ang}
\end{equation} 
where the luminosity distance, $D_L(z)$, is given by the distance duality: 
\begin{equation}  
D_L(z) = (1+z)^2 D_A(z).  
\label{eq:dual} 
\end{equation}
The coupling  between $D_A(z)$ and $D_L(z)$ persists in any metric theory of
gravity as long as photon number is conserved (see Sec. \ref{transparency-and-Etherington-relation} for
cases in which the duality relation is violated). 
BAO yield both $D_A(z)$ and $H(z)$ making use of an almost completely linear
physics (unlike for example SNIa, demanding  complex and poorly understood
mechanisms of explosions). Furthermore, they provide the chance of constraining
the growth rate  through the change in the amplitude of the power spectrum.

The characteristic scale of the BAO is set by the sound horizon at decoupling.
Consequently, one can attain  the angular diameter distance and Hubble parameter
separately. This  scale along the line of sight ($s_{||}(z)$) measures $H(z)$
through 
$H(z) = c\Delta z/s_{||}(z)$, while the tangential mode measures the angular
diameter distance $D_A(z)  = s_{\perp}/\Delta \theta (1+z)$.

One can then use the power spectrum to derive predictions on the parameter
constraining power of the survey (see e.g.
\cite{2005MNRAS.357..429A,2008Natur.451..541G,Wang08,Wang10,diporto10}).

In order to explore the cosmological parameter
constraints from a given redshift survey, one needs to specify
the measurement uncertainties of the galaxy power spectrum.
In general, the statistical error on the measurement of the galaxy
power spectrum $P_{\rm g}(k)$
at a given wave-number bin is \citep{FKP}
\begin{equation}
\left[\frac{\Delta P_{\rm g}}{P_{\rm g}}\right]^2=
\frac{2(2\pi)^2 }{\Vsur k^2\Delta k\Delta \mu}
\left[1+\frac1{n_{\rm g}P_{\rm g}}\right]^2,
\label{eqn:pkerror}
\end{equation}
where $n_{\rm g}$ is the mean number density of galaxies, 
$\Vsur$ is the comoving survey volume of the galaxy survey, and $\mu$
is the cosine of the angle between $\bf{k}$ and the line-of-sight
direction $\mu = \vec{k}\cdot \hat{r}/k$.

In general, the \emph{observed} galaxy power spectrum is different
from the \emph{true} spectrum, and it can be reconstructed approximately
assuming a reference cosmology (which we consider to be our fiducial
cosmology) as (e.g. \citep{seo03})
\begin{equation}
P_{\rm obs}(k_{{\rm ref}\perp},k_{{\rm ref}\parallel},z)
=\frac {\DA _{\rm ref} ^2 \hz}{\DA ^2 \hz _{\rm ref}} P_{\rm g}(k_{{\rm ref}\perp},k_{{\rm ref}\parallel},z)
+P_{\rm shot}\,,
\label{eq:Pobs}
\end{equation}
where
\begin{equation}
P_{\rm g}(k_{{\rm ref}\perp},k_{{\rm
    ref}\parallel},z)=b(z)^2\left[1+\beta(z) 
\frac{k_{{\rm ref}\parallel}^2}{k_{{\rm ref}\perp}^2+k_{{\rm ref}\parallel}^2}\right]^2\times
P_{{\rm matter}}(k,z)\,.
\label{eq:Pg}
\end{equation}
In Eq.~(\ref{eq:Pobs}), $H(z)$ and $D_A(z)$ are the Hubble parameter and the angular
diameter distance, respectively, and the prefactor 
$(\DA _{\rm ref} ^2 \hz)/(\DA ^2 \hz _{\rm ref})$ encapsulates the
geometrical distortions due to the  Alcock-Paczynski
effect \citep{seo03,9605017}.  Their values in the reference cosmology are
distinguished by the subscript `ref', while those in the true cosmology have no
subscript. $k_\perp$ and $k_\parallel$ are the wave-numbers across and along
the line of sight in the true cosmology, and they are related to the 
wave-numbers calculated assuming the reference
cosmology by
$k_{{\rm ref}\perp} = k_\perp D_A(z)/D_A(z)_{\rm ref}$ and
$k_{{\rm ref}\parallel} = k_\parallel H(z)_{\rm ref}/H(z)$. 
$P_{shot}$ is the unknown white shot noise that 
remains even after the conventional shot noise of inverse number density has been 
subtracted \citep{seo03}.
In Eq.~(\ref{eq:Pg}), $b(z)$ is the \emph{linear bias} factor between galaxy and
matter density distributions, $f_g(z)$ is the linear growth\label{symbol:bias}
rate\footnote{In presence of massive neutrinos $f_g$ depends also on
  the scale $k$ \citep{0709.0253}.}, 
and $\beta(z)=f_g(z)/b(z)$ is the linear \label{symbol:beta}
redshift-space distortion parameter \citep{Kaiser:1987}. 
The linear matter power spectrum $P_{{\rm matter}}(k,z)$ in Eq.~(\ref{eq:Pobs}) takes the form
\begin{eqnarray}
P_{{\rm matter}}(k,z)=\frac{8\pi^2c^4k_0\Delta^2_{\cal R}(k_0)}{25
  H_0^4\Omega_{m}^2} T^2(k) \left [\frac{G(z)}{G(z=0)}\right]^2
\left(\frac{k}{k_0}\right)^{n_s}e^{-k^2\mu^2\sigma_r^2},
\label{eq:Pm}
\end{eqnarray}
where $G(z)$ is the usual \emph{scale independent} linear
growth-factor in the absence of massive neutrino free-streaming (see Eq.~(25)
in \cite{Eisenstein_Hu_1997}), whose fiducial value in each redshift bin is computed 
through numerical integration of the differential equations governing the growth
of linear perturbations in presence of dark-energy
\citep{astro-ph/0305286} or employing the approximation of Eq. (\ref{def-growth-rate}). $T(k)$
 depends on matter and baryon densities\footnote{If we assume that
   neutrinos have a non-vanishing mass, then the transfer function is
   also redshift-dependent.} (neglecting dark-energy at early times), 
and is computed in each redshift bin using a Boltzmann code like
{\sc camb}\footnote{http://camb.info/} \citep{CAMB} or {\sc cmbfast}.

In Eq.~(\ref{eq:Pm})  a damping factor
$e^{-k^2\mu^2\sigma_r^2}$ has been added,
due to redshift uncertainties, where $\sigma_r=(\partial r/\partial
z)\sigma_z$, $r(z)$ being the comoving
distance \citep{0904.2218,seo03},
and assumed that the power spectrum of primordial curvature
perturbations, $P_{\cal R}(k)$, is
\begin{equation}
\Delta^2_{\cal R}(k) \equiv \frac{k^3P_{\cal R}(k)}{2\pi^2}
= \Delta^2_{\cal R}(k_0)\left(\frac{k}{k_0}\right)^{n_s},
\label{eq:pR}
\end{equation}
where $k_0=0.002$/Mpc, $\Delta^2_{\cal R}(k_0)|_{\rm fid}=2.45\times
10^{-9}$ is the dimensionless amplitude of the primordial curvature perturbations
evaluated at a pivot scale $k_0$, and $n_s$ is the scalar spectral
index \citep{arXiv:1001.4635}.

In the limit where the survey volume is much larger than the scale of 
any features in $P_{\rm obs}(k)$, it has been shown that
the redshift survey Fisher matrix for a given redshift bin can be approximated as \citep{Tegmark97}
\begin{eqnarray}
F_{ij}^{\rm LSS}
&=&\int_{-1}^{1} \int_{k_{\rm min}}^{\kmax}\frac{\partial \ln
  P_{\rm obs}(k,\mu)}{\partial p_i} \frac{\partial \ln P_{\rm obs}(k,\mu)}{\partial p_j} 
\Veff(k,\mu) \frac{2\pi k^2 dk d\mu}{2(2\pi)^3}
\label{Fisher}                 
\end{eqnarray}
where the derivatives are evaluated at the parameter values $p_i$
of the fiducial model,
and $\Veff$ is the effective volume of the survey:
\begin{eqnarray}
\Veff(k,\mu) =
\left [ \frac{{n_{\rm g}}P_{\rm g}(k,\mu)}{{n_{\rm g}}P_{\rm g}(k,\mu)+1} \right ]^2 \Vsur,
\label{V_eff}                                                                                        
\end{eqnarray}
where the comoving number density
$n_{\rm g}(z)$ is assumed to be spatially constant.
Due to azimuthal symmetry around the line of sight,
the three-dimensional galaxy redshift power spectrum
$P_{\rm obs}(\vec{k})$ depends only on $k$ and $\mu$, i.e. is reduced
to two dimensions by symmetry \citep{SE03}. The total Fisher matrix
can be obtained by summing over the redshift bins.

To minimise nonlinear effects, one should restrict wave-numbers to the 
quasi-linear regime, e.g. imposing that $\kmax$ is given
by requiring that the variance of matter fluctuations in a sphere of
radius $R$ is, for instance, $\sigma^2(R)=0.25$ for $R=\pi/(2\kmax)$. 
Or one could model the non-linear distortions as in \cite{eisenstein07}.
On scales larger than ($\sim 100\, h^{-1}$Mpc) where we focus our analysis,
nonlinear effects can be represented in fact as a displacement field in Lagrangian space
modeled by an elliptical Gaussian function. Therefore, following
\cite{eisenstein07,seo07}, to model nonlinear effect
we multiply $P(k)$ by the factor
\begin{equation}
\exp\left\{ 
-k^{2}\left[\frac{(1-\mu^{2})\Sigma_{\perp}^{\,2}}{2}+\frac{\mu^{2}\Sigma_{
\parallel}^{\,2}}{2}\right]\right\}
\,, \label{eq:damping}
\end{equation}
where $\Sigma_{\perp}$ and $\Sigma_{\parallel}$ represent the
displacement across and along the line of sight, respectively. They  are related
to the growth factor $G$ and to the growth rate
$f_g$ through $\Sigma_{\perp}=\Sigma_{0}G$ and
$\Sigma_{\parallel}=\Sigma_{0}G(1+f_g)$.
The value of $\Sigma_{0}$ is proportional to $\sigma_{8}$. For a
reference cosmology where $\sigma_{8}=0.8$~\citep{wmap7}, we have
$\Sigma_{0}=11\, h^{-1}$Mpc.

Finally, we note that when actual data are available, the usual
way to measure  $\beta=f_g/b$ is  by fitting the measured
galaxy redshift-space correlation function $\xi(\sigma,\pi)$ to a model 
\citep{Peebles80}:
\begin{equation}
\xi(\sigma,\pi)= \int_{-\infty}^{\infty}{\rm d}v\, f(v)\,
\tilde{\xi}(\sigma,\pi-v/H_0),
\end{equation}
where $f(v)$ describes the small-scale random motion (usually modeled by 
a Gaussian that depends on the galaxy pairwise peculiar velocity dispersion),
and $\tilde{\xi}(\sigma,\pi)$ is the model accounting for coherent 
infall velocities\footnote{See \cite{Hamilton92}. $\tilde{\xi}(\sigma,\pi)$ 
is the Fourier transform of $P_s(k)=(1+\beta \mu^2)^2 P_r(k)$
\citep{Kaiser:1987}.}: 
\begin{equation}
\tilde{\xi}(\sigma,\pi)=\xi_0(s) P_0(\mu)
+\xi_2(s) P_2(\mu)+ \xi_4(s) P_4(\mu).
\end{equation}
$P_l(\mu)$ are Legendre polynomials; $\mu=\cos\theta$,\label{symbol:legendre}
where $\theta$ denotes the angle between ${\mbox{\bf r}}$ and $\pi$; 
$\xi_0(s)$, $\xi_2(s)$, and $\xi_4(s)$ depend on
$\beta$ and the real-space correlation function $\xi(r)$.

The bias between galaxy and matter distributions can be estimated from
either galaxy clustering, or weak lensing. 
To determine bias, we can assume that the galaxy density perturbation 
$\delta_g$ is related to the matter density perturbation $\delta_m({\mbox{\bf
x}})$
as \citep{Fry_Gaztanaga93}:
\begin{equation}
\delta_g= b \delta_m({\mbox{\bf x}})+ b_2 \delta_m^2({\mbox{\bf x}})/2.
\end{equation}

Bias can be derived from galaxy clustering by measuring the galaxy bispectrum:
\begin{eqnarray}
\langle \delta_{g{\mbox{\bf k}}_1} \delta_{g{\mbox{\bf k}}_2}
\delta_{g{\mbox{\bf k}}_1}\rangle
&=& (2\pi)^3 \left\{P({\mbox{\bf k}}_1) P({\mbox{\bf k}}_2)\left[J({\mbox{\bf
k}}_1,{\mbox{\bf k}}_2)/b
+b_2/b^2\right] \right. \nonumber\\
& & \left. \hskip 1cm +cyc.\right\} \delta^D({\mbox{\bf k}}_1+{\mbox{\bf
k}}_2+{\mbox{\bf k}}_3),
\end{eqnarray}
where $J$ is a function that depends on the shape of the
triangle formed by (${\mbox{\bf k}}_1$, ${\mbox{\bf k}}_2$, ${\mbox{\bf k}}_3$)
in
${\mbox{\bf k}}$ space, but only depends very weakly on cosmology 
\citep{Matarrese_Verde_Heavens97,Verde02}.

In general, bias can be measured from weak lensing through the comparison of the
shear-shear and shear-galaxy correlations functions.
A combined constraint on bias and the growth factor $G(z)$
can be derived from weak lensing by comparing the cross-correlations
of multiple redshift slices.

Of course, if bias is assumed to be linear ($b_2=0$) and scale independent, or is parametrized in
some simple way, e.g.
with a power law scale dependence, then it is possible to estimate it even from
linear galaxy clustering alone,
as we will see in  Sec. \ref{gamma-bias-forecasts}.

%% file: de_mg/bulk.tex
\subsection{Cosmological Bulk Flows}

As we have seen, the additional redshift induced by the galaxy peculiar velocity field generates the redshift 
distortion in the power spectrum. In this section we discuss
a related effect on the luminosity of the galaxies and on its use
to measure the peculiar velocity in large volumes, the so-called bulk flow.

In the gravitational instability framework, inhomogeneities in the matter distribution 
induce gravitational accelerations ${\bf g}$, which result in galaxies having peculiar velocities 
${\bf v}$ that add to the Hubble flow.
In linear theory the peculiar velocity field is proportional to the peculiar acceleration
\begin{equation}
\label{eq:vg}
{\bf v}({\bf r})=\frac{2f_g}{3H_0\Omega_m}{\bf g}({\bf r})=
\frac{H_0 f_g}{4\pi} \int \delta_m({\bf r^\prime})\frac{({\bf r^\prime} - {\bf r})}{|{\bf r^\prime}-{\bf r}|^3}d^3{\bf r^\prime} \; ,
\end{equation}
and the bulk flow of a spherical region is solely determined by the gravitational
pull of the dipole of the external mass distribution. For this reason, bulk flows
are reliable indicators to deviations from homogeneity and isotropy on 
large scale, should they exist.

Constraints on the power spectrum and growth rate can be obtained by comparing
the bulk flow estimated from
the volume-averaged motion of the sphere of radius $R$:
\begin{equation}
\label{eq:vbulk}
{\bf B}_{R}\equiv \frac{\int {\bf v}({\bf x}) W({\bf x}/R)d^3{\bf x}}{\int W({\bf x}/R)d^3{\bf x}}
\end{equation}
with expected variance:
\begin{equation}
\label{eq:vbexp}
\sigma^2_{{\bf B},{R}}=
\frac{H_0^2 f_g^2}{6\pi^2}\int P(k){\cal W}(kR)^2(k)dk \;,
\end{equation}
where the window function $W({\bf x}/R)$ and its Fourier transform
${\cal W}(kR)$ describe the spatial distribution of the dataset.

Over the years the bulk flows has been estimated from 
the measured peculiar velocities of a large variety of objects ranging from galaxies
\citep{1998ApJ...505L..91G, 1998AJ....116.2632G, 1999ApJ...522....1D, 2000ApJ...544..636C, 2000ApJ...537L..81D,
2007MNRAS.375..691S}
clusters of galaxies \citep{1994ApJ...425..418L, 1996ApJ...461L..17B, 2004MNRAS.352...61H}
and SNIa  \citep{1995ApJ...445L..91R}. Conflicting results triggered by the use of error-prone distance indicators 
have fueled a long lasting controversy on the amplitude and  convergence of the bulk flow that is still on.
For example, the recent claim of a bulk flow of $407\pm81$ km s$^{-1}$ within $R=50$ $h^{-1}$Mpc
\citep{2009MNRAS.392..743W}, inconsistent with expectation from the $\Lambda$CDM model, has been seriously
challenged by the re-analysis of the same data by \cite{2011arXiv1101.1650N} who found a 
bulk flow amplitude consistent  with  $\Lambda$CDM expectations and from which they 
were able to set the strongest constraints on modified gravity models so far. 
On larger scales, \cite{2010ApJ...712L..81K} claimed the detection of a dipole anisotropy
attributed to the kinetic SZ decrement in the WMAP temperature map at the 
position of X-ray galaxy clusters. When interpreted  as a coherent motion, this signal would indicate 
a  gigantic bulk flow of $1028\pm 265$
km s$^{-1}$ within $R=528$ $h^{-1}$Mpc. This highly debated result 
has been  seriously questioned by independent analyses of  WMAP data
(see e.g. \cite{2010arXiv1011.2781O})

The large, homogeneous dataset expected from Euclid has the potential to settle these issues.
The idea is to measure bulk flows in 
large redshift surveys, based on the apparent,  dimming or brightening of galaxies
due to their peculiar motion. The method, originally proposed by  \cite{TYS}, has been recently
extended by \cite{2011arXiv1102.4189N}
 who propose to estimate the bulk flow 
by minimizing systematic variations in galaxy luminosities
with respect to a reference luminosity function measured 
from the whole survey.  It turns out that, if applied to the
photo-z catalog expected from Euclid, this method would be able to 
detect at $\> 5 \sigma$ significance
 a bulk flow like the one of  \citep{2009MNRAS.392..743W} 
over  $\sim 50$ independent spherical volumes at $z \ge 0.2$, provided that the
systematic magnitude offset over the corresponding areas in the sky does not exceed 
the expected random magnitude errors of 0.02-0.04 mag.
Additionally, photo-z or spectral-z  could be used to validate or disproof 
with very large ($\> 7 \sigma$) significance
the claimed bulk flow detection of \cite{2010ApJ...712L..81K} at $z=0.5$.

Closely related to the bulk flow 
is the Local Group peculiar velocity inferred from the observed CMB dipole
\citep{1990ApJ...349..408J}
\begin{equation}
\label{eq:vlg}
{\bf v}_{CMB}={\bf v}_{LG,R}-\frac{H_0f_g}{3}{\bf x}_{c.m.}+{\bf B}_{R} \;,
\end{equation}
where ${\bf v}_{LG,R}$ is the Local Group velocity resulting from the gravitational pull
of all objects in the sample within the radius $R$, ${\bf x}_{c.m.}$ is the position of
the center of mass of the sample and ${\bf v}_{CMB}$ is the LG velocity  inferred from the 
CMB dipole 
\citep{2003ApJS..148....1B}.
The convergence of ${\bf v}_{LG,R}$ with the radius and its alignment with the CMB dipole direction
indicates a crossover to homogeneity
\citep{1991ApJ...376L...1S}
 and allows to constrain the growth rate by comparing 
${\bf v}_{CMB}$ with ${\bf v}_{LG,R}$. The latter can be estimated from the dipole in the distribution 
of objects either using a number-weighting scheme if redshifts are available for all objects of the sample,
or using a flux-weighting scheme. In this second case the fact that both gravitational acceleration and
flux are inversely proportional to the distance allows to compute the dipole from photometric catalogs with no
need to measure redshifts. The drawback is that the information on the convergence scale is lost.

As for the bulk flow case, despite the many measurements of cosmological dipoles using galaxies
\citep{1980ApJ...242..448Y, 1982ApJ...254..437D, 1986AJ.....91..191M, 1992ApJ...397..395S,
1999MNRAS.304..893S, 2006ApJ...645.1043K}
there is still no general consensus on the scale of convergence and even on
the convergence itself. Even the recent analyses of measuring the acceleration of the Local Group from
the 2MASS redshift catalogs provided conflicting results. 
\cite{2006MNRAS.368.1515E} found that the galaxy dipole seems to converge beyond  $R=60$ $h^{-1}$Mpc
whereas \cite{2010ApJ...709..483L} find no convergence within $R=120$ $h^{-1}$Mpc.

Once again, Euclid will be in the position to solve this controversy by measuring the galaxy 
and cluster dipoles 
not only at the LG position and out to very large radii, but also in several independent ad truly
all-sky spherical samples carved out from the the observed areas with $|b|>20^{\circ}$.
In particular, coupling photometry with photo-$z$ one expects to be able to estimate
the convergence scale of the flux-weighted dipole over about 100 independent spheres of radius
$200$ $h^{-1}$Mpc out to $z=0.5$ and, beyond that, to compare number-weighted and flux-weighted
dipoles over a larger number of similar volumes using spectroscopic redshifts.

%% file: de_mg/obs_prospects.tex
\noindent Here we describe forecasts for the constraints on modified gravity
parameters which Euclid observations should be able to achieve. We begin
with reviewing the relevant works in literature. Then, after we define
 our {\color{red}``Euclid model''}, i.e. the main specifics of the redshift
and weak lensing survey,
we illustrate a number of Euclid forecasts obtained through a Fisher matrix approach.

\subsection{A review of forecasts for parametrised  modified gravity with Euclid}

\cite{Heavens/etal:2007} have used Bayesian evidence to distinguish between
models, using the Fisher matrices for the parameters of interest. This study
calculates the ratio of evidences $B$ for a 3D weak lensing analysis of the full
Euclid survey, for a dark energy model with varying equation of state, and
modified gravity with additionally varying growth parameter $\gamma$. They find
that Euclid can decisively distinguish between e.g. DGP and dark energy, with
$|\ln B|\simeq 50$. In addition, they find that it will be possible to
distinguish any departure from GR which has a difference in $\gamma$ greater
than $\simeq 0.03$. A phenomenological extension of the DGP model \citep{Dvali:2003rk, Afshordi:2008rd} has also been tested with Euclid. Specifically, \citet{Camera:2011mg} found that it will be possible to discriminate between this modification to gravity from $\Lambda$CDM at the $3\sigma$ level in a wide range of angular scale, approximately $1000\lesssim\ell\lesssim4000$.

\begin{figure}
 
\vspace{-.1cm}
\begin{center}
\includegraphics[width=9cm]{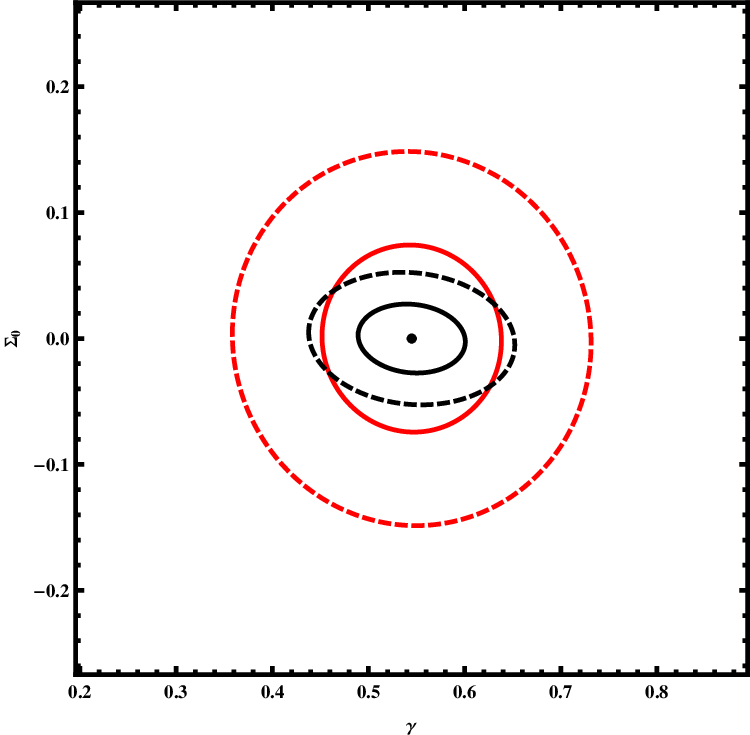}
\caption{Marginalized $\gamma-\Sigma_0$ forecast for weak lensing only analysis
with Euclid. {\color{red}Here $\Sigma_0$ is defined from $\Sigma =
1+\Sigma_0 a$ and $\Sigma, defined via Eq. \ref{Sigma-wl-isw}$, is related
to the WL potential. } Black contours correspond to $\ell_{max}=5000$, demonstrating an
error of 0.089$(1\sigma)$ on $\Sigma_0$, whereas the red contours correspond to
$\ell_{max}=500$ giving an error of 0.034. In both cases, the inner and outer
contours are $1\sigma$ and $2\sigma$ respectively. General Relativity resides at
[0.55, 0], while DGP resides at [0.68, 0]. \label{fig:thomas}}
\end{center}
\end{figure}

\cite{2009MNRAS.395..197T}  construct Fisher matrix forecasts for the Euclid
weak lensing survey, shown in figure \ref{fig:thomas}. The constraints obtained
depend on the maximum wavenumber which we are confident in using; $\ell_{\rm
max}=500$ is relatively conservative as it probes the linear regime where we can
hope to analytically track the growth of structure; $\ell_{\rm max}=10000$ is more
ambitious as it includes nonlinear power, using the \cite{Smith2003}
fitting function. This will not be strictly correct, as the fitting function was
determined in a GR context. Note that $\gamma$ is not very sensitive to $\ell_{\rm
max}$, while $\Sigma_0${\color{red}, defined in \cite{Amendola:2007rr} as $\Sigma = 1 + \Sigma_0 a$ (and where $\Sigma$ is defined in Eq. \ref{Sigma-wl-isw})}   is measured much more accurately in the non-linear
regime.

\cite{Amendola:2007rr} find Euclid weak lensing constraints for a more
general parameterization that includes evolution. In particular, $\Sigma(z)$ is
investigated by dividing the Euclid weak lensing survey into three redshift bins
with equal numbers of galaxies in each bin, and approximating that $\Sigma$ is
constant within that bin. {\color{red}Since $\Sigma_1$, i.e.  the value of $\Sigma$ in the $a=1$ bin (present-day)  is degenerate with the amplitude of matter
fluctuations, it is set to unity.} The study finds that a
deviation from unit $\Sigma$ (i.e. General Relativity) of 3\% can be detected in
the second redshift bin, and a deviation of 10\% is still detected in the
furthest redshift bin.

\cite{2009arXiv0910.1480B} make forecasts for modified gravity with Euclid weak
lensing   including \cite{Hu:2007pj} in interpolating between the linear
spectrum predicted by modified gravity, and GR on small scales as required by
Solar System tests. This requires parameters $A$ (a measure of the abruptness of
transitioning between these two regimes), $\alpha_1$ (controlling the
$k$-dependence of the transition) and $\alpha_2$ (controlling the $z$-dependence
of the transition). 

\begin{figure}

\begin{center}
\includegraphics[width=14cm]{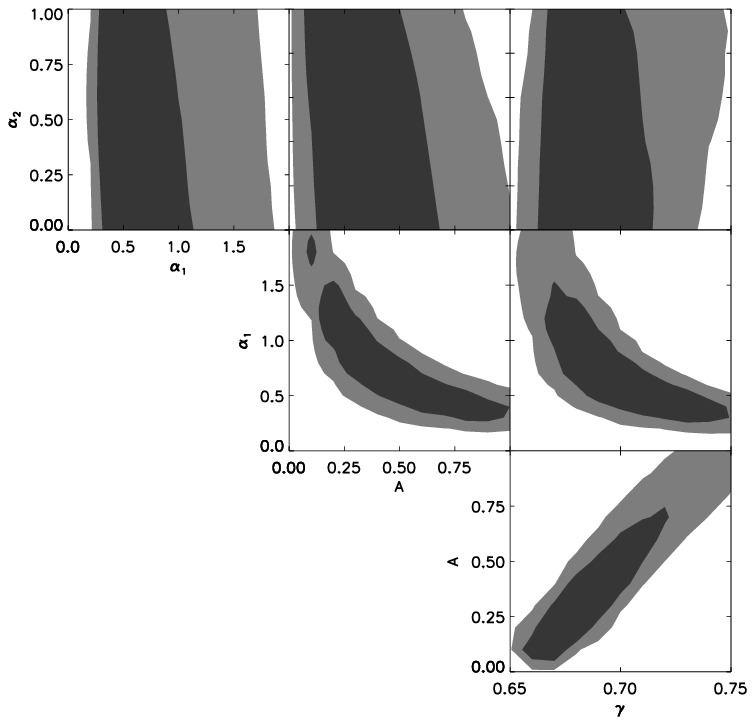}
\caption{Constraints on $\gamma$, $\alpha_1$, $\alpha_2$ and $A$ from Euclid,
using a DGP fiducial model and 0.4 redshift bins between 0.3 and 1.5 for the
central cosmological parameter values fitting WMAP+BAO+SNe. \label{fig:beynon}}
\end{center}
\end{figure}

The forecasts for modified gravity parameters are shown in Figure
\ref{fig:beynon} for the Euclid lensing data. Even with this larger range of
parameters to fit, Euclid provides a measurement of the growth factor $\gamma$
to within 10\%, and also allows some constraint on the $\alpha_1$ parameter,
probing the physics of nonlinear collapse in the modified gravity model.

Finally, Song et al. (2010) have shown forecasts for measuring $\Sigma$ and $\mu$ using
both imaging and spectroscopic surveys. They combine $20,000$
square-degree lensing data (corresponding to \citealt{euclidyellowbook} rather than to the updated \citealt{euclidredbook}) with the peculiar velocity dispersion measured from redshift
space distortions in the spectroscopic survey, together with stringent
background expansion measurements from the CMB and supernovae. They find that
for simple models for the redshift evolution of $\Sigma$ and $\mu$, both
quantities can be measured to 20\% accuracy.

%% file: de_mg/euclid-surveys.tex
\subsection{Euclid surveys}\label{sec:baofm_survey}

The Euclid mission will produce a catalog of up to 100 million galaxy redshifts
and an imaging survey that should allow to estimate the galaxy ellipticity
of up to 2 billion galaxy images. Here we discuss these surveys and fix their 
main properties into a ``Euclid model", i.e. an approximation to the real
Euclid survey that will be used as reference mission in the following.
\\

{\bf Modeling the Redshift Survey.}

The main goals of next generation redshift surveys will be
to constrain the Dark Energy parameters and  to explore models
alternative to standard Einstein Gravity. For these purposes they will
need to consider very large volumes that encompass
$z\sim 1$, i.e. the epoch at which dark energy  started dominating
the energy budget, spanning a range of epochs large enough to
provide a sufficient leverage to discriminate among
competing models at different redshifts.

Here we consider a survey covering a large fraction of the extragalactic corresponding to $\sim 15000$ deg$^2$
capable to measure a large number of galaxy redshifts out to $z\sim 2$.
A promising observational strategy is to target H$\alpha$ emitters at
near-infrared  wavelengths (which implies $z>0.5$) since they guarantee both
relatively dense sampling (the space density of
this population is expected to increase out to $z\sim 2$) and an 
efficient method to measure the redshift of the object.
The limiting flux of the survey should be the tradeoff between
the requirement of minimizing the shot noise, the contamination by
other lines (chiefly among them the [O${\rm II}$] line), and that of 
maximizing the so-called efficiency $\varepsilon$, i.e. the fraction of
successfully measured redshifts.
To minimize shot noise one should obviously strive for a low flux. Indeed, 
\cite{geach10} found that  a limiting flux  $f_{{\rm H} \alpha} \ge
1\times10^{-16}$  $ \textrm{erg } \textrm{cm}^{-2}  \textrm{s}^{-1}$ would be
able to balance shot noise and cosmic variance
out to $z=1.5$. However, simulated observations of mock  H$\alpha$ galaxy
spectra have shown that $\varepsilon$ ranges between 30 \% and 60\% (depending
on the redshift) for a limiting flux $f_{{\rm H}_\alpha}\ge 3\times10^{-16}$ 
$\textrm{erg } \textrm{cm}^{-2}  \textrm{s}^{-1}$~\citep{euclidredbook}. Moreover,
contamination from  [O${\rm II}$] line drops from 12\% to 1\% when the limiting
flux increases from $1\times10^{-16}$ to
$5\times10^{-16}$ $\textrm{erg } \textrm{cm}^{-2}  \textrm{s}^{-1}$~\citep{geach10}. 

Taking all this into account, in order to reach the top-level science requirement on the number density of H$\alpha$ galaxies, the average effective H$\alpha$ line flux limit from a 1-arcsec diameter
source shall be lower than or equal to $3\times10^{-16}$ $\textrm{erg } \textrm{cm}^{-2}  \textrm{s}^{-1}$.
However, a slitless spectroscopic survey has a success rate in measuring redshifts
that is a function of the emission line flux. As such, the Euclid survey cannot be characterized by a
single flux limit, as in conventional slit spectroscopy.

We use the number density of H$\alpha$ galaxies at a given redshift, $n(z)$,
estimated using the latest empirical data (see Fig. 3.2
of~\cite{euclidredbook}), where the values account for redshift - and flux - success rate, to which we refer as our reference efficiency $\varepsilon_r$.

However, in an attempt to bracket current uncertainties in modeling galaxy surveys, we consider two further scenarios, one where the efficiency is only the half of $\varepsilon_r$ and one where it is increased by a factor of 40\%. Then we define the following cases:

\begin{itemize}
\item {\it Reference case (ref.)}. Galaxy number density $n(z)$ which include efficiency $\varepsilon_r$ (column $n_2(z)$ in Tab. \ref{tab:n_z}).
\item {\it Pessimistic case (pess.)}. Galaxy number density $n(z)\cdot 0.5$. i.e. efficiency is $\varepsilon_{r}\cdot 0.5$ (column $n_3(z)$ in Tab. \ref{tab:n_z}).
\item {\it Optimistic case (opt.)}. Galaxy number density $n(z)\cdot 1.4$ i.e. efficiency is $\varepsilon_{r}\cdot 1.4$ (column $n_1(z)$ in Tab. \ref{tab:n_z}).

\end{itemize}

The total number of observed galaxies ranges from $3\cdot10^7$ (pess.) to
$9\cdot10^7$ (opt.).
For all cases we assume that the  error on the measured 
redshift is $\Delta z=0.001(1+z)$, independent of the  limiting flux of the survey. \\

\begin{table}
\centering
\begin{tabular}
{|>{\centering}p{2.5cm}|>{\centering}p{2.5cm}|>{\centering}p{2.5cm}|>{\centering}p{2.5cm}|}
\hline 
$z$  & $n_{1}(z)$ $\times10^{-3}$  & $n_{2}(z)$ $\times10^{-3}$  &
$n_{3}(z)$ $\times10^{-3}$
\tabularnewline
\hline 
0.65-0.75  & $1.75$  & $1.25$  & $0.63$\tabularnewline
0.75-0.85  & $2.68$  & $1.92$  & $0.96$\tabularnewline
0.85-0.95  & $2.56$  & $1.83$  & $0.91$\tabularnewline
0.95-1.05  & $2.35$  & $1.68$  & $0.84$\tabularnewline
1.05-1.15  & $2.12$  & $1.51$  & $0.76$\tabularnewline
1.15-1.25  & $1.88$  & $1.35$  & $0.67$\tabularnewline
1.25-1.35  & $1.68$  & $1.20$  & $0.60$\tabularnewline
1.35-1.45  & $1.40$  & $1.00$  & $0.50$\tabularnewline
1.45-1.55  & $1.12$  & $0.80$  & $0.40$\tabularnewline
1.55-1.65  & $0.81$  & $0.58$  & $0.29$\tabularnewline
1.65-1.75  & $0.53$  & $0.38$  & $0.19$\tabularnewline
1.75-1.85  & $0.49$  & $0.35$  & $0.18$\tabularnewline
1.85-1.95  & $0.29$  & $0.21$  & $0.10$\tabularnewline
1.95-2.05  & $0.16$  & $0.11$  & $0.06$\tabularnewline
\hline
\end{tabular}
\caption{\label{tab:n_z}Expected galaxy number densities in units of
($h/$Mpc)$^{3}$
for Euclid survey. Let us notice that the galaxy number densities $n(z)$ depend on the fiducial cosmology adopted in the computation of the survey volume, needed for the conversion from the galaxy numbers $dN/dz$ to $n(z)$.}
\end{table}

{\bf Modeling the weak lensing survey.}

For the weak lensing survey, we assume again a sky coverage of 15,000 square degrees.  For the number
density we use the common
parameterization
\begin{equation}
n(z) = z^2 \exp(-(z/z_0)^{3/2}),
\end{equation}
where $z_0 =z_\mathrm{mean}/1.412$ is the peak of $n(z)$ and
$z_\mathrm{mean}$ the
median and typically 
we assume $z_{mean}=0.9$ and a surface density of valid images of $n_g=30$ per arcmin$^2$ ~\citep{euclidredbook}).
We also assume that the photometric redshifts give an error of $\Delta z=0.05(1+z)$. Other
specifications will be presented in the relevant sections.

%% file: de_mg/gamma-param.tex
\subsection{Forecasts for the growth rate from the redshift survey}\label{gamma-bias-forecasts}

In this section we forecast the constraints that future observations
can put on the growth rate and on a scale-independent bias, employing the Fisher matrix method
presented in Sec. \ref{dark-energy-and-redshift-surveys}. We use the representative Euclid survey
presented in Sec. \ref{sec:baofm_survey}.   We assess
how well one can constrain the bias function from the analysis
of the power spectrum itself and evaluate the impact that treating
bias as a free parameter has on the estimates of the growth factor.   We
estimate how errors depend on the parametrization of the growth factor and on
the number and type of degrees of freedom in the analysis.
Finally, we explicitly explore the case of coupling between dark energy and
dark matter and assess the ability of measuring the coupling constant.
Our parametrization is defined as follows. More details can be found in \cite{diporto10}.

{\it Equation of state.}
In order to represent the evolution of the
equation of state parameter $w$, we use the popular CPL
parameterization~\citep{chevallier01,Linder03} 
\begin{equation}
w(z)=w_{0}+w_{1}\frac{z}{1+z}\;.\label{eq:w_CPL}
\end{equation}
As a special case we will also consider the case of a constant $w$.
We refer to this as the $w$-parametrization.

{\it Growth Rate.}
Here we assume that the growth rate, $f_g$, is a function of time but not
of scale. As usual, we use the simple prescription ~\citep{peebles76,lahav91,polarski08,linder05,wang98}
\begin{equation}
f_g=\Omega_{m}^{\gamma}\,,\label{eq:standard}
\end{equation}
where $\Omega_{m}(z)$ is the matter density in units of the critical
density as a function of redshift. A value $\gamma\approx0.545$ reproduces
well the $\Lambda$CDM behavior while departures from this value characterize
different models. Here we explore three different parameterizations of $f_g$:

\begin{itemize}

\item {\it $f$-parameterization}. This is in fact a non-parametric 
model in which the growth rate itself is modeled as a step-wise function
$f_g(z)=f_{i}$, specified in different redshift bins. The errors are derived on
$f_{i}$ in each $i$-th redshift bin of the survey. 

\item {\bf {\it $\gamma$-parameterization}}.
As a second case we assume
\begin{equation}
f_g\equiv\Omega_{m}(z)^{\gamma(z)}\;.\label{eq:s_parametriz}
\end{equation}
where the $\gamma(z)$ function is parametrized as 
\begin{equation}
\gamma(z)=\gamma_{0}+\gamma_{1}\frac{z}{1+z}\;.\label{eq:gam_CPL}
\end{equation}
As shown by \cite{wu09,fu09}, this parameterization is more accurate than that
of eq.~(\ref{eq:standard}) for both $\Lambda$CDM and DGP models.
Furthermore, this parameterization is especially effective to distinguish
between a $w$CDM model  (i.e. a dark energy model with a constant equation of
state) that has a negative $\gamma_{1}$ 
($-0.020\lesssim\gamma_{1}\lesssim-0.016$) and a DGP model that instead, has 
a positive $\gamma_{1}$ ($0.035<\gamma_{1}<0.042$).
In addition, modified gravity models show a strongly evolving
$\gamma(z)$~\citep{gannouji09,motohashi10,fu09}, in contrast with conventional
Dark Energy models.
As a special case we also consider $\gamma=$ constant (only when $w$ also is
assumed constant), to compare our results with those of previous works.

\item {\bf {\it $\eta$-parameterization}}.
To explore models in which perturbations grow faster than in the 
$\Lambda$CDM case, like in the case of a coupling between dark energy and dark
matter~\citep{diporto08}, we consider a model in which $\gamma$ is constant and
the growth rate varies as
\begin{equation}
f_g\equiv\Omega_{m}(z)^{\gamma}(1+\eta)\;,\label{eq:eta_paramet}
\end{equation}
where $\eta$ quantifies the strength of the coupling.
The example of the coupled quintessence model worked out by~\cite{diporto08}
illustrates this point. In that model, the numerical solution for the growth
rate can be fitted by the formula (\ref{eq:eta_paramet}), with 
$\eta=c\beta_{c}^{2}$,
where $\beta_{c}$ is the dark energy-dark matter coupling constant
and  best fit values  $\gamma=0.56$ and $c=2.1$.
In this simple case, observational constraints over
$\eta$ can be readily  transformed into constraints over $\beta_{c}$.

\end{itemize}

{\it Reference Cosmological Models.}
 We assume as reference model a ``pseudo'' $\Lambda$CDM, where the growth rate values are obtained from
eq.~(\ref{eq:standard}) with $\gamma=0.545$ and $\Omega_m(z)$ is given by the
standard evolution.
Then $\Omega_m(z)$ is completely specified by setting $\Omega_{m,0}=0.271$, 
$\Omega_k=0$, $w_0=-0.95$, $w_1=0$.
When the corresponding parameterizations are
employed,   we choose as fiducial
values $\gamma_{1}=0$ and $\eta=0$, We also assume a primordial slope $n_s=0.966$ and a present
normalization $\sigma_8=0.809$.

One of the goals of this work is to assess whether the
analysis of the power spectrum in redshift-space
can distinguish the fiducial model from alternative
cosmologies, characterized by their own set of parameters
(apart from $\Omega_{m,0}$ which is set equal to 0.27 for all of them). The
alternative models that we consider in this work are:

\begin{itemize}

\item {\it DGP model}. We consider the flat space case studied
in~\cite{maartens06}.
When we adopt this model then we set
$\gamma_{0}=0.663$,\,$\gamma_{1}=0.041$~\citep{fu09} or
$\gamma=0.68$~\citep{linder07} and $w=-0.8$ when $\gamma$ and $w$ are assumed
constant.

\item {\it $f(R)$ model}. {\color{red}Here we consider different classes of $f(R)$ models: i) the one proposed in 
\cite{Hu07}, depending on two parameters, $n$ and $\lambda$, which we fix to
$n=0.5,1,2$ and $\lambda=3$. For the model with $n=2$ we assume $\gamma_{0}=0.43$,
$\gamma_{1}=-0.2$, values that apply quite generally in the limit of small
scales (provided they are still linear, see \cite{gannouji09}) or $\gamma=0.4$
and $w=-0.99$. Unless differently specified, we will always refer to this specific model when we mention comparisons to a single $f(R)$ model. ii) The model proposed in \cite{Star07} fixing $\lambda=3$ and $n=2$, which shows a very similar behaviour to the previous one. iii) The one proposed in \cite{Tsuji08} fixing $\lambda=1$.}

\item {\it coupled dark energy (CDE) model}.  This is the coupled model 
proposed by \cite{Amendola:1999er,wetterich95}. In this case we assume
$\gamma_{0}=0.56$, $\eta=0.056$ (this value comes from putting $\beta_{c}=0.16$
as coupling, which is of the order of the maximal value allowed by CMB
constraints)~\citep{amendola_quercellini_2003}. As already explained, this model cannot be
reproduced by a constant $\gamma$. Forecasts on coupled quintessence based on \cite{Amendola:2011ie, Amendola:1999er, Pettorino:2008ez} are discussed in more detail in subsec.\ref{cdeforecast}.

\end{itemize}

For the fiducial values of the bias parameters in every bin, we assume 
$b(z)=\sqrt{1+z}$ (already used in~\cite{rassat08}) since this function provides
a good fit to H$_{\alpha}$ line galaxies with luminosity $L_{{\rm
H}_{\alpha}}=10^{42}$ erg$^{-1}$ s$^{-1}$ h$^{-2}$ modeled by~\cite{orsi10}
using  the semi-analytic
$GALFORM$ models of~\cite{baugh05}.
For the sake of comparison, we will also consider the case of constant
$b=1$ corresponding to the rather unphysical case of a redshift-independent
population of unbiased mass tracers.

The fiducial values for $\beta$ are computed through
\begin{equation}
\beta^F(z)
=\frac{\Omega_m^F(z)^{\gamma^F}}{b^F(z)}=\frac{f_g^F}{b^F}\label{eq:beta_sb}
\end{equation}

Now we express the growth function $G(z)$ and the redshift distortion parameter
$\beta(z)$ in terms of the growth rate {\color{red}$f_g$} (see eqs.~(\ref{eq:gengz}),
(\ref{eq:beta_sb})). When we compute the derivatives of the spectrum
in the Fisher matrix $b(z)$ and {\color{red}$f_g(z)$} are considered as independent
parameters in each redshift bin. In this way we can compute the errors
on $b$ (and {\color{red}$f_g$}) self consistently by marginalizing over
all other parameters.

Now we are ready to present the main result of the Fisher matrix analysis .
We note that in all tables below we always quote errors at 68\% probability
level and draw in the plots the probability regions at 68\% and/or 95\% (denoted
for shortness as 1 and 2$\sigma$ values). Moreover, in all figures, all the
parameters that are not shown have been marginalized over or fixed to a fiducial
value when so indicated.

{\it Results for the $f$-parameterization.}

The total number of parameters that enter in the Fisher matrix analysis is 45:
5 parameters that describe the background cosmology
($\Omega_{m,0}h^{2},\Omega_{b,0}h^{2},$ $h$, $n$, $\Omega_{k}$)
plus 5 $z$-dependent parameters specified in 8 redshift bins evenly spaced in
the range $z=[0.5,2.1]$. They are $P_{\textrm{s}}(z)$, $D(z)$, $H(z)$, $f_g(z)$,
$b(z)$. However, since we are not interested in {\color{red}constraining} $D(z)$ and $H(z)$, we always project them to the set of parameters they depend on (as explained in \cite{seo03}) instead of marginalizing over, so extracting more information on the background parameters.

\begin{figure}[t]

\begin{centering}
\includegraphics[width=9cm]{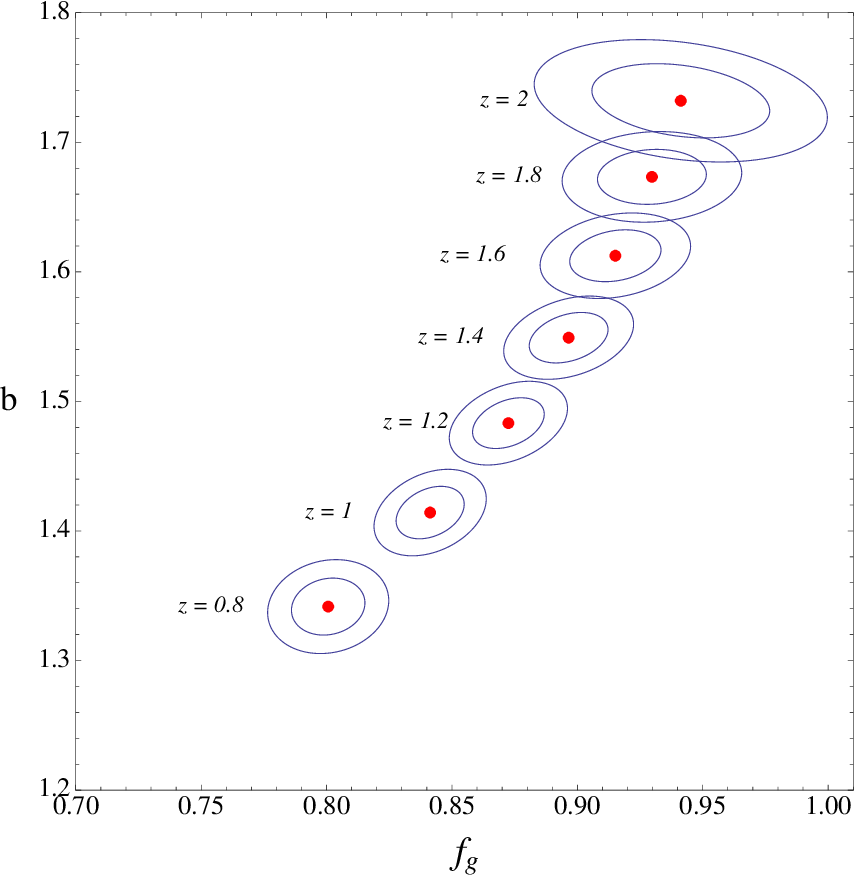} 
\par\end{centering}

\caption{\label{fig:s_b_ref_cp}
Contour plots at 68\% and 98\% of probability for the pairs  $s(z_i)-b(z_i)$ in 7 redshift bins (with $b=\sqrt{1+z}$). The ellipses are centered on the fiducial values of the growth rate and bias parameters, computed  in the central values of the bins, $z_{i}$. }

\end{figure}

The fiducial growth function $G(z)$ in the $(i+1)$-th redshift bin is evaluated
from a step-wise, constant growth rate $f_g(z)$ as
\begin{equation}
G(z)=\exp\left\{\int_{0}^{z}f_g(z)\frac{dz}{1+z}\right\}=\prod_{i}\left(\frac{1+z_
{i}}{1+z_{i-1}}\right)^{f_{i}}\left(\frac{1+z}{1+z_{i}}\right)^{f_{i+1}}\,
.\label{eq:gengz}\end{equation}
To obtain the errors on $s_{i}$ and $b_{i}$ we compute the elements of the
Fisher matrix and marginalize over all other parameters. In this case one is
able to obtain, self-consistently, the error on the bias and on the growth
factor at different redshifts, as detailed in Tab.~\ref{tab:sigma_bias_s_bint}.
In Fig.~\ref{fig:s_b_ref_cp} we show the contour plots at 68\% and 95\% of probability for all the pairs $s(z_i)-b(z_i)$ in several redshift bins (with $b=\sqrt{1+z}$), where $z_{i}$'s are the central values of the bins. We do not show the ellipses for all the 14 bins to avoid overcrowding. 

The table~\ref{tab:sigma_bias_s_bint} illustrates one important result: through the analysis of
the redshift-space galaxy power spectrum in a next-generation
Euclid-like survey, it will be possible to measure galaxy biasing
in $\Delta z=0.1$ redshift bins with less than 1.6\% error, provided
that the bias function is independent {\color{red}of} scale. 
We also tested a different choice for the fiducial form of the bias: $b(z)=1$
finding that the precision in
measuring the bias as well as the other parameters has a very little dependence
on the $b(z)$ form.
Given the robustness of the results on the choice of $b(z)$  
in the following we  only consider the $b(z)=\sqrt{1+z}$ case.

\begin{figure}[t]
\begin{centering}
\includegraphics[width=14cm]{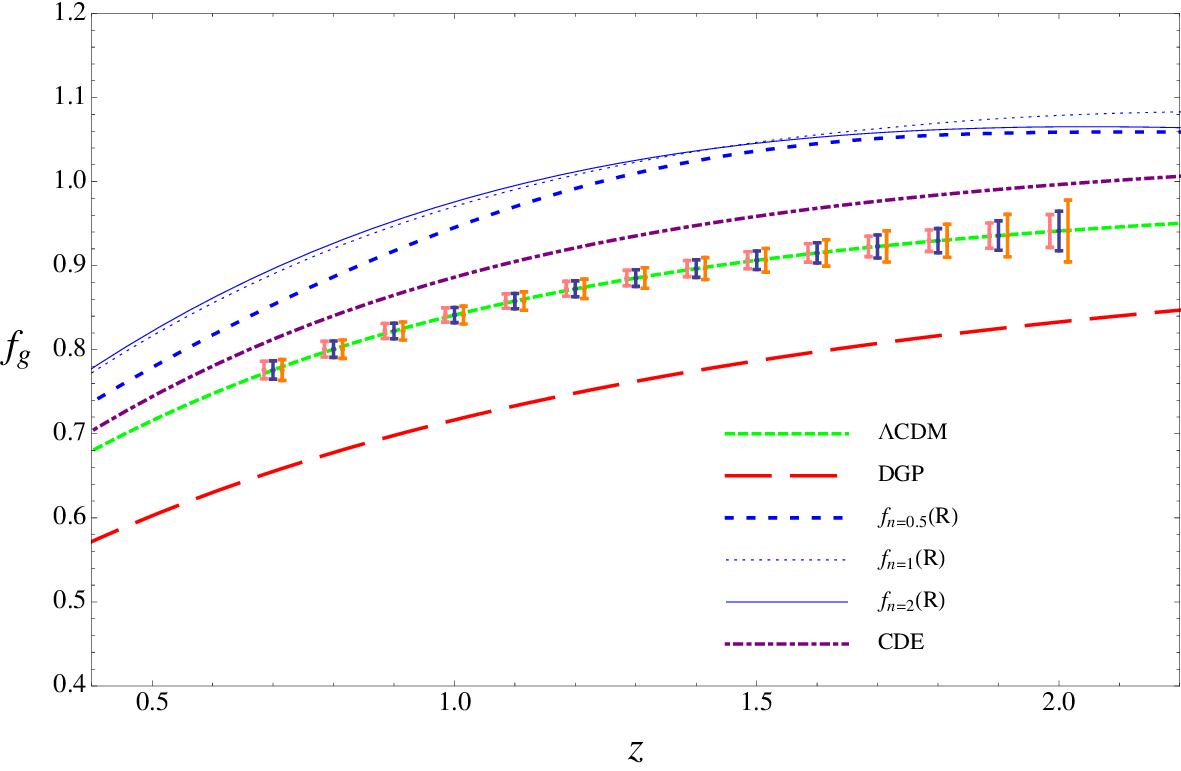} 
\par\end{centering}
\caption{
Expected constraints on the growth rates
 in each redshift bin.
 For each $z$ the central error bars refer to the {\it Reference case} while
 those referring to the {\it Optimistic} and {\it Pessimistic} case have been
 shifted
 by -0.015 and +0.015 respectively. {\color{red}The growth rates for different
 models are also plotted: $\Lambda$CDM (green tight shortdashed curve), flat DGP
(red longdashed curve) and a model
 with coupling between dark energy and dark matter (purple, dot-dashed
 curve).  The blue curves (shortdashed, dotted and solid) represent the $f(R)$ model by \cite{Hu07} with $n=0.5,1,2$ respectively. The plot shows that it will be possible to distinguish these models
 with next generation data.}}
\label{fig:s_err_bint_revised1} 
\end{figure}

In Fig.~\ref{fig:s_err_bint_revised1} we show the errors on
the growth rate $f_g$ as a function of redshift,
overplotted to our fiducial $\Lambda$CDM (green solid
curve).  The three sets of error bars are plotted in correspondence of
the 14 redshift bins and refer (from left to right) to the {\it Optimistic, 
Reference} and {\it Pessimistic}
cases, respectively. {\color{red}The other curves show the expected growth rate in three alternative cosmological
models: flat DGP (red, longdashed curve), CDE (purple, dot-dashed curve) and different $f(R)$ models (see description in the figure caption).}
This plot clearly illustrates the ability of next generation surveys to 
distinguish between alternative models, even in the less favorable choice of
survey parameters.

The main results  can be summarized as follows.
\begin{enumerate}

\item The ability of measuring the biasing function is not too sensitive
to the characteristic of the survey ($b(z)$ can be constrained to within 
1\% in the {\it Optimistic} scenario and up to 1.6\% in the {\it Pessimistic}
one) provided that the bias function is independent {\color{red}of} scale. Moreover, we checked that the
precision in measuring the bias has a very little dependence on the $b(z)$ form.

\item The growth rate $f_g$ can be estimated to within 1-2.5\%
in each bin for the {\it Reference case} survey with
no need of estimating the bias function $b(z)$ from some
dedicated, independent analysis using higher order statistics~\citep{Verde02}
or full-PDF analysis~\citep{sigad00}.

\item The estimated errors on $f_g$ depend weakly on the fiducial
model of $b(z)$.
\end{enumerate}


\begin{figure}[t]
\begin{centering}
\includegraphics[width=14cm]{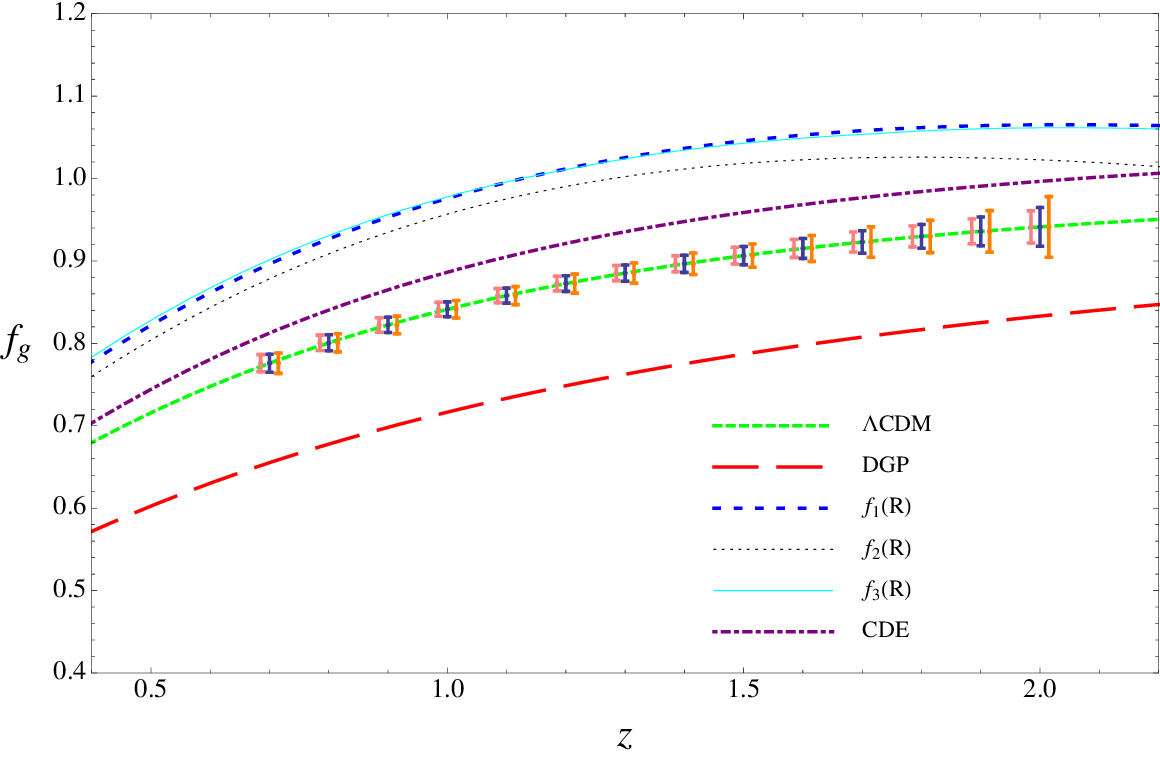}
\par\end{centering}
\caption{\label{fig:s_err_bint_revised2}
 Expected constraints on the growth rates
 in each redshift bin.
 For each $z$ the central error bars refer to the {\it Reference case} while
 those referring to the {\it Optimistic} and {\it Pessimistic} case have been
 shifted
 by -0.015 and +0.015 respectively.  {\color{red}The growth rates for different
 models are also plotted: $\Lambda$CDM (green tight shortdashed curve), flat DGP
(red longdashed curve) and a model
 with coupling between dark energy and dark matter (purple, dot-dashed
 curve).  Here we plot again the $f(R)$ model by \cite{Hu07} with $n=2$ (blu shortdashed curve) together with the model by \cite{Star07} (cyan solid curve) and the one by \cite{Tsuji08} (black dotted curve). Also in this case it will be possible to distinguish these models
 with next generation data.}}
\end{figure}

\begin{table}
\centering
\begin{tabular}{|>{\centering}>{$\color{red}$}p{1cm}|>{\centering}p{1.3cm}|>{\centering}p{1.3cm}|>{\centering}p{1.3cm}|>{\centering}p{1cm}|>{\centering}p{1cm}|>{\centering}p{1cm}|>{\centering}p{1.3cm}|>{\centering}p{1.3cm}|>{\centering}p{1.3cm}|}
\hline 
z &
\multicolumn{1}{|>{\centering}p{1.3cm}}{} &
\multicolumn{1}{>{\centering}p{1.3cm}}{$\sigma_{b}$} &  & $b^F$ & $z$ &  $f_g^F$ &
\multicolumn{1}{>{\centering}p{1.3cm}}{} &
\multicolumn{1}{>{\centering}p{1.3cm}}{$\sigma_{f_g}$} & \tabularnewline
\hline 
&ref.  & opt.  & pess.  & & & & ref.  & opt.  & pess.\tabularnewline
\hline
0.7 & 0.016 & 0.015 & 0.019 & 1.30 & 0.7 & 0.76  & 0.011  & 0.010 & 0.012\tabularnewline
0.8 & 0.014 & 0.014 & 0.017 & 1.34 & 0.8 & 0.80  & 0.010  & 0.009 & 0.011\tabularnewline
0.9 & 0.014 & 0.013 & 0.017 & 1.38 & 0.9 & 0.82  & 0.009  & 0.009 & 0.011\tabularnewline
1.0 & 0.013 & 0.012 & 0.016 & 1.41 & 1.0 & 0.84  & 0.009  & 0.008 & 0.011\tabularnewline
1.1 & 0.013 & 0.012 & 0.016 & 1.45 & 1.1 & 0.86  & 0.009  & 0.008 & 0.011\tabularnewline
1.2 & 0.013 & 0.012 & 0.016 & 1.48 & 1.2 & 0.87  & 0.009  & 0.009 & 0.011\tabularnewline
1.3 & 0.013 & 0.012 & 0.016 & 1.52 & 1.3 & 0.88  & 0.010  & 0.009 & 0.012\tabularnewline
1.4 & 0.013 & 0.012 & 0.016 & 1.55 & 1.4 & 0.89  & 0.010  & 0.009 & 0.013\tabularnewline
1.5 & 0.013 & 0.012 & 0.016 & 1.58 & 1.5 & 0.91  & 0.011  & 0.010 & 0.014\tabularnewline
1.6 & 0.013 & 0.012 & 0.016 & 1.61 & 1.6 & 0.91  & 0.012  & 0.011 & 0.016\tabularnewline
1.7 & 0.014 & 0.013 & 0.017 & 1.64 & 1.7 & 0.92  & 0.014  & 0.012 & 0.018\tabularnewline
1.8 & 0.014 & 0.013 & 0.018 & 1.67 & 1.8 & 0.93  & 0.014  & 0.013 & 0.019\tabularnewline
1.9 & 0.016 & 0.014 & 0.021 & 1.70 & 1.9 & 0.93  & 0.017  & 0.015 & 0.025\tabularnewline
2.0 & 0.019 & 0.016 & 0.028 & 1.73 & 2.0 & 0.94  & 0.023  & 0.019 & 0.037\tabularnewline
\hline
\end{tabular}

\caption{\label{tab:sigma_bias_s_bint}
$1\sigma$ marginalized errors for the bias and the growth rates in each redshift
bin.} 
\end{table}

Next,  we focus on the ability of determining $\gamma_0$
and $\gamma_1$, in the context of the {\it $\gamma$-parameterization}
and $\gamma$, $\eta$ in the {\it $\eta$-parameterization}.
In both cases the Fisher matrix elements have been estimated
by expressing the growth factor as
\begin{eqnarray}
G(z) & = & 
\delta_{0}\exp\left[(1+\eta)\int_{0}^{z}\Omega_{m}(z^{\prime})^{\gamma(z)}\frac{
dz^{\prime}}{1+z^{\prime}}\right]
\, ,
\label{eq:growth_fact_def_gam}\end{eqnarray}
where  for the {\it $\gamma$-parameterization} we fix $\eta=0$.

\begin{figure}[t]
\includegraphics[width=7.6cm]{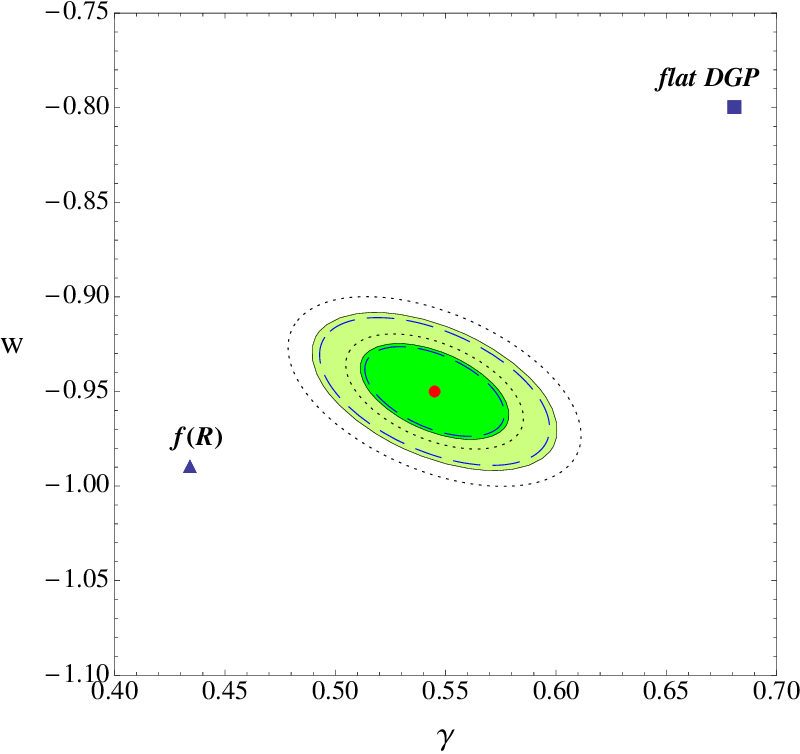} 
\includegraphics[width=7.4cm]{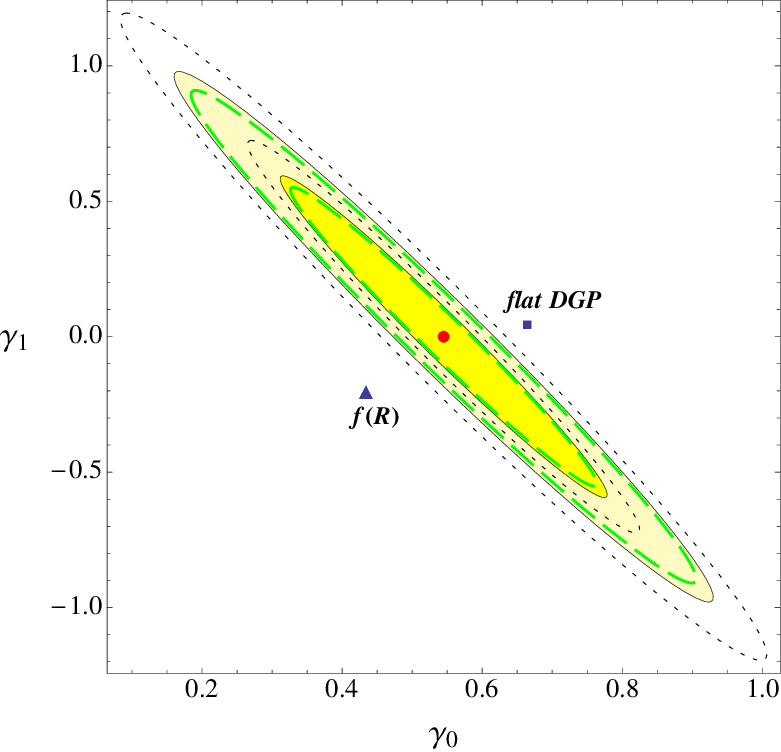} 
\caption{\label{fig:gam_w_b1_dgp}
$\gamma$-parameterization. Left panel: 1 and 2$\sigma$ marginalized probability
regions for constant $\gamma$ and $w$: the green (shaded) regions
are relative to the {\it Reference case}, the blue long-dashed ellipses
to the {\it Optimistic case}, while the black short-dashed ellipses are
the probability regions for the {\it Pessimistic case}. The red dot marks the
fiducial model; two alternative models are also indicated for comparison.
Right panel: 1 and 2$\sigma$ marginalized probability
regions for the parameters $\gamma_{0}$ and $\gamma_{1}$, relative
to the {\it Reference case} (shaded yellow regions), to the {\it Optimistic
case}
(green long-dashed ellipses), and to the {\it Pessimistic case} (black dotted
ellipses). Red dots represent the fiducial model, blue squares
mark the DGP while triangles stand for the $f(R)$ model. Then, in the case of
$\gamma$-parameterization, one
could distinguish these three models (at 95\% probability). } 
\end{figure}

\begin{figure}[t]

\begin{centering}
\includegraphics[width=15cm]{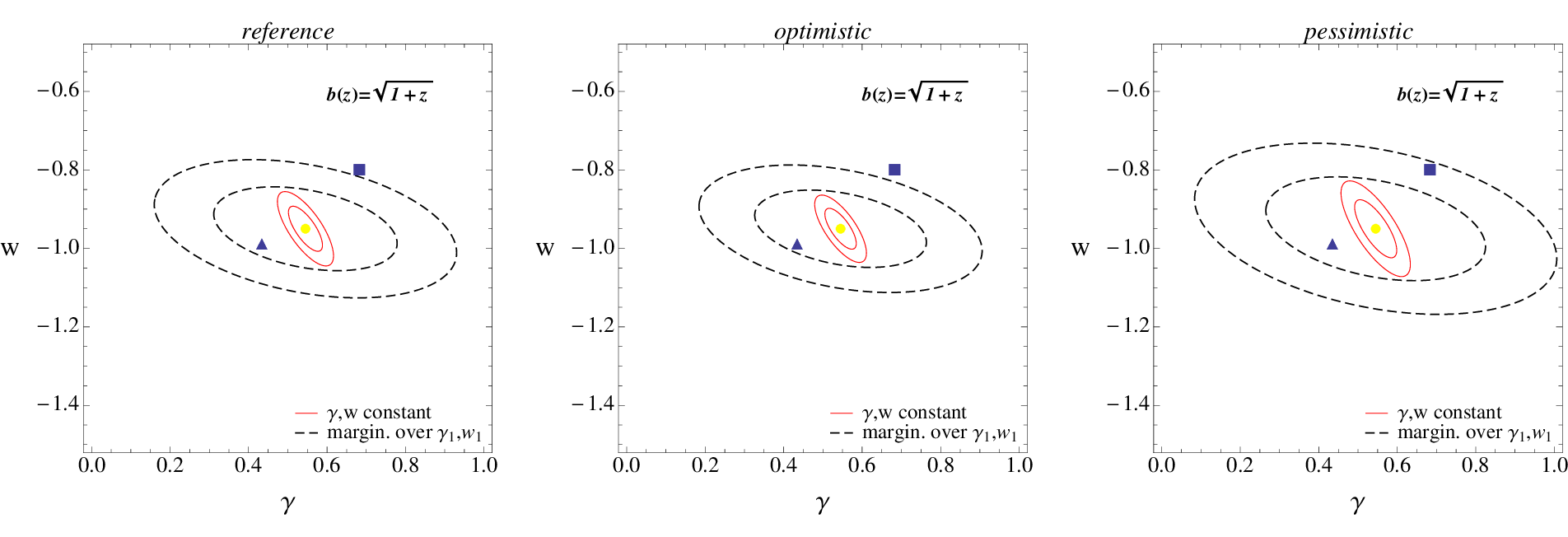} 
\par\end{centering}

\caption{\label{fig:gam_w_margover_gam1w1}
$\gamma$-parameterization. 1 and 2$\sigma$ marginalized probability
regions obtained assuming constant $\gamma$ and $w$ (red solid curves)
or assuming the parameterizations (\ref{eq:gam_CPL}) and (\ref{eq:w_CPL})
and marginalizing over $\gamma_{1}$ and $w_{1}$ (black dashed curves);
marginalized error values are in columns $\sigma_{{\gamma}_{\textrm{marg},1}}$,
$\sigma_{{w}_{\textrm{marg},1}}$ of Tab.~\ref{tab:sigma_gam_w_marg}. Yellow dots
represent the fiducial model, the triangles a $f(R)$ model and the squares mark
the flat DGP.}
\end{figure}

\begin{figure}[t]

\begin{centering}
\includegraphics[width=15cm]{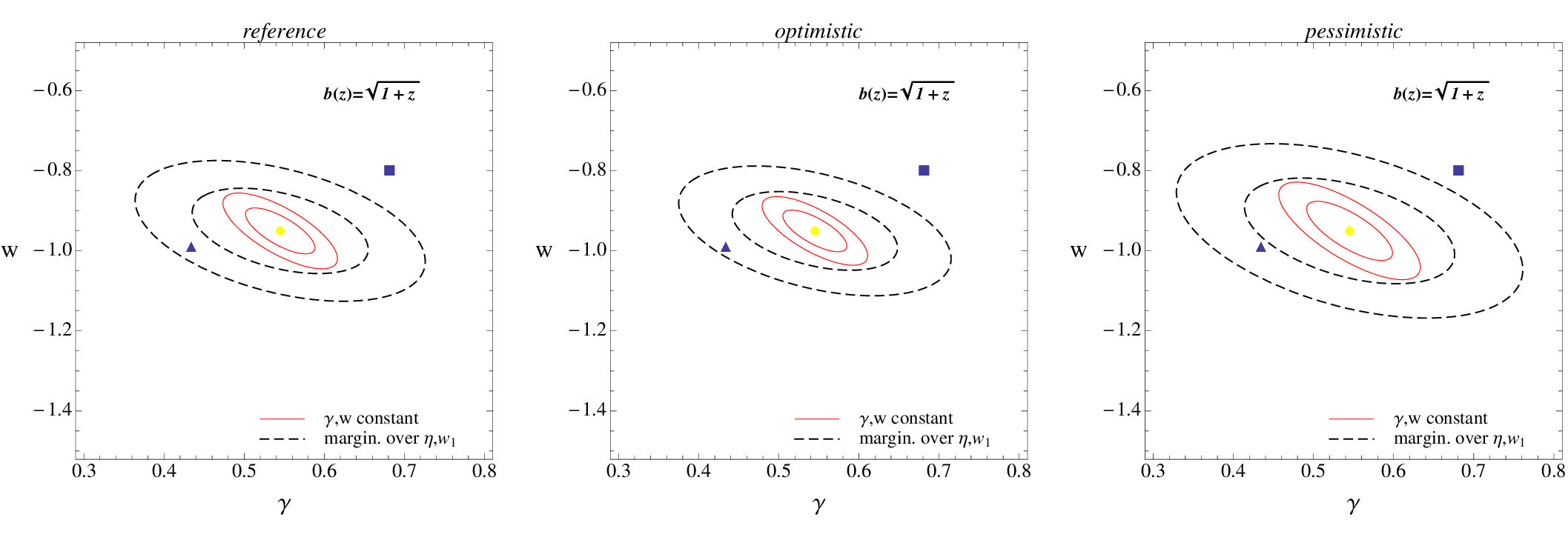} 
\par\end{centering}

\caption{\label{fig:gam_w_margover_etaw1}
$\eta$-parameterization. 1 and 2$\sigma$ marginalized probability
regions obtained assuming constant $\gamma$ and $w$ (red solid curves)
or assuming the parameterizations (\ref{eq:eta_paramet}) and (\ref{eq:w_CPL})
and marginalizing over $\eta$ and $w_{1}$ (black dashed curves);
marginalized error values are in columns $\sigma_{{\gamma}_{\textrm{marg},2}}$,
$\sigma_{{w}_{\textrm{marg},2}}$ of Tab.~\ref{tab:sigma_gam0_gam1}. Yellow dots
represent the fiducial model, the triangles stand for a $f(R)$ model and the
squares mark the flat DGP. }
\end{figure}

\begin{figure}[t]
\!\includegraphics[height=7.cm]{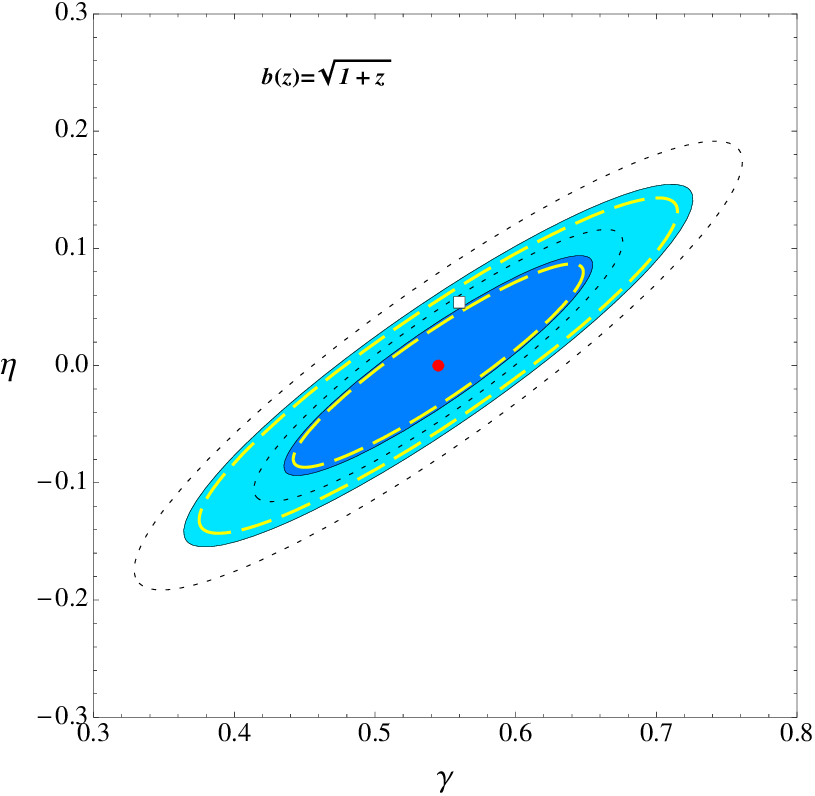}
\includegraphics[height=7.cm]{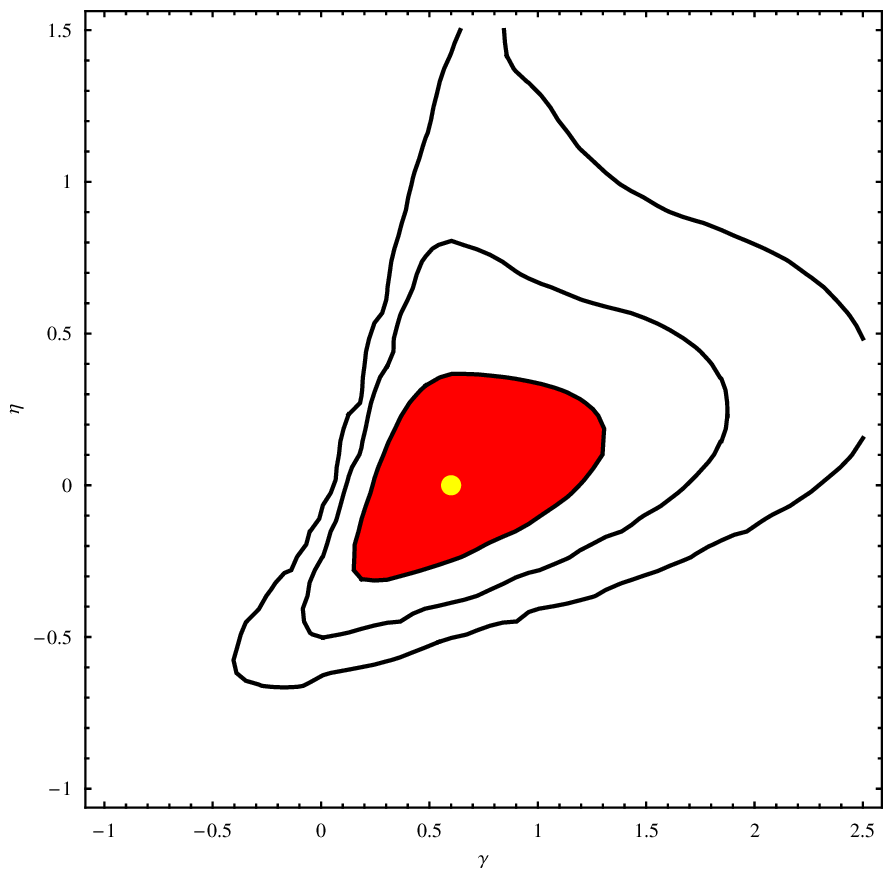}

\caption{\label{fig:gamma_eta_new_past}
$\eta$-parameterization. Left panel: 1 and 2$\sigma$ marginalized probability
regions for the parameters $\gamma$ and $\eta$ in eq.~(\ref{eq:eta_paramet})
relative to the reference case (shaded blue regions), to the optimistic
case (yellow long-dashed ellipses) and to the pessimistic case (black
short-dashed ellipses). The red dot marks the fiducial model while the square
represents the coupling model. Right panel: present constraints on $\gamma$ and
$\eta$ computed
through a full likelihood method (here the red dot marks the likelihood
peak) \protect \citep{diporto08}. }
\end{figure}

\begin{itemize}

\item {\it $\gamma$-parameterization}.
We start by considering the  case
of constant $\gamma$ and $w$ in which we set $\gamma=\gamma^F=0.545$ and
$w=w^F=-0.95$. As we will discuss in the next Section, this simple case will
allow us to cross-check our results with those in the literature.
In  Fig.~\ref{fig:gam_w_b1_dgp} we show the marginalized probability 
regions, at 1 and 2$\sigma$ levels, for  $\gamma$ and $w$.
The regions with different shades of green {\color{red}illustrate} the
{\it Reference case} for the survey whereas the
blue long-dashed and the black short-dashed ellipses
refer to the {\it Optimistic} and {\it Pessimistic} cases, respectively.
Errors on $\gamma$ and $w$ are listed in Tab.~\ref{tab:sigma_gam_w}
together with the corresponding figures of merit [FOM]
defined to be the squared inverse of the Fisher matrix determinant and therefore
equal to the inverse of the
product of the errors in the pivot point, see~\cite{Albrecht2006}.
Contours are centered on the fiducial model. The blue triangle and
the blue square represent the  flat DGP and the $f(R)$ models' predictions,
respectively.
It is clear that, in the case of constant $\gamma$ and $w$, the
measurement of the growth rate in a Euclid-like survey will
allow us to discriminate among these models. These results have been 
obtained by fixing the curvature to its fiducial value $\Omega_k=0$. If instead,
we consider 
curvature as a free parameter and marginalize over, the errors on $\gamma$ and
$w$ increase significantly, as shown in Table~\ref{tab:sigma_gam_w_omk_marg},
and yet the precision is good enough to distinguish the different models.
For completeness, we also computed the fully marginalized errors over the other
Cosmological parameters for the reference survey, given in
Tab.~\ref{tab:cosm_par_errors}.

As a second step we considered the case in which
$\gamma$ and $w$ evolve with redshift according to
eqs.~(\ref{eq:gam_CPL}) and~(\ref{eq:w_CPL})
and then we {\color{red}marginalized} over the parameters $\gamma_{1}$, $w_{1}$ and  
$\Omega_k$.
The marginalized probability contours
are shown  in Fig.~\ref{fig:gam_w_margover_gam1w1} in which we have shown
the three survey setups in three different panels to avoid overcrowding.
Dashed contours refer to the $z$-dependent parameterizations while red,
continuous
contours refer to the case of  constant $\gamma$ and $w$ obtained after
marginalizing over $\Omega_k$.
Allowing for time dependency increases the size of the confidence ellipses since
the Fisher matrix analysis now accounts for the additional uncertainties in the
extra-parameters $\gamma_{1}$ and $w_{1}$; marginalized error values are in
columns $\sigma_{{\gamma}_{\textrm{marg},1}}$,
$\sigma_{{w}_{\textrm{marg},1}}$ of Tab.~\ref{tab:sigma_gam_w_marg}.  
The uncertainty ellipses are now larger and show that DGP and
fiducial models could be distinguished at $>2\sigma$ level only if the redshift survey
parameter will be more favorable than in the {\it Reference case}.

We have also projected the marginalized ellipses for the parameters $\gamma_{0}$
and $\gamma_{1}$  and calculated their marginalized
errors and figures of merit, which are reported in 
Tab.~\ref{tab:sigma_gam0_gam1}.
The corresponding uncertainties contours are shown in the right panel of
Fig.~\ref{fig:gam_w_b1_dgp}. Once again we overplot the expected values
in the $f(R)$ and DGP scenarios to stress the fact that
one is expected to be able to distinguish among competing models, 
irrespective on the survey's precise characteristics.

\item {\it $\eta$-parameterization}.

We have repeated the same analysis as for
the {\it $\gamma$-parameterization} taking into account the 
possibility of coupling between DE and DM
i.e. we have
modeled the growth factor according to eq.~(\ref{eq:eta_paramet}) and 
the dark energy equation of state as in eq.~(\ref{eq:w_CPL}) and 
marginalized over all parameters, including $\Omega_k$. The marginalized errors
are shown in columns $\sigma_{{\gamma}_{\textrm{marg},2}}$,
$\sigma_{{w}_{\textrm{marg},2}}$ of Tab.~\ref{tab:sigma_gam_w_marg} and the
significance contours are shown in the three panels of
Fig.~\ref{fig:gam_w_margover_etaw1} which is analogous to
Fig.~\ref{fig:gam_w_margover_gam1w1}. 
Even if the ellipses are now larger we note that errors are still small enough to distinguish the fiducial model
from the $f(R)$ and DGP scenarios at  $>1\sigma$ and  $>2\sigma$ level respectively.

Marginalizing over all other parameters we can compute the uncertainties in
the $\gamma$ and $\eta$ parameters, as listed in Tab.~\ref{tab:sigma_gam_eta}.
The relative confidence ellipses are shown in the left panel of
Fig.~\ref{fig:gamma_eta_new_past}.
This plot shows that next generation Euclid-like surveys will be able to 
distinguish the reference model with no coupling (central, red dot) to the CDE
model proposed by~\cite{amendola_quercellini_2003} (white square) only at the 1-1.5 $\sigma$
level.

\end{itemize}

\begin{table}
\centering
\begin{tabular}{|>{\centering}p{2.5cm}|>{\centering}p{2cm}|>{\centering}p{2cm}
|c|>{\centering}p{2cm}|}
\hline 
  & case  & $\sigma_{\gamma}$  &
\multicolumn{1}{>{\centering}p{2cm}|}{$\sigma_{w}$} & FOM\tabularnewline
\hline
\hline 
$b=\sqrt{1+z}$ & ref.  & 0.02  & 0.017  & 3052\tabularnewline
 with & opt.  & 0.02  & 0.016  & 3509\tabularnewline
$\Omega_{k}$ fixed & pess.  & 0.026  & 0.02  & 2106\tabularnewline
\hline
\end{tabular}

\caption{\label{tab:sigma_gam_w}Numerical values for 1$\sigma$ constraints
on parameters in Fig.~\ref{fig:gam_w_b1_dgp} and figures of merit. Here we have
fixed $\Omega_{k}$ to its fiducial value, $\Omega_k = 0$.}

\end{table}
\begin{table}
\centering
\begin{tabular}{|>{\centering}p{2.5cm}|>{\centering}p{2cm}|>{\centering}p{2cm}|c|>{\centering}p{2cm}|}
\hline 
bias  & case  & $\sigma_{\gamma}$  &
\multicolumn{1}{>{\centering}p{2cm}|}{$\sigma_{w}$} & FOM\tabularnewline
\hline
\hline 
 & ref.  & 0.03  & 0.04  & 1342\tabularnewline
$b=\sqrt{1+z}$  & opt.  & 0.03  & 0.03  & 1589\tabularnewline
 & pess.  & 0.04  & 0.05  & 864\tabularnewline
\hline
\end{tabular}

\caption{\label{tab:sigma_gam_w_omk_marg}Numerical values for 1$\sigma$
constraints
on parameters $\gamma$ and $w$ (assumed constant), relative to the red
ellipses in Figs~\ref{fig:gam_w_margover_gam1w1},~\ref{fig:gam_w_margover_etaw1}
 and figures of merit. Here we have marginalized over $\Omega_{k}$.}

\end{table}
\begin{table}
\centering
\begin{tabular}{|>{\centering}p{2.5cm}|>{\centering}p{1cm}|>{\centering}p{1cm}|>
{\centering}p{1cm}|>{\centering}p{1cm}|c|>{\centering}p{1cm}|>{\centering}p{1cm}|}
\hline 
  & case & $\sigma_{h}$ & $\sigma_{\Omega_m h^2}$  & $\sigma_{\Omega_b h^2}$ &
$\sigma_{\Omega_k}$ & $\sigma_{n_s}$ & $\sigma_{\sigma_8}$ \tabularnewline
\hline
$b=\sqrt{1+z}$ & ref.  & 0.007  & 0.002  & 0.0004 & 0.008 & 0.03 & 0.006\tabularnewline
\hline
\end{tabular}

\caption{\label{tab:cosm_par_errors}Numerical values for marginalized 1$\sigma$
constraints on Cosmological parameters using constant $\gamma$ and $w$.}

\end{table}
\begin{table}
\centering
\begin{tabular}{|>{\centering}p{2.5cm}|>{\centering}p{1cm}|>{\centering}p{
1.5cm}|c|>{\centering}p{1.5cm}||>{\centering}p{1.5cm}|>{\centering}p{1.5cm}|>{
\centering}p{1.5cm}|}
\hline 
bias  & case  & $\sigma_{\gamma_{marg,1}}$  &
\multicolumn{1}{>{\centering}p{1.5cm}|}{$\sigma_{w_{marg,1}}$} & FOM  &
$\sigma_{\gamma_{marg,2}}$  & $\sigma_{w_{marg,2}}$  & FOM\tabularnewline
\hline
\hline 
 & ref.  & 0.15  & 0.07  & 97  & 0.07  & 0.07  & 216\tabularnewline
$b=\sqrt{1+z}$  & opt.  & 0.14  & 0.06  & 112  & 0.07  & 0.06  & 249\tabularnewline
 & pess.  & 0.18  & 0.09  & 66  & 0.09  & 0.09  & 147\tabularnewline
\hline
\end{tabular}

\caption{\label{tab:sigma_gam_w_marg}1$\sigma$ marginalized errors for
parameters
$\gamma$ and $w$ expressed through $\gamma$ and $\eta$ parameterizations.
Columns $\gamma_{0,marg1},w_{0,marg1}$ refer to marginalization over
$\gamma_{1},w_{1}$ (Fig.~\ref{fig:gam_w_margover_gam1w1}) while
columns $\gamma_{0,marg2},w_{0,marg2}$ refer to marginalization over
$\eta,w_{1}$ (Fig.~\ref{fig:gam_w_margover_etaw1}).}

\end{table}

\begin{table}
\centering
\begin{tabular}{|>{\centering}p{2.5cm}|>{\centering}p{2cm}|>{\centering}p{2cm}
|c|>{\centering}p{2cm}|}
\hline 
bias  & case  & $\sigma_{\gamma_{0}}$  &
\multicolumn{1}{>{\centering}p{2cm}|}{$\sigma_{\gamma_{1}}$} &
FOM\tabularnewline
\hline
\hline 
 & ref.  & 0.15  & 0.4  & 87\tabularnewline
$b=\sqrt{1+z}$  & opt.  & 0.14  & 0.36  & 102\tabularnewline
 & pess.  & 0.18  & 0.48  & 58\tabularnewline
\hline
\end{tabular}

\caption{\label{tab:sigma_gam0_gam1}Numerical values for 1$\sigma$ constraints
on parameters in right panel of Fig.~\ref{fig:gam_w_b1_dgp} and figures of
merit.}

\end{table}

\begin{table}
\centering
\begin{tabular}{|>{\centering}p{2.5cm}|>{\centering}p{2cm}|>{\centering}p{2cm}
|c|>{\centering}p{2cm}|}
\hline 
bias  & case  & $\sigma_{\gamma}$  &
\multicolumn{1}{>{\centering}p{2cm}|}{$\sigma_{\eta}$} & FOM\tabularnewline
\hline
\hline 
 & ref.  & 0.07  & 0.06  & 554\tabularnewline
$b=\sqrt{1+z}$  & opt.  & 0.07  & 0.06  & 650\tabularnewline
 & pess.  & 0.09  & 0.08  & 362\tabularnewline
\hline
\end{tabular}

\caption{\label{tab:sigma_gam_eta}Numerical values for 1$\sigma$ constraints
on parameters in Fig.~\ref{fig:gamma_eta_new_past} and figures of merit.}

\end{table}
\begin{figure}[t]

\begin{centering}
\includegraphics[width=8cm]{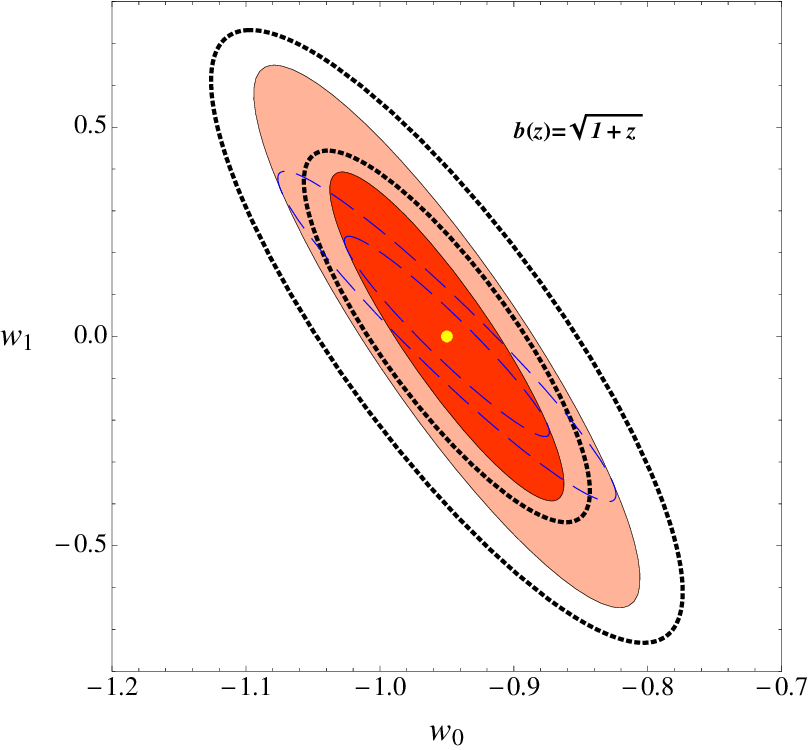} 
\par\end{centering}

\caption{\label{fig:w0_w1}
Errors on the equation of state. 1 and 2$\sigma$ marginalized probability
regions for the parameters $w_{0}$ and $w_{1}$, relative
to the reference case and bias $b=\sqrt(1+z)$. The blue dashed ellipses are
obtained fixing $\gamma_{0}, \gamma_{1}$ and $\Omega_{k}=0$ to their fiducial values and
marginalizing over all the other parameters; for the red shaded ellipses
instead, we also marginalize over 
$\Omega_{k}=0$ but we fix $\gamma_{0}, \gamma_{1}$. Finally, the black dotted ellipses are obtained marginalizing
over all parameters but $w_{0}$ and $w_{1}$. The progressive increase in the
number of parameters reflects in a widening of the ellipses with a consequent
decrease in the figures of merit (see Tab.~\ref{tab:w0_w1}). }
\end{figure}

Finally, in order to explore the dependence on the number of parameters and to
compare our results to previous works, we also draw the confidence ellipses for
$w_0$, $w_1$ with three different methods: $i$) fixing $\gamma_{0}, \gamma_{1}$ and
$\Omega_k$ to their fiducial values and marginalizing over all the other parameters; $ii$)
fixing only {\color{red}$\gamma_{0}$ and $\gamma_{1}$}; $iii$) marginalizing over all parameters but $w_0$, $w_1$. As one
can see in Fig.~\ref{fig:w0_w1} and Tab.~\ref{tab:w0_w1} this progressive
increase in the number of marginalized parameters reflects in a widening of the
ellipses with a consequent decrease in the figures of merit. These results are
in agreement with those of other authors (e.g.~\cite{Wang2010}).

The results obtained {\color{red} in} this Section can be summarized as follows.
\begin{enumerate}

\item If both $\gamma$ and $w$ are assumed to be constant
and setting  $\Omega_k=0$,
then a redshift survey described by our {\it Reference case}
will be able to constrain these parameters to within 4\% and 2\%,
respectively.

\item Marginalizing over $\Omega_{k}$ degrades these constraints to 5.3\% and
4\% respectively.

\item If $w$ and $\gamma$ are considered redshift-dependent and
parametrized according to eqs~(\ref{eq:gam_CPL}) and~(\ref{eq:w_CPL})
then the errors on $\gamma_{0}$ and $w_{0}$ obtained after
marginalizing over  $\gamma_{1}$ and  $w_{1}$ increase by a
factor $\sim 7$, 5. However, with this
precision we will
be able to distinguish the fiducial model from the  DGP and $f(R)$
scenarios with more than 2$\sigma$ and 1$\sigma$ significance, respectively.

\item The ability to discriminate these models with a significance above
2$\sigma$ is confirmed by the confidence contours drawn in the
$\gamma_{0}$-$\gamma_{1}$ plane, obtained after marginalizing over all other
parameters.

\item If we allow for a coupling between dark matter and dark  energy,
and  we marginalize over $\eta$ rather than over $\gamma_{1}$,
then the errors on $w_{0}$ are almost identical
to those obtained in the case of the {\it $\gamma$-parameterization}, while the errors on $\gamma_{0}$ decrease significantly.

However, our ability in separating the fiducial model from the
CDE model is significantly hampered: the confidence contours
plotted in the $\gamma$-$\eta$ plane show that discrimination
can only be performed wit 1-1.5$\sigma$ significance. Yet, this is still a
remarkable improvement over the present situation, as can be appreciated from
Fig.~\ref{fig:gamma_eta_new_past} where we compare the constraints expected by
next generation data to the present ones. Moreover, the {\it Reference} survey
will be able to constrain the parameter $\eta$ to within 0.06. Reminding that we
can write $\eta=2.1 \beta_c^2$~\citep{diporto08}, this means that the coupling
parameter $\beta_c$ between dark energy and dark matter can be constrained to
within 0.14, solely employing the growth rate information. This is comparable to
existing constraints from the CMB but is complementary since obviously it is
obtained at much smaller redshifts. A variable coupling could therefore be
detected by comparing the redshift survey results with the CMB ones.

\end{enumerate}
\begin{table}
\centering
\begin{tabular}{|>{\centering}p{7cm}|>{\centering}m{2cm}|>{\centering}p{2cm}|>{
\centering}p{2cm}|}
\hline 
 & $\sigma_{w_{0}}$ & $\sigma_{w_{1}}$ & FOM\tabularnewline
\hline
\hline 
$\gamma_{0}, \gamma_{1}$, $\Omega_{k}$ fixed & 0.05 & 0.16 & 430\tabularnewline
\hline 
 $\gamma_{0},\gamma_{1}$ fixed & 0.06
& 0.26 & 148\tabularnewline
\hline 
marginalization over all other parameters & 0.07 & 0.3 & 87\tabularnewline
\hline
\end{tabular}\caption{\label{tab:w0_w1}
1  $\sigma$ marginalized errors for the parameters $w_{0}$
and $w_{1}$, obtained with three different methods (reference case, see
Fig.~\ref{fig:w0_w1} \label{tab:sigma_w0_w1}).}
\end{table}

It is worth pointing out that, whenever we have performed statistical tests
similar to those already discussed by other authors in the context of a
Euclid-like survey, we did find consistent results.
Examples of this are the values of FOM and errors for $w_0$, $w_1$, similar to
those in~\cite{Wang10,majerotto11} and the errors on constant $\gamma$ and
$w$~\citep{majerotto11}. However, let us notice that all these values strictly
depend on the parametrizations adopted and on the numbers of parameters fixed
or marginalized over (see e.g. \cite{rassat08}).

%% file: de_mg/hg-weaklensing.tex
\subsection{Weak lensing non-parametric measurement of expansion and growth rate}
\label{weak-lensing-non-parametric}

In this section we apply power spectrum tomography \citep{Hu1999} to the Euclid weak lensing
survey without using any parameterization of the Hubble parameter $H(z)$ as
well as the growth function $G(z)$. Instead, we add the 
fiducial values of those functions at the center of some redshift bins of
our choice to the list of cosmological parameters. Using the Fisher
 matrix formalism, we can forecast the constraints that future
surveys can put on $H(z)$ and $G(z)$. Although such a non-parametric approach is quite common
for as concerns  the equation-of-state ratio $w(z)$ in supernovae surveys (see e.g.
\cite{Albrecht:2009ct}) and also in redshift surveys \citep{seo03}, it has not been
investigated for weak lensing surveys.

The Fisher matrix is given by \citep{1999ApJ...514L..65H}
\begin{equation}
F_{\alpha\beta} = f_\mathrm{sky} \sum_\ell 
\frac{(2\ell+1)\Delta\ell}{2}\frac{\partial P_{ij}(\ell)}{\partial
p_\alpha}C^{-1}_{jk}\frac{\partial P_{km}(\ell)}{\partial
p_\beta}C^{-1}_{mi},
\label{eq:3-fm}
\end{equation}
where {\color{red}$f_\mathrm{sky}$ is the observed fraction of the sky,} $C$ is the covariance matrix, $P(\ell)$ is the convergence
power spectrum and $\mathbf p$ is the vector of the parameters defining our
cosmological model. Repeated indices are being summed over from $1$ to $N$,
the number of redshift bins. The covariance matrix is defined as (no
summation over $j$)
\begin{equation}
C_{jk}=P_{jk} + \delta_{jk}\gamma_\mathrm{int}^2 n^{-1}_j,
\end{equation}
where $\gamma_\mathrm{int}$ is the intrinsic galaxy shear and $n_j$ is the
fraction of galaxies per steradian belonging to the $j$-th redshift bin:
\begin{equation}
n_j = 3600 \left( \frac{180}{\pi} \right)^2 
n_\theta\int_{0}^\infty n_j(z)\mathrm d z
\end{equation}
where $n_\theta$ is the galaxy density per arc minute and $n_j(z)$ the galaxy
density for the $j$-th bin, convolved with a gaussian around $\hat z_j$, the
center of that bin, with a width of $\sigma_z(1+\hat z_j)$ in order to
account for errors in the redshift measurement.

For the matter power spectrum we use the
fitting formulae from \cite{Eisenstein_Hu_1997} and for its non-linear
corrections the results from \cite{Smith2003}. Note that this is where the
growth function enters. The convergence power
spectrum for the $i$-th and $j$-th bin can then be written as 
\begin{equation}
P_{ij}(\ell) = \frac{9H_0^3}{4}\int_0^\infty 
\frac{W_i(z)W_j(z)E^3(z)\Omega_m^2(z)}{(1+z)^4}
P_{\delta_m}\left(\frac{\ell}{\pi r(z)}\right)\mathrm d z.
\label{eq:3-convspec}
\end{equation}
Here we make use of the window function
\begin{equation}
W_i(z) = \int_z^\infty \frac{\mathrm d\tilde z}{H(\tilde z)}\left[
1-\frac{r(z)}{r(\tilde z)}
\right] n_i[r(\tilde z)]
\label{eq:3-wind}
\end{equation}
(with $r(z)$ being the comoving distance)
and the dimensionless Hubble parameter
\begin{equation} 
E^2(z) = \Omega_m^{(0)}(1+z)^3 + (1-\Omega_m^{(0)})
\exp\left[ \int_0^z
\frac{3(1+w(\tilde z))}{1+\tilde z}\mathrm d\tilde z \right].
\label{eq:3-e(z)}
\end{equation}
For the equation-of-state ratio, finally, we use the usual CPL   parameterization.

We determine $N$ intervals in redshift space such that each interval
contains the same amount of galaxies. For this we use the common
parameterization
\begin{equation}
n(z) = z^2 \exp(-(z/z_0)^{3/2}),
\end{equation}
where $z_0 =z_\mathrm{mean}/1.412$ is the peak of $n(z)$ and
$z_\mathrm{mean}$ the
median.
Now we can define $\hat z_i$ as the center of the $i$-th redshift bin and
add $h_i\equiv \log\left({H(\hat z_i)/H_0}\right)$ as well as $g_i\equiv\log
G(\hat z_i)$ to the list of cosmological parameters. The Hubble parameter
and the growth function now become functions of the $h_i$ and $g_i$
respectively:
\begin{align}
H(z;\Omega_m^{(0)},w_0,w_1) &\rightarrow H(z;h_1,\ldots,h_N)\\
G(z;\Omega_m^{(0)},\gamma) &\rightarrow G(z;g_1,\ldots,g_N)
\label{}
\end{align}
This is being done by linearly interpolating the functions through their
supporting points, e.g. $(\hat z_i,\exp( h_i))$ for $H(z)$. Any function
that depends on either $H(z)$ or $G(z)$ hence becomes a function of the 
$h_i$ and $g_i$ as well.

\begin{table}[htb]
\centering
 \begin{tabular}{l|l}
   $\omega_m$ & 0.1341 \\ 
   $\omega_b$ & 0.02258  \\
   $\tau$     & 0.088 \\
   $n_s$      & 0.963 \\
   $\Omega_m$ & 0.266 \\
   $w_0$      & -1  \\
   $w_1$      & 0   \\
   $\gamma$   & 0.547 \\
   $\gamma_{\mathrm{ppn}}$     & 0   \\
   $\sigma_8$  & 0.801 \\
 \end{tabular}
 \hspace{1cm}
\begin{tabular}{l|l}
$f_\mathrm{sky}$ & 0.375 \\
$z_\mathrm{mean}$ & 0.9 \\
$\sigma_z$ & 0.05 \\
$n_\theta$& 30 \\
$\gamma_\mathrm{int}$ & 0.22 \\
\hline
$\ell_\mathrm{max}$ & $5\cdot 10^3$\\
$\Delta \log_{10}\ell$ & 0.02
\end{tabular}
\caption{Values used in our computation. The values of the fiducial model
(WMAP7, on the left) and the survey parameters (on the right)} \label{tab1}
\end{table}
The values for our fiducial model (taken from WMAP 7-year data
\citep{Komatsu2010}) and the survey parameters that we chose
for our computation can be found in table \ref{tab1}.

As for the sum in eq. (\ref{eq:3-fm}), we generally found that with a
realistic upper
limit of $\ell_\mathrm{max} = 5\cdot 10^3$ and a step size of $\Delta \lg \ell =
0.2$ we get the best result in terms of a figure of merit (FOM),
that we defined as 
\begin{equation}
FOM = \sum \sigma_i^{-2}.
\end{equation}

Note that this is a fundamentally different FOM than the one defined by the
Dark Energy Task Force. Our definition allows for a single large error
without influencing the FOM significantly and should stay almost constant
after dividing a bin arbitrarily in two bins, assuming the error scales
roughly as the inverse of the root of the number of galaxies in a given bin.

\begin{figure}[tb] 
\begin{center}
\includegraphics[width=10.cm]{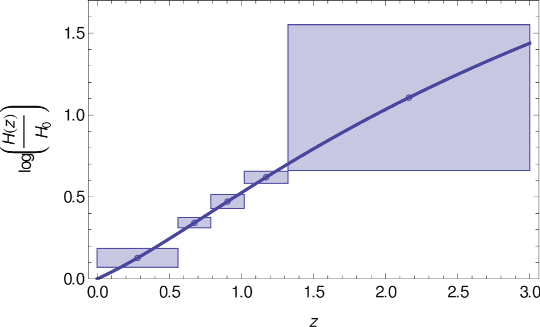}
\end{center}
\vspace{-.5cm}
\caption{Error bars on the Hubble parameter $H(z)$ with five redshift bins.
The exact height of the error bars respectively are $(0.23, 0.072, 0.089, 0.064, 0.76)$.}
\label{fig:hplot}
\end{figure}

\begin{figure}[htb]
\begin{center}
\includegraphics[width=10.cm]{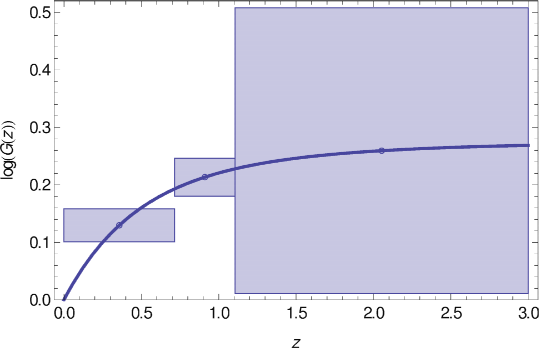}
\end{center}
\vspace{-.5cm}
\caption{Error bars on the growth function $G(z)$ with three redshift bins
while marginalizing over the $h_i$s. The
exact height of the error bars respectively are $(0.029, 0.033, 0.25)$.}
\label{fig:gplot}
\end{figure}

We first did the computation with just binning $H(z)$ and using the common
fit for the growth function slope \citep{wang98}
\begin{equation}
\frac{d \log G(z)}{d\log a} = \Omega_m(z)^\gamma,
\end{equation}
yielding the result in fig.~\ref{fig:hplot}. Binning both $H(z)$ and
$G(z)$ and marginalizing over the $h_i$s yields the plot for $G(z)$ seen in
fig.~\ref{fig:gplot}. 

Notice that here we assumed no prior information. Of course one could improve the FOM by taking into account
some external constraints due to other experiments.

%% file: de_mg/weaklensing-nbody.tex
\subsection{Testing the non-linear corrections for weak lensing forecasts.}
\label{weak-lensing-non-parametric_test}

In order to fully exploit next generation weak lensing survey
potentialities, accurate knowledge of non--linear power spectra up to
$\sim1\%$ is needed~\citep{Huterer:2001yu,Huterer:2004tr}. However,
such precision goes beyond the claimed $\pm 3\, \%$ accuracy of the
popular {\sc halofit} code~\citep{Smith2003}.

 \cite{McDonald:2005gz} showed that, using {\sc halofit} for
non--$\Lambda$CDM models, requires suitable corrections. In spite of
that, {\sc halofit} has been often used to calculate the spectra of
models with non--constant DE state parameter $w(z)$.  This procedure
was dictated by the lack of appropriate extensions of {\sc halofit} to
non--$\Lambda$CDM cosmologies.

In this paragraph we quantify the effects of using the {\sc halofit}
code instead of $N$--body outputs for non--linear corrections for DE
spectra, when the nature of DE is investigated through weak lensing
surveys. Using a Fisher Matrix approach, we evaluate the discrepancies
in error forecasts for $w_{0}$, $w_{a}$ and $\Omega_m$ and compare the
related confidence ellipses. See \cite{casarini:2010} for further details.

The weak lensing survey is as specified in Sec. \ref{sec:baofm_survey}.
Tests are performed assuming three different fiducial cosmologies:
$\Lambda$CDM model ($w_0 = -1$, $w_a = 0$) and two dynamical DE models, still
consistent with the WMAP+BAO+SN combination \citep{Komatsu:2010fb} at
95\% C.L.~. They will be dubbed M1 ($w_0 = -0.67$, $w_a = 2.28$) and
M3 ($w_0 = -1.18$, $w_a = 0.89$). In this way we explore the
dependence of our results on the assumed fiducial model.
For the other parameters we adopt the fiducial cosmology of Sec. \ref{sec:baofm_survey}.

The derivatives to calculate the
Fisher matrix are evaluated by extracting the power spectra from the $N$-body
simulations of models close to the fiducial ones, obtained by
considering parameter increments $\pm 5\, \%$. For the $\Lambda$CDM
case, two different initial seeds were also considered, to test the
dependence on initial conditions, finding that Fisher Matrix results
are almost insensitive to it. For the other fiducial models, only one
seed is used. 

$N$--body simulations are performed by using a modified version of
PKDGRAV \citep{2001PhDT........21S} able to handle any DE state
equation $w(a)$, with $N^3 = 256^{3}$ particles in a box with side $L
= 256\, h^{-1}$Mpc. Transfer functions generated using the CAMB
package are employed to create initial conditions, with a modified
version of the PM software by \cite{Klypin:1997sk}, also able to
handle suitable parameterizations of DE.

Matter power spectra are obtained by performing a FFT (Fast Fourier
Transform) of the matter density fields, computed from the particles
distribution through a Cloud--in--Cell algorithm, by using a regular
grid with $N_{g}=2048$. This allows us to obtain non--linear spectra
in a large $k$--interval. In particular, our resolution allows to work
out spectra up to $k \simeq 10\, h$Mpc$^{-1}$. However, for $k >
2$--$3\, h\, $Mpc$^{-1}$ neglecting baryon physics is no longer
accurate
\citep{Jing:2005gm,Rudd:2007zx,Bonometto:2010kz,Zentner:2007bn,Hearin:2009hz}.
For this reason, we consider WL spectra only up to $\ell_{max} = 2000$.

Particular attention has to be paid to matter power spectra
normalizations. {\color{red}In fact}, we found that, normalizing all models to the
same linear $\sigma_8 (z=0)$, the shear derivatives with respect to
$w_0$, $w_a$ or $\Omega_m$ were largely dominated by the normalization
shift at $z=0$, $\sigma_{8}$ and $\sigma_{8,nl}$ values being quite
different and the shift itself depending on $w_0$, $w_a$ and
$\Omega_m$. This would confuse the $z$ dependence of the growth
factor, through the observational $z$--range. This normalization
problem was not previously met in analogous tests with the Fisher
matrix, as {\sc halofit} does not directly depend on the DE state
equation.

As a matter of fact, one should keep in mind that, observing the
galaxy distribution with future surveys, one can effectively measure
$\sigma_{8,nl}$, and not its linear counterpart. For these reasons, we
choose to normalize matter power spectra to $\sigma_{8,nl}$, assuming
to know it with high precision.

\begin{figure}
\begin{center}
\includegraphics[scale=0.4]{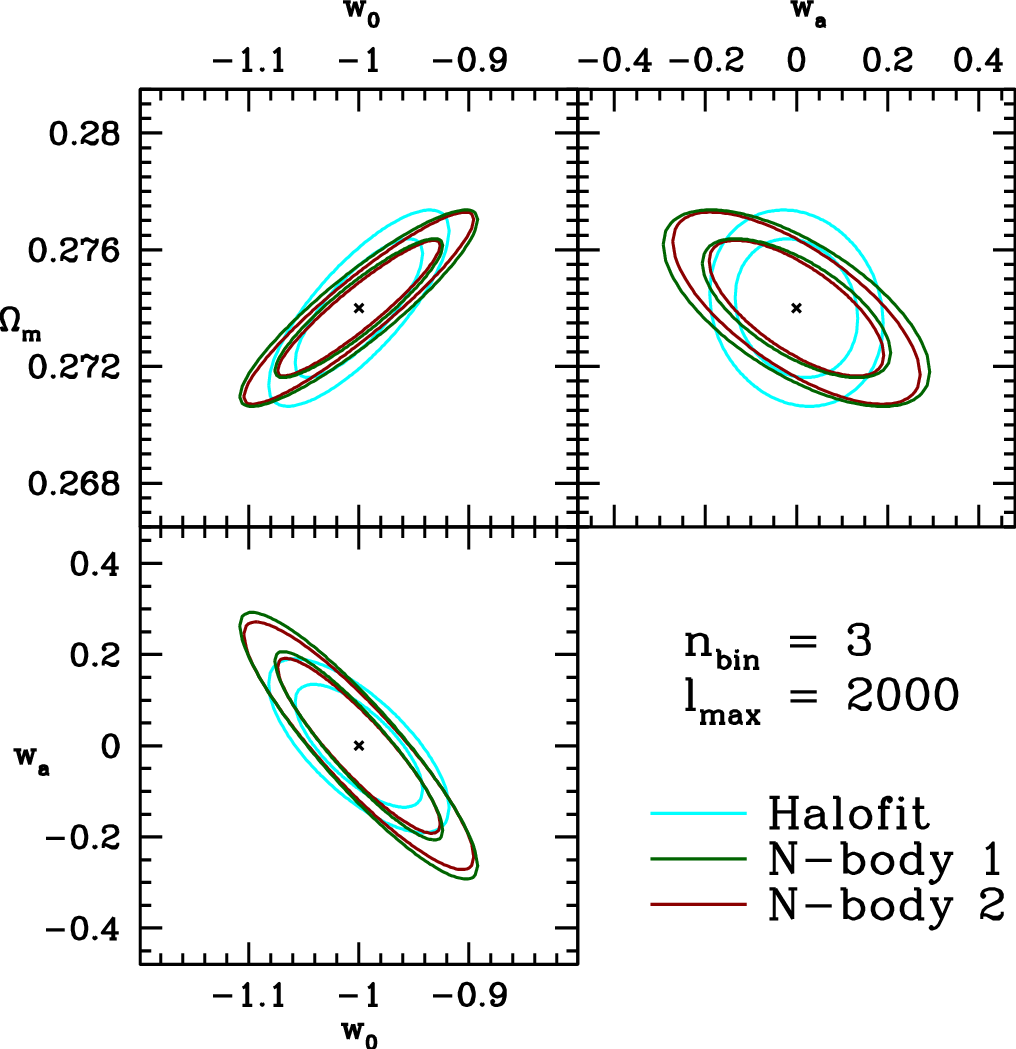}
\hskip1.truecm
\includegraphics[scale=0.4]{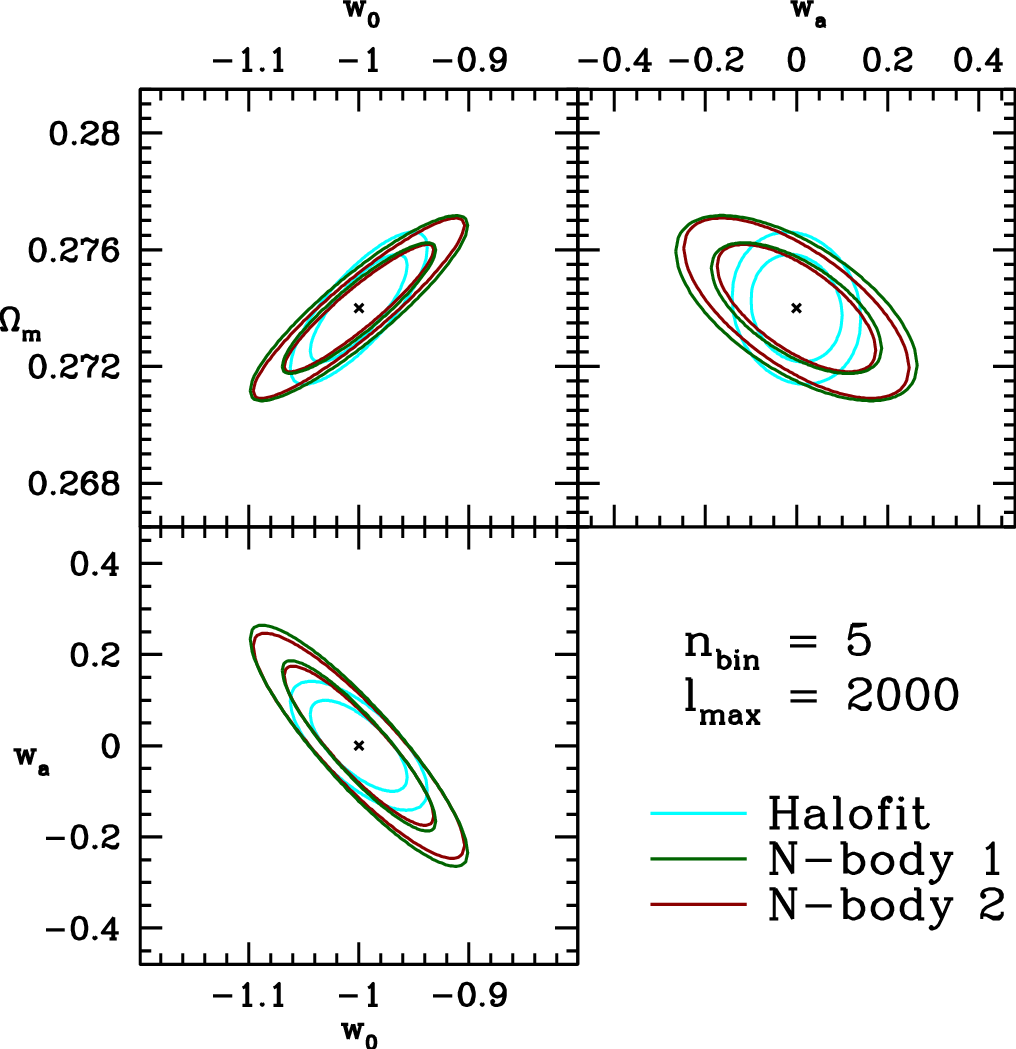}
\end{center}
\caption{Likelihood contours, for 65$\, \%$ and 95$\, \%$ C.L.,
  calculated including signals up to $\ell \simeq 2000$ for the
  $\Lambda$CDM fiducial. Here simulations and {\sc halofit} yield
  significantly different outputs.}
\label{lcdm2000}
\end{figure}
In figures \ref{lcdm2000} we show the confidence ellipses, when the
fiducial model is $\Lambda$CDM, in the cases of 3 or 5 bins and with
$\ell_{max} = 2000$. Since the discrepancy between different seeds are
small, discrepancies between {\sc halofit} and simulations are truly
indicating an underestimate of errors in the {\sc halofit} case.

As expected, the error on $\Omega_m$ estimate is not affected by the
passage from simulations to {\sc halofit}, since we are dealing with
$\Lambda$CDM models only. On the contrary, using {\sc halofit} leads
to underestimates of the errors on $w_0$ and $w_a$, by a substantial
30--40$\, \%$ (see \cite{casarini:2010} for further details).

This confirms that, when considering models different from
$\Lambda$CDM, non linear correction obtained through {\sc halofit} may
be misleading. This is true even when the fiducial model is
$\Lambda$CDM itself and we just consider mild deviations of $w$ from
$-1$.

\begin{figure}
\begin{center}
\includegraphics[scale=0.45]{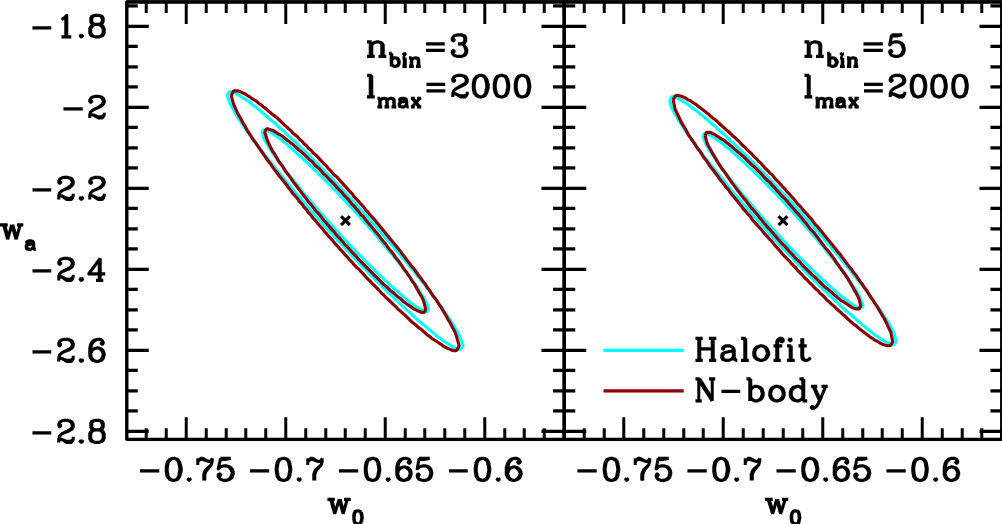}
\hskip1.truecm
\includegraphics[scale=0.45]{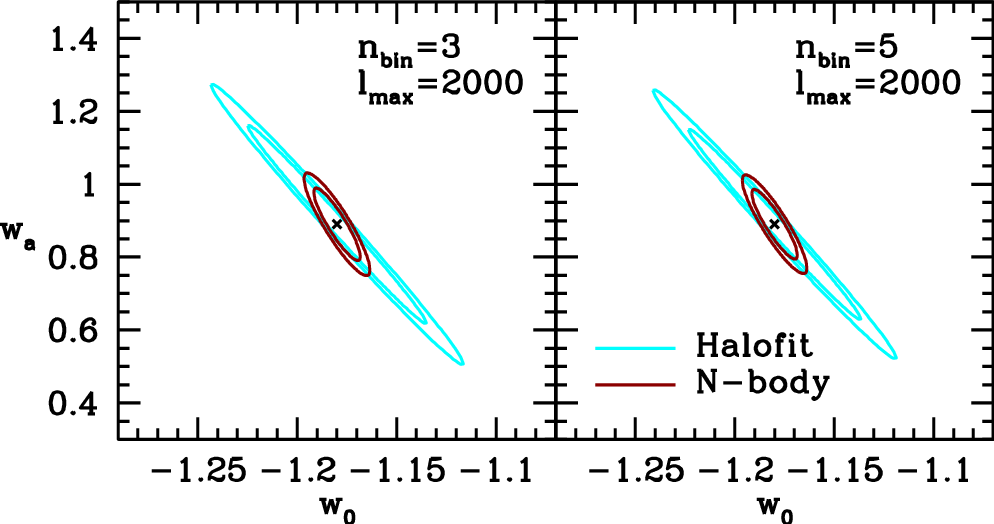}
\end{center}
\caption{On the left (right) panel, 1-- and 2--$\sigma$ contours for
  the M1 (M3) model. The two fiducial {\color{red}models} exhibit quite different
  behaviours.}
\label{mfig}
\end{figure}
Figure \ref{mfig} then show the results in the $w_0$--$w_a$ plane,
when the fiducial models are M1 or M3. It is evident that the two
cases are quite different. In the M1 case, we see just quite a mild
shift, even if they are $\cal O$ $(10 \, \%)$ on error predictions. In
the M3 case, errors estimated through {\sc halofit} exceed simulation
errors by a substantial factor. Altogether, this is a case when
estimates based on {\sc halofit} are  not trustworthy.

{\color{red} The effect of baryon physics is another non-linear correction to be considered. We note that the details of a study on  the impact of baryon physics on the power spectrum and the parameter estimation can be found in \cite{2011MNRAS.417.2020S}}

%% file: de_mg/soundspeed.tex
\subsection{Forecasts for the dark energy sound speed}\label{soundspeed}

As we have seen  in Sec.~\ref{sec:cosmo_perts},  when dark
energy clusters, the standard sub-horizon Poisson equation that links matter
fluctuations to the gravitational potential is modified and $Q\ne 1$. The deviation
from unity will depend on the degree of DE clustering and therefore on the sound
speed $c_s$. \label{symbol:c_s}
In this subsection we try to forecast the constraints that Euclid can put
on a constant $c_s$ by measuring $Q$ both via weak lensing and via redshift clustering. Here
 we assume standard Einstein gravity and zero anisotropic stress (and therefore we have $\Psi=\Phi$) and
 we allow $c_{s}$ to 
assume different values in the range $0 - 1$. 

Generically, while dealing with a non-zero sound speed, 
we have to worry about the sound horizon $k_{sh}=aH/c_{s}$, which 
characterizes the growth of the perturbations; 
then we have at least three regimes with different behaviour of the perturbations: 
\begin{enumerate}
\item perturbations larger than the causal horizon (where perturbations 
are {\color{red}not causally} connected and their growth is suppressed),
\item perturbations smaller than the causal horizon but larger than 
the sound horizon, $k\ll aH/c_{s}$ (this is the only regime where perturbations are free to 
grow because the velocity dispersion, or equivalently the pressure perturbation, 
is smaller than the gravitational attraction), 
\item perturbations smaller than the sound horizon, $k\gg aH/c_{s}$ (here perturbations stop growing 
because the pressure perturbation is larger than the gravitational attraction).
\end{enumerate}

As we have set the anisotropic stress to zero, the perturbations are
fully described by $Q$. 
The main problem is therefore to find an explicit expression that shows
how $Q$ depends on $c_s$. 
\cite{Sapone:2009} have provided 
the following explicit approximate expression for  $Q\left(k,a\right)$ 
which captures the behavior for both super- and sub-horizon scales: 
\begin{equation}
Q(k,a)=1+\frac{1-\Omega_{M,0}}{\Omega_{M,0}}\frac{(1+w)a^{-3w}}{1-3w+\frac{2}{3}\nu(a)^{2}}.
\label{eq:qtot}
\end{equation}
Here $\nu(a)^{2}=k^{2}c_{s}^2 a/\left(\Omega_{M,0}H_{0}^{2}\right)$
which it is defined through $c_{s}k\equiv\nu aH$ so that $\nu$ counts
how deep a mode is inside the sound horizon.

Eq.~(\ref{eq:qtot}) depends substantially on the value of the 
sound speed or, to put it differently, on the scale considered. 
For scales larger than the sound horizon ($\nu\approx0$), Eq.~(\ref{eq:qtot}) 
scales as $a^{-3w}$ and for $\Omega_{m,0}=0.25$ and $w=-0.8$ we have that 
\begin{equation}
Q-1\approx\frac{3}{17}a^{2.4}\simeq0.18a^{2.4}\,.
\end{equation}
This is not a negligible deviation today, but it decreases rapidly
as we move into the past, as the dark energy becomes less important.
\footnote{For this reason, early dark energy models can have a much stronger
impact.} 
As a scale enters the sound horizon, $Q-1$ grows with one power
of the scale factor slower (since $\delta_{DE}$ stops growing), suppressing
the final deviation roughly by the ratio of horizon size to the scale
of interest (as now $\nu^2\gg 1$). In the observable range, $(k/H_{0})^{2}\approx10^{2}-10^{4}$.
Therefore if $c_{s}\approx1$, $Q\to1$ and the dependence on $c_{s}$
is lost. This shows that $Q$ is sensitive to $c_{s}$ only for small
values, $c_{s}^{2}\lesssim10^{-2}$.

We can characterize the dependence of $Q$ on the main perturbation
parameter $c_{s}^2$ by looking at its derivative, a key quantity for
Fisher matrix forecasts: 
\begin{equation}
\frac{\partial\log Q}{\partial\log c_s^2}=-\frac{x}{\left(1+x\right)}\frac{Q-1}{Q}.
\label{eq:Qdercs}
\end{equation}
where $x=\frac{2}{3}\nu(a)^{2}/(1-3w)\simeq0.2\nu(a)^{2}$ (with
the last expression being for $w=-0.8$). For the values we are interested {\color{red} in}
here, this derivative has a peak at the present epoch at the sound
horizon, i.e. for $c_{s}\approx H_{0}/k$, which in the observable
range of $k$ is $c_{s}\approx.01-.001$, and declines rapidly for
larger $c_{s}$.
This means that the sensitivity of $Q$ to the sound speed can be
boosted by several orders of magnitude as the sound speed is decreased.

There are several observables that depend on $Q$:
\begin{itemize} 

\item The growth of matter perturbations.
There are two
ways to influence the growth factor: firstly at background level,
with a different Hubble expansion. Secondly at perturbation level:
if dark energy clusters then the gravitational potential changes because
of the Poisson equation, and this will also affect the growth rate
of dark matter. All these effects can be included in the growth index
$\gamma$ and we therefore expect that $\gamma$ is a function of
$w$ and $c_s^2$ (or equivalently of $w$ and $Q$).

The growth index depends on dark energy
perturbations (through $Q$) as \citep{Sapone:2009}
\begin{equation}
\gamma=\frac{3\left(1-w-A\left(Q\right)\right)}{5-6w}
\label{eq:gamma-Q}
\end{equation} 
where 
\begin{equation}
A\left(Q\right)=\frac{Q-1}{1-\Omega_{M}\left(a\right)}.
\label{eq:A-Q}
\end{equation}
Clearly here, the key quantity is the derivative of the growth 
factor with respect to the sound speed: 
\begin{equation}
\frac{\partial\log G}{\partial \ln c_s^2}\propto \int_{a_0}^{a_1}{\frac{\partial\gamma}{\partial c_{s}^2}{\rm d}a}\propto \int_{a_0}^{a_1}{\frac{\partial Q}{\partial c_{s}^2}{\rm d}a} \propto \int_{a_0}^{a_1}{\left(Q-1\right){\rm d}a}\,.
\label{eq:Gdercs}
\end{equation}
From the above equation we also notice that the derivative of the growth factor  
does not depend on $Q-1$ like the derivative 
$Q$, but on $Q-Q_{0}$ as it is an integral (being $Q_0$ the 
value of $Q$ today). The growth factor is thus not directly probing
the deviation of $Q$ from unity, but rather how $Q$ evolves over
time, see \cite{Sapone:2010} for more details.

\item Redshift space distortions

The distortion induced by redshift can be expressed in linear theory
by the $\beta$ factor, related to the bias factor and the growth
rate via: 
\begin{equation}
\beta(z,k)=\frac{\Omega_{m}\left(z\right)^{\gamma(k,z)}}{b(z)}\,.
\end{equation}
The derivative of the redshift distortion parameter with respect to
the sound speed is: 
\begin{equation}
\frac{\partial\log\left(1+\beta\mu^{2}\right)}{\partial\log c_s^2}= 
-\frac{3}{5-6w}\frac{\beta\mu^{2}}{1+\beta\mu^{2}}\frac{x}{1+x}\left(Q-1\right)\,.
\label{eq:derbetadcs}
\end{equation}
We see that the behavior versus $c_{s}^{2}$ is similar to the one
for the $Q$ derivative, so the same discussion applies. Once again,
the effect is maximized for small $c_{s}$. 
The $\beta$ derivative is {\color{red}comparable to that of} $G$ at $z=0$ but becomes more important 
at low redshifts.

\item Shape of the dark matter power spectrum

{\color{red}Quantifying} the impact of the sound speed on the matter power
spectrum is quite hard as we {\color{red}need} to run 
Boltzmann codes (such as CAMB, \cite{CAMB}) in order to get the full impact of 
dark energy perturbations into the matter power spectrum. 
\cite{Sapone:2010} proceeded in two ways: 
first using the CAMB output and then considering the analytic expression 
from  \cite{Eisenstein_Hu_1997} (which does not include dark 
energy perturbations, i.e. does not include $c_{s}$).

They find that the impact of the derivative of
the matter power spectrum with respect the sound speed on the final
errors is only relevant if high values of $c_s^2$ are considered; 
by decreasing the sound speed, the results are less and less affected. 
The reason is that for low values of the sound
speed other parameters, like the growth factor, start to be the dominant
source of information on $c_{s}^{2}$.

\end{itemize}

{\bf Impact on weak lensing.}

Now it is possible to investigate the response of weak lensing (WL) to the
dark energy parameters. Proceeding with a Fisher matrix as in \cite{Amendola:2007rr}, 
the main difference here being that the parameter $Q$ has an
explicit form. Since $Q$ depends on $w$ and $c_s^2$, we can forecast
the precision with which those parameters can be extracted. We can
also try to trace where the constraints come from.
For a vanishing anisotropic stress the WL potential becomes: 
\begin{equation}
k^{2}\left(\Phi+\Psi\right) = -2Q\frac{3H_{0}^{2}\Omega_{M,0}}{2a}\Delta_{M}
\label{eq:WL-potential-Q}
\end{equation}
which can be written, in linear perturbation theory as: 
\begin{equation}
k^{2}\left(\Phi+\Psi\right)=-3H\left(a\right)^{2}a^{3}Q\left(a,k\right)\Omega_{M}\left(a\right)G\left(a,k\right)\Delta_{M}\left(k\right)\,.
\label{eq:phiwl}
\end{equation}

Hence, the lensing potential contains three conceptually different
contributions from the dark energy perturbations: 
\begin{itemize}
\item The direct contribution of the perturbations to the gravitational
potential through the factor $Q$.
\item The impact of the dark energy perturbations on the growth rate of
the dark matter perturbations, affecting the time dependence of {\color{red}$\Delta_{M}$},
through $G\left(a,k\right)$.
\item A change in the shape of the matter power spectrum $P(k)$, corresponding
to the dark energy induced $k$ dependence of {\color{red}$\Delta_{M}$}.
\end{itemize}
We use the representative Euclid survey presented in Sec.~\ref{sec:baofm_survey} 
and we extend our survey up to three different redshifts: $z_{max}=2,3,4$. 
We choose different values of $c_s^2$ and  $w_0 = -0.8$ in order to 
maximize the impact on $Q$: values closer to $-1$
reduce the effect and therefore increase the errors on $c_{s}$.

\begin{figure}
\centering \includegraphics[width=2.7in]{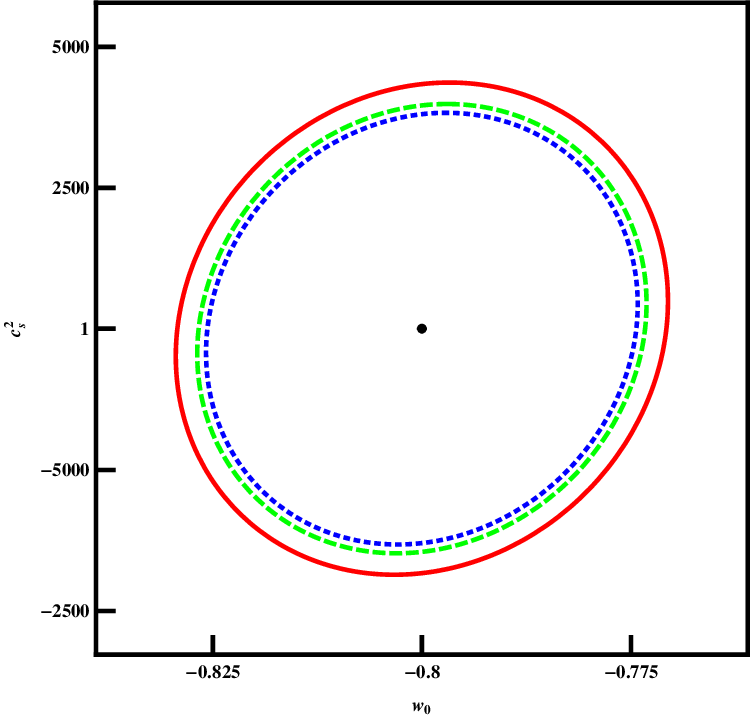}
\hspace{0.1in} \includegraphics[width=2.7in]{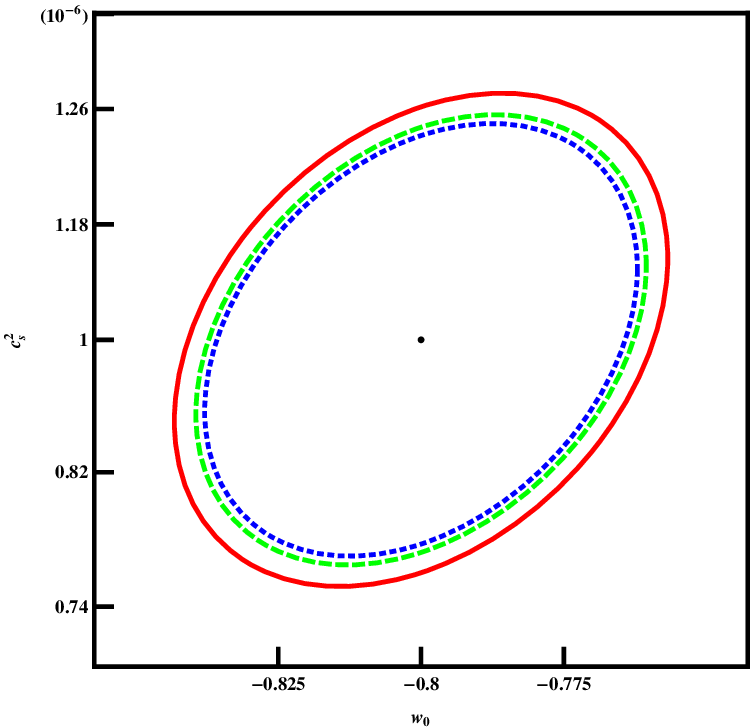}\\
\caption{Confidence region at $68\%$ for three different {\color{red}values} of $z_{max}=2.5,3.5,4$,
red solid,  green long-dashed and blue dashed contour, respectively. The
left panel shows the confidence region when the sound speed is $c_s^2=1$;
the right panel with the sound speed $c_s^2=10^{-6}$. The 
{\color{red}equation of state parameter} is for both cases $w_{0}=-0.8$.}
\label{fig:ellipsesw0cs} 
\end{figure}

In Fig.~\ref{fig:ellipsesw0cs} we report the $1-\sigma$ confidence 
region for $w_{0},c_s^2$ for two different
values of the sound speed and $z_{max}$. For high value of the sound
speed ($c_s^2=1$) we find $\sigma(w_{0})=0.0195$ and the relative
error for the sound speed is $\sigma(c_s^2)/c_s^2=2615$. As expected,
WL is totally insensitive to the clustering properties of quintessence
dark energy models when the sound speed is equal to $1$. The presence
of dark energy perturbations leaves a $w$ and $c_s^2$ dependent signature
in the evolution of the gravitational potentials through {\color{red}$\Delta_{DE}/\Delta_{m}$}
and, as already mentioned, the increase of the $c_s^2$ enhances the
suppression of dark energy perturbations which brings $Q\rightarrow1$.

Once we decrease the sound speed then dark energy perturbations are
free to grow at smaller scales. In Fig.~\ref{fig:ellipsesw0cs} {\color{red}  the confidence region for $w_{0},c_s^2$ for $c_s^2=10^{-6}$ is 
shown;} 
we find $\sigma(w_{0})=0.0286$, $\sigma(c_s^2)/c_s^2=0.132$;
in the last case the error on the measurement {\color{red}of the sound speed  is} reduced
to the 70\% of the total signal.

{\bf Impact on galaxy power spectrum.}

We now explore a second probe of clustering, the galaxy power spectrum. The procedure is the same outlined in
 Sec.~\ref{dark-energy-and-redshift-surveys}.
We use the representative Euclid survey presented in  Sec.~\ref{sec:baofm_survey}. 
Here too we also consider in addition possible extended surveys to $z_{max}=2.5$ and $z_{max}=4$. 

\begin{figure}
\centering \includegraphics[width=2.7in]{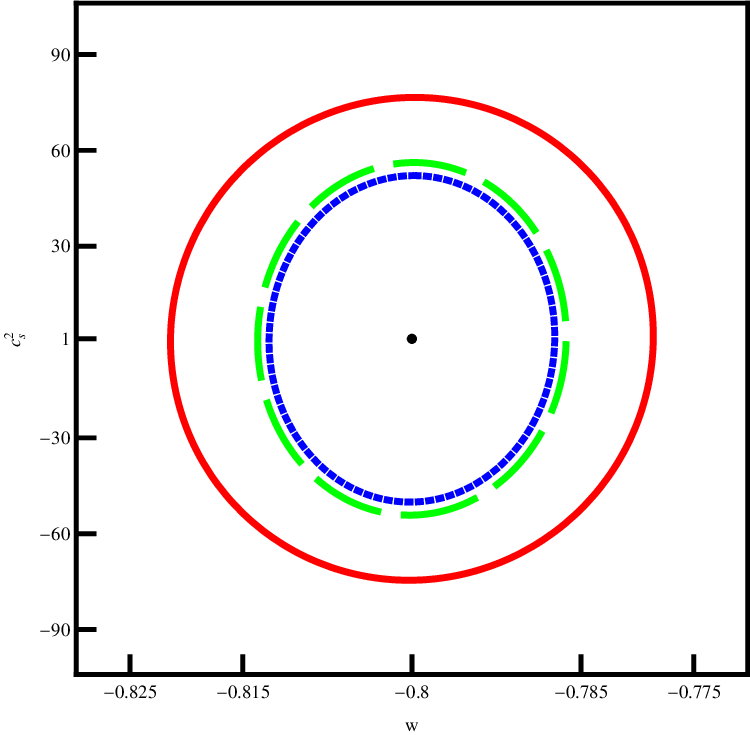}
\hspace{0.1in} \includegraphics[width=2.7in]{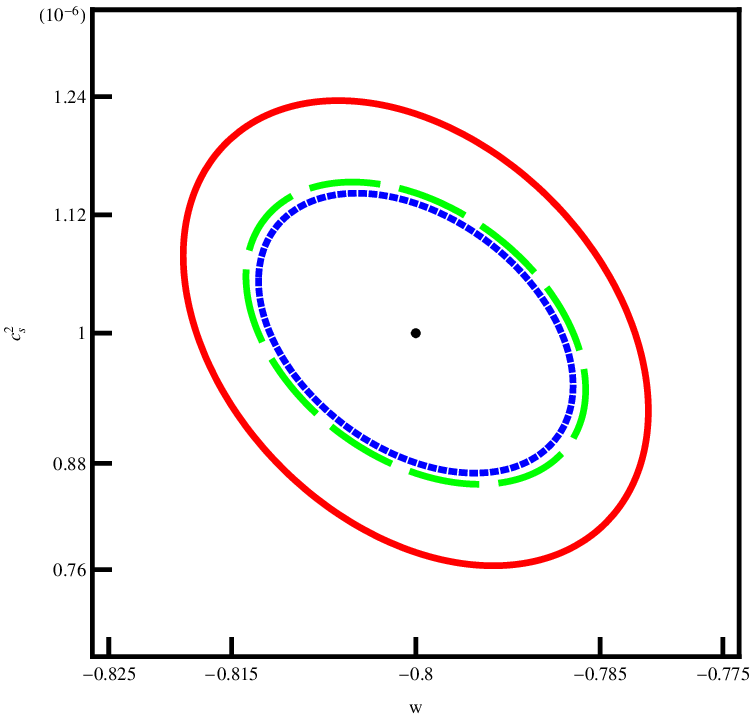}\\

\caption{ Confidence region at $68\%$ for three different {\color{red}values} of $z_{max}=2.5,3.5,4$,
red solid, green long-dashed and blue dashed {\color{red}contour}, respectively. The
left panel shows the confidence region when the sound speed is $c_s^2=1$;
the right panel with the sound speed $c_s^2=10^{-6}$. The 
{\color{red}equation of state parameter} is for both cases $w_{0}=-0.8$.}
\label{fig:ellipsesw0cs-pk} 
\end{figure}

In Fig.~\ref{fig:ellipsesw0cs-pk} we report the confidence region
for $w_{0},c_s^2$ for two different values of the sound speed and $z_{max}$.
For high values of the sound speed ($c_s^2=1$) we find, for our benchmark
survey: $\sigma(w_{0})=0.0133$, and $\sigma(c_s^2)/c_s^2=50.05$. Here
again we find that galaxy power spectrum is not sensitive to the clustering
properties of dark energy when the sound speed is of order unity.
If we decrease the sound speed down to $c_s^2=10^{-6}$ then the errors
are $\sigma(w_{0})=0.0125$, $\sigma(c_s^2)/c_s^2=0.118$.

In conclusion, as perhaps expected, we find that
dark energy perturbations have a very small effect on dark matter
clustering unless the sound speed is extremely small, $c_{s}\le 0.01$. 
Let us remind that in order to boost the observable
effect, we always assumed $w=-0.8${\color{red}; for values closer to $-1$ the
sensitivity to $c_s^2$ is further reduced. A}s a test,  \cite{Sapone:2010} 
performed the calculation for $w=-0.9$ and $c_s^2=10^{-5}$ and found $\sigma_{c_s^2}/c_s^2=2.6$
and $\sigma_{c_s^2}/c_s^2=1.09$ for WL and galaxy power spectrum experiments,
respectively.

Such small sound speeds are not in contrast with the fundamental expectation
of dark energy being much smoother that dark matter: even with $c_{s}\approx0.01$,
dark energy perturbations are more than one order of magnitude weaker
than dark matter ones (at least for the class of models investigated
here) and safely below non-linearity at the present time at all scales.
Models of {}``cold'' dark energy are interesting because they can
cross the phantom divide \citep{Kunz:2006wc} and contribute to the cluster
masses \citep{Creminelli/etal:2010} (see also Sec. \ref{the-spherical-collapse-model} of this review ). 
Small $c_{s}$ could be constructed for instance
with scalar fields with non-standard kinetic energy terms.

%% file: de_mg/fRforecastconstraints.tex
\subsection{Weak lensing constraints on $f(R)$ gravity \label{fRforecastconstraints}}

In this section {\color{red}we} present the Euclid weak lensing forecasts of  a specific, but very popular, class of models, the
so-called $ f(R)$ models of gravity. As we have already seen in Sec. \ref{fr-general} these models
are described by the action

\begin{equation} S_{\rm grav} = \int \sqrt{-g} d^{4}x \left[ \frac{f(R)}{16\pi
G} - {\cal L}_{\rm m} \right] , \end{equation}

\noindent where $f(R)$ is an arbitrary function of the Ricci scalar and ${\cal
L}_{\rm m}$ is the Lagrange density of standard matter and radiation.

In principle one has complete freedom to specify the function $f(R)$, and so any expansion history  can be
reproduced.
However, as discussed in Sec. \ref{fr-general}, those that remain viable are the subset
that very closely mimic the standard $\Lambda$CDM background expansion, as this
restricted subclass of models can evade solar system constraints
\citep{Chiba:2003ir,Tsujikawa:2008uc,Lin:2010hk}, have a standard matter era in
which the scale factor evolves according to $a(t) \propto t^{2/3}$
\citep{APT07} and can also be free of ghost and
tachyon instabilities \citep{Nariai:1973eg,Gurovich:1979xg}.

\begin{figure}
\includegraphics[width=0.9\textwidth]{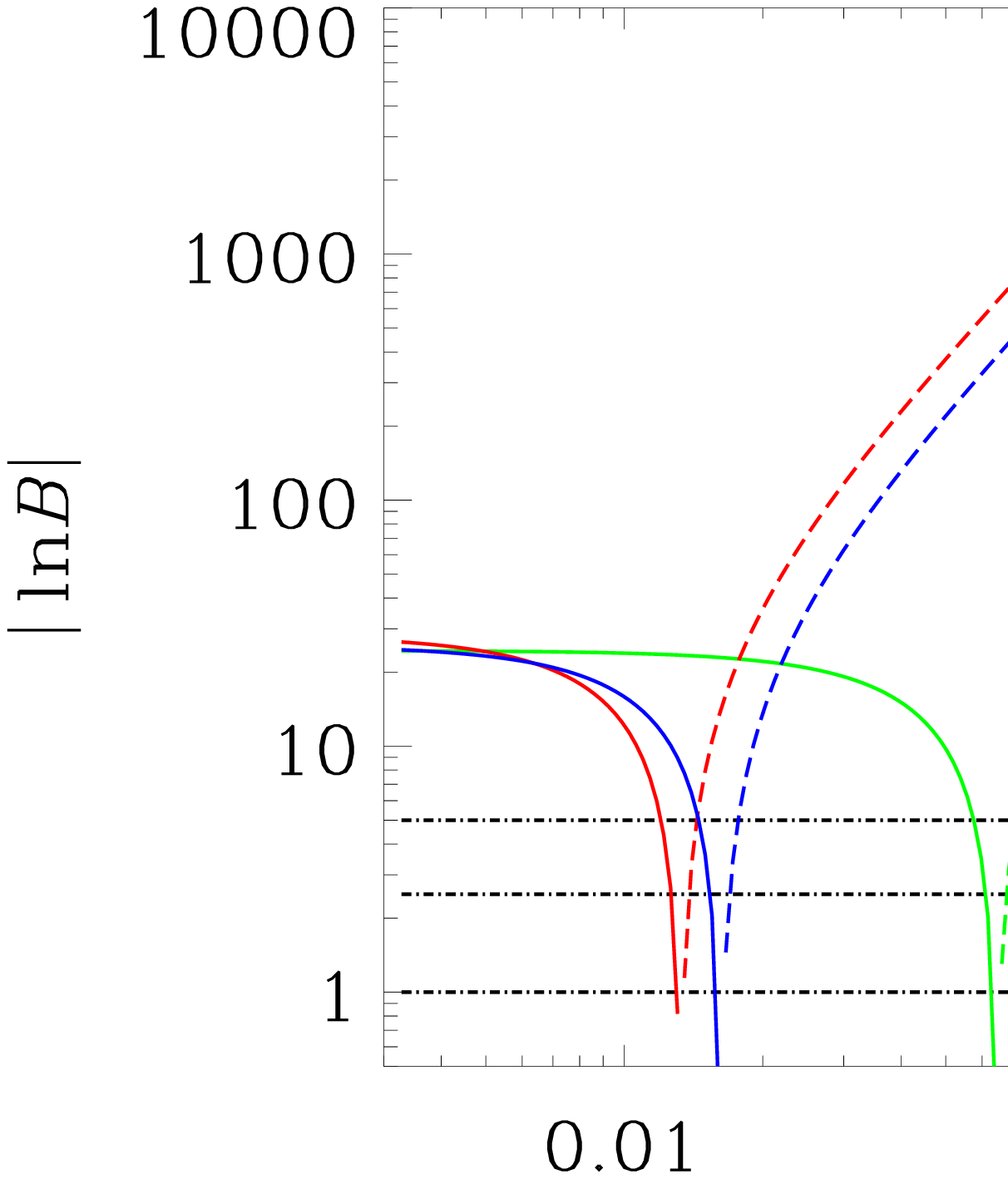}
\caption{The Bayes factor $\ln B$ for the $f(R)$ model of Eq.~(\ref{Bmodel})
over standard $\Lambda$CDM as a function of the extra parameter $n$. The green,
red and blue curves refer to the conservative, bin-dependent and optimistic
$\ell_\mathrm{max}$, respectively. The horizontal lines denote the Jeffreys'
scale levels of significance.}\label{fig:lnB-HS}
\end{figure}

{\color{red}To this subclass belongs  the popular} $f(R)$ model proposed by
\citet{Hu07} (\ref{Bmodel}). \citet{Camera:2011ms} demonstrated that Euclid will
have the {\color{red}power of distinguishing} between it and $\Lambda$CDM with a
good accuracy. They performed a tomographic analysis using several values of the
maximum allowed wavenumber of the Fisher matrices; specifically, a conservative
value of $1000$, an optimistic value of $5000$ and a bin-dependent setting,
which increases the maximum angular wavenumber for distant shells and reduces it
for nearby shells. Moreover, they computed the Bayesian expected evidence for
the model of Eq.~(\ref{Bmodel}) over the $\Lambda$CDM model as a function of the
extra parameter $n$. This can be done because the $\Lambda$CDM model is formally
nested in this $f(R)$ model, and the latter is equivalent to the former when
$n=0$. Their results are shown in Fig.~\ref{fig:lnB-HS}. {\color{red}For another  Bayesian  evidence analysis of $f( R)$ models  and the added value of probing the growth of structure with galaxy surveys  see also \cite{Song:2007da}.}

This subclass of $f(R)$ models can be parameterised solely in terms of the mass
of the scalar field, {\color{red} which as we have seen in  Eq.~(\ref{frr-mass})} is related to the $f(R)$ functional form via the
relation 

\begin{equation} M^{2}(a) = {\frac{1} {3 f_{,RR}[R_{\rm back}(a)]}} \end{equation}

\noindent  {\color{red} where $R$ subscripts denote differentiation with respect to $R$.} The function $f_{,RR}$  
can be approximated by its standard $\Lambda$CDM form,

\begin{equation} {\frac{R_{\rm back}} {H_{0}^{2}}} \simeq  {\frac{3\Omega_{\rm m0}} 
{a^{3}}}  + 12\Omega_{\Lambda} , \end{equation}

\noindent valid for $z \lesssim 1000$. The mass $M(a)$ is typically a function
of redshift which decays from a large value in the early Universe to its present
day value $M_{0}$.

Whilst these models are practically indistinguishable from $\Lambda$CDM at the
level of background expansion, there is a significant difference in the
evolution of perturbations relative to the standard General Relativistic
behaviour.

The evolution of linear density perturbations in the context of $f(R)$ gravity
is markedly different than in the standard $\Lambda$CDM scenario; $\delta_{\rm
m} \equiv \delta \rho_{\rm m} /\rho_{\rm m}$ acquires a nontrivial scale
dependence at late times. This is due to the presence of an additional scale
$M(a)$ in the equations; as any given mode crosses the modified gravity
`horizon' $k = aM(a)$, said mode will feel an enhanced gravitational force due
to the scalar field. This will have the effect of increasing the power of small
scale modes.

Perturbations on sub-horizon scales in the Newtonian gauge evolve approximately
according to

\begin{eqnarray} \label{eq:16} & &  \Psi = \left(1 + {\frac{2\bar{K}^{2}} {3 +
2\bar{K}^{2}}}\right) \Phi , \\ \label{eq:17} & & k^{2}\Phi = -4\pi G \left({\frac{3 +
2\bar{K}^{2}} {3 + 3\bar{K}^{2}}}\right)a^{2}\rho_{\rm m} \delta_{\rm m} , \\
\label{eq:18}  & & \label{eq:p1} \ddot{\delta}_{\rm m} + 2H \dot{\delta}_{\rm m}
- 4\pi G \left({\frac{3 + 4\bar{K}^{2}}{ 3 + 3\bar{K}^{2}}}\right)\rho_{\rm
m}\delta_{\rm m} = 0 , \end{eqnarray}

\noindent where $\bar{K} = k/(aM(a))$. These equations represent a particular
example of a general parameterization introduced in
\cite{Martinelli:2010wn,Bertschinger:2008zb,Zhao:2008bn}. To solve them one
should first parameterize the scalaron mass $M(a)$, choosing a form that broadly
describes the behaviour of viable $f(R)$ models. A suitable functional form,
which takes into account the evolution of $M(a)$ in both the matter era and the
late time accelerating epoch, is given by \cite{Thomas:2011pj}

\begin{equation} \label{eq:107} M^{2} = M_{0}^{2} \left( {\frac{ a^{-3} + 4 a_{*}^{-3}}
{ 1 + 4 a_{*}^{-3}}}\right)^{2\nu} , \end{equation}

\noindent where $a_{*}$ is the scale factor at matter-$\Lambda$ equality; $a_{*}
= (\Omega_{\rm m0}/\Omega_{\Lambda})^{1/3}$.  There are two modified gravity
parameters; $M_{0}$ is the mass of the scalaron at the present time and $\nu$ is
the rate of increase of $M(a)$ to the past.

\begin{figure*}[!ht]
  \begin{flushleft}
    \centering
    \begin{minipage}[c]{1.00\textwidth}
      \centering
      \includegraphics[width=7cm,height=7cm]{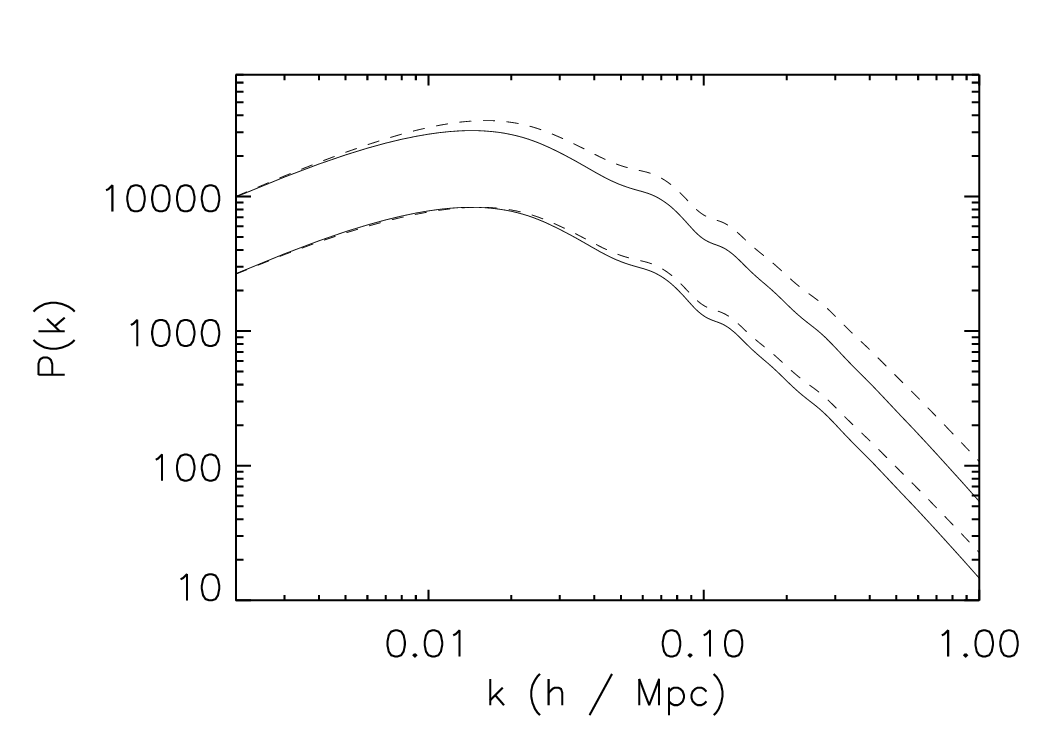}
      \includegraphics[width=7cm,height=7cm]{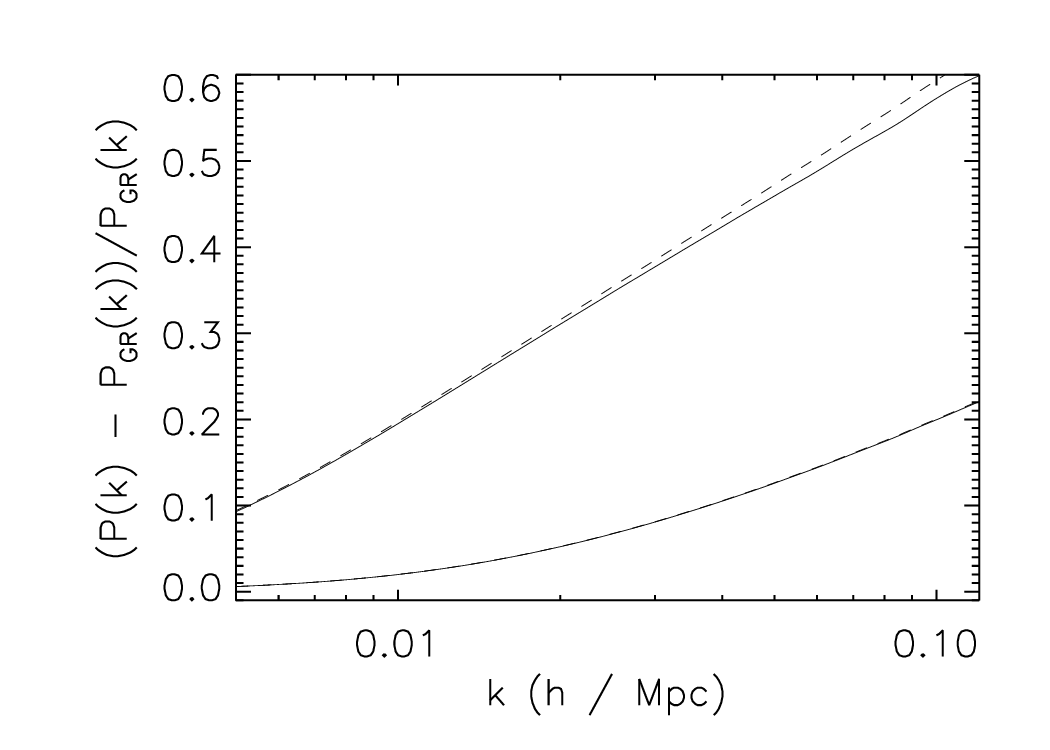}
    \end{minipage}
    \caption{\small{{\it Left Panel:} Linear matter power spectra for
$\Lambda$CDM (solid line; $M_{0}^{-1}=0$, $\nu=1.5$) and scalaron (dashed line;
$M^{-1}_{0}=375 [10^{28}{\rm h^{-1} \hspace{1mm} eV^{-1}}]$, $\nu=1.5$)
cosmologies. The modification to gravity causes a sizeable scale dependent
effect in the growth of perturbations. The redshift dependence of the scalaron
can be seen by comparing the top and bottom pairs of power spectra evaluated at
redshifts $z=0.0$ and $z=1.5$, respectively. {\it Right Panel:} The
environmental dependent chameleon mechanism can be seen in the mildly non linear
regime. We exhibit the fractional difference $(P(k) - P_{\rm GR}(k))/P_{\rm
GR}(k)$ between the f(R) and GR power spectra for the model ($\ref{eq:107}$)
with parameters $M^{-1}_{0}=375 [10^{28}{\rm h^{-1} \hspace{1mm} eV^{-1}}]$ and
$\nu=1.5$. The dashed lines represent linear power spectra ($P(k)$ and $P_{\rm
GR}(k)$ calculated with no higher order effects) and the solid lines are the
power spectra calculated to second order. We see that the nonlinearities
decrease the modified gravity signal. This is a result of the chameleon
mechanism. The top set of lines correspond to $z=0$ and the bottom to $z=0.9$;
demonstrating that the modified gravity signal dramatically decreases for larger
$z$. This is due to the scalaron mass being much larger at higher redshifts.
Furthermore, non linear effects are less significant for increasing $z$. }}
    \label{fig:2b}
  \end{flushleft}
\end{figure*}
\noindent

In fig.\ref{fig:2b} the linear matter power spectrum is exhibited for this
parameterization (dashed line), along with the standard $\Lambda$CDM power
spectrum (solid line). The observed, redshift dependent tilt is due to the
scalaron's influence on small scale modes, and represents a clear modified
gravity signal. Since weak lensing is sensitive to the underlying matter power
spectrum, we expect Euclid to provide  direct constraints on the mass of the
scalar field.

{\color{red}

By performing a Fisher analysis, using the standard Euclid specifications, \cite{Thomas:2011pj} 
calculates the expected $f(R)$ parameter sensitivity of the weak lensing survey. 
By combining Euclid weak lensing and Planck Fisher matrices, both modified gravity parameters $M_{0}$ and $\nu$ 
are shown to be strongly constrained by the growth data in fig. \ref{fig:fr_saa}. The expected 1$\sigma$ bounds on $M_{0}$ and $\nu$ 
are quoted as $M_{0} = 1.34 \pm 0.62 \times 10^{-30} [h eV]$, $\nu = 1.5 \pm 0.18$ when using linear data $l < 400$ only 
and $M_{0} = 1.34 \pm 0.25 \times 10^{-30} [h eV]$, $\nu = 1.5 \pm 0.04$ when utilizing the full set of non-linear modes $l < 10000$.

\begin{figure*}[htbp!]
    \centering
\includegraphics[width=10cm]{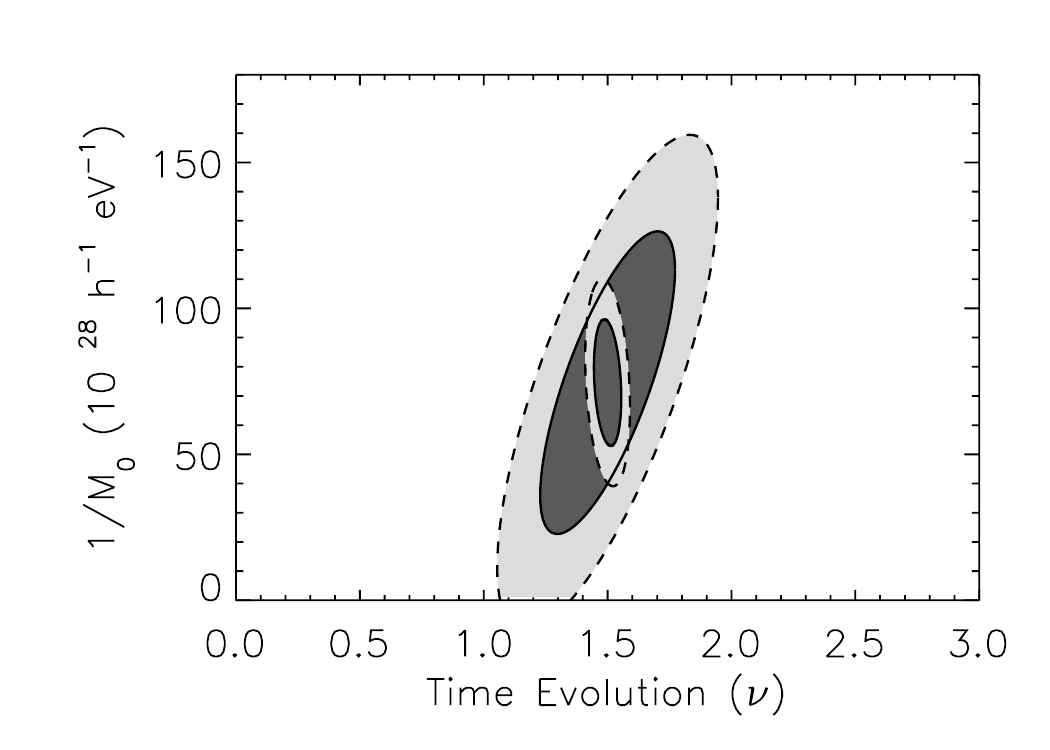}
\caption{68\% (dark grey) and 95\% (light grey) projected bounds on the modified gravity parameters $M_{0}^{-1}$ and $\nu$ for the combined Euclid weak lensing and Planck CMB 
surveys. The smaller (larger) contours correspond to including modes $l = 400 (10000)$ in the weak lensing analysis.} 
\label{fig:fr_saa}
\end{figure*}

}

%% file: de_mg/cdeforecast.tex
\subsection{Forecast constraints on coupled quintessence cosmologies \label{cdeforecast}}

In this section we present forecasts for coupled quintessence cosmologies \cite{Amendola:1999er,wetterich95,Pettorino:2008ez}, obtained when combining Euclid weak lensing, Euclid redshift survey (baryon acoustic oscillations, redshift distortions and full $P(k)$ shape) and CMB as obtained in Planck (see also the next {\color{red}section} for CMB priors). Results reported here were obtained in \cite{Amendola:2011ie} and we refer to it for details on the analysis and Planck specifications 
(for weak lensing and CMB constraints on coupled quintessence with a different coupling see also \cite{martinelli,DeBernardis:2011iw}). 
In \cite{Amendola:2011ie} the coupling is the one described in section \ref{cde_eq}, as induced by a scalar-tensor
model. The slope $\alpha$ of the Ratra-Peebles potential is included as an additional parameter and Euclid specifications refer to the Euclid Definition phase \cite{euclidredbook}.

\begin{figure*}
\centering \includegraphics[width=15cm]{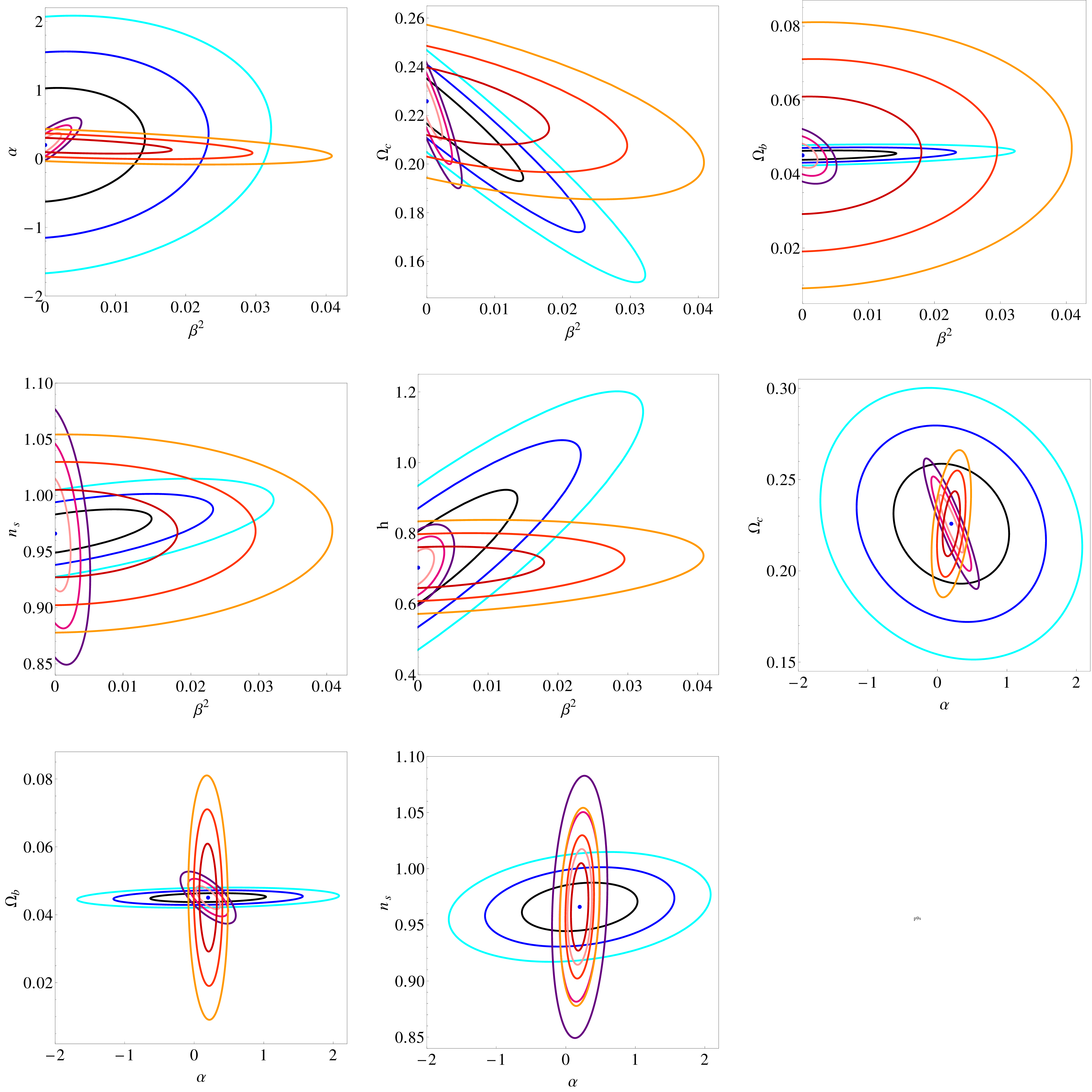}\caption{{\small Comparison among predicted confidence contours for the cosmological
parameter set $\Theta\equiv\{\beta^{2},\alpha,\Omega_{c},h,\Omega_{b},n_{s},\sigma_{8},log(A)\}$
using CMB (Planck, blue contours), $P(k)$ (pink-violet contours) and weak
lensing (orange-red contours) with Euclid-like specifications. [From Amendola, Pettorino, Quercellini, Vollmer 2011].}}
{\small \label{fig:cmbbaowl-cont} } 
\end{figure*}

The combined Fisher confidence regions are plotted in Fig.~\ref{fig:cmbbaowl-cont} 
and the results are in Table \ref{tab:combined}.
The main result is that future surveys can constrain the coupling
of dark energy to dark matter $\beta^{2}$ to less than $3\cdot10^{-4}$. Interestingly, some combinations of parameters (e.g.\ $\Omega_b$ vs $\alpha$) seem to profit {\color{red}at the most} from the combination of the three probes. 

\begin{table*}
\begin{tabular}{lllll}
\begin{minipage}[c]{100pt}
\flushleft Parameter
\end{minipage} &
\begin{minipage}[c]{100pt}
\flushleft $\sigma_{i}$ CMB+$P(k)$
\end{minipage} &
\begin{minipage}[c]{100pt}
\flushleft $\sigma_{i}$ CMB+$P(k)$+WL
\end{minipage} &  & \tabularnewline
\hline 
$\beta^{2}$  & 0.00051  & 0.00032  &  & \tabularnewline
$\alpha$  & 0.055  & 0.032  &  & \tabularnewline
$\Omega_{c}$  & 0.0037  & 0.0010  &  & \tabularnewline
$h$  & 0.0080  & 0.0048  &  & \tabularnewline
$\Omega_{b}$  & 0.00047  & 0.00041  &  & \tabularnewline
$n_{s}$  & 0.0057  & 0.0049  &  & \tabularnewline
$\sigma_{8}$  & 0.0049  & 0.0036  &  & \tabularnewline
$\log(A)$  & 0.0051  & 0.0027  &  & \tabularnewline
\hline 
\end{tabular} 
\caption{1-$\sigma$ errors for the set $\Theta\equiv\{\beta^{2},\alpha,\Omega_{c},h,\Omega_{b},n_{s}\,\sigma_{8},\log(A)\}$
of cosmological parameters, combining CMB+$P(k)$ (left column) and
CMB+$P(k)$+WL (right column).}
\label{tab:combined}
\end{table*}

We can also ask whether a better knowledge of the parameters $\{\alpha,\Omega_{c},h,\Omega_{b},n_{s},\sigma_{8},\log(A)\}$,
obtained by independent future observations, can give us better constraints
on the coupling $\beta^{2}$. In Table \ref{tab:fixedpar}  we show the errors
on $\beta^{2}$ when we have a better knowledge of only one other
parameter, which is here fixed to the reference value. All remaining
parameters are marginalized over. 

It is remarkable to notice that the combination of CMB, power spectrum
and weak lensing is already a powerful tool and a better knowledge
of one parameter doesn't improve much the constraints on $\beta^{2}$.
CMB alone, instead, improves by a factor 3 when $\Omega_{c}$ is known
and by a factor 2 when $h$ is known. The power spectrum is mostly
influenced by $\Omega_{c}$, which allows to improve constraints on
the coupling by more than a factor 2. Weak lensing gains the most
by a better knowledge of $\sigma_{8}$.

\begin{table*}
\begin{tabular}{llllll}
\begin{minipage}[c]{140pt}
\flushleft Fixed parameter
\end{minipage} &
\begin{minipage}[c]{40pt}
\flushleft CMB
\end{minipage} &
\begin{minipage}[c]{40pt}
\flushleft $P(k)$
\end{minipage} &
\begin{minipage}[c]{40pt}
\flushleft WL
\end{minipage} &
\begin{minipage}[c]{90pt}
\flushleft CMB + $P(k)$ + WL
\end{minipage} & \tabularnewline
\hline 
(Marginalized on all params)  & 0.0094  & 0.0015  & 0.012  & 0.00032  & \tabularnewline
$\alpha$  & 0.0093  & 0.00085  & 0.0098  & 0.00030  & \tabularnewline
$\Omega_{c}$  & 0.0026  & 0.00066  & 0.0093  & 0.00032  & \tabularnewline
$h$  & 0.0044  & 0.0013  & 0.011  & 0.00032  & \tabularnewline
$\Omega_{b}$  & 0.0087  & 0.0014  & 0.012  & 0.00030  & \tabularnewline
$n_{s}$  & 0.0074  & 0.0014  & 0.012  & 0.00028  & \tabularnewline
$\sigma_{8}$  & 0.0094  & 0.00084  & 0.0053  & 0.00030  & \tabularnewline
$\log(A)$  & 0.0090  & 0.0015  & 0.012  & 0.00032  & \tabularnewline
\hline 
\end{tabular}\caption{1-$\sigma$ errors for $\beta^{2}$, for CMB, $P(k)$, WL and CMB+$P(k)$+WL.
For each line, only the parameter in the left column has been fixed
to the reference value. The first line corresponds to the case in
which we have marginalized over all parameters. [From Amendola, Pettorino, Quercellini, Vollmer 2011].}
\label{tab:fixedpar} 
\end{table*}

%% file: de_mg/extra_priors.tex
\subsection{Extra-Euclidean data and priors}

Other dark energy projects will enable the cross-check of the dark energy 
constraints from Euclid. These include Planck, BOSS, WiggleZ, HETDEX, DES,
Panstarrs, LSST, BigBOSS {\color{red}and SKA}.

Planck will provide exquisite constraints
on cosmological parameters, but not tight constraints on dark energy
by itself, as CMB data are not sensitive to the nature of dark energy
(which has to be probed at $z<2$, where dark energy becomes increasingly
important in the cosmic expansion history and the growth history of 
cosmic large scale structure). Planck data in combination with Euclid
data provide powerful constraints on dark energy and tests of gravity.
In the next section we will discuss how to create a Gaussian approximation
to the Planck parameter constraints that can be combined with Euclid
forecasts in order to model the expected sensitivity until the actual
Planck data is available towards the end of 2012.

{\color{red}The galaxy redshift surveys BOSS, WiggleZ, HETDEX, and BigBOSS are
complementary to Euclid, since the 
overlap in redshift ranges of different galaxy redshift surveys, both space 
and ground-based, is critical for understanding systematic effects such 
as bias through the use of multiple tracers of cosmic large scale structure. 
Euclid will survey H-alpha emission line galaxies 
at $0.5 < z < 2.0$ over 20,000 square degrees.
The use of multiple tracers of cosmic large scale structure can reduce 
systematic effects and ultimately increase the precision of dark energy 
measurements from galaxy redshift surveys (see, e.g., 
\cite{Seljak_Hamaus_Desjacques09}).}

{\color{red}Currently on-going or recently completed surveys which  cover a sufficiently large volume to measure BAO at several redshifts and thus have  science goals common to  Euclid, are the Sloan Digital Sky Survey III Baryon  Oscillations  Spectroscopic Survey (BOSS for short)  and the WiggleZ survey.

BOSS \footnote{http://www.sdss3.org/surveys/boss.php} maps the redshifts of 1.5 million Luminous Red Galaxies (LRGs) out to $z\sim0.7$ over 10,000 square degrees, measuring the BAO signal, the  large-scale galaxy correlations  and extracting information of the growth from redshift space distortions. A simultaneous survey of $2.2 < z < 3.5$ quasars measures the acoustic oscillations in the correlations of the Lyman-alpha forest. LRGs were chosen for their high bias, their approximately  constant number density and, of course,  the fact that they are bright.  Their spectra and redshift can be measured with  relatively short    exposures in a 2.4m  ground-based telescope.  The data-taking of BOSS will end in 2014. 

The WiggleZ \footnote{http://wigglez.swin.edu.au/site/index.html} survey is now completed, it measured redshifts for almost 240,000 galaxies over 1000 square degrees at $0.2<z<1$. The target  are luminous blue star-forming galaxies with spectra dominated by patterns of strong atomic emission lines. This choice is motivated by the fact that these emission lines  can be used to measure a galaxy redshift in relatively short exposures of a 4m class ground-based telescope.

Red quiescent galaxies inhabit dense clusters environments, while blue star-forming galaxies trace better lower density regions such as sheets and filaments. It is believed that on large cosmological scales these details are unimportant and that galaxies are simply tracers of the underlying dark matter: different galaxy type will only have a different `bias factor'. The fact that so far results from BOSS and WiggleZ agree well confirms this assumption.

Between now and the availability of  Euclid data other  wide-field spectroscopic galaxy redshift surveys  will take place.  
Among them, eBOSS will extend BOSS operations focusing on 3100 square degrees using a variety of tracers.   Emission line galaxies will be targeted in the redshift window $0.6<z<1$. This will  extend to higher redshift and extend the sky coverage of the WiggleZ survey.  Quasars in the redshift range $1<z<2.2$ will be used as  tracers of the BAO feature instead of galaxies.  The BAO LRG measurement will be extended to $z \sim 0.8$, and the quasar number density at $z>2.2$ of BOSS will be tripled, thus improving the BAO lyman alpha forest measure.

HETDEX is expected to begin full science operation is 2014: it  aims at surveying 1 million  Lyman-alpha emitting galaxies at $1.9 < z < 3.5$ over 420 square degrees. The main science goal is to map the BAO feature over this redshift range. 

Further in the future  we highlight here the  proposed BigBOSS survey  and SuMIRe survey  with HyperSupremeCam on  the Subaru telescope. 
The BigBOSS survey will target  [OII] emission line galaxies at $0.6 < z < 1.5$ (and LRGs at $z < 0.6$) over 14,000  square degrees.
The SuMIRe wide survey  proposes to survey $\sim 2000$ square degrees  in the redshift range $0.6<z<1.6$ targeting LRGs  and  [OII] emission-line galaxies. Both these surveys will likely reach full science operations roughly at the same time as  the Euclid launch. 

Wide field photometric surveys  are also being carried out and planned. 
The on-going Dark Energy Survey (DES)\footnote{http://www.darkenergysurvey.org} will cover 5000 square degrees out to $z\sim1.3$ and is expected to complete observations in 2017; The  Panoramic Survey Telescope \& Rapid Response System (Pan-STARRS), on-going  at the single-mirror stage,  
The PanSTARSS survey, which first phase is already on-going, will cover 30000 square degrees with 5 photometry bands  for redshifts up to $z\sim 1.5$. The second pause of the survey is expected to be competed  by the time Euclid launches.
More in the future  the Large Synoptic Survey Telescope (LSST) will cover  redshifts $0.3<z<3.6$ over  20000 square degrees, but is expected to begin operations in 2021,  after Euclid's planned launch date.  
The galaxy imaging surveys DES, Panstarrs, and LSST will complement
Euclid imaging survey in both the choice of band passes, and the
sky coverage. 

{\color{red}SKA (which is expected to begin operations in 2020 and reach full operational capability in 2024)  will  survey neutral atomic hydrogen (HI) through the radio 21 cm line, over  a very wide area of the sky. It is expected to detect HI emitting galaxies out to  $z\sim 1.5$ making it nicely complementary to Euclid. Such galaxy redshift survey will of course offer the opportunity  to measure the galaxy power spectrum (and therefore the BAO feature) out to $z \sim 1.5$. The well behaved point spread function of a synthesis array like the SKA should ensure superb image quality enabling cosmic shear to be accurately measured and tomographic weak lensing  used to constrain cosmology and in particular dark energy. This weak lensing capability also makes SKA and Euclid very complementary. For more information see e.g., \cite{Rawlings:2004wk, Blake:2004pb}.}

\begin{figure}
\centering
\includegraphics[width=0.9\textwidth]{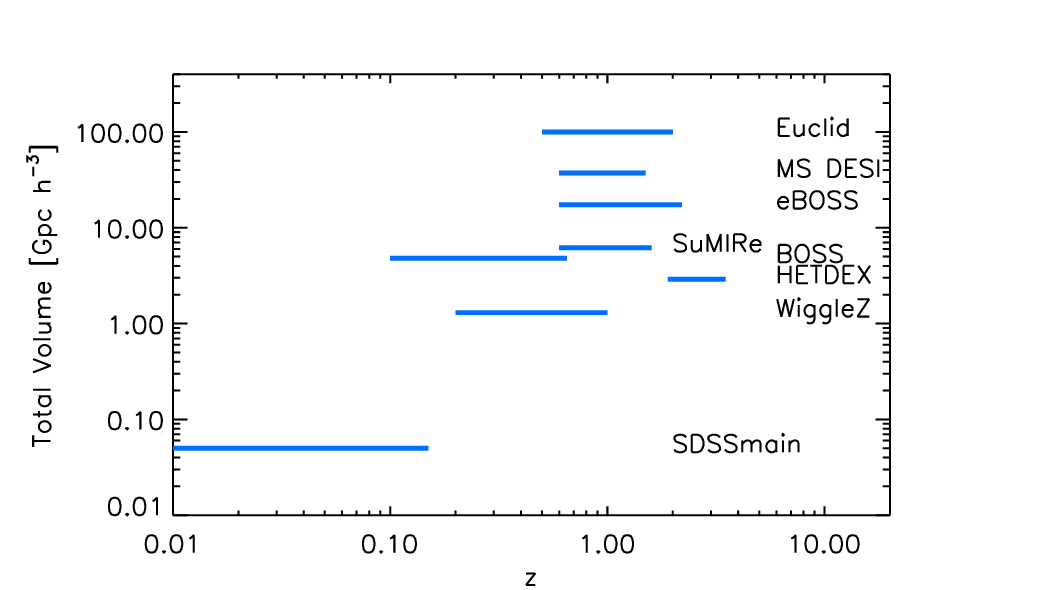}
\caption{{\color{red}Redshift coverage and volume for the surveys mentioned in the text. Spectroscopic surveys only are shown.  Recall that while future and forthcoming photometric surveys focus on weak gravitational lensing, spectroscopic surveys can extract the three dimensional galaxy clustering information and therefore measure radial and tangential BAO signal, the power spectrum shape and the growth of structure via redshift space distortions. The three-dimensional clustering information is crucial for BAO. For example to obtain the same figure of merit  for dark energy properties a photometric survey must cover a volume roughly  ten times bigger than a spectroscopic one.}} \label{fig:future_probes_context}
\end{figure}

The figure \ref{fig:future_probes_context}  puts Euclid into context.   Euclid will survey H-alpha emission line galaxies at $0.5 < z < 2.0$ over 20,000 square degrees.
Clearly Euclid with both spectroscopic and photometric capabilities and wide field coverage  surpasses all surveys that will be carried out by the time it launches.   The  large volume surveyed is crucial as the number of modes to sample for example the power spectrum and the BAO feature scales with the volume. The redshift coverage is also important especially at $z<2$ where the  dark energy contribution to the density pod the Universe  is non-negligible (at $z>2$ for most cosmologies the Universe is effectively Einstein de Sitter therefore high redshifts do not   contribute much to constraints on dark energy). 
Having a single instrument, a uniform target selection and calibration is also crucial to perform precision tests of cosmology without having to build a `ladder' from different surveys selecting different targets.  On the other hand it is  also easy to see the synergy between these ground-based surveys and Euclid: by mapping different targets (over the same sky area and ofter the same redshift range) one can gain better control over issues such as bias factors.  The use of multiple tracers of cosmic large scale structure can reduce systematic effects and ultimately increase the precision of dark energy measurements from galaxy redshift surveys (see, e.g., Seljak et al. (2009)).

Moreover, having both spectroscopic and imaging capabilities Euclid is uniquely poised to explore the clustering with both  the three dimensional distribution of galaxies and weak gravitational lensing.}

\subsubsection{The Planck prior}

Planck will provide highly accurate constraints on many cosmological parameters,
which makes the construction of a Planck Fisher matrix somewhat non-trivial as it
is very sensitive to the detailed assumptions. A relatively robust approach was
used by \cite{Mukherjee:2008kd} to construct a Gaussian approximation to the
WMAP data by introducing two extra parameters,
\begin{equation}
R \equiv \sqrt{\Omega_m H_0^2} \,r(z_{CMB}), \hskip 0.1in
l_a \equiv \pi r(z_{CMB})/r_s(z_{CMB}),
\label{eq:planckfish1}
\end{equation}
where $r(z)$ is the comoving distance from the observer to redshift $z$,
and $r_s(z_{CMB})$ is the comoving size of the sound-horizon at decoupling.

In this scheme, $l_a$ describes the peak location through the angular diameter
distance to decoupling and the size of the sound horizon at that time.
If the geometry changes, either due to non-zero curvature or due to a different
equation of state of dark energy, $l_a$ changes in the same way as the 
peak structure. $R$ encodes similar information, but in addition 
contains the matter density which is connected with the peak height.
In a given class of models (for example, quintessence
dark energy), these parameters are ``observables'' {\color{red}related} to 
the shape of the observed CMB spectrum, and constraints on them 
remain the same independent of (the prescription for) the equation of 
state of the dark energy.

As a caveat we note that if some assumptions regarding the evolution of
perturbations are changed, then the corresponding 
$R$ and $l_a$ constraints and covariance 
matrix will need to be recalculated under each such hypothesis, for 
instance if massive neutrinos were to be included, or even if tensors were 
included in the analysis \citep{Corasaniti:2007rf}. Further $R$ as defined in Eq.~(\ref{eq:planckfish1})
can be badly constrained and {\color{red} is} quite useless if the dark energy clusters
as well, e.g. if it has a low sound speed, as in the model discussed in
\cite{Kunz:2007rk}.

In order to derive a Planck fisher matrix,  \cite{Mukherjee:2008kd}
simulated Planck data as described in \cite{Pahud:2006kv} and derived constraints on
our base parameter set $\{R,l_a,\Omega_b h^2,n_s\}$ with a MCMC based
likelihood analysis. In addition to $R$ and $l_a$ they used the
baryon density $\Omega_bh^2$, and optionally the spectral index of the 
scalar perturbations $n_s$, as these are strongly
correlated with $R$ and $l_a$, which means that we will lose information
if we do not include these correlations. As shown in  \cite{Mukherjee:2008kd},
the resulting Fisher matrix loses some information relative to the full likelihood
when only considering Planck data, but it is very close to the full analysis as soon
as extra data is used. Since this is the intended application here, it is perfectly
sufficient for our purposes.

The following tables, from \cite{Mukherjee:2008kd}, give the covariance matrix
for quintessence-like dark energy (high sound speed, no anisotropic stress)
on the base parameters and the Fisher matrix derived from it. Please consult the
appendix of that paper for the precise method used to compute $R$ and $l_a$
as the results are sensitive to small variations.

\begin{table*}[htb]
\caption{$R$, $l_a$, $\Omega_bh^2$ and $n_s$ estimated from Planck 
simulated data.}
\begin{center}
\begin{tabular}{lll}
\hline
Parameter & mean & rms variance \\
\hline
\hline
& $\Omega_k\neq 0$&\\
\hline
$R$ & 1.7016 & 0.0055\\
$l_a$ & 302.108 & 0.098 \\
$\Omega_b h^2$ & 0.02199 & 0.00017 \\
$n_s$ & 0.9602 & 0.0038\\
 \hline		
\end{tabular}
\end{center}
\end{table*}

\begin{table*}[htb]
\caption{Covariance matrix for $(R, l_a, \Omega_b h^2, n_s)$from Planck.}
\begin{center}
\begin{tabular}{lrrrr}
\hline
\hline
 & $R$ & $l_a$ & $\Omega_b h^2$ & $n_s$ \\
 \hline
 \hline
& & $\Omega_k\neq 0$& &\\
\hline
$R$            & 0.303492E-04  & 0.297688E-03  & $-$0.545532E-06  & $-$0.175976E-04 \\
$l_a$          & 0.297688E-03  & 0.951881E-02  & $-$0.759752E-05  & $-$0.183814E-03 \\
$\Omega_b h^2$ & $-$0.545532E-06 & $-$0.759752E-05 &  0.279464E-07  &  0.238882E-06 \\
$n_s$          & $-$0.175976E-04 & $-$0.183814E-03 &  0.238882E-06  &  0.147219E-04 \\
\hline
\end{tabular}
\end{center}
\end{table*}

\begin{table*}[htb]\scriptsize
\caption{Fisher matrix for ($w_0$, $w_a$, $\Omega_{DE}$, $\Omega_k$, $\omega_m$, 
$\omega_b$, $n_S$) derived from the covariance matrix for
$(R, l_a, \Omega_b h^2, n_s)$ from Planck.}
\begin{center}
\begin{tabular}{lrrrrrrr}
\hline
\hline
 & $w_0$ & $w_a$ & $\Omega_{DE}$ & $\Omega_k$ & $\omega_m$ & $\omega_b$ & $n_S$ \\
 \hline
 \hline
$w_0$ &  .172276E+06 &  .490320E+05 &  .674392E+06 & $-$.208974E+07 &  .325219E+07 & $-$.790504E+07 & $-$.549427E+05 \\
$w_a$ &  .490320E+05 & .139551E+05 & .191940E+06& $-$.594767E+06 & .925615E+06 &$-$.224987E+07 &$-$.156374E+05\\
$\Omega_{DE}$ & .674392E+06 & .191940E+06 & .263997E+07& $-$.818048E+07 & .127310E+08 &-.309450E+08 &$-$.215078E+06\\
$\Omega_k$ & $-$.208974E+07 & $-$.594767E+06 & $-$.818048E+07 &  .253489E+08 & $-$.394501E+08 &  .958892E+08 &  .666335E+06\\
$\omega_m$ &  .325219E+07 &  .925615E+06 &  .127310E+08 & $-$.394501E+08 &  .633564E+08 & $-$.147973E+09 & $-$.501247E+06\\
$\omega_b$ & $-$.790504E+07 &$-$.224987E+07 &$-$.309450E+08 & .958892E+08 &$-$.147973E+09  &.405079E+09 & .219009E+07\\
$n_S$ & $-$.549427E+05 & $-$.156374E+05 & $-$.215078E+06  & .666335E+06 & $-$.501247E+06  & .219009E+07  & .242767E+06\\
\hline 
\end{tabular}
\end{center}
\end{table*}

%% file: de_mg/outlook.tex
\section{Summary and outlook}

This chapter introduced the main features of the most popular
dark energy{\color{red}/modified gravity} models.
Here we  summarize the performance of Euclid with respect to these models.
Unless otherwise indicated, we always assume Euclid with no external priors  and all errors
fully marginalized over the standard cosmological parameters. Here RS denotes the redshift survey, WLS the weak lensing one.

\begin{enumerate}

\item
Euclid (RS)  should be able to measure the main standard cosmological parameters to percent or sub-percent level
as detailed
in Table \ref{tab:cosm_par_errors} (all marginalized errors, including constant equation of state and 
constant growth rate{\color{red}, see Table~\ref{tab:w0_w1} and Fig.~\ref{fig:w0_w1}}).

\item The two CPL parameters $w_0,w_1$ should be measured with {\color{red}errors 0.06 and 0.26}, respectively (fixing the
growth rate to fiducial){\color{red}, see Table~\ref{tab:w0_w1} and Fig.~\ref{fig:w0_w1}}.

\item The equation of state $w$ and the growth rate parameter $\gamma$, both assumed constant,
 should be simultaneously constrained to within 0.04 and 0.03, respectively.

\item The growth function should be constrained to within 0.01-0.02 for each redshift bin from {\color{red}$z=0.7$} to $z=2$ {\color{red}(see Table~\ref{tab:sigma_bias_s_bint})}.

\item A scale-independent bias function $b(z)$ should be constrained to within {\color{red}0.02} for each redshift bin {\color{red}(see Table~\ref{tab:sigma_bias_s_bint})}.

\item The growth rate parameters $\gamma_0,\gamma_1$ defined in Eq. \ref{eq:gam_CPL} 
should be measured to within 0.08, 0.17, respectively.

{\color{red}\item Euclid will achieve an accuracy on measurements of the dark energy sound speed of $\sigma(c_s^2)/c_s^2=2615$ (WLS) and $\sigma(c_s^2)/c_s^2=50.05$ (RS), if $c_s^2=1$, or $\sigma(c_s^2)/c_s^2=0.132$ (WLS) and $\sigma(c_s^2)/c_s^2=0.118$ (RS), if $c_s^2=10^{-6}$.}

{\color{red}\item The coupling $\beta^{2}$ between dark energy and dark matter can be constrained by Euclid (with Planck) to less than $3\cdot10^{-4}$ (see Fig.~\ref{fig:cmbbaowl-cont} and Table~\ref{tab:combined}).}

{\color{red}\item Any departure from GR greater
than $\simeq 0.03$ in the growth index $\gamma$ will be distinguished by the WLS \citep{Heavens/etal:2007}.}

{\color{red}\item Euclid WLS can detect deviations between $3\%$ and $10\%$ from the GR value of the modified-gravity parameter $\Sigma$ (Eq.~\ref{Sigma-wl-isw}), whilst with the RS there will be a $20\%$ accuracy on both $\Sigma$ and $\mu$ (Eq.~\ref{eq:muQeta}).}

\item With the WLS, Euclid should provide an upper limit to the present dimensionless scalaron
inverse mass $\mu \equiv H_{0}/M_{0}$ of the $f(R)$ scalar ({\color{red} where the time dependent scalar field mass is defined} in Eq.~\ref{eq:107}) 
as $\mu = 0.00 \pm 1.10\times 10^{-3}$ for $l < 400$ and $\mu = 0.0 \pm 2.10 \times 10^{-4}$ for $l < 10000$

{\color{red}\item The WLS will be able to rule out the DGP model growth index with a Bayes factor $|\ln B|\simeq 50$ \citep{Heavens/etal:2007}, and viable phenomenological extensions could be detected at the $3\sigma$ level for $1000\lesssim\ell\lesssim4000$ \citep{Camera:2011mg}.}

\end{enumerate}

At the same time, there are  several areas of research that we feel are important for the future of Euclid, both
to improve the current analyses and to
maximize its science return.
Here we provide a preliminary, partial list.

\begin{enumerate}

\item The results of the redshift survey and weak lensing surveys should be combined in a statistically coherent way

\item The set of possible priors to be combined with Euclid data should be better defined

\item The forecasts for the parameters of the modified gravity  and  clustered dark energy models should be extended to include more
general cases

\item We should estimate the errors on a general reconstruction of the modified gravity functions $\Sigma,\mu$ or of
the metric potentials $\Psi,\Phi$ as a function of both scale and time.

\end{enumerate}

%% file: dark_matter/dark_matter.tex

\chapter{Dark matter and neutrinos}\label{dark-matter}



\input{dark_matter/particle_intro}

\input{dark_matter/euclid_halo}


\input{dark_matter/euclid_dm_xray_olehost}

\input{dark_matter/DMmapping}


\input{dark_matter/Scattering}

\input{dark_matter/ClusterCrossSection.tex}

\input{dark_matter/WDM}

\input{dark_matter/textforeuclid_massnu}

\input{dark_matter/de_nu_writeup}


\input{dark_matter/UDMwriteupFinal}

\input{dark_matter/de_dm_writeup}

\input{dark_matter/UltraLightScalarFields}

\input{dark_matter/multifield}


 
\input{dark_matter/Outlook}


%% file: dark_matter/particle_intro.tex
\section{Introduction}

The identification of dark matter is one of the most important open problems in particle
physics and cosmology. In Standard Cosmology, dark matter contributes 85\% of all the
matter in the Universe, but we do not know what it is made of, as we have never observed
dark matter particles in our laboratories. The foundations of the modern dark matter
paradigm were laid in the 1970's and 1980's, after decades of slow accumulation of evidence. 
Back in the 1930's it was noticed that the Coma cluster seemed to contain much more mass
than what could be inferred from visible galaxies \citep{Zwicky:1933gu,Zwicky:1937zza}, and a few years later, it became clear
that the Andromeda galaxy M31 rotates anomalously fast at large radii, as if most of its mass 
resides in its outer regions. Several other pieces of evidence provided further support to the dark matter
hypothesis, including the so called timing-argument. In the 1970's rotation curves were
extended to larger radii and to many other spiral galaxies, proving the presence of large
amounts of mass on scales much larger than the size of galactic disks (\cite{Peacock99}).

{\color{red}We are now in the position of determining} the total abundance
of dark matter relative to normal, baryonic matter, in the Universe with exquisite accuracy; 
we have a much better understanding of how dark matter is distributed in structures ranging from dwarf galaxies
to clusters of galaxies, thanks to gravitational lensing 
observations (see Massey et al., 2010 for a review) and theoretically from 
high-resolution numerical simulations made possible by
modern supercomputers (such as, for example, the Millennium or Marenostrum simulations).

Originally, Zwicky thought of dark matter as most likely baryonic -- missing cold gas, or low mass stars. 
Rotation curve observation could be explained by dark matter in  the form of MAssive Compact Halo
Objects (MACHOs e.g. a halo of {\color{red}black holes} or brown dwarfs). 
However, the MACHO and EROS experiments have shown that dark matter cannot be in 
the mass range $0.6\times10^{-7}$\,M$_\odot<$M$<15$\,M$_\odot$ if it comprises 
massive compact objects \citep{2000ApJ...542..281A,2007A&A...469..387T}. 
Gas measurements are now extremely sensitive, ruling out dark matter as
undetected gas (\bcite{1997ApJ...479..523B}; \bcite{2001ApJ...559...29C};
\bcite{2006A&A...445..827R}; but see \bcite{1994A&A...285...79P}). And the CMB
and Big Bang Nucleosynthesis require the total mass in baryons in the Universe
to be significantly less that the total matter density
\citep{2002NuPhS.110...16R,2002PhRvD..65d3510C,2002ApJ...576L.101T}.

This is one of the most spectacular results in cosmology obtained at the end
of the {\color{red}XX} century: dark matter has to be non-baryonic. 
As a result, our expectation of
the nature of dark matter shifted from an astrophysical explanation to particle
physics,
linking the smallest and largest scales that we can probe. 

During the seventies
the possibility of the neutrino to be the dark matter particle with a mass of
tenth of $\mathrm{eV}$ was explored, but it was realised that such light
particle would erase the primordial fluctuations on small scales, leading
to a lack of structure formation on galactic scales and below. It was therefore
postulated that {\color{red} the dark matter particle} must be cold (low thermal
energy, to allow structures on small scale to
form), collisionless 
(or have a very low interaction cross section, because dark matter is observed
to be pressureless) and  stable 
over a long period of time: such a candidate is referred to as a Weakly Interacting Massive Particle (WIMP). 
This the standard cold dark matter (CDM) picture (Peebles, 1992).

Particle physicists have proposed several possible dark matter candidates. 
Supersymmetry (SUSY) is an attractive extension of the Standard Model of particle physics. The lightest SUSY particle (the LSP) is stable, uncharged, and weakly interacting, providing a perfect WIMP candidate known as a neutralino. Specific realisations of SUSY each provide slightly different dark matter candidates \citep[for a review see][]{jungman1996}. Another distinct dark matter candidate arising from extensions of the Standard Model is the axion, 
a hypothetical pseudo-Goldstone boson whose existence was postulated to solve the so
called strong CP problem in Quantum Chromo-Dynamics \citep{pecceiquinn1977}, also arising generically in string theory \citep{witten1984,witten2006}. They are known to be very well
motivated dark matter candidates \citep[for a review of axions in cosmology see][]{sikivie2008}. Other well-known candidates are sterile neutrinos,
which interact only gravitationally with ordinary matter, apart from a small mixing with
the familiar neutrinos of the Standard Model (which should make them ultimately unstable), and candidates arising from technicolor \citep[see e.g.][]{gudnason2006}. A wide array of other possibilities have been discussed in the literature, and they are currently being searched for with a variety of experimental strategies \citep[for a complete review of dark matter in particle physics see][]{pdg}.

There remain some possible discrepancies in the standard cold dark matter model, such as the missing satellites
problem, and the cusp-core controversy (see below for details and references) that {\color{red}have led} some
authors to question the CDM model and to propose alternative solutions.
The physical mechanism by which one may reconcile the observations with the standard
theory of structure formation is the suppression of the matter power spectrum at small scales.
This can be achieved with dark matter particles with a strong self-scattering cross section, or
with particles with a non-negligible velocity dispersion at the epoch of structure formation,
also referred to as warm dark matter (WDM) particles.

{\color{red}Another possibility is that the extra gravitational degrees of freedom arising in modified theories of gravity play the role of dark matter.  In particular this happens for the Einstein-Aether, TeVeS and bigravity models. These theories were developed following the idea that the presence of unknown dark components in the Universe may be indicating us that it is not the matter component that is exotic but rather that gravity is not described by standard general relativity.}

Finally we note that only from astrophysical probes can any dark matter candidate found in either direct detection experiments or accelerators, such as the LHC, be confirmed. Any direct dark matter candidate discovery will give Euclid a clear goal to verify the existence of this particle on astrophysical scales. Within this context, Euclid can provide precious information on the nature of dark matter. In this chapter, we discuss the most relevant results that can be obtained with Euclid, and that can be summarised as follows:
\begin{itemize}
\item 
The discovery of an exponential suppression in the power spectrum at small scales, that
would rule out CDM and favor WDM candidates, or, in absence of it, the determination
of a lower limit on the mass of the WDM particle, $m_{{\rm WDM}}$, of $2\,\mathrm{keV}$;
\item 
the determination of an upper limit on the dark matter self-interaction cross section $\sigma/m\sim 10^{-27}\,\mathrm{cm}^2\,\mathrm{GeV}^{-1}$ at $68\%$ CL, which represents an improvement of three orders of magnitude compared to the best constraint available today, which arises from the analysis of the dynamics of the bullet cluster;
\item 
the measurement of the slope of the dark matter distribution within galaxies and clusters of galaxies with unprecedented accuracy;
\item 
the determination of the properties of the only known -- though certainly subdominant -- non-baryonic dark matter particle: the standard neutrino, for which Euclid can provide information on the absolute mass scale, its normal or inverted hierarchy, as well as its Dirac or Majorana nature;
\item 
the test of Unified Dark Matter (UDM, or quartessence) models, through the detection of characteristic oscillatory features predicted by these theories on the matter power spectrum, detectable through weak lensing or baryonic acoustic oscillations studies;
\item
a probe of the Axiverse, i.e. of the legacy of string theory through the presence of ultra-light scalar fields that can affect the growth of structure, introducing features in the matter power spectrum and modifying the growth rate of structures.
\end{itemize}
Finally, Euclid will provide, through gravitational lensing measurement, a map of the dark matter distribution over the entire extragalactic sky, allowing us to study the effect of the dark matter environment on galaxy evolution and structure formation as a function of time. This map will pinpoint our place within the dark Universe.

%% file: dark_matter/euclid_halo.tex
\section{Dark matter halo properties}

Dark matter was first proposed by \citet{1937ApJ....86..217Z} to explain the anomalously high velocity of galaxies in galaxy clusters. Since then, evidence for dark matter has been accumulating on all scales. The velocities of individual stars in dwarf galaxies suggest that these are the most dark matter dominated systems in the Universe \citep[e.g.][]{1998ARA&A..36..435M,2001ApJ...563L.115K,2007ApJ...670..313S,2007MNRAS.380..281M,2007ApJ...667L..53W}. Low surface brightness (LSB) and giant spiral galaxies rotate too fast to be supported by their stars and gas alone, indicating the presence of dark matter \citep{2001ApJ...552L..23D,2005ApJ...621..757S, 2001MNRAS.323..285B, 2002ApJ...573..597K}. Gravitationally lensed giant elliptical galaxies and galaxy clusters require dark matter to explain their observed image distributions \citep[e.g.][]{1964MNRAS.128..295R,1975ApJ...195...13B,1979Natur.279..381W,1987A&A...172L..14S,2006ApJ...648L.109C}. Finally, the temperature fluctuations in the cosmic microwave background (CMB) radiation indicate the need for dark matter in about the same amount as that required in galaxy clusters \citep[e.g.][]{1992ApJ...396L...1S,1992ApJ...396L..13W,SpergelEtAl2006}.

While the case for particle dark matter is compelling, until we find direct evidence for such a particle, astrophysics remains a unique dark matter probe. Many varieties of dark matter candidates produce a noticeable change in the growth of structure in the Universe \citep{1996PhR...267..195J,2009EPJC...59..557S}. Warm dark matter (WDM) suppresses the growth of structure in the early Universe producing a measurable effect on the small-scale matter power spectrum \citep{2001ApJ...556...93B,2001ApJ...559..516A,2001ApJ...558..482B}. Self-interacting dark matter (SIDM) changes the expected density distribution \textit{within} bound dark matter structures \citep{2001ApJ...561...35D,2000PhRvD..62f3511H}. In both cases, the key information about dark matter is contained on very small scales. In this section, we discuss previous work that has attempted to measure the small scale matter distribution in the Universe, and discuss how Euclid will revolutionise the field. We divide efforts into three main areas: measuring the halo mass function on large scales, but at high redshift; measuring the halo mass function on small scales through lens \textit{substructures}; measuring the dark matter density profile within galaxies and galaxy clusters.

\begin{figure}
\centering
\includegraphics[height=0.47\textwidth]{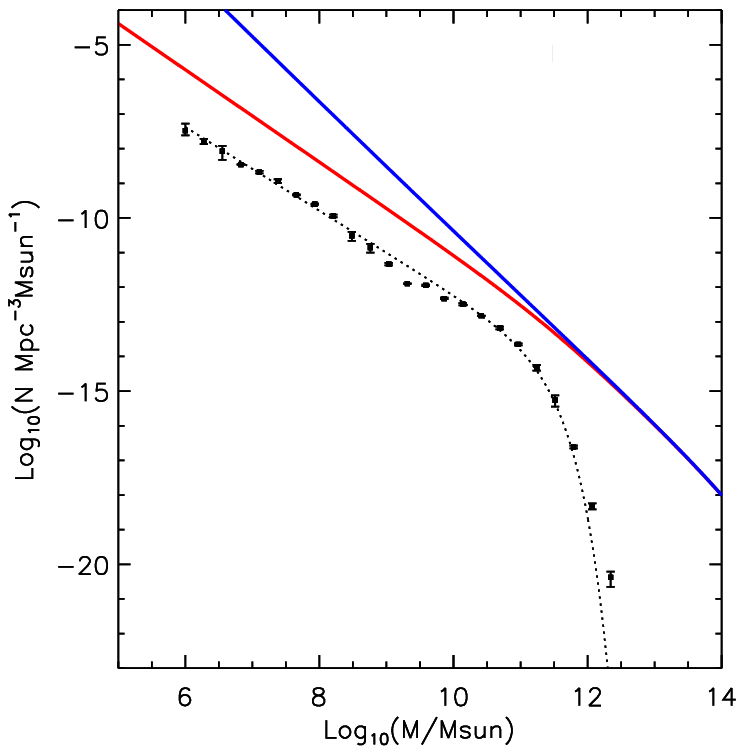}
\caption{The baryonic mass function of galaxies (data points). The dotted line shows a Schechter function fit to the data. The blue line shows the predicted mass function of dark matter haloes, assuming that dark matter is cold. The red line shows the same assuming that dark matter is warm with a (thermal relic) mass of $m_\mathrm{WDM}=1\,\mathrm{keV}$.}
\label{fig:readtrent}
\end{figure}

\subsection{The halo mass function as a function of redshift}\label{sec:matterpower}

Attempts have already been made to probe the small scale power in the Universe through galaxy counts. Fig.~\ref{fig:readtrent} shows the best measurement of the `baryonic mass function' of galaxies to date \citep{2005RSPTA.363.2693R}. This is the number of galaxies with a given total mass in baryons normalised to a volume of $1\,\mathrm{Mpc}$. To achieve this measurement,\citet{2005RSPTA.363.2693R} sewed together results from a wide range of surveys reaching a baryonic mass of just $\sim10^6\,M_\odot$ -- some of the smallest galaxies observed to date.

The baryonic mass function already turns up an interesting result. Over-plotted in blue on Fig.~\ref{fig:readtrent} is the \textit{dark matter} mass function expected assuming that dark matter is `cold' -- i.e. that it has no preferred scale. Notice that this has a different shape. On large scales, there should be bound dark matter structures with masses as large as $10^{14}\,M_\odot$, yet the number of observed galaxies drops off exponentially above a baryonic mass of $\sim 10^{12}\,M_\odot$. This discrepancy is well-understood. Such large dark matter haloes have been observed, but they no longer host a single galaxy; rather they are bound collections of galaxies -- galaxy clusters \citep[see e.g.][]{1937ApJ....86..217Z}. However, there is also a discrepancy at low masses that is not so well understood. There should be far more bound dark matter haloes than observed small galaxies. This is the well-known `missing satellite' problem \citep{1999ApJ...524L..19M,1999ApJ...522...82K}. 

The missing satellite problem could be telling us that dark matter is not cold. The red line on Fig.~\ref{fig:readtrent} shows the expected dark matter mass function for WDM with a (thermal relic) mass of $m_\mathrm{WDM}=1\,\mathrm{keV}$. Notice that this gives an excellent match to the observed slope of the baryonic mass function on small scales. However, there may be a less exotic solution. It is likely that star formation becomes inefficient in galaxies on small scales. A combination of supernovae feedback, reionisation and ram-pressure stripping is sufficient to fully explain the observed distribution assuming pure CDM \citep{2004ApJ...609..482K,2006MNRAS.371..885R,2009arXiv0903.4681M}. Such `baryon feedback' solutions to the missing satellite problem are also supported by recent measurements of the orbits of the Milky Way's dwarf galaxies \citep{2010arXiv1001.1731L}.

\subsubsection{Weak and strong lensing measurements of the halo mass function} 
To make further progress on WDM constraints from astrophysics, we must avoid the issue of baryonic physics by probing the halo mass function \textit{directly}. The only tool for achieving this is gravitational lensing. In weak lensing this means stacking data for a very large number of galaxies to obtain an averaged mass function. In strong lensing, this means simply finding enough systems with `good data.' Good data ideally means multiple sources with wide redshift separation \citep{2009ApJ...690..154S}; combining independent data from dynamics with lensing may also prove a promising route \citep[see e.g.][]{2002MNRAS.337L...6T}. 

Euclid will measure the halo mass function down to $\sim 10^{13}M_{\odot}$ using weak lensing. It will simultaneously find $1000$'s of strong lensing systems. However, in both cases, the lowest mass scale is limited by the lensing critical density. This limits us to probing down to a halo mass of $\sim 10^{11}\,M_\odot$ which gives poor constraints on the nature of dark matter. However, if such measurements can be made as a \textit{function of redshift}, the constraints improve dramatically. We discuss this in the next Section. 

\subsubsection{The advantage of going to high redshift}\label{sec:redhigh}
\begin{figure}
\centering
\includegraphics[height=0.47\textwidth]{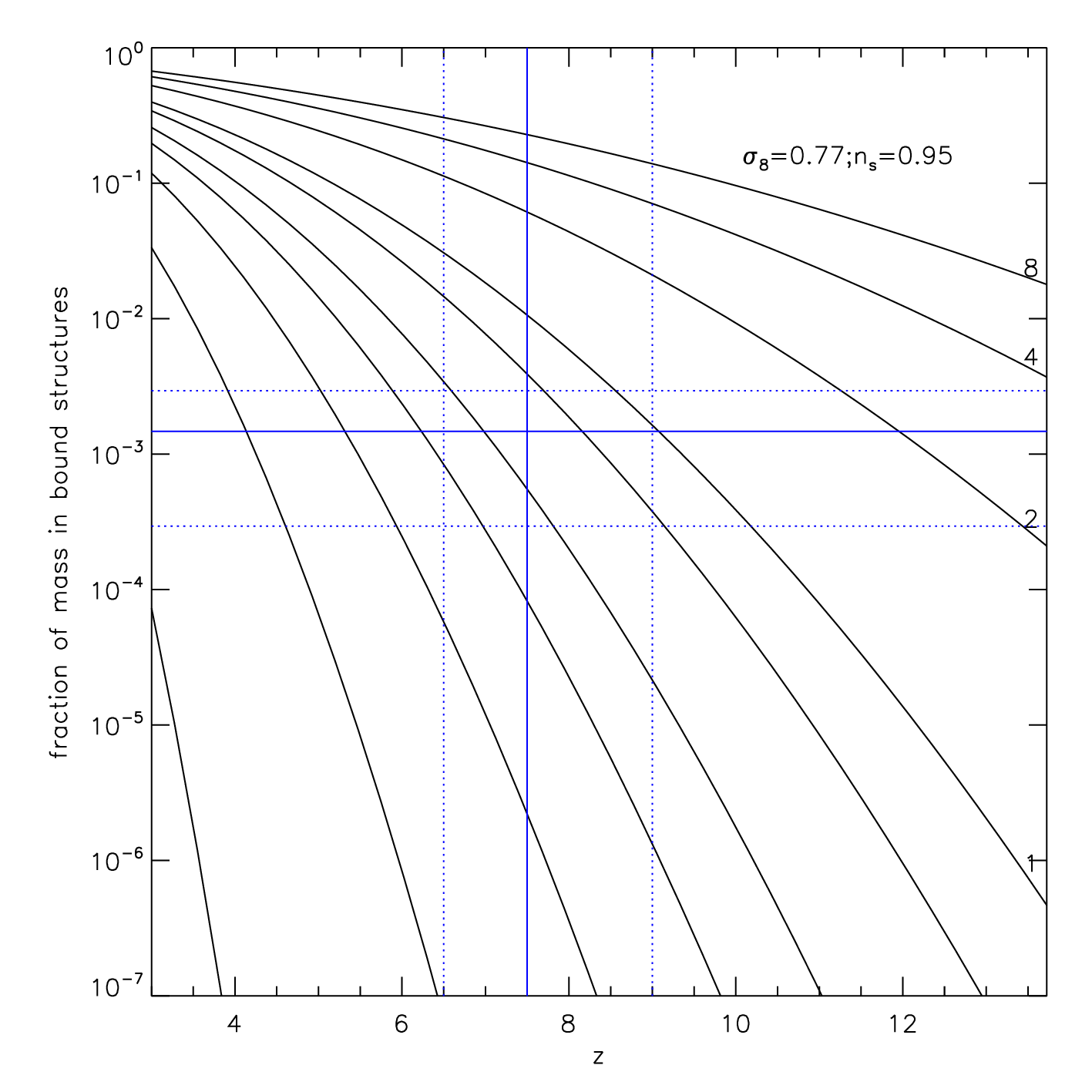}
\caption{The fraction of mass in bound structures as a function of redshift, normalised to a halo of Milky Way's mass at redshift $z=0$. Marked are different masses of thermal-relic WDM particles in $\mathrm{keV}$ (black solid lines). Notice that the differences between different WDM models increases towards higher redshift.}
\label{fig:fractionbound}
\end{figure}

Dark matter constraints from the halo mass function become much stronger if the halo mass function is measured as a function of redshift. This is because warm dark matter \textit{delays} the growth of structure formation as well as suppressing small scale power. This is illustrated in Fig.~\ref{fig:fractionbound}, which shows the fraction of mass in bound structures as a function of redshift, normalised to a halo of Milky Way's mass at redshift $z=0$. Marked are different thermal relic WDM particle masses in $keV$ (black solid lines). Notice that the differences between WDM models increase significantly towards higher redshift at a given mass scale. Thus we can obtain strong constraints on the nature of dark matter by moving to higher $z$'s, rather than lower halo mass.

The utility of redshift information was illustrated recently by observations of the Lyman-$\alpha$ absorption spectra from Quasars \citep{2008PhRvL.100d1304V,2006PhRvL..97s1303S}. Quasars act as cosmic `flashlights' shining light from the very distant Universe. Some of this light is absorbed by intervening neutral gas leading to absorption features in the Quasar spectra. Such features contain rich information about the matter distribution in the Universe at high redshift. Thus, the Lyman-$\alpha$ forest measurements have been able to place a lower bound of $m_\mathrm{WDM}>4\,\mathrm{keV}$ probing scales of $\sim1\,\mathrm{Mpc}$. Key to the success of this measurement is that much of the neutral gas lies in-between galaxies in filaments. Thus, linear approximations for the growth of structures in WDM versus CDM remain acceptable, while assuming that the baryons are a good tracer of the underlying matter field is also a good approximation. However, improving on these early results means probing smaller scales where non-linearities and baryon physics will creep in. For this reason, tighter bounds must come from techniques that either probe even higher redshifts, or even smaller scales. Lensing from Euclid is an excellent candidate since it will achieve both while measuring the halo mass function directly rather than through the visible baryons.

\subsection{The dark matter density profile}\label{sec:darkdist}
\begin{figure}
\centering
\includegraphics[height=0.47\textwidth]{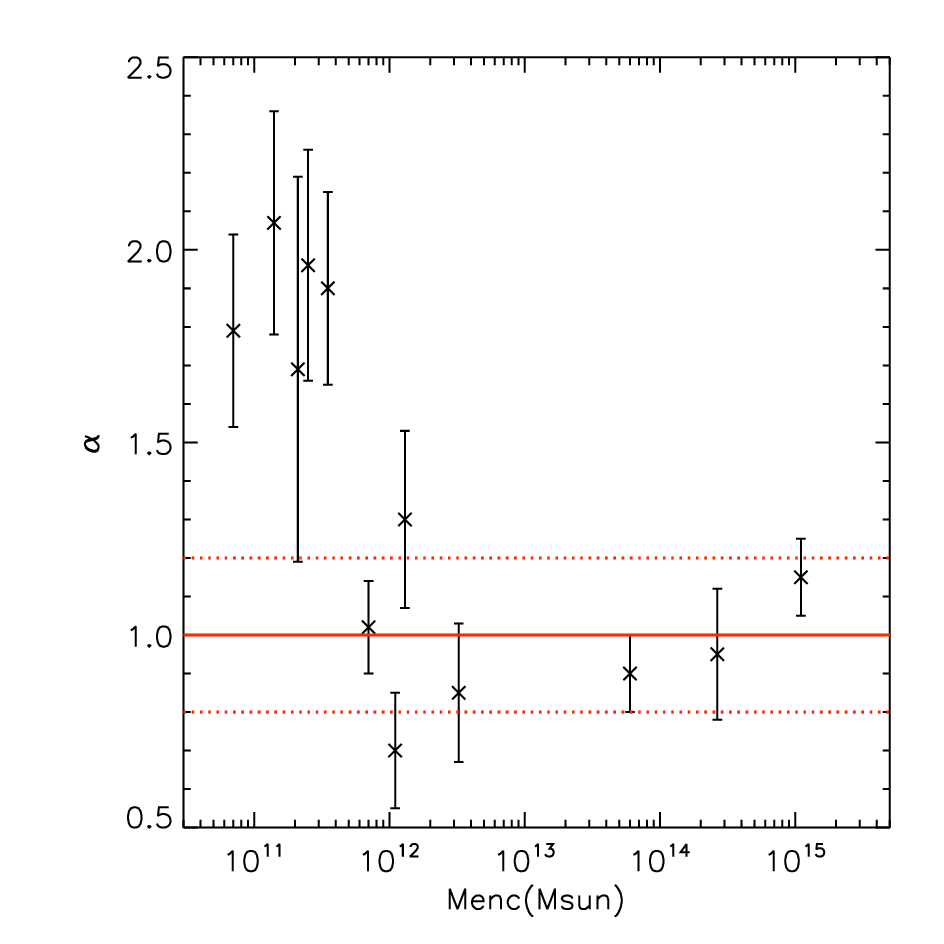}
\caption{The central log-slope {\color{red}$\alpha$} of the density distribution {\color{red}$\rho \propto r^\alpha$} for $9$ galaxies/groups and $3$ lensing clusters as a function of the enclosed lensing mass. Marked in red is the prediction from structure formation simulations of the standard cosmological model, that assume non-relativistic CDM, and that do not include any baryonic matter.}
\label{fig:cuspslope}
\end{figure}

An alternative approach to constraining dark matter models is to measure the distribution of dark matter \textit{within} galaxies. Fig.~\ref{fig:cuspslope} shows the central log-slope of the density distribution for $9$ galaxies/groups and $3$ lensing clusters as a function of the enclosed lensing mass \citep{2006ApJ...652L...5S,2007arXiv0704.3267R,2009ApJ...690..154S}. {\color{red}Over the visible region of galaxies, the dark matter distribution tends towards a single power law: $\rho \propto r^\alpha$.} Marked in red is the prediction from structure-formation simulations of the standard cosmological model, that assume non-relativistic CDM, and that do not include any baryonic matter. Notice that above an enclosed lensing mass of $\sim 10^{12}\,M_\odot$, the agreement between theory and observations is very good. This lends support to the idea that dark matter is cold and not strongly self-interacting. However, this result is based on only a handful of galaxy clusters with excellent data. Furthermore, lower mass galaxies and groups can, in principle, give tighter constraints. In these mass ranges, however ($M_\mathrm{enc}<10^{12}\,M_\odot$), the lensing mass is dominated by the visible stars. Determining the underlying dark matter distribution is then much more difficult. It is likely that the dark matter distribution is also altered from simple predictions by the dynamical interplay between the stars, gas and dark matter during galaxy formation \citep[e.g.][]{2007arXiv0707.0737D}.

%% file: dark_matter/euclid_dm_xray_olehost.tex
\section{Euclid dark matter studies: wide-field X-Ray complementarity}

The predominant extragalactic X-ray sources are AGNs and galaxy clusters. For dark matter studies the latter are the more interesting targets. X-rays from clusters are emitted as thermal bremsstrahlung by the hot intracluster medium (ICM) which contains most of the baryons in the cluster. The thermal pressure of the ICM supports it against gravitational collapse so that measuring the temperature through X-ray observations provides information about the mass of the cluster and its distribution. Hence X-rays form a complementary probe of the dark matter in clusters to Euclid weak lensing measurements.

The ongoing X-ray missions XMM-Newton and Chandra have good enough angular resolution to measure the temperature and mass profiles in $\sim10$ radial bins for clusters at reasonable redshifts, although this requires long exposures. Many planned X-ray missions aim to improve the spectral coverage, spectral resolution, and/or collection area of the present mission, but they are nonetheless mostly suited for targeted observations of individual objects. Two notable exceptions are eROSITA\footnote{http://www.mpe.mpg.de/erosita/}\citep[][launch 2012]{Cappelluti:2010ay} and the Wide Field X-ray Telescope\footnote{http://www.wfxt.eu/home/Overview.html} \citep[WFXT][proposed] {2009astro2010S..90G, Vikhlinin:2009xb, 2010MNRAS.407.2339S, 2011MSAIS..17....3R, 2011MSAIS..17...36B, 2011arXiv1112.0327S} which will both conduct full sky surveys and, in the case of WFXT, also smaller but deeper surveys of large fractions of the sky. 

A sample of high-angular resolution X-ray cluster observations can be used to test the prediction from $N$-body simulations of structure formation that dark matter haloes are described by the NFW profile \citep{1996ApJ...462..563N} with a concentration parameter $c$. This describes the steepness of the profile, which is related to the mass of the halo \citep{ 2007MNRAS.381.1450N}. Weak or strong lensing measurements of the mass profile, such as those that will be provided from Euclid, can supplement the X-ray measurement and have different systematics. Euclid could provide wide field weak lensing data for such a purpose with very good point spread function (PSF) properties, but it is likely that the depth of the Euclid survey will make dedicated deep field observations a better choice for a lensing counterpart to the X-ray observations. However, if the WFXT mission becomes a reality, the sheer number of detected clusters with mass profiles would mean Euclid could play a much more important r\^ole.

X-ray observations of galaxy clusters can constrain cosmology by measuring the geometry of the Universe through the baryon fraction $f_\mathrm{gas}$ \citep{2008MNRAS.383..879A} or by measuring the growth of structures by determining the high-mass tail of the mass function \citep{2010MNRAS.406.1759M}. The latter method would make the most of the large number of clusters detected in full-sky surveys and there would be several benefits by combining an X-ray and a lensing survey. It is not immediately clear which type of survey would be able to better detect clusters at various redshifts and masses, and the combination of the two probes could improve understanding of the sample completeness. An X-ray survey alone cannot measure cluster masses with the required precision for cosmology. Instead, it requires a calibrated relation between the X-ray temperature and the cluster mass. Such a calibration, derived from a large sample of clusters, could be provided by Euclid. In any case, it is not clear yet whether the large size of a Euclid sample would be more beneficial than deeper observations of fewer clusters.

Finally X-ray observations can also confirm the nature of possible 'bullet-like' merging clusters. In such systems the shock of the collision has displaced the ICM from the dark matter mass, which is identified through gravitational lensing. This offers the opportunity to study dark matter haloes with very few baryons and, e.g. search for signatures of decaying or annihilating dark matter.

%% file: dark_matter/DMmapping.tex
\section{Dark matter mapping}
Gravitational lensing offers a unique way to chart dark matter structures in the Universe as it is sensitive to all forms of matter. Weak lensing has been used to map the dark matter in galaxy clusters \citep[see for example][]{2006A&A...451..395C} with high resolution reconstructions recovered for the most massive strong lensing clusters \citep[see for example][]{2006ApJ...652..937B}. Several lensing studies have also mapped the projected surface mass density over degree scale-fields \citep{2007A&A...462..459G,2007A&A...462..875S,2009ApJ...702..980K} to identify shear-selected groups and clusters. The minimum mass scale that can be identified is limited only by the intrinsic ellipticity noise in the lensing analysis and projection effects. Using a higher number density of galaxies in the shear measurement reduces this noise, and for this reason the Deep Field Euclid Survey will be truly unique for this area of research, permitting high resolution  reconstructions of dark matter in the field \citep{2007Natur.445..286M,2008MNRAS.385.1431H} and the study of lenses at higher redshift.

There are several non-parametric methods to reconstruct dark matter in 2D which can be broadly split into two {\color{red}categories}: convergence techniques \citep{1993ApJ...404..441K} and potential techniques \citep{1996ApJ...464L.115B}. In the former one measures the projected surface mass density (or convergence) $\kappa$ directly by applying a convolution to the measured shear under the assumption that $\kappa\ll1$. Potential techniques perform a $\chi^2$ minimisation and are better suited to the cluster regime and can also incorporate strong lensing information \citep{2005A&A...437...39B}. In the majority of methods, choices need to be made about smoothing scales to optimise signal-to-noise whilst preserving reconstruction resolution. Using a wavelet method circumvents this choice \citep{2006A&A...451.1139S, 2008ApJ...684..794K} but makes the resulting significance of the reconstruction difficult to measure.

\subsection{Charting the universe in 3D}
The lensing distortion depends on the total projected surface mass density along the line of sight and a geometrical factor that increases with source distance. This redshift dependence can be used to recover the full 3D gravitational potential of the matter density as described in \citet{2002PhRvD..66f3506H,2003MNRAS.344.1307B} and applied to the COMBO-17 survey in \citet{2004MNRAS.353.1176T} and the COSMOS survey in \citet{2007Natur.445..286M}. This work has been extended in \citet{2009MNRAS.399...48S} to reconstruct the full 3D mass density field and applied to the STAGES survey in  \citet{2011arXiv1109.0932S}.

All 3D mass reconstruction methods require the use of a prior based on the expected mean growth of matter density fluctuations. Without the inclusion of such a prior, \citet{2002PhRvD..66f3506H} have shown that one is unable to reasonably constrain the radial matter distribution, even for densely sampled space-based quality lensing data. Therefore 3D maps cannot be directly used to infer cosmological parameters.

The driving motivation behind the development of 3D reconstruction techniques was to enable an unbiased 3D comparison of mass and light. Dark haloes for example would only be detected in this manner. However the detailed analysis of noise and the radial PSF in the 3D lensing reconstructions presented for the first time in \citet{2011arXiv1109.0932S} show how inherently noisy the process is. Given the limitations of the method to resolve only the most massive structures in 3D the future direction of the application of this method for the Euclid Wide survey should be to reconstruct large scale structures in the 3D density field. Using more heavily spatially smoothed data we can expect higher quality 3D resolution reconstructions as on degree scales the significance of modes in a 3D mass density reconstruction are increased \citep{2009MNRAS.399...48S}. Adding additional information from flexion may also improve mass reconstruction, although using flexion information alone is much less sensitive than shear \citep{2010ApJ...723.1507P}.

%% file: dark_matter/Scattering.tex
\section{Scattering cross sections}
We now move towards discussing the particulate aspects of dark matter, starting with a discussion on the scattering cross-sections of dark matter. At present, many physical properties of the dark matter particle remain highly uncertain. Prospects for studying the scattering of dark matter with each of the three major constituents of the Universe -- itself, baryons, and dark energy -- are outlined below.  

\subsection{Dark matter - dark matter interactions}
Self-interacting dark matter (SIDM) was first postulated by \citet{spergel-steinhardt}, in an attempt to explain the apparent paucity of low-mass haloes within the Local Group. The key characteristic of this model is that CDM particles possess a large scattering cross-section, yet with negligible annihilation or dissipation. The process of elastic scattering erases small substructures and cuspy cores, whilst preserving the density profile of the haloes.

However, as highlighted by \citet{ostriker-cross}, cross-sections large enough to alleviate the structure formation issues would also allow significant heat transfer from particles within a large halo to the cooler sub-haloes. This effect is most prominent close to the centres of clusters. As the sub-halo evaporates, the galaxy residing within the halo would be disrupted. Limiting this rate of evaporation to exceed the Hubble time allows an upper bound to be placed on the scattering cross-section of approximately {\color{red}$\sigma_p/m_p\lesssim0.3\,\mathrm{cm^2\,g^{-1}}$} (neglecting any velocity dependence). Note the dependence on particle mass -- a more massive CDM particle would be associated with a lower number density, thereby reducing the frequency of collisions.

\citet{2001MNRAS.325..435M} have performed ray-tracing through $N$-body simulations, and have discovered that the ability for galaxy clusters to generate giant arcs from strong lensing is compromised if the dark matter is subject to just a few collisions per particle. This constraint translates to an upper bound {$\sigma_p/m_p\lesssim0.1\,\mathrm{cm^2\,g^{-1}}$}. Furthermore, more recent analyses of SIDM models \citep{mark-bullet,randall2008} utilise data from the Bullet Cluster to provide another independent limit on the scattering cross section, though the upper bound remains unchanged. \citet{massey-bulleticity} have proposed that the tendency for baryonic and dark matter to become separated within dynamical systems, as seen in the Bullet Cluster, could be studied in greater detail if the analysis were to be extended over the full sky in Euclid. This concept is explored in further detail in the following section.

How do these cosmological constraints relate to the values anticipated by particle physics? WIMPs are expected to fall in the range of $10\,\mathrm{GeV}$ to a few $\mathrm{TeV}$. The aforementioned values would then correspond to around $\sigma_p\lesssim10^{-24}\,\mathrm{cm^2}$, at least twenty order of magnitudes greater than what one might expect to achieve from neutral current interactions. Therefore in a cosmological context WIMPs are essentially collisionless, as are axions, since they exhibit an even smaller cross section. Any cosmological detection of SIDM would thus point towards the more exotic candidates postulated by particle physicists, particularly those which are not point particles but instead comprise of extended objects such as Q-balls. A measurement of the scattering cross-section would also place an upper bound on the mass of the dark matter particle, since unitarity of the scattering matrix forbids extremely large cross sections \citep{PhysRevLett.86.3467}, i.e.
\begin{equation}
\sigma_\mathrm{tot}\leq1.76\times10^{-17}\,\mathrm{cm^2}\left(\frac{\mathrm{GeV}}{m_\chi}\right)^2\left(\frac{10\,\mathrm{km\,s^{-1}}}{v_\mathrm{rel}}\right)^2
\end{equation}

\subsection{Dark matter - baryonic interactions}
Currently, a number of efforts are underway to directly detect WIMPs via the recoil of atomic nuclei. The underground experiments such as CDMS, CRESST, XENON, EDELWEISS and ZEPLIN have pushed observational limits for the spin-independent WIMP-nucleon cross-section down to the $\sigma\lesssim10^{-43}\mathrm{cm}^2$ r\'egime.\footnote{\textbf{It is anyway worth noticing the controversial results of DAMA/LIBRA, and more recently of CoGeNT.}} A collection of the latest constraints can be found at \texttt{http://dmtools.brown.edu}.

Another opportunity to unearth the dark matter particle lies {\color{red}in accelerators such as the LHC}. By 2018 it is possible these experiments will have yielded mass estimates for dark matter candidates, provided its mass is lighter than a few hundred $\mathrm{GeV}$. However, the discovery of more detailed properties of the particle, which are essential to confirm the link to cosmological dark matter, would have to wait until the International Linear Collider is constructed.

\subsection{Dark matter - dark energy interactions}
Interactions in the dark sector have provided a popular topic for exploration, with a view to building models which alleviate the coincidence and fine-tuning issues associated with dark energy (see Sec.~\ref{mg:cde}). The great uncertainty surrounding the physical nature of dark energy leaves plenty of scope for non-gravitational physics to play a r\^ole. These models are discussed at length in other sections of this review (\ref{models-of-modified-gravity} {\color{red}and} \ref{dms:de_dm}). Here, we only mention that \citet{simpscat} have explored the phenomenology associated with dark matter scattering elastically with dark energy. The growth rate of large-scale structures is artificially slowed, allowing a modest constraint of 
\begin{equation}
{\color{red}\sigma_p/m_p}\lesssim\frac{10}{1+w}\,\mathrm{cm^2\,g^{-1}} \,\,\, .
\end{equation}

It is clear that such dark sector interactions do not arise in the simplest models of dark matter and dark energy. However a rigorous refutation of General Relativity will require not only a robust measure of the growth of cosmic structures, but confirmation that the anomalous dynamics are not simply due to physics within the dark sector.

%% file: dark_matter/ClusterCrossSection.tex
\section{Cross section constraints from galaxy clusters}
Clusters of galaxies present an interesting environment in which the dark matter density is high and where processes such as collisions present the possibility of distinguishing dark matter from baryonic matter as the two components interact differently. For instance, particulate dark matter and baryonic matter may be temporarily separated during collisions between galaxy clusters, such as 1E~0657-56 \citep{2006ApJ...648L.109C,2006ApJ...652..937B} and MACSJ0025-12 \citep{BradacEtAl2008}. These `bullet clusters' have provided astrophysical constraints on the interaction cross-section of hypothesised dark matter particles \citep{randall2008}, and may ultimately prove the most useful laboratory in which to test for any velocity dependence of the cross-section. Unfortunately, the contribution of individual systems is limited by uncertainties in the collision velocity, impact parameter and angle with respect to the plane of the sky. Current constraints are three orders of magnitude weaker than constraints from the shapes of haloes \citep{feng10} and, since collisions between two massive progenitors are rare \citep{2010MNRAS.406.1134S,2010MNRAS.408.1277S}, the total observable number of such systems may be inadequate to investigate a physically interesting regime of dark matter properties.

Current constraints from bullet clusters on the cross-section of particulate dark matter are $\sim18$ orders of magnitude larger than that required to distinguish between plausible particle-physics dark matter candidates (for example from supersymmetric extensions to the standard model). In order to investigate a physically interesting r\'egime of dark matter cross-section, and provide smaller error bars, many more individual bullet clusters are required. However collisions between two massive progenitors are rare and ultimately the total observable number of such systems may be inadequate. 

\subsection{Bulleticity}\label{symbol:bulleticity}
In \citet{2011MNRAS.413.1709M}, a method for using every individual infalling substructure in every cluster has been proposed. For each piece of infalling substructure, a local vector from the dark matter peak (identified using weak lensing analysis) and the baryonic mass peak (from X-rays) -- dubbed `bulleticity' -- can be defined
\begin{equation}
\mathbf b=b_r\er+b_t\et,
\end{equation}
where the radial $b_r$ and azimuthal $b_t$ components are defined relative to unit vector towards the cluster centre and tangentially and $b=|\mathbf b|$. An integrated bulleticity signal of zero would imply an equal cross sections for the dark matter and baryonic matter. By measuring the amplitude of the bulleticity one can empirically measure the ratio between the dark matter and baryonic cross sections.

\begin{figure}
\centering
\includegraphics[width = 0.3\columnwidth]{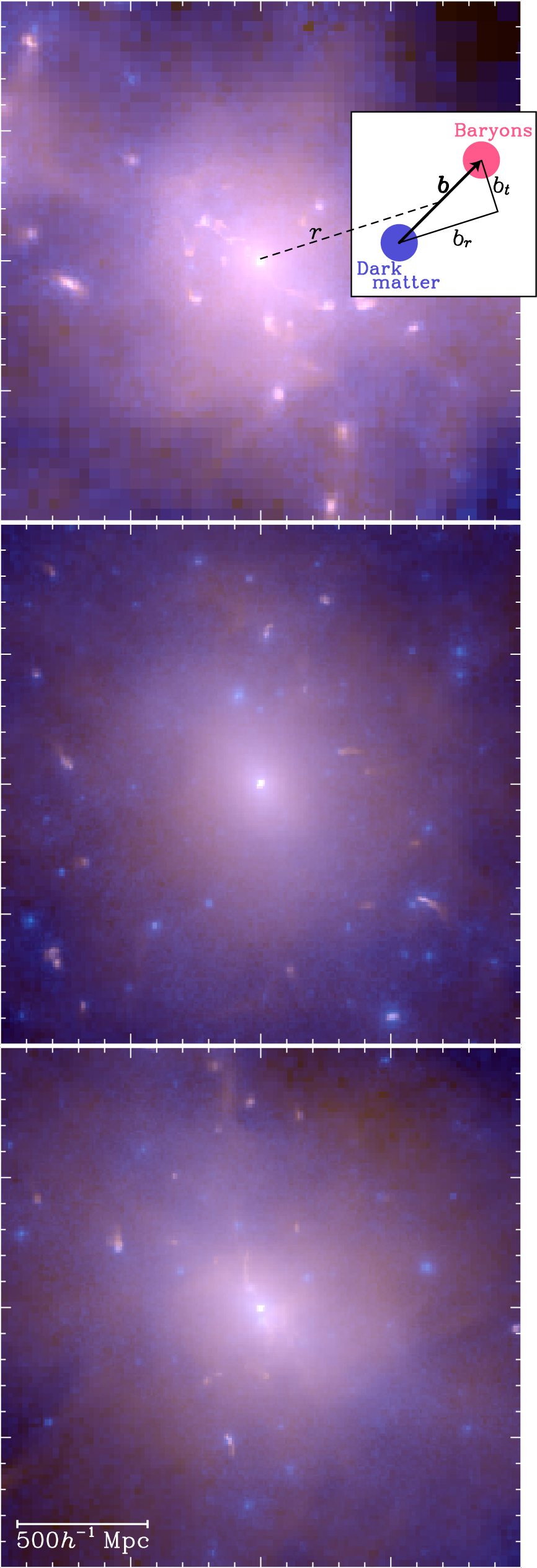}
\caption{Full hydrodynamical simulations of massive clusters at redshift $z=0.6$. Total projected mass is shown in blue, while X-ray emission from baryonic gas is in red. The preferential trailing of gas due to pressure from the ICM, and its consequent separation from the non interacting dark matter, is apparent in much of the infalling substructure.}\label{fig:hydro_images}
\end{figure}

In Fig.~\ref{fig:hydro_images} a result from full hydrodynamical simulations of dark and baryonic matter within clusters in shown. \citet{2011MNRAS.413.1709M} have used these simulations to show that the measurement of a net bulleticity consistent with the cold dark matter used in the simulations is possible. 

Finally a Fisher matrix calculation has shown that, under the assumption that systematic effects can be controlled, Euclid could use such a technique to constrain the relative particulate cross-sections to $6\times10^{-27}\,\mathrm{cm^2\,GeV^{-1}}$.

The raw bulleticity measurement would constrain the relative cross-sections of the baryon-baryon interaction and the dark matter-dark matter interaction. However, since we know the baryonic cross-section relatively well, we can infer the dark matter-dark matter cross-section. The dark matter-dark matter interaction probed by Euclid using this technique will be complementary to the interactions constrained by direct detection and accelerator experiments where the primary constraints will be on the dark matter-baryon interaction.

%% file: dark_matter/WDM.tex
\section{Constraints on warm dark matter} 
$N$-body simulations of large-scale structures that assume a $\Lambda$CDM cosmology appear to over-predict the power on small scales when compared to observations \citep{Primack:2009jr}: `the missing-satellite problem' \citep{Kauffmann:1993gv,Klypin:1999uc,Strigari:2007ma,Bullock:2010uy}, the `cusp-core problem' \citep{Li:2009mp,2005ApJ...621..757S, Zavala:2009ms} and sizes of mini-voids \citep{Tikhonov:2009jq}. These problems may be more or less solved by several different phenomena \citep[e.g.][]{Diemand:2009bm}, however one which could explain all of the above is Warm Dark Matter (WDM) \citep{2001ApJ...556...93B, Colin:2000dn, Boyanovsky:2007ay}. If the dark matter particle  is very light, it can cause a suppression of the growth of structures on small scales via free-streaming of the dark matter particles whilst relativistic in the early universe.

\subsection{Warm dark matter particle candidates}
Numerous WDM particle models can be constructed, but there are two that occur most commonly in literature, because they are most plausible from particle physics theory as well as from cosmological observations:
\begin{itemize}
\item Sterile neutrinos may be constructed to extend the standard model of particle physics. The standard model active (left-handed) neutrinos can then receive the observed small masses through e.g. a see-saw mechanism. This implies that right-handed sterile neutrinos must be rather heavy, but the lightest of them naturally has a mass in the $\mathrm{keV}$ region, which makes it a suitable WDM candidate. The simplest model of sterile neutrinos as WDM candidate assumes that these particles were produced at the same time as active neutrinos, but they never thermalised and were thus produced with a much reduced abundance due to their weak coupling \citep[see][and references therein]{Biermann:2007ap}.
\item The gravitino appears as the supersymmetric partner of the graviton in supergravity models. If it has a mass in the $\mathrm{keV}$ range, it will be a suitable WDM candidate. It belongs to a more general class of \textit{thermalised} WDM candidates. It is assumed that this class of particles achieved a full thermal equilibrium, but at an earlier stage, when the number of degrees of freedom was much higher and hence their relative temperature with respect to the CMB is much reduced. Note that in order for the gravitino to be a good dark matter particle in general, it must be very stable, which in most models corresponds to it being the LSP \citep[e.g.][]{Borgani:1997ds,Cembranos2005}.
\end{itemize}
Other possible WDM candidates exist, for example a non-thermal neutralino \citep{Hisano:2000dz} or a non-thermal gravitino \citep{Baltz:2001rq} etc.

\subsection{Dark matter free-streaming}
The modification of the shape of the linear-theory power spectrum of CDM due to WDM can be calculated by multiplication by a transfer function \citep{2001ApJ...556...93B}
\begin{equation}
T(k)\equiv\sqrt{\frac{P_\mathrm{WDM}(k)}{P_\mathrm{CDM}(k)}}=\left[1+(\alpha k)^{2\mu}\right]^{-5/\mu},
\end{equation}
with suitable parameter $\mu=1.12$ \citep{Viel:2005qj} and with the scale break {\color{red}parameter, $\alpha$} in the case of thermal relic DM
\begin{equation}
\alpha=0.049\left(\frac{m_\mathrm{WDM}}{\mathrm{keV}}\right)^{-1.11}\left(\frac{\Omega_\mathrm{WDM}}{0.25}
\right)^{0.11}\left(\frac{h}{0.7}\right)^{1.22}\,h^{-1}\,\mathrm{Mpc}.
\end{equation}
This is a fit to the solution of the full Boltzman equations.

There is a one-to-one relation between the mass of the thermalised WDM particle $m_\mathrm{WDM}$ (e.g. gravitino), and the mass of the simplest sterile neutrino $m_\mathrm{\nu s}$, such that the two models have an identical impact on cosmology \citep{Viel:2005qj}
\begin{equation}
m_\mathrm{\nu s}=4.43\left(\frac{m_\mathrm{WDM}}{\mathrm{keV}}\right)^{4/3}\left(\frac{\omega_\mathrm{WDM}}{0.1225}\right)^{-1/3}\mathrm{keV}.
\end{equation}
where $\omega=\Omega h^2$. The difference comes from the fact that in the gravitino case the particle is fully thermalised, the number of effective degrees of freedom being determined by mass and energy density of dark matter, while in the simplest sterile neutrino case the number of degrees of freedom is fixed, while abundance is determined by mass and energy density of dark matter.

In order to extrapolate the matter power spectrum to later times one must take into account the non-linear evolution of the matter density field. This is not straightforward in the WDM case \citep{Markovic:2010te} and most likely needs to be explored through further simulations \citep{Zavala:2009ms}.
 
\subsection{Current constraints on the WDM particle from large-scale structure}
Measurements in the particle-physics energy domain can only reach masses uninteresting in the WDM context, since direct {\color{red}detectors} look mainly for a WIMP, whose mass should be in the $\mathrm{GeV}$-$\mathrm{TeV}$ range. However, as described above, cosmological observations are able to place constraints on light dark matter particles. Observation of the flux power spectrum of the Lyman-$\alpha$ forest, which can indirectly measure the fluctuations in the dark matter density on scales between $\sim100\,\mathrm{kpc}$ and $\sim10\,\mathrm{Mpc}$ gives the limits of $m_\mathrm{WDM}>4\,\mathrm{keV}$ or equivalently $m_\mathrm{\nu s}>28\,\mathrm{keV}$ at $95\%$ confidence level \citep{Viel:2007mv,Viel:2005qj,Seljak:2006qw}. For the simplest sterile neutrino model, these lower limits are at odds with the upper limits derived from X-ray observations, which come from the lack of observed diffuse X-ray background from sterile neutrino annihilation and set the limit $m_\mathrm{\nu s}<1.8\,\mathrm{keV}$ at the $95\%$ confidence limit \citep{Boyarsky:2006jm}. However, these results do not rule the simplest sterile neutrino models out. There exist theoretical means of evading small-scale power constraints \citep[see e.g.][and references therein]{Boyarsky:2008mt}. The weak lensing power spectrum from Euclid will be able to constrain the dark matter particle mass to about $m_\mathrm{WDM}>2\,\mathrm{keV}$ \citep{Markovic:2010te}.

%% file: dark_matter/textforeuclid_massnu.tex
\section{Neutrino properties}
The first significant evidence for a finite neutrino mass \citep{SuperK} indicated the incompleteness of the standard model of particle physics. Subsequent experiments have further strengthened this evidence and improved the determination of the neutrino mass splitting required to explain observations of neutrino oscillations.

As a summary of the last decade of neutrino experiments, two hierarchical neutrino mass splittings and {\color{red}three mixing angles have been measured}. Furthermore, the standard model has three neutrinos: the motivation for considering deviations from the standard model in the form of extra sterile neutrinos has disappeared \citep{mena,miniboone}. Of course, deviations from the standard effective numbers of neutrino species could still indicate exotic physics which we will discuss below (\S~\ref{sec:Neff}).

\begin{figure}
\centering
\includegraphics[width=.75\textwidth]{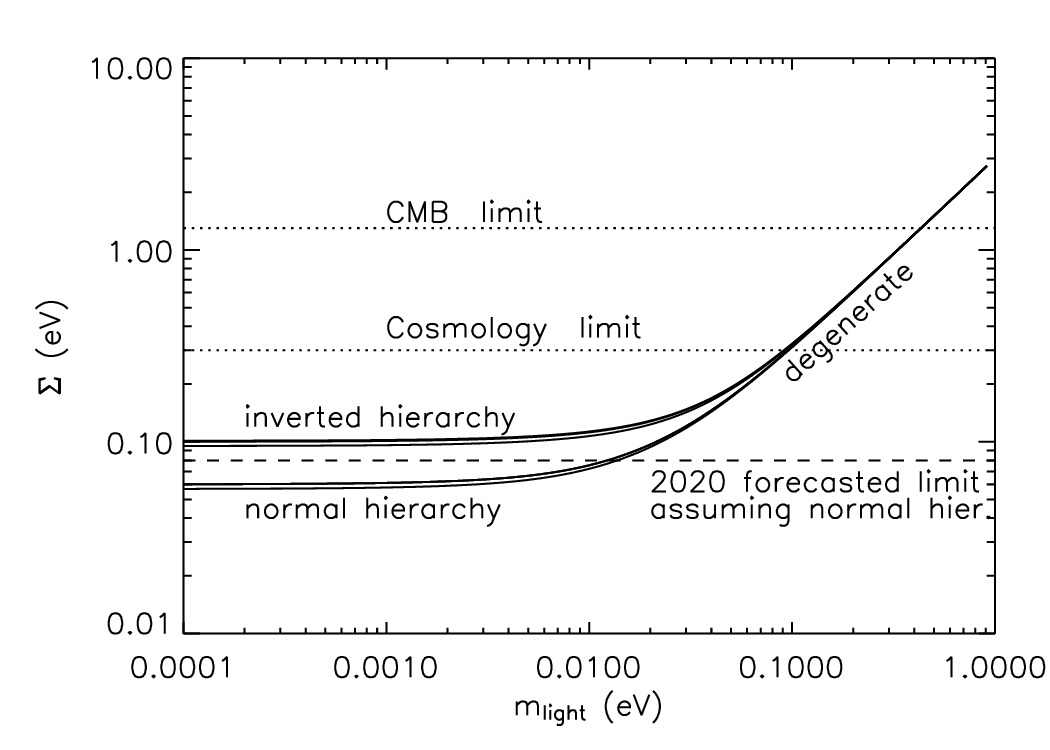}
\caption{Constraints from neutrino oscillations and from cosmology in the $m$-$\Sigma$ plane. From \citet{Jimenez:2010ev}.}\label{fig:hierarchy-massnu}
\end{figure}

New and future neutrino experiments aim to determine the remaining parameters of the neutrino mass matrix and the nature of the neutrino mass. Within three families of neutrinos, and given all neutrino oscillation data, there are three possible mass spectra: a) degenerate, with mass splitting smaller than the neutrino masses, and two non-degenerate cases, b) normal hierarchy (NH), with the larger mass splitting between the two more massive neutrinos and c) inverted hierarchy (IH), with the smaller spitting between the two higher mass neutrinos. Fig.~\ref{fig:hierarchy-massnu} \citep{Jimenez:2010ev} illustrates the currently allowed regions in the plane of total neutrino mass, $\Sigma$, \label{symbol:Sigma} vs. mass of the lightest neutrino, $m$. Note that a determination of $\Sigma<0.1\,\mathrm{eV}$ would indicate normal hierarchy and that there is an expected minimum  mass $\Sigma>0.054\,\mathrm{eV}$. The cosmological constraint is from \citet{Reid/etal:2010}.

Cosmological constraints on neutrino properties are highly complementary to particle physics experiments for several reasons:
\begin{itemize}
\item \textbf{Relic neutrinos}  produced in the early Universe are hardly detectable by weak interactions, making it impossible with foreseeable technology to detect them directly. But new cosmological probes such as Euclid offer the opportunity to detect (albeit indirectly) relic neutrinos, through the effect of their mass on the growth of cosmological perturbations.
\item \textbf{Cosmology remains a key avenue to determine the absolute neutrino mass scale}. Particle physics experiments will be able to place lower limits on the \textit{effective} neutrino mass, which depends on the hierarchy, with no rigorous limit achievable in the case of normal hierarchy \citep{murayama}. Contrarily, neutrino free streaming suppresses the small-scale clustering of large-scale cosmological structures by an amount that depends on neutrino mass.
\item \textbf{``What is the hierarchy (normal, inverted or degenerate)?''} Neutrino oscillation data are unable to resolve whether the mass spectrum consists in two light states with mass $m$ and a heavy one with mass $M$ -- normal hierarchy -- or two heavy states with mass $M$ and a light one with mass $m$ -- inverted hierarchy -- in a model-independent way. Cosmological observations, such as the data provided by Euclid, can determine the hierarchy, {\color{red} complementarily to data from particle physics experiments}.
\item \textbf{``Are neutrinos their own anti-particle?''} If the answer is yes, then neutrinos are Majorana fermions; if not, they are Dirac. If neutrinos and anti-neutrinos are identical, there could have been a process in the early Universe that affected the balance between particles and anti-particles, leading to the matter anti-matter asymmetry we need to exist \citep{leptogenesis}. This question can, in principle, be resolved if neutrino-less double-$\beta$ decay is observed \citep[see][and references therein]{murayama}. However, if such experiments \citep[on going and planned e.g.][]{Cremonesi:2010} lead to a negative result, the implications for the nature of neutrinos depend on the hierarchy. As shown in \citet{Jimenez:2010ev}, in this case cosmology can offer complementary information by  helping determine the hierarchy. 
\end{itemize}

\subsection{Evidence of relic neutrinos}\label{sec:relictnu}
The hot big bang model predicts a background of relic neutrinos in the Universe with an average number density of $\sim100N_{\nu}\,\mathrm{cm}^{-3}$, where $N_{\nu}$ is the number of neutrino species. These neutrinos decouple from the CMB at redshift $z\sim10^{10}$ when the temperature was $T\sim o(\mathrm{MeV})$, but remain relativistic down to much lower redshifts depending on their mass. A detection of such a neutrino background would be an important confirmation of our understanding of the physics of the early Universe.

Massive neutrinos affect cosmological observations in different ways. Primary CMB data alone can constrain the total neutrino mass $\Sigma$, if it is above $\sim1\,\mathrm{eV}$ \citep[][finds $\Sigma<1.3\,\mathrm{eV}$ at $95\%$ confidence]{KomatsuWMAP:2010} because these neutrinos become non-relativistic before recombination leaving an imprint in the CMB. Neutrinos with masses $\Sigma<1\,\mathrm{eV}$ become non-relativistic after recombination altering matter-radiation equality for fixed $\Omega_mh^2$; this effect is degenerate with other cosmological parameters from primary CMB data alone. After neutrinos become non-relativistic, their free streaming damps the small-scale power and modifies the shape of the matter power spectrum below the free-streaming length. The free-streaming length of each neutrino family depends on its mass.

Current cosmological observations do not detect any small-scale power suppression and break many of the degeneracies of the primary CMB, yielding constraints of $\Sigma<0.3\,\mathrm{eV}$ \citep{Reid/etal:2010} if we assume the neutrino mass to be a constant. A detection of such an effect, however, would provide a detection, although indirect, of the cosmic neutrino background. As shown in the next section, the fact that  oscillations predict a  minimum total mass $\Sigma\sim0.054\,\mathrm{eV}$ implies that Euclid has the statistical power to detect the cosmic neutrino background. We finally remark that the neutrino mass may also very well vary in time \cite{Wetterich:2007kr}; this might be tested by comparing (and not combining) measurements from CMB at decoupling with low-z measurements. An inconsistency would point out a direct measurement of a time varying neutrino mass \cite{Wetterich:2009qf}.
 
\subsection{Neutrino mass}\label{sec:numass}
Particle physics experiments are sensitive to neutrino flavours making a determination of the neutrino absolute-mass scales very model dependent. On the other hand, cosmology is not sensitive to neutrino flavour, but is sensitive to the total neutrino mass.

The small-scale power-suppression caused by neutrinos leaves imprints on CMB lensing: forecasts indicate that Planck should be able to constrain the sum of neutrino masses $\Sigma$, with a $1\sigma$ error of $0.13\,\mathrm{eV}$ \citep{Knox/Kaplinghat:2003,Lesgourgues/etal:2006,dePutter/etal:2009}.
 
Euclid's measurement of the galaxy power spectrum, combined with Planck (primary CMB only) priors should yield an error on $\Sigma$ of $0.04\,\mathrm{eV}$  \citep[for details see][]{Carbone/etal:2010} which is in  qualitative agreement with previous work \citep[e.g.][]{Saito/etal:2009}), assuming a minimal value for $\Sigma$ and constant neutrino mass. Euclid's weak lensing should also yield an error on $\Sigma$ of $0.05\,\mathrm{eV}$ \citep{Kitching/etal:2008}. While these two determinations are not fully independent (the  cosmic variance part of the error is in common given that the lensing survey and the galaxy survey cover the same volume of the Universe) the size of the error-bars implies more than $1\sigma$ detection of even the minimum $\Sigma$ allowed by oscillations. Moreover, the two independent techniques will offer cross-checks  and robustness to systematics. The error on $\Sigma$  depends on the fiducial model assumed, decreasing for fiducial models with larger $\Sigma$. Euclid will enable us not only to detect the effect of massive neutrinos on clustering but also to determine the absolute neutrino mass scale.

\subsection{Hierarchy and the nature of neutrinos}\label{sec:hierarchy}
Since cosmology is insensitive to flavour, one might expect that cosmology may not help in determining the neutrino mass hierarchy. However, for $\Sigma<0.1\,\mathrm{eV}$, only normal hierarchy is allowed, thus a mass determination can help disentangle the hierarchy. There is however another effect: neutrinos of different masses become non-relativistic at slightly different epochs; the free streaming length is sightly different for the different species and thus the detailed shape of the small scale power suppression depends on the individual neutrino masses and not just on their sum. As discussed in \citet{Jimenez:2010ev}, in cosmology one can safely neglect the impact of the solar mass splitting. Thus, two masses characterise the neutrino mass spectrum: the lightest $m$, and the heaviest $M$. The mass splitting can be parameterised by $\Delta=(M-m)/\Sigma$ for normal hierarchy and $\Delta=(m-M)/\Sigma$ for inverted hierarchy. The absolute value of $\Delta$ determines the mass splitting, whilst the sign of $\Delta$ gives the hierarchy. Cosmological data are very sensitive to $|\Delta|$; the direction of the splitting -- i.e. the sign of $\Delta$ -- introduces a sub-dominant correction to the main effect. Nonetheless, \citet{Jimenez:2010ev} show that weak gravitational lensing from Euclid data will be able to determine the hierarchy (i.e. the mass splitting and its sign) if far enough away from the degenerate hierarchy (i.e. if $\Sigma<0.13$). 
 \begin{figure}
\centering
\includegraphics[width=.45\textwidth]{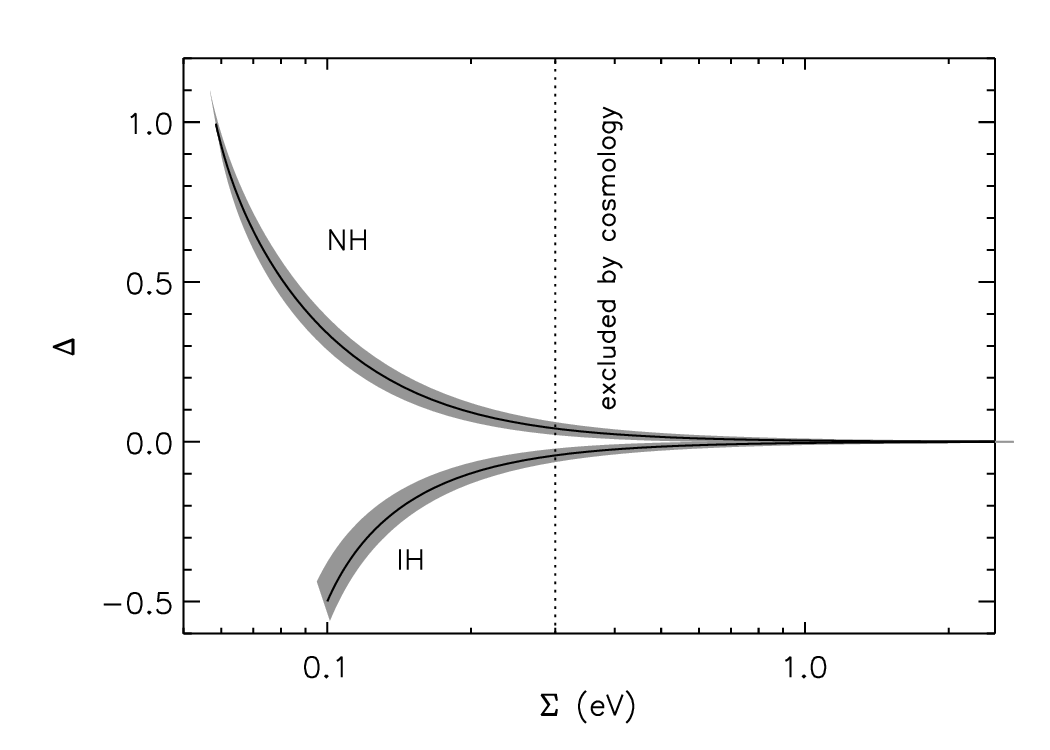}
\includegraphics[width=.43\textwidth,height=0.4\textwidth]{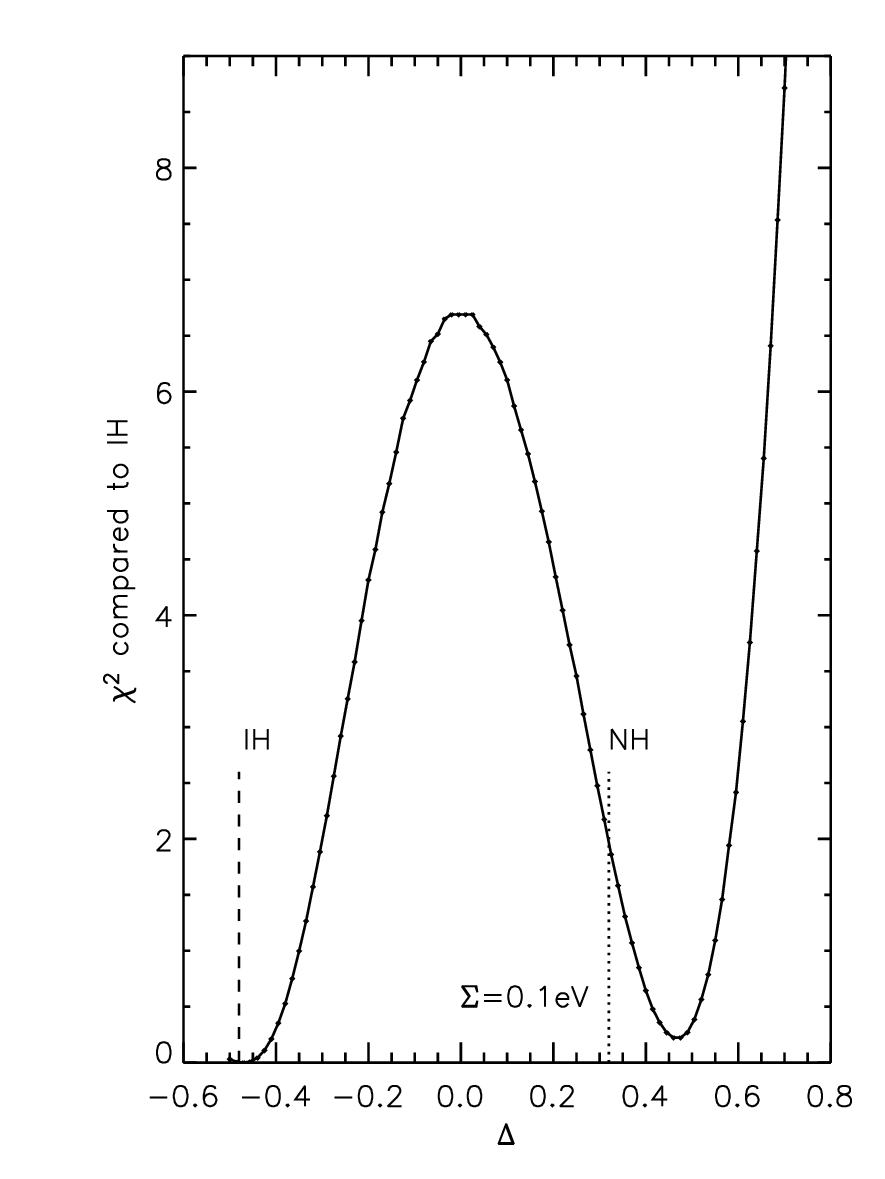}
\caption{Left: region in the $\Delta$-$\Sigma$ parameter space allowed  by oscillations data. Right: Weak lensing forecasts. The dashed and dotted vertical lines correspond to the central value for $\Delta$ given by oscillations data. In this case Euclid could discriminate NI from IH with a $\Delta \chi^2=2$. From \citet{Jimenez:2010ev}.}\label{fig:hierarchy}
\end{figure}

A detection of neutrino-less double-$\beta$ decay from the next generation experiments would indicate that neutrinos are Majorana particles. A null result of such double-$\beta$ decay experiments would lead to a definitive result pointing to the Dirac nature of the neutrino only for degenerate or inverted mass spectrum. This information can be obtained from large-scale structure cosmological data, improved data on  the tritium beta decay, or the long-baseline neutrino oscillation experiments. If the small mixing in the neutrino mixing matrix is negligible, cosmology might be the most promising arena to help in this puzzle.

\subsection{Number of neutrino species}\label{sec:Neff}
Neutrinos decouple early in cosmic history and contribute to a relativistic energy density with an effective number of species $N_{\nu,\mathrm{eff}}=3.046$. Cosmology is sensitive to the physical energy density in relativistic particles in the early Universe, which in the standard cosmological model includes only photons and neutrinos: $\omega_\mathrm{rel}=\omega_{\gamma}+N_{\nu,\mathrm{eff}}\omega_{\nu}$, where $\omega_{\gamma}$ denotes the energy density in photons and is exquisitely constrained from the CMB, and $\omega_{\nu}$ is the energy density in one neutrino. Deviations from the standard value for $N_{\nu,\mathrm{eff}}$ would signal non-standard neutrino features or additional relativistic species. $N_{\nu,\mathrm{eff}}$ impacts the big bang nucleosynthesis epoch through its effect on the expansion rate; measurements of primordial light element abundances can constrain $N_{\nu,\mathrm{eff}}$ and rely on physics at $T\sim MeV$ \citep{Bowen:2001in}. In several non-standard models -- e.g. decay of dark matter particles, axions, quintessence -- the energy density in relativistic species can change at some later time. The energy density of free-streaming relativistic particles alters the epoch of matter-radiation equality and leaves therefore a signature in the CMB and in the matter-transfer function. However, there is a degeneracy between $N_{\nu,\mathrm{eff}}$ and $\Omega_m h^2$ from CMB data alone (given by the combination of these two parameters that leave matter-radiation equality unchanged) and between $N_{\nu,\mathrm{eff}}$ and $\sigma_8$ and/or $n_s$. Large-scale structure surveys measuring the shape of the power spectrum at large scale can constrain independently the combination $\Omega_m h$ and $n_s$, thus breaking the CMB degeneracy. Furthermore, anisotropies in the neutrino background affect the CMB anisotropy angular power spectrum at a level of $\sim20\%$ through the gravitational feedback of their free streaming damping and anisotropic stress contributions. Detection of this effect is now possible by combining CMB and large-scale structure observations. This yields an indication at more than $ 2\sigma$ level that there exists a neutrino background with characteristics compatible with what is expected under the cosmological standard model \citep{Trotta:2004ty,DeBernardis:2008ys}.

The forecasted errors on $N_{\nu,\mathrm{eff}}$ for Euclid (with a Planck prior) are $\pm0.1$ at $1\sigma$ level \citep{Kitching/etal:2008}, which is a factor $\sim5$ better than current constraints from CMB and LSS and about a factor $\sim2$ better than constraints from light element abundance and nucleosynthesis.
 
\subsection{Model dependence}\label{Model_dependence}
A recurring question is how much model dependent will the neutrino constraints be. It is important to recall that usually parameter-fitting is done within the context of a $\Lambda$CDM model and that the neutrino effects are seen indirectly in the clustering. Considering more general cosmological models, might degrade neutrino constraints, and vice versa, including neutrinos in the model might degrade dark energy constraints. Here below we discuss the two cases of varying the total neutrino mass $\Sigma$ and the number of relativistic species $N_{\rm eff}$, separately.

\subsection{$\Sigma$ forecasted error bars and degeneracies}
\label{Mnu_cosmology}

In \citet{Carbone_etal2010b} it is shown that, for a general model which allows for a non-flat Universe, and a redshift dependent dark energy equation of state, the $1\sigma$ spectroscopic errors on the neutrino mass $\Sigma$ are in the range $0.036-0.056$ eV, depending on the fiducial total neutrino mass $\Sigma$, for the combination Euclid+Planck.

On the other hand, looking at the effect that massive neutrinos have on the dark energy parameter constraints, it is shown that the total CMB+LSS dark-energy FoM decreases only by $\sim 15\%-25\%$ with respect to the value obtained if neutrinos are supposed to be massless, when the forecasts are computed using the so-called ``$P(k)$--method marginalised over growth-information'' (see Methodology chapter),  which therefore results to be quite robust in constraining the dark-energy equation of state.

For what concerns the parameter correlations, at the LSS level, the total neutrino mass $\Sigma$ is correlated with all the cosmological parameters affecting the galaxy power spectrum shape and BAO positions. When Planck priors are added to the Euclid constraints, all degeneracies are either resolved or reduced, and the remaining dominant correlations among $\Sigma$ and the other cosmological parameters are $\Sigma$-$\Omega_{de}$, $\Sigma$-$\Omega_m$, and $\Sigma$-$w_a$, with the $\Sigma$-$\Omega_{de}$ degeneracy being the largest one.

\subsubsection{Hierarchy dependence}
\label{Hierarchy_dependence}
In addition, the neutrino mass spectroscopic constraints depend also on the neutrino hierarchy. In fact, the $1\sigma$ errors on total neutrino mass for normal hierarchy are $\sim 17\%-20\%$ larger than for the inverted one. It appears that the matter power spectrum is less able to give information on the total neutrino mass when the normal hierarchy is assumed as fiducial neutrino mass spectrum. This is similar to what found in Ref.~\citep{Jimenez:2010ev} for the constraints on the neutrino mass hierarchy itself, when a normal hierarchy is assumed as the fiducial one. On the other hand, when CMB information are included, the $\Sigma$-errors decrease by $\sim$35\% in favour of the normal hierarchy, at a given fiducial value $\Sigma|_{\rm fid}$. This difference arises from the changes in the free-streaming effect due to the assumed mass hierarchy, and is in agreement with the results of Ref.~\citep{0403296}, which confirms that the expected errors on the neutrino masses depend not only on the sum of neutrino masses, but also on the order of the mass splitting between the neutrino mass states.

\subsubsection{Growth and incoherent peculiar velocity dependence}
\label{growth+FoG}
$\Sigma$ spectroscopic errors stay mostly unchanged whether growth--information are included or marginalised over, and decrease only by $10\%$--$20\%$ when adding $f_g\sigma_8$ measurements. This result is expected, if we consider that, unlike dark energy parameters, $\Sigma$ affects the shape of the power spectrum via a redshift-dependent transfer function $T(k,z)$, which is sampled on a very large range of scales including the $P(k)$ turnover scale, therefore this effect dominates over the information extracted from measurements of $f_g\sigma_8$. This quantity, in turn, generates new correlations with $\Sigma$ via the $\sigma_8$-term, which actually is anti-correlated with $M_\nu$ \citep{Marulli2010_inprep}. On the other hand, if we suppose that early dark-energy is negligible, the dark-energy parameters $\Omega_{de}$, $w_0$ and $w_a$ do not enter the transfer function, and consequently growth information have relatively more weight when added to constraints from $H(z)$ and $D_A(z)$ alone. Therefore, the value of the dark-energy FoM does increase when growth-information are included, even if it decreases by a factor $\sim 50\%$--$60\%$ with respect to cosmologies where neutrinos are assumed to be massless, due to the correlation among $\Sigma$ and the dark-energy parameters. As confirmation of this degeneracy, when growth-information are added and if the dark-energy parameters $\Omega_{de}$, $w_0$, $w_a$ are held fixed to their fiducial values, the errors $\sigma({\Sigma})$ decrease from $0.056$ eV to $0.028$ eV, for Euclid combined with Planck.

We expect that dark-energy parameter errors are somewhat sensitive also to the effect of incoherent peculiar velocities, the so-called ``Fingers of God'' (FoG). This can be understood in terms of correlation functions in the redshift-space; the stretching effect due to random peculiar velocities contrasts the flattening effect due to large-scale bulk velocities. Consequently, these two competing effects act along opposite directions on the dark-energy parameter constraints (see Methodology chapter).

On the other hand, the neutrino mass errors are found to be stable again at $\sigma({\Sigma})=0.056$, also when FoG effects are taken into account by marginalising over $\sigma_v(z)$; in fact, they increase only by $10\%$--$14\%$ with respect to the case where FoG are not taken into account.

Finally, in Table~\ref{summary} we summarise the dependence of the $\Sigma$-errors on the model cosmology, for Euclid combined with Planck\footnote{In this case we have added the contribution from BOSS at redshifts $0.1<z<z_{min}$, where $z_{min}=0.5$ is the minimum redshift of the Euclid spectroscopic survey.}. We conclude that, if $\Sigma$ is $>0.1$ eV, spectroscopy with Euclid will be able to determine the neutrino mass scale independently of the model cosmology assumed. If $\Sigma$ is $<0.1$ eV, the sum of neutrino masses, and in particular the minimum neutrino mass required by neutrino oscillations, can be measured in the context of a $\Lambda$CDM model.

\subsection{$N_{\rm eff}$ forecasted errors and degeneracies}
\label{Neff_cosmology}
Regarding the $N_{\rm eff}$ spectroscopic errors, Ref.~\citep{Carbone_etal2010b} finds $\sigma(N_{\rm eff})\sim 0.56$ from Euclid, and $\sigma(N_{\rm eff})\sim 0.086$, for Euclid+Planck. Concerning the effect of $N_{\rm eff}$ uncertainties on the dark-energy parameter errors, the CMB+LSS dark-energy FoM decreases only by $\sim 5\%$ with respect to the value obtained holding $N_{\rm eff}$ fixed at its fiducial value, meaning that also in this case the ``$P(k)$--method marginalised over growth--information'' is not too sensitive to assumptions about  model cosmology when constraining the dark-energy equation of state.

About the degeneracies between $N_{\rm eff}$ and the other cosmological parameters, it is necessary to say that the number of relativistic species gives two opposite contributions to the observed power spectrum $P_{\rm obs}$ (see Methodology chapter), and the total sign of the correlation depends on the dominant one, for each single cosmological parameter. In fact, a larger $N_{\rm eff}$ value suppresses the transfer function $T(k)$ on scales $k\leq \kmax$. On the other hand, a larger $N_{\rm eff}$ value also increases the Alcock-Paczynski prefactor in $P_{\rm obs}$. For what concerns the dark-energy parameters $\Omega_{de}$, $w_0$, $w_a$, and the dark-matter density $\Omega_m$, the Alcock-Paczynski prefactor dominates, so that $N_{\rm eff}$ is positively correlated to $\Omega_{de}$ and $w_a$, and anti-correlated to $\Omega_m$ and $w_0$. In contrast, for the other parameters, the $T(k)$ suppression produces the larger effect and $N_{\rm eff}$ results to be anti-correlated to $\Omega_b$, and positively correlated to $h$ and $n_s$. The degree of the correlation is very large in the $n_s$-$N_{\rm eff}$ case, being of the order $\sim 0.8$ with and without Planck priors. For the remaining cosmological parameters, all the correlations are reduced when CMB information are added, except for the covariance $N_{\rm eff}$-$\Omega_{de}$, as happens also for the $M_\nu$--correlations. To summarise, after the inclusion of Planck priors, the remaining dominant degeneracies among $N_{\rm eff}$ and the other cosmological parameters are $N_{\rm eff}$-$n_s$, $N_{\rm eff}$-$\Omega_{de}$, and $N_{\rm eff}$-$h$, and the forecasted error is $\sigma(N_{\rm eff})\sim 0.086$, from Euclid+Planck. Finally, if we fix to their fiducial values the dark energy parameters $\Omega_{de}$, $w_0$ and $w_a$, $\sigma(N_{\rm eff})$ decreases from $0.086$ to $0.048$, for the combination Euclid+Planck.
\begin{table*}
\caption{$\sigma(M_\nu)$ and $\sigma(N_{\rm eff})$ marginalised errors                                                                                                                                                                                   
  from LSS+CMB}
\setlength{\tabcolsep}{0.3pt}
\begin{tabular}{|l c c c c c c|}
\hline
\hline
{}&{}&{}&\footnotesize{{\small General cosmology}}&{}&{}&{}
\\
\hline
\footnotesize{{\small {\rm fiducial}$\to$ }}&
\footnotesize{{\small $\Sigma$=0.3 eV}}${}^a$&
\footnotesize{{\small $\Sigma$=0.2 eV}}${}^a$ &
\footnotesize{{\small $\Sigma$=0.125 eV}}${}^b$ &
\footnotesize{{\small $\Sigma$=0.125 eV}}${}^c$ &
\footnotesize{{\small $\Sigma$=0.05 eV}}${}^b$ &
\footnotesize{{\small $N_{\rm eff}$=3.04}}${}^d$
\\
\hline
\footnotesize{{\small EUCLID+Planck}} & $0.0361$ & $0.0458$ &
$0.0322$ & $0.0466$ & $0.0563$ & $0.0862$\\
\hline
\hline
{}&{}&{}&\footnotesize{{\small $\Lambda$CDM cosmology}}&{}&{}&{}
\\
\hline
\footnotesize{{\small EUCLID+Planck}} & $0.0176$ & $0.0198$ &$0.0173$ & $0.0218$ & $0.0217$ & $0.0224$\\
\hline
\end{tabular}
\begin{flushleft}
${}^a$\footnotesize{for degenerate spectrum: $m_1\approx m_2\approx
  m_3$};
${}^b$\footnotesize{for normal hierarchy: $m_3\neq 0$, $m_1\approx
  m_2\approx 0$}\\
${}^c$\footnotesize{for inverted hierarchy: $m_1\approx m_2$,
  $m_3\approx 0$};
${}^d$\footnotesize{fiducial cosmology with massless neutrinos}
\end{flushleft}
\label{summary}
\end{table*}

\subsection{Non-linear effects of massive cosmological neutrinos on bias and {\color{red}RSD}}
In general, forecasted errors are obtained using techniques, like the Fisher-matrix approach, that are not particularly well suited to quantify systematic effects. These techniques forecast only statistical errors, which are meaningful as long as they dominate over systematic errors. It is therefore important to consider sources of systematics and their possible effects on the recovered parameters. Possible sources of systematic errors of major concern are the effect of non-linearities and the effects of galaxy bias.

The description of non-linearities in the matter power spectrum in the presence of massive neutrinos has been addressed in several different ways: \citet{2008JCAP...10..035W,Saito1,Saito2,Saito3} have used perturbation theory, \citet{Pietroni} the time-RG flow approach and \citet{Brandbyge1,Brandbyge2,Brandbyge3,Viel_etal2010} different schemes of $N$-body simulations.  {\color{red} Another nonlinear scheme that has been examined in the literature is the halo model.  This has been applied to massive neutrino cosmologies in \cite{2005PhRvD..71d3507A,2005JCAP...09..014H,2006JCAP...06..025H}}.

On the other hand, galaxy/halo bias is known to be almost scale-independent only on large, linear  scales, but to become non-linear and scale-dependent for small scales and/or for very massive haloes. From the above discussion and references, it is clear that the effect of massive neutrinos on the galaxy power spectrum  in the non-linear regime must be explored via $N$-body simulations to encompass all the relevant effects.

Here below we focus on the behaviour of the DM-halo mass function (MF), the DM-halo bias, and the redshift-space distortions ({\color{red}RSD}), in the presence of a cosmological background of massive neutrinos. To this aim,{\color{red}\citet{Brandbyge3}} and \citet{Marulli2010_inprep} have analysed a set of large $N$-body hydrodynamical simulations, developed with an extended version of the code GADGET-3 \citep{Viel_etal2010}, which take into account the effect of massive free-streaming neutrinos on the evolution of cosmic structures.

\begin{figure}
\center
\includegraphics[width=0.78\textwidth]{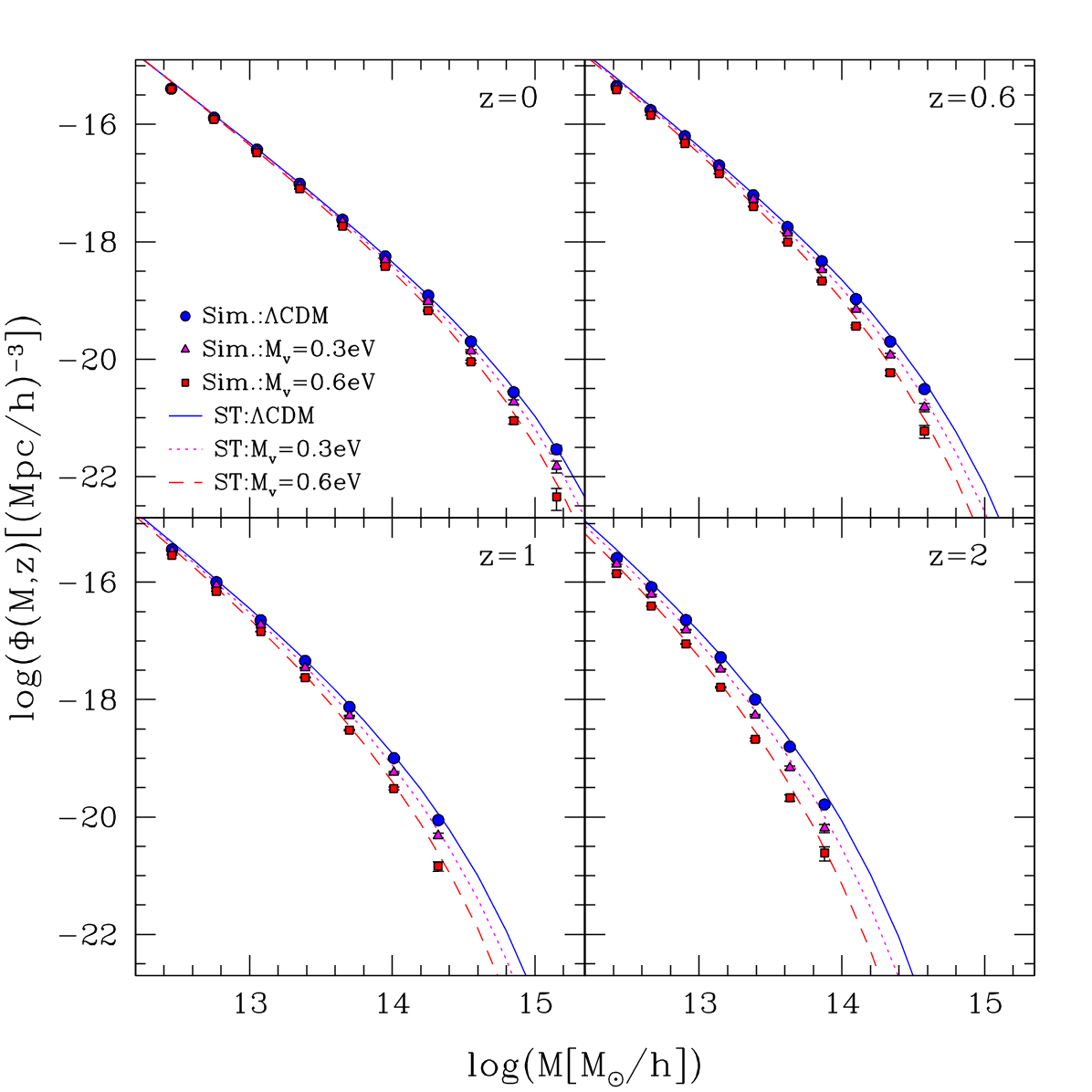}
\caption{DM halo mass function (MF) as a function of $\Sigma$ and redshift. MF of the SUBFIND haloes in the $\Lambda$CDM $N$-body simulation (blue circles) and in the two simulations with $\Sigma=0.3\,\mathrm{eV}$ (magenta triangles) and $\Sigma=0.6\,\mathrm{eV}$ (red squares). The blue, magenta and red lines show the halo MF predicted by \citet{sheth2002}, where the variance in the density fluctuation field, $\sigma(M)$, at the three cosmologies, $\Sigma=0,0.3,0.6\,\mathrm{eV}$, has been computed with the software CAMB \citep{CAMB}.}\label{fig:MF_xireal_left}
\end{figure}

The pressure produced by massive neutrino free-streaming contrasts the gravitational collapse which is the basis of cosmic structure formation, causing a significant suppression in the average number density of massive structures. This effect can be observed in the high mass tail of the halo MF in Fig.~\ref{fig:MF_xireal_left}, as compared with the analytic predictions of \citet{sheth2002} (ST), where the variance in the density fluctuation field, $\sigma(M)$, has been computed via CAMB \citep{CAMB}, using the same cosmological parameters of the simulations. In particular, {\color{red}here the MF of sub-structures is shown,} identified using the SUBFIND package \citep{Springel_etal2001}, while the normalisation of the matter power spectrum is fixed by the dimensionless amplitude of the primordial curvature perturbations $\Delta^2_{\cal R}(k_0)|_{\rm fid}=2.3\times 10^{-9}$, evaluated at a pivot scale $k_0=0.002$/Mpc \citep{arXiv:1001.4635}, which has been chosen to have the same value both in the $\Lambda$CDM$\nu$ and in the $\Lambda$CDM cosmologies.

\begin{figure}
\center
\includegraphics[width=0.78\textwidth]{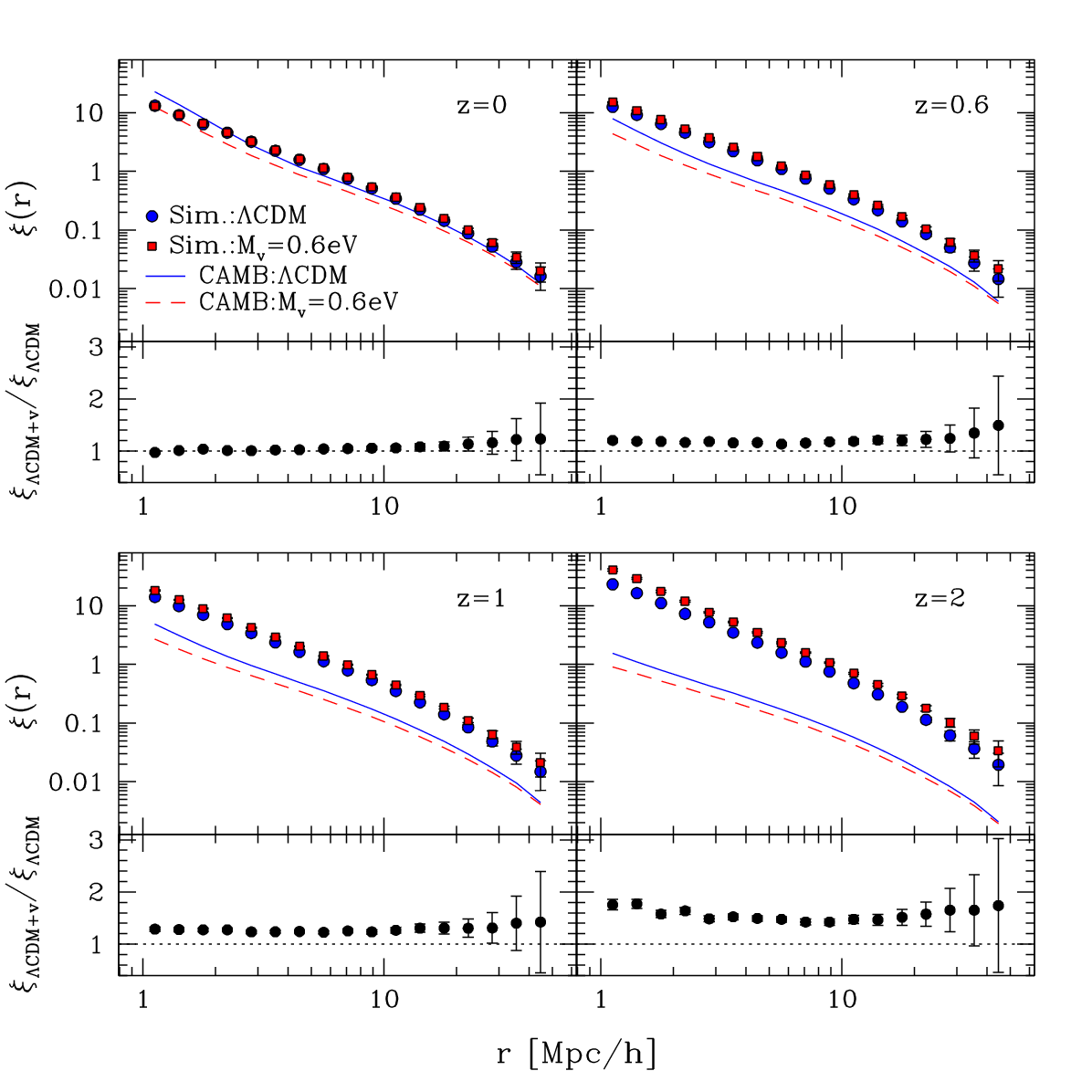}
\caption{DM halo mass function (MF) as a function of $\Sigma$ and redshift. Real space two-point auto-correlation function of the DM haloes in the $\Lambda$CDM $N$-body simulation (blue circles) and in the simulation with $\Sigma=0.6\,\mathrm{eV}$ (red squares). The blue and red lines show the DM correlation function computed using the CAMB matter power spectrum with $\Sigma=0$ and $\Sigma=0.6\,\mathrm{eV}$, respectively. The bottom panels show the ratio between the halo correlation function extracted from the simulations with and without massive neutrinos.}\label{fig:MF_xireal_right}
\end{figure}

In Figs.~\ref{fig:MF_xireal_left} and ~\ref{fig:MF_xireal_right}, two fiducial neutrino masses have been considered, $\Sigma=0.3$ and $\Sigma=0.6\,\mathrm{eV}$. From the comparison of the corresponding MFs, we confirm the theoretical predictions, i.e. that the higher the neutrino mass is, the larger the suppression in the comoving number density of DM haloes becomes.

As is well known, massive neutrinos also strongly affect the spatial clustering of cosmic structures. A standard {\color{red}statistics} generally used to quantify the degree of clustering of a population of sources is the two-point auto-correlation function. Although the free-streaming of massive neutrinos causes a suppression of the matter power spectrum on scales $k$ larger than the neutrino free-streaming scale, the halo bias is significantly enhanced. This effect can be physically explained thinking that, due to neutrino structure suppression, the same halo bias would correspond, in a $\Lambda$CDM cosmology, to more massive haloes (than in a $\Lambda$CDM$\nu$ cosmology), which as known are typically more clustered.

\begin{figure}
\center
\includegraphics[width=0.78\textwidth]{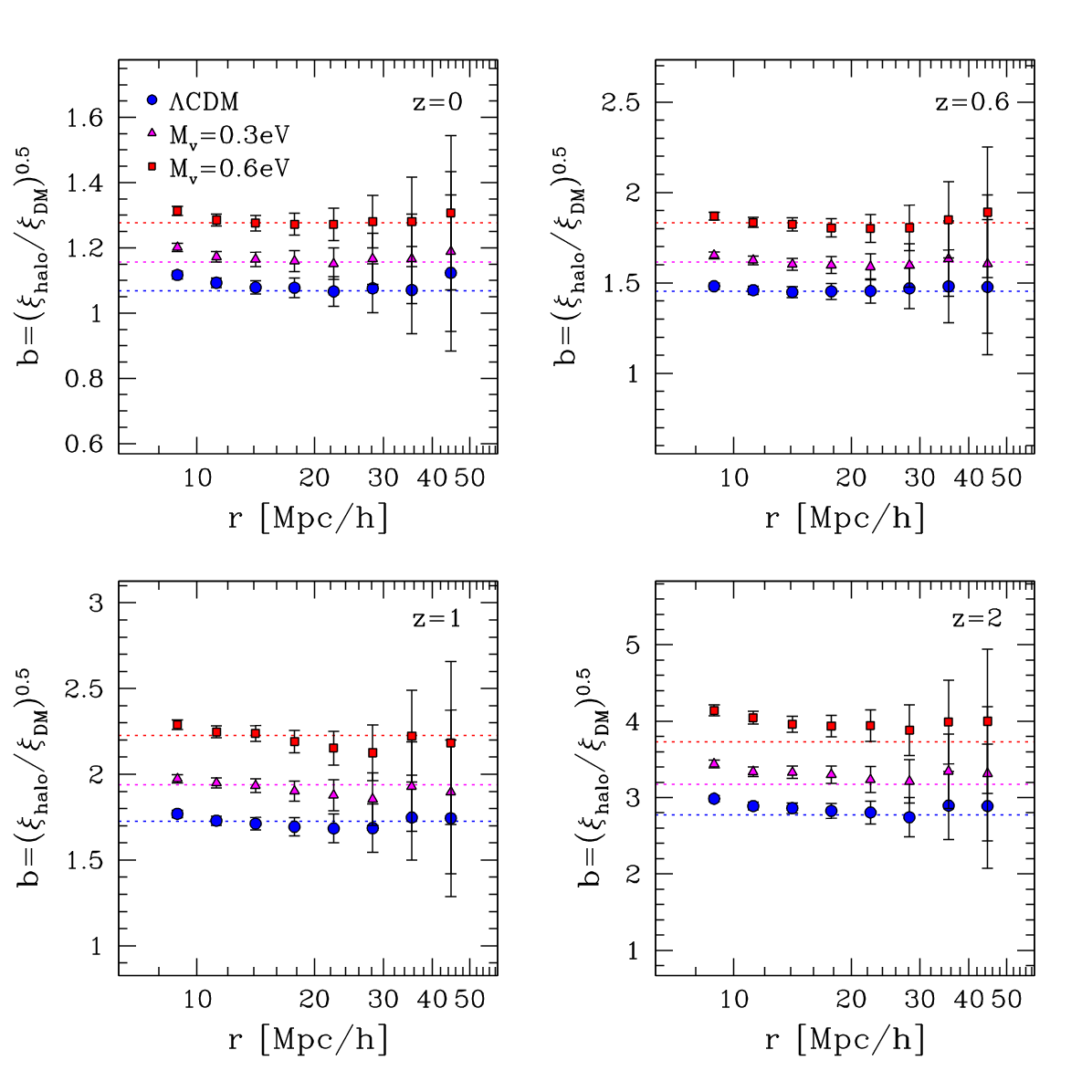}
\caption{Real space two-point auto-correlation function of the DM haloes in the $\Lambda$CDM $N$-body simulation (blue circles) and in the simulation with $\Sigma=0.6\,\mathrm{eV}$ (red squares). The blue and red lines show the DM correlation function computed using the CAMB matter power spectrum with $\Sigma=0$ and $\Sigma=0.6\,\mathrm{eV}$, respectively. The bottom panels show the ratio between the halo correlation function extracted from the simulations with and without massive neutrinos.}\label{fig:bias_left}
\end{figure}

\begin{figure}
\center
\includegraphics[width=0.78\textwidth]{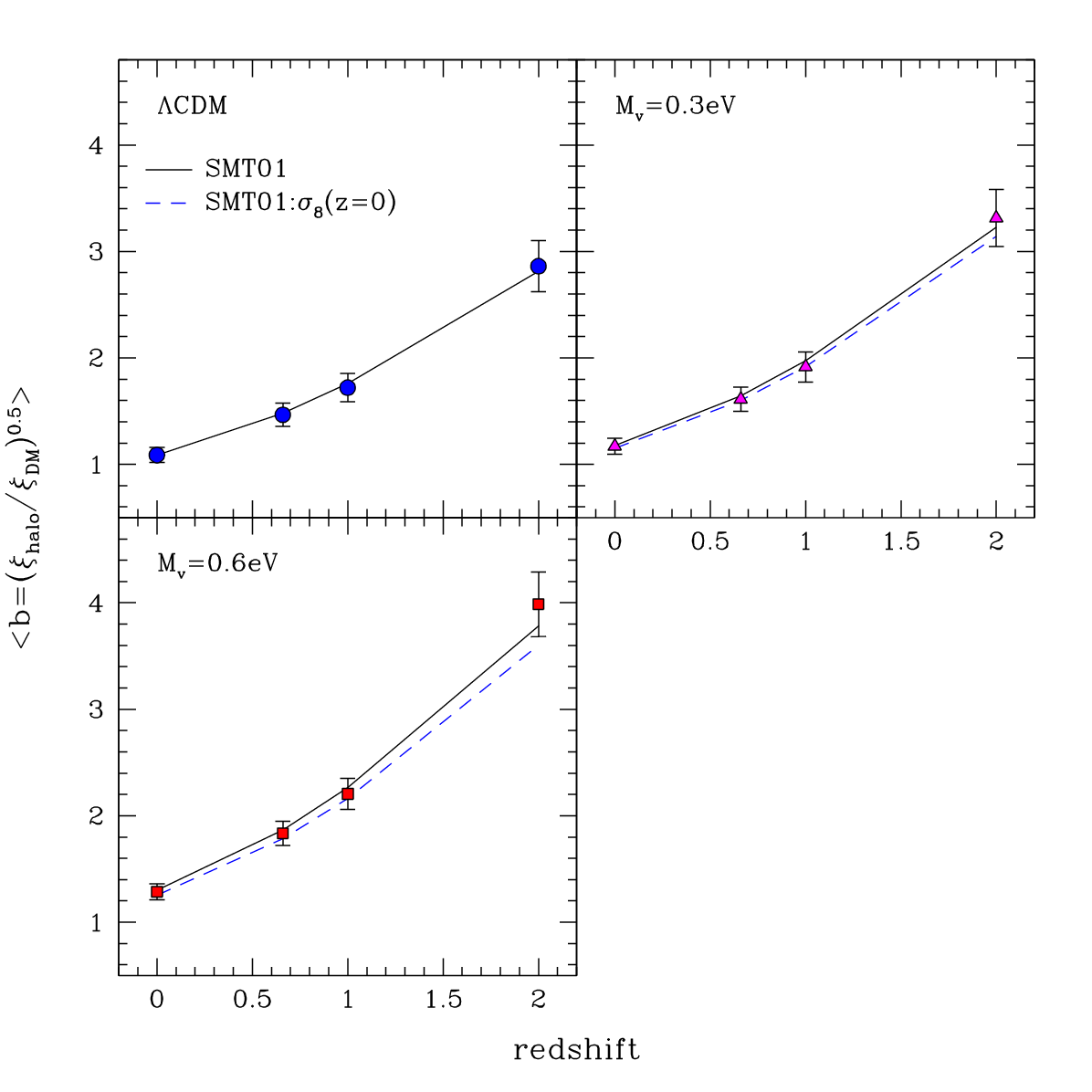}
\caption{Mean bias (averaged in $10<r\,[\mathrm{Mpc}/h]<50$) as a function of redshift compared with the theoretical predictions of \citet{sheth2002}. Here the dashed lines represent the theoretical expectations for a LCDM cosmology renormalized with the $\sigma_8$ value of the simulations with a massive neutrino component.}\label{fig:bias_right}
\end{figure}

This effect is evident in Fig.~\ref{fig:MF_xireal_right} which shows the two-point DM-halo correlation function measured with the Landy and Szalay (1993) estimator, compared to the matter correlation function. In particular, the clustering difference between the $\Lambda$CDM and $\Lambda$CDM$\nu$ cosmologies increases at higher redshifts, as it can be observed from Figs.~\ref{fig:bias_left} and  ~\ref{fig:bias_right} and the windows at redshifts $z>0$ of Fig.~\ref{fig:MF_xireal_left}. Note also the effect of non-linearities on the bias, which clearly starts to become scale-dependent for separations $r<20$ Mpc/$h$.

\begin{figure}
\center
\includegraphics[width=0.78\textwidth]{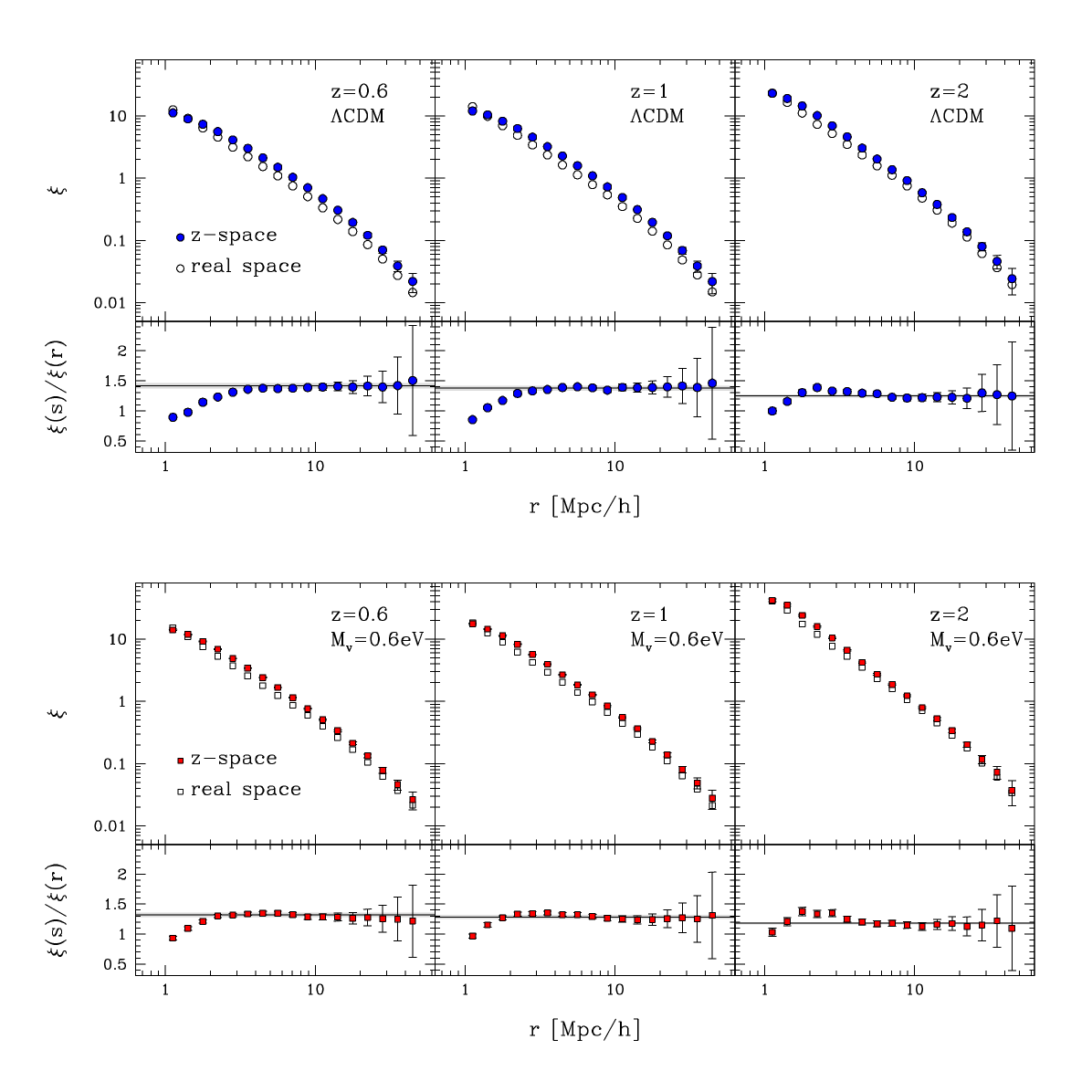}
\caption{Two-point auto-correlation function in real and redshift space of the DM-haloes in the $\Lambda$CDM $N$-body simulation (blue circles) and in the simulation with $\Sigma=0.6\,\mathrm{eV}$ (red squares). The bottom panels show the ratio between them, compared with the theoretical expectation.}\label{fig:xiZspace_beta_left}
\end{figure}

\begin{figure}
\center
\includegraphics[width=0.78\textwidth]{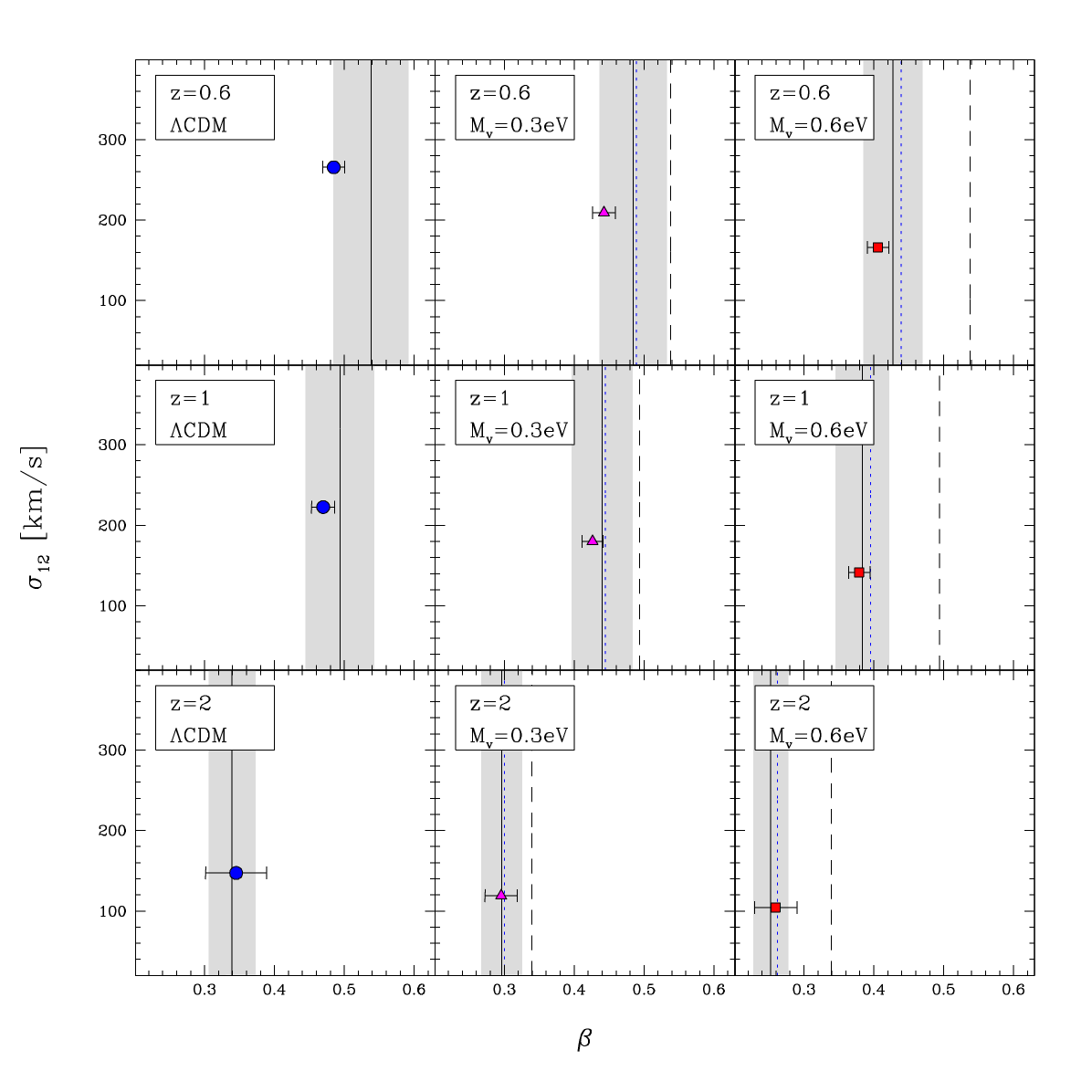}
\caption{Best-fit values of $\beta$-$\sigma_{12}$, as a function of $\Sigma$ and redshift (points), compared with the theoretical prediction (grey shaded area). The blue dotted lines show the theoretical prediction for $\Sigma=0$ and with $\sigma_8(z=0)$.}\label{fig:xiZspace_beta_right}
\end{figure}

As it happens for the MF and clustering, also {\color{red}RSD} are strongly affected by massive neutrinos. Fig.~\ref{fig:xiZspace_beta_left} shows the real and redshift space correlation functions of DM haloes as a function of the neutrino mass. The effect of massive neutrinos is particularly evident when the correlation function is measured as a function of the two directions perpendicular and parallel to the line of sight. As a consequence, the value of the linear growth rate that can be derived by modelling galaxy clustering anisotropies can be greatly suppressed with respect to the value expected in a $\Lambda$CDM cosmology. Indeed, neglecting the cosmic relic massive neutrino background in data analysis might induce a bias in the inferred growth rate, from which a potentially fake signature of modified gravity might be inferred. Fig.~\ref{fig:xiZspace_beta_right} demonstrates this point, showing the best-fit values of $\beta$ and $\sigma_{12}$, as a function of $\Sigma$ and redshift, where $\beta = {\frac{f(\Omega_{\rm M})} {b_\mathrm{eff}}}$, $b_\mathrm{eff}$ being the halo effective linear bias factor, $f(\Omega_{\rm M})$ the linear growth rate and $\sigma_{12}$ the pairwise velocity dispersion.  \label{symbol:halobias}

%% file: dark_matter/de_nu_writeup.tex
\section{Coupling between dark energy and neutrinos}\label{dms:de_nu}
As we have seen in section (\ref{mg:cde}), it is interesting to consider the possibility that dark energy, seen as a dynamical scalar field (quintessence), may interact with other components in the Universe. In this section we focus on the possibility that a coupling may exist between dark energy and neutrinos.

The idea of such a coupling has been addressed and developed by several authors within MaVaNs theories first \citep{Fardon:2003eh,Peccei:2004sz,Bi:2004ns,Afshordi:2005ym,Weiner:2005ac,Das:2006ht, Takahashi:2006jt,Spitzer:2006hm,Bjaelde:2007ki,Brookfield:2005td,Brookfield:2005bz} and more recently within growing neutrino cosmologies \citep{Amendola2008b,Wetterich:2007kr,Mota:2008nj,Wintergerst:2009fh,Wintergerst:2010ui,Pettorino:2010bv,Brouzakis:2010md}. It has been shown that neutrinos can play a crucial role in cosmology, setting naturally the desired scale for Dark Energy. Interestingly, a coupling between neutrinos and dark energy may help solving the `why now' problem, explaining why dark energy dominates only in recent epochs. The coupling follows the description illustrated in section (\ref{mg:cde}) for a general interacting dark energy cosmology, where now $m_\nu=m_\nu(\phi)$. 

Typically, in growing neutrino cosmologies, the function $m_\nu(\phi)$ is such that the neutrino mass grows with time from low, nearly massless values (when neutrinos are non-relativistic) up to present masses in a range in agreement with current observations (see the previous section of this review for latest bounds on neutrino masses). The key feature of growing neutrino models is that the amount of dark energy today is triggered by a cosmological event, corresponding to the transition from relativistic to non-relativistic neutrinos at redshift $z_\mathrm{NR}\sim5\div10$. As long as neutrinos are relativistic, the coupling plays no role on the dynamics of the scalar field, which follows attractor solutions of the type described in section \ref{mg:cde}. From there on, the evolution of dark energy resembles that of a cosmological constant, plus small oscillations of the coupled dark energy-neutrino fluid. As a consequence, when a coupling between dark energy and neutrinos is active, the amount of dark energy and its equation of state today are strictly connected to the present value of the neutrino mass.

The interaction between neutrinos and dark energy is a nice and concrete example of the significant imprint that dynamical coupled dark energy can leave on observables and in particular on structure formation and on the cosmic microwave background. This is due to the fact that the coupling, playing a role only after neutrinos become non-relativistic, can reach relatively high values as compared to gravitational attraction. Typical values of $\beta$ are order $50\div100$ or even more such that even the small fraction of cosmic energy density in neutrinos can have a substantial influence on the time evolution of the quintessence field. During this time the fifth force can be of order $10^2\div10^4$ times stronger than gravity. The neutrino contribution to the gravitational potential influences indirectly also dark matter and structure formation, as well as CMB, via the Integrated Sachs Wolfe effect and the non-linear Rees-Sciama effect, which is non-negligible at the scales where neutrinos form stable lumps. Furthermore, backreaction effects can substantially modify the growth of large scale neutrino lumps, with effects which are much larger than in the dark matter case. The presence of a fifth force due to an interaction between neutrinos and dark energy can lead to remarkably peculiar differences with respect to a cosmological constant scenario. 

Here, we just recall some of the typical features that can arise when such an interaction is active:
\begin{itemize}
\item existence of very large structures, order $10\div500\,\mathrm{Mpc}$ \citep{Afshordi:2005ym,Mota:2008nj,Wintergerst:2009fh,Wintergerst:2010ui,Pettorino:2010bv};
\item enhanced ISW effect, drastically reduced when taking into account non-linearities \citep{Pettorino:2010bv}: information on the gravitational potential is a good mean to constrain the range of allowed values for the coupling $\beta$;
\item large-scale anisotropies and enhanced peculiar velocities \citep{Watkins:2008hf,Ayaita:2009qz};
\item the influence of the gravitational potential induced by the neutrino inhomogeneities can affect BAO in the dark-matter spectra \citep{Brouzakis:2010md}.
\end{itemize}

Investigation of structure formation at very large scales (order $1\div100\,\mathrm{Mpc}$) as well as cross correlation with CMB are crucial in order to disentangle coupled neutrino-quintessence cosmologies from a cosmological constant scenario. Detection of a population of very large-scale structures could pose serious difficulties to the standard framework and open the way to the existence of a new cosmological interaction stronger than gravity.

%% file: dark_matter/UDMwriteupFinal.tex
\section{Unified Dark Matter}

The appearance of two unknown components in the Standard Cosmological Model, dark matter and dark energy, has prompted discussion of whether they are two facets of a single underlying dark component. This concept goes under the name of quartessence \citep{Makler2003PhLB}, or Unified Dark Matter (UDM). \emph{A priori} this is attractive, replacing two unknown components with one, and in principle it might explain the `why now?' problem of why the energy densities of the two components are similar (also referred to as the coincidence problem). Many UDM models are characterised by a sound speed, whose value and evolution imprints oscillatory features on the matter power spectrum, which may be detectable through weak lensing or BAO signatures with Euclid.

The field is rich in UDM models \citep[see][for a review and for references to the literature]{2010AdAst2010E..78B}. The models can grow structure, as well as providing acceleration of the Universe at late times. In many cases, these models have a non-canonical kinetic term in the Lagrangian, e.g. an arbitrary function of the square of the time derivative of the field in a homogeneous and isotropic background. Early models with acceleration driven by kinetic energy \citep[$k$-inflation][]{1999PhLB..458..209A,1999PhLB..458..219G,2009PhRvD..80j3508B} were generalised to more general Lagrangians \citep[$k$-essence; e.g.][]{2000PhRvL..85.4438A,2001PhRvD..63j3510A,2004PhRvL..93a1301S}. For UDM, several models have been investigated, such as the generalised Chaplygin gas \citep{2001PhLB..511..265K,2002PhRvD..66d3507B,2002PhLB..535...17B,2006JCAP...01..003Z,2010PhLB..686..211P}, although these may be tightly constrained due to the finite sound speed \citep[e.g.][]{2003JCAP...07..005A,2003GReGr..35.2063B,2004PhRvD..69l3524S,2004A&A...423..421Z}. Vanishing sound speed models however evade these {\color{red}constraints} \citep[e.g. the Silent Chaplygin gas of][]{2005JCAP...11..009A}. Other models consider a single fluid with a two-parameter equation of state \citep[e.g][]{2007PhRvD..76j3519B}), models with canonical Lagrangians but a complex scalar field \citep{2006PhRvD..74d3516A}, models with a kinetic term in the energy-momentum tensor \citep{2010PhRvD..81d3520G,2008PhLB..666..205C}, models based on a DBI action \citep{2010GReGr..42.1189C}, models which violate the weak equivalence principle \citep{2007PhRvD..75l3007F} and models with viscosity \citep{2011AdAst2011E...4D}. Finally, there are some models which try to unify inflation as well as Dark Matter and Dark Energy \citep{2006PhLB..632..597C,2008arXiv0801.4843N,2008PhRvD..77l1301L,2009arXiv0906.5021L,2009PhRvD..79j3522H}.

A requirement for UDM models to be viable is that they must be able to cluster to allow structure to form. A generic feature of the UDM models is an effective sound speed, which may become significantly non-zero during the evolution of the Universe, and the resulting Jeans length may then be large enough to inhibit structure formation. The appearance of this sound speed leads to observable consequences in the CMB as well, and generally speaking the speed needs to be small enough to allow structure formation and for agreement with CMB measurements. In the limit of zero sound speed, the standard cosmological model is recovered in many models. Generally the models require fine-tuning, although some models have a fast transition between a dark matter only behaviour and $\Lambda$CDM. Such models \citep{2010JCAP...01..014P} can have acceptable Jeans lengths even if the sound speed is not negligible.

\subsection{Theoretical Background}
\begin{figure}[t]
\centering
\includegraphics[width=10truecm]{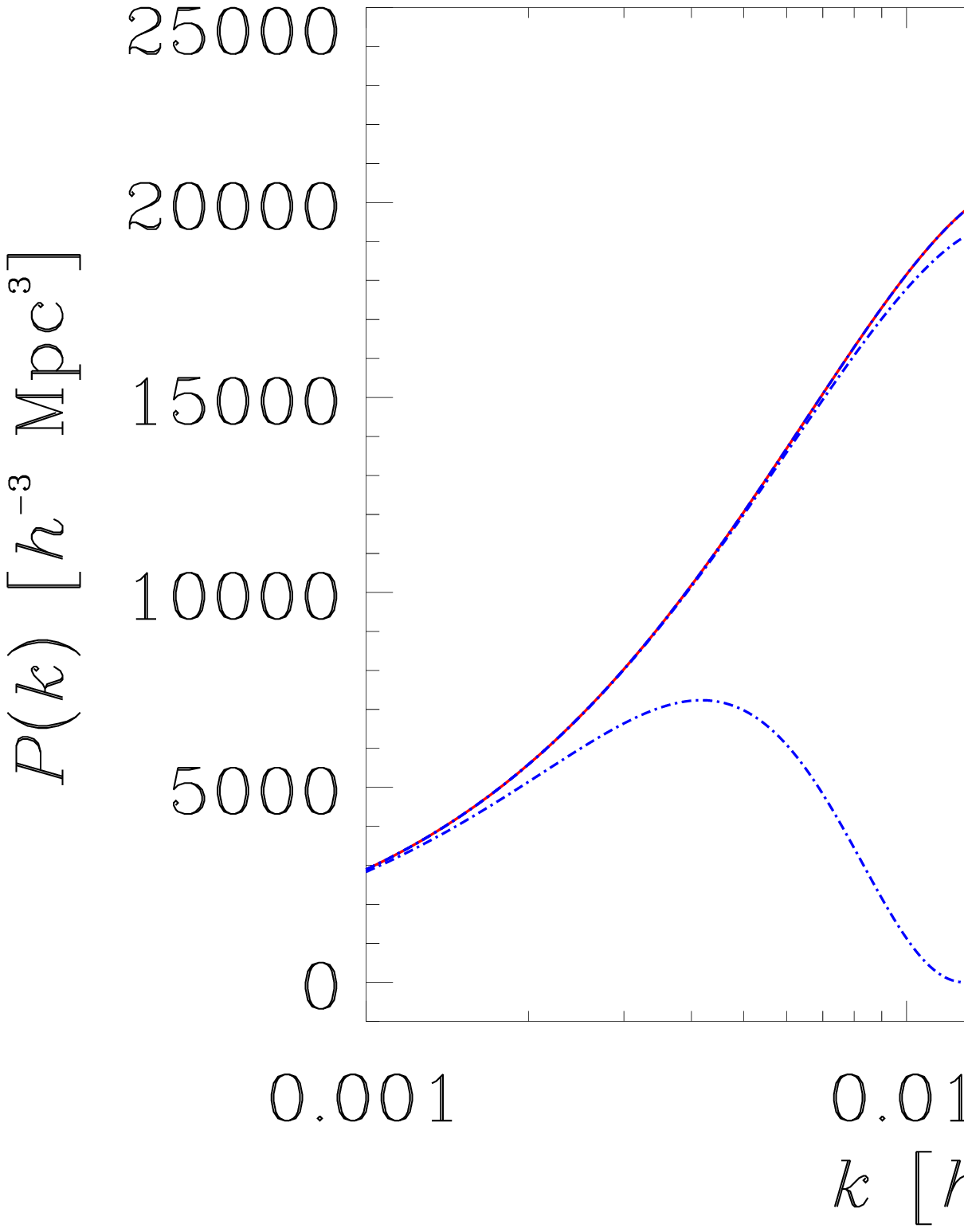}
\caption{The $z=0$ matter power spectrum arising in UDM models with a Lagrangian given by equation (\ref{UDML}). $\Lambda$CDM is solid, and UDM models with $c_\infty=10^{-1}, \, 10^{-2},\,10^{-3}$ are shown from bottom to top. From \citet{Camera:2010wm}.}\label{UDMPk}
\end{figure}

An action which is applicable for most UDM models, with a single scalar field $\varphi$, is
\begin{equation}
S=\int\,d^4x\sqrt{-g}\left[\frac{R}{2}+{\cal L}(\varphi,X)\right],
\end{equation}
where 
\begin{equation}
X\equiv-\frac{1}{2}\nabla_\mu\varphi \nabla^\mu \varphi
\end{equation}
and $\nabla$ indicates covariant differentiation. This leads to an energy density which is $\rho=2X\,\partial p/\partial X-p$, and hence an equation-of-state parameter $w \equiv p/\rho$ (in units of $c=1$) given by
\begin{equation}
w=\frac{p}{2X\,\partial p/\partial X-p},
\end{equation}
and $p={\cal L}$. A full description of the models investigated and Lagrangians considered is beyond the scope of this work; the reader is directed to the review by \citet{2010AdAst2010E..78B} for more details. Lagrangians of the form
\begin{equation}
{\cal L}(\varphi,X) = f(\varphi)g(X)-V(\varphi), 
\label{UDML}
\end{equation}
where $g(X)$ is a Born-Infeld kinetic term, were considered in a Euclid-like context by \citet{Camera:2010wm}, and models of this form can avoid a strong ISW effect which is often a problem for UDM models \citep[see][and references therein]{2008JCAP...10..023B}. This model is parameterised by a late-time sound speed, $c_\infty$, and its influence on the matter power spectrum is illustrated in Fig. \ref{UDMPk}. For zero sound speed $\Lambda$CDM is recovered.

\subsection{Euclid Observables}
Of interest for Euclid are the weak lensing and BAO signatures of these models, although the supernova Hubble diagram can also be used \citep{2009MNRAS.397.1935T}. The observable effects come from the power spectrum and the evolution of the equation-of-state parameter of the unified fluid, which affects distance measurements. The observational constraints of the generalised Chaplygin gas have been investigated \citep{2010PhRvD..81f3532P}, with the model already constrained to be close to $\Lambda$CDM with SDSS data and the CMB. The effect on BAO measurements for Euclid has yet to be calculated, but the weak lensing effect has been considered for non-canonical UDM models \citep{2010arXiv1002.4740C}. The change in shape and oscillatory features introduced in the power spectrum allow the sound speed parameter to be constrained very well by Euclid, using 3D weak lensing \citep{2003MNRAS.343.1327H,2007MNRAS.376..771K} with errors $\sim 10^{-5}$ \citep[see also][]{Camera:2009uz}.

%% file: dark_matter/de_dm_writeup.tex
\section{Dark energy and dark matter}\label{dms:de_dm}
In section \ref{models-of-modified-gravity}, we have illustrated the possibility
that dark energy, seen as a dynamical scalar field (quintessence), may interact
with other components in the Universe. When starting from an action such as
Eq.(\ref{mg:cde:action}), the species which interact with quintessence are
characterised by a mass function that changes in time
\citep{Kodama:1985bj,Amendola:1999er,Amendola:2003wa,Pettorino:2008ez}. Here, we
consider the case in which the evolution of cold dark matter (CDM) particles
depends on the evolution of the dark energy scalar field. In this case the
general framework seen in section \ref{models-of-modified-gravity} is specified
by the choice of the function $m_c=m_c(\phi)$. The coupling is not constrained
by tests of the equivalence principle and solar system constraints, and can
therefore be stronger than the coupling with baryons. Typical values of $\beta$
presently allowed by observations (within current CMB data) are within the range
$0< \beta < 0.06$ at $95\%$ CL for a constant coupling and an exponential
potential,
\citep{Bean:2008ac,amendola_etal_2003,Amendola:2003wa,amendola_quercellini_2003}
, or possibly more if neutrinos are taken into account or more realistic
time-dependent choices of the coupling are used
\citep{LaVacca:2009yp,Kristiansen:2009yx}. As mentioned in section \ref{mg:cde},
this framework is generally referred to as `coupled quintessence' (CQ). Various
choices of couplings have been investigated in the literature, including
constant $\beta$
\citep{Amendola:1999er,Mangano:2002gg,Amendola:2003wa,Koivisto:2005nr,Guo:2007zk
,Quartin:2008px,quercellini_etal_2008,Pettorino:2008ez} and varying couplings
\citep{Baldi:2010vv}, with effects on Supernov\ae, CMB and cross-correlation of
the CMB and LSS
\citep{Bean:2008ac,amendola_etal_2003,Amendola:2003wa,amendola_quercellini_2003,
LaVacca:2009yp,Kristiansen:2009yx,Mainini:2010ng}.

The presence of a coupling (and therefore, of a fifth force acting among dark
matter particles) modifies the expansion of the universe, linear perturbations
and most relevantly, structure formation. Coupled quintessence is a concrete
model in which a non-negligible amount of dark energy is present at early times.
The presence of such an early dark energy component is accompanied specific
features, as illustrated in section \ref{models-of-modified-gravity} for a
general framework:
\begin{enumerate}
\item a fifth force ${\bf{\nabla }} \left[\Phi_\alpha + \beta \phi \right]$ with
an effective $\tilde{G}_{\alpha} = G_{N}[1+2\beta^2(\phi)]$;
\item a velocity-dependent term $\tilde{H}{\bf{v}}_{\alpha} \equiv H \left(1 -
{\beta(\phi)} \frac{\dot{\phi}}{H}\right) {\bf{v}}_{\alpha}$;
\item a time-dependent mass for each particle $\alpha$, evolving according to
Eq.(\ref{mass_def}).
\end{enumerate} 
All these effects, and in particular the first two, contribute significantly to
structure formation. Note that the second and third terms are not independent of
each other as they are a direct consequence of momentum conservation. Depending
on the function $m_c(\phi)$, and therefore $\beta(\phi)$, the first two terms
can partially balance: the fifth force increases gravitational attraction whilst
the velocity-dependent term, if the CDM mass decreases with time, tries to
dilute the concentration of the virialised haloes. In particular, a striking
difference between constant and variable-coupling models concerning the
interplay of all these three effects has been highlighted in
\citet{Baldi:2010vv}: whilst for constant couplings only the latter two effects
can alter the virial equilibrium of an already-collapsed object, for the case of
a variable coupling the time evolution of the effective gravitational constant
can also modify the virial status of a halo, and can either enhance or
counteract the effect of reducing halo concentrations (for decreasing and
increasing couplings, respectively). Non-linear evolution within coupled
quintessence cosmologies has been addressed using various methods of
investigation, such as spherical collapse
\citep{Mainini:2006zj,Wintergerst:2010ui,Manera:2005ct,Koivisto:2005nr,
Sutter:2007ky,Abdalla:2007rd,Bertolami:2007tq} and alternative semi-analytic
methods \citep{Saracco_etal_2010,amendola_quercellini_2004}. $N$-body and
hydro-simulations have also been done
\citep{maccio_etal_2004,Baldi_etal_2010,Baldi:2010vv,Baldi_Pettorino_2010,
Baldi:2010ks,Li:2010zw,Li:2010eu,Baldi:2010pq,Zhao:2010dz}.

We list here briefly the main observable features typical of this class of
models:
\begin{itemize}
\item enhanced ISW effect
\citep{Amendola:1999er,Amendola:2003wa,Mainini:2010ng}; such effects may be
partially reduced when taking into account non-linearities, as described in
\citet{Pettorino:2010bv};
\item increase in the number counts of massive clusters at high redshift
\citep{Baldi_Pettorino_2010};
\item scale-dependent bias between baryons and dark matter, which behave
differently if only dark matter is coupled to dark energy
\citep{Baldi_etal_2010,Baldi:2010pq};
\item less steep inner core halo profiles (depending on the interplay between
fifth force and velocity-dependent terms)
\citep{Baldi_etal_2010,Baldi:2010vv,Li:2010zw,Li:2010eu,Baldi:2010pq};
\item lower concentration of the halos
\citep{Baldi_etal_2010,Baldi:2010vv,Li:2010eu};
\item voids are emptier when a coupling is active \citep{Baldi:2010ks}.
\end{itemize}

As discussed in subsection \ref{Nbody_sims}, when a variable coupling $\beta (\phi )$ 
is active the relative balance of the fifth-force and other dynamical effects depends on
 the specific time evolution of the coupling strength. Under such conditions, certain cases
 may also lead to the opposite effect of larger halo inner overdensities and higher concentrations, 
as in the case of a steeply growing coupling function \citep[see][]{Baldi:2010vv}. Alternatively, 
the coupling can be introduced by choosing directly a covariant stress-energy tensor, treating dark 
energy as a fluid in the absence of a starting action \citep{Mangano:2002gg,Valiviita:2008iv,CalderaCabral:2008bx,Schaefer:2008ku,Valiviita:2009nu,
Majerotto:2009np,Gavela:2009cy,CalderaCabral:2009ja,Gavela:2010tm}. For an illustration of non-linear 
effects in the presence of a coupling see section \ref{non-linear-aspects}.

%% file: dark_matter/UltraLightScalarFields.tex
\section{Ultra-light scalar fields}
Ultra-light scalar fields arise generically in high energy physics, most commonly as axions or other axion-like particles (ALPs). They are the Pseudo-Goldstone bosons (PGBs) of spontaneously broken symmetries. Their mass remains protected to all loop orders by a shift symmetry, which is only weakly broken to give the fields a mass and potential, through non perturbative effects. Commonly these effects are presumed to be caused by instantons, as in the case of the QCD axion, but the potential can also be generated in other ways that give potentials that are useful, for example, in the study of quintessence \citep{panda2010}. Here we will be considering a general scenario, motivated by the suggestions of \citet{axiverse2009} and \citet{hu2000}, where an ultralight scalar field constitutes some fraction of the dark matter, and we make no detailed assumptions about its origin.

Axions arise generically in string theory \citep{witten2006}. They are similar to the well known QCD axion \citep{pecceiquinn1977,thooft1976a,thooft1976b,dine1981,preskill1983,steinhardt1983,turner1983,abbott1983,dine1983,turner1986,visinelli2009}, and their cosmology has been extensively studied \citep[see, for example,][]{banks1996}. String axions are the Kaluza-Klein zero modes of anti-symmetric tensor fields, the number of which is given by the number of closed cycles in the compact space: for example a two-form such as $B_{MN}$\footnote{\color{red}$B_{MN}$ is the antisymmetric partner of the metric, which in heterotic string theory gives rise to the Òmodel-independent axionÓ. The indices $M, N$ run over the spacetime dimensions, $0, \dots, D - 1$.} 
%
 has a number of zero modes coming from the number of closed two-cycles. In any realistic compactification giving rise to the Standard Model of particle physics the number of closed cycles will typically be in the region of hundreds. Since such large numbers of these particles are predicted by String Theory, we are motivated to look for their general properties and resulting cosmological phenomenology.

The properties of the axion $\theta$ are entirely determined by its potential $U$, whose specific form depends on details in string theory that will not concern us, and two parameters in the four-dimensional Lagrangian
\begin{equation}
\mathcal{L} = \frac{f_a^2}{2}(\partial \theta)^2 - \Lambda^4 U(a),
\end{equation}
where $f_a$ is the scale at which the Peccei-Quinn-like symmetry -- an additional global $U(1)$ symmetry -- is broken, also referred to as the axion decay constant, and $\Lambda$ is the overall scale of the potential. In terms of the canonically normalised field $\phi = f_a \theta$, we find that the mass is given by
\begin{equation}
m=\frac{\Lambda^2}{f_a}.
\end{equation}
The values of these parameters are determined by the action $S$ of the non-perturbative physics that generates the potential for a given axion, and it was argued in \citet{axiverse2009} that this scales with the volume/area of the closed cycle giving rise to that axion, $S\sim A$. $f_a$ and $S$ are related by
\begin{equation}
f_a\sim\frac{\Mp}{S}.
\end{equation}
$f_a$ is typically of order $10^{16}$GeV and can be considered constant for all string axions \citep{witten2006}. However, the mass of each axion depends exponentially on $S$ from
\begin{equation}
\Lambda^4=\mu^4e^{-S},
\end{equation} 
where $\mu$ sets the scale of the non-perturbative physics (essentially, the Planck Scale in the string case), and so, as $S$ varies from axion to axion depending on the cycle areas in the compact space, we expect axion masses to evenly distribute on a logarithmic mass scale all the way down to the Hubble scale today, $H_0\sim 10^{-33}$eV \citep{axiverse2009}.

There will be a small thermal population of ALPs, but the majority of the cosmological population will be cold and non-thermally produced. Production of cosmological ALPs proceeds by the vacuum realignment mechanism. When the Peccei-Quinn-like $U(1)$ symmetry is spontaneously broken at the scale $f_a$ \label{symbol:f_a} the ALP acquires a vacuum expectation value, the misalignment angle $\theta_i$, uncorrelated across different causal horizons. However, provided that inflation occurs after symmetry breaking, and with a reheat temperature $T\lesssim f_a$, then the field is homogenised over our entire causal volume. This is the scenario we will consider. The field $\theta$ is a PGB and  evolves according to the potential $U$ acquired at the scale $\mu$. However, a light field will be frozen at $\theta_i$ until the much later time when the mass overcomes the Hubble drag and the field begins to roll towards the minimum of the potential, in exact analogy to the minimum of the instanton potential restoring $\mathcal{CP}$ invariance in the Peccei-Quinn mechanism for the QCD axion. Coherent oscillations about the minimum of $U$ lead to the production of the weakly coupled ALPs, and it is the value of the misalignment angle that determines the cosmological density in ALPs \citep{linde1991, hertzberg2008, sikivie2008}.

The underlying shift symmetry restricts $U$ to be a periodic function of $\theta$ for true axions, but since in the expansion all couplings will be suppressed by the high scale $f_a$, and the specific form of $U$ is model-dependent, we will make the simplification to consider only the quadratic mass term as relevant in the cosmological setting, though some discussion of the effects of anharmonicites will be made. In addition, \citet{panda2010} have constructed non-periodic potentials in string theory.

Scalar fields with masses in the range $10^{-33}$eV$<m<10^{-22}$eV are also well-motivated dark matter candidates independently of their predicted existence in string theory, and constitute what Hu has dubbed ``fuzzy cold dark matter'', or FCDM \citep{hu2000}. The Compton wavelength of the particles associated to ultra-light scalar fields, $\lambda_c = 1/m$ in natural units, is of the size of galaxies or clusters of galaxies, and so the uncertainty principle prevents localisation of the particles on any smaller scale. This naturally suppresses formation of structure and serves as a simple solution to the problem of ``cuspy halos'' and the large number of dwarf galaxies, which are not observed and are otherwise expected in the standard picture of CDM. Sikivie has argued \citep{sikivie2010b} that axion dark matter fits the observed caustics in dark matter profiles of galaxies, which cannot be explained by ordinary dust CDM.

The large phase space density of ultralight scalar fields causes them to form Bose-Einstein condensates \citep[see][and references therein]{sikivie2009} and allows them to be treated as classical fields in a cosmological setting. This could lead to many interesting, and potentially observable phenomena, such as formation of vortices in the condensate, which may effect halo mass profiles \citep{silverman2002,kain2010}, and black hole super radiance \citep{axiverse2009,arvanitaki2010,rosa2010}, which could provide direct tests of the ``string axiverse'' scenario of \citet{axiverse2009}. In this summary we will be concerned with the large-scale indirect effects of ultra-light scalar fields on structure formation via the matter power spectrum in a cosmology where a fraction $f=\Omega_a/\Omega_m$ of the dark matter is made up of such a field, with the remaining dark matter a mixture of any other components but for simplicity we will here assume it to be CDM so that $(1-f)\Omega_m = \Omega_c$.

If ALPs exist in the high energy completion of the standard model of particle physics, and are stable on cosmological time scales, then regardless of the specifics of the model \citet{tegmark2006} have argued that on general statistical grounds we indeed expect a scenario where they make up an order one fraction of the CDM, alongside the standard WIMP candidate of the lightest supersymmetric particle. However, it must be noted that there are objections when we consider a population of light fields in the context of inflation \citep{mack2009a,mack2009b}. The problem with these objections is that they make some assumptions about what we mean by ``fine tuning'' of fundamental physical theories, which is also related to the problem of finding a measure on the landscape of string theory and inflation models \citep[see, for example,][]{linde2010}, the so-called ``Goldilocks problem.'' Addressing these arguments in any detail is beyond the scope of this summary.

We conclude with a summary of the most important equations and properties of ultra-light scalar fields.
\begin{itemize}
\item In conformal time and in the synchronous gauge with scalar perturbation $h$ as defined in \citet{bertschinger1995}, a scalar field with a quadratic potential evolves according to the following equations for the homogeneous, $\phi_0(\tau)$, and first order perturbation, $\phi_1(\tau,\vec{k})$, components
\begin{align}
\ddot{\phi}_0 + 2\mathcal{H}\dot{\phi}_o + m^2 a^2 \phi_0 &= 0 \label{eqn:phi0}, \\
\ddot{\phi}_1 +2\mathcal{H} \dot{\phi}_1 +(m^2 a^2 + k^2)\phi_1 &= -\frac{1}{2}\dot{\phi}_0\dot{h} \label{eqn:phi1};
\end{align}
\item In cosmology we are interested in the growth of density perturbations in the dark matter, and how they effect the expansion of the universe and the growth of structure. The energy-momentum tensor for a scalar field is
\begin{equation}
T^{\mu}_{\ \ \nu} = \phi^{;\mu}\phi_{;\nu} - {\frac{1}{2}}(\phi^{;\alpha} \phi_{;\alpha} +2V)\delta^{\mu}_{\: \nu}
\end{equation}
which to first order in the perturbations has the form of a perfect fluid and so we find the density and pressure components in terms of $\phi_0$, $\phi_1$,
\begin{align}
\rho_a = &\frac{a^{-2}}{2}\dot{\phi}_0^2 + \frac{m^2}{2}\phi_0^2 \label{eqn:rhoa}, \\
\delta\rho_a =& a^{-2}\dot{\phi}_0\dot{\phi}_1 + m^2 \phi_0 \phi_1 \label{eqn:deltarho},\\
P_a =& \frac{a^{-2}}{2}\dot{\phi}_0^2 - \frac{m^2}{2}\phi_0^2 \label{eqn:pa},\\
\delta P_a =& a^{-2}\dot{\phi}_0 \dot{\phi}_1 - m^2\phi_0 \phi_1 \label{eqn:deltap}, \\
(\rho + P)\theta_a =&a^{-2}k^2 \dot{\phi}_0 \phi_1;
\end{align}
\item The scalar field receives an initial value after symmetry breaking and at early times it remains frozen at this value by the Hubble drag. A frozen scalar field behaves as a cosmological constant; once it begins oscillating it will behave as matter. A field begins oscillating when
\begin{equation}
H(t) < m;
\end{equation}
\item Do oscillations begin in the radiation or matter dominated era? The scale factor at which oscillations begin, $a_{osc}$, is given by
\begin{align}
a_{osc} &= \left ( \frac{t_{eq}}{t_0} \right)^{1/6} \left( \frac{1}{m t_0} \right)^{1/2},\qquad m \gtrsim 10^{-27}\mathrm{eV} \nonumber, \\
a_{osc}&= \left ( \frac{1}{m t_0} \right)^{2/3}, \qquad m \lesssim 10^{-27}\mathrm{eV} \nonumber; \\
\end{align}
\item If oscillations begin in the matter-dominated era then the epoch of equality will not be the same as that inferred from the matter density today. Only CDM will contribute to the matter density at equality, so that the scale factor of equality is given by
\begin{equation}
a_{eq} \simeq \frac{\Omega_{r}}{\Omega_m}\frac{1}{(1-f)};
\end{equation}
\item The energy density today in such an ultralight field can be estimated from the time when oscillations set in and depends on its initial value as
\begin{equation}
\Omega_a = \frac{1}{6} \left( \frac{1}{t_0} \right)^2 \phi_0(t_i)^2,
\end{equation}
while fields that begin oscillations in the radiation era also have a mass dependence in the final density as $\sim m^{1/2}$;
\item In the context of generalized dark matter \citep{hu1998b} we can see the effect of the Compton scale of these fields through the fluid dynamics of the classical field. The sound speed of a field with momentum $k$ and mass $m$ at a time when the scale factor of the FLRW metric is $a$ is given by
\begin{align}
c_s^2 &= \frac{k^2}{4m^2 a^2}, \quad k<2ma \nonumber, \\
c_s^2 &= 1, \quad k>2ma \nonumber .\\
\label{eqn:cssquared}
\end{align}
On large scales the pressure becomes negligible, the sound speed goes to zero and the field behaves as ordinary dust CDM and will collapse under gravity to form structure. However on small scales, set by $\lambda_c$, the sound speed becomes relativistic, suppressing the formation of structure;
\item This scale-dependent sound speed will affect the growth of overdensities, so we ask: are the perturbations on a given scale at a given time relativistic? The scale
\begin{equation}
k_R=ma(t)
\end{equation}
separates the two regimes. On small scales: $k>k_R$ the sound speed is relativistic. Structure formation is suppressed in modes that entered the horizon whilst relativistic.
\item Time dependence of the scale $k_R$ and the finite size of the horizon mean that suppression of structure formation will accumulate on scales larger than $k_R$. For the example of ultralight fields that began oscillations in the matter-dominated regime, we calculate that suppression of structure begins at a scale
\begin{equation}
k_m \sim \left( \frac{m}{10^{-33}\mathrm{eV}} \right)^{1/3}\left( \frac{100\mathrm{km}\mathrm{s}^{-1}}{c} \right) h\mathrm{Mpc}^{-1},
\end{equation}
which is altered to $k_m \sim m^{1/2}$ for heavier fields that begin oscillations in the radiation era \citep{2006PhLB..642..192A};
\item The suppression leads to steps in the matter power spectrum, the size of which depends on $f$. The amount of suppression can be estimated, following \citet{2006PhLB..642..192A}, as
\begin{equation}
S(a)=\left( \frac{a_{osc}}{a} \right)^{2(1-1/4(-1+\sqrt{25-24f}))}.
\end{equation}
As one would expect, a larger $f$ gives rise to greater suppression of structure, as do lighter fields that free-stream on larger scales.
\end{itemize}

Numerical solutions to the perturbation equations indeed show that the effect of ultralight fields on the growth of structure is approximately as expected, with steps in the matter power spectrum appearing. However, the fits become less reliable in some of the most interesting regimes where the field begins oscillations around the epoch of equality, and suppression of structure occurs near the turnover of the power spectrum, and also for the lightest fields that are still undergoing the transition from cosmological constant to matter-like behaviour today \citep{marsh2010}. These uncertainties are caused by the uncertainty in the background expansion during such an epoch. In both cases a change in the expansion rate away from the expectation of the simplest $\Lambda$CDM model is expected. During matter and radiation eras the scale factor grows as $a\sim \tau^p$ and $p$ can be altered away from the $\Lambda$CDM expectation by $\mathcal{O}(10)\%$ by oscillations caused during the scalar field transition, which can last over an order of magnitude in scale factor growth, before returning to the expected behaviour when the scalar field is oscillating {\color{red}sufficiently} rapidly and behaves as CDM.

The combined CMB-large scale structure  likelihood analysis of \citet{2006PhLB..642..192A} has shown that ultralight fields with mass around $10^{-30}-10^{-24}$eV might account for up to $10\%$ of the dark matter abundance.

\subsection{Requirements}
Ultralight fields are similar in many ways to massive neutrinos \citep{2006PhLB..642..192A}, the major difference being that their non-thermal production breaks the link between the scale of suppression, $k_m$, and the fraction of dark matter, $f_{ax}$, through the dependence of $f_{ax}$ on the initial field value $\phi_i$. Therefore an accurate measurement of the matter power spectrum in the low-$k$ region where massive neutrinos corresponding to the WMAP limits on $\Omega_\nu$ are expected to suppress structure will determine whether the expected relationship between $\Omega_\nu$ and $k_m$ holds. These measurements will limit the abundance of ultralight fields that begin oscillations in the matter-dominated era.

Another powerful test of the possible abundance of ultralight fields beginning oscillations in the matter era will be an accurate measure of the position of the turn over in the matter power spectrum, since this gives a handle on the species present at equality. Ultralight fields with masses in the regime such that they begin oscillations in the radiation-dominated era may suppress structure at scales where the BAO are relevant, and thus distort them. An accurate measurement of the BAO that fits the profile in $P(k)$ expected from standard $\Lambda$CDM would place severe limits on ultralight fields in this mass regime.

\begin{figure}
\centering
\includegraphics[width=0.75\textwidth]{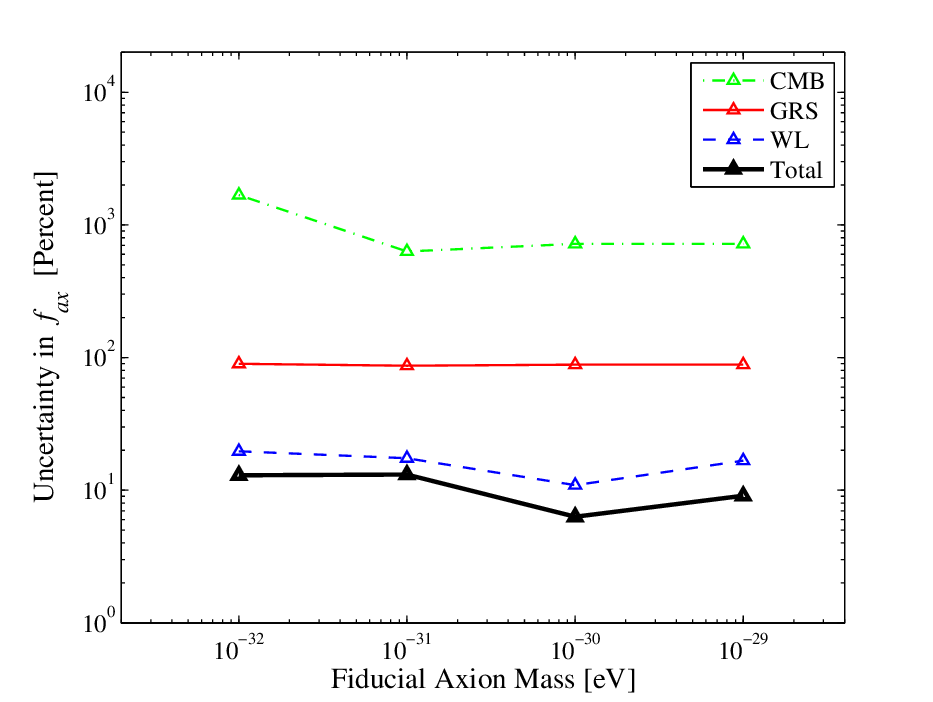}
\caption{Marginalised uncertainty in $f_{ax}$ for CMB (green), a galaxy redshift survey (red), weak lensing (blue) and the total (black) evaluated for four different fiducial axion masses, for the cosmology $\Lambda$CDM+$f_{ax}$+$\nu$ \citep[from][]{Marsh:2011bf}.}\label{fig:axion_constraints}
\end{figure}

{\color{red} Recently, \citet{Marsh2012} showed that with current and next generation galaxy surveys alone it should be possible to unambiguously detect a fraction of dark matter in axions of the order of $1\%$ of the total. Furthermore, they demonstrated that the tightest constraints on the axion fraction $f_{ax}$ come from weak lensing; when combined with a galaxy redshift survey, constraining $f_{ax}$ to $0.1\%$ should be possible, see Fig.\ref{fig:axion_constraints}. 
The strength of the weak lensing constraint depends on the photometric redshift measurement, i.e. on tomography. 
Therefore, lensing tomography will allow Euclid -- {\color{red}through} the measurement of the growth rate -- {\color{red}to resolve} the redshift evolution of the axion suppression of small scale convergence power. Further details can be found in \citet{Marsh2012, Marsh:2011bf}.}

Finally, the expected suppression of structure caused by ultralight fields should be properly taken into account in $N$-body simulations. The non-linear regime of $P(k)$ needs to be explored further both analytically and numerically for cosmologies containing exotic components such as ultralight fields, especially to constrain those fields which are heavy enough such that $k_m$ occurs around the scale where non-linearities become significant, i.e. those that begin oscillation deep inside the radiation-dominated regime. For lighter fields the effects in the non-linear regime should be well-modelled by using the linear $P(k)$ for $N$-body input, and shifting the other variables such as $\Omega_c$ accordingly.

%% file: dark_matter/multifield.tex
\section{Dark-matter surrogates in theories of modified gravity}
\subsection{Extra fields in modified gravity}
The idea that the dark Universe may be a signal of modified gravity has led to the development of a plethora of theories. From polynomials in curvature invariants, preferred reference frames, UV and IR modifications and extra dimensions, all lead to significant modifications to the gravitational sector. A universal feature that seems to emerge in such theories is the existence of fields that may serve as a proxy to dark matter. This should not be unexpected. On a case by case basis, one can see that modifications to gravity generically lead to extra degrees of freedom. 

For example, polynomials in curvature invariants lead to higher derivative theories which inevitably imply extra (often unstable) solutions that can play the role of dark matter. This can be made patently obvious when mapping such theories onto the Einstein frame with an addition scalar field (Scalar-Tensor theories). Einstein-Aether theories \citep{ZlosnikFerreiraStarkman2008} explicitly introduce an extra time-like vector field. The time-like constraint locks the background, leading to modifications to the background expansion; perturbations in the vector field can, under certain conditions, lead to growth of structure, mimicking the effect of pressureless dark matter. The vector field plays the same role in TeVeS \citep{Bekenstein2004a}, where two extra fields are introduced to modify the gravitational dynamics. And the same effects come into play in bigravity models \citep{Banadosetal2009} where two metrics are {\color{red}explicitly} introduced -- the scalar modes of the second metric can play the role of dark matter.

In what follows we briefly focus on three of the above cases where extra gravitational degrees of freedom play the role of dark matter: Einstein-Aether models, TeVeS models and bigravity models. We will look at the Einstein-Aether model more carefully and then discuss briefly the other two cases.

\subsection{Vector dark matter in Einstein-Aether models}
As we have seen in a previous section, Einstein-Aether models introduce a time-like vector field $A^{a}$ into gravitational dynamics. The four vector $A^{a}$ can be expanded as $A^{\mu}=(1+\epsilon X,\epsilon\partial^{j}Z)= (1+\epsilon X,\frac{\epsilon}{a^{2}}\partial_{j}Z)$ \citep{ZlosnikFerreiraStarkman2008}. In Fourier space we have $A^{\mu}=(1-\epsilon \Psi,i\frac{\epsilon}{a} k_{j}V)$, where, for computational convenience, we have defined $V\equiv Z / a$ and have used the fact that the constraint fixes $X=-\Psi$.

The evolution equation for the perturbation in the vector field becomes (where primes denote derivatives with respect to conformal time)
\begin{align}
\label{vecp}
0 &= c_{1}[V''+k^{2}V+2{\cal H}V'+2{\cal H}^{2}V+\Psi'+\Phi '+2{\cal H}\Psi] \\
\nonumber & +c_2[k^{2}V+6{\cal H}^{2}V-3\frac{a''}{a}V+3\Phi'+3{\cal H}\Psi]\\
\nonumber & +c_3[k^{2}V+2{\cal H}^{2}V-\frac{a''}{a}V+\Phi'+{\cal H}\Psi]\\
\nonumber & +\frac{F_{KK}}{F_{K}}[-K^{\epsilon}\alpha{\cal H}-K^{0'}(-c_{1}(V'+\Psi)+3c_{2}{\cal H} V+c_{3}{\cal H} V)].
\end{align}
The perturbation in the vector field is sourced by the two gravitational potentials $\Phi$ and $\Psi$ and will in turn source them through Einstein's equations. The Poisson equation takes the form
\begin{align}
\label{eqn:Poisson}
k^{2}\Phi&=-\frac{1}{2}F_{K}c_{1}k^{2}[V'+\Psi+(3+2\tilde{c}_{3}){\cal H}V]\\
\nonumber & -4\pi Ga^{2}\sum_{a}(\bar{\rho}_{a}\delta_{a}+3(\bar{\rho}_{a}+\bar{P}_{a}){\cal H}\frac{\theta_{a}}{k^{2}}).
\end{align}

To understand why the vector field can play the role of dark matter it is instructive to study the effect of the vector field during matter domination. It should give us a sense of how in the generalised Einstein-Aether case, the growth of structure is affected. Let us consider the simplest case in which the the dominant remaining contribution to the energy density is baryonic, treated as a pressureless perfect fluid with energy-momentum tensor $\mathbf{T}$ and let us introduce the variable $V'\equiv E$. For ease of illustration we will initially consider only the case where $V$ is described by a growing monomial, i.e. $V=V_{0}(\eta/\eta_0)^p$. During the matter era we have 
\begin{align}
a^{2}\delta T^{0}_{\phantom{0}0} &\simeq -l_{E}\xi(k)k^{2}\eta^{5+p-6n} \label{too} \\ 
k^{2}(\Psi-\Phi) &\simeq -l_{S}\xi(k)k^{2}\eta^{5+p-6n} 
\end{align}
with $l_{E} \equiv -(c_{1}(2+p)n+2\alpha(1-2n)n)$, $l_{S} \equiv -(c_{1}+c_{3})n(6n-p-10)$, and
\begin{equation}
\xi(k) \sim \gamma V_{0}(k)\eta_0^{-p}k_\mathrm{hub}^{6-6n}\left[3\alpha\Omega_{m}\left(\frac{H_{0}}{M}\right)^{2}\right]^{n-1} \label{xixi}
\end{equation}
where $k_\mathrm{hub}\equiv1/\eta_\mathrm{today}$. Hence, the vector field affects our evolution equations for the matter and metric perturbations only through its contribution to the energy density and its anisotropic stress. On large scales, $k \eta \ll1$, and assuming adiabatic initial conditions for the fields $\delta$, $\Phi$ and $\theta$, this leads to 
\begin{equation}
\delta= C_{1}(k)+\frac{6l_{S}\xi(k)}{(10+p-6n)}\eta^{5+p-6n} 
\end{equation}
where $C_{1}$ is a constant of integration and we have omitted the decaying mode. Therefore, even before horizon crossing, the anisotropic stress term due to the vector field can influence the time evolution of the baryon density contrast.

On small scales ($k\eta \gg1$), we find
\begin{equation}
\delta(k,\eta) = C_{2}(k)\eta^{2} +\frac{(\frac{1}{2}l_{E}+l_{S})}{(5+p-6n)(10+p-6n)}\xi(k)(k\eta)^{2}\eta^{5+p-6n}
\end{equation}
where $C_{2}(k)$ is a constant of integration. Hence, for sub-horizon modes, the influence of the vector field on the evolution of $\delta$ is a combination of the effect of the energy density and anisotropic stress contributions though both, in this limit, result in the same contributions to the scale dependence and time evolution of the density contrast. The net effect is that, for particular choices of parameters in the action, the perturbations in the vector field can enhance the growth of the baryon density contrast, very much along the lines of dark matter in the dark matter dominated scenario.

\subsection{Scalar and tensors in TeVeS}
We have already come across the effect of the extra fields of TeVeS. Recall that, in TeVeS, as well as a metric (tensor) field, there is a time-like vector field and a scalar field both of which map the two frames on to each other. While at the background level the extra fields contribute to modifying the overall dynamics, they do not contribute significantly to the overall energy density. This is not so at the perturbative level. The field equations for the scalar modes of all three fields can be found in the conformal Newtonian gauge in \citet{SkordisEtAl2006}. While the perturbations in the scalar field will have a negligible effect, the space-like perturbation in the vector field has an intriguing property: it leads to growth. \citet{DodelsonLiguori2006} have shown that the growing vector field  feeds into the Einstein equations and gives rise to a growing mode in the gravitational potentials and in the baryon density. Thus, baryons will be aided by the vector field leading to an effect akin to that of pressureless dark matter. The effect is very much akin to that of the vector field in Einstein-Aether models -- in fact it is possible to map TeVeS models onto a specific subclass of Einstein-Aether models. Hence the discussion above for Einstein-Aether scenarios can be used in the case of TeVeS.

\subsection{Tensor dark matter in models of bigravity}
In bigravity theories \citep{Banadosetal2009}, one considers two metrics: a dynamical metric $g_{\mu\nu}$ and a background metric, ${\tilde g}_{\alpha\beta}$. As in TeVeS, the dynamical metric is used to construct the energy-momentum tensor of the non-gravitational fields and is what is used to define the geodesic equations of test particles. The equations that define its evolution are usually not the Einstein field equations but may be defined in terms of the background metric.

Often one has that ${\tilde g}_{\alpha\beta}$ is dynamical, with a corresponding term in the gravitational action. It then becomes necessary to link ${\tilde g}_{\alpha\beta}$ to $g_{\mu\nu}$ with the background metric determining the field equations of the dynamical metric through a set of interlinked field equations. In bigravity models both metrics are used to build the Einstein-Hilbert action even though only one of them couples to the matter content. A complete action is of the form
\begin{equation}
S=\frac{1}{16 \pi G}\int d^4 x\left [\sqrt{-g}(R-2\Lambda)+\sqrt{-{\tilde g}}
({\tilde R}-2{\tilde \Lambda})-\sqrt{-{\tilde g}}\frac{1}{\ell^2}({\tilde g}^{-1})^{\alpha\beta}g_{\alpha\beta}\right ]
\end{equation}
where $\Lambda$ and ${\tilde \Lambda}$ are two cosmological constant terms and $\ell^2$ defines the strength of the linking term between the two actions. The cosmological evolution of perturbations in these theories has been worked out in some detail. It turns out that perturbations in the auxiliary field can be rewritten in the form of a generalised dark matter fluid \citep{Hu1998} with fluid density, momentum, pressure and shear that obey evolution equations which are tied to the background evolution. As a result, it is possible to work out cosmological observables such as perturbations in the CMB and large scale structure. If we restrict ourselves to a regime in which ${\tilde \rho}$ simply behaves as dark matter, then the best-fit bimetric model will be entirely indistinguishable from the standard CDM scenario.

%% file: dark_matter/Outlook.tex
\section{Outlook}
Dark matter dominates the matter content of the Universe, and only through astrophysical and cosmological observations can the nature of dark matter on large scales be determined. In this review, we have discussed a number of observational techniques available to Euclid: dark matter mapping, complementarity with other astronomical observations (e.g. X-ray and CMB experiments); cluster and galaxy scale dark matter halo mapping; and power spectrum analyses. The techniques described will allow Euclid to constrain a variety of dark matter candidates and their microphysical properties. We have discussed Warm Dark Matter scenarios, axion-like dark matter, scalar field dark matter models (as well as the possible interactions between dark energy and scattering with ordinary matter) and massive neutrinos (the only known component of dark matter). 

{\color{red} Here, we briefly list the main dark matter constraints so far forecasted for Euclid:
\begin{itemize}
\item The weak lensing power spectrum from Euclid will be able to constrain warm dark matter particle mass to about $m_\mathrm{WDM}>2\,\mathrm{keV}$ \citep{Markovic:2010te};
\item The galaxy power spectrum, with priors from Planck (primary CMB only), will yield an error on the sum of neutrino masses $\Sigma$ of $0.04\,\mathrm{eV}$  (see Table~\ref{summary}; \citealt{Carbone/etal:2010});
\item Euclid's weak lensing should also yield an error on $\Sigma$ of $0.05\,\mathrm{eV}$ \citep{Kitching/etal:2008};
\item \citet{Jimenez:2010ev} have shown that weak gravitational lensing from Euclid data will be able to determine neutrino hierarchy (if $\Sigma<0.13$);
\item The forecasted errors on the effective number of neutrino species $N_{\nu,\mathrm{eff}}$ for Euclid (with a Planck prior) are $\pm0.1$ \citep[for weak lensing][]{Kitching/etal:2008} and $\pm0.086$ \citep[for galaxy clustering][]{Carbone/etal:2010};
\item The sound speed of unified dark energy-dark matter can be constrained with errors $\sim10^{-5}$ by using 3D weak lensing \citep{2010arXiv1002.4740C};
\item Recently, \citet{Marsh2012} showed that with current and next generation galaxy surveys alone it should be possible to unambiguously detect a fraction of dark matter in axions of the order of $1\%$ of the total;
\end{itemize}
}

We envisage a number of future scenarios, all of which give Euclid an imperative to confirm or identify the nature of dark matter. In the event that a dark matter candidate is discovered in direct detection experiments or an accelerator (e.g. LHC) a primary goal for Euclid will be to confirm, or refute, the existence of this particle on large scales. In the event that no discovery is made directly, then astronomical observations will remain our only way to determine the nature of dark matter.

%% file: ini_cond/ICmerged.tex
\chapter{Initial conditions}\label{ini-cond}



 \section{Introduction}
The exact origin of the primordial perturbations that seeded the formation of the large-scale 
structure in the Universe is still unknown. 
Our current understanding of the initial conditions is based on inflation, a phase of accelerated expansion preceding the standard evolution of the Universe (\cite{Guth:1981,Starobinsky:1979,Starobinsky:1982,Sato:1981}). In particular, inflation explains why the Universe is so precisely flat, homogeneous and isotropic.
During this phase, scales much smaller than the Hubble radius are inflated to super-horizon sizes, so that regions appearing today as causally disconnected were in fact very close in the past. This mechanism is also at the origin of the cosmic large scale structure. Vacuum quantum fluctuations of any light field present during inflation are amplified by the accelerated expansion and {\em freeze-out} on super-Hubble scales acquiring a quasi-scale invariant spectrum (\cite{Mukhanov/Chibisov1981,Hawking:1982,Starobinsky:1982,Guth/Pi:1982,Bardeen/Steinhardt/Turner:1983}). 

From the early development of inflation, the simplest proposal based on a weakly-coupled single field rolling along its potential (\cite{Linde:1982,Albrecht/Steinhardt:1982}) has gained strength and many models have been built based on this picture (see for instance \cite{Linde:2008} for a review). Although some inflationary potentials are now excluded by current data (see for instance \cite{KomatsuWMAP7}), this scenario has been extremely successful in passing many observational tests: it predicts perfectly adiabatic and almost Gaussian fluctuations with a quasi scale-invariant spectrum and a small amount of gravitational waves. 

While current data have ruled out some classes of inflationary models, the next qualitative step 
forward is investigating the physics responsible for inflation: we still lack a complete understanding of the high energy physics describing it.   In fact, most likely the physics of inflation is far out of reach of terrestrial experiments, many orders of magnitude  
larger than the centre-of-mass energy at the Large Hadron Collider.
Thus, cosmological tests of inflation offer a unique  opportunity to  learn about ultra-high energy physics.
We can do this by targeting observations which directly probe the dynamics of inflation. One route is  to accurately measure the shape of the primordial power spectrum of scalar perturbations produced during the phase of accelerated
expansion, which is directly related to the shape of the inflaton potential, and to constrain the amplitude of the corresponding stochastic gravitational-wave background, which is related instead to the energy-scale of inflation.

A complementary approach is offered by constraining -- or exploring -- how much the distribution of primordial density perturbations departs from Gaussian statistics and purely adiabatic fluctuations. Indeed, future large-scale structure surveys like Euclid can probe these features with an unprecedented accuracy, thus providing a way to test aspects of inflationary physics that are not easily accessible otherwise. Non-Gaussianity is a very sensitive probe of self-couplings and interactions between the fields generating the primordial perturbations, whereas the presence of isocurvature modes can teach us about the number of fields present during inflation and their role in reheating and generating the matter in the Universe. 

Furthermore, non-minimal scenarios or proposals even radically different
from single-field inflation are still compatible with the data. In order to
learn something about the physics of the early Universe we need to rule out
or confirm the conventional slow-roll scenario and possibly discriminate
between non-conventional models. Non-Gaussianities and isocurvature
perturbations currently represent the best tools that we have to accomplish
this task. Any deviation from the conventional Gaussian and adiabatic
initial perturbations would represent important breakthroughs in our
understanding of the early Universe. In this chapter we are going to review
what we can learn by constraining the initial conditions with a large-scale
structure survey such as {\color{red}like } Euclid.

 \section{Constraining inflation}

The spectrum of cosmological perturbations represents an important source of information on the early Universe. During inflation scalar (compressional) and tensor (purely gravitational) fluctuations are produced. The shape and the amplitude of the power spectrum of scalar fluctuations can be related to the dynamics of the inflationary phase, providing a window on the inflaton potential. Inflation generically predicts a deviation from a purely scale-invariant spectrum. Together with future CMB experiments such as Planck, Euclid will improve our constraints on the scalar spectral index and its running, helping to pin down the model of inflation.

\subsection{Primordial perturbations from inflation}

It is convenient to describe primordial perturbations using the so-called curvature perturbation on uniform density hypersurfaces $\zeta$ introduced in \cite{Bardeen/Steinhardt/Turner:1983}. An important property of this quantity is that for adiabatic perturbations -- i.e.~in absence of isocurvature perturbations, discussed in Sec.~\ref{sec:isocurvature} --  it remains constant on super-Hubble scales, allowing us to connect the early inflationary phase to the late time Universe observations, regardless of the details of reheating. In a gauge where the energy density of the inflaton vanishes, we can define $\zeta$ from the spatial part of the metric (assuming a flat FRW Universe), as \citep{Salopek/Bond:1990,Maldacena:2003}
\begin{equation}\label{eq:zeta}
g_{ij} = a^2(t) \exp \left( 2 \zeta \right) \delta_{ij}\;.
\end{equation}
This definition, where $\zeta$ enters the metric in the exponential form, has the advantage that it is valid also beyond linear order and can be consistently used when discussing non-Gaussian fluctuations, such as in Sec.~\ref{sec:NG-TH}. 

The power spectrum of primordial perturbations is given by 
\begin{equation}
\langle \zeta_{\bf{k}} \zeta_{{\bf k}'} \rangle = (2 \pi)^3 \delta({\bf k}+{\bf k}') P_{\zeta}(k)\;,
\end{equation}
where $\langle \ldots \rangle$ denotes the average over an ensemble of
realizationso {\color{red}realizations}. It is useful to define a dimensionless spectrum as ${\cal P}_s(k) \equiv \frac{k^3}{2\pi^2} P_\zeta (k)\;$, where the index $s$ stands for scalar, to distinguish it from the spectrum of tensor perturbations, defined below.
The deviation from scale-invariance of the scalar spectrum is characterized by the spectral index $n_s$, defined by  (see for instance \cite{Liddle/Lyth:2000})
\begin{equation}
\label{eq:index_ns}
n_s \equiv 1 + \frac{d  \ln {\cal P}_s}{d \ln k}\;,
\end{equation}
where $n_s=1$ denotes a purely scale-invariant spectrum. We also define the running of the spectral index $\alpha_s$ as
\begin{equation}
\label{eq:running}
\alpha_s \equiv \frac{d n_s}{d \ln k}\;.
\end{equation}
These quantities are taken at a particular pivot scale. For our analysis we chose it to be $k_* \equiv 0.05 \, {\rm Mpc}^{-1}$. Thus, with these definitions the power spectrum can be written as
\begin{equation}
P_\zeta(k) = \frac{2 \pi^2}{k^3} A_s(k_*) (k/k_*)^{n_s(k_*) -1 + \frac12 \alpha_s(k_*) \ln (k/k_*) }\;,
\label{eq:shape}
\end{equation}
where $A_s$ is the normalization parameterizing the amplitude of the fluctuations. 

During inflation tensor modes are also generated. They are described by the gauge invariant metric perturbation $h_{ij}$, defined from the spatial part of the metric as
\begin{equation}
g_{ij} = a^2(t) \left( \delta_{ij} + h_{ij} \right)\;, \quad h^j_{i,j} = 0 = h_{i}^i\;.
\end{equation}
Each mode has $2$ polarizations, $h_+$ and $h_\times$, each with power spectrum given by 
\begin{equation}
\langle h_{\bf{k}} h_{{\bf k}'} \rangle = (2 \pi)^3 \delta({\bf k}+{\bf k}') P_h(k)\;.
\end{equation}
Defining the dimensionless power spectrum of tensor fluctuations as ${\cal P}_t(k) \equiv 2 \frac{k^3}{2\pi^2} P_h (k)\;$, where the factor of $2$ comes from the two polarizations, it is convenient to define the ratio of tensor to scalar fluctuations as 
\begin{equation}
\label{eq:r_ts}
r\equiv {{\cal P}_t (k_*)}/{{\cal P}_s (k_*)}\;.
\end{equation}

The form of the power spectrum given in eq.~\eqref{eq:shape} approximates very well power spectra of perturbations generated by slow-roll models. In particular, the spectrum of scalar fluctuations  is given in terms of the Hubble rate $H$ and the first slow-roll parameter $\epsilon \equiv - \dot H/H^2$, both evaluated at the time when the comoving scale $k$ crosses the Hubble radius during inflation,
\begin{equation}
{\cal P}_s(k) = \left.\frac{1}{8 \pi^2 \epsilon}\frac{H^2}{M_{\rm Pl}^2} \right|_{k=aH}\;.
\end{equation}
During slow-roll, $\epsilon$ is related to the first derivative of the inflaton potential $V(\phi)$, $\epsilon \approx \frac{M_{\rm Pl}^2}{2} \left( \frac{V'}{V} \right)^2$, where the prime denotes differentiation with respect to $\phi$. As $H$ and $\epsilon$ vary slowly during inflation, this spectrum is almost scale-invariant. Indeed, the scalar spectral index $n_s$ in eq.~\eqref{eq:index_ns} reads
\begin{equation}
n_s=1 - 6 \epsilon + 2 \eta_V \;,
\end{equation}
where the second slow-roll parameter $\eta_V \equiv M_{\rm Pl}^2 \frac{V''}{V} $ must be small for inflation to yield a sufficient number of $e$-foldings. The running of the spectral index defined in eq.~\eqref{eq:running} is even smaller, being second-order in the slow-roll parameters. It is given by $\alpha_s = 16\epsilon \eta_V - 24 \epsilon^2- 2\xi_V$ where we have introduced the third slow-roll parameter $\xi_V \equiv M_{\rm Pl}^4 \frac{V' V''' }{V^2} $.

The spectrum of tensor fluctuations is given by
\begin{equation}
{\cal P}_t(k) = \left.\frac{2}{\pi^2 }\frac{H^2}{M_{\rm Pl}^2} \right|_{k=aH}\;,
\end{equation}
which shows that the ratio of tensor to scalar fluctuations in eq.~\eqref{eq:r_ts} is simply related to the first slow-roll parameter by $r =  16 \epsilon$.

As a fiducial model, in the next section we will consider chaotic inflation \citep{Linde:1983}, based on the quadratic inflaton potential $V = \frac12 m^2 \phi^2$. In this case, the first two slow-roll parameters are both given in terms of the value of the inflaton field at Hubble crossing $\phi$ or, equivalently, in terms of number of $e$-folds from Hubble crossing to the end of inflation $N$, as $\epsilon=\eta_V = 2 M_{\rm Pl}^2 /\phi^2 = 1/2 N$, while $\xi_V =0$. This implies \label{symbol:efolds}
\begin{equation}
n_s = 1 - 2/N_*\;, \quad r = 8/N_*\;, \quad \alpha_s = -2 /N_*^2\;,
\end{equation}
where the star denotes Hubble crossing of the pivot scale $k_*$.
Choosing $N_*=62.5$, this yields $n_s= 0.968$, $r=0.128$ and $\alpha_s =0$ as our fiducial model.

\subsection{Forecast constraints on the power spectrum}

We will now study how much Euclid will help in improving the already very
tight constraints on the power spectrum given by the Planck satellite. Let
us start discussing the forecast for Planck. We assume 2.5 years (5 sky
surveys) of multiple CMB channel data, with instrument characteristics for
the different channels listed in Tab.~\ref{tbl:PlanckSpec}. We take the
detector sensitivities and the values of the full width half
maxima {\color{red}maximum} from the Planck ``Blue Book'' \citep{PlanckBlueBook}. In this analysis we use three channels for Planck mock data and we assume that the other channels are used for foreground removal and thus do not provide cosmological information.
\begin{table}
\caption{Instrument specifics for the Planck satellite with 30 months of integration.\label{tbl:PlanckSpec}}
\centering
\begin{tabular}{llll}
\hline
Channel Frequency (GHz) & 70 &100 & 143 \\
\hline
\hline
Resolution (arcmin) & 14 & 10 & 7.1 \\
Sensitivity  - intensity ($\mu K$)&  8.8 & 4.7 & 4.1 \\
Sensitivity - polarization ($\mu K$)& 12.5 & 7.5  & 7.8  \\
\hline
\end{tabular}
\end{table}
For a nearly full-sky CMB experiment (we use $f_{\rm sky}=0.75$), the likelihood ${\cal L}$ can be approximated by \citep{Verde/Peiris/Jimenez:2006}
\begin{equation}
\begin{split}
-2 \ln{{\cal L}} =& \sum_{\ell=\ell_{\rm min}}^{\ell_{\rm max}} (2\ell+1)f_{\rm sky}\, \left[ -3 + \frac{\hat{C}_\ell^{BB}}{C_\ell^{BB}}  + \ln\left(\frac{C_\ell^{BB}}{\hat{C}_\ell^{BB}}\right) \right. \\
& \left.  + \frac{\hat{C}_\ell^{TT}C_\ell^{EE} + \hat{C}_\ell^{EE}C_\ell^{TT} - 2\hat{C}_\ell^{TE}C_\ell^{TE}}{C_\ell^{TT}C_\ell^{EE}-(C_\ell^{TE})^2}  +\ln{\left(\frac{C_\ell^{TT}C_\ell^{EE}-(C_\ell^{TE})^2}{\hat{C}_\ell^{TT}\hat{C}_\ell^{EE}-(\hat{C}_\ell^{TE})^2}\right)}\right] \ , 
\end{split}
\end{equation}
where we assume $l_{\rm min}=3$ and $l_{\rm max}=2500$. \label{symbol:Cell}
Here, $C_\ell$ is the sum of the model-dependent theoretical power spectrum $C_\ell^{\rm theory}$ and of the noise spectrum $N_\ell$, which we assume perfectly known. The mock data $\hat{C}_\ell$ is $C_\ell$ for the fiducial model, with $C_\ell^{\rm theory}$ calculated using the publicly available code CAMB \citep{Lewis/etal:2000} and $N_\ell$ calculated assuming a Gaussian beam.
We use the model described in \cite{Verde/Peiris/Jimenez:2006,Baumann/etal:2009} to propagate the effect of polarization foreground residuals into the estimated uncertainties on the cosmological parameters. For simplicity, in our simulation we consider only the dominating components in the frequency bands that we are using, i.e. the synchrotron and dust signals. The fraction of the residual power spectra are all assumed to be 5\%.

Let us turn now to the Euclid forecast based on the spectroscopic redshift survey. We will model the galaxy power spectrum in redshift space as (\cite{Kaiser:1987, Peacock:1992, Peacock/Dodds:1994}; see also discussion in Sec.~\ref{dark-energy-and-redshift-surveys})
\begin{equation}
P_g(k,z,\mu) = \left(b+ f_g \mu^2\right)^2 G^2(z)P_{\rm matter}(k;z=0) e^{-k^2\mu^2\sigma_r^2}, \label{eq:Pgini}
\end{equation}
where $\mu$ is the cosine of the angle between the wavenumber $\mathbf{k}$ 
and the line of sight, $G(z)$ is the linear growth factor defined in eq.~\eqref{def_gf},
 $f_g \equiv {d\ln G}/{d\ln a}$ is the linear growth rate (see eq.~\eqref{def-growth-rate}) 
and $P_{\rm matter}(k;z=0)$ is the matter power spectrum at redshift $0$. The term $f_g \mu^2$  
comes for the redshift distortions due to the large-scale peculiar velocity field \citep{Kaiser:1987}, 
which is correlated with the matter density field. The factor $e^{-k^2\mu^2\sigma_r^2}$ accounts for 
the radial smearing due to the redshift distortions that are uncorrelated with the large scale structure. 
We consider two contributions. The first is due to the redshift uncertainty of the spectroscopic galaxy 
samples. Assuming a typical redshift uncertainty $\sigma_z=0.001(1+z)$, this turns into a contribution 
to $\sigma_r$ given by $\partial r /\partial z \, \sigma_z = H^{-1} \, \sigma_z
$ where $r(z)=\int_0^z cdz'/H(z')$ is the comoving distance of a flat FRW
Universe and $H$ is the Hubble parameter as a function of the redshift. The
second contribution comes from the Doppler shift due to the
virialized {\color{red}virialised} motion of
galaxies within clusters, which typically have a pairwise velocity dispersion
$v_p$ of the order of few hundred kilometers per second. This term can be
parameterized as $\frac{v_p}{\sqrt{2}}H^{-1} (1+z)$ \citep{Peacock/Dodds:1994}.
Taking the geometric mean of the two contributions, we obtain 
\begin{equation}
\sigma_r^2 = \frac{(1+z)^2}{H^2}  \left(10^{-6} + {v_p^2}/{2} \right)\;, \label{eq:sigma2}
\end{equation}
where the two velocities in the parenthesis contribute roughly the same.
Practically neither the redshift measurement nor the virialized
{\color{red}virialised} motion of 
galaxies can be precisely quantified. In particular, the radial smearing due 
to peculiar velocity is not necessarily close to Gaussian. Thus, eq.~(\ref{eq:Pgini}) 
should not be used for wavenumbers  $k>\frac{H(z)}{v_p(1+z)}$, where the radial smearing effect is important. 

On large scales the matter density field has, to a very good approximation, Gaussian statistics and uncorrelated Fourier modes. Under the assumption that the positions of observed galaxies are generated by a random Poissonian point process, the band-power uncertainty is given by (\cite{Tegmark/etal:1998}; see also eq.~\eqref{eqn:pkerror} in Sec.~\ref{dark-energy-and-redshift-surveys})
\begin{equation}
\Delta  P_g = \left[ \frac{2 (2\pi)^3}{(2\pi k^2dk d\mu) (4\pi r^2f_{\rm sky} dr)}\right]^{1/2}\left(P_g+\frac{1}{\bar{n}}\right). \label{eq:dPg}
\end{equation}
Here $f_{\rm sky}$ is the observed fraction of sky, $r$ the comoving distance defined above, and $\bar{n}$ is the expected number density of galaxies that can be used.

Finally, we ignore the band-band correlations and write the likelihood as 
\begin{equation}
-2 \ln {\cal L} =  \sum_{k,\mu,z\ \rm bins} \left(\frac{ P_g^{\rm model} - P_g^{\rm fiducial}  }{\Delta P_g^{\rm fiducial} }\right)^2\ .
\end{equation}

To produce the mock data we use a fiducial $\Lambda$CDM model with $\Omega_ch^2=0.1128$, $\Omega_bh^2 = 0.022$, $h=0.72$, $\sigma_8 = 0.8$ and $\tau=0.09$, where $\tau$ is the reionization optical depth.  As mentioned above, we take the fiducial value for the spectral index, running and tensor to scalar ratio, defined at the pivot scale $k_*=0.05\,{\rm Mpc}^{-1}$, as given by chaotic inflation with quadratic potential, i.e.~$n_s = 0.968$, $\alpha_s=0$ and $r=0.128$. We have checked that for Planck data $r$ is almost orthogonal to $n_s$ and $\alpha_s$. Therefore our result is not sensitive to the fiducial value of $r$. 

The fiducial Euclid spectroscopically selected galaxies are split into 14
redshift bins. The redshift ranges and expected numbers of observed galaxies per
unit volume $\bar{n}_{\rm obs}$ are taken from \cite{euclidredbook} and shown in the third column 
of Tab.~\ref{tab:n_z} in Sec.~\ref{sec:baofm_survey} ($n_2(z)$). The number density of
galaxies that can be used is $\bar{n}=\varepsilon\bar{n}_{\rm obs}$, where
$\varepsilon$ is the fraction of galaxies with measured redshift. 
The boundaries of the wavenumber range used in the analysis, labeled $k_{\min}$ and $k_{\max}$, vary  in the ranges (0.00435-0.00334)$h{\rm Mpc}^{-1}$ and (0.16004-0.23644)$h{\rm Mpc}^{-1}$ respectively, for $0.7\leq z\leq 2$. The IR cutoff $k_{\min}$ is
chosen such that $k_{\min}r = 2\pi$, where $r$ is the comoving distance of the
redshift slice. The UV cutoff is the smallest between $\frac{H}{v_p(1+z)}$ and
$\frac{\pi}{2R}$. Here $R$ is chosen such that the r.m.s.~linear density
fluctuation of the matter field in a sphere with radius $R$ is $0.5$. In each
redshift bin we use 30 $k$-bins uniformly in $\ln k$ and 20 uniform $\mu$-bins.

For the fiducial value of the bias, in each of the 14 redshift bins of
width $\Delta z=0.1$ in the range (0.7-2), we use those derived from
\cite{orsi10},
i.e.\ (1.083, 1.125, 1.104, 1.126, 1.208, 1.243, 1.282, 1.292, 1.363, 1.497, 1.486, 1.491, 1.573, 1.568), 
and we assume that $v_p$ is redshift dependent
choosing $v_p=400\,{\rm km/s}$ as the fiducial value in each redshift bin. Then
we marginalize over $b$ and $v_p$ in the 14 redshift bins, for a total of 28
nuisance parameters. 

\begin{table}[ht]
\caption{Cosmological Parameters \label{tbl:cosmoparams}}
\centering
\begin{tabular}{lll}
\hline
\hline
parameter & Planck  constraint & Planck + Euclid constraint\\
\hline
$\Omega_bh^2$ & $0.02227^{+0.00011}_{-0.00011}$ & $0.02227^{+0.00008}_{-0.00008}$ \\
$\Omega_ch^2$ & $0.1116^{+0.0008}_{-0.0008}$ & $0.1116^{+0.0002}_{-0.0002}$ \\
$\theta$ & $1.0392^{+0.0002}_{-0.0002}$ & $1.0392^{+0.0002}_{-0.0002}$ \\
$\tau_{\rm re}$ & $0.085^{+0.004}_{-0.004}$ & $0.085^{+0.003}_{-0.003}$ \\
$n_s$ & $0.966^{+0.003}_{-0.003}$ & $0.966^{+0.002}_{-0.002}$ \\
$\alpha_s$ & $-0.000^{+0.005}_{-0.005}$ & $-0.000^{+0.003}_{-0.003}$ \\
$\ln(10^{10}A_s)$& $3.078^{+0.009}_{-0.009}$ & $3.077^{+0.006}_{-0.006}$ \\
$r$ & $0.128^{+0.018}_{-0.018}$ & $0.127^{+0.019}_{-0.018}$ \\
$\Omega_m$ & $0.271^{+0.005}_{-0.004}$ & $0.271^{+0.001}_{-0.001}$ \\
$\sigma_8$ & $0.808^{+0.005}_{-0.005}$ & $0.808^{+0.003}_{-0.003}$ \\
$h$ & $0.703^{+0.004}_{-0.004}$  & $0.703^{+0.001}_{-0.001}$ \\
\hline
\end{tabular}
\end{table}
In these two cases, we consider the forecast constraints on eight cosmological parameters, i.e.~$\Omega_bh^2$, $\Omega_ch^2$, $\theta$, $\tau$, $\ln A_s$, $n_s$, $\alpha_s$, and $r$. Here $\theta$ is the angle subtended by the sound horizon on the last scattering surface, rescaled by a factor $100$. We use the publicly available code CosmoMC \citep{Lewis/Bridle:2002} to perform Markov Chain Monte Carlo calculation.  The nuisance parameters are marginalized over in the final result. The marginalized 68.3\% confidence level (CL) constraints on cosmological parameters for Planck forecast only, and Planck and Euclid forecast are  listed in the second and third columns of Tab.~\ref{tbl:cosmoparams}, respectively. 

\begin{figure}
\centering
\includegraphics[width=0.5\textwidth]{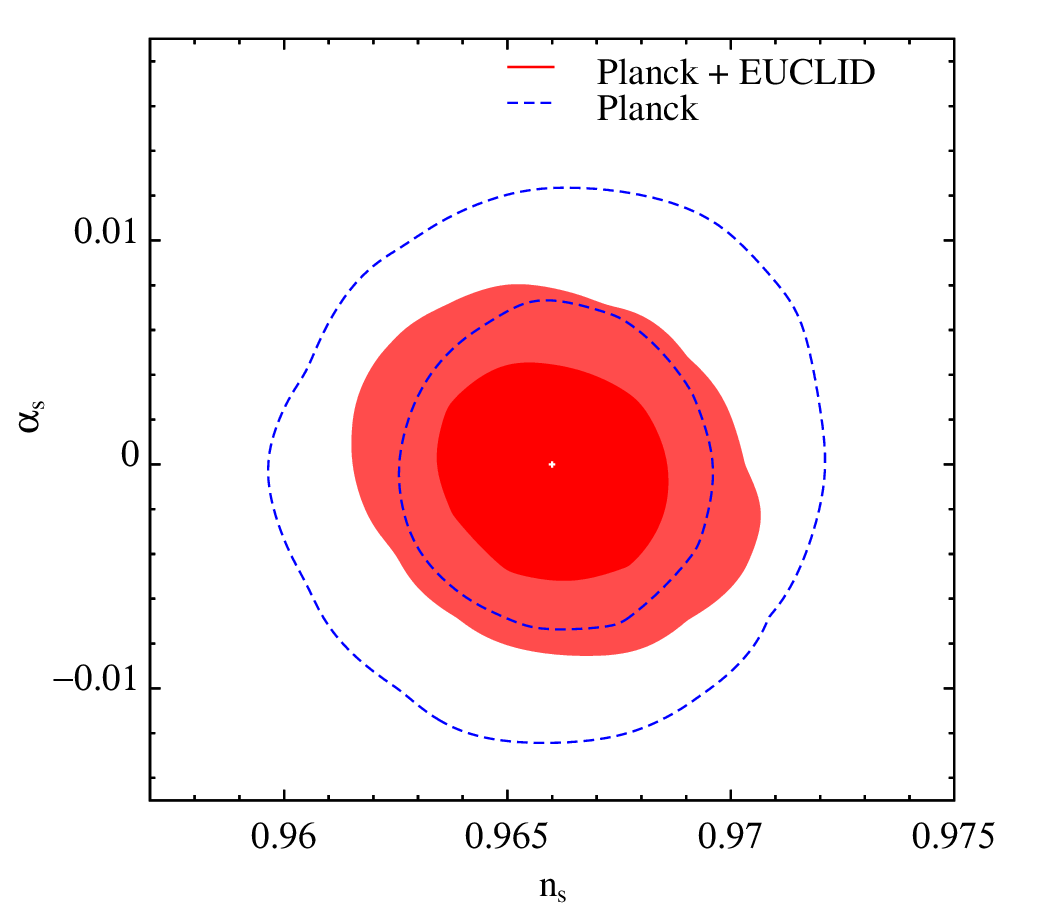}%
\includegraphics[width=0.5\textwidth]{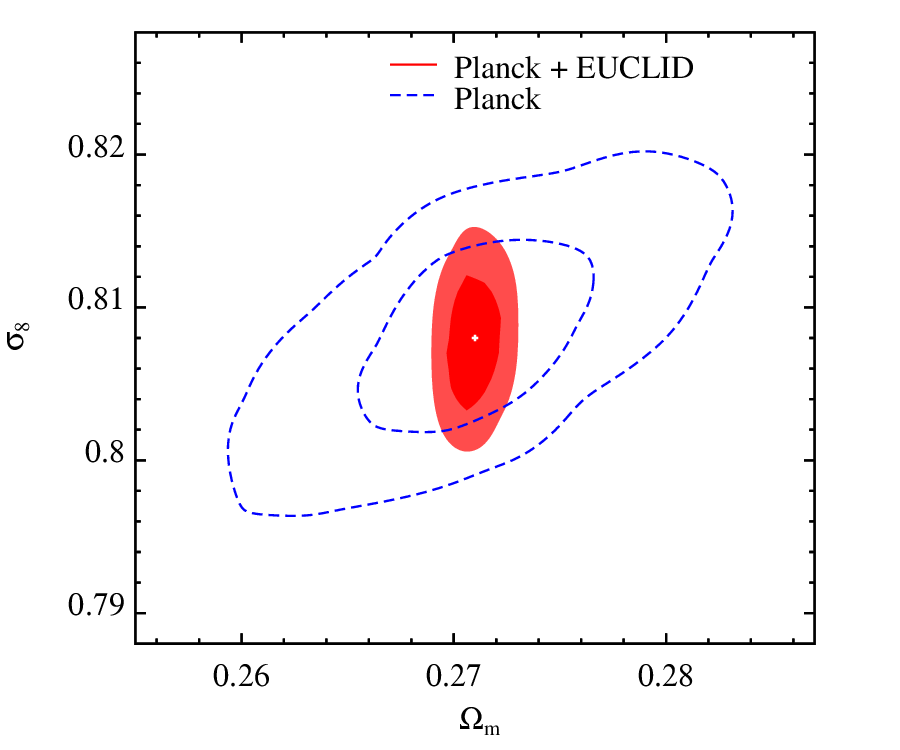}
\caption{The marginalized likelihood contours (68.3\% and 95.4\% CL) for Planck forecast only (blue dashed lines) and {\color{red}Planck plus Euclid pessimistic (red filled contours).} The white points correspond to the fiducial model. \label{fig:2D_Planck_EUCLID}}
\end{figure}
Euclid can improve the `figure of merit' on the $n_s$-$\alpha_s$ plane by a factor of $2.2$, as shown in the left panel of Figure~\ref{fig:2D_Planck_EUCLID}. Because the bias is unknown, the LSS data do not directly measure $A_s$ or $\sigma_8$. However, Euclid can measure $\Omega_m$ to a much better accuracy, which can break the degeneracy between $\Omega_m$ and $\sigma_8$ that one typically finds using CMB data alone. This is shown in the right panel of Figure~\ref{fig:2D_Planck_EUCLID}.

{\color{red}A more extensive and in depth analysis of what constraints on inflationary models a survey like Euclid can provide is presented in \cite{Huang:2012mr}.  In particular they find that  for models  where the primordial power spectrum  is not featureless (i.e. close to  a power law  with small running) a survey like Euclid will be crucial to detect and measure features. Indeed, what we measure with the CMB is the angular power spectrum of the anisotropies in the 2-D multipole space, which is a projection of the power spectrum in the 3-D momentum space. Features at large $\ell$'s and for small width in momentum space get smoothed during this projection but  this does not happen for large-scale structure surveys.  The main limitation on the width of features measured using  large-scale structure comes from the size of the volume of the survey:  the smallest detectable feature being of the order of the inverse cubic root of this volume and the error  being determined by number of modes contained in this volume. Euclid, with the large volume surveyed and the  sheer number of modes that are sampled and cosmic variance dominated   offers a unique  opportunity to  probe inflationary models where the potential is not featureless. In addition the increased statistical power would  enable  us to perform a Bayesian  model selection on the space of inflationary models (e.g., \cite{Easther:2011yq, Norena:2012rs} and references therein).}

 \section{Probing the early Universe with non-Gaussianities}
 \label{sec:NG-TH}
 
 

The workhorse for primordial non-Gaussianity  has been so far the so-called ``local model''
\citep{Salopek/Bond:1990, Gangui/etal:1994, Verde/etal:2000a, Komatsu/Spergel:2001,Bartolo/etal:2004}:
\begin{equation}
\Phi=\phi+f_{\rm NL}\left( \phi^2-\langle \phi^2\rangle \right)\,.
\label{eq:fnl}
\end{equation}
Here $\phi$ is a Gaussian random field while  $\Phi$ denotes 
Bardeen's gauge-invariant potential, which, 
on sub-Hubble scales reduces to the usual Newtonian peculiar 
gravitational potential, up to a minus sign. On large scales it is related to the conserved variable $\zeta$ by
\begin{equation}
\zeta = \frac{5 + 3 w}{3 + 3w} \Phi\;,
\end{equation}
where $w$ is the equation of state of the dominant component in the Universe.
The amount of primordial non-Gaussianity is quantified by the non-linearity parameter
$f_{\rm NL}$. Note that, since $\Phi\simeq \phi\simeq 10^{-5}$, $f_{\rm NL}\sim 100$ corresponds
to relative non-Gaussian corrections of order $10^{-3}$.
While $\zeta$ is constant on large scales, $\Phi$ is not. For this reason, 
in the literature there are two conventions for 
eq.~(\ref{eq:fnl}): the large-scale structure (LSS) and the Cosmic Microwave Background (CMB) 
one. In the LSS  convention, $\Phi$ is linearly extrapolated at $z=0$; 
in the CMB convention $\Phi$ is instead primordial:  thus $f^{\rm LSS}_{\rm
NL}=g(z=\infty)/g(0) f_{\rm NL}^{\rm CMB}\sim 1.3 \,f^{\rm CMB}_{\rm NL}$, 
where $g(z)$ denotes the linear growth suppression factor relative to an 
Einstein-de-Sitter Universe. In the past few years it has become customary to always report 
$ f_{\rm NL}^{\rm CMB}$ values even though, for simplicity as it will be clear below, 
one carries out the calculations with $f_{\rm NL}^{\rm LSS}$.

In this section we review the theoretical motivations and implications for looking into primordial non-Gaussianity; the readers less theoretically oriented can go directly to Sec. \ref{sec:NG-LSS}.

 \subsection{Local non-Gaussianity}

The non-Gaussianities generated in the conventional scenario of inflation (single-field with standard kinetic term, in slow-roll, initially in the Bunch-Davies vacuum) are predicted to be extremely small. Earlier calculations showed that $f_{\rm NL} $ would be of the order of the slow-roll parameters (\cite{Salopek/Bond:1990,Falk/etal:1993,Gangui/etal:1994}). More recently, with an exact calculation \cite{Maldacena:2003} confirmed this result and showed that the dominant contribution to non-Gaussianity comes from gravitational interaction and it is thus independent of the inflaton potential. More precisely, in the squeezed limit, i.e.~when one of the modes is much smaller than the other two, the bispectrum of the primordial perturbation $\zeta$ is given by
\begin{equation}
B_\zeta(k_1 \ll k_2,k_3) = 4 f_{\rm NL}^{\rm local} P_\zeta(k_2) P_\zeta(k_3)\;, \label{local_sl}
\end{equation}
where $f_{\rm NL}^{\rm local}$ is proportional to the tilt of scalar fluctuations, $f_{\rm NL}^{\rm local} = -(5/12) (n_s-1)$, a value much too small to be observable. Thus, any deviation from this prediction would rule out a large class of models based on the simplest scenario. 

Furthermore, \cite{Creminelli/Zaldarriaga:2004} showed that irrespective of slow-roll and of the particular inflaton Lagrangian or dynamics, in single-field inflation, or more generally when only adiabatic fluctuations are present, there exists a consistency relation involving the 3-point function of scalar perturbations in the squeezed limit (see also \cite{Seery/Lidsey:2005,Chen/etal:2007,Cheung/etal:2008a}). In this limit, when the short wavelength modes are inside the Hubble radius during inflation, the long mode is far out of the horizon and its only effect on the short modes is to rescale the unperturbed history of the Universe. This implies that the 3-point function is simply proportional to the 2-point function of the long wavelength modes times the 2-point function of the short wavelength mode times its deviation from scale invariance. In terms of local non-Gaussianity this translates into the same $f_{\rm NL}^{\rm local}$ found in \cite{Maldacena:2003}. Thus, a convincing detection of local non-Gaussianity would rule out all classes of inflationary single-field models.
%
%

To overcome the consistency relation and produce large local non-Gaussianity
one can go beyond the single-field case and consider scenarios where a
second field plays a role in generating perturbations. In this case, because
of non-adiabatic fluctuations, scalar perturbations can evolve outside the
horizon invalidating the argument of the consistency relation and possibly
generating a large $f_{\rm NL}^{\rm local}$ as in
\cite{Linde/Mukhanov:1997}. The curvaton scenario is one of such mechanisms.
The curvaton is a light scalar field that acquires scale-invariant
fluctuations during inflation and decays after inflation but well before
nuclesynthesis {\color{red}nucleosynthesis} (\cite{Mollerach:1990,Moroi/Takahashi:2001,Lyth/Wands:2002,Enqvist/Sloth:2002}). During the decay it dominates the Universe affecting its expansion history thus imprints its perturbations on super-horizon scales. The way the expansion history depends on the value of the curvaton field at the end of the decay can be highly non-linear, leading to large non-Gaussianity. Indeed, the non-linear parameter $f_{\rm NL}^{\rm local}$ is inversely proportional to the curvaton abundance before the decay  (\cite{Lyth/Ungarelli/Wands:2003}). 

Models exists where both curvaton and inflaton fluctuations contribute to cosmological perturbations (\cite{Langlois/Vernizzi:2004}). Interestingly, curvaton fluctuations could be negligible in the 2-point function but detectable through their non-Gaussian signature in the 3-point function, as studied in \cite{Boubekeur/Lyth:2006}. We shall come back on this point when discussing isocurvature perturbations. Other models generating local non-Gaussianities are the so called modulated reheating models, in which one light field modulates the decay of the inflaton field (\cite{Dvali/Gruzinov/Zaldarriaga:2004,Kofman:2003}). Indeed, non-Gaussianity could be a powerful window into the physics of reheating and preheating, the phase of transition from inflation to the standard radiation dominated era (see e.g.~\cite{Bond/etal:2009,Chambers/Nurmi/Rajantie:2010}).

In the examples above only one field is responsible for the dynamics of inflation, while the others are spectators. When the inflationary dynamics is dominated by several fields along the $\sim 60$ e-foldings of expansion from Hubble crossing to the end of inflation we are truly in the multi-field case. For instance, a well-studied model is double inflation with two massive non-interacting scalar fields (\cite{Polarski/Starobinky:1992}). In this case, the overall expansion of the Universe is affected by each of the field while it is in slow-roll; thus, the final non-Gaussianity is slow-roll suppressed, as in single field inflation (\cite{Rigopoulos/Shellard/vanTent:2006,Alabidi/Lyth:2006,Vernizzi/Wands:2006}). 

Because the slow-roll conditions are enforced on the fields while they dominate the inflationary dynamics, it seems difficult to produce large non-Gaussianity in multi-field inflation; however, by tuning the initial conditions it is possible to construct models leading to an observable signal (see \cite{Byrnes/Choi/Hall:2008,Tanaka/Suyama/Yokoyama:2010}). 
Non-Gaussianity can be also generated at the end of inflation, where large-scale perturbations may have a non-linear dependence on the non-adiabatic modes, especially if there is an abrupt change in the equation of state (see e.g.~\cite{Bernardeau/Uzan:2002,Lyth:2005}). Hybrid models (\cite{Linde:1994}), where inflation is ended by a tachyonic instability triggered by a waterfall field decaying in the true vacuum, are natural realizations of this mechanism (\cite{Enqvist/Vaihkonen:2004,Barnaby/Cline:2006}).

\subsection{Shapes: what do they tell us?}
\label{nongaussianshapes}

As explained above, local non-Gaussainity {\color{red}non-Gaussianity} is expected for models where non-linearities develop outside the Hubble radius. However, this is not the only type of non-Gaussianity. Single-field models with derivative interactions
yield a negligible 3-point function in the squeezed limit, yet leading to possibly observable non-Gaussianities. Indeed, as the interactions contain time derivatives and gradients, they vanish outside the horizon and are unable to produce a signal in the squeezed limit. Correlations will be larger for other configurations, for instance between modes of comparable wavelength. In order to study the observational signatures of these models we need to go beyond the local case and study the {\em shape} of non-Gaussianity (\cite{Babich/Creminelli/Zaldarriaga:2004}).

Because of translational and rotational invariance, the 3-point function is characterized by a function of the modulus of the three wave-vectors, also called the bispectrum $B_{\zeta}(k_1,k_2,k_3)$, defined as
\begin{equation}\label{eq:bispectrumdef}
\langle \zeta_{\bf k_1} \zeta_{\bf k_2} \zeta_{\bf k_3} \rangle = (2 \pi)^3 \delta({\bf k_1}+{\bf k_2}+{\bf k_3}) B_{\zeta}(k_1,k_2,k_3)\;.
\end{equation}
Relaxing the assumption of a local $f_{\rm NL}$, this function is a rich object which can contain a wealth of information, depending on the size and shape of the triangle formed by $k_1$, $k_2$ and $k_3$. 
Indeed, the dependence of the bispectrum on configuration in momentum space is related to the particular inflationary model generating it. Namely, each third-order operator present in the field action gives rise to a particular shape of the bispectrum.

An example of models containing large derivative interactions has been proposed by \cite{Silverstein/Tong:2004,Alishshiha/Silverstein/Tong:2004}. Based on the Dirac-Born-Infeld Lagrangian,
${\cal L} = f(\phi)^{-1} \sqrt{1- f(\phi) X} +V(\phi)$, with $X= -g^{\mu \nu} \partial_\mu \phi \partial_\nu \phi$, it is called DBI inflation. 
This Lagrangian is string theory-motivated and $\phi$ describes the low-energy radial dynamics of a D3-brane in a warped throat: $f(\phi)^{-1}$ is the warped brane tension and $V(\phi)$ the interaction field potential. In this model the non-Gaussianity is dominated by derivative interactions of the field perturbations so that we do not need to take into account mixing with gravity. 
An estimate of the non-Gaussianity is given by the ratio between the third-order and the second order Lagrangians, respectively ${\cal L}_3$ and ${\cal L}_2$, divided by the amplitude of scalar fluctuations. This gives $f_{\rm NL} \sim  ({{\cal L}_3}/{{\cal L}_2}) \Phi^{-1} \sim -1/{c_s^2}$, where $c_s^2 = [1+ 2 X (\partial^2 {\cal L} / \partial X^2) /(\partial {\cal L} / \partial X)]^{-1} <1$ is the speed of sound of linear fluctuations and we have assumed that this is small, as it is the case for DBI inflation.
Thus, the non-Gaussianity can be quite large if $c_s \ll 1$. 

However, this signal vanishes in the squeezed limit due to the derivative interactions. More precisely, the particular momentum configuration of the bispectrum is very well described by
\begin{equation}
B_{\zeta}(k_1,k_2,k_3) =   6 f_{\rm NL}^{\rm equil}  \bigg( \frac{P_\zeta(k_1) P_\zeta(k_2)}{2 }  + \frac{\left[ P_\zeta(k_1) P_\zeta(k_2) P_\zeta(k_3) \right]^{\frac{2}{3}}}{3}  - P_\zeta(k_1)^{\frac{1}{3}} P_\zeta(k_2)^{\frac{2}{3}} P_\zeta(k_3) + 5 \ {\rm perms.} \bigg)\;,
\end{equation}
where, up to numerical factors of order unity, $f_{\rm NL}^{\rm equil}
\simeq -1/ c_s^2 $. The function of momenta inside the parenthesis is the
so-called {\em equilateral} shape (\cite{Creminelli/etal:2006}), a {\em
template} used to approximate a large class of inflationary models. It is
defined in such a way as to be factorizable {\color{red}factorisable}, maximized for equilateral configurations and vanishing  in the squeezed limit faster than the local shape, see eq.~(\ref{local_sl}).

To compare  two shapes $F_1$ and $F_2$, it is useful to define a 3-dimensional scalar product between them as (\cite{Babich/Creminelli/Zaldarriaga:2004})
\begin{equation}
F_1 \cdot F_2 = \sum F_1 (k_1,k_2,k_3) F_2 (k_1,k_2,k_3) / (P_\zeta(k_1)P_\zeta(k_2)P_\zeta(k_3))\;,
\end{equation}
where the sum is over all configurations forming a triangle. Then, $\cos \theta = F_1 \cdot F_2/ \sqrt{(F_1 \cdot F_1)(F_2 \cdot F_2)}$ defines a quantitative measure of how much two shapes ``overlap'' and their signal is correlated. The cosine is small between the local and equilateral shapes. Two shapes with almost vanishing cosine are said to be orthogonal and any estimator developed to be sensitive to a particular shape will be completely blind to its orthogonal one. Note that the observable signal could actually be a combination of different shapes. For instance, multi-field models base on the DBI action (\cite{Langlois/etal:2008}) can generate a linear combination of local and equilateral non-Gaussianities (\cite{Renaux-Petel:2009}).

The interplay between theory and observations, reflected in the relation
between derivative interactions and the shape of non-Gaussianity, has
motivated the study of inflation according to a new approach, the so-called
effective field theory of inflation (\cite{Cheung/etal:2008}; see also
\cite{Weinberg:2008}). Inflationary models can be viewed as effective field
theories in presence of symmetries. Once symmetries are defined, the
Lagrangian will contain each possible operator respecting such symmetries.
As each operator leads to a particular non-Gaussian signal, constraining
non-Gaussianity directly constrains the coefficients in front of these
operators, similarly to what {\color{red}is } done in high-energy physics with particle accelerators. For instance, the operator ${\cal L}_3$ discussed in the context of DBI inflation leads to non-Gaussianity  controlled by the speed of sound of linear perturbations. This operator can be quite generic in single field models. Current constraints on  non-Gaussianity allow to constrain the speed of sound of the inflaton field during inflation to be $c_s \ge 0.01$ (\cite{Cheung/etal:2008,Senatore/Smith/Zaldarriaga:2010}).
Another well-studied example is ghost inflation
(\cite{Arkani-Hamed/etal:2004b}), based on the ghost condensation, a model
proposed by \cite{Arkani-Hamed/etal:2004} to modified {\color{red}modify} gravity in the infrared. This model is motivated by shift symmetry and exploits the fact that in the limit where this symmetry is exact, higher derivative operators play an important role in the dynamics, generating large non-Gaussianity with approximately equilateral shape. 

Following this approach has allowed to construct operators or combination of operators leading to new shapes, orthogonal to the equilateral one. An example of such a shape is the {\em orthogonal} shape proposed in \cite{Senatore/Smith/Zaldarriaga:2010}. This shape is generated by a particular combination of two operators already present in DBI inflation. It is peaked both on equilateral-triangle configurations and on flattened-triangle configurations (where the two lowest-$k$ sides are equal exactly to half of the highest-$k$ side) -- the sign in this two limits being opposite. The orthogonal and equilateral are not an exhaustive list. For instance, \cite{Creminelli/etal:2010} have shown that the presence in the inflationary theory of an approximate Galilean symmetry (proposed by \cite{Nicolis/Rattazzi/Trincherini:2009} in the context of modified gravity) generates third-order operators with two derivatives on each field. A particular combination of these operators produces a shape that is approximately orthogonal to the three shapes discussed above.

Non-Gaussianity is also sensitive to deviations from the initial adiabatic Bunch-Davies vacuum of inflaton fluctuations. Indeed, considering excited states over it, as done in {\color{red}\cite{Chen/etal:2007}}  \cite{Holman/Tolley:2008,Meerburg/vanderSchaar/Corasaniti:2009}, leads to a shape which is maximized in the collinear limit, corresponding to enfolded or squashed triangles in momentum space, although one can show that this shape can be written as a combination of the equilateral and orthogonal ones (\cite{Senatore/Smith/Zaldarriaga:2010}).

{\color{red}
\subsection{Beyond shapes: scale dependence and  the squeezed limit}

There is a way out to generate large non-Gaussianity in single-field inflation. Indeed, one can temporarily break scale-invariance, for instance by introducing features in the potential as in \cite{Chen/Easther/Lim:2007}. This can lead to large non-Gaussianity typically associated to scale-dependence. These signatures could even teach us something about string theory. Indeed, in axion monodromy, a model recently proposed by \cite{Silverstein/Westphal:2008} based on a particular string compactification mechanism,  the inflaton potential is approximately linear, but periodically modulated. These modulations lead to tiny oscillations in the power spectrum of cosmological fluctuations and to large non-Gaussianity (see for instance \cite{Flauger:2010J}).

This is not the only example of scale dependence.
While in general the amplitude of the non-Gaussianity signal  is considered constant, there are  several models, beside the above example,  which predict a scale-dependence. For example models like the Dirac-Born-Infeld (DBI) inflation e.g., \cite{Alishahiha/Silverstein/Tong:2004,Chen:2005a,Chen2005b,Bean/etal:2008} can be characterized by a primordial bispectrum whose amplitude varies significantly over the range of scales accessible by cosmological probes.  

In view of measurements from observations it is  also worth considering the so-called squeezed limit of non-Gaussianity that is the limit in which one of the momenta is much smaller than the other two. Observationally this is because some probes (like for example the halo bias \S\ref{sec:halobias},  accessible by large-scale structure surveys like Euclid) are sensitive to this limit. Most importantly, from the theoretical point of view, there are consistency relations valid in this limit that identify different classes of inflation e.g., \cite{Creminelli:2012ed,Creminelli:2011rh} and references therein. 

The scale dependence of non-gaussianity, the shapes of non-gaussianity and the  behaviour of the squeezed limit are all  promising avenues, where the combination of CMB data and large scale structure surveys such as Euclid can provide powerful constraints as illustrated e.g., in \cite{Sefusattietal:2009, Norena/etal:2012, Sefusatti/etal:2012}.}

\subsection{Beyond inflation}


As explained above, the search of non-Gaussianity could represent a unique
way to rule out the simplest of the inflationary models and distinguish
between different scenarios of inflation. Interestingly, it could also open
up a window on new scenarios, alternative to inflation. There have been
numerous attempts to construct  models alternative to inflation able to
explain the initial conditions of our Universe. In order to solve the
cosmological problems and generate large-scale primordial fluctuations, most
of them require a phase during which observable scales today have exited the
Hubble size. This can happen in bouncing cosmologies, in which the present
era of expansion is preceeded {\color{red}preceded} by a contracting phase. Examples are the pre-big bang (\cite{Gasperini/Veneziano:1993}) and the ekpyrotic scenario (\cite{Khoury/etal:2001}). 

In the latter, the 4-d effective dynamics corresponds to {\color{red}to} a cosmology driven by a scalar field with a steep exponential potential $V(\phi) = \exp(-c \phi)$, with $c \gg 1$.
Leaving aside the problem of the realization of the bounce, it has been
shown that the adiabatic mode in this models {\color{red}model} generically leads to a steep blue spectrum for the curvature perturbations (\cite{Lyth:2002,Creminelli/Zaldarriaga:2005}). Thus, at least a second field is required to generate an almost scale-invariant spectrum of perturbations (\cite{Finelli:2002,Creminelli/Senatore:2007,Buchbinder/Khoury/Ovrut:2007,Koyama/Wands:2007}). If two fields are present, both with exponential potentials and steepness coefficients $c_1$ and $c_2$, the non-adiabatic component has negative mass and acquires a quasi invariant spectrum of fluctuations with tilt $n_s-1 = 4(c_1^{-2} + c_2^{-2})$, with $c_1,c_2 \gg 1$. Then one needs to convert the non-adiabatic fluctuation into curvature perturbation, similarly to what the curvaton mechanism does.

As the Hubble rate increases during the collapse, one expects non-linearities in the fields to become more and more important, leading to non-Gaussianity in the produced perturbations. As non-linearities grow larger on super-Hubble scales, one expects the signal to be of local type. The particular amplitude of the non-Gaussianity in the observable curvature perturbations depends on the conversion mechanism from the non-adiabatic mode to the observable perturbations. The tachyonic instability itself can lead to a phase transition to an ekpyrotic phase dominated by just one field $\phi_1$. In this case \cite{Koyama/etal:2007} have found that $f_{\rm NL}^{\rm local} = -(5/12)c_1^2$. 
Current constraints on $f_{\rm NL}^{\rm local}$ (WMAP7 year data imposes $-10<f_{\rm NL}^{\rm local}< 74$ at 95\% confidence) gives an unacceptably large value for the scalar spectral index.  In fact in this model, even  for $f_{\rm NL}=-10$, $c_2\simeq 5$ which implies a too large value of the scalar spectral index  ($n_s-1 > 0.17$) which is  excluded by  observations (recall that WMAP7 year data implies $n_s=0.963\pm 0.014$ at 68\% confidence). Thus, one needs to modify the potential to accommodate a red spectrum or consider alternative conversion mechanisms to change the value of the generated non-Gaussianity (\cite{Buchbinder/Khoury/Ovrut:2008,Lehners/Steinhardt:2008}).

\section{Primordial Non-Gaussianity and Large-Scale Structure}
\label{sec:NG-LSS}
As we have seen, even the simplest inflationary models predict deviations from Gaussian initial conditions. 
Confirming or ruling out the simplest inflationary model 
is an important goal and in this section we will show how Euclid can help achieving this.
Moreover, Euclid data (alone or in combination with CMB experiments like Planck) can be used to 
explore the primordial bispectrum and thus explore the interaction of the fields during inflation.

\subsection{Constraining primordial non-Gaussianity and {\color{red}gravity} from 3-point statistics}

Contrary to CMB research which mainly probes the high-redshift Universe, 
current studies of the LSS focus on data at much lower redshifts and are 
more heavily influenced by cosmic evolution.
Even if the initial conditions were Gaussian,
non-linear evolution due to gravitational instability generates a non-zero bispectrum
for the matter distribution.  
The  first non-vanishing term in perturbation theory (e.g.~
\cite{Catelan/etal:1995}) gives
\begin{equation}
B(\vk_1,\vk_2,\vk_3)=2(P(k_1) P(k_2)J(\vk_1,\vk_2)+  {\rm cyclic\,\,
permutations})
\end{equation}
where   $J(\vk_1,\vk_2)$ is the gravitational instability ``kernel" which depends very weakly on 
cosmology and for an Einstein-de-Sitter Universe assumes the form:
\begin{equation}
J(\vk_1,\vk_2)=\frac{5}{7}+\frac{\vk_1\cdot \vk_2}{2k_1k_2}\left(\frac{k_1}{k_2}+\frac{k_2}{k_1}\right)+\frac{2}{7}\left(\frac{\vk_1\cdot \vk_2}{k_1 k_2}\right)^2\,.
\end{equation}
This  kernel represents  the ``signature" of gravity as we know it on the large-scale structure
of the Universe. Either a modification of the gravitational law or the introduction of 
a coupling between dark matter and another component (say dark energy) 
would alter the bispectrum shape from the standard form.
The volume covered by Euclid will enable us to exploit this.

It was recognized a decade ago \citep{Verde/etal:2000} that the contribution to the matter
bispectrum generated by graviational {\color{red}gravitational} instability
is large compared to the fossil signal due to primordial non-Gaussianity
and that the primordial signal ``redshifts away''  compared to the gravitational signal. 
In fact, primordial non-Gaussianity of the local type
would affect the late-time dark matter density bispectrum with a 
contribution of the form
\begin{equation}
B^{\fnl\, \rm local}(\vk_1,\vk_2,\vk_3,z)=
2(\fnl P(k_1)P(k_2)\frac{{\cal F}(\vk_1,\vk_2)}{D(z)/D(z=0)}+  {\rm
cyclic\,\,
permutations}).\end{equation}
where $D(z)$ is the linear growth function which in an Einstein-de-Sitter Universe goes like $(1+z)^{-1}$ and 
\begin{equation}
{\cal F}= \frac{{\cal M}(k_3)}{{\cal M}(k_1){\cal M}(k_2)}\,; \,\,\,{\cal M}(k)=\frac{2}{3}\frac{k^2 T(k)}{H_0^2\Omega_{m,0}}\,,
\end{equation}
$T(k)$ denoting the matter transfer function, \label{symbol:Tk} 
$H_0$ the Hubble constant and $\Omega_{m,0}$ the matter density parameter.
Clearly the contributions due to primordial non-Gaussianity and gravitational instability 
have different scale and redshift dependence and the two 
kernel shapes in configuration space are different, thus,  making the two components, at least 
in principle and for high signal-to-noise, separable.
This is particularly promising for high-redshift probes of the matter distribution
like the 21-cm background which should simultaneously provide competing measures of
$f_{\rm NL}$ and a test of the gravitational law \citep{Pillepichetal2007}.
Regrettably,
these studies require using a long-wavelength radio telescope 
above the atmosphere (e.g.~on the Moon) and will certainly come well after Euclid.

Galaxy surveys do not observe the dark matter distribution directly. However,
dark matter halos are believed to host galaxy formation, and different galaxy types at different 
redshifts are expected to populate halos in disparate ways 
\citep{Magliocchetti/Porciani:2003,Zehavietal2005}.
A simple (and approximate) way to account for galaxy biasing is to assume that
the overdensity in galaxy counts can be written as a truncated power expansion
in terms of the mass overdensity (smoothed on some scale):
$\delta_g(x)=b_1\delta_{\rm DM}(x)+b_2(\delta_{\rm DM}^2-\langle \delta_{\rm DM}^2 \rangle)$.
The linear and quadratic bias coefficient $b_1$ and $b_2$ are assumed to be scale-independent 
(although this assumption must break down at some point) but they can vary with redshift
and galaxy type. Obviously, a quadratic bias will introduce non-Gaussianity even on an initially 
Gaussian field. In summary, for local non-Gaussianity and scale-independent quadratic bias we 
have \citep{Verde/etal:2000}:
\begin{equation}
B(\vk_1,\vk_2,\vk_3,z)=2 P(k_1)P(k_2) b_1(z)^3\times  \left[ \fnl \frac{{\cal F}(\vk_1,\vk_2)}{D(z)} + J(\vk_1,\vk_2) +\frac{b_2(z)}{2 b_1(z)}\right]+ cyc.\\ 
\end{equation}
Before the above expression can be compared against observations, it needs to be further 
complicated to account for redshift-space distortions and shot noise. 
Realistic surveys use galaxy redshifts as a proxy for distance, but gravitationally-induced 
peculiar velocities distort the redshift-space galaxy distribution. 
At the same time, the discrete nature of galaxies gives rise to corrections 
that should be added to the bispectrum computed in the continuous limit.
We will not discuss these details here as including redshift space distortions and shot noise 
will not change the gist of the message.

From the observational point of view, it is important to note that photometric surveys
are not well suited for extracting a primordial signal out of the galaxy bispectrum.
Although in general they can cover larger volumes that {\color{red}than} spectroscopic surveys, 
the projection effects due to the photo-z smearing along the line-of-sight is 
expected to suppress significantly the sensitivity of the measured bispectrum to the shape of 
the primordial one (see e.g.~\cite{Verde/Heavens/Matarrese:2000}).
\cite{Sefusatti/Komatsu:2007}
have shown that, if the evolution of the bias parameters
is known a priori, spectroscopic surveys like Euclid would be able to give 
constraints on the $f_{\rm NL}$ parameter that are competitive with CMB studies.
%
%
While the gravitationally-induced non-Gaussian signal in the bispectrum has been detected  
to high statistical significance (see  e.g., \cite{Verde2df, Kulkarni/etal:2007} and references 
therein),  the identification of non-linear biasing (i.e. $b_2\neq 0$) is still controversial, 
and there has been so far no detection of any extra (primordial) bispectrum contributions.

Of course, one could also consider higher-order correlations. One of the advantages of considering
 e.g. the trispectrum is that, contrary to the bispectrum, it has very weak non-linear growth 
\cite{Verde/Heavens:2001}, but it has the disadvantage that the signal is de-localized: the number
of possible configurations grows fast with the dimensionality $n$ of the $n$-point function!  

Finally, it has been proposed to 
measure the level of primordial non-Gaussianity using 
Minkowski functionals applied either to the galaxy distribution or the weak
lensing maps (see e.g.~\cite{Hikageetal2008, Munshietal2011} and references therein).
The potentiality of this approach compared to more traditional methods
needs to be further explored in the near future.



\subsection{Non-Gaussian halo bias}\label{sec:halobias}
The discussion above neglects an important fact which went unnoticed until year 2008:
the presence of small non-Gaussianity can have a large effect on the clustering of dark matter 
halos \citep{Dalal/etal:2008,Matarrese/Verde:2008}. The argument goes as follows.
The clustering of the peaks in a Gaussian random field is completely specified by the field power 
spectrum. Thus, assuming that halos form out of linear density peaks, 
for Gaussian initial conditions the clustering of the dark matter halos is 
completely specified by the linear matter power spectrum.
On the other hand, for a non-Gaussian field, the clustering of the peaks depends on all 
higher-order correlations, not just on the power spectrum. Therefore, for non-Gaussian initial 
conditions,  the clustering of dark matter halos  depends on the 
linear bispectrum (and higher-order moments).

One can also understand the effect in the peak-background-split framework:
overdense patches of the (linear) Universe collapse to form dark matter halos if their 
overdensity lies above a critical collapse threshold.  
Short-wavelength modes define the overdense patches while the long-wavelength modes 
determine the spatial distribution of the collapsing ones by modulating their height above and 
below the critical threshold.  
In the  Gaussian case, long- and short-wavelength modes are uncorrelated, 
yielding  the well  known linear, scale-independent peak bias. In the non-Gaussian case, 
however, long and short wavelength modes are coupled, yielding a different spatial pattern of 
regions that cross the collapse threshold.

In particular, for primordial non-Gaussianity of the local type, 
the net effect is that the halo distribution on very large scales relates to the underlying
dark matter in a strongly scale-dependent fashion.
For $k\lesssim 0.02\,h$ Mpc$^{-1}$,
the effective linear bias parameter scales as $k^{-2}$. 
\citep{Dalal/etal:2008,Matarrese/Verde:2008,Giannantonio/Porciani:2010}.
This is because the halo overdensity depends not only
on the underlying matter density but also on the value of the auxiliary Gaussian potential $\phi$
(\cite{Giannantonio/Porciani:2010}).

\begin{figure}[htbp!]
    \centering
\includegraphics[width=.75\textwidth]{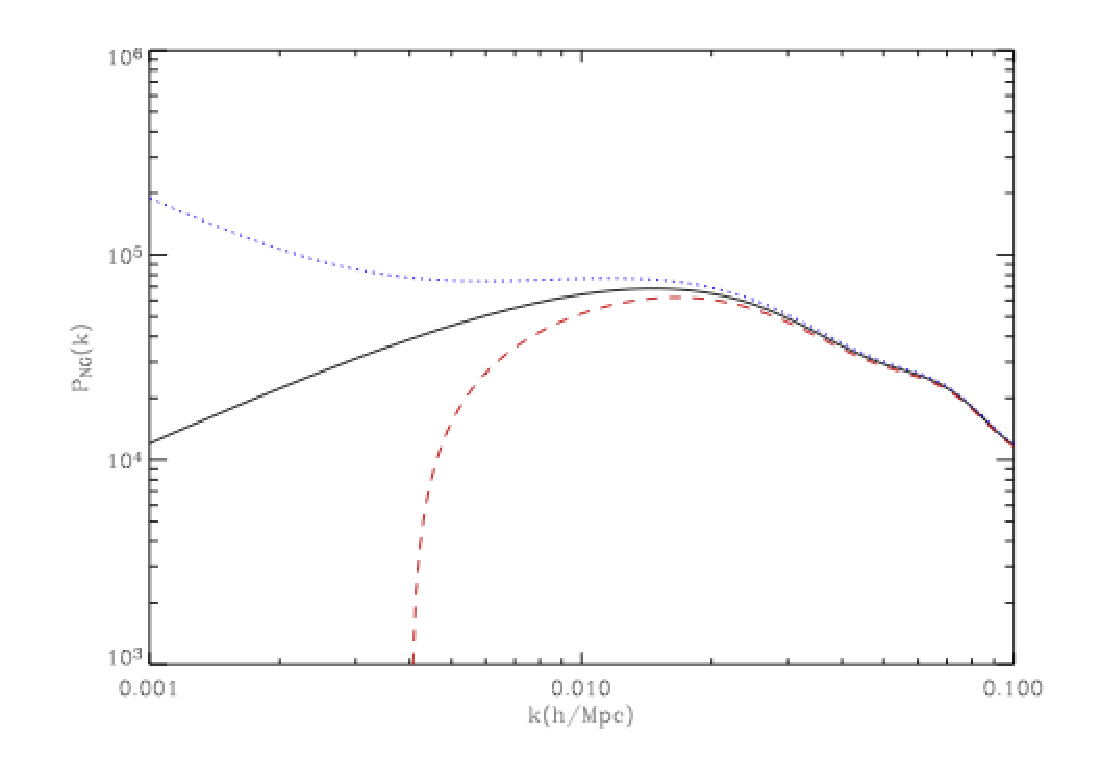}        
               \caption{For illustration purposes this is the
 effect of a local $f_{\rm NL}$ of $\pm 50$ on the $z=0$ power spectrum of 
halos with mass above $10^{13}$ M$_{\odot}$. 
}
                 \label{fig:Pkng}
\end{figure}


The presence of this effect is extremely important for observational studies 
as it allows to detect primordial non-Gaussianity from 2-point
statistics of the galaxy distribution like the power spectrum.
Combining current LSS data gives constraints on $f_{\rm NL}$ which are comparable to the CMB ones
 \citep{Slosar/etal:2008,Xia/etal:2010}. Similarly, planned galaxy surveys are expected
to progressively improve upon existing limits
\citep{Carbone/Verde/Matarrese:2008,Carbone/Mena/Verde:2010,Giannantonioetal2010}. 
For example Euclid could reach an error on $\fnl$ of $\sim 5$ (see below for further details)
which is comparable with the BPol forecasted {\color{red}forecast} errors.  

The scale dependence of the halo bias changes considering different
shapes of primordial non-Gaussianity \citep{Schmidt/Kamion:2010,Wagneretal2010}. 
For instance, orthogonal and folded models produce an effective bias that scales as $k^{-1}$
while the scale dependence becomes extremely weak for equilateral models.
Therefore, measurements of the galaxy power spectrum on the largest possible scales
have the possibility to constrain the shape and the amplitude of primordial non-Gaussianity
and thus shed new light on the dynamics of inflation.

On scales comparable with the Hubble radius, matter and halo clustering
are affected by general-relativity effects: 
the Poisson equation gets a quadratic correction that acts effectively as a non-zero local
$\fnl$ \citep{Bartoloetal2005,Pillepichetal2007}.
This contribution is peculiar to the inflationary initial conditions because it 
requires perturbations on super-horizon scales and it is mimicked in the halo bias by a 
local $\fnl=-1.6$ \citep{Verde/Matarrese:2009}. This is at the level of detectability by 
a survey like Euclid.
 
\subsection{Number counts of non-linear structures}\label{sec:massfn}
Even a small deviation from Gaussianity in the initial conditions can have a strong impact on those 
statistics which probe the tails of the linear density distribution. 
This is the case for the abundance of the most extreme  
non-linear objects existing at a given cosmic epoch, massive dark matter halos and voids,
as they correspond to the highest and lowest density peaks (the rarest events) in the
underlying linear density field.

Thus small values of $f_{\rm NL}$ are potentially detectable by measuring 
the abundance of massive dark matter halos as traced by galaxies and galaxy clusters
at $z \gtrsim 1$ 
\citep{Matarrese/Verde/Jimenez:2000}. This approach has recently received  
renewed attention, e.g.~\cite{Loverde/etal:2008,Grossi/etal:2009,Pillepich/Porciani/Hahn:2010, 
Maggiore/Riotto:2010,Damico/etal:2010,Verdereview:2010,Pillepich/Porciani/Reiprich2011} and references therein)
and might represent a promising tool for Euclid science.
In Euclid, galaxy clusters at high redshift can be identified either by lensing studies or by
building group catalogs based on the spectroscopic and photometric galaxy data. The main
challenge here is to determine the corresponding halo mass with sufficient accuracy to allow
comparison with the theoretical models. 

While galaxy clusters form at the highest overdensities of the primordial density field and 
probe the high-density tail of the PDF, voids form in the low-density regions and thus probe the 
low-density tail of the PDF. Most of the volume of the evolved Universe is underdense, so 
it seems interesting to pay attention to the distribution of underdense regions.  
For the derivation of the non-Gaussian void probability function one proceeds in parallel to the 
treatment for halos with the only subtlety that the critical threshold is not negative and that 
its numerical value depends on the precise definition of a void (and may depend on the 
observables used to find voids), e.g.~\cite{Kamionkowski/Verde/Jimenez:2009}.
Note that while a positive skewness ($\fnl>0$) boosts the number of halos at the high mass end 
(and slightly suppress the number of low-mass halos), it is a negative skewness that will 
increase the voids size distribution at  the largest voids end (and slightly decrease it for 
small void sizes).  In addition voids may probe slightly larger scales than halos, making the 
two approaches highly complementary.

Even though a number of observational techniques to detect voids in galaxy surveys 
have been proposed
(see e.g.~\citet{Colbergetal2008} and references therein), the challenge here is to match
the theoretical predictions to a particular void-identification criterion based on a specific
galaxy sample. We envision that mock galaxy catalogs based on numerical simulations will
be employed to calibrate these studies for Euclid.

\subsection{Forecasts for Euclid}
A number of authors have used the Fisher-matrix formalism to
explore the potentiality of Euclid in determining the level and
the shape of primordial non-Gaussianity \citep{Carbone/Verde/Matarrese:2008,Carbone/Mena/Verde:2010,Giannantonioetal2010}. 
In what follows, unless specifically mentioned, we will focus on the local 
type of non-Gaussianity which has been more widely studied so far. 

The most promising avenue is exploiting the scale-dependent bias on very large scales in
studies of galaxy clustering at the two-point level.
Early Fisher forecasts for the Euclid redshift survey found that,
for a fiducial model with $f_{\rm NL}=0$, this gives a marginalized $1\sigma$ error
 on the non-linearity parameter of $\Delta f_{\rm NL}\simeq 2$ \citep{Carbone/Verde/Matarrese:2008,Carbone/Mena/Verde:2010}.
Forecasts based on the most recent specifics for the Euclid surveys 
(see Tab.~\ref{tbl:EucludSpecfnl})
are presented
in \cite{Giannantonioetal2010} and summarized in Tab.~\ref{tbl:EUCLIDFNL} 
below.
Updated values of the galaxy number counts and of the efficiency in measuring 
spectroscopic redshifts correspond to a marginalized $1\sigma$ error
of $f_{\rm NL} \simeq 4-5$ 
(depending a little on the detailed assumptions of the Fisher
matrix calculation), 
with a slightly better result obtained using the 
Euclid spectroscopic sample rather than the
photometric one
(complemented with multi-band ground-based photometry), 
at least for a fiducial value of $f_{\rm NL}=0$ \citep{Giannantonioetal2010}.
The forecasted {\color{red}forecast} errors further improve by nearly a few per cent 
using Planck priors on the cosmological
parameters determined with the power spectrum of CMB temperature anisotropies.  

\begin{table}
\caption{Specifications of the surveys used in the Euclid forecasts given
in Tab.~\ref{tbl:EUCLIDFNL}. The redshift distributions of the different 
galaxy samples are as in Chapter 1 (see also \cite{Giannantonioetal2010}).}
\label{tbl:EucludSpecfnl}
\centering
\begin{tabular}{ccc}
\hline
 & Photometric survey  & Spectroscopic survey \\
\hline
\hline
Surveyed area (deg$^2$)& 15,000 & 15,000 \\
Galaxy density (arcmin$^{-2}$)& 30 & 1.2 \\
Median redshift & 0.8 & 1.0 \\
Number of redshift bins & 12 & 12 \\
Redshift uncertainty $\sigma_z/(1+z)$ & 0.05 & 0.001 \\
Intrinsic ellipticity noise $\gamma$ & - & 0.247 \\
Gaussian linear bias param. & $\sqrt{1+z}$ & $\sqrt{1+z}$ \\
\hline
\end{tabular}
\end{table}

The amplitude and shape of the matter power spectrum in the mildly 
non-linear regime depend (at a level of a few per cent)
on the level of primordial non-Gaussianity \citep{Taruyaetal2008,Pillepich/Porciani/Hahn:2010,Giannantonio/Porciani:2010}. Measuring this signal with the Euclid weak-lensing survey 
gives $\Delta f_{\rm NL}\simeq 70$ (30 with Planck priors) \citep{Giannantonioetal2010}. 
On the other hand, counting non-linear structures in terms of 
peaks in the weak-lensing maps (convergence or shear)
should give limits in the same ballpark 
(\cite{Marianetal2010} find $\Delta f_{\rm NL}=13$
assuming perfect knowledge of all the cosmological parameters).

Finally, by combining lensing and angular power spectra 
(and accounting for all possible cross-correlations)
one should achieve $\Delta f_{\rm NL}\simeq 5$
(4.5 with Planck priors) \citep{Giannantonioetal2010}. 
This matches what is expected from both the Planck mission and 
the proposed BPol satellite.

Note that the forecasted {\color{red}forecast} errors on $f_{\rm NL}$ are somewhat sensitive to 
the assumed fiducial values of the galaxy bias. 
In our study we have adopted the approximation 
$b(z)=\sqrt{1+z}$ \citep{rassat08}. On the other hand, using 
semi-analytic models of galaxy formation, \citet{orsi10} found
bias values which are nearly 10-15\% lower at all redshifts.
Adopting this slightly different bias, 
the constraint on $f_{\rm NL}$ already degrades by 50\% with respect to our 
fiducial case. 

Euclid data can also be used to constrain the scale dependence of the 
non-linearity parameter (see Tab.~\ref{tbl:Euclidfnlrun}).
To this purpose, we consider a local model of primordial non-Gaussianity where
\begin{equation}
f_{\rm NL}=f_{\rm NL}^{\rm (piv)}\cdot \left( \frac{k}{k_{\rm piv}}\right)^{n_{f_{\rm NL}}}\;,
\end{equation}
 with fiducial values $k_{\rm piv}=0.02\, h$ Mpc$^{-1}$, 
$f_{\rm NL}^{\rm (piv)}=30$, and $n_{f_{\rm NL}}=0$.
In this case, the combination of lensing and clustering data gives 
$\Delta \alpha_{\rm s,m}=0.18$ (0.14 with Planck priors) 
and $\Delta f_{\rm NL}^{\rm (piv)}\simeq 9$ (7 with Planck priors)
\citep{Giannantonioetal2010}. These constraints are similar to what is
expected from future studies of the CMB bispectrum with Planck 
\citep{Sefusattietal2009}.

\begin{table}
\caption{Forecasted  {\color{red}Forecast} 1$\sigma$ errors for the non-linearity parameter 
$f_{\rm NL}$ based on two-point statistics (power spectra)
of the Euclid redshift and weak-lensing surveys. 
Results are obtained using the
Fisher-matrix formalism and marginalizing over eight cosmological 
parameters ($\Omega_\Lambda$, $\Omega_m$, $\Omega_b$, $h$, $n_s$, $\sigma_8$,
$w_0$, $w_a$) plus a large number of nuisance parameters to account for galaxy 
biasing, non-linear redshift-space distortions and shot noise
(see \cite{Giannantonioetal2010} for details).
Results within parentheses include the forecasted  {\color{red}forecast} priors for the cosmological
parameters from the power spectrum of CMB temperature anisotropies measured
with the Planck satellite (note that no prior is assumed on $f_{\rm NL}$).
The label ``Galaxy clustering'' refers to the anisotropic power
spectrum $P(k_\parallel,k_\perp)$ for spectroscopic data and to 
the angular power spectrum $C_\ell$ for photometric data.
The combined analysis of clustering and lensing data is based on
angular power spectra and includes all possible cross-correlations between
different redshift bins and probes.
Non-linear power spectra are computed using the halo model.
This introduces possible inaccuracies in the forecasts for weak lensing
data in the equilateral and orthogonal shapes (see main text for details).
\label{tbl:EUCLIDFNL}}
\centering
\begin{tabular}{ccccc}
\hline
Bispectrum shape  & local & orthogonal & equilateral \\
Fiducial $f_{\rm NL}$ & 0 & 0 & 0 \\
\hline
\hline
Galaxy clustering (spectr. $z$) & 4.1 (4.0) & 54 (11) & 220 (35) \\
Galaxy clustering (photom. $z$) & 5.8 (5.5) & 38 (9.6)& 140 (37) \\
Weak lensing                    & 73 (27)   & 9.6 (3.5) & 34 (13) \\
Combined                        & 4.7 (4.5) & 4.0 (2.2) & 16 (7.5)\\
\end{tabular}
\end{table}

\begin{table}
\caption{Forecasted {\color{red}Forecast} 1$\sigma$ errors for a scale-dependent local model
of primordial non-Gaussianity 
\citep{Giannantonioetal2010}. 
Details of the forecasts are as in the previous table.
\label{tbl:Euclidfnlrun}}
\centering
\begin{tabular}{ccc}
\hline
& $\Delta f_{\rm NL}^{\rm (piv)}$ & $\Delta n_{f_{\rm NL}}$ \\
\hline
\hline
Galaxy clustering (spectr. $z$) & 9.3 (7.2) & 0.28 (0.21) \\
Galaxy clustering (photom. $z$) & 25 (18)   & 0.38 (0.26) \\
Weak lensing                    & 134 (82)  & 0.66 (0.59)   \\
Combined                        & 8.9 (7.4) & 0.18 (0.14) \\
\hline
\end{tabular}
\end{table}

In the end, we briefly comment on how well Euclid data could constrain
the amplitude of alternative forms of primordial non-Gaussianity than
the local one. In particular, we consider the equilateral and orthogonal shapes
introduced in Section \ref{nongaussianshapes}.
Tab.~\ref{tbl:EUCLIDFNL} summarizes the resulting constraints on 
the amplitude of the primordial bispectrum, $f_{\rm NL}$. 
The forecasted {\color{red}forecast} errors from galaxy clustering grow larger and larger
when one moves from the local to the orthogonal and finally to the equilateral
model.
This reflects the fact that the scale-dependent part of the galaxy bias
for $k\to 0$
approximately scales as $k^{-2}$, $k^{-1}$, and $k^0$ 
for the local, orthogonal, and equilateral shapes, respectively  
\citep{Schmidt/Kamion:2010,Wagneretal2010,Scoccimarroetal2011,DJS2011,DJSb2011}. 
On the other hand, 
the lensing constraints (that, in this case, come from the 
very non-linear scales) 
appear to get much stronger for the non-local shapes.
A note of caution is in order here.
In \cite{Giannantonioetal2010}, the non-linear matter power spectrum is
computed used {\color{red}using} a halo model which has been tested against N-body simulations
only for non-Gaussianity of the local type.
\footnote{Very few N-body simulations of the non-local models 
are currently available and none of them has very high spatial resolution.}
In consequence,
the weak-lensing forecasts might be less reliable
than in the local case (see the detailed discussion in \citet{Giannantonioetal2010}). This does not apply for the forecasts based
on galaxy clustering which are always robust as they are based
on the scale dependence of the galaxy bias on very large scales.

\subsection{Complementarity} 

The CMB bispectrum is very sensitive to the shape of non-Gaussianity; 
halo bias and mass function, the most promising approaches to constrain $\fnl$ with a survey 
like Euclid, are much less sensitive.
However, it is the complementarity between CMB and LSS that matters. 
One could  envision different scenarios. 
If non-Gaussianity is local with negative $\fnl$ and CMB 
obtains a detection, then the halo bias approach should 
also give a high-significance detection (GR correction and 
primordial contributions add up), while if it is local but 
with positive $\fnl$, the halo-bias approach could give a 
lower statistical significance as the GR correction contribution has the opposite sign. 
If CMB detects $\fnl$ at the level of  10 and a form 
that is close to local, but halo bias does not detect it, 
then the CMB bispectrum is given by secondary effects  (e.g.~\cite{Mangilli/Verde:2009}). 
If CMB detects non-Gaussianity that is not of the local 
type, then halo bias can help discriminate between equilateral and enfolded shapes: if halo bias sees a signal, it  indicates the enfolded type, and if halo bias does not see  a signal, it indicates the equilateral type. Thus even a  non-detection of the halo-bias effect, in combination with 
CMB constraints, can have an important discriminative  power.

\section{Isocurvature modes}
\label{sec:isocurvature}

At some time well after inflation but deep into the radiation era the Universe
is filled with several components. For instance, in the standard picture right
before recombination there are four components: baryons, cold dark matter,
photons and neutrinos. One can study the distribution of {\em super-Hubble}
fluctuations between different species, which represent the initial conditions
for the subsequent evolution. So far we have investigated mostly the adiabatic initial conditions;
in this section we explore more generally the possibility of isocurvature initial conditions.
Although CMB data are the most sensitive to constrain isocurvature perturbations,
we discuss here the impact on Euclid results.

\subsection{The origin of isocurvature perturbations}

Let us denote by $\rho_\alpha$ the energy density of the component $\alpha$. Perturbations are purely adiabatic when for each component $\alpha$ the quantity $\zeta_\alpha \equiv - 3 H \delta \rho_\alpha/\dot \rho_\alpha$ is the same (\cite{Weinberg:2003,Malik/Wands/Ungarelli:2003}). Let us consider for instance cold dark matter and photons. When fluctuations are adiabatic it follows that $\zeta_{\rm cdm} = \zeta_\gamma$. Using the energy conservation equation, $\dot \rho_\alpha = - 3 H (\rho_\alpha + p_\alpha)$ with $p_{\rm cdm}=0$ and $p_\gamma = \rho_\gamma/3$, one finds that the density contrasts of these species are related by
\begin{equation}
\frac{\delta \rho_{\rm cdm}}{\rho_{\rm cdm}} = \frac34 \frac{\delta \rho_\gamma}{\rho_\gamma} \;.
\end{equation}
Using that $n_{\rm cdm} \propto \rho_{\rm cdm}$ and $n_\gamma \propto \rho_\gamma^{3/4}$, this also implies that particle number ratios between these species is fixed, i.e.~$\delta(n_{\rm cdm}/n_\gamma) = 0$. 

When isocurvature perturbations are present, the condition described above is not satisfied.\footnote{Strictly speaking, isocurvature perturbations are defined by the condition that their total energy density in the total comoving gauge vanishes, i.e.~$\sum_\alpha \delta \rho^{(\rm com.)}_\alpha=0$. Using the relativistic Poisson equation, one can verify that this implies that they do not contribute to the ``curvature'' potential.}
In this case one can define a non-adiabatic or entropic perturbation between two components $\alpha$ and $\beta$ as ${\cal S}_{\alpha, \beta} \equiv \zeta_\alpha -\zeta_\beta$,  
so that, for the example above one has
\begin{equation}
{\cal S}_{{\rm cdm},r } = \frac{\delta \rho_{\rm cdm}}{ \rho_{\rm cdm}} -   \frac34 \frac{\delta \rho_\gamma}{ \rho_\gamma} = \frac{\delta (n_{\rm cdm}/n_\gamma)}{n_{\rm cdm}/n_\gamma} \;.
\end{equation}

A sufficient condition for having purely adiabatic perturbations is that all
the components in the Universe where {\color{red} were} created by a single degree of freedom, such as during reheating after single field inflation.\footnote{In this case in the flat gauge one finds, for each species $\alpha$, $\zeta_\alpha = \zeta$, where $\zeta$ is the Bardeen curvature perturbation conserved on super-Hubble scales.} Even if inflation has been driven by several fields, thermal equilibrium may erase isocurvature perturbations if it is established before any non-zero conserving quantum number was created (see \cite{Weinberg:2004}). 
Thus, a detection of non-adiabatic fluctuations would imply that several
scalar fields where present during inflation {\em and} that either some of
the species where {\color{red}were} not in thermal equilibrium afterwards or that some non-zero conserving quantum number was created before thermal equilibrium. 

The presence of many fields is not unexpected. Indeed, in all the extension of the Standard Model scalar fields are rather ubiquitous. In particular, in String Theory dimensionless couplings are functions of moduli, i.e.~scalar fields describing the compactification. Another reason to consider the relevant role of a second field other than the inflaton is that this can allow to circumvent the necessity of slow-roll (see e.g.~\cite{Dvali/Kachru:2003}) enlarging the possibility of inflationary models. 

Departure from thermal equilibrium is one of the necessary conditions for the generation of baryon asymmetry and thus of the matter in the Universe. Interestingly, the oscillations and decay of a scalar field requires departure from thermal equilibrium. Thus, baryon asymmetry can be generated by this process; examples are the decay of a right-handed sneutrino (\cite{Hamaguchi/Murayama/Yanagida:2002}) or the \cite{Affleck/Dine:1985}  scenario. If the source of the baryon-number asymmetry in the Universe is the condensation of a scalar field after inflation, one expects generation of baryon isocurvature perturbations (\cite{Moroi/Takahashi:2001}). This scalar field can also totally or partially generate adiabatic density perturbations through the curvaton mechanism.

In summary, given our ignorance about inflation, reheating, and the generation of matter in the Universe, a discovery of the presence of isocurvature initial conditions would have radical implications on both the inflationary process and on the mechanisms of generation of matter in the Universe. 

Let us concentrate on the non-adiabatic perturbation between cold dark matter (or baryons, which are also non-relativistic) and radiation ${\cal S} = {\cal S}_{{\rm cdm}, \gamma}$. Constraints on the amplitude of the non-adiabatic component are given in terms of the parameter $\alpha$, defined at a given scale $k_0$, by $P_{\cal S}/P_\zeta \equiv \alpha/(1-\alpha)$, see e.g.~\cite{Beltran/etal:2004,Bean/Dunkley/Pierpaoli:2006,KomatsuWMAP7}. As discussed in \cite{Langlois:1999}, adiabatic and entropy perturbations may be correlated. To measure the amplitude of the correlation one defines a cross-correlation coefficient, $\beta \equiv  - P_{{\cal S}, \zeta} / \sqrt{P_{\cal S} P_\zeta}$. Here $P_{{\cal S}, \zeta}$ is the cross-correlation power-spectrum between ${\cal S}$ and $\zeta$ and for the definition of $\beta$ we have adopted the sign convention of \cite{KomatsuWMAP7}.
Observables, such as for instance the CMB anisotropies, depend on linear combinations of $\zeta$ and ${\cal S}$. Thus, constraints on $\alpha$ will considerably depend on the cross-correlation coefficient $\beta$ (see for instance discussion in \cite{Gordon/Lewis:2003}).

If part of the cold dark matter is created out of equilibrium from a field other than the inflaton, totally uncorrelated isocurvature perturbations, with $\beta=0$, are produced, as discussed for instance in \cite{Efstathiou/Bond:1986,Linde/Mukhanov:1997}. The axion is a well-known example of such a field. 
The axion is the Nambu-Goldstone boson associated to the
\cite{Peccei/Quinn:1977} mechanism to solve the strong-CP problem in QCD. As
it acquires a mass through QCD non-perturbative effects, when the Hubble
rate drops below its mass the axion starts oscillating coherently, behaving
as cold dark matter (\cite{Preskill/Wise/Wilczek:1983,abbott1983,dine1983}).
During inflation, the axion is practically massless and acquires fluctuations which are totally uncorrelated from photons, produced by the inflaton decay (\cite{Seckel/Turner:1985,Linde:1985,Linde:1991,Turner/Wilczek:1991}). As constraints on $\alpha_{\beta=0}$ are currently very strong (see e.g.~\cite{Beltran/etal:2007,Komatsuetal2010}), axions can only represent a small fraction of the total dark matter. 

Totally uncorrelated isocurvature perturbations can also be produced in the curvaton mechanism, if the dark matter or baryons are created from inflation, before the curvaton decay, and remain decoupled from the product of curvaton reheating (\cite{Langlois/Vernizzi/Wands:2008}). This scenario is ruled out if the curvaton is entirely responsible for the curvature perturbations. However, in models when the final curvature perturbation is a mix of the inflaton and curvaton perturbations (\cite{Langlois/Vernizzi:2004}), such an entropy contribution is still allowed.

When dark matter or baryons are produced solely from the curvaton decay,
such as discussed by \cite{Lyth/Ungarelli/Wands:2003}, the isocurvature
perturbations are totally anit-corrlated {\color{red}anti-correlated}, with $\beta=-1$. For instance, some fraction of the curvaton decays to produce CDM particles or the out-of-equilibrium curvaton decay generates the primordial baryon asymmetry (\cite{Hamaguchi/Murayama/Yanagida:2002,Affleck/Dine:1985}).

If present, isocurvature fields are not constrained by the slow-roll conditions imposed on the inflaton field to drive inflation. Thus, they can be highly non-Gaussian \cite{Linde/Mukhanov:1997,Bernardeau/Uzan:2002}. Even though negligible in the two-point function, their presence could be detected in the three-point function of the primordial curvature and isocurvature perturbations and their cross-correlations, as studied in \cite{Kawasaki/etal:2008,Langlois/Vernizzi/Wands:2008}.

\subsection{Constraining isocurvature perturbations}

Even if pure isocurvature models have been ruled out, current observations allow for mixed adiabatic and isocurvature contributions (e.g.~\cite{CrottyEtAl2003,Trotta:2006ww,KomatsuWMAP7, Valiviita/Giannantonio:2009}). 
As shown in \cite{Trotta/Riazuelo/Durrer:2001,2002PhRvL..88u1302A,Valiviita/Giannantonio:2009,Langlois/Riazuelo:2000, Bucher/Moodly/Turok:2001,Sollom/Challinor/Hobson:2009}, the initial conditions issue is a very delicate problem: in fact, for current cosmological data, relaxing the assumption of adiabaticity reduces our ability to do precision cosmology since it compromises the accuracy of parameter constraints.   
 Generally, allowing for isocurvature modes introduces new degeneracies in
 the parameters {\color{red}parameter} space which weaken  constraints considerably.
  
 The Cosmic Microwave Background radiation (CMB), being our window on the early Universe, is the preferred 
 data set to learn about initial conditions. 
 Up to now, however, the CMB temperature power spectrum alone, which is the CMB observable better constrained so far, has not been able to  break the degeneracy between the nature of initial  perturbations (i.e. the amount  and properties of an  isocurvature component) and cosmological parameters,  e.g.~\cite{Kurki/etal:2005,Trotta/Riazuelo/Durrer:2001}.
 Even if the precision measurement of the CMB first acoustic peak at $\ell \simeq 220$  ruled out the possibility of a dominant isocurvature mode, allowing for isocurvature perturbations together with the adiabatic ones introduce additional degeneracies in the interpretation of the CMB data that current experiments could not break.
 Adding external data sets somewhat alleviates the issue for some degeneracy
 directions, e.g.~\cite{Trotta:2002iz,Beltran/etal:2004,Dunkley/etal:2005}.
  As shown in \cite{Bucher/Moodly/Turok:2001}, the precision polarization measurement of the next CMB experiments like Planck will be crucial to lift such degeneracies, i.e. to distinguish the effect of the isocurvature modes from those due to the variations of the cosmological parameters.
  
It is important to keep in mind that analyzing the CMB data with the
prior assumption of purely
adiabatic initial conditions when the real Universe contains even a
small isocurvature contribution,
could lead to an incorrect determination of the cosmological
parameters and on the inferred value
of the sound horizon at radiation drag. The sound horizon at radiation
drag is the standard ruler
that is used to extract information about the expansion history of the
Universe from measurements
of the baryon acoustic oscillations.
Even for a CMB experiment like Planck, a small but non-zero
isocurvature contribution, still allowed by Planck data, if ignored,
can introduce a systematic error in the interpretation of the BAO signal that is comparable if not
larger than the statistical errors.
In fact, \cite{Mangilli/Verde/Beltran:2010} shows that even a tiny amount of
isocurvature perturbation, if not accounted for, could  affect
standard rulers calibration from  CMB observations such as those
provided by the Planck mission, 
affect BAO interpretation, and introduce biases in the recovered dark energy
properties that are larger than forecasted {\color{red}forecast}  statistical errors from future surveys.
In addition it will introduce a mismatch of the expansion history as
inferred from CMB and as masured {\color{red}measured} by BAO surveys. The mismatch between
CMB predicted and the measured expansion histories has been proposed
as a signature for deviations from a DM cosmology in the form of
deviations from Einstein's gravity (e.g.~\cite{Acquaviva/Verde:2007, Ishak/etal:2006}), couplings in 
the dark sector (e.g.~\cite{Lopez/etal:2010}) or time-evolving dark energy.

For the above reasons, extending on the work of \cite{Mangilli/Verde/Beltran:2010} , \cite{Carbone_etal2011} adopted a general fiducial cosmology
which includes a varying dark energy equation of state parameter and
curvature. Beside {\color{red}In addition to} BAO measurements, in this case the information from the shape of the galaxy power spectrum
are included and a joint analysis of a Planck-like CMB probe and a
Euclid-type survey is considered.
This allows one to break the degeneracies that affect the
CMB and BAO combination. As a result, most of the cosmological
parameter systematic biases arising from an incorrect assumption on
the isocurvature fraction parameter $f_{\rm iso}$, become negligible with
respect to the statistical errors.   The combination of CMB and LSS gives a statistical
error $\sigma(f_{\rm iso}) \sim 0.008$,
even when curvature and a varying dark energy  equation of state are
included, which is smaller than the error obtained from CMB  alone when flatness and cosmological constant are
assumed. These results confirm the synergy and complementarity
between  CMB and LSS, and the great potential of   future  and planned galaxy surveys.

\section{Summary and Outlook}
\noindent 
We have  summarized  aspects of the initial conditions  for the growth of
cosmological perturbations that Euclid will enable us to probe. In particular we
have considered the shape of the primordial power spectrum and its connection to
inflationary models, primordial non-Gaussianity and isocurvature perturbations.

A survey like Euclid will greatly improve our knowledge of the initial
conditions for the growth of perturbations and  will help shed light on the
mechanism  for the generation of primordial perturbations.
The addition of Euclid data will improve the Planck satellite's Cosmic Microwave
Background  constraints on parameters describing  the shape of the primordial
power spectrum by a factor of 2-3.  

Primordial non-Gaussianity can be tested by Euclid in three different and
complementary ways: via the galaxy bispectrum, number counts of non-linear
structures and the non-Gaussian halo bias.  These approaches are also highly 
competitive with and complementary to CMB constraints. In combination with
Planck, Euclid will not only test a possible scale-dependence of non-Gaussianity
but also its shape. The shape of non-Gaussianity  is the  key to constrain and
classify possible deviations for the simplest single-field slow roll inflation. 

Isocurvature modes affect the  interpretation of large-scale structure
clustering in two ways.
The power spectrum shape is modified on small scales due to the extra
perturbations although this effect however can be mimicked by  scale-dependent
bias. More importantly  isocurvature modes can lead to an incorrect inferred
value for the sound horizon at radiation drag from CMB data. This then predicts
an incorrect   location of the baryon acoustic feature. It is through this
effect that  Euclid BAO measurements improve constraints on isocurvature modes.






%% file: structure/structure.tex
\chapter{Testing the basic cosmological hypotheses}\label{testing}

\section{Introduction}

The standard cosmological analyses implicitly make several assumptions, none of which are seriously challenged by current data.  Nevertheless, Euclid offers the possibility of testing some of these basic hypotheses.  Examples of the standard assumptions are that
photon number is conserved, that the Copernican  principle holds 
(i.e. we are not at a special place in the Universe)  and that  the Universe 
is homogeneous and isotropic,  at least on large enough scales. 
These are the pillars on which  standard cosmology is built, 
 so it is important to take the opportunity offered by Euclid observations to test these basic hypotheses.

\section{Transparency and Etherington relation}\label{transparency-and-Etherington-relation}

The Etherington relation \citep{Etherington:1933} implies that, in a cosmology based on a metric theory 
of gravity, distance measures are unique: the luminosity distance is $(1 + z)^2$ times the 
angular diameter distance. This is valid in any cosmological background where photons 
travel on null geodesics and where, crucially, photon number is conserved. 
There are several scenarios in which the Etherington relation would be violated: for 
instance we can have deviations from a metric theory of gravity, photons not
traveling {\color{red}travelling} 
along unique null geodesics, variations of fundamental constants, etc.
We follow here the approach of  \cite{Avgoustidis/etal:2010}.\\

\subsection{Violation of photon conservation}

A change in the photon flux during propagation towards the Earth will affect
the Supernovae (SNe) luminosity distance measures $D_L(z)$
\label{symbol:luminosity}but not the determinations of the angular diameter
distance. BAO will not be affected so $D_A(z)$ and $H(z)$   measurements
from BAO  could be combined with Supernovae measurements of $D_L(z)$ to
constrain deviations from photon number conservation. Photon conservation
can be violated by simple astrophysical effects or by exotic physics.
Amongst the former we find, for instance, attenuation due to interstellar
dust, gas and/or plasmas. Most known sources of attenuation are expected to
be clustered and can be typically constrained down to the 0.1\% level
\citep{Menard/etal:2008, More/etal:2009}. Unclustered sources of attenuation
are however much more difficult to constrain. For  example, gray
{\color{red}grey} dust \citep{Aguirre:1999} has been invoked to explain the
observed dimming of Type Ia Supernovae without resorting to cosmic
acceleration. More exotic sources of photon conservation violation involve a
coupling of photons  to particles beyond the standard model of particle
physics. Such couplings would mean that, while passing through the
intergalactic medium, a photon could disappear or even (re)appear!
Interacting with such exotic particles, modifying the apparent  luminosity
of sources. Recently,   \cite{Avgoustidis/etal:2010} considered the mixing of photons with scalars, known as axion-like particles, chameleons,  and the possibility of mini-charged particles which have a tiny, and unquantized electric charge. In particular, the implications of these particles on the SN luminosity have been described in a number  of publications~\citep{Csaki/etal:2002, Mortsell/etal:2002,Burrage:2008, Ahlers:2009} and a detailed discussion of the proposed approach can be found  in Refs.~\cite{BassettKunz1:2004,BassettKunz2:2004,Avgoustidis/etal:2009,Avgoustidis/etal:2010}.

\begin{figure}[htbp!]
    \centering
       
\includegraphics[width=.75\textwidth]{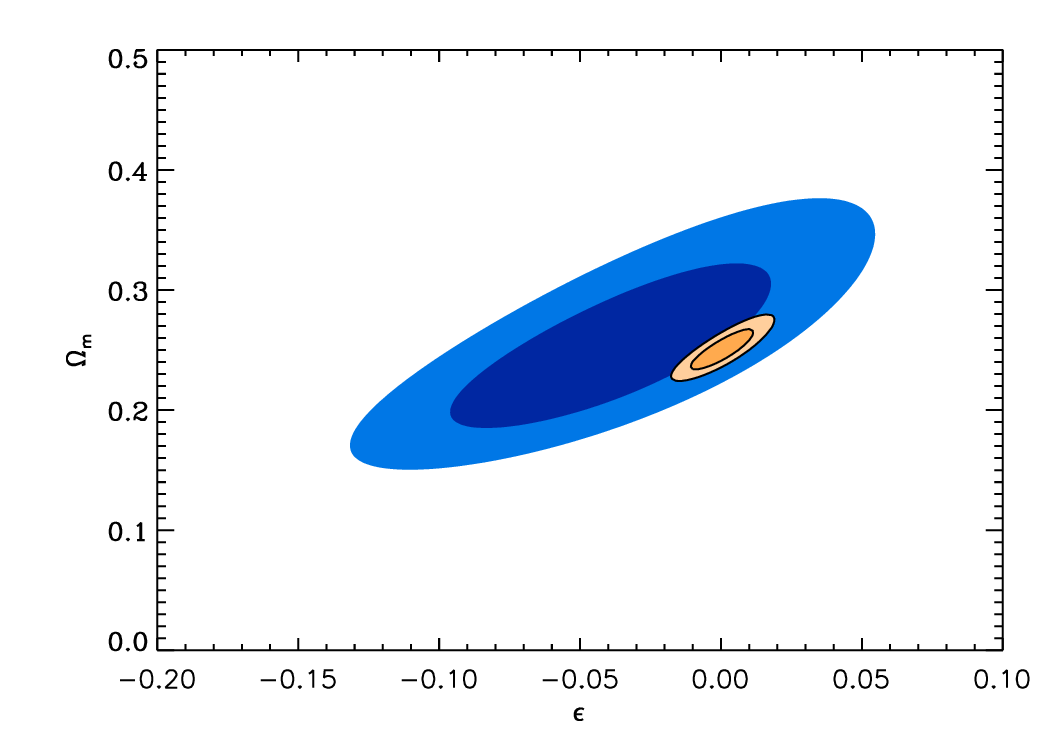}
          
               \caption{Constraints on possible violation of the Etherington
relation in the form of deviations from a perfectly transparent Universe
($\epsilon=0$). Blue regions represent current constraints while orange are
forecasted {\color{red}forecast} Euclid constraints assuming it is accompanied by  a dark energy task
force stage IV supernovae sample.}
                 \label{fig:transparency}

\end{figure}

Any systematic violations in  photon conservation can then be interpreted as
an opacity effect in the observed luminosity distance, parametrised through
a generic opacity parameter,$\tau(z)$, as: $$
D_{L,\rm obs}^2=D^2_{L,\rm true}\exp[\tau(z)]\,.$$ Note that a negative $\tau(z)$
allows for apparent brightening of light sources, as would be the case, for
example, if exotic particles were also emitted from the source and converted
into photons along the line of sight \citep{Burrage:2008}. Following
\cite{Avgoustidis/etal:2009} generic deviations from the Etherington
relation  can be parameterized {\color{red}parametrized} as: \label{symbol:tau}
$$D_L(z) = D_A(z)(1 + z)^{2+\epsilon}\,.$$ 
Forecasted {\color{red}Forecast} Euclid constraints are shown in Fig. \ref{fig:transparency}, taken from \cite{Avgoustidis/etal:2010}. This assumes that  Euclid is accompanied by a supernova sample with the characteristic of a dark energy task force stage IV survey

For particular models of exotic matter-photon 
coupling, namely axion-like particles (ALPs), chameleons, and mini-charged
particles (MCPs), the appropriate parameterization
{\color{red}parametrization} of $\tau(z)$ is used instead. 

\subsection{Axion-like particles}

Axion-like particles (ALP) can arise from field theoretic extensions of the standard model as Goldstone bosons when a global shift symmetry, present in the high energy sector, is spontaneously broken. Interestingly, these fields  also arise naturally in string theory (for a review see \cite{SvrcekWitten:2006}). Chameleon scalar fields are another very interesting type of ALPs \citep{Brax/etal:2010}. They were originally invoked to explain the current accelerated expansion of the Universe with a quintessence field which can couple to matter without giving rise to large fifth forces or unacceptable violations of the weak equivalence principle.  A chameleon  model with only matter couplings will induce a coupling to photons.

\begin{figure}[htbp!]
    \centering
        \includegraphics[width=.75\textwidth]{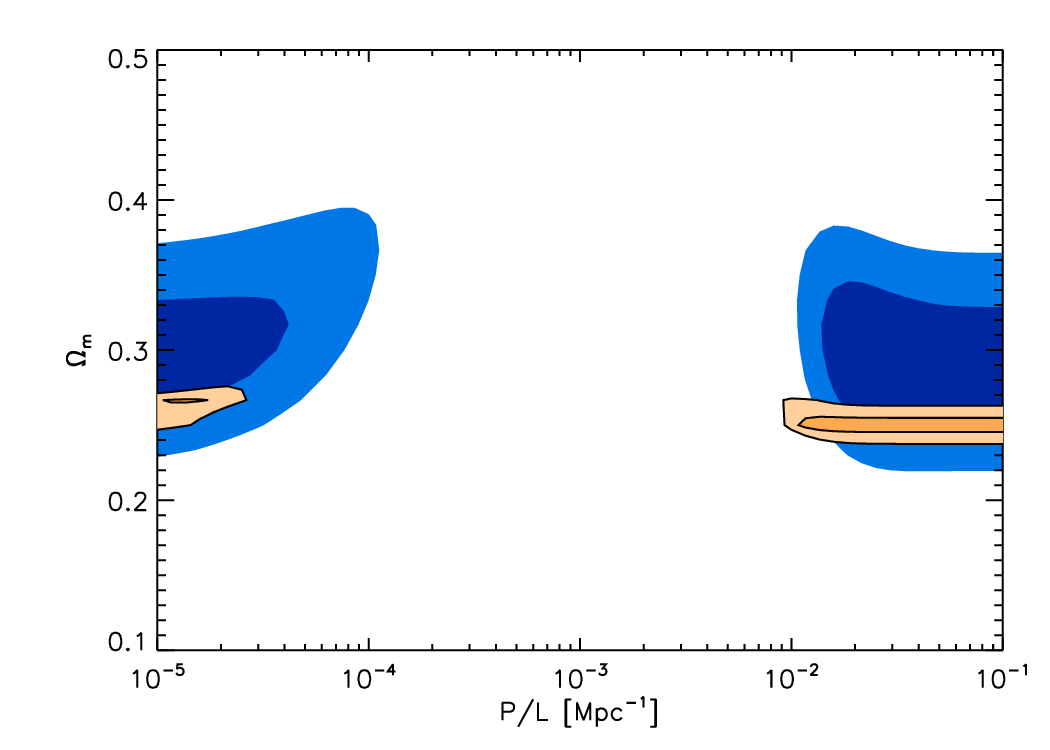}
   \caption{Constraints on the simplest Axion-like particles models. 
Blue regions represent current constraints while orange are forecasted
{\color{red}forecast} 
Euclid constraints assuming it is accompanied by  a dark energy task force stage IV supernovae sample.
Here $P/L$ is the conversion probability per unit length
and is the relevant parameter for $\tau(z)$ (see \protect{\cite{Avgoustidis/etal:2010}}).}
    \label{fig:axion2}
\end{figure}

The presence of ALPs will have an impact on observations of SNe if their
observed light passes through (intergalactic) magnetic fields. The  net
effect depends on the ratio of the transition probability to the length
traveled {\color{red}travelled} through a magnetic field, and a parameter
$A$ describing the degree of thermalization of the initial flux ($A=1$ means
thermalized flux where the photon to ALP transition is compensated by  the
inverse ALP to photon, making the photon number constant). For the simplest
ALP model $A=2/3$, the present and forecasted {\color{red}forecast} constraints are shown in Fig. \ref{fig:axion2} taken from~\cite{Avgoustidis/etal:2010}. 

\subsection{Mini-charged particles}
New particles with a small unquantized charge have been investigated in several 
extensions of the standard model \citep{Holdom:1986, BatellGherghetta:2006}. In particular, they arise naturally in 
extensions of the standard model which contain at least one additional U(1) hidden 
sector gauge group \citep{Holdom:1986, Bruemmer/etal:2009}. The gauge boson of this additional U(1) is known as 
a hidden photon, and hidden sector particles, charged under the hidden U(1), get an 
induced electric charge proportional to the small mixing angle between the kinetic terms 
of the two photons. In string theory, such hidden U(1)s and the required kinetic mixing 
are a generic feature \citep{Abel/etal:2008a, Abel/etal:2008b, Dienes/etal:1997, AbelSchofield:2004, Goodsell/etal:2009}. Hidden photons are not necessary however to 
explain mini-charged particles, and explicit brane-world scenarios have been constructed 
\citep{BatellGherghetta:2006} where MCPs arise without the need for hidden photons. 

\begin{figure}[htbp!]
    \centering
        \includegraphics[width=.75\textwidth]{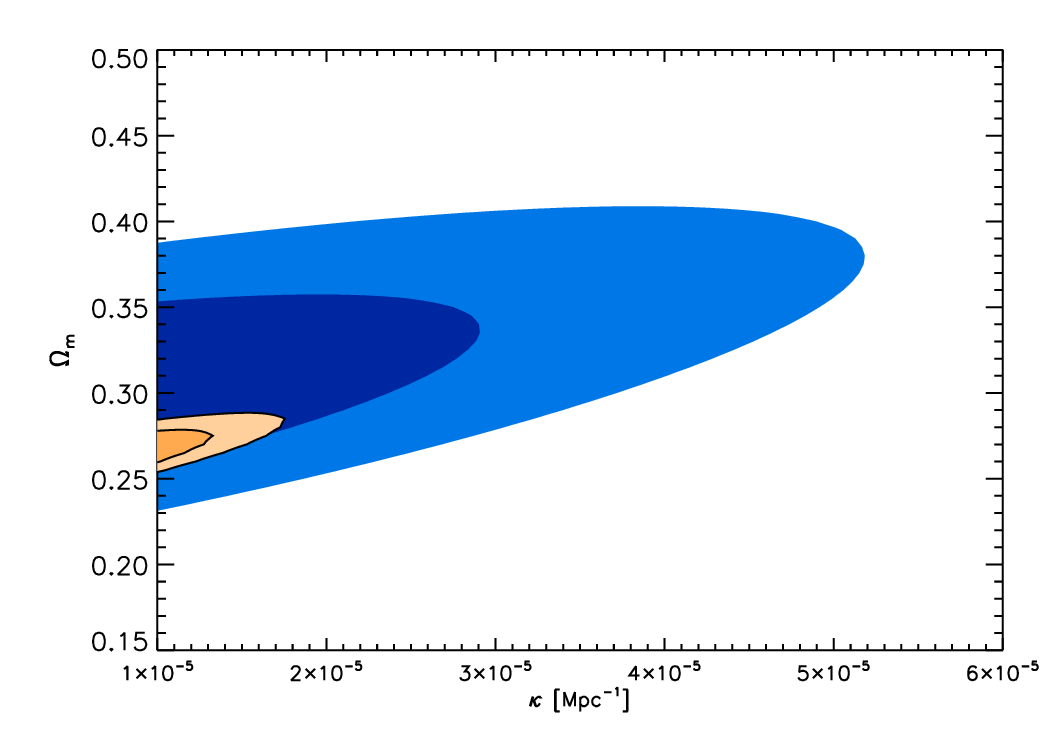}
   \caption{Constraints on MCP models. Blue regions represent current
   constraints while orange are forecasted {\color{red}forecast} Euclid 
constraints assuming it is accompanied by  a dark energy task force stage IV supernovae sample.}
    \label{fig:mcp}
\end{figure}

More interestingly, \citet{Ahlers:2009,Gies/etal:2006,Ahlers/etal:2008} pointed
out that photons propagating in a background magnetic field can actually
pair-produce 
MCPs without the need for a second photon in the initial state.
The opacity in this case is  parameterized {\color{red}parametrized} by $\kappa y(z)$ where $y$ is the 
comoving distance to the source and $\kappa$ encloses information on the MCP electric
charge and the intervening magnetic field strength. 
Fig.~\ref{fig:mcp} shows current and forecasted {\color{red}forecast} Euclid's constraints, taken from
\cite{Avgoustidis/etal:2010} assuming Euclid is accompanied by a supernova
sample with the characteristic of a dark energy task force stage IV survey. 

\section{Beyond homogeneity and isotropy}

The crucial ingredient that kickstarted dark energy research was the interpretation in 1998 of standard candle observations in terms of cosmic acceleration required to explain the data in the context of the FLRW metric. What we observe is however merely that distant sources ($z>0.3$) are dimmer than we would predict in  
a matter-only Universe
calibrated through {}``nearby'' sources. That is, we observe a different evolution of luminosity rather than directly an increase in the expansion rate. Can this be caused by a strong inhomogeneity rather than by an accelerating Universe?

In addition, cosmic acceleration seems to be a recent phenomenon at least for standard dark energy models, which gives rise to the
coincidence problem. The epoch in which dark energy begins to play a role is close to the epoch in which most of the cosmic structures formed out of the slow linear gravitational growth. We are led to ask again: can the acceleration be caused by strong inhomogeneities rather than by a dark energy component?

Finally, one must notice that in all the standard treatment of dark energy one always assumes a perfectly isotropic expansion. Could it be that some of the properties of acceleration depends critically on this assumption?

In order to investigate these issues, in this section we explore radical deviations from homogeneity and isotropy and see how Euclid can test them.

\subsection{Anisotropic models}
\label{sec:anisotropicmodels}
In recent times, there has been a resurgent interest towards anisotropic
cosmologies, classified in terms of Bianchi solutions to general relativity.
This has been mainly motivated by hints of anomalies in the cosmic
microwave background (CMB) distribution observed on the full sky by
the WMAP satellite
\citep{de-Oliveira:2004,2004ApJ...609...22V,2005MNRAS.356...29C,
2004ApJ...605...14E}.
While the CMB is very well described as a highly isotropic (in a statistical
sense) Gaussian random field, {\color{red}and the anomalies are {\it a posteriori} statistics and therefore their statistical significance should due corrected  at least for the so-called {\it look elsewhere effect} (see e.g., \cite{Pontzen:2010yw, Bennett:2010jb} and references therein)} recent analyses have shown that local
deviations from Gaussianity in some directions (the so called cold
spots, see \cite{2005MNRAS.356...29C}) cannot be excluded at high
confidence levels. Furthermore, the CMB angular power spectrum extracted
from the WMAP maps has shown in the past a quadrupole power lower than expected
from the best-fit cosmological model \citep{2004MNRAS.348..885E}.
Several explanations for this anomaly have been proposed (see e.g.
\cite{2003PhLB..574..141T,2003JCAP...09..010C,2003PhRvD..67j3509D,
2007PhRvD..76f3007C,2007PhRvD..76h3010G})
including the fact that the Universe is expanding with different velocities
along different directions. While deviations from homogeneity and
isotropy are constrained to be very small from cosmological observations,
these usually assume the non-existence of anisotropic sources in the
late Universe. Conversely, as suggested in
~\cite{2008JCAP...06..018K,2008ApJ...679....1K,2006PhRvD..74d1301B,
2006PhRvD..73f3502C,2008arXiv0812.0376C},
dark energy with anisotropic pressure acts as a late-time source of
anisotropy. Even if one considers no anisotropic pressure fields,
small departures from isotropy cannot be excluded, and it is interesting
to devise possible strategies to detect them.

The effect of assuming an anisotropic cosmological model on the CMB
pattern has been studied by
\cite{1973MNRAS.162..307C,1985MNRAS.213..917B,1995A&A...300..346M,
1996A&A...309L...7M,1996PhRvL..77.2883B,1997PhRvD..55.1901K}.
The Bianchi solutions describing the anisotropic line element were
treated as small perturbations to a Friedmann-Robertson-Walker (FRW)
background. Such early studies did not consider the possible presence
of a non-null cosmological constant or dark energy and were upgraded
recently by \cite{2006MNRAS.369.1858M,2006ApJ...644..701J}.

One difficulty with the anisotropic models that have been shown to fit
the large-scale CMB pattern, is that they have to be produced according
to very unrealistic choices of the cosmological parameters. For example,
the Bianchi VIIh template used in \cite{2006ApJ...644..701J} requires
an open Universe, an hypothesis which is excluded by most cosmological
observations. An additional problem is that an inflationary phase
-- required to explain a number of feature of the cosmological model
-- isotropizes the Universe very efficiently, leaving a residual anisotropy
that is negligible for any practical application. These difficulties
vanish if an anisotropic expansion takes place only well after the
decoupling between matter and radiation, for example at the time of
dark energy domination
~\citep{2008JCAP...06..018K,2008ApJ...679....1K,2006PhRvD..74d1301B,
2006PhRvD..73f3502C,2008arXiv0812.0376C}.

Bianchi models are described by homogeneous and anisotropic metrics. If
anisotropy is  slight, the dynamics of any Bianchi model can be decomposed  into
an isotropic FRW background linearly perturbed to break  isotropy; on the other
side,  homogeneity is maintained with respect to three Killing vector fields. 

The geometry of Bianchi models is set up by the structure constants ${\it
C}^k_{ij}$, defined by the commutators of the {\color{red}(these)} three Killing fields
$\vec{\xi}_i$: 
\begin{equation}\label{eq:killing}
\left[ \vec{\xi}_i, \vec{\xi}_j \right] = {\it C}^k_{ij} \vec{\xi}_k.
\end{equation}
The structure constants are subjected {\color{red}subject} to the antisymmetry relation ${\it
C}^k_{ij} = - {\it C}^k_{ji}$ and the Jacobi identities ${\it C}^a_{[bc} {\it
C}^d_{e]a}=0$. As a consequence,  their attainable values are restricted to only
four of the initial $27$ necessary to  describe a given space. In
\cite{1969CMaPh..12..108E} these four values are dubbed as  $n_1, n_2, n_3$ and
$a_1$.  The categorization of Bianchi models into different types relies  on
classifying the inequivalent sets of these four constants. In
table~\ref{tab:bianchi} the subclass of interest containing the FRW limit is
shown. Bianchi types VIIh and IX contain the open and closed FRW model,
respectively. Type VII$_{0}$ contains the flat FRW; types I and V are just
particular subcases of the VII$_{0}$ and VIIh. In type I  no vortical
{\color{red}vertical} motions are
allowed and the only extension with respect to the FRW case is that there are 
three different scale factors. 
The metric in general can be written as
\begin{equation}
\label{eq:bianmetr}
g_{\mu\nu}=-n_{\mu}n_{\nu}+g_{ab}\xi_{\mu}^{a}\xi_{\nu}^{b},
\end{equation}
where $g_{ab}$ is a $3\times3$ metric depending on t {\color{red}$t$}. It can be decomposed as
$g_{ab}=e^{2\alpha}[e^{2\beta}]_{ab}$, where the first term represents the
volumetric expansion and the second term includes the anisotropy.
\begin{table}
\begin{center}
\begin{tabular}{c|ccccl}
Type & $a$ & $n_1$ & $n_2$ & $n_3$  \\
\hline
I & 0 & 0 & 0 & 0 \\
V & $1$ & 0 & 0 & 0  \\
VII$_{0}$ & 0 & 0 & 1 & 1  \\
VII$_h$ & $\sqrt{h}$ & 0 & 1 & 1  \\
IX & 0 & 1 & 1 & 1  \\
\end{tabular}
\caption{Bianchi models containing FRW limit and their structure
constants.}\label{tab:bianchi}
\end{center}
\end{table}

\subsubsection{Late time anisotropy}
\label{sec:ltanisotr}

 While deviations from homogeneity and isotropy are constrained to be very small
from cosmological observations,
these usually assume the non-existence of anisotropic sources in the
late Universe. The CMB provides very tight constraints on Bianchi models
at the time of
recombination~\citep{1996PhRvL..77.2883B,1997PhRvD..55.1901K,1995A&A...300..346M}
of order of the quadrupole value, i.e. $\sim10^{-5}$. Usually, in
standard cosmologies with a cosmological constant the anisotropy parameters
scale as the inverse of the comoving volume. This implies an isotropization
of the expansion from the recombination up to the present, leading
to the typically derived constraints on the shear today, namely
$\sim10^{-9}\div10^{-10}$.
However, this is only true if the anisotropic expansion
is not generated by any anisotropic source arising after decoupling,
e.g. vector fields representing anisotropic dark energy
~\citep{2008ApJ...679....1K}.

As suggested in
~\cite{2008JCAP...06..018K,2008ApJ...679....1K,2006PhRvD..74d1301B,
2006PhRvD..73f3502C,2008arXiv0812.0376C},
dark energy with anisotropic pressure acts as a late-time source of
anisotropy. 
An additional problem is that an inflationary phase -- required to explain a
number of feature of the cosmological model
-- isotropizes the Universe very efficiently, leaving a residual anisotropy that
is negligible for any practical application. These difficulties
vanish if an anisotropic expansion takes place only well after the decoupling
between matter and radiation, for example at the time of
dark energy domination
~\citep{2008JCAP...06..018K,2008ApJ...679....1K,2006PhRvD..74d1301B,
2006PhRvD..73f3502C,2008arXiv0812.0376C}.

For example the effect of cosmic parallax~\citep{2008arXiv0809.3675Q} 
has been
recently proposed as a tool to assess the presence of an anisotropic
expansion of the Universe. It is essentially the change in angular
separation in the sky between far-off sources, due to an anisotropic
expansion.

A common parameterization of an anisotropically distributed dark energy
component  is studied in  a class of Bianchi I type, where the line element is
 \begin{equation}
ds^{2}=-dt^{2}+a^{2}(t)dx^{2}+b^{2}(t)dy^{2}+c^{2}(t)dz^{2}.\label{metric}
\end{equation}
 The expansion rates in the three Cartesian directions $x$, $y$
and $z$ are defined as $H_{X}=\dot{a}/a$, $H_{Y}=\dot{b}/b$ and
$H_{Z}=\dot{c}/c$, where the dot denotes the derivative with respect
to coordinate time. In these models they differ from each other, but
in the limit of $H_{X}=H_{Y}=H_{Z}$ the flat FRW isotropic expansion
is recovered. Among the Bianchi classification models the type I exhibits
flat geometry and no overall vorticity; conversely, shear components
$\Sigma_{X,Y,Z}=H_{X,Y,Z}/H-1$ are naturally generated, where $H$
is the expansion rate of the average scale factor, related to the
volume expansion as $H=\dot{A}/A$ with $A=(abc)^{1/3}$.

The anisotropic expansion is caused by the anisotropically stressed dark energy
fluid whenever
its energy density contributes to the global energy budget. If the
major contributions to the overall budget come from matter and dark
energy, as after recombination, their energy-momentum tensor can be
parametrized as: \begin{eqnarray}
T_{(m)\nu}^{\mu} & = & \mbox{diag}(-1,w_{m},w_{m},w_{m})\rho_{m}\\
T_{({\rm DE)\nu}}^{\mu} & = & \mbox{diag}(-1,w,w+3\delta,w+3\gamma)\rho_{{\rm
DE}},\label{eq:Tmunu-de}\end{eqnarray}
 respectively, where $w_{m}$ and $w$ are the equation of state parameters
of matter and dark energy and the skewness parameters $\delta$ and
$\gamma$ can be interpreted as the difference of pressure along the
$x$ and $y$ and $z$ axis. Note that the energy-momentum tensor~\eqref{eq:Tmunu-de}
is the most general one compatible with the
metric~\eqref{metric}~\citep{2008ApJ...679....1K}.
Two quantities are introduced to define the degree of anisotropic
expansion: \begin{equation}
\begin{aligned}R &
\,\equiv\,(\dot{a}/a-\dot{b}/b)/H\;=\;\Sigma_{x}-\Sigma_{y}\,,\\
S & \,\equiv\,(\dot{a}/a-\dot{c}/c)/H\;=\;2\Sigma_{x}+\Sigma_{y}\,.\end{aligned}
\label{dom}\end{equation}

Considering the generalized Friedmann equation, the continuity equations
for matter and dark energy and no coupling between the two fluids,
the derived autonomous system reads 
 \citep{2008JCAP...06..018K,2008ApJ...679....1K}: 
 \begin{equation}
\begin{aligned}U'= & U(U-1)[\gamma(3+R-2S)\;+\,\delta(3-2R+S)\,+\,3(w-w_{m})]\\
S'= &
\frac{1}{6}(9-R^{2}+RS-S^{2})\big\{S[U(\delta+\gamma+w-w_{m})+w_{m}-1]-6\,
\gamma\, U\big\}\\
R'= &
\frac{1}{6}(9-R^{2}+RS-S^{2})\big\{R[U(\delta+\gamma+w-w_{m})+w_{m}-1]-6\,
\delta\, U\big\},
\end{aligned}
\label{sys}
\end{equation}
 where $U\equiv\rho_{{\rm DE}}/(\rho_{{\rm DE}}+\rho_{m})$ and the
derivatives are taken with respect to $\log(A)/3$. 
System~(\ref{sys}) exhibits many different fixed points,
defined as the solutions of the system $S'=R'=U'=0$. Beside the Einstein-de
Sitter case ($R_{*}=S_{*}=U_{*}=0$), the most physically interesting
for our purposes are the dark energy dominated solution \begin{equation}
R_{*}\,=\,\frac{6\delta}{\delta+\gamma+w-1},\;\;\;
S_{*}\,=\,\frac{6\gamma}{\delta+\gamma+w-1},\;\;\;
U_{*}=1,\label{eq:de-domination}\end{equation}
 and the scaling solution \begin{equation}
\begin{aligned} &
R_{*}\,=\,\frac{3\delta(\delta+\gamma+w)}{2(\delta^{2}-\delta\gamma+\gamma^{2})}
,\quad
S_{*}\,=\,\frac{3\gamma(\delta+\gamma+w)}{2(\delta^{2}-\delta\gamma+\gamma^{2})}
,\;\;\;
U_{*}\,=\,\frac{w+\gamma+\delta}{w^{2}-3(\gamma-\delta)^{2}+2w(\gamma+\delta)},
\end{aligned}
\label{scal}\end{equation}
 in which $\rho_{{\rm DE}}/\rho_{m}=\rm const.$, i.e., the fractional
dark energy contribution to the total energy density is constant.

Anisotropic distribution of sources in Euclid survey might constrain the
anisotropy at present, when the
dark energy density is of order $74\%$, hence not yet in the final
dark energy dominant attractor phase~(\ref{eq:de-domination}). 

\subsection{Late-time inhomogeneity}

Inhomogeneity is relatively difficult to determine, as observations are
typically made on our past light cone, but some methods exist (e.g.
\cite{Clarkson/Maartens:2010,Ellis:2011,Maartens:2011}).  However,
homogeneity may be tested by exploring the interior of the past light cone
by using the fossil record of galaxies to probe along the past world line of
a large number of galaxies~\citep{Heavens:2011}.  One can use the average
star formation rate at a fixed lookback time as a diagnostic test for
homogeneity.  The lookback time has two elements to it -- the lookback time of the emission of the light, plus the time along the past world line.  The last of these can be probed using the integrated stellar spectra of the galaxies, using a code such as VESPA~\citep{Tojeiro:2007}, and this is evidently dependent only on atomic and nuclear physics, independent of homogeneity.  The lookback time can also be computed, surprisingly simply, without assuming homogeneity from 
\begin{equation}
\Delta t = \int_0^z \frac{dz'}{(1+z')H_r(z')}
\end{equation}
\citep{Heavens:2011} where $H_r$ is the radial Hubble constant.  In principle this can be obtained from radial BAOs, assuming early-time homogeneity so that the physical BAO scale is fixed.   The spectroscopic part of Euclid could estimate both the star formation histories from stacked spectra, and the radial expansion rate.

\subsection{Inhomogeneous models: Large Voids}

Non-linear inhomogeneous models are traditionally studied either with
higher-order perturbation theory or with $N$-body codes. Both approaches
have their limits. A perturbation expansion obviously break
{\color{red}breaks} down when the perturbations are deeply in the non-linear regime. $N$-body codes, on the other hand, are intrinsically Newtonian and, at the moment, are unable to take into account full relativistic effects. Nevertheless, these codes can still account for the general relativistic behaviour of gravitational collapse in the case of inhomogeneous large void models, as shown recently in ~\cite{AGBTV2010}, where the growth of the void follows the full  non-linear GR solution down to large density contrasts (of order one).

A possibility to make progress is to proceed with the most extreme simplification: radial symmetry. By assuming that the inhomogeneity is radial (i.e. we are at the center of a large void or halo) the dynamical equations can be solved exactly and one can make definite observable predictions. 

It is however clear from the start that these models are highly controversial, since  the observer needs to be located at the center of the void with a tolerance of  about few percent of the void scale radius, see~\cite{Blomqvist/Mortsell:2010,Clarkson/Maartens:2010}, disfavoring the long-held Copernican Principle (CP).  Notwithstanding this, the idea that we live near the center of  a huge  void is attractive for another important reason: a void creates an apparent acceleration field that could in principle match the supernovae observations~\citep{tomita00,tomita01,cele00,iguchi02}. Since we observe that nearby SN Ia recede faster than the $H(z)$ predicted by the Einstein-de Sitter Universe, we could assume that we live in the middle of a huge spherical region which is expanding faster because it is emptier than the outside. The transition redshift $z_{e}$, i.e. the void edge, should be located around 0.3-0.5, the value at which in the standard interpretation we observe the beginning of acceleration.

The consistent way to realize such a spherical inhomogeneity has been studied since the 1930s in the relativistic literature: the Lema\^{i}tre-Tolman-Bondi (LTB)\index{Lemaitre-Tolman-Bondi (LTB) metric@Lema\^{i}tre-Tolman-Bondi (LTB)
metric} metric. This is the generalization of a FLRW metric in which the expansion factor along the radial coordinate $r$ is different relative to the surface line element $\rd\Omega^{2}=\rd\theta^{2}+\sin^{2}\theta\,\rd\phi^{2}$. If we assume the inhomogeneous metric (this subsection follows closely the treatment in \cite{Amendola2010})
\begin{equation}
{\textrm{d}}s^{2}=-{\textrm{d}}t^{2}+X^{2}(t,r)\,{\textrm{d}}r^{2}
+R^{2}(t,r)\,{\textrm{d}}\Omega^{2}\,,\label{eq:inh-met}
\end{equation}
and solve the $(0,1)$ Einstein equation for a fluid at rest 
 we find that the LTB metric is given by 
\begin{equation}
{\textrm{d}}s^{2}=-{\textrm{d}}t^{2}+\frac{\left[R'(t,r)\right]^{2}}
{1+\beta(r)}{\textrm{d}}r^{2}+R^{2}(t,r){\textrm{d}}\Omega^{2}\,,\label{eq:LTB}
\end{equation}
where $R(t,r),\beta(r)$ are arbitrary functions. Here primes and dots refer
to partial space and time derivatives, respectively. The function $\beta(r)$
can be thought of as a position-dependent spatial curvature. If $R$ is
factorized so that $R(t,r)=a(t)f(r)$ and $\beta(r)=-Kf^{2}(r)$, then we
recover the FLRW metric (up to a redefinition of $r$: from now on when we
seek the FLRW limit we put $R=a(t)r$ and $\beta=-Kr^{2}$). Otherwise, we
have a metric representing a spherical inhomogeneity centered
{\color{red}centred} on the origin. An observer located at the origin will observe an isotropic Universe. We can always redefine $r$ at the present time to be $R_{0}\equiv R(t_{0},r)=r$, so that the metric is very similar  to a FLRW today.

Considering the infinitesimal radial proper length $D_{||}=R'\rd r/\sqrt{1+\beta}$, 
we can define the \emph{radial} \emph{Hubble function} \index{Hubble
parameter!radial}as
\begin{equation}
H_{||}\equiv\dot{D}_{||}/D_{||}=\dot{R}'/R'\,,\end{equation}
 and similarly the \emph{transverse Hubble function}:\index{Hubble
parameter!transverse}
\begin{equation}
H_{\perp}=\dot{R}/R\,.\end{equation}
Of course the two definitions coincide for the FLRW metric.
The non-vanishing components of the Ricci tensor for the LTB 
metric\index{Ricci tensor!for LTB metric} are 
\begin{eqnarray}
& & R_{0}^{0}=\frac{2\ddot{R}}{R}+\frac{\ddot{R}'}{R'}\,,\\
& & R_{1}^{1}= \frac{2\dot{R}\dot{R'}+R\ddot{R}'-\beta'}{RR'}\,,\\
& & R_{2}^{2}=R_{3}^{3}= \frac{\dot{R}^{2}-\beta}{R^{2}}+
\frac{\dot{R}\dot{R}'+R'\ddot{R}-\beta'/2}{RR'}\,.
\end{eqnarray}

In terms of the two Hubble functions, we find that the Friedmann equations for the pressureless matter density $\rho_{m}(t,r)$ are given by
\cite{alnes-amarz} \begin{eqnarray}
H_{\perp}^{2}+2H_{||}H_{\perp}-\frac{\beta}{R^{2}}-\frac{\beta'}{RR'} & = & 8\pi
G\rho_{m}\,,\label{eq:ltb-fri1}\\
6\frac{\ddot{R}}{R}+2H_{\perp}^{2}-2\frac{\beta}{R^{2}}-2H_{||}H_{\perp}+\frac{
\beta'}{RR'} & = & -8\pi G\rho_{m}\,.\label{eq:ltb-fri2}\end{eqnarray}
 Adding eqs.~(\ref{eq:ltb-fri1}) and (\ref{eq:ltb-fri2}), it follows
that $2R\ddot{R}+\dot{R}^{2}=\beta$. Integrating this equation, we
obtain a Friedmann-like equation \begin{equation}
H_{\perp}^{2}=\frac{\alpha(r)}{R^{3}}+\frac{\beta(r)}{R^{2}}\,,\label{eq:hperpeq}\end{equation}
 where $\alpha(r)$ is a free function that we can use along with
$\beta(r)$ to describe the inhomogeneity. From this we can define
an effective density parameter $\Omega_{m}^{(0)}(r)=\Omega_{m}(r,t_{0})$
today: \begin{equation}
\Omega_{m}^{(0)}(r)\equiv\frac{\alpha(r)}{R_{0}^{3}H_{\perp0}^{2}}\,,\end{equation}
 where $R_{0}\equiv R(r,t_{0})=r,\, H_{\perp0}\equiv H_{\perp}(r,t_{0})$
(the superscript $(0)$ denotes the present value)
and an effective spatial curvature
\begin{equation}
\Omega_{K}^{(0)}(r)=1-\Omega_{m}^{(0)}(r)=\frac{\beta(r)}{R_{0}^{2}H_{\perp0}^{2
}}\,.\label{eq:ltb-curv}
\end{equation}
Hence we see that the initial condition at some time $t_{0}$ (which here we take as the present time) must specify two free functions of $r$, for instance $\alpha(r),\beta(r)$ or $\Omega_{m}^{(0)}(r),H_{\perp0}^{}(r)$. The latter choice shows that the inhomogeneity can be in the matter distribution or in the expansion rate or in both. This freedom can be used to fit simultaneously for any expansion rate (and therefore luminosity and angular diameter distances \citep{romano09}) and for
any source number density \citep{musta97}.

If one imposes the additional constraint that the age of the Universe is the same for every observer, then only one free function is left \citep{garcia-haug08a}. The same occurs if one chooses $\Omega_{m}^{(0)}(r)={\rm constant}$ (notice that this is different from $\rho_{m}^{(0)}(r)={\rm constant}$, which is another possible choice) i.e. if the matter density fraction is assumed homogeneous today (and only today) \citep{enq07}. The choice of a homogeneous Universe age guarantees against the existence of diverging inhomogeneities in the past. However, there is no compelling reason to impose such restrictions.

Eq.~(\ref{eq:hperpeq}) is the classical cycloid equation\index{cycloid
equation} whose solution for $\beta>0$ is given parametrically by\begin{align}
R(r,\eta)=\,\frac{\alpha(r)}{2\beta(r)} &
(\cosh\eta-1)=\frac{R_{0}\Omega_{m}^{(0)}(r)}{2[1-\Omega_{m}^{(0)}(r)]}
(\cosh\eta-1)\,,\label{eq:sol-R}\\
t(r,\eta)-t_{B}(r)=\, &
\frac{\alpha(r)}{2\beta^{3/2}(r)}(\sinh\eta-\eta)=\frac{\Omega_{m}^{(0)}(r)}{2[
1-\Omega_{m}^{(0)}(r)]^{3/2}H_{\perp0}}(\sinh\eta-\eta)\,\,,\label{eq:sol-beta-t}\end{align}
 where $t_{B}(r)=t(r,\eta=0)$ is the inhomogeneous {}``big-bang''
time, i.e. the time for which $\eta=0$ and $R=0$ for a point at
comoving distance $r$. 
This can be put to zero in all generality by a redefinition of time.
The {}``time'' variable $\eta$ is defined
by the relation 
\begin{equation}
\eta=\int^{t}_0\frac{\beta(r)^{1/2}}{R(\tilde{t},r)}\rd \tilde{t}\,.
\end{equation}
Notice that the {}``time'' $\eta$ that corresponds to a given $t$
depends on $r$; so $R(r,t)$ is found by solving numerically $\eta(t,r)$
from eq.~(\ref{eq:sol-beta-t}) and then substituting $R[r,\eta(r,t)]$.
The present epoch $\eta_{0}(r)$ is defined by the condition $R=R_{0}$.
In the problem {[}10.2{]} we will derive the age of the Universe 
$t_{{\rm age}}(r)=t(r,\eta_{0})-t_{B}(r)$
in terms of $\Omega_{m}^{(0)},H_{\perp0}$. For $\beta<0$ the $\eta$
functions in eqs.\ (\ref{eq:sol-R}-\ref{eq:sol-beta-t}) become
$(1-\cos\eta)$ and $(\eta-\sin\eta)$ for $R$ and $t$, respectively,
while for $\beta=0$ they are $\eta^{2}/2$ and $\eta^{3}/6$: we
will not consider these cases further. 

As anticipated, since we need to have a faster expansion inside some distance to mimic
cosmic acceleration, we need to impose to our solution the structure
of a void. An example of the choice of
$\Omega_{m}^{(0)}(r)\equiv\Omega_{m}(r,t_{0}),h^{(0)}(r)\equiv
H_{\perp0}/(100$\,km\,sec$^{-1}$\,Mpc$^{-1}$)
is \cite{garcia-haug08} \begin{eqnarray}
\Omega_{m}^{(0)}(r) & = & \Omega_{{\rm out}}+(\Omega_{{\rm in}}-\Omega_{{\rm
out}})f(r,r_{0},\Delta)\,,\\
h^{(0)}(r) & = & h_{{\rm out}}+(h_{{\rm in}}-h_{{\rm
out}})f(r,r_{0},\Delta)\,,\end{eqnarray}
 with \begin{equation}
f(r,r_{0},\Delta)=\frac{1-\tanh[(r-r_{0})/2\Delta]}{1+\tanh(r_{0}/2\Delta)}\,,
\end{equation}
 representing the transition function of a shell of radius $r_{0}$
and thickness $\Delta$. The six constants $\Omega_{{\rm in}},\Omega_{{\rm
out}},h_{{\rm in}},h_{{\rm out}},r_{0},\Delta$
completely fix the model. If $h_{{\rm in}}>h_{{\rm out}}$ we can
mimic the accelerated expansion.

In order to compare the LTB model to observations we need to generalize
two familiar concepts: redshift and luminosity distance. The redshift can be
calculated
through the equation  \citep{alnes2}
 \begin{equation}
\frac{{\rm d} z}{{\rm d}
r}=(1+z)\frac{\dot{R}'}{\sqrt{1+\beta}}\,,\label{eq:ltb-red}\end{equation}
 where $R(t,r)$ must be calculated on the trajectory $t_{p}(r)$
and we must impose $z(r=0)=0$. Every LTB function, e.g., $H_{\perp}(t,r),R(t,r)$
etc. can be converted into line-of-sight functions of redshift by
evaluating the arguments $r_{p}(z),t_{p}(z)$ along the past light
cone.

The proper area of an infinitesimal surface at $r,t={\rm constant}$
is given by $A=R^{2}(r,t)\sin\theta\,\rd\theta\,\rd\phi$. The angular
diameter distance is the square root of $A/(\sin\theta\,\rd\theta\,\rd\phi)$
so that $d_{A}(z)=R(t_{p}(z),r_{p}(z))$. 
Since the Etherington duality relation\index{duality relation|see{Etherington
relation}}
$d_{L}=(1+z)^{2}d_{A}$ remains valid in inhomogeneous models, we
have \citep{sachs66} \begin{equation}
d_{L}(z)=(1+z)^{2}R(t_{p}(z),r_{p}(z))\,.\label{eq:dlvoid}\end{equation}
 This clearly reduces to $d_{L}=(1+z)r(z)$ in the FLRW background.
Armed with these observational tools, we can compare any LTB model
to the observations.

Besides matching the SN Ia Hubble diagram, we do not want to spoil
the CMB acoustic peaks and we also need to impose a local density
$\Omega_{{\rm in}}$ near 0.1-0.3, a flat space outside (to fulfill
inflationary {\color{red}fulfil} predictions), i.e. $\Omega_{{\rm out}}=1$, and finally
the observed local Hubble value $h_{{\rm in}}\approx0.7\pm0.1$. The
CMB requirement can be satisfied by a small value of $h_{{\rm out}}$,
since we know that to compensate for $\Omega_{{\rm out}}=1$ we need
a small Hubble rate (remember that the CMB essentially constrains
$\Omega_{m}^{(0)}h^{2}$). This fixes $h_{{\rm out}}\approx0.5$.
So we are left with only $r_{0}$ and $\Delta$ to be constrained
by SN Ia. As anticipated we expect $r_{0}$ to be near $z=0.5$, which
in the standard $\Lambda$CDM model gives a distance $r(z)\approx2$\,Gpc.
An analysis using SN Ia data \citep{garcia-haug08c} finds that
$r_{0}=2.3\pm0.9$\,Gpc
and $\Delta/r_{0}>0.2$. Interestingly, a
{}``cold spot'' in the CMB sky could be attributed to a void of
comparable size \citep{coldspot,coldspot-notari}.

There are many more constraints one can put on such large inhomogeneities.
Matter inside the void moves with respect to CMB photons coming from
outside. So the hot intracluster gas will scatter the CMB photons
with a large peculiar velocity and this will induce a strong kinematic
Sunyaev-Zel'dovich effect\index{Sunyaev-Zel'dovich effect} \citep{garcia-haug08b}.
Moreover, secondary photons scattered towards us by reionized matter
inside the void should also distort the black-body spectrum due to
the fact that the CMB radiation seen from anywhere in the void (except
from the center {\color{red}centre}) is anisotropic and therefore at different temperatures
\citep{cald-steb08}. These two constraints require the voids not to
exceed 1 or 2 Gpc, depending on the exact modeling {\color{red}modelling} and are therefore
already in mild conflict with the fit to supernovae.

Moreover, while in the FLRW background the function $H(z)$ fixes
the comoving distance $\chi(z)$ up to a constant curvature (and consequently
also the luminosity and angular diameter distances), in the LTB model
the relation between $\chi(z)$ and $H_{\perp}(z)$ or $H_{\Vert}(z)$
can be arbitrary. That is, one can choose the two spatial free functions
to be for instance $H_{\perp}(r,0)$ and $R(r,0)$, from which the
line-of-sight values $H_{\perp}(z)$ and $\chi(z)$ would also be
arbitrarily fixed. This shows that the {}``consistency'' FLRW relation
between $\chi(z)$ and $H(z)$ is violated in the LTB model, and in
general in any strongly inhomogeneous Universe.

Further below we discuss how this consistency test can be exploited by Euclid to test for large scale inhomogeneities. 
Recently, there has been an implementation of LTB models in large scale structure
N-body simulations~\citep{AGBTV2010}, where inhomogeneities grow in the presence 
of a large-scale void and seen to follow the predictions of linear perturbation theory.

An interesting class of tests on large-scale inhomogeneities involve probes of 
the growth of structure. However, progress in making theoretical predictions 
has been hampered by the increased complexity of cosmological perturbation 
theory in the LTB spacetime, where scalar and tensor perturbations couple,
see for 
example \cite{Clarkson:2009sc}. Nevertheless, a number of promising tests 
of large-scale inhomogeneity using the growth of structure have been proposed. 
\cite{Alonso:2012ds} used N-body simulations to modify the Press-Schechter halo 
mass function, introducing a sensitive dependence on the background shear. The 
shear vanishes in spatially-homogeneous models, and so a direct measurement of 
this quantity would put stringent constraints on the level of background 
inhomogeneity, independent of cosmological model assumptions. Furthermore, 
recent upper limits from the ACT and SPT experiments on the linear, all-sky 
kinematic Sunyaev-Zel'dovich signal at $\ell=3000$, a probe of the peculiar 
velocity field, appear to put strong constraints on voids (\cite{Moss:2011ze}). 
This result depends sensitively on theoretical uncertainties on the matter 
power spectrum of the model, however.

Purely geometric tests involving large-scale structure have been proposed, 
which neatly side-step the perturbation theory issue. The Baryon Acoustic 
Oscillations (BAO) measure a preferred length scale, $d(z)$, which is a 
combination of the acoustic length scale, $l$, set at matter-radiation 
decoupling, and projection effects due to the geometry of the Universe, 
characterized by the volume distance, $D_V(z)$. In general, the volume distance 
in an LTB model will differ significantly from that in the standard model, even 
if the two predict the same SN Ia Hubble diagram and CMB power spectrum. 
Assuming that the LTB model is almost homogeneous around the decoupling epoch, 
$l$ may be inferred from CMB observations, allowing the purely geometric volume 
distance to be reconstructed from BAO measurements. It has been shown by 
\cite{Zumalacarregui:2012pq} that, based on these considerations, recent BAO 
measurements effectively rule out giant void models, independent of other 
observational constraints.

The tests discussed so far have been derived under the assumption of a 
homogeneous Big Bang (equivalent to making a particular choice of the bang time 
function). Allowing the Big Bang to be inhomogeneous considerably loosens or 
invalidates some of the constraints from present data. It has been shown 
(\cite{Bull:2011wi}) that giant void models with inhomogeneous bang times can 
be constructed to fit the SN Ia data, WMAP small-angle CMB power spectrum, and 
recent precision measurements of $h$ simultaneously. This is contrary to claims 
by, e.g. \cite{Riess:2011yx}, that void models are ruled out by this 
combination of observables. However, the predicted kinematic Sunyaev-Zel'dovich 
signal in such models was found to be severely incompatible with existing 
constraints. When taken in combination with other cosmological observables, 
this also indicates a strong tension between giant void models and the data, 
effectively ruling them out.

\subsection{Inhomogeneous models: Backreaction}

In general, we would like to compute directly the impact of the inhomogeneities,
without requiring an exact and highly symmetric solution of Einstein's equations
like FLRW or even LTB. Unfortunately there is no easy way to approach this
problem. One ansatz tries to construct average quantities that follow
equations similar to those of the traditional FLRW model, see e.g. 
\citet{Buchert:1999er,Rasanen:2003fy,Rasanen:2006kp,Buchert:2007ik}.
This approach is often called {\em backreaction} as the presence of the 
inhomogeneities acts on the background evolution and changes it. In this
framework, it is possible to obtain a set equations, often called Buchert equations,
that look surprisingly similar to the Friedmann equations for the averaged
scale factor $a_\CD$, with extra contributions:
\begin{eqnarray}
3\left( \frac{{\dot a}_\CD}{a_\CD}\right)^2 - 8\pi G \average{\varrho}-\Lambda &=& - \frac{\average{\CR}+{\CQ}_\CD }{2} \;, \\
3\frac{{\ddot a}_\CD}{a_\CD} + 4\pi G \average{\varrho} -\Lambda &=& {\CQ}_\CD\;,
\end{eqnarray}
Here $\CR$ is the 3--Ricci scalar of the spatial hypersurfaces and
$\CQ$ is given by
\begin{equation}
\label{Q} 
{\cal Q}_\CD =
\frac{2}{3}\average{\left(\theta - \average{\theta}\right)^2 } - 
2\average{\sigma^2}\;,
\end{equation}
i.e. it is a measure of the variance the expansion rate $\theta$ and of the shear $\sigma_{ij}$. We see that
this quantity, if it {\color{red}is} positive, can induce an accelerated growth of $a_\CD$, which suggests
that observers would conclude that the Universe is undergoing accelerated expansion.

However,  it is not possible to directly link this formalism to observations. A first step can
be done by imposing by hand an effective, average geometry with the help of a template
metric that only holds on average. The probably simplest first choice is to impose on each
spatial hypersurface a spatial metric with constant curvature, by imagining that the inhomogeneities
have been smoothed out. But in general the degrees of freedom of this metric (scale factor
and spatial curvature) will not evolve as in the FLRW case, since the evolution is given by
the full, inhomogeneous Universe, and we would not expect that the smoothing of the
inhomogeneous Universe follows exactly the evolution that we would get for a smooth (homogeneous)
Universe. For example, the average curvature could grow over time, due to the collapse of
overdense structure and the growth (in volume) of the voids. Thus, unlike in the FRLW case,
the average curvature in the template metric should be allowed to evolve. This is the
case that was studied in \cite{Larena:2008be}. 

While the choice of template metric and the Buchert equations complete the set of equations,
there are unfortunately further choices that need to be made. Firstly, although there is an
integrability condition linking the evolution of $\average{\CR}$ and $\CQ_\CD$ and in addition a consistency
requirement that the effective curvature $\kappa(t)$ in the metric is related to $\average{\CR}$, we still need to impose
an overall evolution by hand as it was not yet possible to compute this from first principles.
Larena assumed a scaling solution $\average{\CR}\propto a_\CD^n$, with $n$ 
a free exponent. In a dark energy context, this scaling exponent $n$ corresponds to an
effective dark energy with $w_\CD = -(n+3)/3$, but in the backreaction case with the template
metric the geometry is different from the usual dark energy case. A perturbative analysis \citep{Li:2007ci}
found $n=-1$, but of course this only an indication of the possible behaviour as the situation
is essentially non-perturbative.

\begin{figure}[htbp]
\begin{center}
\includegraphics[width=8cm]{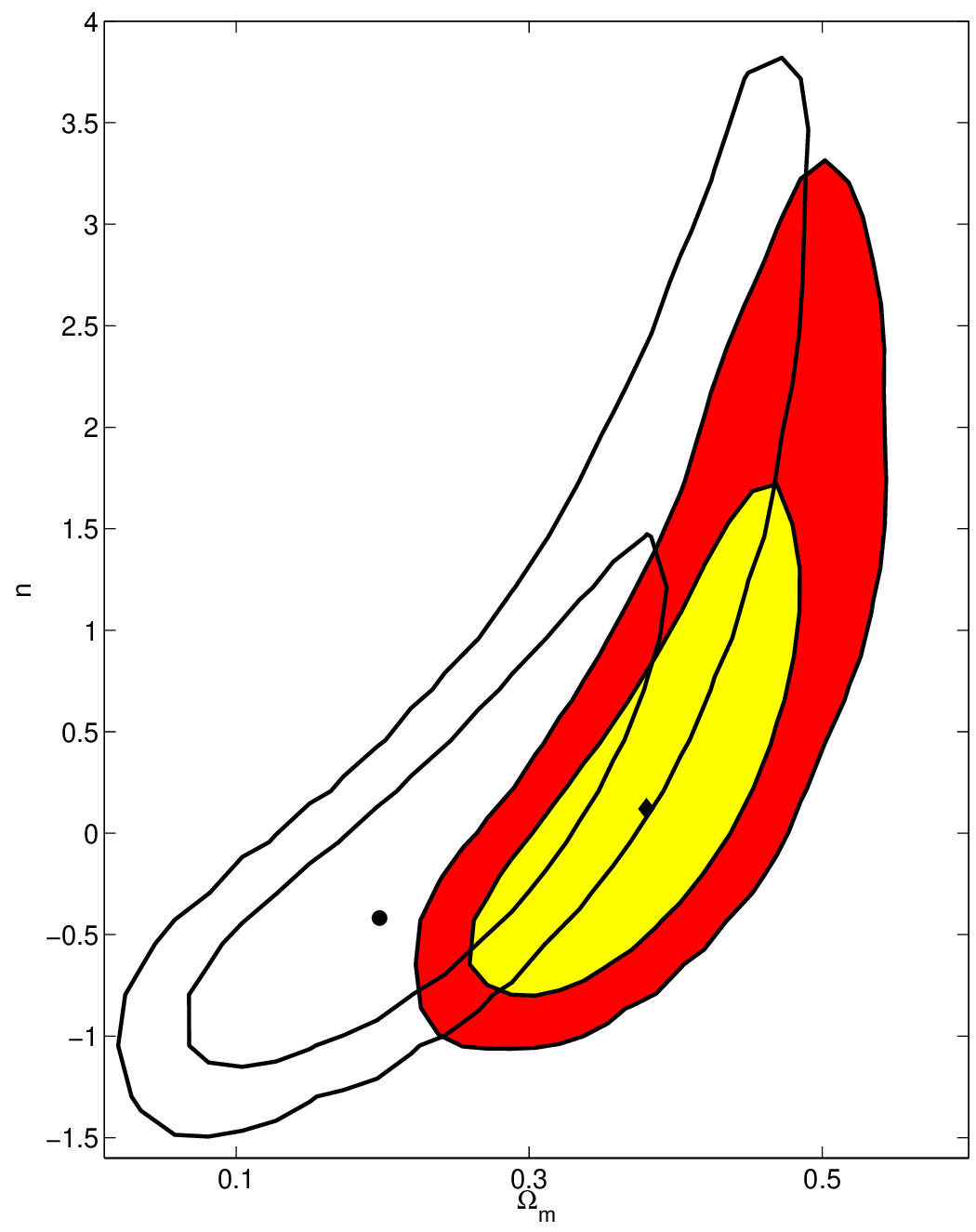}
\caption{\label{fig:larena1}Supernovae and CMB constraints in the $(\Omega^{\CD_0}_{m}$,n) plane for the averaged effective model with zero Friedmannian curvature (filled ellipses) and for a standard flat FLRW model with a quintessence
field with constant equation of state $w=-(n+3)/3$ (black ellipses). The disk and diamond represent the absolute best--fit models respectively for the standard FLRW model and the averaged effective model.}
\end{center}
\end{figure}

The second choice concerns the computation of observables. \cite{Larena:2008be} studied
distances to supernovae and the CMB peak position, effectively another distance. The assumption
taken was that distances could be computed within the averaged geometry as if this was the
true geometry, by integrating the equation of radial null geodesics. In other words, the effective
metric was taken to be the one that describes distances correctly. 
The resulting constraints are shown in Fig.~\ref{fig:larena1}. We see that the leading perturbative mode ($n=1$) is
marginally consistent with the constraints. These contours should be regarded as an indication of what kind
of backreaction is needed if it is to explain the observed distance data.

One interesting point, and maybe the main point in light of the discussion in the following section, is that the averaged curvature
needs to become necessarily large at late times due to the link between it and the backreaction term $\CQ$, in order
to explain the data. Just as in the case of a huge void, this effective curvature makes the backreaction scenario testable
to some degree with future large surveys like Euclid.

\begin{figure}[htbp]
\begin{center}
\includegraphics[width=8cm,height=12cm]{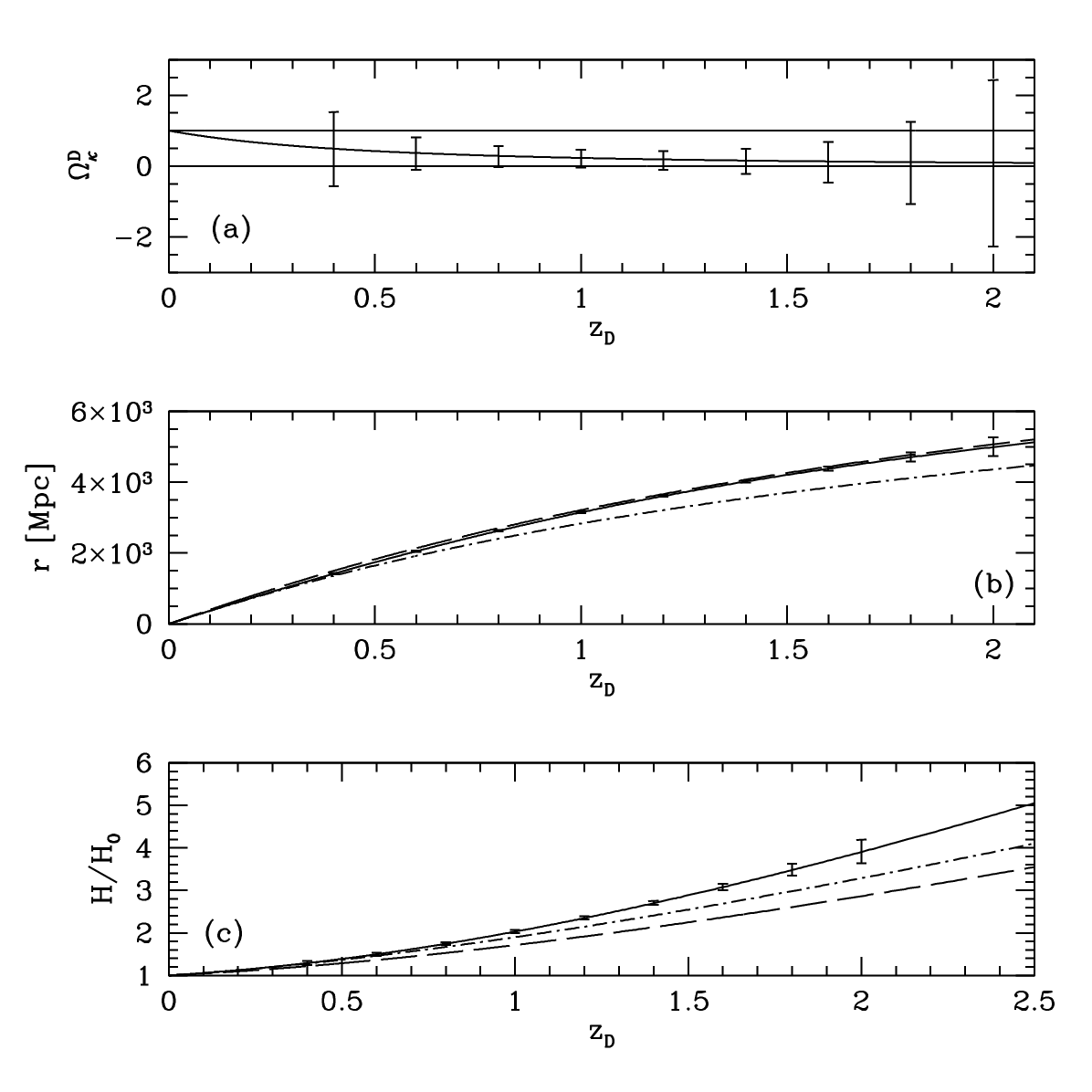}
\caption{\label{fig:larena2}Upper panel: Evolution of $\Omega_{k}(z_{\CD})$ as a function of redshift for the absolute best--fit averaged model represented by the diamond in Fig.~\ref{fig:larena1}. One can see that all positively curved FLRW models ($\Omega_{k,0}<0$) and only highly negatively curved FLRW models ($\Omega_{k,0}>0.5$) can be excluded by the estimation of $\Omega_{k}(z_{\CD})$. Central panel: Evolution of the coordinate distance for the best--fit averaged model (solid line), for a $\Lambda$CDM model with $\Omega_{m,0}=0.277$, $\Omega_{\Lambda}=0.735$ and $H_{0}=73\mbox{ km/s/Mpc}$ (dashed line), and for the FLRW model with the same parameters as the best--fit averaged model (dashed-dotted line). Lower panel: Evolution of the Hubble parameter $H/H_{0}$ for the best--fit averaged model (solid line), the FLRW model with the same parameters as the averaged best--fit model (dashed-dotted line), and for the same $\Lambda$CDM model as in the central panel (dashed line). The error bars in all panels correspond to the expectations for future large surveys like Euclid.}
\end{center}
\end{figure}

\section{Reconstructing the global curvature at different redshifts}

\cite{Clarkson/Bassett/Lu:2008} presented an observational test for the 
Copernican Principle which relies on the consistency relation between
expansion rate and angular diameter distance. Here we discuss the implications
for Euclid.

Let us recall that  the angular diameter distance in a FLRW model can
be written as:
\begin{equation}
D_{A}(z)=\frac{1}{1+z}\frac{1}{H_{0}\sqrt{-\Omega^{(0)}_{K}}}\sin\left(\sqrt{-\Omega^{(0)}_{K}}\int_{0}^{z}{dz'\frac{H_{0}}{H(z')}}\right)\,.
\label{eq:distanza-angolare-curv}
\end{equation}
where $\Omega^{(0)}_{K}$ is the curvature parameter {\it today}.

We can invert the last equation to obtain an expression for the curvature parameter that 
depends on the Hubble parameter $H$ and comoving angular diameter distance 
$D\left(z\right)=\left(1+z\right)D_{A}\left(z\right)$ only, 
see \cite{Clarkson/Bassett/Lu:2008}:
\begin{equation}
\Omega^{(0)}_{K}=\frac{\left[H\left(z\right)D'\left(z\right)\right]^{2}-1}{\left[H_{0}D\left(z\right)\right]^{2}}\label{eq:omegak}
\end{equation}
where here the prime refers to the derivative with respect the redshift.
Then eq.~(\ref{eq:omegak}) tells us how the curvature parameter 
can be measured from the distance and the Hubble rate observations, 
in a model-independent way.

The idea is then  to measure the curvature 
parameter $\Omega^{(0)}_{K}$ at different redshifts. 
Let us consider again eq.~(\ref{eq:omegak}); if we are in a FLRW Universe then 
$\Omega^{(0)}_K$ should be  independent of redshift, i.e. its  derivative 
with respect to $z$ should be zero 
\begin{equation}
\mathcal{C}(z) = \frac{{\rm d}\Omega^{(0)}_{K}}{{\rm d}z}=0\,.
\label{eq:domegak}
\end{equation}
If it happens that $\mathcal{C}(z)\neq 0$ even at a single redshift then 
this means the large-scale Universe is not homogeneous.

A possible test to measure $\Omega^{(0)}_K$ at various redshifts is provided by 
baryon acoustic oscillations. Observing the features 
of BAO in the galaxy power spectrum in both angular (orthogonal to 
the line of sight $L_{\perp}$) and radial direction (along the 
line of sight $L_{\parallel}$)  allows us to measure with a 
great accuracy both $D_{A}(z)$ and $H(z)$, respectively. 

If the geometry is not  FLRW, then the standard BAO will be deformed in three different 
ways:
\begin{enumerate}
\item The sound horizon scale, which is the characteristic ruler, will be different 
in the $\perp$ and $\parallel$ directions and it will be also different from that for
 the FLRW Universe.
\item Even if the sound horizon were isotropic at decoupling, the subsequent 
expansion in the $\perp$ and $\parallel$ directions will be different just 
because they will be governed by two distinct Hubble parameters: $H_{\perp}$ 
and $H_{\parallel}$.
\item The redshift distortion parameter will be different  because 
it will depend on the background expansion.
\end{enumerate} 

Also the growth factor will be modified, perhaps in a scale dependent way.
If the true underlying model is radically inhomogeneous, but we assume a
FLRW in interpreting the observations, the derived cosmological parameters will
be biased (or unphysical) and the parameters derived from  BAO data  will be
different from those 
measured by SNIa and/or lensing. As argued also in different contexts, a
mismatch on the value of  one of more  parameters may  indicate that  we are
assuming  a wrong model.

We show here the sensitivity that can be reached with an experiment 
like Euclid for the curvature parameter $\Omega^{(0)}_{K}$ (L. Amendola and 
D. Sapone in preparation).
We choose a redshift survey with a depth of $z=1.6$ and consider  different redshift bins.

\begin{figure}
\centering\includegraphics[width=2.4in]{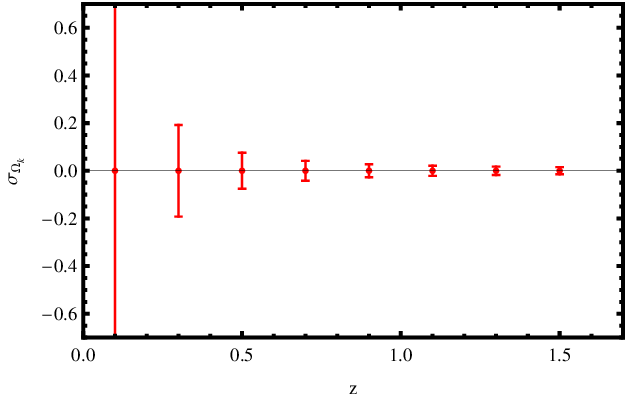}
\caption{Relatives {\color{red}Relative} errors on $\Omega_{K}$ for our benchmark survey for different 
redshifts.}
\label{fig:omk}
\end{figure}

In Fig.~(\ref{fig:omk}) we show the first $1$-$\sigma$ absolute errors on the curvature 
parameter for different redshift bins that can be obtained measuring the Hubble parameter and 
the angular diameter distance.   In obtaining these errors we used Fisher-based forecasts for
 the radial and angular BAO signal following \cite{seo03,eisenstein07}, as
 dicussed {\color{red}discussed}
in Sec.~\ref{dark-energy-and-redshift-surveys}.

\begin{figure}
\centering\includegraphics[width=2.4in]{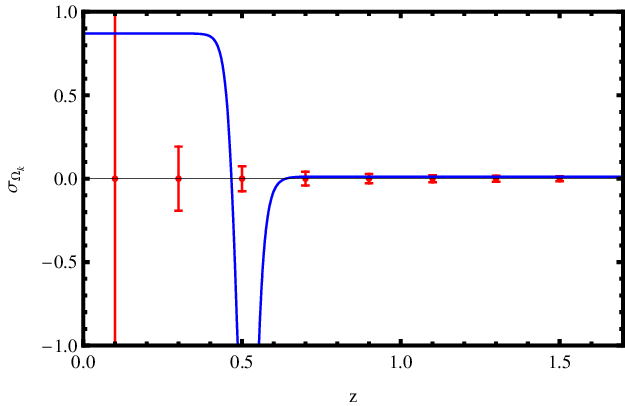}
\includegraphics[width=2.4in]{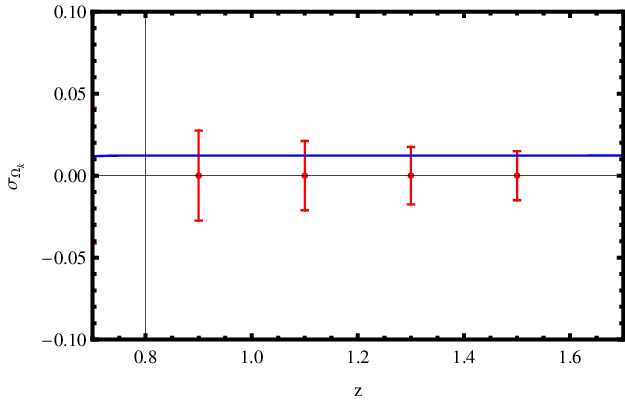}
\caption{Left: same as Fig.~\ref{fig:omk} but now with superimposed the prediction
for  the Lema\^itre-Tolman-Bondi model  considered by
\cite{GarciaBellido/Haugbolle:2008}. Right: zoom in the high-redshift range.}
\label{fig:newfigomk}
\end{figure}

The sensitivity that can be reached with an experiment like Euclid is extremely high; 
we can measure the curvature parameter better than $0.02$ at redshift of 
the order of $z\simeq 1$. This will allow us to discriminate between FLRW 
and averaged cosmology as for example illustrated in Fig.~\ref{fig:newfigomk}.

 An alternative to measuring the global curvature is to measure the shear of
 the background geometry. If there is {\color{red}a} large inhomogeneous void then a congruence of geodesics will not only expand but also suffer shear~\citep{garcia-haug08c}. The amount of shear will depend on the width and magnitude of the transition between the interior of the void and the asymptotic Einstein-de Sitter Universe. Normalizing the shear w.r.t. the overall expansion, one finds~\citep{garcia-haug08c}
\begin{equation}\label{eq:normshear}
\varepsilon = \frac{H_\perp(z)-H_{||}(z)}{2H_\perp+H_{||}} \simeq
\frac{1 - H_{||}(z)\,\partial_z\Big[(1+z)\,D_A(z)\Big]}{3H_{||}(z)D_A(z) + 2\Big(1 - H_{||}(z)\,\partial_z\Big[(1+z)\,D_A(z)\Big]\Big)}\,.
\end{equation}
Clearly, in homogeneous FRW Universes the shear vanishes identically since $H_\perp = H_{||} = H$. Also note that the function
$H_{||}(z)D_A(z)$ is nothing but the Alcock-Paczynski factor, which is normally used as a geometric test for the existence of vacuum energy in $\Lambda$CDM FRW models.

\section{Speculative avenues: non-standard models of primordial fluctuations}

In this section we explore other non-conventional scenarions
{\color{red}scenarios} that challenge our understanding of the Universe. Here we
present models that include mechanisms for primordial anisotropy in the fluctuation spectrum, due to
space-time non-commutativity, to inflationary vector fields or to
superhorizon {\color{red}super-horizon} fluctuations. Since inflation can occur at high energies 
for which we lack robust direct 
experimental probes, it is reasonable to pay attention on possible deviations from some standard properties of low 
energy physics.  We review these here
and point out possible observables for the Euclid project.

\subsection{Probing the quantum origin of primordial fluctuations}

Conventionally, the 2-point correlation function of a random variable $X(\vec k,t)$ is regarded as a classical object,  related to the power spectrum $P_{X}$ via the relation
\ba
\langle X(\vec k,t)X(\vec k',t)\rangle=(2\pi)^{3}\delta(\vec k-\vec k')P_{X}(k)\ ,
\ea
where $k=|\vec k|$.

When we look at $X(\vec k,t)$ in terms of a \emph{quantum field} in momentum
space, we need to reinterpret the average $\langle\ldots \rangle$ as the
expectation value of the 2-point function over a determined quantum state.
This raises several issues that are usually ignored in a classical analysis.
For instance, the value of the expectation value depends in the algebra of
the annihilation and creation operators that compose the field operator.
Non-trivial algebra such as non-commutative, {\color{red}Any non-trivial
algebra such as a non-commutative one,} leads to non-trivial power spectra. Also, the quantum expectation value depends on the state of the field, and different choices can lead to radically different results.

Suppose that $\varphi(\vec x,t)$ represents a perturbation propagating on an inflationary background. Upon quantization, we have
\ba
\hat\varphi(\vec x,t)=(2\pi)^{-3/2}\int d^{3}k\left[\varphi_{k}(t)\hat a_{\vec k}\,e^{i\vec k\cdot t}+\varphi_{k}^{*}(t)\hat a_{\vec k}^{\dagger}\,e^{-i\vec k\cdot t}\right]\ ,
\ea
where $\hat a_{\vec k}$ is the usual annihilation operator. When calculated
in the limit $\vec k\rightarrow \vec k'$, the expectation value of the
two-point function in coordinate space diverges, signaling
{\color{red}signalling} the breakdown of the theory at short distances. From
the quantum field theory perspective, this means that the expectation value
needs to be regularized in the ultraviolet (UV). It has been argued that
this has in specific scenarios sizable {\color{red}sizeable} effects on the
observable spectrum {\color{red}--} see e.g. \cite{renorm1}, see however e.g. \cite{DMR} for a contrary viewpoint.

In addition to UV divergences, there are infrared (IR) ones in long-range
correlations. Usually, one tames these by putting the Universe in
{\color{red} a} box and cutting off super-horizon correlations. However,
several authors have recently proposed more sensible IR regulating
techniques, see e.g. \cite{sloth,Koivisto:2010pj}. Very natural way
{\color{red} ways} to obtain IR finite results are to take into account the presence of tiny spatial curvature or a pre-inflationary phase which alters the initial conditions \cite{proko,Koivisto:2010pj}. In principle these regularizations will leave an imprint in the large-scale structure data, in the case that regularization scale is not too far beyond the present horizon scale. If this pre-inflationary phase is characterized by modified field theory, such as modified dispersion relations or lower dimensional effective gravity,  the scalar and tensor power spectra show a modification whose magnitude is model-dependent, see e.g. \cite{maxnew}.

The two-point function of a scalar field is constructed from basic quantum
field theory, according to a set of rules determined in the context of
relativistic quantum mechanics. In particular, the usual commutation rules
between position and momentum are promoted to commutation rules between the
field and its canonical conjugate. A modification of the fundamental quantum
mechanical commutation rules can be easily generalized to field theory. The
most popular case is represented by non-commutative geometry, which implies that coordinate operators do not commute, i.e.
\ba \label{canonical_nc}
[\hat x^{\mu},\hat x^{\nu}]=i\theta^{\mu\nu}\ ,
\ea
where $\theta^{\mu\nu}$ is an anti-symmetric matrix, usually taken to be
constant, see e.g. \cite{ncfund1,ncfund2}. There are many fundamental
theories that phenomenologically reduce to an ordinary field theory over a
non-commutative manifold, from string theory to quantum gravity. It is
therefore important to consider the possibility that non-commutative effects  took place during the inflationary era and try to extract some prediction.

One can construct models where the inflationary expansion of the Universe is
driven by non-commutative effects, as in \cite{mag,ncmax}. In this kind of
models, there is no need for an inflaton field  and non-commutativity modifies the equation of state in the radiation-dominated Universe in a way that it generates a quasi-exponential expansion. The initial conditions are thermal and not determined by a quantum vacuum. For the model proposed in \cite{mag}, the predictions for the power spectra have been worked out in \cite{koh}. Here, Brandenberger and Koh find that the spectrum of fluctuations is nearly scale invariant, and shows a small red tilt, the magnitude of which is different from what is obtained in a usual inflationary model with the same expansion rate.

On the other hand, non-commutativity could introduce corrections to standard
inflation. Such {\color{red} a}  perhaps less radical approach consists in assuming the usual
inflaton-driven background, where scalar and tensor perturbations propagate
with a Bunch and Davies vacuum as initial condition, but are subjected to
non-commutativity at short distance. It turns out that the power spectrum is modified according to (see e.g. \cite{Koivi1}, and references therein)
\ba \label{nc_spectrum}
P=P_{0}\,e^{H\vec\theta\cdot \vec k}\ ,
\ea
where $H$ is the Hubble parameter, $P_{0}$ is the usual commutative
spectrum, and $\vec \theta$ is the vector formed by the $\theta^{0i}$
components of $\theta^{\mu\nu}$. This prediction can be obtained by using a
deformation of statistics in non-commutative spacetime on the usual
inflationary computation. It can be also derived in an alternative way
beginning from an effective deformation of the Heisenberg algebra of the
inflaton field. The most important aspect {\color{red} of} the result is that the spectrum
becomes direction-dependent. The perturbations thus distinguish a preferred
direction given by the vector $\vec\theta$ that specifies the non-commutativity between space and time.

Furthermore, it is interesting that the violation of isotropy can also
violate parity. This could {\color{red}provide what seems a quite} unique property of
possible signatures in the CMB and large scale structure. However, these
{\color{red}there} is
also an ambiguity with the predictions of the simplest models, which is
related to interpretation {\color{red}interpretations} of non-commuting quantum observables at the
classical limit. This is evident from the fact that {\color{red}one} has to consider an
effectively imaginary $\vec{\theta}$ in the above formula
(\ref{nc_spectrum}). Reality of physical observables requires the odd parity
part of the spectrum (\ref{nc_spectrum}) to be imaginary. The appearance of
this imaginary parameter $\vec \theta$ into the theory may signal the
unitary violation that has been reported in theories of time-space
non-commutativity. It is known that the Seiberg-Witten map to string theory
applies only for space-space non-commutativity \cite{Seiberg:1999vs}.
Nevertheless, the phenomenological consequence that the primordial
fluctuations can distinguish handedness, seems in principle a physically
perfectly plausible -- though speculative -- possibility, and what ultimately
renders it very interesting is that we can test by cosmological
observations. Thus, while lacking the completely consistent and unique
non-commutative field theory, we can parameterize {\color{red}parametrize} the ambiguity by a phenomenological parameter whose correct value is left to be determined observationally. The parameter $\alpha \in [0,1]$ can be introduced \cite{Koivi1} to quantify the relative amplitude of odd and even contributions in such a way that $P = \alpha P^+  + i(1-\alpha)P^-$, where $P^\pm = (P(\vec k)\pm P(-\vec k))/2$.

The implications of the anisotropic power spectra, such as
(\ref{nc_spectrum}), to {\color{red}for} the large scale structure measurements, is discussed
below in subsection \ref{anisotropicconstraints_cmb_pk}. Here we proceed to
analyze {\color{red}analyse } some consequences of the non-commutativity
relation  (\ref{canonical_nc}) to the higher order correlations of
cosmological perturbations. We find that they can violate both isotropy and
parity symmetry of the FRW background. In particular, the latter effect
persists also in the case $\alpha=1$. The $\alpha=1$ {\color{red}This case}
corresponds to the prescription in ref.~\cite{Akofor:2007fv} and in the
remainder of this subsection we restrict to this case for simplicity.  Thus,
even when we choose this special prescription where the power spectrum is
even, higher order correlations will violate parity. This realizes the
possibility of {\color{red}an} odd bispectrum that was recently contemplated
upon in ref.~\cite{Kamionkowski:2010rb}.

More precisely, the functions $B$ defined in eq.~(\ref{eq:bispectrumdef}) for the three-point function of the curvature perturbation can be shown to have the form
\ba \label{bispectrum}
B_\Phi(\vec{k}_1,\vec{k}_2,\vec{k}_3)   & = &
2\cos{\left( \vec{k}_1\wedge\vec{k}_2 \right)}\Big(\cosh(2H\vec{\theta}\cdot\vec{k}_3)P_0(\vec{k}_1)P_0(\vec{k}_2)f_s(\vec{k}_3) +   2\,{\rm  perm.}\Big)  \nonumber \\
& - & 2i\sin{\left( \vec{k}_1\wedge\vec{k}_2 \right)}\Big(\sinh(2H\vec{\theta}\cdot\vec{k}_3)P_0(\vec{k}_1)P_0(\vec{k}_2)f_s(\vec{k}_3)  +   2\,{\rm  perm.}\Big)\,,
\ea
where the function $f_s(k)$ is
\be
f_s(k)=\frac{N''}{2N'^2}\left(1+{n_{f_{{\rm NL},0}}}\,\ln\frac{k}{k_p}\right)\,,
\ee
$k_p$ being a pivot scale and primes denoting derivatives with respect to the inflaton field. The quantity ${n_{f_{{\rm NL},0}}}$ is the scale dependence in the commutative case explicitly given by
\be \label{nf}
{n_{f_{{\rm NL},0}}}
  =\frac{N'}{N''}\left(
  -3\eta+
  \frac{V'''}{3H^2}\right)\,.
\ee
The spatial components of the non-commutativity
matrix $\theta_{ij}$ enter the bispectrum through the phase $\vec{k}_1\wedge\vec{k}_2= k_1^i k_2^j\,\theta_{ij}$.
They do not appear in the results for the spectrum and therefore affect only
the non-Gaussian statistics of primordial perturbations.

We now focus on this part in the following only and set all components of $\vec{\theta}$ equal to
zero. This gives
  \ba
  \label{fnl_theta_ij}
  f_{{\rm NL},\theta}
  &=&\frac{5}{3}\cos{\left( \vec{k}_1\wedge\vec{k}_2
  \right)}\frac{P_{0}(k_1)P_{0}(k_2)f_s(k_3)+2\,{\rm
  perm.}}{P_{0}(k_1)P_{0}(k_2)+2\,{\rm perm.}}\ ,
  \ea
where the only contribution from the non-commutativity is the pre-factor
involving the wedge product. This affects the scale dependence of
${n_{f_{{\rm NL},\theta}}}$ and can hence be constrained observationally. For example,
computing the scale-dependence for shape preserving variations of
the momentum space triangle, $\vec{k}_i\rightarrow \lambda \vec{k}_i$,
defined as
\be
{n_{f_{{\rm NL},\theta}}} =  \frac{\partial {\rm ln}\,|f_{{\rm NL},\theta}(\lambda\vec{k}_1,\lambda\vec{k}_2,\lambda\vec{k}_3)|}{\partial {\rm ln}\,
  \lambda}\Big|_{\lambda=1}\,,
\ee
we find, in the present case
\be \label{nfnl_result}
{n_{f_{{\rm NL},\theta}}} =
 -2k_1^ik_2^j\theta_{ij}\tan(k_1^ik_2^j\theta_{ij})+{n_{f_{{\rm NL},0}}}\,,
\ee
where ${n_{f_{{\rm NL},0}}}$ given by (\ref{nf}) is the result in the commuting
case. The part dependent on $\theta_{ij}$ arises purely from
non-commutative features.
The Euclid data can be used to constrain the scale dependence of the non-linearity parameter $f_{{\rm NL},\theta}$, and the scale dependence could therefore place interesting bounds on $\theta_{ij}$.
We note however that the amplitude of the non-linearity is not enhanced by
the purely spatial non-commutativity, but is given by the underlying
inflationary model. The amplitude on the other hand is exponentially
enhanced by the possible timespace non-commutativity.

Moreover, it is worth noting
that the result (\ref{nfnl_result}) depends on the wave vectors
$\vec{k}_1$ and $\vec{k}_2$ and hence on the shape of the momentum space
triangle. This is in contrast with the commutative case, where the
scale dependence is given by the same result (\ref{nf})for all shape
preserving variations, $\vec{k}_i\rightarrow \lambda \vec{k}_i$, regardless of
triangle shape. This allows, in principle, to distinguish between
the contributions arising from the non-commutative properties of the
theory and from the standard classical inflationary physics or gravitational clustering.

To recapitulate, parity violations in the statistics of large-scale
structures would be a smoking gun signature of timespace non-commutativity
at work during inflation. Moreover, purely spatial non-commutativity
predicts peculiar features in the higher order correlations of the
perturbations, and in particular these can be most efficiently detected by
combining information of the scale- and shape-dependence of non-Gaussianity.
As discussed earlier in this document, these {\color{red}this} information
are {\color{red}is} extractable from the Euclid data.

\subsection{Early-time anisotropy}

Besides the non-commutative effects seen in the  previous section, anisotropy can be generated by the presence of
anisotropic fields at inflation. Such could be spinors, vectors or higher order
forms which modify the properties of fluctuations in a direction-dependent
{\color{red}way} 
either directly through perturbation dynamics or by causing the background to
inflate slightly anisotropically. The most common alternative is vector
fields (see Sec.~\ref{VSect}).
Whereas a canonical scalar field easily inflates the Universe if suitable
initial conditions are chosen, it turns out that it much less straightforward to
construct vector field alternatives. In particular, one must maintain a
sufficient level of isotropy of the Universe, achieve slow roll and keep
perturbations stable. Approaches to deal with the anisotropy have
{\color{red}been} based on a
''triad'' of three identical vectors aligned with the three axis
\citep{ArmendarizPicon:2004pm}, a large number of randomly oriented fields
averaging to isotropy \citep{Golovnev:2008cf}, time-like \citep{Koivisto:2008xf}
or sub-dominant  \citep{Dimopoulos:2008} fields. There are many variations of
inflationary scenarios involving vector fields, and in several cases the
predictions to {\color{red}of} the primordial spectra of perturbations have been worked out in
detail, see e.g. \cite{Watanabe:2010fh}. The generic prediction is that the
primordial perturbation spectrum {\color{red}spectra} become statistically anisotropic,  see
e.g. \cite{Ackerman:2007}.

Anisotropy could be also regarded simply as a trace of the initial conditions
set before inflation. One then assumes that inflation has lasted just about the
60 e-folds so that the largest observable scales were not yet smoothed out, or
isotropized, by the early inflationary expansion \citep{Pitrou:2008gk}. Such
{\color{red}a}
scenario can also be linked to various speculative ideas of pre-inflationary
physics such as gravitational tunneling {\color{red}tunnelling} into an anisotropic Universe, see e.g.
\cite{Adamek:2010sg}.

Also in this case the interest in such possibilities has been stimulated by several anomalies observed in the temperature 
WMAP maps, see ~\cite{Copi:2010} for a recent review 
(some of them were also present in the COBE maps). Their statistical evidence is quite robust w.r.t. the increase of the 
signal-to-noise ratio over the years of 
the WMAP mission and to independent tests by the international scientific community, although the a posteriori choice of statistics could make their 
interpretation difficult, see ~\cite{Bennet:2011}. Apart from those already mentioned in Sec.~\ref{sec:anisotropicmodels},  
these anomalies include
an alignment between the harmonic quadrupole and octupole modes in the temperature anisotropies (\cite{de-Oliveira:2004}), 
an asymmetric distribution of CMB power between two hemispheres, or dipole asymmetry~(\cite{Eriksen:2004}), the lack of power 
of the temperature two-point correlation function 
on large angular scales ($> 60^\circ$), asymmetries in the even vs. odd multipoles of the CMB power spectra (parity symmetry breaking), both at large 
(\cite{Kim:2010a}, \cite{Gruppuso:2010}) and intermediate angular scales (\cite{Bennet:2011}). 
Some of the anomalies could be connected among each other, e.g. the CMB parity breaking has been recently linked to the lack of large-scale power
(\cite{Maris:2010}, \cite{Copi:2007}, \cite{Kim:2010b}).

\subsubsection{Vector field models} \label{VSect}

Various inflationary models populated by vector fields can be described with a Lagrangian of the following form 
\begin{equation}
L_{\rm vector}=-\frac{1}{4} f(\varphi) F_{\mu\nu}F^{\mu\nu}+\frac{1}{2}m^2 B_{\mu}B^{\mu}\, ,
\end{equation}
where  $F_{\mu\nu}\equiv \partial_{\mu}B_{\nu}-\partial_{\nu}B_{\mu}$, and $f(\varphi)$ is a suitable function of the inflaton field. 
A Lagrangian containing just the standard kinetic term $F_{\mu\nu}F^{\mu\nu}$ would be conformally invariant 
thus preventing fluctuations of the vector field 
$B_{\mu}$ to be excited on super-horizon scales. Contrary to the case of a light scalar 
field, large-scale primordial perturbations of the vector field can be generated during inflation if the vector field 
is sufficiently massive (with $m^2\approx -2H^2$). 
This Lagrangian includes the case of a massive (curvaton) vector field (when $f\equiv 1$) studied by~\cite{Dimopoulos:2006,Dimopoulos:2008} and  
where the mass of the vector field is acquired via a non-minimal coupling to gravity to break conformal invariance. For some of these models 
there are actually some 
instability issues about the evolution of the primordial longitudinal perturbation modes of the vector field~(\cite{Himmetoglou:2009a}, 
\cite{Himmetoglou:2009b}). 
The models with varying kinetic function (when $f(\varphi)$ is switched on) allows to overcome these difficulties, 
since in this case the longitudinal mode is gauged away. They have been
studied in various contexts 
(e.g. \cite{Soda:2008}, \cite{Dimopoulos:2010a}). 
The Ackerman-Carroll-Wise models, (\cite{Ackerman:2007}), employ a different Lagrangian of the form 
$L_{\rm vector}=-\frac{1}{4} F_{\mu\nu}F^{\mu\nu}+\lambda(B^\mu B_\mu-m^2)$, so that the norm of the vector field is fixed by the Lagrangian 
multiplier $\lambda$. 
In these models (where inflation is driven by an inflaton field) the main effect of the vector field is a slightly anisotropic  
background evolution described by a metric,  with $c(t)=b(t)$) with a backreaction on the inflaton 
field fluctuations, rather than the vector field perturbations themselves. Another possibility that has been 
explored is based on a non-Abelian gauge $SU(2)$ vector multiplet (\cite{Bartolo:2009a}, \cite{Bartolo:2009b}), 
providing  a realistic model of gauge interactions neglected so far.  

A general prediction from all these scenarios is that the power spectrum of primordial perturbations can be written as 
\begin{equation}
\label{Panis}
P({\bf k})=P(k)\left[1+g(k) (\hat{\bf k} \cdot \hat{\bf n})^2 \right]\, ,
\end{equation}
where $g(k)$ is the amplitude of the rotational invariance breaking (statistical isotropy breaking) induced by a 
preferred direction ${\bf n}$. Thus, the power spectrum 
is not just a function of $k$ but it depends on the wave vector ${\bf k}$. 
Usually the preferred direction is related to the vector fields $n^i \propto B^i$ 
while the amplitude is related to the contribution of the vector field perturbations 
to the total curvature perturbation $g \sim P_{\zeta_B}/P_\zeta$. 

However, beyond the various concrete realizations, the expression~(\ref{Panis}), 
first introduced in~\cite{Ackerman:2007}, provides a robust and useful way to study observable 
consequences of a preferred direction during inflation and also a practical
template for comparsion {\color{red}comparison} with observations (see below). 
Usually the amplitude $g(k)$ is set to a 
constant $g_*$. A generalization of the above 
parametrization is $P({\bf k})=P(k)\left[1+\sum_{LM} g_{\rm LM}(k) Y_{LM}({\hat{\bf k}}) \right]$, 
where $Y_{\rm LM}({\hat{\bf k}})$ are spherical harmonics with only even multipoles $L \ge 2$ ~(\cite{Pullen:2007}).
Interestingly enough, inflationary models with vector fields can also generate 
higher-order correlators, such as bispetrum and trispectrum, which display
anisotropic fetaures {\color{red}features} as well 
(e.g.,\cite{Soda:2008}, \cite{Kar:2009}, \cite{Bartolo:2009a}, \cite{Bartolo:2009b}).

\subsubsection{Modulated perturbations}

The alignment of low CMB multipoles and the hemispherical power asymmetry observed in the CMB anisotropies can find an explanation in some models 
where the primordial 
gravitational perturbation is the result of fluctuations  within our Hubble volume, modulated by super-horizon fluctuations. The primordial 
gravitational perturbation can thus be thought of as a product of two fields
$\Phi_1({\bf x})$ and $\Phi_2({\bf x})$ (\cite{Dvorkin:2008}, and refs.
{\color{red}references}
therein)   
\begin{equation}
\Phi({\bf x})=\Phi_1({\bf x})\left[1+\Phi_2({\bf x}) \right]\, , 
\label{modulated}
\end{equation}
with $\Phi_2({\bf x})$ which {\color{red}where $\Phi_2({\bf x})$} has only super-horizon fluctuations, so that within a given Hubble volume it takes a fixed value, while 
$\Phi_1({\bf x})$ has sub-horizon stochastic fluctuations within that volume. The result is that an observer within our Hubble volume 
would see broken statistical 
homogeneity from the modulation on large scales of $\Phi_1({\bf x})$, 
and also broken statistical isotropy from the gradient of the modulating field $\Phi_2({\bf x})$. 
The dipole modulation $\delta T (\hat{\bf p})/T = S(\hat{\bf p})\left[1+A(\hat{\bf p} \cdot \hat{\bf n}) \right]$
used for CMB by, e.g., \cite{Eriksen:2007pc} and \cite{Hanson:2009} (or for LSS \cite{Hirata:2009ar}) to explain the hemispherical asymmetry
falls within the parametrization of eq.~(\ref{modulated}).
A scenario with a dipole modulation has been realized in some concrete and detailed models, 
such as those involving adiabatic and isocurvature modulating perturbations from 
a curvaton field (\cite{Erickcek:2008} and \cite{Erickcek:2009}).

\subsection{Current and future constraints from CMB and LSS on an anisotropic power spectrum} \label{anisotropicconstraints_cmb_pk}

\cite{Groeneboom:2009}, using WMAP5 year data (up to multipoles $\ell =400$), claimed a detection of 
a quadrupolar power spectrum of the form of eq.~(\ref{Panis}) at more than $3 \sigma$ ($g_*=0.15 \pm 0.039$) with preferred direction 
$(l,b)=(110^\circ, 10^\circ)$. Subsequently this result has been put under further check. \cite{Hanson:2009} confirmed 
this effect at high statistical 
significance, pointing out however that beam asymmetries could be a strong contaminant (see also~\cite{Hanson:2010}). 
The importance of this systematic effect is 
somewhat debated: \cite{Groeneboom:2010}, including polarization and beam asymmetries analysis excluded that the latter 
can be responsible for the observed effect. 
Their claim is a $9 \sigma$ detection with $g_*=0.29 \pm 0.031$. However, the preferred direction shifted much closer to 
the ecliptic poles, which is probably 
an indication that some unknown systematic is  involved and must be corrected in order to obtain true constraints on any primordial modulation. 
Foregrounds and noise are disfavoured as possible systematic effects~\citep{Bennet:2011,Groeneboom:2009}. Thus the cause of this kind of 
asymmetry is not definitely known. Planck should be able to detect a power quadrupole as 
small as $2\%$ (at $3 \sigma$)~\citep{Pullen:2007,Groeneboom:2009,Groeneboom:2010}. It is of course desirable to test this (and other anisotropic effects) with other techniques. 

What about Large-Scale Structure Surveys? Up to now there are just
{\color{red}a} few analyses testing anisotropies in large-scale structure surveys,  but all of them have been crucial, indicating that  large-scale structure surveys such as Euclid  offer a promising avenue to constrain these features.

\cite{Hirata:2009ar} used high-redshift quasars from the Sloan Digital Sky Survey to rule out the simplest version of dipole modulation of the 
primordial power spectrum. In comparison  the Planck mission  using the CMB hemispherical asymmetry  would only  marginally distinguish it from the standard case  \cite{Eriksen:2007pc}. 
The constraints obtained by high-redshift quasars require an amplitude for the dipole modulation $6$ times smaller than the one required by CMB.
This would disfavour the simple curvaton spatial 
gradient scenario \citep{Erickcek:2008} proposed to generate this dipole modulation. 
Only a curvaton scenario with a non-negligible fraction of isocurvature perturbations at late 
times could avoid this constraint from current high-redshift quasars \citep{Erickcek:2009}.

\cite{Pullen:2010} considered a sample of photometric luminous red galaxies from the SDSS survey to assess the quadrupole anisotropy 
in the primordial power spectrum
of the type described by eq.~(\ref{Panis}). The sample is divided into eight redshift slices 
(from $z=0.2$ up to $z=0.6$), 
and within each slice the galaxy angular power spectrum is analyzed
{\color{red}analysed}. They also accounted for 
possible systematic effects (such as a modulation of the signal and noise due to a slow variation 
of the photometric calibration errors across the survey) and redshift-space
distorsion {\color{red}distortion} effects.
In this case \citep{Pullen:2010}
\begin{eqnarray}
\label{Cg}
C_g({\bf n},{\bf n'}) = \langle \delta_g({\bf n}) \delta_g({\bf n'}) \rangle &=& \sum_l \frac{2l+1}{4\pi} C_{g,l}P_l({\bf n} \cdot {\bf n'})
\nonumber \\ &&
+\sum_{LM}\sum_{lml'm'}D_{g,ll'}^{LM}X_{lml'm'}^{LM}R_{lm}({\bf n})R_{l'm'}({ \bf n'})\, .
\end{eqnarray}
Here, the set of $C_{g,l}$s are given by the usual galaxy angular power spectrum for the case of statistical isotropy.  
Statistical anisotropy produces the second term 
\begin{eqnarray}
\label{Dg}
D_{g,ll'}^{LM}={\rm i}^{l-l'}\frac{2}{\pi}\int_0^\infty {\rm d}k\,k^2 P_g(k)g_{LM}W_l(k)W_{l'}(k)\, ,
\end{eqnarray}
where $X_{lml'm'}^{LM}$ are geometric coefficients related to Wigner 3-j
{\color{red}$3-j$} symbols,  $R$ denote {\color{red}denotes} the real spherical harmonics (see eqs.~(3) and (13) of \cite{Pullen:2007} for more details), $P_g(k)=b^2_g P(k)$ is the isotropic galaxy power spectrum and 
$W_l(k) = \int {\rm d}\chi f(\chi)j_l(k\chi)$ 
is the window function ($\chi$ is the comving {\color{red}comoving}
distance, and $f(\chi)$ is the selection fucntion {\color{red}function}, i.e. the normalized redshift distribution for 
a redshift slice). 
 
Assuming the same preferred direction singled out by \cite{Groeneboom:2009}, they derive a constraint on the anisotropy amplitude 
$g_*=0.006 \pm 0.036$ ($1 \sigma$), 
thus finding no evidence for anisotropy. 
Marginalizing over ${\bf n}$ with a uniform prior they find $-0.41 < g_* < 0.38$ at $95\%$ C.L. These results could confirm that the signal 
seen in CMB data is of systematic nature. However, it must be stressed that CMB and LSS analyses probe different scales, 
and in general the amplitude of the anisotropy is 
scale dependent $g=g(k)$, as in the model proposed in \cite{Erickcek:2009}. 
An estimate for what an experiment like Euclid can achieve is to consider how the uncertainty in $g_*$ 
scale in terms of number of modes measured 
and the number of redshift slices. Following the arguments of \cite{Pullen:2010}, the uncertainty will scale roughly as $\ell_{\rm max}^{-1} N_z^{-1/2}$, 
where $\ell_{\rm max}$ is the maximum multipole at which the galaxy angular
power spectrum is probed, and $N_z$ is the number of redshit
{\color{red}redshift} slices. 
Considering  that the redshift survey of Euclid will cover 
redshifts $0.4 < z <2$, there is an increase by a factor of $3$ in distance of the survey and hence a factor $3$ increase in $l_{\rm max}$ 
($l_{\rm max} \sim k_{\rm max} \chi(z)$, see the expression for the selection function after eq.~(\ref{Dg})). 
Taking $k_{\rm max}= 0.2h \rm{Mpc}^{-1}$ the effective number of redshift slices 
is also increased of a factor of $\sim 3$ ($N_z \sim k_{\rm max} \
\Delta \chi /\pi$, with $\Delta \chi$ the radial width of the survey). Therefore one could expect that for a mission like 
Euclid one can achieve an uncertainty  (at 1 $\sigma$)
$\sigma_{g_*} \sim 10^{-3}-10^{-2}$ or $\sigma_{g_*} \sim 10^{-2}$, for a fixed anisotropy axis or marginalizing over ${\bf n}$, respectively. 
This will be competitive with Planck measurements and highly complementary to it \citep{Paci:2010wp,Gruppuso:2010}.
Notice that these constraints apply to  an analysis of the galaxy angular  
power spectrum. An analysis of the 3-dimensional power spectrum $P({\bf k})$ could improve the sensitivity further. 
In this case the uncertainty would scale as $\Delta g_* \sim N^{-1/2}_{\rm modes}$, where $N_{\rm modes}$ is the number of independent Fourier modes.

%% file: methodology/methodology.tex
\chapter{Statistical methods for performance forecasts}\label{statistical}


\section{Introduction}

As cosmology becomes nowadays increasingly dominated by results emerging from
large-scale observational programmes, it is imperative to be able to
justify that resources are being deployed as effectively as possible.
In recent years it has become standard to quantify the expected
outcome of cosmological surveys to enable comparison, a procedure
exemplified by the Figure of Merit (FoM) introduced by~\cite{Huterer:2000mj} and later used in the influential dark energy
task force (DETF) report about dark energy surveys~\citep{Albrecht2006,Albrecht:2009ct}. 

The idea is to be able to capture in one single number the scientific return of a future mission, in order to be able to rank competing proposals and to forecast their ability to answer relevant scientific questions, such as: is dark energy a cosmological constant or does it evolve with time? Is it an expression of modified gravity? How well can a time-evolution of dark energy be constrained? 

Encapsulating the entire value of a proposed cosmological survey in one single number is of course highly reductive, and the ensuing conclusions should therefore be taken with a large grain of salt. Having said that, work in recent years has focused on attempts to devise Figures of Merit (FoMs) that represent in an increasingly realistic way future missions. It is perhaps obvious that, to a certain extent, the assessment of a future probe will depend on the scientific question one is most interested in: parameter constraints, model selection, robustness to systematics are but a few examples of the different levels on which a proposed mission can be evaluated and optimized.

This chapter gives an overview of some of the approaches recently adopted in the field, and used elsewhere in this document to produce forecasts for Euclid. Useful references and background material to some of the concepts discussed below are: \cite{Trotta:2008qt,BMIC} for an introduction to Bayesian methods in cosmology,  \cite{Sivia:1996, MacKay:2003} for introductions to the Bayesian approach in data analysis, \cite{MCMC:1996} for an introduction to Markov Chain Monte Carlo (MCMC) methods.


\section{Predicting the science return of a future experiment}

\begin{figure}
\centering
 \includegraphics[angle=0,width=0.8\columnwidth]{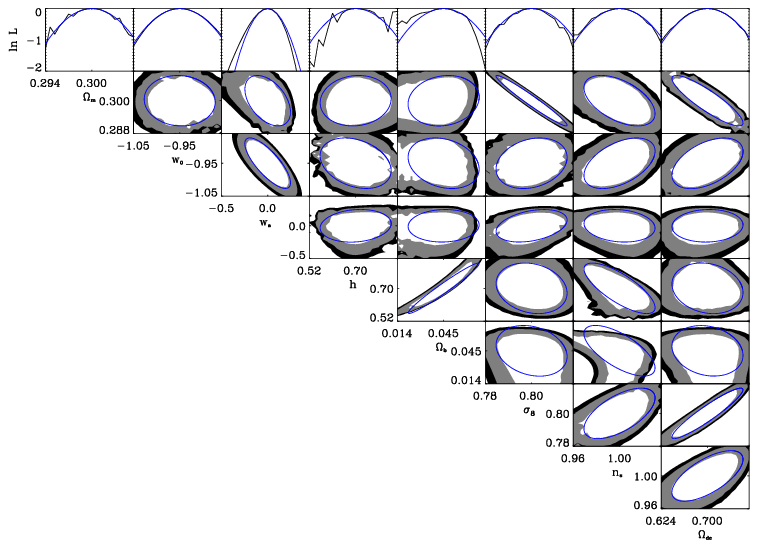}
 \caption{Projected 
   cosmological $8$-parameter space for a
   20,000 square degrees, median redshift of $z=0.8$,
   10 bin tomographic cosmic shear survey. Specifications are based on Euclid Yellow book \cite{euclidyellowbook} as this figure is representative of a method, rather than on forecast analysis; the discussion is still valid with more updated \cite{euclidredbook} Euclid specifications.   The upper panel shows the 1D parameter
   constraints using analytic marginalization (black) and the Gaussian approximation (Fisher
   matrix, blue, dark gray {\color{red}grey}). The other panels show the
   2D parameter constraints. Grey contours are 1- 2-
   and 3-$\sigma$ levels using analytic marginalization over the extra
   parameters,
   solid blue ellipses are the 1-$\sigma$ contours using the
   Fisher-matrix approximation to the projected
   likelihood surface, solid red ellipses are the 1-$\sigma$ fully
   marginalized. From~\cite{2010MNRAS.408..865T}.}
 \label{fig:Gauss_1}
\end{figure}

\begin{figure}
\centering
\includegraphics[width=0.8\linewidth]{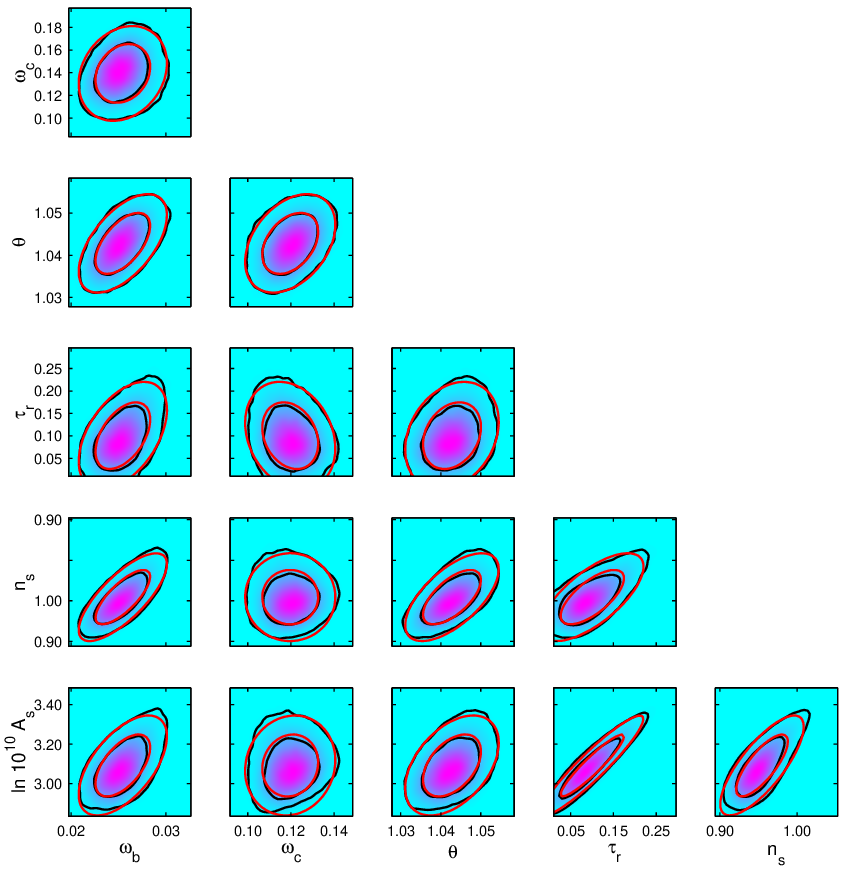}
\caption{Gaussian approximation (Laplace approximation) to a 6-dimensional posterior distribution for cosmological parameters,
from WMAP1 and SDSS data.
For all couples of parameters, panels show contours enclosing 68\%
and 95\% of joint probability from $2\cdot10^5$ MC samples (black
contours), along with the Laplace approximation (red ellipses). It
is clear that the Laplace approximation captures the bulk of the
posterior volume in parameter space in this case where there is
little non-Gaussianity in the posterior PDF. From~\cite{Trotta:2005ar}.} \label{fig:Gauss_2}
\end{figure}

We consider a toy Gaussian linear model in order to illustrate the different approaches to performance forecast. We notice that, although motivated by computational simplicity and the ability to obtain analytical results, a Gaussian model is actually a fairly close representation of many cases of interest. In Fig.~\ref{fig:Gauss_1} we illustrate this point by plotting the parameter constraints expected from a Euclid-like survey and the corresponding Gaussian approximation in the Fisher matrix approach to the likelihood (described below). In these cases, it seem clear that the Gaussian model captures fairly well the full probability distribution. Another example shown in Fig.~\ref{fig:Gauss_2} are cosmological constraints from  WMAP and SDSS data, where a Gaussian approximation to the likelihood (so-called Laplace approximation) is seen to give an excellent description of the full distribution obtained numerically via MCMC. 

\subsection{The Gaussian linear model}

Suppose we have $N$ cosmological probes, whose likelihood function is assumed to be a multi-dimensional Gaussian, given by:
$L_{i}$ ($i=1,\dots, N$), i.e. 
\begin{equation} \label{eq:likelihood_i}
\mathcal{L}_i (\Theta) \equiv p(D_i |\Theta)= \mathcal{L}_{0}^{i}\exp\left(-\frac{1}{2}(\mu_{i}-\Theta)^{t}L_{i}(\mu_{i}-\Theta)\right).
\end{equation}
where $\Theta$ are the parameters one is interested in constraining, $D_i$ are the available data from probe $i$ and $\mu_i$ is the location of the maximum likelihood value in parameter space. The matrix $L_i$ is the inverse of the covariance matrix of the parameters.


The posterior distribution for the parameters from each probe, $p(\Theta|D_i)$, is obtained by Bayes' theorem as 
\begin{equation} \label{eq:bayes}
p(\Theta|D_i)=\frac{p(\Theta)p(D_i|\Theta)}{p(D_i)},
\end{equation}
where and $p(\Theta)$ is the prior and $p(D_i)$ is a normalizing constant (the Bayesian evidence). 
If we assume a Gaussian prior centered {\color{red}centred} on the origin with inverse covariance
matrix $\Sigma$, the posterior from each probe is also a Gaussian, with
inverse covariance matrix \label{symbol:invcov}
\begin{equation}
F_i = L_i + \Sigma \quad (i=1,\dots, N)
\end{equation}
and posterior mean 
\begin{equation}
\overline{\mu}_{i} = F_{i}^{-1}(L_{i}\mu_{i}).
\end{equation}
Tighter constraints on the parameters can be usually obtained by combining all available probes together (provided there are no systematics, see below). If we combine all probes together, we obtain a Gaussian posterior with inverse covariance matrix
\begin{equation} \label{eq:post_Fisher}
F = \sum_{i=1}^N L_i  + \Sigma
\end{equation}
and mean  
\begin{equation}
\overline{\mu} =F^{-1}\sum_{i=1}^N L_{i}\mu_{i}.  \label{eq:post_mean}
\end{equation}
 Notice that the precision of the posterior (i.e., the inverse covariance matrix) does not
 depend on the degree of overlap of the likelihoods from the individual
 probes. This is a property of the Gaussian linear model. 
 
For future reference, it is also useful to write down the general expression
for the Bayesian evidence. For a Normal {\color{red}normal} prior
$p(\Theta)\sim{\mathcal{N}}(\theta_{\pi},\Sigma)$ and a
likelihood 
\begin{equation}
  \mathcal{L}(\Theta)=\mathcal{L}_{0}
  \exp\left(-\frac{1}{2}(\theta_{0}-\Theta)^{t}L(\theta_{0}-\Theta)\right),\end{equation}
the evidence for data $d$ is given by \begin{equation}
\begin{split}p(d) & \equiv\int{\rm {d}}\Theta p(d|\Theta)p(\Theta)=
  \mathcal{L}_{0}\frac{|\Sigma|^{1/2}}{|F|^{1/2}}\\
 & \exp\left[-\frac{1}{2}\left(\theta_{0}^{t}L\theta_{0}+
    \theta_{\pi}^{t}\Sigma\theta_{\pi}-\overline{\theta}^{t}F\overline{\theta}\right)\right],\end{split}
\label{eq:evidence}
\end{equation}
where $F$ is given by eq.~\eqref{eq:post_Fisher} with $N=1$ and
$\overline{\theta} = F^{-1}L\theta_0$.

\subsection{Fisher matrix error forecast} \label{Fisher_matrix}

A general likelihood function for a future experiment (subscript $i$) can be Taylor-expanded around its maximum--likelihood value, $\mu_i$. By definition, at the maximum the first derivatives vanish, and the shape of the log--likelihood in parameter space is approximated by the Hessian matrix
$H_i$,
 \begin{equation}
 \label{eq:like_second_order_expansion}
 \lnlike_i(\params) \approx \lnlike_i(\mu_i)
 + \frac{1}{2}(\params-\mu_i)^t H_i (\params-\mu_i),
 \end{equation}
where $H_i$ is given by
 \begin{equation}
 \left(H_i\right)_{\alpha \beta} \equiv
\frac{\partial^2 \lnlike_i}{\partial \params_\alpha \partial \params_\beta} {{\Big\arrowvert}_{\mu_i}} ,
 \end{equation}
and the derivatives are evaluated at the maximum--likelihood
point. By taking the expectation of equation
\eqref{eq:like_second_order_expansion} with respect to many data
realizations, we can replace the maximum--likelihood value $\mu_i$ with
the true value, $\fid$, as the maximum--likelihood estimate is
unbiased (in the absence of systematics), i.e. $\langle \mu_i \rangle = \fid$. We then define
the Fisher information matrix as the expectation value of the
Hessian,
 \begin{equation} \label{eq:Fisher_Matrix}
 F_i \equiv
\langle H_i \rangle. 
 \end{equation}
The inverse of the Fisher matrix, $F^{-1}$, is an estimate of the
covariance matrix for the parameters, and it describes how fast
the log--likelihood falls (on average) around the maximum
likelihood value, and we recover the Gaussian expression for the likelihood, eq.~\eqref{eq:likelihood_i}, with the maximum likelihood value replaced by the true value of the parameters and the inverse covariance matrix given by the Fisher matrix, $L_i = F_i^{-1}$~\citep{KendallStuart}.
In general, the derivatives depend on where in parameter space we
take them (except for the simple case of linear models), hence it
is clear that $F_i$ is a function of the fiducial parameters.

Once we have the Fisher matrix, we can give estimates for the
accuracy on the parameters from a future measurement, by computing the posterior as in eq.~\eqref{eq:bayes}. If we are only interested in a subset of the parameters, then
we can marginalise easily over the others: computing the Gaussian
integral over the unwanted parameters is the same as inverting the
Fisher matrix, dropping the rows and columns corresponding to
those parameters (keeping only the rows and columns containing the
parameters of interest) and inverting the smaller matrix back. The
result is the marginalised Fisher matrix $\mathcal{F}_i$. For example,
the 1 sigma error for parameter $\alpha$ from experiment $i$, marginalized over all other
parameters, is simply given by $\sigma_\alpha =
\sqrt{(F_i^{-1})_{\alpha \alpha}}$. 

It remains to compute the Fisher matrix for the future experiment.
This can be done analytically for the case where the likelihood
function is approximately Gaussian in the data, which is a good
approximation for many applications of interest. We can write for
the log--likelihood (in the following, we drop the subscript $i$ denoting the experiment under consideration for simplicity of notation)
 \begin{equation}
 -2 \lnlike = \ln |C| + (D-\mu)^t C^{-1} (D-\mu),
 \end{equation}
where $D$ are the (simulated) data that would be observed by the
experiment and in general both the mean $\mu$ and covariance
matrix $C$ may depend on the parameters $\params$ we are trying to
estimate. The expectation value of the data corresponds to the
true mean, $\langle D \rangle = \mu$, and similarly the
expectation value of the data matrix $\Delta \equiv
(D-\mu)^t(D-\mu)$ is equal to the true covariance, $\langle \Delta
\rangle = C$. Then it can be shown (see e.g.~\cite{Tegmark:1996bz})
that the Fisher matrix is given by
\begin{equation}\label{eq:Falphabeta}
    F_{\alpha \beta} = \frac{1}{2} {\rm tr}\left[ A_{\alpha} A_{\beta} +
    C^{-1} \langle \Delta_{,\alpha\beta} \rangle \right],
\end{equation}
where $A_\alpha \equiv C^{-1} C_{,\alpha}$ and the comma denotes a
derivative with respect to the parameters, for example  $C_{,\alpha}
\equiv
\partial C/\partial \params_\alpha$. The fact that this
expression depends only on {\em expectation values} and not on the
particular data realization means that the Fisher matrix can be
computed from knowledge of the noise properties of the experiment
without having to go through the step of actually generating any
simulated data. The specific form of the Fisher matrix then
becomes a function of the type of observable being considered and
of the experimental parameters.

Explicit expressions for the Fisher matrix for cosmological
observables can be found in~\cite{Tegmark:1996bz} for cosmic
microwave background data, in~\cite{Tegmark:1997rp} for the matter power
spectrum from galaxy redshift surveys (applied to baryonic
acoustic oscillations in~\cite{Seo:2003pu} and in~\cite{Hu:2003pt} for weak lensing. 
These approaches have been discussed in Sec.~\ref{observational-properties-of-modified-gravity}.
A useful summary of Fisher matrix
technology is given in the Dark Energy Task Force report~\citep{Albrecht2006} and 
in \cite{Verde:2009tu}. A useful numerical package which includes several of the above 
calculations is the publicly available Matlab
 code\footnote{Available from \url{http://www.cosmology.org.za}}
 \texttt{Fisher4Cast}~\citep{Bassett:2009uv,Bassett:2009tw}. 
Attempts to include systematic errors modelling in this framework can 
be found in \cite{Kitching/Taylor:2010,Taylor/Kitching:2010, Kitching/etal:2009}.

\subsection{Figure of Merits}

It has become customary to describe the statistical power of a future dark
energy probe by the inverse area enclosed by the 68\% covariance ellipse marginalized down to the dark energy parameter space. This measure of
statistical performance for probe $i$ (widely known as the DETF
FoM~\cite{Albrecht2006,Huterer:2000mj}) is usually defined (up to multiplicative
constants) as 
\begin{equation}
\text{FoM} = |F_i|^{-1/2}.  
\end{equation}
where the Fisher matrix $F_i$ is given in eq.~\eqref{eq:Fisher_Matrix}. 
\cite{Albrecht2006} suggested to use the inverse area of the $95\%$ error ellipse of $w_0-w_a$ (where $w_0$ and $w_a$ are defined in \cite{Linder:2002et}, \cite{chevallier01}). This definition was inspired by \cite{Huterer:2000mj}. In \cite{Albrecht:2009ct} it is suggested to model $w(a)$ as piecewise constant values of $w(a)$ defined in many small redshift bins ($\Delta a = 0.025$). The suggestion is then to apply a principal component approach (\cite{Huterer:2002hy}) in order to understand the redshifts at which each experiment has the power to constrain $w$.

A closely related but more statistically motivated measure of the information gain is the Kullback-Leibler divergence (KL) between the
posterior and the prior, representing the information gain
obtained when upgrading the prior to the posterior via Bayes' theorem:
\begin{equation} 
D_{KL} \equiv\int
  p(\Theta|D)\ln\frac{p(\Theta|D)}{p(\Theta)}d\Theta.
\label{eq:def_KL}
\end{equation}
The KL divergence measures the relative entropy between the two distributions:
it is a dimensionless quantity which expressed the information gain obtained via
the likelihood. 
For the Gaussian likelihood and prior introduced above, the information gain
(w.r.t. the prior $\Sigma$) from the combination of all probes is given
by~\cite{BMICRTetal}
\begin{equation} \label{eq:DKL}
  D_{KL}=\frac{1}{2}\left(\ln|F|-\ln|\Sigma|-{\rm {tr}[1-\Sigma
      F^{-1}]}\right).
\end{equation} 

A discussion of other, alternative FoMs (D-optimality, A-optimality) can be
found in~\cite{Bassett05}. In \cite{Wang:2008zh} a different FoM for Dark Energy
is suggested. For a set of DE parameters $\params$, the FoM is defined as $FoM =
1/\sqrt{\text{Cov}(\params)}$, where $Cov(\params)$ is the covariance matrix of
$\params$. This definition is more flexible since one can use it for any DE
parametrisation~\citep{Wang:2010gq}. 

Given that Euclid can constrain both the expansion history and the growth of
structure, it is also useful to introduce a new FoM for the growth of
perturbations. Similarly to the DETF FoM, one can define this new FoM as the
inverse area of the $95\%$ error ellipse of $\Omega_m-\gamma$, where $\gamma$ is
the growth index, defined starting from the growth rate $f_G(z) \equiv \frac{d\ln
G(z)}{d\ln a} = \Omega_m^\gamma$, or as $1/\sqrt{Cov(w_0,w_a,\gamma)}$ or similar variants
\citep{Maj-in_preparation,diporto10}. Instead of $\gamma$, other
parameters describing the growth can also be employed.

A FoM targeted at evaluating the robustness of a future probe to potential systematic 
errors has been introduced in \cite{MarchRobustness}. The robustness of a
future probe is defined via the degree of overlap between the posterior
distribution from that probe and the posterior from other, existing probes.
The fundamental notion is that maximising statitical {\color{red}statistical} power (e.g. by designing a future probe to deliver orthogonal constraints w.r.t.~current probes) will in general reduce its robustness (by increasing the probability of an incompatible results, for example because of systematic bias). Thus in evaluating the strength of a probe, both its statistical power and its resilience to plausible systematics ought to be considered.

\subsection{The Bayesian approach} 

When considering the capabilities of future experiments, it
is common stance to predict their performance in terms of
constraints on relevant parameters, assuming a fiducial point in
parameter space as the true model (often, the current best--fit
model), as explained above. While this is a useful indicator for parameter inference
tasks, many 
questions in cosmology fall rather in the model
comparison category. Dark energy is a case in point, where
the science driver for many future probes (including Euclid)
is to detect possible departures from a cosmological constant,
hence to gather evidence in favour of an evolving dark energy
model. It is therefore preferable to assess the capabilities of
future experiments by their ability to answer model selection
questions.

The procedure is as follows (see~\cite{Mukherjee:2005tr} for
details and the application to dark energy scenarios). At every
point in parameter space, mock data from the future observation
are generated and the Bayes factor between the competing models is
computed, for example between an evolving dark energy and a
cosmological constant. Then one delimits in parameter space the
region where the future data would {\em not} be able to deliver a
clear model comparison verdict, for example $\vert \ln B_{01} \vert <
5$ (evidence falling short of the ``strong'' threshold). Here, $B_{01}$ is the Bayes factor, which
is formed from the ratio of the Bayesian evidences of the two models
being considered:
\begin{equation}
B_{01} = \frac{p(\data | \mdl_0)}{p(\data | \mdl_1)},
\label{eq:bayesfactor}
\end{equation}
where the Bayesian evidence is the average of the likelihood under the prior in each model (denoted by a subscript $m$):
\begin{equation}
p(\data | \mdl_m) = \int \dr\params_m p(\data | \params_m, \mdl_m)p(\params_m | \mdl_m).
\end{equation}
 The Bayes factor updates the prior probability ratio
of the models to the posterior one, indicating the extent to which the
data have modified one's original view on the relative probabilities
of the two models. The
experiment with the smallest ``model--confusion'' volume in
parameter space is to be preferred, since it achieves the highest
discriminative power between models. An application of a related
technique to the spectral index from the Planck satellite is
presented in~\cite{Pahud:2007gi,Pahud:2006kv}.

Alternatively, we can investigate the full probability
distribution for the Bayes factor from a future observation. This
allows to make probabilistic statements regarding the outcome of a
future model comparison, and in particular to quantify the
probability that a new observation will be able to achieve a
certain level of evidence for one of the models, given current
knowledge. This technique is based on the {\em predictive
distribution} for a future observation, which gives the expected
posterior for an observation with a certain set of experimental
capabilities (further details are given in~\cite{Trotta:2007hy}). This method
is called PPOD, for {\em predictive posterior odds distribution}
and can be useful in the context of experiment design and
optimization

Hybrid approaches have also been attempted, i.e., to defined model-selection
oriented FoMs while working in the Fisher Matrix
framework, such as the expected Bayesian evidence ratio~\cite{Heavens/etal:2007, Amara/Kitching:2010}.

The most general approach to performance forecasting involves the use of a
suitably defined utility function, and it has recently been presented in
\cite{Trotta:2010ug}. Consider the different levels of uncertainty that are
relevant when
predicting the probability of a certain model selection outcome from a
future probe, which can be summarized as follows:
\begin{itemize}
\item {\bf Level 1:} current uncertainty about the correct model
  (e.g., is it a cosmological constant or a dark energy model?). 
\item {\bf Level 2:} present-day uncertainty in the value of the
  cosmological parameters for a given model (e.g., present error on
  the dark energy equation of state parameters assuming an evolving
  dark energy model). 
\item {\bf Level 3:} realization noise, which will be present in
  future data even when assuming a model and a fiducial choice
  for its parameters.    
\end{itemize}
The commonly-used Fisher matrix forecast ignores the uncertainty arising from Levels
1 and 2, as it assumes a fiducial model (Level 1) and fiducial
parameter values (Level 2). It averages over realization noise (Level
3) in the limit of an infinite number of realizations. Clearly, the Fisher
matrix procedure provides a very limited assessment of what we can expect for
the scientific return of a future probe, as it ignores the uncertainty
associated with the choice of model and parameter values. 

The Bayesian framework allows improvement on the usual Fisher matrix error
forecast thanks to a general 
procedure which fully accounts for all three levels of uncertainty given above. 
Following~\cite{Loredo:2003nm}, we think of
future data $\Df$ as {\em outcomes}, which arise as consequence of our
choice of experimental parameters $e$ ({\em actions}). For each action
and each outcome, we define a utility function $\Uf(\Df, e)$. Formally, the
utility only depends on the future data realization $\Df$. However, as will
become clear below, the data $\Df$ are realized from a fiducial model and model
parameter values. Therefore, the utility function implicitly depends on the
assumed model and parameters from which the data $\Df$ are generated. The best
action is the one that maximizes the expected utility, i.e. the
utility averaged over possible outcomes:
\begin{equation} \label{def:EU}
\EU (e) \equiv \int \dr \Df p(\Df | e, \data) \Uf(\Df, e).
\end{equation}
Here, $p(\Df | e, \data) $ is the predictive distribution
for the future data, conditional on the experimental setup ($e$) and
on current data ($\data$). For a single fixed model the
predictive distribution is given by 
\be
\begin{aligned} \label{eq:predictive_general}
p(\Df | e, \data) &  = \int \dr\params \, p(\Df, \params | e, \data) \\
 			 & =  \int \dr\params \, p(\Df | \params ,  e,
 			 \data) p(\params|e,\data) \\ 
			 & = \int \dr\params \, p(\Df | \params ,  e)
 			 p(\params|\data), 
\end{aligned}
\ee
where the last line follows because $p(\Df | \params ,  e,  \data) =
p(\Df | \params ,  e)$ (conditioning on current data is irrelevant
once the parameters are given) and  $p(\params|e,\data)  = p(\params|
\data)$ (conditioning on future experimental parameters is irrelevant
for the present-day posterior). So we 
can predict the probability distribution for future data $\Df$ 
by averaging the likelihood function for the future measurement
(Level 3 uncertainty) over the current posterior on the parameters
(Level 2 uncertainty). The expected utility then becomes
\begin{equation} \label{eq:exp_utility_1}
\EU (e) = \int \dr \params p(\params| o, \data) \int \dr \Df p(\Df |
\params, e) \Uf(\Df, e). 
\end{equation}

So far, we have tacitly assumed that only one model was being considered for the
data. In practice, there will be several models that one is interested in
testing (Level 1 uncertainty), and typically there is uncertainty over which one
is best. This is in fact one of the main motivations for designing a new dark
energy probe. If $M$ models $\{ \mdl_1, \dots, \mdl_M \}$ are being considered,
each one with parameter vector $\params_m$ ($m=1,\dots, M$), the current
posterior can be further extended in terms of model
averaging (Level 1), weighting each model by its current model
posterior probability, $p(\mdl_m | \data)$, obtaining from
eq.~\eqref{eq:exp_utility_1} the model-averaged expected utility
\begin{equation} \label{eq:exp_utility_2}
\begin{aligned}
\EU (e) & = \sum_{m=1}^M p(\mdl_m| \data) \int \dr \params_m p(\params_m|
\data ,\mdl_m) \\ & \times \int \dr \Df p(\Df | \params_m,
e,\mdl_m) \Uf(\Df, e,\mdl_m).
\end{aligned}
\end{equation}
This expected utility is the most general definition of a FoM for a future
experiment characterized by experimental parameters $e$. The usual Fisher matrix
forecast is recovered as a special case of eq.~\eqref{eq:exp_utility_2}, as are
other {\em ad hoc} FoMs that have been defined in the literature. Therefore
eq.~\eqref{eq:exp_utility_2} gives us a formalism to define in all generality
the scientific return of a future experiment. This result clearly accounts for
all three levels of uncertainty in making our predictions: the utility function
$\Uf(\Df, e,\mdl_m)$ (to be specified below) depends on the future data
realization, $\Df$, (Level 3), which in turn is a function of the fiducial
parameters value, $\params_m$, (Level 2), and is averaged over present-day model
probabilities (Level 1).

This approach is used in \cite{Trotta:2010ug} to define two model-selection
oriented Figures of Merit: the decisiveness $\mathcal D$, which quantifies the
probability that a probe will deliver a decisive result in favour or against the
cosmological constant, and the expected strength of evidence, $\mathcal E$, that
returns a measure of the expected power of a probe for model selection.

\section{Survey design and optimization}

Although the topic of survery {\color{red}survey} design is still in its infancy, the basic idea is to carry
out an optimization of survey parameters (such as for example choise
{\color{red}choice} of targets,
depth of field, number of spectroscopic fibers, etc.) in order to identify the
configuration that is more likely to return a high FoM for the scientific
question being considered. Example of this approach applied to dark energy parameters can be found
in~\cite{Bassett05,Parkinson:2007cv,Parkinson:2009zi,Bassett:2004st,
Bassett:2004np}, while \cite{Loredo:2003nm} discussed a more general methodology.  In \cite{Bassett:2004st} a method is defined to
optimise future surveys, in the framework of Bayesian statistics and without
necessarily assuming a dark energy model. In \cite{Bassett:2004np},
\cite{Parkinson:2007cv} and \cite{Parkinson:2009zi} this method is used to
produce forecasts for future weak lensing and galaxy redshift surveys.

The optimization process is carried out subject to constraints, such as for
example design parameter ranges and/or cost constraints. This is generally a
numerically complex and computationally expensive procedure. It typically
requires to explore the design parameters space (e.g. via MCMC), generating at
each point a set of pseudo-data that are analyzed {\color{red}analysed} as real data would, in order
to compute their FoM. Then the search algorithm moves on to maximise the FoM. 

In order to carry out the optimization procedure, it might be useful to
adopt a principal component analysis (PCA) to determine a suitable
parameterization {\color{red}parametrization } of $w(z)$~\citep{Huterer:2002hy,Simpson:2006bd}. The redshift range of the survey can be split into $N$ bins, with the equation of state taking on a value $w_i$ in the $i$-th bin:
\begin{equation}
w(z) = \sum_{i = 1}^N w_i b_i(z) \,.
\end{equation}
where the basis functions $b_i$ are top-hats of value 1 inside the bin, and 0 elsewhere. If $F$ is the Fisher matrix for the $N$ parameters $w_i$, one can diagonalize it by writing $F = W^T \Lambda W$, where $\Lambda$ is a diagonal matrix, and the rows of $W$ are the eigenvectors $e_i(z)$ or the so-called principal components. These define a new basis (in which the new coefficients $\alpha_i$ are uncorrelated) so the equation of state can be written as
\begin{equation}
w(z) = \sum_{i = 1}^N \alpha_i e_i(z) \,.
\end{equation}
The diagonal elements of $\Lambda$ are the eigenvalues $\lambda_i$ and define the variance of the new parameters, $\sigma^2(\alpha_i)  = 1/\lambda_i$.

One can now reconstruct $w(z)$ by keeping only a certain number of the most accurately determined modes, i.e., the ones with largest eigenvalues. The optimal number of modes to retain can be estimated by minimizing the risk, defined as the sum of the bias squared (how much the reconstructed equation of state departs from the true one by neglecting the more noisy modes) plus the variance of the estimate~\citep{Huterer:2002hy}.

\section{Future activities and open challenges}

As outlined in the previous sections, several approaches are available to capture the expected scientific performance of Euclid. As part of future theoretical activities, it will be necessary to build on the above concepts in 
order to obtain a realistic assessment of the science return of Euclid. Operationally, this means that the following tasks will need to be carried out:
\begin{itemize}
\item Estimation of likelihood contours around the maximum likelihood peak beyond the Fisher matrix approach. We envisage here a programme where  
simulated mock data will be generated and then used to blindly reconstruct the likelihood surface to sufficient accuracy. 

\item Estimation of Bayesian posterior distributions and assessment of impact of various priors. Bayesian inference is a mature field in cosmology and we now have at our disposal a number of efficient and reliable numerical algorithms based on Markov Chain Monte Carlo or nested sampling methods.

\item Comparison of Bayesian inferences with inferences based on profile likelihoods. Discrepancies might occur in the presence of large ``volume effects'' arising from insufficiently constraining data sets and highly multi-modal likelihoods~\citep{Trotta:2008bp}. Based on our experience so far, this is unlikely to be a problem for most of the statistical quantities of interest here but we recommend to check this explicitly for the more complicated distributions.
  
\item Investigation of the coverage properties of Bayesian credible and frequentist confidence intervals. Coverage of intervals is a fundamental property in particle physics, but rarely discussed in the cosmological setting. We recommend a careful investigation of coverage from realistically simulated data sets (as done recently in \cite{March:2011xa}). Fast neural networks techniques might be required to speed up the inference step by several orders of magnitude in order to make this kind of studies computationally feasible~\citep{shaw07,Bridges:2010de}.

\item Computation of the Bayesian evidence to carry out Bayesian model
selection~\citep{Trotta:2008qt,Mukherjee:2005tr}. Algorithms based on nested
sampling, and in particular, MultiNest~\citep{feroz08}, seem to be ideally
suited to this task, but other approaches are available, as well, such as
population Monte Carlo~\citep{2011arXiv1101.0950K} and semi-analytical
ones~\citep{Trotta:2005ar,Heavens/etal:2007}. A robust Bayesian model
selection will require a careful assessment of the impact of priors.
Furthermore, the outcome of Bayesian model selection is dependent on the
chosen parameterization {\color{red}parametrization}, if different
non-linearly related reparameterizations {\color{red}reparametrization} can
equally plausibly be chosen from physical consideration (relevant examples
include parameterizations {\color{red}parametrizations} of the isocurvature fraction~\citep{Beltran:2005xd}, the tensor-to-scalar ratio~\citep{Parkinson:2006ku} and the inflaton potential~\citep{Martin:2010hh}). It will be important to cross check results with frequentist hypothesis testing, as well. The notion of Bayesian doubt, introduced in~\cite{March:2010ex}, can also be used to extend the power of Bayesian model selection to the space of unknown models in order to test our paradigm of a $\Lambda$CDM cosmological model.

\item Bayesian model averaging~\citep{Liddle:2006kn,Parkinson:2010zr} can also be used to obtain final inferences parameters which take into account the residual model uncertainty. Due to the concentration of probability mass onto simpler models (as a consequence of Occam's razor), Bayesian model averaging can lead to tighter parameter constraints than non-averaged procedures, for example on the curvature parameter~\citep{Vardanyan:2011in}.

\end{itemize}

%% file: ack.tex
\chapter*{\centering Acknowledgments}
\addcontentsline{toc}{chapter}{Acknowledgments}
It is a pleasure to thank the Euclid theory science working group, all
science working group members, the Euclid Consortium Board and the whole
Euclid consortium for fruitful discussions. We also kindly thank Rene
Laureijs, Ana Heras, Philippe Gondoin, Ludovic Duvet and Marc Sauvage for
their {\color{red} continuous} work and support, and Thomas Buchert for comments on the draft.

%% file: journals.tex
\def\ref@jnl#1{{#1}}

\def\aj{\ref@jnl{AJ}}                   
\def\actaa{\ref@jnl{Acta Astron.}}      
\def\araa{\ref@jnl{ARA\&A}}             
\def\apj{\ref@jnl{ApJ}}                 
\def\apjl{\ref@jnl{ApJ}}                
\def\apjs{\ref@jnl{ApJS}}               
\def\ao{\ref@jnl{Appl.~Opt.}}           
\def\apss{\ref@jnl{Ap\&SS}}             
\def\aap{\ref@jnl{A\&A}}                
\def\aapr{\ref@jnl{A\&A~Rev.}}          
\def\aaps{\ref@jnl{A\&AS}}              
\def\azh{\ref@jnl{AZh}}                 
\def\baas{\ref@jnl{BAAS}}               
\def\bac{\ref@jnl{Bull. astr. Inst. Czechosl.}}
\def\caa{\ref@jnl{Chinese Astron. Astrophys.}}
\def\cjaa{\ref@jnl{Chinese J. Astron. Astrophys.}}
\def\icarus{\ref@jnl{Icarus}}           
\def\jcap{\ref@jnl{J. Cosmology Astropart. Phys.}}
\def\jrasc{\ref@jnl{JRASC}}             
\def\memras{\ref@jnl{MmRAS}}            
\def\mnras{\ref@jnl{MNRAS}}             
\def\na{\ref@jnl{New A}}                
\def\nar{\ref@jnl{New A Rev.}}          
\def\pra{\ref@jnl{Phys.~Rev.~A}}        
\def\prb{\ref@jnl{Phys.~Rev.~B}}        
\def\prc{\ref@jnl{Phys.~Rev.~C}}        
\def\prd{\ref@jnl{Phys.~Rev.~D}}        
\def\pre{\ref@jnl{Phys.~Rev.~E}}        
\def\prl{\ref@jnl{Phys.~Rev.~Lett.}}    
\def\pasa{\ref@jnl{PASA}}               
\def\pasp{\ref@jnl{PASP}}               
\def\pasj{\ref@jnl{PASJ}}               
\def\rmxaa{\ref@jnl{Rev. Mexicana Astron. Astrofis.}}%
\def\qjras{\ref@jnl{QJRAS}}             
\def\skytel{\ref@jnl{S\&T}}             
\def\solphys{\ref@jnl{Sol.~Phys.}}      
\def\sovast{\ref@jnl{Soviet~Ast.}}      
\def\ssr{\ref@jnl{Space~Sci.~Rev.}}     
\def\zap{\ref@jnl{ZAp}}                 
\def\nat{\ref@jnl{Nature}}              
\def\iaucirc{\ref@jnl{IAU~Circ.}}       
\def\aplett{\ref@jnl{Astrophys.~Lett.}} 
\def\apspr{\ref@jnl{Astrophys.~Space~Phys.~Res.}}
\def\bain{\ref@jnl{Bull.~Astron.~Inst.~Netherlands}} 
\def\fcp{\ref@jnl{Fund.~Cosmic~Phys.}}  
\def\gca{\ref@jnl{Geochim.~Cosmochim.~Acta}}   
\def\grl{\ref@jnl{Geophys.~Res.~Lett.}} 
\def\jcp{\ref@jnl{J.~Chem.~Phys.}}      
\def\jgr{\ref@jnl{J.~Geophys.~Res.}}    
\def\jqsrt{\ref@jnl{J.~Quant.~Spec.~Radiat.~Transf.}}
\def\memsai{\ref@jnl{Mem.~Soc.~Astron.~Italiana}}
\def\nphysa{\ref@jnl{Nucl.~Phys.~A}}   
\def\physrep{\ref@jnl{Phys.~Rep.}}   
\def\physscr{\ref@jnl{Phys.~Scr}}   
\def\planss{\ref@jnl{Planet.~Space~Sci.}}   
\def\procspie{\ref@jnl{Proc.~SPIE}}   

\let\astap=\aap
\let\apjlett=\apjl
\let\apjsupp=\apjs
\let\applopt=\ao